\documentclass[11pt,letter]{article}

\usepackage[letterpaper, top=1in, bottom=1in, left=1in, right=1in, includefoot]{geometry}

\everymath{\color{MidnightBlack}}

\usepackage{titling}
\thanksmarkseries{arabic}
\newcommand*\samethanks[1][\value{footnote}]{\footnotemark[#1]}

\RequirePackage[T1]{fontenc}
\RequirePackage[utf8]{inputenc}
\usepackage[greek,russian,english]{babel}
\usepackage{alphabeta}

\usepackage[pdftex,
backref=true,
plainpages = false,
pdfpagelabels,
hyperfootnotes=true,
pdfpagemode=FullScreen,
bookmarks=true,
bookmarksopen = true,
bookmarksnumbered = true,
breaklinks = true,
hyperfigures,
pagebackref,
urlcolor = magenta,
urlcolor = MidnightBlack,
anchorcolor = green,
hyperindex = true,
colorlinks = true,
linkcolor = black!30!blue,
citecolor = black!30!green
]{hyperref}

\newcommand{\myskip}{\vspace{0mm}}
\newcommand{\mysubsection}[1]{\myskip\myskip\subsection{#1}}

\usepackage[svgnames]{xcolor}
\usepackage{bm,aliascnt,amsthm,amsmath,amsfonts,rotate,datetime,amsmath,ifthen,eurosym,wrapfig,cite,url,subcaption,cite,amsfonts,amssymb,ifthen,color,wrapfig,rotate,lmodern,aliascnt,datetime,graphicx,algorithmic,algorithm,enumerate,enumitem,todonotes,fancybox,cleveref,bm,subcaption,tabularx,colortbl,xspace,graphicx,algorithmic,algorithm}
\usepackage{stmaryrd,lmodern}

\newcommand{\bound}[1]{\textbf{#1}}
\newcommand{\tw}{{\bf tw}}
\newcommand{\tritri}{\mbox{\footnotesize $\,\blacktriangleright\,$}}
\newcommand{\torso}{{\sf torso}}
\newcommand{\p}{{\bf p}}

\newcommand{\N}{\mathbb{N}}
\newcommand{\Mod}{{\rm Mod}}
\newcommand{\MSOL}{\mbox{\sf CMSOL}\xspace}
\newcommand{\FOL}{{\sf FOL}\xspace}
\newcommand{\DP}{{\sf DP}}
\newcommand{\FOLDP}{\FOL[τ{\normalfont+}\DP]}
\newcommand{\NTMC}{{\sf EM}}
\newcommand{\bool}{{\bf PB}}
\newcommand{\hh}{\end{document}}
\renewcommand{\O}{{\cal O}}
\newcommand{\labels}[1]{\label{#1}}
\newcommand{\hw}{{\bf hw}}

\newcommand{\bd}{{\sf bd}}
\newcommand{\yes}{{\sf yes}}

\newcommand{\remove}[1]{}

\newcommand{\bigmid}{\;\big|\;}
\newcommand{\cupall}{\pmb{\pmb{\bigcup}}}

\newcommand{\pretp}{\preceq_{\sf tm}}
\newcommand{\prem}{\preceq_{\sf m}}

\newcommand{\NP}{{\sf NP}\xspace}

\newcommand{\frR}{{\mathfrak{R}}}
\newcommand{\obs}{{\sf obs}}
\newcommand{\excl}{{\sf excl}}

\RequirePackage{stmaryrd}
\usepackage{textcomp}
\DeclareUnicodeCharacter{2286}{\subseteq}
\DeclareUnicodeCharacter{2192}{\ifmmode\to\else\textrightarrow\fi}
\DeclareUnicodeCharacter{2203}{\ensuremath\exists}
\DeclareUnicodeCharacter{183}{\cdot}
\DeclareUnicodeCharacter{2200}{\forall}
\DeclareUnicodeCharacter{2264}{\leq}
\DeclareUnicodeCharacter{2265}{\geq}
\DeclareUnicodeCharacter{8614}{\mathbin{\mapsto}}
\DeclareUnicodeCharacter{8656}{\Leftarrow}
\DeclareUnicodeCharacter{8657}{\Uparrow}
\DeclareUnicodeCharacter{8658}{\Rightarrow}
\DeclareUnicodeCharacter{8659}{\Downarrow}
\DeclareUnicodeCharacter{8669}{\rightsquigarrow}
\newcommand{\eqdef}{\stackrel{{\scriptsize\rm def}}{=}}
\DeclareUnicodeCharacter{8797}{\eqdef}
\DeclareUnicodeCharacter{8870}{\vdash}
\DeclareUnicodeCharacter{8873}{\Vdash}
\DeclareUnicodeCharacter{22A7}{\models}
\DeclareUnicodeCharacter{9121}{\lceil}
\DeclareUnicodeCharacter{9123}{\lfloor}
\DeclareUnicodeCharacter{9124}{\rceil}
\DeclareUnicodeCharacter{2208}{\in}
\DeclareUnicodeCharacter{9126}{\rfloor}
\DeclareUnicodeCharacter{9655}{\triangleright}
\DeclareUnicodeCharacter{9665}{\triangleleft}
\DeclareUnicodeCharacter{9671}{\diamond}
\DeclareUnicodeCharacter{9675}{\circ}
\DeclareUnicodeCharacter{10178}{\bot}
\DeclareUnicodeCharacter{10214}{} 
\DeclareUnicodeCharacter{10215}{} 
\DeclareUnicodeCharacter{10229}{\longleftarrow}
\DeclareUnicodeCharacter{10230}{\longrightarrow}
\DeclareUnicodeCharacter{10231}{\longleftrightarrow}
\DeclareUnicodeCharacter{10232}{\Longleftarrow}
\DeclareUnicodeCharacter{10233}{\Longrightarrow}
\DeclareUnicodeCharacter{10234}{\Longleftrightarrow}
\DeclareUnicodeCharacter{10236}{\longmapsto}
\DeclareUnicodeCharacter{10238}{\Longmapsto} 
\DeclareUnicodeCharacter{10503}{\Mapsto}    
\DeclareUnicodeCharacter{10971}{\mathrel{\not\hspace{-0.2em}\cap}}
\DeclareUnicodeCharacter{65294}{\ldotp}
\DeclareUnicodeCharacter{65372}{\mid}

\usepackage{color}
\definecolor{MidnightBlack}{rgb}{0.1,0.1,.34}
\definecolor{MidnightBlue}{rgb}{0.1,0.1,0.44}
\definecolor{Black}{rgb}{0,0, 0}
\definecolor{Blue}{rgb}{0, 0 ,1}
\definecolor{Red}{rgb}{1, 0 ,0}
\definecolor{White}{rgb}{1, 1, 1}
\definecolor{Grey}{rgb}{.6, .6, .6}
\definecolor{Mygreen}{rgb}{.0, .7, .0}
\definecolor{Yellow}{rgb}{.55,.55,0}
\definecolor{Mustard}{rgb}{1.0, 0.86, 0.35}
\definecolor{applegreen}{rgb}{0.55, 0.71, 0.0}
\definecolor{darkturquoise}{rgb}{0.0, 0.81, 0.82}
\definecolor{celestialblue}{rgb}{0.29, 0.59, 0.82}
\definecolor{green_yellow}{rgb}{0.68, 1.0, 0.18}
\definecolor{crimsonglory}{rgb}{0.75, 0.0, 0.2}
\definecolor{darkmagenta}{rgb}{0.30, 0.0, 0.30}
\definecolor{internationalorange}{rgb}{1.0, 0.31, 0.0}
\definecolor{darkorange}{rgb}{1.0, 0.55, 0.0}
\definecolor{ao}{rgb}{0.0, 0.5, 0.0}
\definecolor{awesome}{rgb}{1.0, 0.13, 0.32}

\newcommand{\midnightblue}[1]{{\color{MidnightBlue}#1}}

\newcommand{\black}[1]{{\color{Black}#1}}

\newcommand{\darkmagenta}[1]{{\color{darkmagenta}#1}}

\newcommand{\blue}[1]{{\color{Blue}#1}}
\newcommand{\red}[1]{{\color{Red}#1}}
\newcommand{\green}[1]{{\color{Mygreen}#1}}

\newcommand{\darkcyan}[1]{{\color{darkcyan}#1}}
\newcommand{\darkgreen}[1]{{\color{darkgreen}#1}}
\newcommand{\darkorange}[1]{{\color{darkorange}#1}}

\usepackage{tikz}

\usetikzlibrary{arrows.meta}

\makeatletter

\pgfdeclarearrow{
  name = ipe _linear,
  defaults = {
    length = +1bp,
    width  = +.666bp,
    line width = +0pt 1,
  },
  setup code = {
    \pgfarrowssetbackend{0pt}
    \pgfarrowssetvisualbackend{
      \pgfarrowlength\advance\pgf@x by-.5\pgfarrowlinewidth}
    \pgfarrowssetlineend{\pgfarrowlength}
    \ifpgfarrowreversed
      \pgfarrowssetlineend{\pgfarrowlength\advance\pgf@x by-.5\pgfarrowlinewidth}
    \fi
    \pgfarrowssettipend{\pgfarrowlength}
    \pgfarrowshullpoint{\pgfarrowlength}{0pt}
    \pgfarrowsupperhullpoint{0pt}{.5\pgfarrowwidth}
    \pgfarrowssavethe\pgfarrowlinewidth
    \pgfarrowssavethe\pgfarrowlength
    \pgfarrowssavethe\pgfarrowwidth
  },
  drawing code = {
    \pgfsetdash{}{+0pt}
    \ifdim\pgfarrowlinewidth=\pgflinewidth\else\pgfsetlinewidth{+\pgfarrowlinewidth}\fi
    \pgfpathmoveto{\pgfqpoint{0pt}{.5\pgfarrowwidth}}
    \pgfpathlineto{\pgfqpoint{\pgfarrowlength}{0pt}}
    \pgfpathlineto{\pgfqpoint{0pt}{-.5\pgfarrowwidth}}
    \pgfusepathqstroke
  },
  parameters = {
    \the\pgfarrowlinewidth,%
    \the\pgfarrowlength,%
    \the\pgfarrowwidth,%
  },
}

\pgfdeclarearrow{
  name = ipe _pointed,
  defaults = {
    length = +1bp,
    width  = +.666bp,
    inset  = +.2bp,
    line width = +0pt 1,
  },
  setup code = {
    \pgfarrowssetbackend{0pt}
    \pgfarrowssetvisualbackend{\pgfarrowinset}
    \pgfarrowssetlineend{\pgfarrowinset}
    \ifpgfarrowreversed
      \pgfarrowssetlineend{\pgfarrowlength}
    \fi
    \pgfarrowssettipend{\pgfarrowlength}
    \pgfarrowshullpoint{\pgfarrowlength}{0pt}
    \pgfarrowsupperhullpoint{0pt}{.5\pgfarrowwidth}
    \pgfarrowshullpoint{\pgfarrowinset}{0pt}
    \pgfarrowssavethe\pgfarrowinset
    \pgfarrowssavethe\pgfarrowlinewidth
    \pgfarrowssavethe\pgfarrowlength
    \pgfarrowssavethe\pgfarrowwidth
  },
  drawing code = {
    \pgfsetdash{}{+0pt}
    \ifdim\pgfarrowlinewidth=\pgflinewidth\else\pgfsetlinewidth{+\pgfarrowlinewidth}\fi
    \pgfpathmoveto{\pgfqpoint{\pgfarrowlength}{0pt}}
    \pgfpathlineto{\pgfqpoint{0pt}{.5\pgfarrowwidth}}
    \pgfpathlineto{\pgfqpoint{\pgfarrowinset}{0pt}}
    \pgfpathlineto{\pgfqpoint{0pt}{-.5\pgfarrowwidth}}
    \pgfpathclose
    \ifpgfarrowopen
      \pgfusepathqstroke
    \else
      \ifdim\pgfarrowlinewidth>0pt\pgfusepathqfillstroke\else\pgfusepathqfill\fi
    \fi
  },
  parameters = {
    \the\pgfarrowlinewidth,%
    \the\pgfarrowlength,%
    \the\pgfarrowwidth,%
    \the\pgfarrowinset,%
    \ifpgfarrowopen o\fi%
  },
}

\newdimen\ipeminipagewidth

\tikzstyle{ipe import} = [
  x=1bp, y=1bp,
  ipe node stretch/.store in=\ipenodestretch,
  ipe stretch normal/.style={ipe node stretch=1},
  ipe stretch normal,
  ipe node/.style={
    anchor=base west, inner sep=0, outer sep=0, scale=\ipenodestretch
  },
  ipe mark scale/.store in=\ipemarkscale,
  ipe mark tiny/.style={ipe mark scale=1.1},
  ipe mark small/.style={ipe mark scale=2},
  ipe mark normal/.style={ipe mark scale=3},
  ipe mark large/.style={ipe mark scale=5},
  ipe mark normal, 
  ipe circle/.pic={
    \draw[line width=0.2*\ipemarkscale]
      (0,0) circle[radius=0.5*\ipemarkscale];
    \coordinate () at (0,0);
  },
  ipe disk/.pic={
    \fill (0,0) circle[radius=0.6*\ipemarkscale];
    \coordinate () at (0,0);
  },
  ipe fdisk/.pic={
    \filldraw[line width=0.2*\ipemarkscale]
      (0,0) circle[radius=0.5*\ipemarkscale];
    \coordinate () at (0,0);
  },
  ipe box/.pic={
    \draw[line width=0.2*\ipemarkscale, line join=miter]
      (-.5*\ipemarkscale,-.5*\ipemarkscale) rectangle
      ( .5*\ipemarkscale, .5*\ipemarkscale);
    \coordinate () at (0,0);
  },
  ipe square/.pic={
    \fill
      (-.6*\ipemarkscale,-.6*\ipemarkscale) rectangle
      ( .6*\ipemarkscale, .6*\ipemarkscale);
    \coordinate () at (0,0);
  },
  ipe fsquare/.pic={
    \filldraw[line width=0.2*\ipemarkscale, line join=miter]
      (-.5*\ipemarkscale,-.5*\ipemarkscale) rectangle
      ( .5*\ipemarkscale, .5*\ipemarkscale);
    \coordinate () at (0,0);
  },
  ipe cross/.pic={
    \draw[line width=0.2*\ipemarkscale, line cap=butt]
      (-.5*\ipemarkscale,-.5*\ipemarkscale) --
      ( .5*\ipemarkscale, .5*\ipemarkscale)
      (-.5*\ipemarkscale, .5*\ipemarkscale) --
      ( .5*\ipemarkscale,-.5*\ipemarkscale);
    \coordinate () at (0,0);
  },
  /pgf/arrow keys/.cd,
  ipe arrow normal/.style={scale=1},
  ipe arrow tiny/.style={scale=.4},
  ipe arrow small/.style={scale=.7},
  ipe arrow large/.style={scale=1.4},
  ipe arrow normal,
  /tikz/.cd,
  ipe arrows/.style={
    ipe normal/.tip={
      ipe _pointed[length=1bp, width=.666bp, inset=0bp,
                   quick, ipe arrow normal]},
    ipe pointed/.tip={
      ipe _pointed[length=1bp, width=.666bp, inset=0.2bp,
                   quick, ipe arrow normal]},
    ipe linear/.tip={
      ipe _linear[length = 1bp, width=.666bp,
                  ipe arrow normal, quick]},
    ipe fnormal/.tip={ipe normal[fill=white]},
    ipe fpointed/.tip={ipe pointed[fill=white]},
    ipe double/.tip={ipe normal[] ipe normal},
    ipe fdouble/.tip={ipe fnormal[] ipe fnormal},
    ipe arc/.tip={ipe normal},
    ipe farc/.tip={ipe fnormal},
    ipe ptarc/.tip={ipe pointed},
    ipe fptarc/.tip={ipe fpointed},
  },
  ipe arrows, 
]

\tikzset{
  rgb color/.code args={#1=#2}{%
    \definecolor{tempcolor-#1}{rgb}{#2}%
    \tikzset{#1=tempcolor-#1}%
  },
}

\makeatother

\usetikzlibrary{patterns}
\usetikzlibrary{calc,3d}
\usetikzlibrary{intersections,decorations.pathmorphing,shapes,decorations.pathreplacing,fit}
\usetikzlibrary{shapes.geometric,hobby,calc}
\usetikzlibrary{decorations}
\usetikzlibrary{decorations.pathmorphing}
\usetikzlibrary{decorations.text}
\usetikzlibrary{shapes.misc}
\usetikzlibrary{decorations,shapes,snakes}

\newcounter{func}
\newcommand{\newfun}[1]{f_{\refstepcounter{func}\label{#1}\thefunc}}
\newcommand{\funref}[1]{\hyperref[#1]{f_{\ref*{#1}}}} 
\newcounter{con}

\newcommand{\conref}[1]{\hyperref[#1]{c_{\ref*{#1}}}} 

\newcommand{\mynewtheorem}[2]{
	\newaliascnt{#1}{dummy}
	\newtheorem{#1}[#1]{#2}
	\aliascntresetthe{#1}
}

\theoremstyle{plain}
\mynewtheorem{theorem}{Theorem}
\mynewtheorem{proposition}{Proposition}
\mynewtheorem{corollary}{Corollary}
\mynewtheorem{lemma}{Lemma}
\theoremstyle{definition}
\mynewtheorem{definition}{Definition}
\theoremstyle{remark}
\mynewtheorem{sublemma}{Sublemma}
\mynewtheorem{claim}{Claim}
\mynewtheorem{remark}{Remark}
\mynewtheorem{fact}{Fact}
\mynewtheorem{conjecture}{Conjecture}
\mynewtheorem{question}{Question}
\mynewtheorem{observation}{Observation}
\mynewtheorem{problem}{Problem}
\mynewtheorem{note}{Note}

\addto\extrasenglish{

}

\makeatletter
\newtheoremstyle{caja1}
  {\topsep}
  {\topsep}
  {\itshape}
  {}
  {}
  {}
  {.5em}
  {\blue{\fbox{\black{\thmname{#1}~\thmnumber{#2}\@ifempty{#3}{.}{}\thmnote{ (#3).}}}}}
\makeatother
\theoremstyle{caja1}
\newtheorem{theoone}{Claim}

\makeatletter
\newtheoremstyle{caja2}
  {\topsep}
  {\topsep}
  {\itshape}
  {}
  {}
  {}
  {.5em}
  {\green{\fbox{\black{\thmname{#1}~\thmnumber{#2}\@ifempty{#3}{.}{}\thmnote{ (#3).}}}}}
\makeatother
\theoremstyle{caja2}
\newtheorem{theotwo}{Claim}

\makeatletter
\newtheoremstyle{caja3}
  {\topsep}
  {\topsep}
  {\itshape}
  {}
  {}
  {}
  {.5em}
  {\red{\fbox{\black{\thmname{#1}~\thmnumber{#2}\@ifempty{#3}{.}{}\thmnote{ (#3).}}}}}
\makeatother
\theoremstyle{caja3}
\newtheorem{theothree}{Claim}

\newtheorem{innercustomgeneric}{\customgenericname}
\providecommand{\customgenericname}{}

\begin{document}

\pagenumbering{roman}
\title{\Large {Compound} Logics for Modification Problems\thanks{The first two authors where supported by 
 the Research Council of Norway  via the project BWCA (314528). The last three authors where supported by the  ANR projects DEMOGRAPH (ANR-16-CE40-0028),  ESIGMA (ANR-17-CE23-0010), and the French-German Collaboration ANR/DFG Project UTMA (ANR-20-CE92-0027). The third author was also supported by  the  ANR project ELIT (ANR-20-CE48-0008-01). }\,\,$^{,}$\thanks{Emails:  \texttt{fedor.fomin@uib.no}, \texttt{petr.golovach@uib.no}, \texttt{ignasi.sau@lirmm.fr}, \texttt{giannos.stamoulis@lirmm.fr}, \texttt{sedthilk@thilikos.info}. }}

{
\author{\bigskip 
Fedor V. Fomin%
\thanks{Department of Informatics, University of Bergen, Norway.} \and
Petr A. Golovach%
\samethanks[3] \and
Ignasi Sau%
\thanks{LIRMM, Univ Montpellier, CNRS, Montpellier, France.}\and
Giannos Stamoulis\samethanks[4]
\and
Dimitrios  M. Thilikos\samethanks[4]}
}
\date{}
\maketitle
\vspace{-7mm}
\begin{abstract}
\noindent
We introduce a novel model-theoretic framework inspired from graph modification and based on the interplay between model theory and algorithmic graph minors.
The core of our framework is  a  new {\sl compound logic}  operating with two types of sentences, expressing graph modification: the {\sl modulator sentence},  defining some property of  the modified part of the graph, and  the {\sl target sentence}, defining some property of the resulting graph. In our framework, modulator sentences are in  counting monadic second-order logic ({\sf CMSOL})  and have models of bounded treewidth, while target sentences express first-order logic ({\sf FOL}) properties along with minor-exclusion.
 Our logic captures  problems that are not definable in first-order logic and, moreover, may  have instances of unbounded treewidth. Also, it permits the modeling of wide families of  problems involving vertex/edge removals, alternative modulator measures (such as elimination distance or ${\cal G}$-treewidth), multistage modifications, and various cut problems.
Our main result is  that, for this compound logic, model-checking can be done in quadratic time.   All derived algorithms are  constructive and this, as a byproduct,  extends the constructibility horizon of the algorithmic applications of the Graph Minors theorem of  Robertson and Seymour.
The proposed logic can be seen as a general framework to capitalize on the potential of the {\sl irrelevant vertex technique}. It gives a way to deal with problem instances of unbounded treewidth, for which Courcelle's theorem does not apply. 
The proof of our meta-theorem combines  novel combinatorial results related to the  {Flat Wall theorem} along with elements of the proof of Courcelle's theorem and Gaifman's theorem. We finally prove extensions where the target property is expressible in {\sf FOL\!+DP}, i.e., the enhancement of {\sf FOL} with disjoint-paths predicates.
Our algorithmic meta-theorems encompass, unify, and extend {all} known meta-algorithmic results on minor-closed graph classes.
\end{abstract}

\medskip\medskip\medskip\medskip

\newpage
\tableofcontents
\newpage
\pagenumbering{arabic}
\thispagestyle{empty}
\newpage

\setcounter{page}{1}
 \myskip\section{Introduction}\myskip
 \label{sec_intro}

Our work is kindled by the current algorithmic advances in graph modification. The core of our approach is a novel model-theoretic framework that is  based on the interplay between model theory and algorithmic graph minors.
Departing from this new perspective, we obtain an algorithmic meta-theorems that encompass, unify, and extend {all}
 known meta-algorithmic results on minor-closed graph classes.

\myskip

\mysubsection{State of the art and our contribution}
\vspace{1mm}
 \label{sec_state-art}

\paragraph{Modification problems.}
 A {\sl graph modification problem} asks whether it is possible to apply a series of modifications to a graph in order  to transform it to a graph with  some desired target property.
Such problems have been the driving force of Parameterized Complexity where parameterization quantifies the concept of ``distance from triviality'' \cite{GuoHN04astru} and measures the amount of the applied modification. Classically, modification operations may be vertex or edge deletions, edge additions/contractions, or combinations of them like taking a minor. In their generality, such problems are \NP-complete \cite{LewisY80theno,Yannakakis81edged} and much research in Parameterized Complexity is on the design of algorithms in time $f(k)\cdot n^{\O(1)},$ where the parameter $k$ is some measure of the modification operation \cite{cygan2015parameterized}.
The target property may express desired structural properties that respond to certain algorithmic or combinatorial demands. A widely studied family of target properties are minor-closed graph classes such as edgeless graphs \cite{ChenKX06impr}, forests \cite{ChenFLLV08,KociumakaP14fast}, bounded treewidth
 graphs \cite{FominLMS12plan,KimLPRRSS16line,FominLMPS16hitti},
 planar graphs \cite{JansenLS14anea,MarxS07obta,Kawarabayashi09}, bounded genus graphs \cite{KociumakaP19}, or, most generally,
 minor-excluding graphs~\cite{SauST20anftp,AdlerGK08comp,SauST21kapiI}. However, other families of target properties have also been considered, such as those that exclude an odd cycle \cite{FioriniHRV05plana}, a topological minor \cite{FominLP0Z20hitti}, an (induced) subgraph~\cite{Cai96,SauS20hitti,CyganMPP17hitti}, an immersion~\cite{GiannopoulouPRT21linea}, or an induced minor~\cite{GolovachKP13}.
A broad class of graph modification problems concerns cuts. In a typical cut problem, one wants to find a minimum-size set of edges or vertices $X$ in a graph $G$ such that in the new graph~$G\setminus X,$ obtained by deleting~$X$ from~$G,$ some terminal-connectivity conditions are satisfied.  For example, the condition can be that a set of specific terminals becomes separated or that at least one connected component in the new graph is of a specific size.
The development of parameterized algorithms for cut problems is a popular trend in parameterized algorithms \cite{marx-razgon-stoc2011-multicut,BousquetDT18,kawarabayashi2011minimum,CyganLPPS19,DBLP:conf/icalp/KleinM12,Lokshtanov0S20,GuptaLL18}.
 More involved modification measures of  vertex set removals, related to  treewidth or treedepth, have been considered  very recently \cite{BulianD16graph,EibenGHK21,JansenK021verte,BulianD17fixe,AgrawalKLPRSZ22delet}.

\myskip

\paragraph{Algorithmic meta-theorems.}
A vibrant line of research in Logic and Algorithms is the development of {\sl algorithmic meta-theorems}. According to Grohe and Kreutzer \cite{GroheK09},
algorithmic meta-theorems state that certain families of algorithmic problems, typically defined by some logical and some combinatorial condition, can be solved ``efficiently'', under some suitable definition of this term.
Algorithmic meta-theorems play an important role in the theory of algorithms as they reveal deep interplays between Algorithms, Logic, and Combinatorics. One of the most celebrated meta-theorems is Courcelle's theorem
asserting that graph properties definable in \MSOL\  (counting monadic second-order logic) are decidable in linear time on graphs of bounded treewidth \cite{Courcelle90them,Courcelle97,Courcelle92}; see also \cite{BoriePT92auto,ArnborgLS91easy}. Another stream of research concerns identifying wide combinatorial structures where model-checking for \FOL \ (first-order logic) can be done in polynomial time. This includes graph classes of bounded degree~\cite{Seese96line}, graph classes of bounded local treewidth~\cite{FrickG01decid}, {\sl minor-closed graph classes} \cite{FlumG01fixe}, graph classes locally excluding a minor~\cite{DawarGK07loca}, and more powerful concepts of sparsity, such as having bounded expansion
\cite{DvorakKT13firs,NesetrilM12spar,NesetrilO08,NesetrilO08I,NesetrilO08II}, nowhere denseness~\cite{GroheKS17dec}, or having bounded twin-width \cite{Bonnet0TW20twinw}. {(See \cite{Kreutzer11algo,Grohe07logi} for surveys. Also for results on the combinatorial horizon
of \FOL\ and \MSOL\ (and its variants) see  \cite{GroheKS17dec,Bonnet0TW20twinw,BonnetGMSTT22twin}  and \cite{BojanczykP16defi,BojanczykGP21} respectively.)}

{Another  line of research, already mentioned in \cite{Grohe07logi}, is to prove algorithmic meta-theorems for extensions of \FOL of greater expressibility. Two such extensions have been recently  presented.
The first one consists in
enhancing \FOL with predicates that can express $k$-connectivity for every $k \geq 1$. This extension of \FOL, was introduced independently
by  Schirrmacher, Siebertz, and Vigny in \cite{SchirrmacherSV22first} (under the name {\sf FOL\!+conn}) and
by   Bojańczyk in \cite{Bojanczyk21separ} (under the name {\sl separator logic}).
The second and more expressive extension, also introduced by Schirrmacher, Siebertz, and Vigny in \cite{SchirrmacherSV22first}, is {\sf FOL\!+DP}, that enhances \FOL with predicates expressing the existence of disjoint paths between certain pairs of vertices. For {\sf FOL\!+conn}, an algorithmic meta-theorem for model-checking on graphs excluding a topological minor has been very recently given by Pilipczuk, Schirrmacher, Siebertz,  Torunczyk, and Vigny~\cite{PilipczukSSTV22algor}. For the more expressive {\sf FOL\!+DP}, an algorithmic meta-theorem for model-checking on graphs excluding a minor has been very recently given by
 Golovach, Stamoulis, and Thilikos in~\cite{GolovachST22model} (see~\cite{GolovachST22model_arXiv} for the full version).
 
 Research on the meta-algorithmics of \FOL\ is  quite active and has moved to several directions such 
as the study of \FOL-interpretability \cite{Bonnet22model,PilipczukOS22transd,NesetrilMS22stru,NesetrilMPRS21rankw,NesetrilRMS20linea,GajarskyKNMPST20first} or the enhancement of \FOL\ with  counting/numerical predicates~\cite{KuskeS17first,KuskeS18gaifm,DreierR21appro,GroheS18fisrt} (see also \cite{HeuvelKPQRS17model,Grange21succe,EickmeyerEH17succi,GroheS00local} for other extensions).
\medskip



{
In this paper, we initiate an alternative approach consisting in combining the expressive power of \FOL and
\MSOL. A typical family of problems where such an approach becomes relevant is the one of modification
problems.}
Courcelle's theorem implies that if the target property
corresponds to a  class of bounded treewidth and the modification conditions are definable in
{\MSOL}, then such modification problems are fixed-parameter tractable when parameterized by the length of
the sentence and the treewidth of the graph. However, when the target class graph is of unbounded treewidth,
none of the aforementioned algorithmic meta-theorems encompasses broad families of modification problems.
As an illustrative example, consider the {\sc Planarization}
problem, which consists in deciding whether at most  $k$ vertices can be removed from an input graph to make
it planar (or equivalently,  minor-excluding $K_{5}$ and $K_{3,3}$). While this problem is definable in {\MSOL},
Courcelle's theorem cannot be applied as we cannot assume that  \yes-instances are of bounded treewidth. On
the other hand, we can easily assume that \yes-instances minor-exclude $K_{k+6}.$ However, all known meta-
theorems whose combinatorial condition encompasses the minor-exclusion are about \FOL, and \FOL\ cannot
express the {\sc Planarization} problem. On the positive side, an algorithm in time $f(k)\cdot n^{2}$  for {\sc
Planarization} is an algorithmic consequence of  Robertson-Seymour's theorem~\cite{RobertsonS04GMXX}
(combined with \cite{KawarabayashiKR12thedis,RobertsonS95XIII}). This automatic implication follows directly
(albeit non-constructively) for a wide family of modification problems whose \yes-instances are minor-closed.
There is  a long line of research in parameterized algorithms towards providing constructive and reasonable
estimations of $f(k)$
\cite{JansenLS14anea,AdlerGK08comp,MarxS07obta,Kawarabayashi09,SauST20anftp,SauST21kapiI}. Note
that  Robertson-Seymour's theorem,  besides not being constructive in general, automatically  offers results
only for problems whose \yes-instances are minor-closed.

\myskip

\paragraph{Our contribution.}
We introduce a {\sl compound logic} that models computational problems  through the lens of the ``modulator vs target'' duality of graph modification problems.
Each sentence of this logic is a composition of two types of sentences. The first one, called the {\em modulator sentence}, models a modification operation, while the second one, called the {\em target sentence}, models a target property. Informally, our result, in its simplest form, asserts that if some appropriate version of the modulator sentence meets the meta-algorithmic assumptions of Courcelle's theorem~\cite{Courcelle90them} (i.e.,
\MSOL-definability and bounded treewidth) and the target sentence meets the meta-algorithmic assumptions of the theorem of Flum and Grohe~\cite{FlumG01fixe}
(i.e., \FOL-definability and minor-exclusion), then model-checking for the composed compound sentence can be done, constructively, in quadratic time.
Our main result (\autoref{@decendientes1}) can be seen as a ``two-dimensional product'' of the two aforementioned meta-algorithmic results, contains both of them as special cases,  and automatically implies the tractability of  wide families of problems that  {\sl neither}    are \FOL -definable {\sl nor}  have instances of bounded treewidth (see~\autoref{sec_applications} for the meta-algorithmic applications).
\myskip

\mysubsection{Our results}
 \label{sec_our-results}
\myskip

In this subsection we give formal statements of our results. We need first some definitions.
\myskip

\paragraph{Preliminaries on graphs.} Most of our graph definitions are compatible with Diestel's book \cite{Diestel10grap}.
Given a graph $G,$ we denote by ${\sf cc}(G)$ the set of all connected components of  $G.$
For a graph $G$ and a set $X\subseteq V(G),$ the {\em stellation} of $X$ in $G$ is the  graph ${\sf stell}(G,X)$  obtained from $G$
if, for every $C∈{\sf cc}(G\setminus X),$ we contract all the edges of $C$ to a single vertex $v_C.$
The {\em torso} of $X$ in $G$ is the graph $\torso(G,X)$ obtained from ${\sf stell}(G,X)$ if, for every $v_{C}$ where $C∈{\sf cc}(G\setminus X),$  we add
all edges between neighbors of $v_{C}$
and  finally  remove all $v_{C}$'s from the resulting graph.

Given a family of graphs ${\cal H},$ we define ${\sf excl}({\cal H})$ as the class of all graphs
minor-excluding the graphs in ${\cal H}$ and note that ${\sf excl}({\cal H})$ is a minor-closed class
(see \autoref{@aproximadament} for the definition of minor relation, minor closeness, and minor-exclusion).
The {\em Hadwiger number} of a graph $G,$ denoted by $\hw(G),$  is the minimum $k$ where $G∈ {\sf excl}(\{K_{k}\})$ and $K_{k}$ is the complete graph on $k$ vertices.
We also use the well-known parameter of {\em treewidth} of a graph $G,$ denoted by $\tw(G),$ that is defined in~\autoref{@aproximadament}.
Given a class of graphs ${\cal G}$ we define
$\tw({\cal G})=\max\{\tw(G)\mid G∈ {\cal G}\}.$ We define $\hw({\cal G})$ analogously. We  use ${\cal G}_{\rm all}$ for the set of all graphs.
\myskip

\paragraph{Preliminaries on logic.}
We use \MSOL\ (resp. \FOL) for the set of sentences in counting monadic second-order logic (resp. first-order logic) -- see \autoref{sec_prelim_logic} for the definitions.
Given some vocabulary $τ$ and a sentence $φ∈ \MSOL[τ],$
we denote by $\Mod(φ)$ the set of
all {finite}
models of $φ,$ i.e., all structures  that are models of $φ.$
In this introduction, in order to simplify our presentation, all structures that we consider are either graphs or annotated graphs, i.e., pairs $(G,X)$ where $G$ is a graph and $X\subseteq V(G).$ In the first case $τ=\{{\sf E}\},$
and in the second $τ=\{{\sf E},{\sf X}\}.$

Given a $φ∈ \MSOL[\{{\sf E}\}],$ we define the {\em connectivity extension} $φ^{({\sf c})}$ of  $φ$ so that $G\models φ^{({\sf c})}$  if   $\forall C∈ {\sf cc}(G), C\models φ.$
Similarly, for every ${\cal L}\subseteq \MSOL[\{{\sf E}\}],$ we define ${\cal L}^{({\sf c})}={\cal L}\cup \{φ^{({\sf c})}\mid φ∈ {\cal L}\}.$
Notice that $\{φ\}^{({\sf c})}=\{φ,φ^{({\sf c})}\}$.
Also by $\bool({\cal L})$ we denote the set of all positive Boolean combinations {(i.e., using only the Boolean connectives $\lor$ and $\land$)} of sentences in ${\cal L}.$
We next define the following  sets of sentences:

\begin{itemize}
\item   The set $\MSOL^{\sf tw}[\{{\sf E},{\sf X}\}]$ contains  every sentence   $β∈\MSOL[\{{\sf E},{\sf X}\}]$ for which there exists some $c_{β}$ such that the torsos of
all the models of $β$ have treewidth at most $c_β.$ Formally,

$\MSOL^{\sf tw}[\{{\sf E},{\sf X}\}]=\{β∈\MSOL[\{{\sf E},{\sf X}\}]\mid \exists c_{β}:  \tw\{{\sf torso}(G,X)\mid (G,X)\models φ\}≤ c_β\}.$
\item The set $\NTMC[\{{\sf E}\}]$ is the set of all sentences  in $\MSOL[\{{\sf E}\}]$ that express the minor-exclusion of a non-empty set of graphs.
Formally,

 $\NTMC[\{{\sf E}\}]=\{μ∈\MSOL[\{{\sf E}\}]\mid \exists {\cal H}\subseteq{\cal G}_{\rm all}, {\cal H}\neq \emptyset: \Mod(μ)={\sf excl}({\cal H})\}.$
\item $Θ_{0}[\{{\sf E}\}]$ contains every sentence $σ\wedge μ$
where $σ ∈ {\sf FOL}[\{{\sf E}\}]$ and $μ∈ {\NTMC}[\{{\sf E}\}].$
\end{itemize}

For simplicity,  we use $\MSOL^{\sf tw},$ $ \NTMC,$ and $Θ_{0}$ as shortcuts  for $\MSOL^{\sf tw}[\{{\sf E},{\sf X}\}],$ $\NTMC[\{{\sf E}\}],$ and $Θ_{0}[\{{\sf E}\}],$ respectively.
\myskip

\paragraph{Algorithmic meta-theorems.}
We are now in position to restate  three  major meta-algorithmic results that were mentioned  in the previous subsection.

%


\begin{proposition}[Courcelle's \cite{Courcelle90them}]
\label{@originallypublishedin}
For every  $β∈ \MSOL^{\sf tw},$ there is  an  algorithm deciding $\Mod(β)$ in linear time.
\end{proposition}

\begin{proposition}[Robertson and Seymour~\cite{RobertsonS95XIII,RobertsonS04GMXX}  and Kawarabayashi,  Kobayashi, and Reed \cite{KawarabayashiKR12thedis}]
\label{@circumscribed} For every minor-closed  graph class ${\cal G}$ deciding membership in ${\cal G}$ can be done in    quadratic time.
\end{proposition}

\begin{proposition}[Flum and  Grohe \cite{FlumG01fixe}]
\label{@universities}
For every  $γ∈ Θ_0,$  there is an  algorithm deciding  $\Mod(γ)$ in  quadratic time.
\end{proposition}

Some comments are in order. The statements of \autoref{@originallypublishedin} and \autoref{@universities}
 have been adapted so to
incorporate the combinatorial
demands in the logical condition. While they can both be stated  for structures, we state \autoref{@originallypublishedin} for annotated graphs and \autoref{@universities}  for graphs in order to facilitate our presentation.
In the classic formulation  of Courcelle's theorem, we are given a sentence $β∈\MSOL$ and a tree decomposition of bounded treewidth. As such a decomposition can be found in linear time, using e.g., \cite{Bodlaender93aline,BodlaenderDDFLP16axnap,DBLP:journals/corr/abs-2104-07463}, the linearity in the running time of Courcelle's theorem is preserved  when it is stated in the form of \autoref{@originallypublishedin}.
For the theorem of Flum and Grohe, the situation is different
as  the combinatorial demand is minor-exclusion of a clique,  which  is not definable is \FOL. For this reason we state \autoref{@universities} using the logic $Θ_0$
that contains {\sl compound} sentences of the form $σ\wedge μ,$  where
$σ∈\FOL$ and $μ$ expresses  minor-exclusion.
For the running time of the algorithm of \autoref{@universities}, we also need
to take into account \autoref{@circumscribed}.
%
As we already mentioned, \autoref{@originallypublishedin} and \autoref{@universities} cannot deal, in general,
with modification problems to properties of unbounded treewidth.  Moreover,  recall that \autoref{@circumscribed} applies only to problems whose  \yes-instances are minor-closed. 

{
We stress that
\autoref{@originallypublishedin}, \autoref{@circumscribed}, and \autoref{@universities}
are non-constructive. In order to construct the algorithms promised by \autoref{@originallypublishedin},
one should also know the bound $c_{β}$ on the treewidth of the models of $β\in \MSOL^{\sf tw}$.
Similarly, for \autoref{@circumscribed} (resp. \autoref{@universities}), one should have an upper bound on the Hadwiger number of the graphs in ${\cal G}$ (resp. the models of $γ$).
}

\myskip
\paragraph{A logic for modification problems.}

As a key ingredient of our result, we define the following  operation between sentences.
Let $β∈  \MSOL[\{{\sf E},{\sf X}\}]$ and $γ∈ \MSOL[\{{\sf E}\}].$
We refer to $β$ as the {\em modulator} sentence on annotated graphs and to $γ$
as the {\em target} sentence on graphs.
We define
$β\triangleright γ$ so that
\begin{eqnarray}
\text{$G\models β\triangleright γ$ if there is   $X\subseteq V(G)$ such that $({\sf stell}(G,X),X)\models β$
and $G\setminus X\models γ$}.\label{@surprenantes}
\end{eqnarray}
In other words, $G\models β\triangleright γ$ means that  the stellation of $X$ in $G,$ along with $X,$ is a model of the modulator sentence $β$
and the $G\setminus X$
is a model of the target sentence $γ.$
That way, $β$ implies the modification operation and $γ$ expresses the target  graph property.
It is easy to see and we prove formally in \autoref{@confiscating} that $β\triangleright γ∈ \MSOL[\{{\sf E}\}].$
This will allow us to apply the operation $\triangleright$ iteratively.

As an example, the problem of removing a set $X$ of $k$ vertices so that  $G\setminus X$ is a triangle-free planar graph  could be expressed by $β\triangleright γ$  if $β$ asks that $X$ has $k$ vertices  and $γ=σ\wedge  μ,$
where $σ$ expresses triangle-freeness and $μ$ expresses planarity {by the exclusion of $K_{3,3}$ and $K_5$}.

Before we present our result in full generality, we give first  the following  indicative  special case, that already expresses the conditions of \autoref{@originallypublishedin} and \autoref{@universities}.

\begin{theorem}
\label{@inflammatory}
For every  $β∈ \MSOL^{\sf tw}$ and every  $γ∈ Θ_{0},$   there is an algorithm deciding $\Mod(β\triangleright γ)$ in quadratic time.
\end{theorem}


Indeed, \autoref{@originallypublishedin} follows\footnote{In particular, \autoref{@inflammatory} contains~\autoref{@originallypublishedin} as a linear-time black-box procedure for deciding models of bounded treewidth.}
if $β$ expresses that $X=V(G)$ {and $γ$ demands that $G\setminus X$ is the empty graph}
and \autoref{@universities} follows
if  $β$  demands that $X=\emptyset.$
In other words,  \autoref{@originallypublishedin} follows if the {target} sentence becomes void  while \autoref{@universities} follows if the {modulator} sentence is void.

As a first step towards a more general statement, \autoref{@inflammatory} also holds
if we replace  $γ∈ Θ_{0}$ by  $γ∈ Θ_{0}^{({\sf c})}$ or even by positive Boolean combinations of sentences in $Θ_{0}^{({\sf c})},$ i.e., $γ∈ \bool(Θ_{0}^{({\sf c})}).$ Moreover, in order to present our result in full generality, we recursively define, for every $i≥ 1,$
\begin{eqnarray}
Θ_{i} &  = & \{β\triangleright γ\mid β∈ \MSOL^{\sf tw}\text{ and } γ∈ \bool(Θ_{i-1}^{({\sf c})})\}.\label{@nichtbestehens}
\end{eqnarray}

Notice that the sentences of \autoref{@inflammatory} (hence also  of \autoref{@originallypublishedin} and \autoref{@universities}) are already contained in $Θ_{1}.$ We set $Θ=\bigcup_{i≥ 1}Θ_{i}.$
The full strength of our results, stated in the vocabulary of graphs, is given by our main theorem.

\begin{theorem}
\label{@decendientes1}
For every  $θ∈ Θ,$ model-checking for   ${θ}$
can be done in quadratic  time.
\end{theorem}
\myskip

\paragraph{An alternative statement.}
 Our results can also be seen under the  typical meta-algorithmic
 framework where a logical and a combinatorial condition are given.
For this, consider an alternative  of  $Θ,$ called $\tilde{Θ},$ that
is defined as in (\ref{@nichtbestehens}) by  taking
$\tilde{Θ}_{0}=\FOL$  as  the base case, i.e.,
by discarding the minor-exclusion   from the definition of ${Θ}_0.$
Notice that $\tilde{Θ}$ contains \FOL\ and can be seen as  a natural extension of it.
A direct consequence of \autoref{@decendientes1} is the following.

\begin{figure}[ht]
	\begin{center}
\tikzstyle{ipe stylesheet} = [
  ipe import,
  even odd rule,
  line join=round,
  line cap=butt,
  ipe pen normal/.style={line width=0.4},
  ipe pen heavier/.style={line width=0.8},
  ipe pen fat/.style={line width=1.2},
  ipe pen ultrafat/.style={line width=2},
  ipe pen normal,
  ipe mark normal/.style={ipe mark scale=3},
  ipe mark large/.style={ipe mark scale=5},
  ipe mark small/.style={ipe mark scale=2},
  ipe mark tiny/.style={ipe mark scale=1.1},
  ipe mark normal,
  /pgf/arrow keys/.cd,
  ipe arrow normal/.style={scale=7},
  ipe arrow large/.style={scale=10},
  ipe arrow small/.style={scale=5},
  ipe arrow tiny/.style={scale=3},
  ipe arrow normal,
  /tikz/.cd,
  ipe arrows, 
  <->/.tip = ipe normal,
  ipe dash normal/.style={dash pattern=},
  ipe dash dotted/.style={dash pattern=on 1bp off 3bp},
  ipe dash dashed/.style={dash pattern=on 4bp off 4bp},
  ipe dash dash dotted/.style={dash pattern=on 4bp off 2bp on 1bp off 2bp},
  ipe dash dash dot dotted/.style={dash pattern=on 4bp off 2bp on 1bp off 2bp on 1bp off 2bp},
  ipe dash normal,
  ipe node/.append style={font=\normalsize},
  ipe stretch normal/.style={ipe node stretch=1},
  ipe stretch normal,
  ipe opacity 10/.style={opacity=0.1},
  ipe opacity 30/.style={opacity=0.3},
  ipe opacity 50/.style={opacity=0.5},
  ipe opacity 75/.style={opacity=0.75},
  ipe opacity opaque/.style={opacity=1},
  ipe opacity opaque,
]
\definecolor{red}{rgb}{1,0,0}
\definecolor{blue}{rgb}{0,0,1}
\definecolor{green}{rgb}{0,1,0}
\definecolor{yellow}{rgb}{1,1,0}
\definecolor{orange}{rgb}{1,0.647,0}
\definecolor{gold}{rgb}{1,0.843,0}
\definecolor{purple}{rgb}{0.627,0.125,0.941}
\definecolor{gray}{rgb}{0.745,0.745,0.745}
\definecolor{brown}{rgb}{0.647,0.165,0.165}
\definecolor{navy}{rgb}{0,0,0.502}
\definecolor{pink}{rgb}{1,0.753,0.796}
\definecolor{seagreen}{rgb}{0.18,0.545,0.341}
\definecolor{turquoise}{rgb}{0.251,0.878,0.816}
\definecolor{violet}{rgb}{0.933,0.51,0.933}
\definecolor{darkblue}{rgb}{0,0,0.545}
\definecolor{darkcyan}{rgb}{0,0.545,0.545}
\definecolor{darkgray}{rgb}{0.663,0.663,0.663}
\definecolor{darkgreen}{rgb}{0,0.392,0}
\definecolor{darkmagenta}{rgb}{0.545,0,0.545}
\definecolor{darkorange}{rgb}{1,0.549,0}
\definecolor{darkred}{rgb}{0.545,0,0}
\definecolor{lightblue}{rgb}{0.678,0.847,0.902}
\definecolor{lightcyan}{rgb}{0.878,1,1}
\definecolor{lightgray}{rgb}{0.827,0.827,0.827}
\definecolor{lightgreen}{rgb}{0.565,0.933,0.565}
\definecolor{lightyellow}{rgb}{1,1,0.878}
\definecolor{black}{rgb}{0,0,0}
\definecolor{white}{rgb}{1,1,1}
\begin{tikzpicture}[ipe stylesheet]
  \fill[darkorange, nonzero rule, ipe opacity 75]
    (176, 480)
     -- (192, 480)
     -- (192, 416)
     -- (192, 416)
     -- (192, 400)
     -- (176, 400)
     -- cycle;
  \draw[->]
    (272, 400)
     -- (288, 400);
  \node[ipe node, font=\footnotesize]
     at (184, 388) {{\sf FOL}};
  \node[ipe node, font=\small]
     at (220, 388) {$\tilde{ \Theta}$};
  \fill[red, ipe opacity 75, ipe opacity 30]
    (176, 416)
     -- (192, 416)
     -- (192, 416)
     -- (256, 416)
     -- (256, 400)
     -- (176, 400) -- (176, 416)
     -- cycle;
  \draw[orange, ipe pen fat]
    (188, 480) rectangle (188, 480);
  \draw
    (176, 400)
     -- (272, 400);
  \draw[<-]
    (176, 512)
     -- (176, 400);
  \node[ipe node, font=\footnotesize]
     at (240, 388) {{\sf CMSOL}};
  \node[ipe node, text=seagreen]
     at (280.554, 388) {{\small Logic}};
  \node[ipe node, font=\footnotesize]
     at (196, 473.059) {[Grohe, Kreutzer, \& Siebertz] /  [Bonnet, Kim, Thomassé, \& Watrigant]};
  \node[ipe node, font=\footnotesize]
     at (262, 417.059) {[Courcelle], [Borie, Parker, \& Tovey],};
  \node[ipe node, font=\footnotesize]
     at (262, 405.059) {and [Arnborg, Lagergren, \&  Seese]};
  \node[ipe node, font=\small]
     at (228, 445.059) {\blue{Theorem {6}}};

     \filldraw[blue, ipe pen ultrafat, ipe opacity 30]
      (176, 400) -- (176, 448)
      -- (176, 448)
     -- (224, 448) --  (224, 448)
     -- (224, 416)
     -- (256, 416) -- (256, 400)
      -- (176, 400)
      ;
  \node[ipe node, font=\footnotesize]
     at (61, 447) {{\sf bounded Hadwiger number}};
  \node[ipe node, font=\footnotesize]
     at (92, 412) {{\sf bounded treewidth}};
  \node[ipe node, font=\footnotesize]
     at (20, 473.059) {{\sf nowhere dense /  bounded twin-width}}; \node[ipe node, text=seagreen, font=\footnotesize]
     at (130, 505) {{Structure}};
\draw[blue, ipe pen ultrafat]
    (176, 400) -- (176, 448)
      -- (176, 448)
     -- (224, 448) --  (224, 448)
     -- (224, 416)
     -- (256, 416) -- (256, 400)
      -- (176, 400)
      ;
\end{tikzpicture}
	\end{center}
\myskip
	\caption{\autoref{@pertrechando1}  in the current meta-algorithmic landscape. The vertical axis is the combinatorial one and is marked by four different types of (structural) sparsity, while the horizontal one is the logical one and is marked with \FOL, $\tilde{Θ},$ and \MSOL.}
	\label{@resolveremos}
	\myskip
\end{figure}
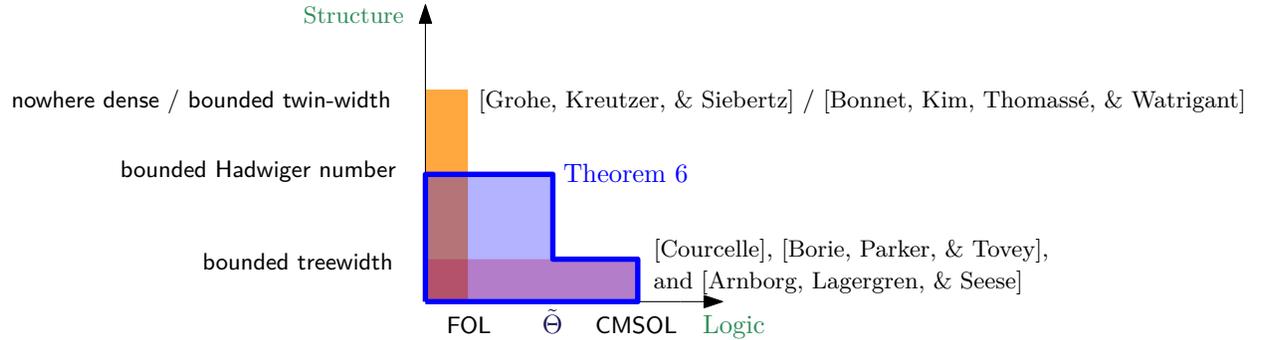

\begin{theorem}
\label{@pertrechando1}
For every $\tilde{θ}∈\tilde{Θ},$ model-checking for   $\tilde{θ}$
can be done
in quadratic  time on every graph class of bounded Hadwiger number.
\end{theorem}

 \autoref{@pertrechando1} is a corollary of \autoref{@decendientes1} and provides an alternative meta-algorithmic  set up between
 the logical and the combinatorial condition (see \autoref{@resolveremos}):
for each sentence $θ$ in $Θ,$ one may consider
a sentence $\tilde{θ}$ in $\tilde{Θ}$ where
we discard minor-exclusion from all its target sentences and then consider the problem
of deciding $\Mod(θ)$ on some minor-excluding graph class. This correspondence
is many-to-one, as many different $θ∈ Θ$ correspond
to the same $\tilde{θ}∈ \tilde{Θ}.$
We opted for presenting and proving our results in the form of \autoref{@decendientes1}, as it is more general and more versatile in expressing modification problems.

%
%

In \autoref{our_logic_stories} we define  $Θ$ on general structures. Under this general setting,  \autoref{@decendientes1} and \autoref{@pertrechando1} will be stated as  \autoref{@decendientes} and \autoref{@pertrechando}.

\paragraph{Compound logics based on {\sf FOL\!+DP}.}  In \autoref{sec_thetadp}, by combining our proofs with
the meta-algorithmic results of \cite{GolovachST22model,GolovachST22model_arXiv}, we extend \autoref{@decendientes1} (resp. \autoref{@pertrechando1})
in the cases of the logic  $Θ^{\sf DP}$ (resp. $\tilde{Θ}^{\sf DP}$) that are obtained if in the definition of  $Θ$ (resp. $\tilde{Θ}$)
we now  consider  the (more expressive) logic {\sf FOL\!+DP} instead of  {\sf FOL} in the target sentences.
That way, the derived extensions of \autoref{@decendientes1} and \autoref{@pertrechando1} (that is \autoref{thm_thetadp} and \autoref{thm_tildethetadp}) encompass,  as special cases,
all  results and applications in  \cite{GolovachST22model,GolovachST22model_arXiv} (see \autoref{@resolvesssremos} for a visualization of the overall
state-of-the-art on the related algorithmic meta-theorems).

While presenting our results and techniques, for the sake of simplicity,  we choose to   focus on the statement and the proof of our meta-theorems for  $Θ$  (\autoref{@decendientes1}) and  $\tilde{Θ}$ (\autoref{@pertrechando1})
and then, in  \autoref{sec_thetadp}, present the modifications that should be applied in order to extend them
for  $Θ^{\sf DP}$ (\autoref{sec_thetadp}) and  $\tilde{Θ}^{\sf DP}$ (\autoref{thm_tildethetadp}).

\paragraph{Constructibility.}

While   Robertson-Seymour's theorem (\autoref{@circumscribed})
implies the existence of an algorithm, its proof is not constructive and cannot be used to construct such an algorithm~\cite{FellowsL88nonc}.
An extra feature of the proof of  \autoref{@decendientes1} (as well as of  its corollary~\autoref{@pertrechando1}) is that it is {\sl constructive}, in the sense that
the implied algorithms can  be constructed if we are given some bound on the Hadwiger number of the models of $θ.$ This considerably extends the constructibility horizon of \autoref{@circumscribed} for graph classes that are not necessarily minor-closed or even hereditary
 (see~\autoref{section_consrtr} for more on the constructibility of our results).
\myskip

\paragraph{Techniques.}
The algorithm and the proofs of~\autoref{@decendientes1} use
as departure point core techniques
{
from the proofs of Propositions~\ref{@originallypublishedin},  \ref{@universities}, and \ref{@circumscribed} such as Courcelle's theorem for dealing with \MSOL-sentences, the use of Gaifman's theorem for dealing with \FOL-sentences,
}
 and an extended version of the {\sl irrelevant vertex technique},
 introduced
by Robertson and Seymour in~\cite{RobertsonS95XIII}, along with some suitable version of the Flat Wall theorem which appeared recently in~\cite{SauST21amor,KawarabayashiTW18anew} (see also~\cite{BasteST20acom,SauST20anftp,SauST21kapiI,SauST21kapiII}). The algorithm produces equivalent
and gradually ``strictly simpler'' instances of an annotated version of the problem. Each equivalent instance is produced in linear time and this simplification is repeated until the graph has bounded treewidth (here we may apply Courcelle's theorem, that is~\autoref{@originallypublishedin}).
This yields a (constructive) quadratic-time algorithm. We  stress that our approach avoids
techniques that have been recently used for this type of problems such as {\sl recursive understanding} (in~\cite{AgrawalKLPRSZ22delet}) or the use of {\sl important separators} (in~\cite{JansenK021verte}) that give worst running times in $n$.  For a more detailed discussion on the results of \cite{AgrawalKLPRSZ22delet,JansenK021verte} and their relation to our results, see the applications section  (\autoref{sec_applications}).
\myskip

\paragraph{Organization of the paper.}
In~\autoref{sec_overview} we provide an overview of our proof.
In~\autoref{sec_applications}, we discuss some applications of our results
in modification problems.
In~\autoref{@aproximadament} we provide some basic definitions that
will be used throughout the paper and in~\autoref{our_logic_stories} we give the formal definition of our logic.
To describe the algorithm for~\autoref{@decendientes1},
we first introduce an annotated version of the problem; this is done in~\autoref{sec_equivalent_version}.
Then, in~\autoref{sec_main_tools}, we give some preliminary concepts and results
and, in~\autoref{sec_scheme}, we present the general scheme of the algorithm for~\autoref{@decendientes1}.
Sections~\ref{sec_first_floor},~\ref{the_second_level},~and~\ref{sec_final_combo} are devoted to the gradual presentation of the {main subroutine} of the algorithm of~\autoref{@decendientes1} and its correctness.
Next, in~\autoref{section_consrtr},
we discuss the constructibility of our results and present some consequences of our results concerning the constructibility of Robertson-Seymour's theorem.
In \autoref{sec_thetadp}, we explain how to modify our proofs so that they also work for the more expressive logics $Θ^{\sf DP}$ and $\tilde{Θ}^{\sf DP}$.
We conclude the paper with~\autoref{sec_conclusions} by mentioning the limitations of our approach, possible extensions, and open research directions. {In~\autoref{@consideracions} we present the flat wall framework that we use in this paper, which was introduced in~\cite{SauST21amor}.
Also, in~\autoref{second_level_more}, we provide all the details of the omitted proofs of~\autoref{the_second_level}.
}

%
%

\myskip\myskip
\myskip\section{Overview of the proof}
\label{sec_overview}
\myskip

In this section we summarize some of the main ideas involved in the proof of \autoref{@decendientes1} (stated as \autoref{@decendientes} in its full {versatility} on structures), while keeping the description at an intuitive level. We would like to stress that some of the informal definitions given in this section are deliberately {\sl imprecise}, since providing the precise ones would result in a huge overload of technicalities that would hinder the flow of the proof.

Our algorithms consider as input a general structure~$\mathfrak{A}$
(not necessarily a graph), and most of the arguments in the proofs concern its Gaifman graph $G_{\mathfrak{A}}$ (see~\autoref{sec_prelim_logic} for the definition). Dealing with general structures, besides making our results more {versatile},
turns out to be useful in the proofs, in particular for using tools such as the {\sl Backwards Translation Theorem}~\cite{CourcelleE12grap,BojanczykP16defi} (see \autoref{@imposibilitada}), or for extending our results to other modification operations beyond vertex removal (see \autoref{thm_grammar} in \autoref{sec_applications_other_variants}). Since the Gaifman graph of a graph is the graph itself, in this overview we will assume for simplicity that the input of our algorithms is a graph~$G,$ instead of a general structure~$\mathfrak{A}.$

%
%

In \autoref{sec_overview_1} we present the general scheme of the algorithm (see \autoref{sec_scheme}, in particular \autoref{fig_general_sch_alg}, for a more detailed presentation), common to the distinct cases presented in \autoref{sec_first_floor}, \autoref{the_second_level}, and \autoref{sec_final_combo}, corresponding to different fragments of our logic~$Θ.$ In \autoref{sec_overview_2} we present a simplified and illustrative setting, where the input sentence $θ$ belongs to the fragment $\bar{Θ}_1$
 (see~\autoref{sec_first_floor}). This (very) particular case of \autoref{@decendientes1} is helpful to illustrate our main conceptual ideas, and after sketching its proof, in \autoref{sec_overview_3} we discuss how to integrate new technical ingredients, step by step, from this particular case up to the general compound logic $Θ$ considered in \autoref{@decendientes1} (see~\autoref{the_second_level} and~\autoref{sec_final_combo} for the details).

\myskip

\mysubsection{General scheme of the algorithm}
\label{sec_overview_1}
%
\myskip

We use the {\sl irrelevant vertex technique} introduced by Robertson and Seymour~\cite{RobertsonS95XIII}. Our overall strategy is the ``typical'' one when using this technique: if the treewidth of the input graph $G$ is bounded by an appropriately chosen function,   depending only on the sentence  $θ∈ Θ,$ then we use Courcelle's theorem~\cite{Courcelle90them,Courcelle92,Courcelle97} and solve the problem in linear time, using the fact that our compound logic $Θ$ is a fragment of counting monadic second-order logic (see \autoref{sec_ourlogic}). Otherwise, we identify an irrelevant vertex in linear time, that is, a vertex whose removal produces an equivalent instance. Naturally, the latter case concentrates all our efforts and, in what follows, we sketch the main ingredients that we use in order to identify such an irrelevant vertex. In a nutshell, our approach is based on introducing a robust combinatorial framework for finding irrelevant vertices. In fact,
what we find is {\sl annotation-irrelevant flat territories}, building on our previous recent work~\cite{BasteST20acom,SauST21amor,FominGST20analgo,SauST20anftp,SauST21kapiI,SauST21kapiII,BasteST20acom},
which is formulated with enough generality so as to allow for the application of powerful tools such as  Gaifman's locality theorem (see~\autoref{lem_gaifman})
 or a variant of Courcelle's theorem on boundaried graphs (see~\autoref{cou_more}).\myskip

\paragraph{Flat walls.} An essential tool of our approach is the notion of {\sl flat wall}, originating in the work of Robertson and Seymour \cite{RobertsonS95XIII}.
Informally speaking, a flat wall $W$ is a structure made up of (non-necessarily planar) pieces, called {\sl flaps}, that are glued together in a bidimensional grid-like way defining the so-called {\sl bricks} of the wall (see~\autoref{label_exhalaciones}).
While such a structure may not be planar, it enjoys topological properties similar to those of planar graphs, in the sense that two paths that are not routed entirely inside a flap cannot ``cross'', except at a constant-sized vertex set $A$ whose vertices are called {\sl apices}. Hence,
 flat walls are only ``locally non-planar'', and after removing apices we can apply useful locality arguments, in the sense that two vertices that are in ``distant'' flaps should also be ``distant'' in the whole graph without the apices.
 One of the most celebrated results in the theory of Graph Minors by Robertson and Seymour~\cite{RobertsonS04GMXX,RobertsonS95XIII}, known as the Flat Wall theorem (see~\autoref{label_proletarians} for a variant recently proved in~\cite{SauST21amor,KawarabayashiTW18anew}), informally states that graphs of large treewidth contain either a large clique minor or a large flat wall. In this article we use the framework recently introduced in~\cite{SauST21amor} that provides a more accurate view of some previously defined notions concerning flat walls, particularly in~\cite{KawarabayashiTW18anew}. We provide these precise definitions in
\autoref{label_exceptionalness}, including the concepts of {\sl flatness pair}, {\sl homogeneity}, {\sl regularity}, {\sl tilt}, and {\sl influence}, and we stress that they are {\sl not} critical in order to understand the main technical contributions of the current article (however, they are critical for their formal correctness).
 In what follows, when considering a flat wall $W$ with an apex set $A$ in a graph $G,$ for simplicity we refer to $W$ by using indistinguishably the terms ``wall'' and ``compass of a wall'', which can be roughly described as the component containing $W$ in the graph obtained from $G$ by removing $A$ and the ``boundary'' of $W$ (see  \autoref{label_exceptionalness} for the formal definition).

\myskip

\paragraph{Working with an annotated version of the problem.} We start by defining a convenient equivalent version of the problem (see \autoref{sec_equivalent_version}), by replacing our sentence $θ ∈ Θ$ with an equivalent enhanced sentence $θ_{{\sf R},{\bf c}}.$ This is done in two steps, presented in \autoref{@vorstellende} and \autoref{sec_equivalent-versions_first-floor}.

Assuming the existence of a flat wall and an apex set in our input graph $G,$ we first transform  (see~\autoref{@vorstellende}) the question $θ$ on $G$ to a question on a structure obtained from $G$ by ``neutralizing'' the apex set (\autoref{@escrostonades}).
The goal of this step is to ask the final \FOL-sentences $σ$
of our sentence $θ$ in a ``flattened'' structure, where apices can no longer ``bring close'' any distant parts of the wall. This transformation of the problem, which we call \emph{apex-projection}, will allow for the application of the locality-based strategy discussed in the definition of the {\sl in-signature} of a wall in \autoref{sec_overview_2}.
To do this, we introduce some additional constant symbols ${\bf c}$ to our vocabulary that will be interpreted as the apex vertices.

The second step (\autoref{sec_equivalent-versions_first-floor}) consists in defining an equivalent \emph{annotated} version of the problem in order to deal with the \FOL-sentences  of $θ,$ inspired by the approach of~\cite{FominGST20analgo}. To do so, we introduce a vertex set $R \subseteq V(G),$ and require, for each \FOL-sentence $σ$ of $θ,$ that the vertices interpreting the variables of (the equivalent Gaifman sentence of) $σ$ belong to the annotated set $R.$
We prove that the initial sentence $θ$ and the obtained sentence, denoted by $θ_{{\sf R},{\bf c}}$ and called an \emph{enhanced sentence},
are equivalent for {\sl any} choice of the apex set interpreting ${\bf c}$ and when ${\sf R}$ is interpreted as the whole vertex set of the graph (see~\autoref{obs_addingR} and \autoref{lem_no_matter_which_apex}). This independence of the choice of the apex set is strongly used in the proofs since, as discussed below, we will consider a number of different flat walls, each of which associated with a different apex set.

Our algorithms will work with the enhanced sentence $θ_{{\sf R},{\bf c}}.$ Starting with the input graph $G$ with $V(G)$ as the annotated set $R,$ we will create successive equivalent annotated instances, in which vertices from $G$ are removed and such that the annotated set $R$ is only reduced.
\myskip

\paragraph{Zooming inside a flat wall.} Our next step is to find, in $G,$ a large flat wall $W_0$  to work with. The definition of our logic $Θ$ implies (see \autoref{@disillusioned}) that models of~$θ$ exclude a fixed complete graph $K_c$ as a minor, where $c$ depends only on $θ.$ Therefore,
we can apply \autoref{label_proletarians} to the input graph $G$ and, assuming that the treewidth of $G$ is large enough, we can find in linear time a flat wall $W_0$ and an apex set $A$ in $G$ such that the height of $W_0$ is a sufficiently large function of $θ.$ Moreover, another crucial property guaranteed by \autoref{label_proletarians} is that the treewidth of  $W_0$ is bounded from above by a function of $θ.$ This will be exploited  in \autoref{sec_overview_2} in order to compute the so-called  {\sl $θ$-characteristic} of a wall. We will now apply a series of  ``zooming'' arguments to the wall $W_0,$ which are illustrated in \autoref{figure_walls_zoom} (see the proof of~\autoref{lemma_irrele_flat} for the precise constants).

 \begin{figure}[h!]
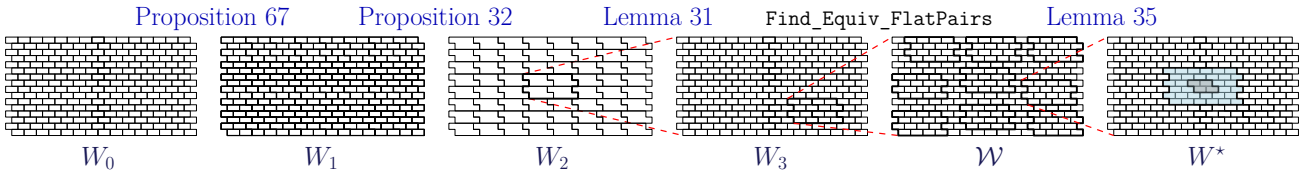

\hspace{-2.5cm}\scalebox{0.58}{%
\tikzstyle{ipe stylesheet} = [
  ipe import,
  even odd rule,
  line join=round,
  line cap=butt,
  ipe pen normal/.style={line width=0.4},
  ipe pen heavier/.style={line width=0.8},
  ipe pen fat/.style={line width=1.2},
  ipe pen ultrafat/.style={line width=2},
  ipe pen normal,
  ipe mark normal/.style={ipe mark scale=3},
  ipe mark large/.style={ipe mark scale=5},
  ipe mark small/.style={ipe mark scale=2},
  ipe mark tiny/.style={ipe mark scale=1.1},
  ipe mark normal,
  /pgf/arrow keys/.cd,
  ipe arrow normal/.style={scale=7},
  ipe arrow large/.style={scale=10},
  ipe arrow small/.style={scale=5},
  ipe arrow tiny/.style={scale=3},
  ipe arrow normal,
  /tikz/.cd,
  ipe arrows, 
  <->/.tip = ipe normal,
  ipe dash normal/.style={dash pattern=},
  ipe dash dotted/.style={dash pattern=on 1bp off 3bp},
  ipe dash dashed/.style={dash pattern=on 4bp off 4bp},
  ipe dash dash dotted/.style={dash pattern=on 4bp off 2bp on 1bp off 2bp},
  ipe dash dash dot dotted/.style={dash pattern=on 4bp off 2bp on 1bp off 2bp on 1bp off 2bp},
  ipe dash normal,
  ipe node/.append style={font=\normalsize},
  ipe stretch normal/.style={ipe node stretch=1},
  ipe stretch normal,
  ipe opacity 10/.style={opacity=0.1},
  ipe opacity 30/.style={opacity=0.3},
  ipe opacity 50/.style={opacity=0.5},
  ipe opacity 75/.style={opacity=0.75},
  ipe opacity opaque/.style={opacity=1},
  ipe opacity opaque,
]
\definecolor{red}{rgb}{1,0,0}
\definecolor{blue}{rgb}{0,0,1}
\definecolor{green}{rgb}{0,1,0}
\definecolor{yellow}{rgb}{1,1,0}
\definecolor{orange}{rgb}{1,0.647,0}
\definecolor{gold}{rgb}{1,0.843,0}
\definecolor{purple}{rgb}{0.627,0.125,0.941}
\definecolor{gray}{rgb}{0.745,0.745,0.745}
\definecolor{brown}{rgb}{0.647,0.165,0.165}
\definecolor{navy}{rgb}{0,0,0.502}
\definecolor{pink}{rgb}{1,0.753,0.796}
\definecolor{seagreen}{rgb}{0.18,0.545,0.341}
\definecolor{turquoise}{rgb}{0.251,0.878,0.816}
\definecolor{violet}{rgb}{0.933,0.51,0.933}
\definecolor{darkblue}{rgb}{0,0,0.545}
\definecolor{darkcyan}{rgb}{0,0.545,0.545}
\definecolor{darkgray}{rgb}{0.663,0.663,0.663}
\definecolor{darkgreen}{rgb}{0,0.392,0}
\definecolor{darkmagenta}{rgb}{0.545,0,0.545}
\definecolor{darkorange}{rgb}{1,0.549,0}
\definecolor{darkred}{rgb}{0.545,0,0}
\definecolor{lightblue}{rgb}{0.678,0.847,0.902}
\definecolor{lightcyan}{rgb}{0.878,1,1}
\definecolor{lightgray}{rgb}{0.827,0.827,0.827}
\definecolor{lightgreen}{rgb}{0.565,0.933,0.565}
\definecolor{lightyellow}{rgb}{1,1,0.878}
\definecolor{black}{rgb}{0,0,0}
\definecolor{white}{rgb}{1,1,1}

}
\caption{Sequence of walls considered in the general scheme of our algorithm, along with the results used to obtain them, where the first wall is obtained by applying \autoref{label_proletarians} to the input graph $G.$}\label{figure_walls_zoom}
\end{figure}
\myskip

Starting from $W_0$ and its associated apex set $A,$ we apply \autoref{label_highlighting} and find, in linear time, a large (again, as a function of $θ$) subwall $W_1$ that is $λ$-\emph{homogeneous}, where $λ$ depends only on~$θ.$ The definition of a homogenous flat wall can be found in \autoref{sec_homogeneouswalls}, and roughly means that each of its bricks can route the same set of partial minors of the graphs corresponding to the minor-exclusion part of the sentence $θ.$ We now apply \autoref{icalp_irrelevancy} to $W_1,$ which is a core result of \cite{SauST21kapiI} (see also~\cite{SauST20anftp}), and obtain in linear time a large subwall $W_2$ that is {\sl irrelevant} with respect to the minor-exclusion part of $θ$ after the removal of a vertex set $X \subseteq V(G)$ of small enough \emph{bidimensionality} (see \autoref{sec_overview_2}). Intuitively, working ``inside'' $W_2$ allows us to ``forget'' the minor-exclusion part of $θ$ in what follows. As our next step, we apply \autoref{lemma_many_apices} to $W_2,$ and obtain in linear time a still large subwall $W_3$ such that its associated apex set $A_3$ is ``tightly tied'' to $W_3,$ in the sense that the neighbors in $W_3$ of every vertex in $A_3$ are spread in a ``bidimensional'' way. This combinatorial technical condition is critically used in the proof of \autoref{@desmembramientos}.

\medskip

\noindent
\textbf{Finding an irrelevant subwall.} So far, we have found a large wall $W_3$ that satisfies the conditions listed in the statement of
 \autoref{@desmembramientos}. Now, in order to identify an irrelevant vertex inside $W_3,$ we proceed as follows (see the algorithm ${\tt Find\_Equiv\_FlatPairs}$ discussed informally in \autoref{sec_what_3_sec} and presented with all details in~\autoref{@inhumainement}). The strategy of the proof is to find, inside the wall $W_3,$ a collection ${\cal W}$ of pairwise disjoint subwalls, and to associate each of these subwalls with an appropriately defined {\sl $θ$-characteristic} that captures its behavior with respect to the partial satisfaction of the sentence $θ.$ Then the idea is that, if there are sufficiently many subwalls in ${\cal W}$ with the same $θ$-characteristic (called \emph{$θ$-equivalent}), then some subwall in the interior of one of them can be declared {\sl annotation-irrelevant} and this implies some progress in simplifying the current problem instance.

  The above strategy is formalized in~\autoref{lemma_irrele_flat}, which allows to identify a subwall $W^{\star}$ inside ${\cal W}$ such that its central part can be removed from the annotated set $R,$ and such that a smaller central part can be removed from $G$ (the blue and grey subwalls in the rightmost wall of \autoref{figure_walls_zoom}, respectively). The proof of \autoref{lemma_irrele_flat} is based on \autoref{@desmembramientos}, which is the main technical part of this paper, and whose full proof
 is postponed to Sections~\ref{sec_first_floor},~\ref{the_second_level}, and~\ref{sec_final_combo} for different fragments of the logic $Θ.$ The proof is based on the algorithm ${\tt Find\_Equiv\_FlatPairs}$ mentioned above, which is in turn based on an appropriate definition of the {\sl $θ$-characteristic} of a wall. A brief explanation of the proof strategy of \autoref{@desmembramientos} is given in~\autoref{sec_what_3_sec} (see \autoref{figu_levels_1and2}), and in what follows we sketch the main ingredients and key ideas.
\myskip

\mysubsection{A simplified and illustrative setting}
\label{sec_overview_2}
\myskip

In order to provide some intuition of the proof of \autoref{@desmembramientos}, in this subsection we
focus on {formulas $θ\in Θ$ of a particular form, i.e., belonging to $\bar{Θ}_1,$  a set of formulas formally defined in \autoref{sec_first_floor} which we proceed to define informally in a semantical level:
Given a general graph $G$ as input}, we seek for a vertex set $X \subseteq V(G),$ called {\sl modulator}, such that, using the notation defined in the introduction, ${\sf stell}(G,X)$ satisfies the so-called {\sl modulator sentence} $β,$ and either every connected component $C$ of $G \setminus X,$ or the whole graph $G \setminus X,$ satisfies the so-called {\sl target sentence} $γ,$ where $γ = σ\wedge μ$ with $σ$ being an arbitrary \FOL-sentence and $μ$ expressing the property of belonging to a {proper} minor-closed graph class.

Note that when $θ ∈ \bar{Θ}_1,$ the target sentence~$γ$ needs to be satisfied either by each of the resulting connected components separately, or jointly by their union. We deal with this easily,  by introducing a $\circ/\bullet$ -flag into the corresponding sentences that distinguishes both cases. The latter case is simpler, but in
this description, in order to better illustrate our techniques, we assume the former.

\myskip

\paragraph{Identifying the privileged component.} A very useful tool in our algorithms is to identify, {for every given $X$}, a {\sl unique} connected component among those of $G \setminus X,$ which we call the \emph{privileged component}, that contains ``most'' of the wall $W_3.$
Let us formalize a bit this idea. For a positive integer $q,$ a \emph{pseudogrid} ${\bf W}_q,$ defined in~\cite{KunOPS21polyn}, is a collection of $q$ ``vertical'' and $q$ ``horizontal'' paths that intersect in a ``grid-like'' way, as illustrated in \autoref{fig_pseudo}. Note that the considered wall $W_3$ naturally defines a (large, as a function of $θ$) pseudogrid. A connected component  $C$ of a graph $G$ is \emph{privileged} with respect to a set $X \subseteq V(G)$ and a pseudogrid ${\bf W}_q$ if $C$ is a connected component of $G\setminus X$ that contains entirely at least one vertical and one horizontal path of ${\bf W}_q.$ It is easy to see (\autoref{@vuitcentistes}) that such a privileged component, if it exists, is unique.

Moreover, when $X$ is a modulator, the fact that ${\sf torso}(G,X)$ has bounded treewidth implies that every connected component of $G\setminus X$ has a ``small interface'' to $X$ and thus the flat wall $W_0$ (and any large subwall of it) is not significantly ``damaged'' by $X,$ which we formalize via the notion of having small \emph{bidimensionality} (see \autoref{@congregation}). Intuitively (see~\autoref{sec_bidimensionality} for the definition), this means that  $X$ intersects a small number of so-called ``bags'' of the wall. Informally, the \emph{bags} of a wall $W$ in a graph $G$ with apex set $A$ define a partition of $G \setminus A$ into connected sets, such that each bag, except the external one, contains the part of the wall $W$ between two neighboring degree-3 vertices of the wall, as illustrated in \autoref{label_bitternesses} (see~\autoref{sec_canonical} for the definition).
This property is used extensively in the proofs and, in particular, it defines, assuming the existence of a large flat wall $W_0$ and a modulator $X,$ a {\sl unique} privileged component $C$  in $G \setminus X$ (regardless of the $\circ/\bullet$-flag).
 In our sentences, in order to identify such a component, we need to integrate the ``recognition''  of a pseudogrid ${\bf W}_q$ and its associated privileged component with respect to a modulator $X$:
 it is easy to see that these properties can be defined in \MSOL\ (see~\autoref{@pejroanlmaaragall}).
\myskip

\paragraph{Splitting the sentence $θ_{{\sf R},{\bf c}}.$}
The existence of a privileged component $C$ allows us to see    the sentence $θ_{{\sf R},{\bf c}}$ as a conjunction of two subsentences: one that concerns  the privileged component $C$ (where we will find the irrelevant vertex) and another one concerning the modulator $X$ and the other (non-privileged) components of $G \setminus X.$ 
Namely, in \autoref{@reconquistasen}
we define a sentence $\tilde{θ}_{{q}},$ called the {\em split version} of~$θ_{{\sf R},{\bf c}},$ that allows us to ``break'' $θ$ into two questions: one denoted by $θ^{\sf out}_{{q}}$ that is the conjunction of the
modulator sentence $β$ and the target sentence $γ$ in the non-privileged components of $G \setminus X$ and another one that concerns the target sentence $γ$ in the privileged component $C.$ This latter question is composed of two subsentences (see~\autoref{@desembarazaron}), namely one about the satisfaction of the \FOL-sentence $σ$ and another one about the minor-exclusion given by~$μ.$ Given this decomposition of $θ$ into three questions (one ``external'' and two ``internal'' ones), our ``irrelevancy'' arguments also decompose into three parts. Concerning the ``irrelevancy'' for minor-exclusion, as discussed above, the fact that the whole wall $W_2$ is irrelevant with respect to~$μ$ allows us to focus on the other two questions.  For this, we need to define the {\sl characteristic} of a wall with respect to $θ,$ denoted by $θ\text{-}{\sf char}$ (see~\autoref{eq_char_first-floor}).
This characteristic is composed of two parts: the \emph{out-signature} (see \autoref{sec_out-sig_first-floor}) corresponding to the satisfiability of the sentence $θ^{\sf out}_{{q}},$ and the \emph{in-signature} (see \autoref{sec_in-sig_first-floor}) corresponding to the \FOL-sentence $σ.$
Let us now explain how we define the out-signature and the in-signature, and sketch why we can eventually declare a subwall irrelevant.

\myskip

\paragraph{Defining the out-signature of a wall.} Dealing with the irrelevancy with respect to the ``external'' sentence $θ^{\sf out}_{{q}}$ turns out to be the most interesting part of the proof of \autoref{@desmembramientos}, and we introduce several ideas which are, in our opinion, one of the main conceptual contributions of this article.  The goal is, for each wall $W$ in the collection ${\cal W},$ to encode all the necessary information that concerns the satisfiability of $θ^{\sf out}_{{q}}$ in the ``non-privileged'' part of the graph and the modulator~$X.$ To do this, for each $W ∈ {\cal W}$ with apex set $A,$ we define a set of {\sl $\ell$-boundaried graphs} (i.e., graphs in which $\ell$ ``boundary'' vertices are equipped with labels), constructed as we describe below, and where $\ell$ depends only on $θ.$
The boundary corresponds to where the sentence has been  ``split'' and we need to ``guess'' how to complement this boundary by the part of the modulator that is not inside the wall. Note that, since $θ^{\sf out}_{{q}}$ is a \MSOL-sentence, by a variant of Courcelle's theorem for boundaried graphs~\cite{Courcelle90them,Courcelle92,Courcelle97} (see~\autoref{cou_more}), there exists a {\sl finite} collection
${\sf rep}^{(\ell)}(θ^{\sf out}_{{q}})$ of sentences on $\ell$-boundaried graphs that are ``representatives'' of the sentence $θ^{\sf out}_{{q}}$ and that can be effectively constructed. We next described how these $\ell$-boundaried graphs are constructed.

We observe that, by \autoref{@congregation} (which uses the bounded-treewidth property of the modulator sentence $β$), there exists a ``buffer'' $I$ in $W,$ consisting of a set of consecutive layers of the wall, which is disjoint from a hypothetical modulator $X.$  We guess with an integer $d$ where this ``buffer'' $I$ is placed in the wall and we denote its inner part by $I^{(d)}.$ This naturally induces a partition of $X$ into $X_{\sf in}$ and $X_{\sf out},$ with $X_{\sf in}$ being the part of $X$ that is inside $I^{(d)}$ (see~\autoref{figure_xinxout}). We also guess which subset of the apex set $A$ will belong to the modulator $X$ and we denote it by $V_L ({\bf a}),$ where $L$ is the set containing the indices of the corresponding apex vertices. Since parts of the ``non-privileged'' vertex set of the graph may lie outside the considered wall,  we need to guess the part of the modulator (namely, its boundary towards the component) that lies outside the wall. More precisely, we need to guess as well
which subset $F'$ of $X_{\sf out},$ other than $V_L ({\bf a}),$ will belong to the neighborhood of the privileged component. This is achieved by guessing all ways an (abstract) graph $F'$ with a bounded number of vertices can extend the boundary (see~\autoref{figure_extension}). We let $F$ be the graph obtained from the union of $V_L ({\bf a})$ and $F'.$ Finally, we also need to consider a set $Z$ that corresponds to $X_{\rm in}$ together with the part inside $I^{(d)}$ that has been ``chopped off''
by the modulator $X,$ that is, the part of $W$ inside $I^{(d)}$ that will not belong to the privileged component after the removal of the modulator $X.$
We denote by $\partial(Z)$ the set of vertices in $Z$ that have a neighbor in $I^{(d)}.$
Altogether, these guesses result in the $\ell$-boundaried graph $K^{(d,Z,L,F)}$ obtained from the graph induced by $I^{(d)}$ and the set $F,$ whose boundary is the set $\partial(Z) \cup F$;
see \autoref{@incomportable} and \autoref{@governorship} for an illustration of $K^{(d,Z,L,F)}$ and of how this graph lies with respect to the privileged component $C,$ respectively.

With each such a guess $(R,d,L,Z)$ we associate the out-signature defined as follows and denoted by {\sf out\mbox{-}sig} (see \autoref{eq_out-sig_first-floor}). Its elements are pairs $({\bf H},\bar{θ}),$ where ${\bf H}$ encodes how the set $V_L ({\bf a})$ in the boundary has been extended by the ``abstract'' graph $F',$ and $\bar{θ} ∈ {\sf rep}^{(\ell)}(θ^{\sf out}_{{q}})$ prescribes the equivalence class, within the set of Courcelle's representatives mentioned above, of the considered $\ell$-boundaried graph. This concludes the description of the out-signature.

While this out-signature indeed encodes the behavior of the considered wall with respect to the ``external'' sentence $θ^{\sf out}_{{q}},$ a crucial issue has been overlooked so far: in order to be able to identify an irrelevant subwall inside the collection ${\cal W} $ within the claimed running time, we need to be able to {\sl compute} the (in- and out-) signature of a wall in linear time. To do this using Courcelle's theorem, we need to consider a graph that has treewidth bounded by a function of $θ.$
Recall that $θ^{\sf out}_{{q}}$ is the conjunction of the modulator sentence $β$ (which is evaluated in the graph ${\sf stell}(G,X)$) and the target sentence $γ$ in the ``non-privileged'' components of $G \setminus X.$ By the condition guaranteed by \autoref{label_proletarians} discussed in the paragraph above \autoref{figure_walls_zoom}, we have that the treewidth of  $W$ is bounded by a function of $θ,$
hence the treewidth of the $\ell$-boundaried ``subwall''  $K^{(d,Z,L,F)},$ for which we want to compute the out-signature, is also bounded
by a function of $θ.$ However, the graph $K^{(d,Z,L,F)}\setminus V(F)$ ``lives'' inside the {\sl whole} privileged component $C,$ and we cannot guarantee that the treewidth of $C$ is bounded by a function of $θ.$ We overcome this problem with the following trick, which is an important tool in the proof of \autoref{claim_1}. We observe that the satisfaction of $θ^{\sf out}_{{q}}$ is preserved if, instead of the whole privileged component $C,$ we consider the graph  $K^{(d,Z,L,F)} ,$ which is obtained
by ``shrinking'' $C$ to the subwall $I^{(d)},$ and which  has bounded treewidth as we need (compare the left part of \autoref{figure_boundariedgraph1} with \autoref{figure_boundariedgraph2}). Indeed, this modification does not change any of the non-privileged components in which the target sentence $γ$ is evaluated and, by adding edges from the ``guessed extended boundary'' $F'$ to $I^{(d)}$ in order to preserve connectivity (see~\autoref{@incomportable}), the resulting graph ${\sf stell}(G,X)$ remains unchanged with this transformation, and therefore
the satisfaction of the modulator sentence $β$ is also preserved.
\myskip

\paragraph{Defining the in-signature of a wall.} To deal with the irrelevancy with respect to the \FOL-sentence~$σ,$ we use arguments strongly inspired by those of~\cite{FominGST20analgo}. The core tool here is Gaifman's locality theorem (see \autoref{lem_gaifman}), which states  that every \FOL-sentence $σ$ is a Boolean combination of basic local sentences $σ_1,\ldots,σ_p,$ in the sense that the satisfaction of each $σ_i$ depends only on the satisfaction of a set of sentences $ψ_1, \ldots, ψ_{\ell_i}$ evaluated on single vertices that can be assumed to be pairwise far apart (see \autoref{sec_equivalent-versions_first-floor}). As discussed before, taking care of the domain of these vertices is the main reason why we consider a annotated version of the problem, corresponding to the enhanced sentence $θ_{{\sf R},{\bf c}}.$
Extending the approach of~\cite{FominGST20analgo} (which does not deal with apices), the in-signature of a wall, denoted by {\sf in\mbox{-}sig}, encodes all (partial) sets of variables, one set for each basic local sentence of the so-called  Gaifman sentence $\breve{σ},$ such that these variables lie inside an ``inner part'' of the wall, they are scattered in the ``apex-projection'' of this inner part, and they satisfy the local sentences $ψ_i$; see \autoref{eq_in-sig_first-floor}
for the formal definition.

\myskip

\paragraph{Declaring a subwall irrelevant.} We now sketch the remaining of the proof of \autoref{@desmembramientos} for sentences in $\bar{Θ}_1,$ presented in \autoref{sec_proof_correctness} (see~\autoref{figu_levels_1and2}).
As mentioned above, suppose that we have already found, inside the collection ${\cal W},$ a large (as a function of $θ$) subcollection ${\cal W}' \subseteq {\cal W}$ of walls all having the same $θ$-characteristic. We pick one of these walls, say $W^{\star} ∈ {\cal W}',$ and we declare its central part irrelevant (see \autoref{figure_walls_zoom}). We need to prove that, if the input graph $G$ satisfies $θ,$ then the graph $G'$ obtained from $G$ by removing the central part of  $W^{\star},$ also satisfies $θ.$ That is, given a modulator $X$ in the original instance $G,$ we need to construct another set $X' \subseteq V(G)$ that is disjoint from $W^{\star}$ and that is a modulator in $G'.$  For this, we proceed as follows.


The cardinality of ${\cal W}'$ and the fact that $X$ intersects few bags of the wall $W_3$ (see~\autoref{@congregation}) imply that there exists a large (again, as a function of $θ$) subcollection ${\cal W}'' \subseteq {\cal W}'$ of walls that are disjoint from $X.$ We take such a wall $\hat{W} ∈ {\cal W}''$ and, using the fact that $W^{\star}$  and $\hat{W}$ have the same $θ$-characteristic, we show that we can ``replace'' the part of the modulator $X$ that intersects $W^{\star}$ with another part in $\hat{W}$ (see~\autoref{figure_boundariedgraph5}), together with an alternative assignment of variables that satisfies the corresponding sentences. This results in another set $X'$ that is a modulator in $G',$ hence yielding the annotation irrelevancy of (the central part of) $W^{\star}.$

Showing these facts is far from being easy and we need a number of technical details
that are structured into three parts, corresponding to \autoref{claim_1}, \autoref{claim_2}, and \autoref{claim_3}. Each of these claims deals, respectively, with the irrelevancy with respect to $θ^{\sf out}_q$ (which incorporates $β),$ $σ,$ and $μ.$ In particular, an important idea in the proof of \autoref{claim_1} is that, changing from $X$ to $X',$ we obtain a new boundaried graph, which is in fact the same graph but with a new boundary (see~\autoref{figure_boundariedgraph5}). In the proof of \autoref{claim_2}, the replacement arguments for the in-signature work because of the aforementioned distance-preservation property of the apex-projection.

\myskip

\mysubsection{How to deal with the general logic $Θ$}
\label{sec_overview_3}
\myskip

In this subsection we sketch how to generalize the ideas presented in \autoref{sec_overview_2} to the general compound logic $Θ.$ We do this in two steps, corresponding to \autoref{the_second_level} and \autoref{sec_final_combo}, respectively.
\myskip\myskip\myskip\myskip\myskip

\paragraph{A less particular case: allowing for recursion.} Once the fragment $\bar{Θ}_1$ of $Θ$ is proved, the next step is to consider  the fragment $\bar{Θ}$ presented in  \autoref{the_second_level}. Namely,  the problem is defined by a sentence $θ∈\bar{Θ}$ composed of $r$ sentences $θ_1, \ldots, θ_r$ with $r$ modulator sentences $β_1, \ldots, β_r$ and one target sentence $θ_0 = σ\wedge μ,$ which are defined recursively. That is, starting with $i=r,$ the sentence $θ_i,$ evaluated on a (dynamic) graph $G,$ asks for the existence of a vertex set $X_i \subseteq V(G)$ that satisfies the modulator sentence $β_i$ and such that either each connected component of $G \setminus X,$ or the whole graph $G \setminus X,$ satisfies the ``next'' sentence $θ_{i-1}$ (see~\autoref{eq_bis}).

The strategy of the proof is essentially the same as in the previous case, and the main extra technical issue is to deal with what we call the $\circ/\bullet$-\emph{scenarios}, which capture whether, in each level of the recursion defined by $θ_i,$ the next sentence $θ_{i-1}$ needs to be satisfied by each connected component of $G \setminus X,$ or by the whole graph $G \setminus X.$ Such a scenario $w$ of length $r$ gives rise to, instead of a privileged component $C,$ to the notion of \emph{$w$-privileged sequence} $C_1,\ldots,C_{r+1}$ with respect to a pseudogrid ${\bf W}_q$ and a collection ${\cal X} = \{X_1,\ldots, X_r\}$ of pairwise disjoint subsets of $V(G).$ A $w$-privileged sequence contains, for $i ∈ [r],$ the privileged component $C_i$ (which is again unique) of the ``current'' graph after the removal of $X_i \cup \ldots \cup X_r,$ defined according to the scenario $w.$ Note that $C_1\subseteq \ldots \subseteq C_{r},$ and we call $C_1$ the \emph{$w$-privileged set}. See~\autoref{sec_scenarios} for the details and \autoref{@corazoncillo} for an illustration of a $\circ/\bullet$-scenario. It is worth observing that a $w$-privileged set is a, possibly disconnected, graph (see~\autoref{@corazoncillo}), and in the proof we always work in the privileged component with respect to ${\bf W}_q$ and ${\cal X}.$

It is again easy to see that the notion of $w$-privileged set can be defined in ${\MSOL}$ (see~\autoref{@assimilation}), which allows to integrate this information in our sentences. We then split the sentence $θ_{{\sf R},{\bf c}}$
 in a similar way as we did in the previous case (see~\autoref{@gaspilleront}), separating the questions that concern the $w$-privileged sequence, in which we aim to find an irrelevant vertex, namely in the $w$-privileged set $C_1$ that contains the large portion of the wall $W_3,$ as illustrated in \autoref{@corazoncillo}.

The proof of \autoref{@desmembramientos} for the case $θ ∈ \bar{Θ}$ follows closely the one for $θ ∈ \bar{Θ}_1$ discussed in \autoref{sec_overview_2}, and is summarized in \autoref{sec_what_3_sec} and presented in detail in \autoref{second_level_more}. One of the main differences is that, when defining the out-signature of a wall (see \autoref{@enthaltenden}), the boundaried graph $K^{(d,{\cal Z},{\cal L},{\cal F})}$ that we consider (see \autoref{fig_contr_mult}) is defined with respect to {\sl collections} ${\cal Z},{\cal L},{\cal F},$ in order to deal with the boundaries that are created recursively by the sentence $θ.$ The roles of \autoref{claim_1}, \autoref{claim_2}, and \autoref{claim_3} is now replaced by \autoref{claim_4}, \autoref{claim_5}, and \autoref{claim_6}, respectively (see \autoref{figu_levels_1and2}).
\myskip

\paragraph{The general case: allowing for Boolean combinations.} Finally, we deal in \autoref{sec_final_combo} with the general logic $Θ,$ corresponding to \autoref{@decendientes1}. The difference with respect to the previous case is that now, within each level of the recursion, we allow for Boolean combinations of sentences in a lower level (see~\autoref{sec_ourlogic}). Since the strategy of the proof is again the same as in the previous cases, in \autoref{sec_final_combo} we do not repeat it again, and we rather focus on the local changes that need to be done with respect to the proof presented so far, which mostly concern the definition of the in-signature and the out-signature of a wall, and provide  a sketch of how to integrate these changes in the previous proof.

Our approach first considers a restricted version of $Θ,$ the logic $\hat{Θ}$ defined in~\autoref{@determinadas} where the only positive Boolean combination of sentences that we allow in each recursive level  is the {\sl conjunction} of a finite number of sentences.
Then, in~\autoref{@insinuations} we show how to insert also disjunctions to our arsenal of
positive Boolean combinations, in order to achieve the generality of~$Θ.$
This approach is based on the fact (see~\autoref{@inexpressibly}) that every sentence $φ$ that is a positive Boolean combination of some set of sentences $Φ$ has an equivalent sentence $φ'$ that is a disjunction of conjunctions of sentences in $Φ.$

In \autoref{@determinadas}, the idea is to associate each sentence $θ∈\hat{Θ}$  with a rooted tree expressing its recursive definition, where the root corresponds to $θ$ and every conjunction to a bifurcation of the tree.
Under the presence of a large enough pseudogrid, we  use this tree to define an equivalent version of the problem, where $θ$ is ``focused'' towards the privileged connected component occurring each time in the leaves of the tree.
Under this scope, every root-to-leaf path of the tree  corresponds to a sentence in $\bar{Θ}.$
Performing this modification, we have to keep track of the bifurcations of the tree and ask the modulators that correspond to each such a bifurcation to be the same sets in all paths that contain this ``bifurcated'' node.
These equalities  have to be respected when searching for an equivalent ``solution-certificate'' that comes with the application of the irrelevant vertex technique.
For this reason, we have to (further) modify the definition of signatures and characteristics given in~\autoref{@enthaltenden} so as to add one ``extra dimension'' to them (corresponding to the swift from the ``path-like'' structure of sentences in $\bar{Θ}$ to the ``tree-like'' structure of sentences in $\hat{Θ}$), while respecting the equalities obtained from above.
At the end of \autoref{@determinadas} we sketch how to prove \autoref{@desmembramientos} for a sentence in~$\hat{Θ}.$

In~\autoref{@insinuations}, when disjunctions are allowed,  we define a notion of a \emph{conjunctive scenario}~${\cal S}$ of a sentence $θ∈ Θ$ (expressed in terms of a tree representation as in \autoref{@determinadas})
and prove (see~\autoref{@envelopperait}) that $θ$ is satisfied by a graph $G$ if and only if $G$ satisfies at least one of the conjunctive scenarios of $θ.$ This results in a redefinition of the characteristic of a wall with respect to $θ,$ as the product of the characteristics of all these possible scenarios (see~\autoref{eq_product_char}). Then, we follow the same arguments as sketched at the end of~\autoref{@determinadas}.
In the current case, when considering a collection of $θ$-equivalent walls, following the definition of the $θ$-characteristic given by \autoref{eq_product_char}, these walls are
$θ_S$-equivalent for {\sl every} conjunctive scenario $S∈ {\cal S}.$
Therefore, when we find a part of a wall and declare it ``irrelevant'', it is ``irrelevant'' for any possible conjunctive scenario of $θ,$ and the proof is complete.

%
%
\myskip

\myskip\section{Applications}
\label{sec_applications}

Before we proceed to a discussion on the consequences  of \autoref{@decendientes1} (stated as \autoref{@decendientes} {for general structures})
Let ${\bf t}=(x_{1},\ldots,x_{l})∈ \N^l$ and $\chi,ψ: \N
\rightarrow \N.$
We say that $\chi(n)=\O_{{\bf t}}(ψ(n))$ if there exists a computable
function $φ:\N^{l} \rightarrow \N$
such that  $\chi(n)=\O(φ({\bf t})\cdot ψ(n)).$
Given a graph $G,$ we define its {\em size} $|G|$ as the number of its vertices and when we present running times of algorithms we always use $n=|G|.$
Given a finite set of graphs ${\cal H},$ we denote by $|{\cal H}|$ the biggest size of a graph in ${\cal H}.$ Given a graph class ${\cal G}$
we define its {\em minor-obstruction} set (or simply, {\em obstruction set}) as the set  $\obs({\cal G})$ of minor-minimal graphs not in ${\cal G}.$ According to  Robertson-Seymour's theorem, $\obs({\cal G})$ is a finite set for every graph class ${\cal G}$ (see~\autoref{@insuperables}).


\myskip\subsection{Vertex removal problems}
\label{sec_applications_vertex-removal}

 In this subsection we consider several modification problems based on vertex removals.

\myskip\paragraph{Classic modification problems.}

The first wide family of modification problems modeled by
$Θ_{1}$ asks whether the {\em vertex removal distance} of a graph $G$ to  ${\cal G}$ is at most $k,$ that is,
whether there is set of at most $k$ vertices in $G$ whose removal gives a graph in ${\cal G}.$
 In many cases, ${\cal G}=\excl({\cal H})$ where ${\cal H}$ is some (finite) set of graphs
and/or satisfies some \FOL-definable property. In the case where  the  \FOL\ demand is void, this problem admits a time $\O_{k,|{\cal H}|}(n^{2})$ algorithm because of \autoref{@circumscribed}, as its \yes-instances are minor-closed for every $k.$ Instantiations of this general problem
occupied a lot of research in the last years~\cite{JansenLS14anea,SauST21kapiI,AdlerGK08comp,MarxS07obta,Kawarabayashi09}.
However, for arbitrary \FOL\ properties, the
\yes-instances of this problem are not minor-closed anymore and cannot be treated by any of Propositions~\ref{@originallypublishedin},~\ref{@universities}, or~\ref{@circumscribed}. The first result
 in this direction appeared in~\cite{FominGST20analgo} and treats  the case where ${\cal H}=\{K_{5},K_{3,3}\},$ i.e., when the target property
 is being planar and satisfying some \FOL\ property. Therefore, \autoref{@decendientes1} constitutes a far-reaching generalization of the results of \cite{FominGST20analgo} in the following sense: while it extends the tractability of modification problems to a much more general property than planarity, it also is applicable to a much broader class of  modulators. We note that the very recent results of~\cite{GolovachST22model,GolovachST22model_arXiv} can also treat certain modulators of bounded size on minor-closed graph classes.

\myskip\paragraph{Alternative measures of the modulator.}
In a recent wave of results, alternative quality measures of the ``modulator'' to some graph class were considered, other than just its size \cite{AgrawalKLPRS21elim,BulianD16graph,BulianD17fixe,BougeretJS20bridg,DLindermayrSV20elimi,EibenGHK21,FominGT22param,SridharanPASK21anftp,JansenK021verte}.
The first step in this direction was
done by Bulian and Dawar~\cite{BulianD16graph,BulianD17fixe}
who considered the {\em elimination distance} to some  graph class ${\cal G}.$  We  assume that the target graph  class is ${\cal G}=\Mod(γ_0),$ for some $γ_0∈\MSOL.$
Also, for $k≥ 1,$ let   $γ_{k}=β\triangleright γ_{k-1},$ where $β$ demands than ${\sf torso}(G,X)$
is edgeless.
Using our notation, a graph $G$ has {\em elimination distance} to ${\cal G}$ at most $k$ if
$G\models γ_{k}.$
Alternatively, one may observe that $γ_k=β_k\triangleright γ_0$ if $β_k$ expresses the fact that ${\sf torso}(G,X)$  has  tree-depth
at most $k.$
If ${\cal G}=\excl({\cal H}),$ where
${\cal H}$ is a finite set of graphs, Bulian and Dawar~\cite{BulianD17fixe}
proved that
the corresponding modification problem
can  be solved by a (constructive) algorithm running in  time  $\O_{k,|{\cal H}|}(n^{2}).$  Elimination distance to ${\cal G}$  has been also studied in~\cite{FominGT22param} for the case where ${\cal G}$ is some \FOL-definable graph class. {We note that the result in~\cite{PilipczukSSTV22algor} implies that  deciding whether the elimination distance to some \FOL-definable graph class is at most $k$ can be done in time $\O_{k,|{\cal H}|}(n^{3})$ on ${\cal H}$-topological-minor-free graphs.}

An alternative elimination distance measure was recently given by the parameter of {\em bridge-depth} to ${\cal G}.$ Using our notation, a graph $G$ has {\em bridge-depth} to ${\cal G}$ at most $k$ if $G∈\Mod(γ_k)$ where $γ_{k}$ is
is defined as above with the difference that now $β$ demands that   $\torso^+(G,X)$ is acyclic.\footnote{\label{ftmlpo}The graph $\torso^{+}(G,X)$ is defined as $\torso(G,X)$ with the difference that now we do not remove the contracted vertices $v_{C}$ in the end.}
Bridge-depth was introduced in \cite{BougeretJS20bridg} for the case where the target graph class ${\cal G}$ is edgeless and has been used on the the study of the existence of polynomial kernels for structural parameterizations.

Another modulator measure is the notion of \emph{${\cal G}$-treewidth}, recently introduced by Eiben, Ganian, Hamm, and Kwon \cite{EibenGHK21}. Having  ${\cal G}$-treewidth at most $k$ is equivalent to asking  that the ${\sf torso}(G,X)$ has treewidth at most $k$ and the target property is containment in ${\cal G}.$ Jansen, de Kroon,  and Włodarczyk proposed in~\cite{JansenK021verte} a time $\O_{k,|{\cal H}|}(n^{\O(1)})$  algorithm for this problem in the case when ${\cal G}=\excl({\cal H}),$ for some finite set of graphs
${\cal H}.$ The algorithms in~\cite{JansenK021verte} are strongly based on the computation of important separators~\cite{Marx06param} and the contribution of $k,|{\cal H}|$ on their running times is explicit.

Recently,  Agrawal,  Kanesh,  Lokshtanov,  Panolan, Ramanujan,  Saurabh, and  Zehavi proved in~\cite{AgrawalKLPRSZ22delet} that, under certain assumptions on ${\cal G},$  ${\cal G}$-treewidth,  {elimination distance} to ${\cal G},$ and vertex deletion to ${\cal G}$ are all  {\sf FPT}-equivalent. The techniques used in~\cite{AgrawalKLPRSZ22delet} are heavily based on the meta-algorithmic result of~\cite{LokshtanovR0Z18redu} (based on the recursive understanding technique) that is non-constructive and implies  {\sf FPT}-algorithms whose running time is worse than quadratic. \smallskip

%
%

%
%

All   problems above are  $Θ$-definable even if, apart from asking containment in $\excl({\cal H}),$ we additionally impose some \FOL-definable demand to the target property (or even a demand definable in $Θ$ itself).
To formalize this, in what follows we give a theoretical framework that comprises, by \autoref{@decendientes1}, all the aforementioned measures on the modulator.

\myskip\paragraph{A parametric variant of our results.}
A {\em graph parameter} is a  function $\p:{\cal G}_{\rm all}\to\mathbb{N}.$ We say that $\p$ is {\em  treewidth-bounded}
if there is a function $f:\mathbb{N}\to\mathbb{N}$ such that
for each $G∈{\cal G}_{\rm all},$  $\p(G)≤ f({\sf tw}(G)).$
We say that $\p$ is {\em \MSOL-definable}
if for every $k∈\mathbb{N}$ there is a \MSOL-sentence (on graphs) $β_{k}$
such that the set of all models of $β_k$ is  $\Mod(β_{k})=\{G\mid \p(G)≤ k\}.$
Clearly, if $\p$ is treewidth-bounded then we can also
assume that each $β_{k}$ is a sentence in $\MSOL^{\sf tw}$ and in this case we say that $\p$ is $\MSOL^{\sf tw}$-definable.
There are several known graph parameters that are $\MSOL^{\sf tw}$-definable, such as treewidth, pathwidth, tree-depth, bridge-depth, block tree-depth, vertex cover, feedback vertex set, branch-width, carving-width, or cutwidth.

For a graph parameter $\p$ and a graph class ${\cal G},$
we define the new graph parameter $\p_{\cal G}:{\cal G}_{\rm all}\to\mathbb{N}$
such that
\begin{eqnarray}
\p_{\cal G}(G) & =  & \min\{k\mid \exists X\subseteq V(G)\mid \p(\torso(G,X))≤ k~\wedge~ G\setminus X∈ {\cal G}\}.
\label{@successfully}
\end{eqnarray}
Thus $\p_{\cal G}$  measures by $\p$ the quality of a modulator $X$ to property   ${\cal G}.$ For example, when  $\p$ is the size of the modulator, then  this is just the {\sl vertex deletion distance to ${\cal G}$}, that is,  the minimum number of vertices $X$ such that $G\setminus X∈ {\cal G}.$
When $\p$ is the tree-depth of a graph, then  $\p_{\cal G}$ is the elimination distance to ${\cal G}.$ Or when $\p$ is the treewidth of a graph, then  $\p_{\cal G}$ corresponds to ${\cal G}$-treewidth.
We consider the general setting  where $\p$ is a $\MSOL^{\sf tw}$-definable graph parameter and ${\cal G}$ is a $Θ$-definable graph class, that is, $\Mod(θ)={\cal G}$ for some $θ∈ Θ.$ By setting  $θ_{k}=β_k\triangleright θ∈ Θ,$ we have that
 $\Mod(θ_k)=\{G\mid \p_{\cal G}(G)≤ k\}.$ Then the following theorem is a direct consequence of~\autoref{@decendientes1}
 and~\autoref{@pertrechando1}.

\begin{theorem}
\label{@consequential}
Let $\p$ be a $\MSOL^{\sf tw}$-definable graph parameter and ${\cal G}=\Mod(θ)$ for some $θ∈ Θ.$  Then there is an algorithm
that, with input a graph $G$ and   $k∈ \mathbb{N},$ checks whether $\p_{\cal G}(G)≤ k$  in time $\O_{k,|θ|}(n^2).$
Moreover, if  ${\cal G}=\Mod(\tilde{θ})$ for some $\tilde{θ}∈ \tilde{Θ},$
then there is an algorithm
that, with the same input, checks whether $\p_{\cal G}(G)≤ k$  in time $\O_{k,|θ|,{\sf hw}(G)}(n^2).$
\end{theorem}

All the results mentioned in this subsection, in what concerns minor-excluded graphs, are subsumed
by \autoref{@consequential}. Moreover, by allowing \FOL-definability in the target sentence and \mbox{$\MSOL^{\sf tw}$}-definability in the modulator sentence, we vastly extend  \autoref{@circumscribed}
to graph classes and parameters that are not necessarily minor-closed or  hereditary. {We stress that none of the results in~\cite{PilipczukSSTV22algor,GolovachST22model} is able to deal with the problems captured by \autoref{@consequential} in their full generality.}

\myskip\subsection{Other variants of modifications problems}
\label{sec_applications_other_variants}

\autoref{@decendientes1}
deals  with  vertex modulators. On the other hand,  there is a lot of literature on graph modification problems involving several types of local operations (see e.g., \cite{CrespelleDFG2020surve} for a survey).
\autoref{@decendientes1} in its full model-theoretic power is stated on structures   (see \autoref{@decendientes}  in \autoref{our_logic_stories}). This permits us to
encode way more entangled modification operations. For instance, one may consider annotated graphs and modification in several steps with different constraints, including colored graphs  and demands on the distribution on the colors in the modulator and the target part.
The systematic study of the power of definability of  $Θ$ goes beyond the scope of this paper. However, in what follows we give  three  indicative examples
of problems that can be expressed via reductions to structures.

\myskip\paragraph{A grammar for modification problems.}
A  compact way to extend the applicability of  \autoref{@decendientes} is by defining the context-free language ${\cal M}$  for graph modifications.
We  define ${\cal M}$ as the context-free language  generated by  specific rules. The alphabet of ${\cal M}$ consists of all sentences of $Θ_0[\{{\sf E}\}]$ plus nine additional letters  $\{{\sf M},{\sf F},{\sf (},{\sf )}, \vee, \wedge, {\sf n},{\sf e},{\sf c}\}.$
We use  ${\sf n},$ and ${\sf e}$
to encode vertex removal, and edge removal, respectively. We also need  ${\sf c}$
to work with connected components.  ${\cal M}$ is generated by the following rules. The terminal strings of ${\cal M}$ are the sentences in $ Θ_0[\{{\sf E}\}],$ thus
 ${\sf F}\rightarrow  Θ_0[\{{\sf E}\}].$ The rest of the production rules are:   ${\sf M}\rightarrow ({\sf F}),$ ${\sf M}\rightarrow ({\sf n}{\sf M}),$ ${\sf M}\rightarrow ({\sf e}{\sf M}),$
${\sf M}\rightarrow ({\sf c}{\sf M}),$ ${\sf M}\rightarrow ({\sf M}\wedge {\sf M}),$ and  ${\sf M}\rightarrow ({\sf M}\vee {\sf M}).$
For each  string ${\sf w}∈ {\cal M},$
we define the class of graphs ${\cal G}_{\sf w}$ as follows.

\begin{itemize}

\item If $|{\sf w}|=1$  then ${\cal G}_{\sf w}=\Mod({\sf w}).$

\item If ${\sf w}=({\sf n} {\sf w}'),$  then ${\cal G}_{\sf w}=\{G\mid \exists v∈ V(G), G\setminus v∈ {\cal G}_{{\sf w}'}\}.$

\item If ${\sf w}=({\sf e} {\sf w}'),$  then ${\cal G}_{\sf w}=\{G\mid \exists e∈ V(G), G\setminus e∈ {\cal G}_{{\sf w}'}\}.$


\item If ${\sf w}=({\sf c} {\sf w}'),$  then ${\cal G}_{\sf w}=\{G\mid \forall C∈ {\sf cc}(G),\ C∈ {\cal G}_{{\sf w}'}\}.$

\item If ${\sf w}=({\sf w}'\vee {\sf w}''),$  then ${\cal G}_{\sf w}={\cal G}_{{\sf w}'}\cup{\cal G}_{{\sf w}''}.$

\item If ${\sf w}=({\sf w}'\wedge {\sf w}''),$  then ${\cal G}_{\sf  w}={\cal G}_{{\sf w}'}\cap{\cal G}_{{\sf w}''}.$
\end{itemize}
\noindent
 Let us remind that we use $G\setminus v$ for vertex deletion and  $G\setminus e$ for edge deletion.

If in the definition of ${\cal M}$ we demand that terminal strings of ${\cal M}$ are the sentences in $\FOL,$ then we define the context-free language  $\tilde{\cal M}.$  That way, the definition of ${\cal G}_{\tilde{\sf w}}$ can also be extended for every string $\tilde{\sf w}∈ \tilde{\cal M}.$
By making use of \autoref{@decendientes}, we prove the following theorem about modifications defined by ${\cal M}.$

\begin{theorem}\label{thm_grammar}
For every ${\sf w}∈ {\cal M},$ ${\cal G}_{\sf w}$ is decidable in quadratic  time. 
Moreover, for every $\tilde{\sf w}∈ \tilde{\cal M},$ ${\cal G}_{\tilde{\sf w}}$ is decidable in quadratic time on graphs of bounded Hadwiger number.
\end{theorem}

\begin{proof}
We start with the proof of the first part of the theorem.
Let $c$ be the maximum Hadwiger number of the models of the target sentences of ${\sf w}.$

The proof is based on a transformation of $G$ to a doubly annotated graph $(G',R,B)$
and the construction of some $θ∈ Θ[\{{\sf E},{\sf X},{\sf X}'\}]$
such that $G∈ {\cal G}_{\sf w}$ if and only if $(G',R,B)\models θ.$ This fact, together with~\autoref{@decendientes}
and the fact that the transformation that we will give can be done in linear time, yield the correctness of the theorem.

The transformation of $G$
creates a doubly annotated graph $(G',R,B)$ whose  annotated vertices  make possible to simulate the edge removals by vertex removals.
The new graph $G'$ is constructed by subdividing once each  edge $e$ of $G$ (we call $v_{e}$ the subdivision vertex) and for each vertex $v$ of $G,$
we construct a clique $K_{v}$ on $c+1$ vertices and identify one of its vertices with $v$
  We also define  $R=\bigcup_{v∈ V(G)}V(K_{v}\setminus\{v\})$ and $B=\{v_{e}\mid e∈ E(G)\}.$ We call the vertices of $R$ (resp. $B$) {\em red vertices} (resp. {\em blue vertices}) of $G'$ and we call the rest of the vertices {\em white vertices} of $G'.$

We next proceed with the definition of $θ∈ Θ[\{{\sf E},{\sf X},{\sf X}'\}].$ Keep in mind that in $G'$ each blue vertex $v_{e}$ of $G'$ corresponds to  an edge $e∈ E(G)$  and each ``red-white'' clique $K_{v}$ of $G'$ corresponds to a vertex  $v∈ V(G).$
To define $θ,$ we use the parsing tree of ${\sf w}$ so that (i) each production rule ${\sf M}\rightarrow ({\sf n}{\sf M})$ corresponds to
a $β∈ \MSOL^{\sf tw}[\{{\sf E},{\sf X},{\sf X}'\}]$ asking for a  clique consisting of one white vertex and $c$ red vertices simulating the removal of a vertex in $G$ (ii) each production rule ${\sf M}\rightarrow ({\sf e}{\sf M})$ corresponds to
a $β∈ \MSOL^{\sf tw}[\{{\sf E},{\sf X},{\sf X}'\}]$ asking for a blue vertex whose removal
corresponds to the removal of an edge in $G,$
(iii) each production rule ${\sf M}\rightarrow ({\sf c}{\sf M})$
is simulated by the application of the $^{\sf c}$ operation (notice that the previous operations maintain the same components in
both $G$ and $G'$), and (iv) each production rule ${\sf M}\rightarrow ({\sf M}\wedge {\sf M})$ or  ${\sf M}\rightarrow ({\sf M}\vee {\sf M})$ is simulated by the \bool\ operation. We also modify each target sentence $γ=σ\wedge μ ∈ Θ_0[\{{\sf E}\}]$
to a sentence $γ'=σ'\wedge μ' ∈ Θ_0[\{{\sf E},{\sf X}\}]$ as follows: (i) $μ'$ is defined so that each obstruction $H'$ in $\obs(\Mod(μ'))$
is created by an obstruction $H$ in $\obs(\Mod(μ))$ by identifying each of its vertices with a vertex of a  clique of size $c+1$ and (ii)
$σ'$ is defined using $σ$ so that the
all quantifications are restricted to white vertices
and the adjacency predicate between two vertices $x$ and $y$
is replaced by the existence of a blue vertex adjacent to both $x$ and $y.$
The definition of $σ'$ guarantees
that $σ$ is simulated on the ``terminal'' graphs resulting after the vertex removals and the edge removals. The definition of $μ'$ implies that if $Q$ and $Q'$ are the terminal graphs corresponding to the evaluation of  $θ$ and $θ',$ respectively,
then $Q$ contains some minor in $\obs(\Mod(μ'))$ if and only if $Q'$ contains some minor
in $\obs(\Mod(μ)).$

For the second part of the theorem, we have the promise that
$\hw(G)≤ c$ for some $c∈ \mathbb{N}.$
Under this promise, if  $G∈ {\cal G}_{\tilde{\sf w}},$ then
the models of the target sentences occurring after the  application of every modification scenario  encoded in $\tilde{\sf w}$ also have  Hadwiger number at most $c.$ We now consider ${\sf w}∈ {\cal M}$ where each target sentence $σ∈\FOL$ in $\tilde{\sf w}$ is replaced by $σ\wedge μ$ where
$\Mod(μ)=\excl(K_{c}).$ Then, under the promise assumption that
$\hw(G)≤ c,$ it holds that $G∈ {\cal G}_{\tilde{\sf w}}\iff G∈ {\cal G}_{{\sf w}}$
and the result follows as, by the first part of the theorem,  ${\cal G}_{{\sf w}}$ can be decided in time $\O_{|{\sf w}|}(n^2)=\O_{|\tilde{\sf w}|,\hw(G)}(n^2).$
\end{proof}

\autoref{thm_grammar} is  able to model
complex {\sl hierarchical modifications}.
That is, it permits to ask for an
iterative removal of  sets of edges/vertices
and,  in each step, apply a different set of modifications to the resulting connected
components.
This multistage modification can be further enhanced by different modification scenarios using the disjunction/conjunction connectors.

Before we give some examples of results that become special cases of \autoref{thm_grammar},
we need some notation. First of all, in the strings of ${\cal M}$
we may omit parenthesis  by agreeing that $\wedge$ has priority over $\vee.$
Also, for a finite set of graphs ${\cal H},$ let  $μ_{\cal H}$ be a {\MSOL}-sentence expressing  ${\cal H}$-minor-exclusion,
and we use $π=μ_{\{K_{5},K_{3,3}\}}$ for expressing planarity.
Then,~\cite{SauST20anftp,SauST20anftp,SauST21kapiI} treat ${\sf w}={\sf n}^kμ_{\cal H},$
~\cite{Grohe04compu,KawarabayashiR2007compu} treat ${\sf w}={\sf e}^kπ,$
\cite{FominGST20analgo} treats (among others) ${\sf w}={\sf n}^k(σ\wedge π)$ and ${\sf w}={\sf e}^k(σ\wedge π),$
for every $σ∈\FOL,$
and \cite{BulianD16graph} treats $({\sf cd})^k{\sf c}μ_{\cal H}.$
Αs an example of the second part
of \autoref{thm_grammar} for $\tilde{M},$
we mention the result of~\cite{DLindermayrSV20elimi}
computing the elimination distance to graphs of bounded degree
in $K_{5}$-minor-free graphs. This problem  corresponds to   $\tilde{\sf w}=({\sf cd})^k{\sf c}σ_d,$ where $σ_d$  expresses degree bounded by $d.$
Finally, we wish to mention the result of Agrawal,  Kanesh,  Panolan, Ramanujan, and  Saurabh in~\cite{SridharanPASK21anftp}
where they consider the problem of checking whether the elimination distance of a graph $G$ to $\Mod(σ_{\cal F})$ is at most $k,$ where
$σ_{\cal F}$ expresses minor-exclusion of some finite set of graphs ${\cal F}$ as induced subgraphs. According to~\cite{SridharanPASK21anftp}, this problem can be solved
in time $\O_{k}(n^{\O_{|σ_{\cal F}|}(1)})$ {for general graphs}.
This problem corresponds to the string  $\tilde{\sf w}=({\sf cd})^k{\sf c}σ_{\cal F}$ and, because of  the second part
of \autoref{thm_grammar}, it can be solved
in time $\O_{k,h,|σ_{\cal F}|}(n^2)$ on
{the special case of}
graphs of Hadwiger number bounded by $h.$

\myskip\paragraph{Variants of {\sc Multiway Cut} and {\sc Multicut}.}
Consider first the {\sc Multiway Cut to ${\cal G}$} problem: the input is
an annotated graph $(G,T),$ where $T$ is a set of terminals. We  ask for a set of edges (or vertices) that, when removed from $G,$ leaves each of the terminals in a separate component of the remaining graph, and each such a component  belongs to ${\cal G}.$
If ${\cal G}=\Mod(σ)$ for some $σ∈ \FOL\subseteq \tilde{Θ},$
then, because of~\autoref{@pertrechando} (that is \autoref{@pertrechando1} stated on structures), this problem can be solved in time $\O_{k,|σ|,h}(n^2)$  on
$K_{h}$-minor-free graphs.
The reduction is a simple version of the proof of~\autoref{thm_grammar}
with the only difference that we add an extra target \FOL-sentence asking that after the removal of the edges the remaining graph should  have only one terminal vertex.
Alternatively, we may move the minor-exclusion to the target property
and ask that ${\cal G}=\Mod(σ)\cap \excl({\cal H})$ for some (finite) set of graphs ${\cal H}.$ In this latter case, the problem can again be solved in time $\O_{k,σ,|{\cal H}|} (n^2)$  without any promise  assumption on its inputs, because of \autoref{@decendientes} (that is \autoref{@decendientes1} stated on structures).
Another application  is the following extension of the well-studied  {\sc Multicut }  problem. We define  {\sc Multicut to ${\cal G}$} as follows.
 For an  input   graph $G$ and a collection $\{(s_i,t_i),i∈[r]\}$ of $r$ pairs of terminals,  the question is whether we may remove $k$ edges (or vertices) from $G$ such that for each $i∈[r],$
$s_i$ and $t_i$ are in different components of the remaining graph, and each of these components belongs to ${\cal G}.$ When ${\cal G}$ is the class of all graphs, this is  {\sc Multicut}.
Because of  \autoref{@pertrechando} and  \autoref{@decendientes}, the previous results about {\sc Multiway Cut to ${\cal G}$} translate to
{\sc Multicut to ${\cal G}$}: when ${\cal G}=\Mod(σ)$ for some $σ∈\FOL,$  {\sc Multiway Cut to ${\cal G}$} can be solved in time $\O_{k,|σ|,h,r}(n^2)$  in $K_{h}$-minor-free graphs,
while, when ${\cal G}=\Mod(σ)\cap \excl({\cal H}),$ {\sc Multiway Cut to ${\cal G}$} can be solved in time $\O_{k,|σ|,|{\cal H}|,r} (n^2)$.

\myskip\paragraph{Removing edges of prescribed adjacency.}

While above we introduced ways to express modifications
involving edge removals, we may further ask for prescribed adjacencies between them. Given a graph $G$ and an edge set $F\subseteq E(G),$ we denote by $V(F)$ the set of all the endpoints of the edges in $F,$ i.e., $V(F)=\bigcup_{e∈ F} e.$ We also define the subgraph
of $G$ {\em spanned} by the edge set $F$ as the graph
$(V(F),E).$

Let $H$ be a graph. We
say that the graph $G'$ is an {\em $H$-modification} of $G$
if $H$ is a subgraph of $G$ and $G'$ is obtained from  $G$ if we remove the edges corresponding to the edges of $H.$
For example, if we  want to remove a matching of $k$ edges in $G$ so that the resulting graph is $G',$ then we ask for an
 $H$-modification of $G$
 where $H=([2k],\{\{2i-1,2i\}\mid i∈[k]\}).$
Given a graph $H$   and a $θ∈ Θ,$
we define the $(H,θ)$-{\sc Modification} problem that, with
input a graph $G,$ asks whether  there is an $H$-modification of $G$ that is a model of $θ.$

\begin{theorem}\label{@commensurate}
For every graph $H$ and every $θ∈Θ,$
there is an algorithm solving the $(H,θ)$-{\sc Modification} problem
in  time $f(|H|,|θ|)\cdot n^2$ . Moreover, if $\tilde{θ}∈\tilde{Θ}$  the same problem can be solved in time $\O_{|H|,|θ|,\hw(G)}(n^2).$
\end{theorem}

\begin{proof}
Let $k=|E(H)|.$
We provide a transformation of $G$ to a multi-annotated graph $(G',R,B,Y_{1},\ldots,Y_{k})$ 
and the construction of some $θ∈ Θ[\{{\sf E},{\sf X}_{1},\ldots,{\sf X}_{k+2}\}]$
such that $G∈ {\cal G}_{\sf w}$ if and only if $(G',R,B,Y_{1},\ldots,Y_{k})\models θ.$
The transformation is similar to the one of \autoref{thm_grammar}. However, instead of subdividing each edge $e=\{x,y\}$ of $G,$
we replace it by the following gadget, called $D_{x,y}$:
take a path of length two between $x$ and $y,$ whose internal vertex is $z,$ increase the multiplicity of each of its two edges to $k$ and, in the resulting multigraph, subdivide each edge and, for  $i∈[k],$ include the $i$-th subdivision vertex into $Y_{i}.$ Also include $z$ in $B$
and identify each vertex $v∈ V(G)$ with
a clique $K_v$ of size $\hw(\Mod(θ))+k+1$ whose
all vertices except from $v$ are included in $R,$ as in the construction of \autoref{thm_grammar}. We call the vertices in $R$ (resp. $B$) \emph{red} (resp. \emph{blue}) vertices of $G'$ and we also call the vertices of $G'$
that correspond to the original vertices of $G$ {\em white}. We set $Y=Y_1 \cup \ldots \cup Y_{k}$ and we call $Y$ the set of  the {\em yellow} vertices of $G'.$
We now consider a sentence $θ^+=β'\triangleright θ'$ defined as follows.
The modulator sentence $β'$ removes $k$ blue vertices $z_{1},\ldots,z_{k}$ and, for each such $z_{i},i∈[k],$ it  also removes all $2(k-1)$ adjacent yellow
vertices that do not belong to $Y_{i}.$ This pair of ``surviving'' yellow vertices will be used to encode the edges of $H.$
Moreover,  $θ'$ is defined so that  (i) each modulator $\MSOL^{\sf tw}$-sentence $β$ composing $θ$ is modified so to exclude from any quantification the red  and the yellow vertices
and (ii) each target sentence $γ=σ\wedge μ ∈ Θ_0[\{{\sf E}\}]$
is modified to $γ'=σ'\wedge μ' ∈ Θ_0[\{{\sf E}\}]$ such that $μ'$ is defined exactly as in the proof of \autoref{thm_grammar} and $σ'= σ_{1}\wedge σ_{2},$ defined as follows. Here $σ_{1}$ is obtained from $σ$ if we
restrict any quantification to the white vertices  and simulate
the adjacency predicate $x\sim y$ by the existence of the gadget $D_{z,y}$ between $x$ and $y.$ While, by the construction presented so far, we are able to simulate the edge removals by the removals of blue vertices, we also need to simulate the adjacencies of $H$ and this is done by the \FOL-sentence $σ_{2},$ that  checks
the existence of an isomorphism between $H$ and the
graph whose vertices are the white vertices that are neighbours of degree-1 yellow vertices, and where two white vertices are adjacent if they have neighboring degree-1 yellow vertices in the same  set $\{Y_{1},\ldots,Y_{k}\}.$
\end{proof}

%

\myskip\section{Basic definitions}
\labels{@aproximadament}
This section as well as Sections 5–11 are devoted to the formal statement and proof of our results.
We present here some basic definitions.

\myskip\subsection{Integers, sets, and tuples}
We denote by $\mathbb{N}$ the set of non-negative integers.
Given two integers $p$ and
$q,$ the set $[p,q]$ refers to the set of every integer $r$ such that $p ≤ r ≤ q.$
For an integer $p≥ 1,$ we set $[p]=[1,p]$ and $\mathbb{N}_{≥ p}=\mathbb{N}\setminus [0,p-1].$
Given a non-negative integer $x,$
we denote by ${\sf odd}(x)$ the minimum odd number that is not smaller than $x.$
For a set $S,$ we denote by $2^{S}$ the set of all subsets of $S$ and, given an integer $r∈[|S|],$
we denote by $\binom{S}{r}$ the set of all subsets of $S$ of size $r.$
Given two sets $A,B$ and a function $f: A\to B,$
for a subset $X\subseteq A$ we use $f(X)$ to denote the set $\{f(x)\mid x∈ X\}.$

Let ${\cal S}$ be a collection of objects where the operations $\cup$ and $\cap$ are defined.
Given two tuples ${\bf x} = (x_1, \ldots, x_l)$ and ${\bf y} = (y_1, \ldots, y_l),$ where $x_i, y_i ∈ {\cal S},$ we denote ${\bf x}\cup{\bf y} = (x_1 \cup y_1, \ldots, x_l \cup y_l)$ and ${\bf x}\cap{\bf y} = (x_1 \cap y_1, \ldots, x_l \cap y_l).$
Also, we denote $\cupall {\cal S} = \bigcup_{X∈ {\cal S}}X.$

\myskip\subsection{Graphs}
\labels{sec_prelim_graphs}

\myskip\paragraph{Basic concepts on graphs.}\labels{label_ricominciavan}
All graphs considered in this paper are undirected, finite, and without loops or multiple edges.
We use standard graph-theoretic notation and we refer the reader to
\cite{Diestel10grap} for any undefined terminology.
Let $G$ be a graph. We say that a pair $(L,R)∈ 2^{V(G)}\times 2^{V(G)}$ is a {\em separation} of $G$
if $L\cup R=V(G)$ and there is no edge in $G$ between a vertex in $L\setminus R$ and a vertex in $R\setminus L.$
Given a vertex $v∈ V(G),$ we denote by $N_{G}(v)$ the set of vertices of $G$ that are adjacent to $v$ in $G.$
Also, given a set $S\subseteq V(G),$ we set $N_G (S) = \bigcup_{v∈ S} N_G (v).$
For $S \subseteq V(G),$ we set $G[S]=(S,E\cap{S \choose 2} )$ and use the shortcut $G \setminus S$ to denote $G[V(G) \setminus S].$
Given a graph $G$ and a set $X\subseteq V(G),$ we denote by $\partial_G (X)$ the set of vertices in $X$ that are adjacent to vertices of $G\setminus X.$

A path $P$ is a {\em $v_1 v_r$-path} if $V(P) = \{v_1, \ldots, v_r\}$ for distinct $v_1, \ldots, v_r$ and $E(P) = \{\{v_1, v_2\}, \{v_2, v_3\}, \ldots, \{v_{r-1}, v_r\}\}.$ We denote this by $P= v_1, \ldots, v_r.$ Given two disjoint paths $P = v_1, \ldots, v_r$ and $Q = u_1, \ldots, u_q$ such that $v_r$ is adjacent to $u_1,$ we say that the path $P\cdot Q= v_1, \ldots, v_r, u_1,\ldots, u_q$ is the {\em concatenation} of $P$ and $Q.$

Given a graph $G$ and a set $S\subseteq V(G),$ we define ${\sf cc}(G,S)$ to be the set of the vertex sets of the connected components of $G\setminus S.$

\myskip\paragraph{Treewidth.}
A \emph{tree decomposition} of a graph~$G$
is a pair~$(T,\chi)$ where $T$ is a tree and $\chi: V(T)\to 2^{V(G)}$
such that
\begin{itemize}
	\item $\bigcup_{t ∈ V(T)} \chi(t) = V(G),$
	\item for every edge~$e$ of~$G$ there is a $t∈ V(T)$ such that
	      $\chi(t)$
	      contains both endpoints of~$e,$ and
	\item for every~$v ∈ V(G),$ the subgraph of~${T}$
	      induced by $\{t ∈ V(T)\mid {v ∈ \chi(t)}\}$ is connected.
\end{itemize}
The {\em width} of $(T,\chi)$ is equal to $\max\big\{\left|\chi(t)\right|-1 \bigmid t∈ V(T)\big\}$ and the {\em treewidth} of $G$ is the minimum width over all tree decompositions of $G.$

\myskip\paragraph{Contractions and minors.}
The \emph{contraction} of an edge $e = \{u,v\}$ of a simple graph $G$ results in a simple graph $G'$
obtained from $G \setminus \{u,v\}$ by adding a new vertex $uv$ adjacent to all the vertices
in the set $N_G(u) \cup N_G(v)\setminus \{u,v\}.$
A graph $G'$ is a \emph{minor} of a graph $G,$ denoted by $G'\prem G,$
if $G'$ can be obtained from $G$ by a sequence of vertex removals, edge removals, and edge contractions.
Given a finite collection of graphs ${\cal F}$ and a graph $G,$ we use the notation ${\cal F}\prem G$ to denote that some graph in ${\cal F}$ is a minor of $G.$
Given a set of graphs ${\cal F},$ we denote by $\excl({\cal F})$ the set containing every graph that excludes all graphs in ${\cal F}$ as minors.
A graph class ${\cal G}$ is {\em minor-closed} if every minor of a graph in ${\cal G}$ is also a member of ${\cal G}.$

\myskip\subsection{First-order logic and monadic second-order logic}
\labels{sec_prelim_logic}

In this subsection, we present some basic notions on logical structures, we define first-order logic and counting monadic second-order logic on structures, and present Gaifman's locality theorem.
We refer the reader to~\cite{CourcelleE12grap} for a broader discussion on logical structures and monadic second-order logic, from the viewpoint of graphs (see also~\cite{Libkin04elem}).

\myskip\paragraph{Structures.}
A {\em vocabulary} is a finite set of relation and constant symbols (we do not use function symbols).
Every relation symbol ${\sf R}$ is associated with a positive integer that is called the {\em arity} of ${\sf R},$ which we denote ${\sf ar}({\sf R}).$
A {\em structure $\mathfrak{A}$ of vocabulary $τ$}, in short a {\em $τ$-structure}, consists of a non-empty set $V(\mathfrak{A}),$ called the {\em universe} of $\mathfrak{A},$
 an $r$-ary relation ${\sf R}^{\mathfrak{A}}\subseteq {V(\mathfrak{A})}^r$ for each relation symbol ${\sf R}∈ τ$ of arity $r≥ 1,$
and an element\footnote{{We stress that we allow constant symbols to be interpreted as the element $\varnothing$, where $\varnothing$ is an element that is not in $V(\mathfrak{A})$.
Throughout this paper, we assume that the universe of every given structure is extended by adding the extra element $\varnothing$, while all relation symbols are interpreted as tuples of elements of $V(\mathfrak{A})$, not containing $\varnothing$.
Moreover, we assume that for every formula that we consider, quantified first order variables are interpreted as elements of the original universe of the structure (and not $\varnothing$).}}
 ${\sf c}^{\mathfrak{A}}∈ \{\varnothing\}\cup V(\mathfrak{A})$ for each constant symbol ${\sf c}∈τ$.
We refer to ${\sf R}^\mathfrak{A}$ (resp. ${\sf c}^{\mathfrak{A}}$) as the {\em interpretation of the symbol ${\sf R}$ (resp. ${\sf c}$) in the structure $\mathfrak{A}$}.
A structure $\mathfrak{A}$ is {\em finite} if its universe $V(\mathfrak{A})$ is a finite set.
We denote by $\mathbb{STR}[τ]$ the set of all finite $τ$-structures.

Let $\mathfrak{A}$ and $\mathfrak{B}$ be $τ$-structures {(both containing $\varnothing$ to their universe)}.
We say that $\mathfrak{A}$ is a {\em substructure} of $\mathfrak{B},$
and we write $\mathfrak{A}\subseteq \mathfrak{B},$ if $V(\mathfrak{A})\subseteq V(\mathfrak{B}),$
for every constant symbol ${\sf c}∈ τ,$ ${\sf c}^{\mathfrak{B}} = {\sf c}^{\mathfrak{A}}$ if ${\sf c}^{\mathfrak{A}}∈ V(\mathfrak{B})$ and ${\sf c}^{\mathfrak{B}}=\varnothing$ otherwise,
and
for every relation symbol ${\sf R}∈ τ$ of arity $r≥ 1$ we have ${\sf R}^{\mathfrak{A}}\subseteq {\sf R}^{\mathfrak{B}}\cap {V(\mathfrak{A})}^r.$
We also say that $\mathfrak{A}$ is an {\em induced substructure} of $\mathfrak{B},$ if $\mathfrak{A}\subseteq \mathfrak{B}$ and
for every relation symbol ${\sf R}∈ τ$ of arity $r≥ 1$ we have ${\sf R}^{\mathfrak{A}}={\sf R}^{\mathfrak{B}}\cap {V(\mathfrak{A})}^r.$
Given a set $S\subseteq V(\mathfrak{A}),$ we use $\mathfrak{A}[S]$ to denote
the $τ$-structure with universe $S,$ where ${\sf R}^{\mathfrak{A}[S]} = {\sf R}^{\mathfrak{A}}\cap S^r$
for each relation symbol ${\sf R}∈ τ$ of arity $r≥ 1$ and for each constant symbol ${\sf c}∈ σ,$
${\sf c}^{\mathfrak{A}[S]} = {\sf c}^{\mathfrak{A}},$ if $ {\sf c}^{\mathfrak{A}} \in S$ and $ {\sf c}^{\mathfrak{A}}=\varnothing.$
Let $σ\subseteq τ$ be a vocabulary.
The {\em $σ$-reduct} of a $τ$-structure $\mathfrak{A}$ is the $σ$-structure $\mathfrak{A}\!\upharpoonright\!σ$ with universe $V(\mathfrak{A})$ such that ${\sf R}^{\mathfrak{A}\upharpoonrightσ} = {\sf R}^{\mathfrak{A}}$ for each relation symbol ${\sf R}∈ σ$ and
 ${\sf c}^{\mathfrak{A}\upharpoonrightσ} = {\sf c}^{\mathfrak{A}}$ for each constant symbol ${\sf c}∈ σ.$

We say that a $τ$-structure $\mathfrak{A}$ is {\em isomorphic} to a $τ$-structure $\mathfrak{B}$ if there is a bijection  $V(\mathfrak{A})\cup\{\varnothing\}$ to $V(\mathfrak{B})\cup\{\varnothing\},$ such that $π(\varnothing) = \varnothing$ and for every $k≥ 1,$ every relation symbol ${\sf R}∈ τ$ of arity $k,$ and every $(a_1, \ldots, a_k)∈ {V(\mathfrak{A})}^k,$ it holds that $(a_1,\ldots, a_k)∈ {\sf R}^{\mathfrak{A}} \iff (π(a_1), \ldots, π(a_k))∈ {\sf R}^{\mathfrak{B}}$ and for every constant symbol ${\sf c}∈ τ,$ it holds that $π({\sf c}^{\mathfrak{A}}) = {\sf c}^{\mathfrak{B}}.$
{Given two $τ$-structures $\mathfrak{A}$ and $\mathfrak{B}$,
where for every constant symbol ${\sf c}\in \tau$ either ${\sf c}^{\mathfrak{A}}={\sf c}^{\mathfrak{B}}$ or ${\sf c}^{\mathfrak{A}} = \varnothing \lor {\sf c}^{\mathfrak{B}} = \varnothing$, we define the {\em disjoint union} of $\mathfrak{A}$ and $\mathfrak{B},$ and we denote it by $\mathfrak{A}\dot\cup\mathfrak{B},$ as
the $τ$-structure where $V(\mathfrak{A}\dot\cup\mathfrak{B})$ is the disjoint union of $V(\mathfrak{A})\setminus\{\varnothing\}$, $V(\mathfrak{B})\setminus\{\varnothing\}$ and $\{\varnothing\}$, for every relation symbol ${\sf R}∈ τ,$ ${\sf R}^{\mathfrak{A}\dot\cup\mathfrak{B}}= {\sf R}^{\mathfrak{A}}\cup {\sf R}^{\mathfrak{B}},$
and
for every constant symbol ${\sf c}\in \tau$, if  ${\sf c}^{\mathfrak{A}}={\sf c}^{\mathfrak{B}}$, then
${\sf c}^{\mathfrak{A}\dot\cup\mathfrak{B}}= {\sf c}^{\mathfrak{A}} ={\sf c}^{\mathfrak{B}}$, and if ${\sf c}^{\mathfrak{A}} = \varnothing$ (resp. ${\sf c}^{\mathfrak{B}} = \varnothing$), then ${\sf c}^{\mathfrak{A}\dot\cup\mathfrak{B}} = {\sf c}^{\mathfrak{B}}$ (resp. ${\sf c}^{\mathfrak{A}\dot\cup\mathfrak{B}} = {\sf c}^{\mathfrak{A}}$).}

An undirected graph  without loops can be seen as an $\{{\sf E}\}$-structure $\mathfrak{G} = (V(\mathfrak{G}),{\sf E}^{\mathfrak{G}}),$ where ${\sf E}^{\mathfrak{G}}$ is a binary relation that is symmetric and anti-reflexive.

\myskip\paragraph{First-order and monadic second-order logic.}
We now define the syntax and the semantics of first-order logic and monadic second-order logic of a vocabulary $τ.$
We assume the existence of a countable infinite set of {\em first-order variables},
usually denoted by lowercase symbols ${\sf x}_1,{\sf x}_2,\ldots,$
and of a countable infinite set of {\em set variables},
usually denoted by uppercase symbols ${\sf X}_1,{\sf X}_2, \ldots.$
A {\em first-order term} is either a first-order variable or a constant symbol.
A {\em first-order logic formula}, in short {\em \FOL-formula}, of vocabulary $τ$ is built from atomic formulas ${\sf x}={\sf y}$
and $({\sf x}_1, \ldots, {\sf x}_r)∈ {\sf R},$ where ${\sf R}∈ τ$ and has arity $r≥ 1,$ on first-order terms ${\sf x},{\sf y},{\sf x}_1,\ldots, {\sf x}_r,$
by using the logical connectives $\vee,$ $\wedge,$ ig$\neg$ and the
quantifiers $\forall, \exists$ on first-order variables.
We denote by $\FOL[τ]$ the set of all \FOL-formulas of vocabulary $τ.$

A {\em monadic second-order logic formula}, in short {\em {\sf MSOL}-formula}, of vocabulary $τ$ is obtained by enhancing
the syntax of \FOL-formulas by allowing the atomic formulas ${\sf x}∈ {\sf X},$
for some first-order term ${\sf x}$ and some set variable ${\sf X},$ and allowing quantification both on first-order and set variables.
We denote by ${\sf MSOL}[τ]$ the set of all {\sf MSOL}-formulas of vocabulary $τ.$
{We make clear that what we call here {\sf MSOL} is what is commonly referred in the literature as {\sf MSO}$_1$, in which, for the vocabulary of graphs, first-order variables are interpreted as vertices and set variables are interpreted as sets of vertices. Our approach uses Courcelle's theorem for bounded treewidth
structures (\autoref{@originallypublishedin}) as a black-box, which applies for a more general logic than  {\sf MSO}$_1$, that is {\sf MSO}$_2$.
For the vocabulary of graphs, {\sf MSO}$_2$ extends  {\sf MSO}$_1$ by also allowing quantification over edges and edge sets (see~\cite[Subsection 9.2]{CourcelleE12grap} for formal definition of  {\sf MSO}$_2$ for general relational vocabularies).
Using this fact, our results hold also in the case we define {\sf MSOL} to be {\sf MSO}$_2$.
}

A {\em counting monadic second-order logic formula}, in short {\em \MSOL-formula}, of vocabulary $τ$ is obtained by enhancing the syntax of {\sf MSOL}-formulas by allowing predicates
of the form ${\sf Card}_p({\sf X}),$ expressing that $|{\sf X}|$ is a multiple of an integer $p>1.$
We denote by $\MSOL[τ]$ the set of all \MSOL-formulas of vocabulary $τ.$

The formulas in $\FOL[τ]$ and $\MSOL[τ]$ are evaluated on $τ$-structures by interpreting every
symbol in $τ$ as its interpretation in the structure and every first-order (resp. set) variable as an element
(resp. set of elements) of the universe of the structure.
Given a formula $φ,$ the {\em free variables} of $φ$ are its variables
that are not in the scope of any quantifier.
We write $φ({\sf x}_1,\ldots, {\sf x}_k)$ to indicate that the free variables of the formula $φ$ are ${\sf x}_1, \ldots,{\sf x}_k.$
A {\em sentence} is a formula without free variables.

Given a $τ$-structure $\mathfrak{A},$ a formula $φ({\sf x}_1, \ldots, {\sf x}_k)∈ \FOL[τ],$
and $a_1,\ldots, a_k$ in $V(\mathfrak{A}),$ we write $\mathfrak{A}\models φ(a_1, \ldots, a_k)$ to denote that $φ({\sf x}_1,\ldots, {\sf x}_k)$ holds in $\mathfrak{A}$ if, for every $i∈[k],$ the variable ${\sf x}_i$ is interpreted as $a_i.$
Two formulas $φ({\sf x}_1, \ldots, {\sf x}_k),$ $ψ({\sf x}_1,\ldots, {\sf x}_k)∈ \FOL[τ]$ are {\em equivalent} if for every $τ$-structure $\mathfrak{A}$ and every $a_1, \ldots, a_k∈ V(\mathfrak{A}),$ we have $\mathfrak{A}\models φ(a_1, \ldots, a_k) \iff \mathfrak{A}\models ψ(a_1, \ldots, a_k).$
We call the set $\{\mathfrak{A}∈ \mathbb{STR}[τ]\mid \mathfrak{A}\modelsφ\}$ the set of {\em models of $φ$} and we denote it by ${\rm Mod}(φ).$

\myskip\paragraph{Gaifman's locality theorem.}
We now aim to present one of the key tools of our proofs, {\sl Gaifman's locality theorem}.
For this, we first give some definitions.
The {\em Gaifman graph} $G_{\mathfrak{A}}$ of a $τ$-structure $\mathfrak{A}$ is the graph with vertex set $V(\mathfrak{A})$
and an edge between two distinct vertices $a,b∈ V(\mathfrak{A})$
if there is an ${\sf R}∈ τ$ of arity $r∈ \mathbb{N}_{≥ 1}$
and a tuple $(a_1,\ldots, a_r)∈ {\sf R}^{\mathfrak{A}}$ such that
$a,b∈ \{a_1, \ldots, a_r\}.$
Notice that in the particular case of graphs (seen as structures), the original graph and its Gaifman graph are the same.

The {\em distance} $d^{\mathfrak{A}}(a,b)$ in $\mathfrak{A}$ between
two elements $a,b∈ V(\mathfrak{A})$ is the length of a shortest path
in $G_{\mathfrak{A}}$ connecting $a$ and $b.$
Given an $r≥ 1$ and an $a∈ V(\mathfrak{A}),$
we define the {\em $r$-neighborhood} of $a$ in $\mathfrak{A}$
to be the set $N^{\mathfrak{A}}_{r} (a) = \{b∈ V(\mathfrak{A})\mid d^{\mathfrak{A}}(a,b)≤ r\}.$
We use $d(a,b)$ instead of $d^{\mathfrak{A}}(a,b)$ and $N_{r} (a)$ instead of $N^{\mathfrak{A}}_{r}(a)$
when $\mathfrak{A}$ is clear from the context.
A first-order formula $ψ({\sf x})$ with one free variable ${\sf x}$
is called {\em $r$-local} if its validity at an element $a$ in the universe of a structure $\mathfrak{A}$ only depends
on the $r$-neighborhood of $a$ in $\mathfrak{A},$ that is $\mathfrak{A}\models
ψ(a) \iff \mathfrak{A}[N_r^{\mathfrak{A}}(a)] \models ψ(a).$

Observe that, for every $r∈ \mathbb{N},$ there is a first-order formula $δ_r ({\sf x},{\sf y})$ such that
for every $τ$-structure $\mathfrak{A}$ and $a,b∈ V(\mathfrak{A})$ we have $\mathfrak{A}\models δ_r (a,b)$
if and only if $d^{\mathfrak{A}}(a,b)≤ r$ (see~\cite[Lemma 2.4.2]{Siebertz16nowh} for a proof).
In what follows, we write $d({\sf x},{\sf y})≤ r$ instead of $δ_r ({\sf x},{\sf y})$ and $d({\sf x},{\sf y})>r$
instead of $\neg δ_r ({\sf x},{\sf y}).$
Let $\ell, r∈ \mathbb{N}_{≥ 1}.$
A {\em basic local sentence with parameters $\ell$ and $r$} is a first-order sentence of the form
\begin{eqnarray*}
\exists {\sf x}_1 \ldots \exists {\sf x}_\ell \Big( \bigwedge_{1≤ i<j≤ \ell} d({\sf x}_i, {\sf x}_j)> 2r \wedge \bigwedge_{i=1}^{\ell} ψ({\sf x}_i)\Big)
\end{eqnarray*}
where $ψ$ is $r$-local. A {\em Gaifman sentence} is a Boolean combination of basic local sentences.

\begin{proposition}[Gaifman's locality theorem~\cite{Gaifman82onlo}]\labels{lem_gaifman}
Every first-order sentence $σ$ is equivalent to a Gaifman sentence $\breve{σ}.$
Moreover, $\breve{σ}$ can be computed effectively from $σ$.
\end{proposition}

For every sentence $σ∈ \FOL[τ],$ we will always denote by $\breve{σ}$ a Gaifman sentence that is equivalent to $σ.$

\myskip\paragraph{Tree decompositions of structures.}
Let $τ$ be a vocabulary.
A {\em tree decomposition} of a $τ$-structure $\mathfrak{A}$ is a pair $(T, \chi),$ where $T$ is a tree and $\chi: V(T) \to 2^{V(\mathfrak{A})}$ such that
\begin{itemize}
\item $\bigcup_{t∈ V(T)} \chi(t) = V(\mathfrak{A}),$
\item for every relation symbol ${\sf R}∈ τ$ of arity $r≥ 1$ and every tuple $(a_1, \ldots, a_r)∈ {\sf R}^{\mathfrak{A}},$
there exists a $t∈ V(T)$ such that $a_1, \ldots, a_r∈ \chi(t),$ and
\item for every $a∈ V(\mathfrak{A}),$ the subgraph of $T$ induced by the set $\{t∈ V(T)\mid a∈ \chi(t)\}$ is connected.
\end{itemize}
The {\em width} of $(T,\chi)$ is equal to $\max\{|\chi(t)|-1\mid t∈ V(T)\}$ and the {\em treewidth} of $\mathfrak{A}$ is the minimum width over all tree decompositions of $\mathfrak{A}.$
Since for every tuple $(a_1, \ldots, a_r)∈ {\sf R}^{\mathfrak{A}},$ for some relation symbol ${\sf R}∈ τ$ of arity $r≥ 1,$ the graph $G_{\mathfrak{A}}[\{a_1, \ldots, a_r\}]$ is a complete graph on $r$ vertices and in a tree decomposition of a graph, every clique is contained in some bag, we have that the treewidth of $\mathfrak{A}$ is the same as the treewidth of $G_{\mathfrak{A}}.$
Therefore, a pair $(T,\chi)$ is a tree decomposition of $\mathfrak{A}$ if and only if it is a tree decomposition of $G_{\mathfrak{A}}.$

\myskip\section{Definition of our logic}
\label{our_logic_stories}
In this section we aim to define our compound logic $Θ.$
This is a logic that is evaluated on structures and its definition is based on a series of modifications done in the initial structure and some questions on ``parts'' of the structure.
To be able to express these modifications, in~\autoref{sec_ope_str}, we define some operations on structures.
Then, in~\autoref{sec_trans}, we show that these operations can be expressed as transductions between structures.
This allows us to do the following: Given a transduction~${\sf f}$ that relates a structure $\mathfrak{A}$ with the structure ${\sf f}(\mathfrak{A}),$
we can ``back-translate'' questions on ${\sf f}(\mathfrak{A})$ to questions on $\mathfrak{A}$ (\autoref{@imposibilitada}).
This implies that we can express properties of ${\sf f}(\mathfrak{A})$ as properties of $\mathfrak{A}.$
Then, in~\autoref{sec_form}, we define some classes of formulas that will be used to define our logic.
Our compound logic $Θ$ is finally defined in~\autoref{sec_ourlogic}.
Our main result is that model-checking of sentences in $Θ$ can be done in quadratic time (\autoref{@decendientes}).

\myskip\subsection{Operations on structures}\labels{sec_ope_str}
In this subsection we define some operations on structures.
All operations defined below are applied on $(τ \cup \{{\sf X}\})$-structures, where $τ$ is a vocabulary and ${\sf X}$ is a unary relation symbol not contained in $τ.$
Intuitively, given a $τ$-structure $\mathfrak{A},$ the interpretation of ${\sf X}$ is a set $X$ of elements of $V(\mathfrak{A})$ (a vertex set, in the case of graphs), which can be considered as elements of a specific ``color''.
Having such a ``colored'' set $X,$ we can define the induced substructure of $\mathfrak{A}$ with respect to this set, denoted by ${\sf ind}_{\sf X} (\mathfrak{A},X),$ and the structure obtained by removing $X$ from the universe of the structure, denoted by ${\sf rm}_{\sf X} (\mathfrak{A},X).$
Also, we define an operation ${\sf cl}_{\sf X}$ (that stands for ``cliquing with respect to ${\sf X}$''), where given a $τ$-structure
$\mathfrak{A}$ and an interpretation $X\subseteq V(\mathfrak{A})$ of ${\sf X},$ we relate (add an edge, in the case of graphs) every two elements of $V(\mathfrak{A})$ that are related (adjacent, in the case of graphs) to a common element of $X.$
Finally, we define the operation ${\sf star}_{\sf X},$ where given a $τ$-structure $\mathfrak{A}$ and an interpretation $X\subseteq V(\mathfrak{A})$ of ${\sf X},$ we replace each vertex set $C$ of the connected components of $G_{\mathfrak{A}}\setminus X$ with a single vertex $v_C$ and we ``project'' every relation of $\mathfrak{A}$ that contains an element of $C$ to a relation that contains $v_C$ (in the case of graphs, this corresponds to contracting each connected component of $G\setminus X$ to a single vertex).
\medskip

Let us now give a formal definition of the above.
Let $τ$ be a vocabulary
and
${\sf X}\notinτ$ be a unary relation symbol.

\myskip\paragraph{Induced structures.}
We define the function ${\sf ind}_{\sf X}: \mathbb{STR}[τ\cup\{{\sf X}\}] \to \mathbb{STR}[τ]$ that maps
every $(τ\cup\{{\sf X}\})$-structure $\mathfrak{A}$ to the $τ$-structure $\mathfrak{A}[{\sf X}^{\mathfrak{A}}]\upharpoonright{τ}.$

\myskip\paragraph{Substructures.}
Also, we define the function ${\sf rm}_{\sf X}: \mathbb{STR}[τ\cup\{{\sf X}\}] \to \mathbb{STR}[τ]$ that maps
every $(τ\cup\{{\sf X}\})$-structure $\mathfrak{A}$ to the $τ$-structure $\mathfrak{A}[V(\mathfrak{A})\setminus {\sf X}^{\mathfrak{A}}]\upharpoonright{τ}.$

\myskip\paragraph{The functions ${\sf cl}_{\sf X}$ and ${\sf star}_{\sf X}.$}
Let ${\sf E}\notinτ$ be a binary relation symbol.
We define the function ${\sf cl}_{\sf X}:\mathbb{STR}[τ\cup \{{\sf X}\}]\to \mathbb{STR}[τ\cup \{{\sf E},{\sf X}\}]$
that maps every $(τ\cup\{{\sf X}\})$-structure $\mathfrak{A}$ to the $(τ\cup\{{\sf E},{\sf X}\})$-structure $\mathfrak{B},$ where
\begin{itemize}
\item $V(\mathfrak{B}) = V(\mathfrak{A}),$
\item  for every ${\sf R}∈ τ\cup \{{\sf X}\},$ ${\sf R}^{\mathfrak{B}} = {\sf R}^{\mathfrak{A}},$
and
\item ${\sf E}^{\mathfrak{B}} =\{(x,y)∈ V(\mathfrak{B})\mid \exists z∈ {\sf X}^{\mathfrak{A}} \ \exists {\sf R}∈ τ\ \exists {\sf Q}∈ τ\ \exists {\bf a}∈ {\sf R}^{\mathfrak{A}}\ \exists {\bf b}∈ {\sf Q}^{\mathfrak{A}} \big(\{x,z\}\subseteq {\bf a} \wedge \{y,z\}\subseteq {\bf b} \big)\},$
\end{itemize}
where by ``$\exists {\bf a}∈ {\sf R}^{\mathfrak{A}}$'' we mean ``$\exists a_1,\ldots, a_r\ (a_1,\ldots, a_{{\sf ar}({\sf R})})∈ {\sf R}^{\mathfrak{A}}$'' and by ``$\{x,z\}\subseteq {\bf a}$'' we mean ``$\{x,z\}\subseteq \{a_1, \ldots, a_{{\sf ar}({\sf R})}\}$''.

For example, in the particular case of graphs (i.e., of structures $\mathfrak{G} = (V(\mathfrak{G}), {\sf E}^\mathfrak{G})$),
for every $X\subseteq V(\mathfrak{G}),$ ${\sf cl}_{\sf X}(\mathfrak{G}, X)$ is the graph obtained from $\mathfrak{G}$ after
transforming the neighborhood of every vertex in $X$ to a clique (see~\autoref{fig_cliquing} for an example).

\begin{figure}[ht]
\centering
\tikzstyle{ipe stylesheet} = [
  ipe import,
  even odd rule,
  line join=round,
  line cap=butt,
  ipe pen normal/.style={line width=0.4},
  ipe pen heavier/.style={line width=0.8},
  ipe pen fat/.style={line width=1.2},
  ipe pen ultrafat/.style={line width=2},
  ipe pen normal,
  ipe mark normal/.style={ipe mark scale=3},
  ipe mark large/.style={ipe mark scale=5},
  ipe mark small/.style={ipe mark scale=2},
  ipe mark tiny/.style={ipe mark scale=1.1},
  ipe mark normal,
  /pgf/arrow keys/.cd,
  ipe arrow normal/.style={scale=7},
  ipe arrow large/.style={scale=10},
  ipe arrow small/.style={scale=5},
  ipe arrow tiny/.style={scale=3},
  ipe arrow normal,
  /tikz/.cd,
  ipe arrows, 
  <->/.tip = ipe normal,
  ipe dash normal/.style={dash pattern=},
  ipe dash dotted/.style={dash pattern=on 1bp off 3bp},
  ipe dash dashed/.style={dash pattern=on 4bp off 4bp},
  ipe dash dash dotted/.style={dash pattern=on 4bp off 2bp on 1bp off 2bp},
  ipe dash dash dot dotted/.style={dash pattern=on 4bp off 2bp on 1bp off 2bp on 1bp off 2bp},
  ipe dash normal,
  ipe node/.append style={font=\normalsize},
  ipe stretch normal/.style={ipe node stretch=1},
  ipe stretch normal,
  ipe opacity 10/.style={opacity=0.1},
  ipe opacity 30/.style={opacity=0.3},
  ipe opacity 50/.style={opacity=0.5},
  ipe opacity 75/.style={opacity=0.75},
  ipe opacity opaque/.style={opacity=1},
  ipe opacity opaque,
]
\definecolor{red}{rgb}{1,0,0}
\definecolor{blue}{rgb}{0,0,1}
\definecolor{green}{rgb}{0,1,0}
\definecolor{yellow}{rgb}{1,1,0}
\definecolor{orange}{rgb}{1,0.647,0}
\definecolor{gold}{rgb}{1,0.843,0}
\definecolor{purple}{rgb}{0.627,0.125,0.941}
\definecolor{gray}{rgb}{0.745,0.745,0.745}
\definecolor{brown}{rgb}{0.647,0.165,0.165}
\definecolor{navy}{rgb}{0,0,0.502}
\definecolor{pink}{rgb}{1,0.753,0.796}
\definecolor{seagreen}{rgb}{0.18,0.545,0.341}
\definecolor{turquoise}{rgb}{0.251,0.878,0.816}
\definecolor{violet}{rgb}{0.933,0.51,0.933}
\definecolor{darkblue}{rgb}{0,0,0.545}
\definecolor{darkcyan}{rgb}{0,0.545,0.545}
\definecolor{darkgray}{rgb}{0.663,0.663,0.663}
\definecolor{darkgreen}{rgb}{0,0.392,0}
\definecolor{darkmagenta}{rgb}{0.545,0,0.545}
\definecolor{darkorange}{rgb}{1,0.549,0}
\definecolor{darkred}{rgb}{0.545,0,0}
\definecolor{lightblue}{rgb}{0.678,0.847,0.902}
\definecolor{lightcyan}{rgb}{0.878,1,1}
\definecolor{lightgray}{rgb}{0.827,0.827,0.827}
\definecolor{lightgreen}{rgb}{0.565,0.933,0.565}
\definecolor{lightyellow}{rgb}{1,1,0.878}
\definecolor{black}{rgb}{0,0,0}
\definecolor{white}{rgb}{1,1,1}
\begin{tikzpicture}[ipe stylesheet]
  \filldraw[draw=darkblue, fill=lightblue, ipe opacity 50]
    (177.925, 493.2568)
     .. controls (186.7433, 488.457) and (204.1777, 486.977) .. (215.2473, 488.6833)
     .. controls (226.317, 490.3897) and (231.022, 495.2823) .. (232.6598, 500.8623)
     .. controls (234.2977, 506.4423) and (232.8683, 512.7097) .. (221.6975, 515.0632)
     .. controls (210.5267, 517.4167) and (189.6143, 515.8563) .. (179.2593, 511.0163)
     .. controls (168.9043, 506.1763) and (169.1067, 498.0567) .. cycle;
  \filldraw[draw=darkblue, fill=lightblue, ipe opacity 50]
    (79.041, 493.1158)
     .. controls (87.8593, 488.316) and (105.2937, 486.836) .. (116.3633, 488.5423)
     .. controls (127.433, 490.2487) and (132.138, 495.1413) .. (133.7758, 500.7213)
     .. controls (135.4137, 506.3013) and (133.9843, 512.5687) .. (122.8135, 514.9222)
     .. controls (111.6427, 517.2757) and (90.7303, 515.7153) .. (80.3753, 510.8753)
     .. controls (70.0203, 506.0353) and (70.2227, 497.9157) .. cycle;
  \draw[darkorange]
    (79.7092, 535.0027)
     .. controls (76.75, 533.0967) and (74.336, 530.2213) .. (80.604, 528.159)
     .. controls (86.872, 526.0967) and (101.822, 524.8473) .. (111.187, 525.559)
     .. controls (120.552, 526.2707) and (124.332, 528.9433) .. (124.8415, 531.564)
     .. controls (125.351, 534.1847) and (122.59, 536.7533) .. (116.1842, 537.9477)
     .. controls (109.7783, 539.142) and (99.7277, 538.962) .. (92.9502, 538.4037)
     .. controls (86.1727, 537.8453) and (82.6683, 536.9087) .. cycle;
  \filldraw[darkorange, ipe opacity 10]
    (79.7092, 535.0027)
     .. controls (76.75, 533.0967) and (74.336, 530.2213) .. (80.604, 528.159)
     .. controls (86.872, 526.0967) and (101.822, 524.8473) .. (111.187, 525.559)
     .. controls (120.552, 526.2707) and (124.332, 528.9433) .. (124.8415, 531.564)
     .. controls (125.351, 534.1847) and (122.59, 536.7533) .. (116.1842, 537.9477)
     .. controls (109.7783, 539.142) and (99.7277, 538.962) .. (92.9502, 538.4037)
     .. controls (86.1727, 537.8453) and (82.6683, 536.9087) .. cycle;
  \filldraw[fill=lightyellow, ipe opacity 30]
    (142.6666, 519.801)
     -- (154.6666, 519.801)
     -- (154.6666, 523.801)
     -- (162.6666, 515.801)
     -- (154.6666, 507.801)
     -- (154.6666, 511.801)
     -- (142.6666, 511.801)
     -- (142.6666, 519.801);
  \filldraw[draw=darkblue, fill=lightblue, ipe opacity 50]
    (78.732, 493.2048)
     .. controls (87.5503, 488.405) and (104.9847, 486.925) .. (116.0543, 488.6313)
     .. controls (127.124, 490.3377) and (131.829, 495.2303) .. (133.4668, 500.8103)
     .. controls (135.1047, 506.3903) and (133.6753, 512.6577) .. (122.5045, 515.0112)
     .. controls (111.3337, 517.3647) and (90.4213, 515.8043) .. (80.0663, 510.9643)
     .. controls (69.7113, 506.1243) and (69.9137, 498.0047) .. cycle;
       \draw[draw=darkblue]
    (78.732, 493.2048)
     .. controls (87.5503, 488.405) and (104.9847, 486.925) .. (116.0543, 488.6313)
     .. controls (127.124, 490.3377) and (131.829, 495.2303) .. (133.4668, 500.8103)
     .. controls (135.1047, 506.3903) and (133.6753, 512.6577) .. (122.5045, 515.0112)
     .. controls (111.3337, 517.3647) and (90.4213, 515.8043) .. (80.0663, 510.9643)
     .. controls (69.7113, 506.1243) and (69.9137, 498.0047) .. cycle;
  \pic[ipe mark small]
     at (111.9196, 504.0652) {ipe disk};
  \pic[ipe mark small]
     at (121.7071, 509.7401) {ipe disk};
  \pic[ipe mark small]
     at (119.6508, 502.0089) {ipe disk};
  \pic[ipe mark small]
     at (129.4383, 507.6838) {ipe disk};
  \pic[ipe mark small]
     at (75.9629, 501.9946) {ipe disk};
  \pic[ipe mark small]
     at (82.7913, 506.1628) {ipe disk};
  \pic[ipe mark small]
     at (86.9594, 499.3344) {ipe disk};
  \pic[ipe mark small]
     at (94.2656, 494.3603) {ipe disk};
  \pic[ipe mark small]
     at (93.0208, 502.2629) {ipe disk};
  \pic[ipe mark small]
     at (100.3009, 507.459) {ipe disk};
  \pic[ipe mark small]
     at (102.7906, 491.6539) {ipe disk};
  \pic[ipe mark small]
     at (109.4483, 500.8013) {ipe disk};
  \node[ipe node, text=darkorange]
     at (131.051, 530.109) {$\darkorange{X}$};
  \pic[ipe mark small]
     at (100.5173, 531.7866) {ipe disk};
  \pic[ipe mark small]
     at (115.2506, 532.1518) {ipe disk};
  \pic[ipe mark small]
     at (86.0278, 531.5702) {ipe disk};
  \draw
    (142.6666, 519.801)
     -- (154.6666, 519.801)
     -- (154.6666, 523.801)
     -- (162.6666, 515.801)
     -- (154.6666, 507.801)
     -- (154.6666, 511.801)
     -- (142.6666, 511.801)
     -- (142.6666, 519.801);
  \draw
    (85.9791, 531.555)
     -- (76.1037, 502.086);
  \draw
    (82.9, 506.179)
     -- (86.2168, 531.6);
  \draw
    (86.3028, 531.625)
     -- (87.0932, 499.326);
  \draw
    (100.608, 531.881)
     -- (93.1322, 502.45);
  \draw
    (100.57, 531.739)
     -- (100.419, 507.435);
  \draw
    (100.692, 531.622)
     -- (94.3467, 494.391);
  \draw
    (102.815, 491.486)
     -- (100.718, 532.052);
  \draw
    (109.584, 500.703)
     -- (100.659, 531.564);
  \draw
    (129.752, 507.734)
     -- (115.251, 532.221);
  \draw
    (115.307, 532.361)
     -- (112.084, 503.991);
  \draw
    (119.743, 502.027)
     -- (115.39, 532.055);
  \draw
    (121.822, 509.814)
     -- (115.308, 532.282);
  \draw
    (82.7367, 506.308)
     -- (86.9943, 499.408);
  \draw
    (86.9943, 499.408)
     -- (75.9743, 501.885);
  \draw
    (93.2252, 502.188)
     -- (109.48, 500.649);
  \draw
    (100.378, 507.595)
     -- (94.5064, 494.727);
  \draw
    (94.5064, 494.727)
     -- (102.762, 491.553);
  \draw
    (111.909, 504.106)
     -- (121.483, 509.642);
  \draw
    (122.023, 509.871)
     -- (119.642, 502.158);
  \draw
    (119.642, 502.158)
     -- (111.701, 504.002);
  \draw[darkorange]
    (178.5932, 535.1437)
     .. controls (175.634, 533.2377) and (173.22, 530.3623) .. (179.488, 528.3)
     .. controls (185.756, 526.2377) and (200.706, 524.9883) .. (210.071, 525.7)
     .. controls (219.436, 526.4117) and (223.216, 529.0843) .. (223.7255, 531.705)
     .. controls (224.235, 534.3257) and (221.474, 536.8943) .. (215.0682, 538.0887)
     .. controls (208.6623, 539.283) and (198.6117, 539.103) .. (191.8342, 538.5447)
     .. controls (185.0567, 537.9863) and (181.5523, 537.0497) .. cycle;
  \filldraw[darkorange, ipe opacity 10]
    (178.5932, 535.1437)
     .. controls (175.634, 533.2377) and (173.22, 530.3623) .. (179.488, 528.3)
     .. controls (185.756, 526.2377) and (200.706, 524.9883) .. (210.071, 525.7)
     .. controls (219.436, 526.4117) and (223.216, 529.0843) .. (223.7255, 531.705)
     .. controls (224.235, 534.3257) and (221.474, 536.8943) .. (215.0682, 538.0887)
     .. controls (208.6623, 539.283) and (198.6117, 539.103) .. (191.8342, 538.5447)
     .. controls (185.0567, 537.9863) and (181.5523, 537.0497) .. cycle;
  \filldraw[draw=darkblue, fill=lightblue, ipe opacity 50]
    (177.616, 493.3458)
     .. controls (186.4343, 488.546) and (203.8687, 487.066) .. (214.9383, 488.7723)
     .. controls (226.008, 490.4787) and (230.713, 495.3713) .. (232.3508, 500.9513)
     .. controls (233.9887, 506.5313) and (232.5593, 512.7987) .. (221.3885, 515.1522)
     .. controls (210.2177, 517.5057) and (189.3053, 515.9453) .. (178.9503, 511.1053)
     .. controls (168.5953, 506.2653) and (168.7977, 498.1457) .. cycle;
  \draw[draw=darkblue]
    (177.616, 493.3458)
     .. controls (186.4343, 488.546) and (203.8687, 487.066) .. (214.9383, 488.7723)
     .. controls (226.008, 490.4787) and (230.713, 495.3713) .. (232.3508, 500.9513)
     .. controls (233.9887, 506.5313) and (232.5593, 512.7987) .. (221.3885, 515.1522)
     .. controls (210.2177, 517.5057) and (189.3053, 515.9453) .. (178.9503, 511.1053)
     .. controls (168.5953, 506.2653) and (168.7977, 498.1457) .. cycle;
  \node[ipe node, text=darkorange]
     at (229.935, 530.25) {$\darkorange{X}$};
  \pic[ipe mark small]
     at (199.4013, 531.9266) {ipe disk};
  \pic[ipe mark small]
     at (214.1343, 532.2928) {ipe disk};
  \pic[ipe mark small]
     at (184.9118, 531.7112) {ipe disk};
  \draw
    (184.8628, 531.6962)
     -- (174.9874, 502.2272);
  \draw
    (181.7837, 506.3202)
     -- (185.1005, 531.7412);
  \draw
    (185.1865, 531.7662)
     -- (185.9769, 499.4672);
  \draw
    (199.4917, 532.0222)
     -- (192.0159, 502.5912);
  \draw
    (199.4537, 531.8802)
     -- (199.3027, 507.5762);
  \draw
    (199.5757, 531.7632)
     -- (193.2304, 494.5322);
  \draw
    (201.6987, 491.6272)
     -- (199.6017, 532.1932);
  \draw
    (208.4677, 500.8442)
     -- (199.5427, 531.7052);
  \draw
    (228.6357, 507.8752)
     -- (214.1347, 532.3622);
  \draw
    (214.1907, 532.5022)
     -- (210.9677, 504.1322);
  \draw
    (218.6267, 502.1682)
     -- (214.2737, 532.1962);
  \draw
    (220.7057, 509.9552)
     -- (214.1917, 532.4232);
  \draw[red]
    (181.6204, 506.4492)
     -- (185.878, 499.5492);
  \draw[red]
    (185.878, 499.5492)
     -- (174.858, 502.0262);
  \draw[red]
    (199.2617, 507.7362)
     -- (193.3901, 494.8682);
  \draw[red]
    (193.3901, 494.8682)
     -- (201.6457, 491.6942);
  \draw[red]
    (210.7927, 504.2472)
     -- (220.3667, 509.7832);
  \draw[red]
    (218.5257, 502.2992)
     -- (210.5847, 504.1432);
  \draw[red]
    (220.9067, 510.0122)
     -- (218.5257, 502.2992);
  \draw[red]
    (192.1089, 502.3292)
     -- (208.3637, 500.7902);
  \draw[red]
    (181.704, 506.279)
     -- (174.971, 502.174);
  \draw[red]
    (199.302, 507.811)
     -- (208.366, 501.032);
  \draw[red]
    (199.332, 507.705)
     -- (192.091, 502.467);
  \draw[red]
    (191.926, 502.706)
     -- (193.107, 494.731);
  \draw[red]
    (201.935, 491.728)
     -- (208.56, 500.906);
  \draw[red]
    (199.038, 507.727)
     -- (201.374, 491.767);
  \draw[red]
    (193.244, 495.151)
     -- (208.26, 500.696);
  \draw[red]
    (192.146, 502.338)
     -- (201.465, 491.958);
  \draw[red]
    (220.983, 509.882)
     -- (228.533, 507.712);
  \draw[red]
    (228.541, 507.712)
     -- (218.426, 502.122);
  \draw[red]
    (210.707, 504.265)
     -- (228.378, 507.667);
  \pic[ipe mark small]
     at (174.8469, 502.1356) {ipe disk};
  \pic[ipe mark small]
     at (181.6753, 506.3038) {ipe disk};
  \pic[ipe mark small]
     at (185.8434, 499.4754) {ipe disk};
  \pic[ipe mark small]
     at (191.9048, 502.4029) {ipe disk};
  \pic[ipe mark small]
     at (199.1849, 507.599) {ipe disk};
  \pic[ipe mark small]
     at (210.8033, 504.2062) {ipe disk};
  \pic[ipe mark small]
     at (208.3323, 500.9413) {ipe disk};
  \pic[ipe mark small]
     at (193.1496, 494.5003) {ipe disk};
  \pic[ipe mark small]
     at (201.6746, 491.7939) {ipe disk};
  \pic[ipe mark small]
     at (218.5345, 502.1499) {ipe disk};
  \pic[ipe mark small]
     at (220.5908, 509.8811) {ipe disk};
  \pic[ipe mark small]
     at (228.322, 507.8248) {ipe disk};
\end{tikzpicture}
\caption{Left: The Gaifman graph of a structure $(\mathfrak{G},X),$ where ${\sf X}^{\mathfrak{A}} = X.$
Right: The Gaifman graph of the structure ${\sf cl}_{\sf X}(\mathfrak{G},X).$}\labels{fig_cliquing}
\end{figure}
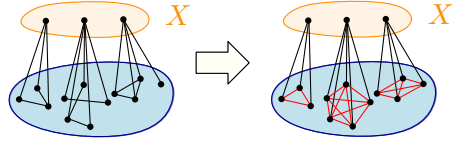

Let $τ$ be a vocabulary without constant symbols
and
${\sf X}\notinτ$ be a unary relation symbol.
We also define the function ${\sf star}_{\sf X}:\mathbb{STR}[τ\cup \{{\sf X}\}]\to \mathbb{STR}[τ\cup \{{\sf X}\}]$
that maps every $(τ\cup\{{\sf X}\})$-structure $\mathfrak{A}$ to the $(τ\cup\{{\sf X}\})$-structure $\mathfrak{B},$ where
\begin{itemize}
\item $V(\mathfrak{B}) = {\sf X}^{\mathfrak{A}} \cup {\sf cc}(G_{\mathfrak{A}},{\sf X}^{\mathfrak{A}}),$
\item for every relation symbol ${\sf R}∈ τ$ of arity $m≥ 1,$
\begin{eqnarray*}
{\sf R}^{\mathfrak{B}} = \bigcup_{(a_1,\ldots, a_m)∈ {\sf R}^{\mathfrak{A}}}\{(z_1, \ldots, z_m)∈ V(\mathfrak{B})^{m}\mid \forall i∈ [m], &  z_i = a_i, & \mbox{ if }a_i∈ {\sf X}^{\mathfrak{A}}, \mbox{~ while~}\\
&  z_i = Y, & \mbox{ if  there is a $Y∈ {\sf cc}(G_{\mathfrak{A}},{\sf X}^{\mathfrak{A}})$}\\
&  \ \ \ \ & \mbox{\ such that $a_i∈ Y$}\}, \text{ and}
\end{eqnarray*}
\item ${\sf X}^{\mathfrak{B}} = {\sf cc}(G_{\mathfrak{A}},{\sf X}^{\mathfrak{A}}).$
\end{itemize}

\begin{figure}[ht]
\centering
\tikzstyle{ipe stylesheet} = [
  ipe import,
  even odd rule,
  line join=round,
  line cap=butt,
  ipe pen normal/.style={line width=0.4},
  ipe pen heavier/.style={line width=0.8},
  ipe pen fat/.style={line width=1.2},
  ipe pen ultrafat/.style={line width=2},
  ipe pen normal,
  ipe mark normal/.style={ipe mark scale=3},
  ipe mark large/.style={ipe mark scale=5},
  ipe mark small/.style={ipe mark scale=2},
  ipe mark tiny/.style={ipe mark scale=1.1},
  ipe mark normal,
  /pgf/arrow keys/.cd,
  ipe arrow normal/.style={scale=7},
  ipe arrow large/.style={scale=10},
  ipe arrow small/.style={scale=5},
  ipe arrow tiny/.style={scale=3},
  ipe arrow normal,
  /tikz/.cd,
  ipe arrows, 
  <->/.tip = ipe normal,
  ipe dash normal/.style={dash pattern=},
  ipe dash dotted/.style={dash pattern=on 1bp off 3bp},
  ipe dash dashed/.style={dash pattern=on 4bp off 4bp},
  ipe dash dash dotted/.style={dash pattern=on 4bp off 2bp on 1bp off 2bp},
  ipe dash dash dot dotted/.style={dash pattern=on 4bp off 2bp on 1bp off 2bp on 1bp off 2bp},
  ipe dash normal,
  ipe node/.append style={font=\normalsize},
  ipe stretch normal/.style={ipe node stretch=1},
  ipe stretch normal,
  ipe opacity 10/.style={opacity=0.1},
  ipe opacity 30/.style={opacity=0.3},
  ipe opacity 50/.style={opacity=0.5},
  ipe opacity 75/.style={opacity=0.75},
  ipe opacity opaque/.style={opacity=1},
  ipe opacity opaque,
]
\definecolor{red}{rgb}{1,0,0}
\definecolor{blue}{rgb}{0,0,1}
\definecolor{green}{rgb}{0,1,0}
\definecolor{yellow}{rgb}{1,1,0}
\definecolor{orange}{rgb}{1,0.647,0}
\definecolor{gold}{rgb}{1,0.843,0}
\definecolor{purple}{rgb}{0.627,0.125,0.941}
\definecolor{gray}{rgb}{0.745,0.745,0.745}
\definecolor{brown}{rgb}{0.647,0.165,0.165}
\definecolor{navy}{rgb}{0,0,0.502}
\definecolor{pink}{rgb}{1,0.753,0.796}
\definecolor{seagreen}{rgb}{0.18,0.545,0.341}
\definecolor{turquoise}{rgb}{0.251,0.878,0.816}
\definecolor{violet}{rgb}{0.933,0.51,0.933}
\definecolor{darkblue}{rgb}{0,0,0.545}
\definecolor{darkcyan}{rgb}{0,0.545,0.545}
\definecolor{darkgray}{rgb}{0.663,0.663,0.663}
\definecolor{darkgreen}{rgb}{0,0.392,0}
\definecolor{darkmagenta}{rgb}{0.545,0,0.545}
\definecolor{darkorange}{rgb}{1,0.549,0}
\definecolor{darkred}{rgb}{0.545,0,0}
\definecolor{lightblue}{rgb}{0.678,0.847,0.902}
\definecolor{lightcyan}{rgb}{0.878,1,1}
\definecolor{lightgray}{rgb}{0.827,0.827,0.827}
\definecolor{lightgreen}{rgb}{0.565,0.933,0.565}
\definecolor{lightyellow}{rgb}{1,1,0.878}
\definecolor{black}{rgb}{0,0,0}
\definecolor{white}{rgb}{1,1,1}
\begin{tikzpicture}[ipe stylesheet]
  \filldraw[draw=navy, fill=lightblue, ipe opacity 50]
    (82.6667, 741.3333)
     .. controls (80, 733.3333) and (88, 722.6667) .. (102, 717.3333)
     .. controls (116, 712) and (136, 712) .. (146.6667, 720)
     .. controls (157.3333, 728) and (158.6667, 744) .. (151.3333, 752)
     .. controls (144, 760) and (128, 760) .. (113.3333, 757.3333)
     .. controls (98.6667, 754.6667) and (85.3333, 749.3333) .. cycle;
  \draw[navy]
    (82.6667, 741.3333)
     .. controls (80, 733.3333) and (88, 722.6667) .. (102, 717.3333)
     .. controls (116, 712) and (136, 712) .. (146.6667, 720)
     .. controls (157.3333, 728) and (158.6667, 744) .. (151.3333, 752)
     .. controls (144, 760) and (128, 760) .. (113.3333, 757.3333)
     .. controls (98.6667, 754.6667) and (85.3333, 749.3333) .. cycle;
  \draw[navy]
    (214.6667, 741.3333)
     .. controls (212, 733.3333) and (220, 722.6667) .. (234, 717.3333)
     .. controls (248, 712) and (268, 712) .. (278.6667, 720)
     .. controls (289.3333, 728) and (290.6667, 744) .. (283.3333, 752)
     .. controls (276, 760) and (260, 760) .. (245.3333, 757.3333)
     .. controls (230.6667, 754.6667) and (217.3333, 749.3333) .. cycle;
  \filldraw[draw=navy, fill=lightblue, ipe opacity 50]
    (214.6667, 741.3333)
     .. controls (212, 733.3333) and (220, 722.6667) .. (234, 717.3333)
     .. controls (248, 712) and (268, 712) .. (278.6667, 720)
     .. controls (289.3333, 728) and (290.6667, 744) .. (283.3333, 752)
     .. controls (276, 760) and (260, 760) .. (245.3333, 757.3333)
     .. controls (230.6667, 754.6667) and (217.3333, 749.3333) .. cycle;
  \draw
    (74, 762)
     .. controls (74.6667, 758.6667) and (81.3333, 757.3333) .. (88, 760)
     .. controls (94.6667, 762.6667) and (101.3333, 769.3333) .. (100.6667, 772.6667)
     .. controls (100, 776) and (92, 776) .. (85.3333, 773.3333)
     .. controls (78.6667, 770.6667) and (73.3333, 765.3333) .. cycle;
  \draw
    (140.6667, 776)
     .. controls (141.3333, 772) and (146.6667, 764) .. (152, 762)
     .. controls (157.3333, 760) and (162.6667, 764) .. (162, 768)
     .. controls (161.3333, 772) and (154.6667, 776) .. (149.3333, 778)
     .. controls (144, 780) and (140, 780) .. cycle;
  \draw
    (100.6667, 694.6667)
     .. controls (97.3333, 690.6667) and (102.6667, 685.3333) .. (110.6667, 684)
     .. controls (118.6667, 682.6667) and (129.3333, 685.3333) .. (132.6667, 689.3333)
     .. controls (136, 693.3333) and (132, 698.6667) .. (124, 700)
     .. controls (116, 701.3333) and (104, 698.6667) .. cycle;
  \draw
    (80, 764)
     -- (92, 740);
  \draw
    (80, 764)
     -- (100, 748);
  \draw
    (96, 772)
     -- (108, 748);
  \draw
    (88, 768)
     -- (100, 740);
  \draw
    (148, 772)
     -- (128, 748);
  \draw
    (152, 768)
     -- (136, 740);
  \draw
    (136, 748)
     -- (152, 768);
  \draw
    (148, 772)
     -- (136, 748);
  \draw
    (108, 692)
     -- (112, 724);
  \draw
    (116, 692)
     -- (120, 716);
  \draw
    (120, 732)
     -- (112, 692);
  \draw
    (124, 692)
     -- (128, 720);
  \draw
    (120, 692)
     -- (128, 728);
  \pic[ipe mark small]
     at (108, 692) {ipe disk};
  \pic[ipe mark small]
     at (112, 692) {ipe disk};
  \pic[ipe mark small]
     at (116, 692) {ipe disk};
  \pic[ipe mark small]
     at (120, 692) {ipe disk};
  \pic[ipe mark small]
     at (124, 692) {ipe disk};
  \pic[ipe mark small]
     at (80, 764) {ipe disk};
  \pic[ipe mark small]
     at (88, 768) {ipe disk};
  \pic[ipe mark small]
     at (96, 772) {ipe disk};
  \pic[ipe mark small]
     at (148, 772) {ipe disk};
  \pic[ipe mark small]
     at (152, 768) {ipe disk};
  \pic[ipe mark small]
     at (92, 740) {ipe disk};
  \pic[ipe mark small]
     at (100, 748) {ipe disk};
  \pic[ipe mark small]
     at (100, 740) {ipe disk};
  \pic[ipe mark small]
     at (108, 748) {ipe disk};
  \pic[ipe mark small]
     at (128, 748) {ipe disk};
  \pic[ipe mark small]
     at (136, 748) {ipe disk};
  \pic[ipe mark small]
     at (136, 740) {ipe disk};
  \pic[ipe mark small]
     at (128, 728) {ipe disk};
  \pic[ipe mark small]
     at (128, 720) {ipe disk};
  \pic[ipe mark small]
     at (120, 716) {ipe disk};
  \pic[ipe mark small]
     at (120, 732) {ipe disk};
  \pic[ipe mark small]
     at (112, 724) {ipe disk};
  \draw
    (220, 768)
     -- (224, 740);
  \draw
    (220, 768)
     -- (232, 748);
  \draw
    (220, 768)
     -- (240, 748);
  \draw
    (220, 768)
     -- (232, 740);
  \draw
    (248, 692)
     -- (252, 716);
  \pic[ipe mark small]
     at (248, 692) {ipe disk};
  \pic[ipe mark small]
     at (220, 768) {ipe disk};
  \pic[ipe mark small]
     at (224, 740) {ipe disk};
  \pic[ipe mark small]
     at (232, 748) {ipe disk};
  \pic[ipe mark small]
     at (232, 740) {ipe disk};
  \pic[ipe mark small]
     at (240, 748) {ipe disk};
  \pic[ipe mark small]
     at (260, 748) {ipe disk};
  \pic[ipe mark small]
     at (268, 748) {ipe disk};
  \pic[ipe mark small]
     at (268, 740) {ipe disk};
  \pic[ipe mark small]
     at (260, 728) {ipe disk};
  \pic[ipe mark small]
     at (260, 720) {ipe disk};
  \pic[ipe mark small]
     at (252, 716) {ipe disk};
  \pic[ipe mark small]
     at (252, 732) {ipe disk};
  \pic[ipe mark small]
     at (244, 724) {ipe disk};
  \pic[ipe mark small]
     at (284, 768) {ipe disk};
  \draw
    (260, 748)
     -- (284, 768);
  \draw
    (268, 748)
     -- (284, 768)
     -- (284, 768);
  \draw
    (268, 740)
     -- (284, 768);
  \draw
    (244, 724)
     -- (248, 692);
  \draw
    (252, 732)
     -- (248, 692);
  \draw
    (260, 728)
     -- (248, 692);
  \draw
    (260, 720)
     -- (248, 692);
  \node[ipe node, text=blue]
     at (140, 732) {$\blue{X}$};
  \node[ipe node, text=blue]
     at (276, 732) {$\blue{X}$};
  \filldraw[fill=lightyellow, ipe opacity 30]
    (176, 744)
     -- (188, 744)
     -- (188, 748)
     -- (196, 740)
     -- (188, 732)
     -- (188, 736)
     -- (176, 736)
     -- (176, 744);
  \node[ipe node, font=\footnotesize]
     at (220, 772) {$v_{C_1}$};
  \node[ipe node, font=\footnotesize]
     at (284, 772) {$v_{C_2}$};
  \node[ipe node, font=\footnotesize]
     at (244, 684) {$v_{C_3}$};
  \node[ipe node]
     at (160, 776) {$C_2$};
  \node[ipe node]
     at (132, 680) {$C_3$};
  \node[ipe node]
     at (82.269, 778.628) {$C_1$};
  \draw
    (176, 744)
     -- (188, 744)
     -- (188, 748)
     -- (196, 740)
     -- (188, 732)
     -- (188, 736)
     -- (176, 736)
     -- (176, 744);
\end{tikzpicture}
\caption{Left: The structure $(\mathfrak{G},X),$ where ${\sf X}^{\mathfrak{A}} = X$ and $C_1, C_2,$ and $C_3$ are the vertex sets of the connected components of $\mathfrak{G}\setminus X.$
Right: The structure $(\mathfrak{G}',Y) = {\sf star}_{\sf X}(\mathfrak{G},X),$ where $Y = \{v_{C_1}, v_{C_2}, v_{C_3}\}.$}\labels{fig_staring}
\end{figure}
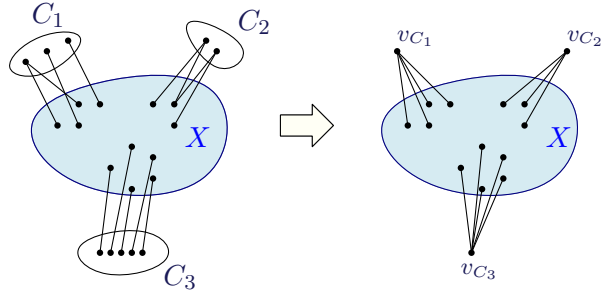

For example, in the particular case of graphs, given a graph $\mathfrak{G}$
and a set $X\subseteq V(\mathfrak{G}),$ ${\sf star}_{\sf X} (\mathfrak{G}, X)$ is
the pair $(\mathfrak{G}', Y),$ where $\mathfrak{G}'$ is the graph obtained from $\mathfrak{G}$ after contracting each
connected component of $\mathfrak{G}\setminus X$ it to a single vertex and $Y$ is the set of vertices
of $\mathfrak{G}'$ corresponding to the connected components of $\mathfrak{G}\setminus X$ (see~\autoref{fig_staring} for an example).
We avoid to define ${\sf star}_{\sf X}$ when $τ$ contains constant symbols, since we will always apply it to structures without constants.

\myskip\subsection{Transductions}\labels{sec_trans}

In this subsection we define (a particular type of) transductions between structures.
The definitions presented here are taken from~\cite{BojanczykP16defi} (see also~\cite{CourcelleE12grap}).

Let $τ$ and $σ$ be two vocabularies without constant symbols\footnote{In this paper, we define transductions between structures without constants. We can extend this definition to transductions between structures with constants with the additional ``promise'' that these transductions do not change the constants.}.
We define a {\em transduction} with input vocabulary $τ$ and output vocabulary $σ$ to be
a set of pairs $(\mathfrak{A}, \mathfrak{B}),$ where $\mathfrak{A}$ is a $τ$-structure and $\mathfrak{B}$ is a $σ$-structure.
Given a transduction ${\cal I}$ with input vocabulary $τ$ and output vocabulary $σ$ and
a $τ$-structure $\mathfrak{A},$ we denote by ${\cal I}(\mathfrak{A})$ the set
of all $σ$-structures $\mathfrak{B}$ such that $(\mathfrak{A},\mathfrak{B})∈ {\cal I}.$
Notice that a transduction is a binary relation between structures that is not necessarily a function.
All the transductions that we will use in our algorithms,
are {\em deterministic}, in the sense that
they are partial functions (up to isomorphism).

\myskip\paragraph{\MSOL-transductions.}
We now define \MSOL-transductions, which are a special case of transductions that can be defined using \MSOL.
We begin by defining three types of transductions:

\begin{itemize}

\item {\bf Copying.}  Let $τ$ be a vocabulary and $k$ be a positive integer.
We define {\em $k$-copying} to be the transduction with input vocabulary $τ$ and output vocabulary $
σ = τ \cup\{ {\sf copy}, {\sf layer}_1, \ldots, {\sf layer}_k\},$ where ${\sf copy}$ is a binary relation symbol,
${\sf layer}_i, i∈ [k]$ is a unary relation symbol,
and for every $τ$-structure $\mathfrak{A},$ outputs a $σ$-structure $\mathfrak{B},$ where
\begin{itemize}
\item $V(\mathfrak{B})$ is the disjoint union of $k$ copies of $V(\mathfrak{A}),$
\item for every ${\sf R}∈ τ$ or arity $r≥ 1,$ ${\sf R}^{\mathfrak{B}}$ is the set of all $r$-tuples over $V(\mathfrak{B})$ such that all the elements of the tuple are in the same copy of $V(\mathfrak{A})$ and the original elements of the copies are in ${\sf R}^{\mathfrak{A}},$

\item ${\sf copy}^{\mathfrak{B}}$ is the set of all pairs of elements in $V(\mathfrak{B})$ that are copies of the same element of $V(\mathfrak{A}),$ and

\item for $i∈ [k],$ ${\sf layer}_i^{\mathfrak{B}}$ is the set of all elements that belong to the $i$-th copy of $V(\mathfrak{A}).$
\end{itemize}

\item {\bf Coloring.}
Let $τ$ be a vocabulary and ${\sf C}\notin τ$ be a unary relation symbol.
We define {\em coloring} to be the transduction with input vocabulary $τ$ and output vocabulary  $σ = τ \cup \{{\sf C}\}$ that, for every $τ$-structure $\mathfrak{A}$ and every $S\subseteq V(\mathfrak{A}),$ outputs the $σ$-structure $\mathfrak{B}_S,$ where $V(\mathfrak{B}) = V(\mathfrak{A}),$ for every ${\sf R}∈ τ,$ ${\sf R}^{\mathfrak{B}} = {\sf R}^{\mathfrak{A}},$ and ${\sf C}^{\mathfrak{B}} = S.$

\item {\bf Interpreting.}
Let $τ$ and $σ$ be two vocabularies.
We define {\em interpretation} to be the transduction with input vocabulary $τ$ and output vocabulary $σ$ as follows:
We consider a family of $\MSOL[τ]$ formulas $$\{φ_{\sf dom}, φ_{\sf univ}\} \cup \{φ_{\sf R}\}_{{\sf R}∈ σ},$$
where the formula $φ_{\sf dom}$ is a sentence (i.e., it has no free variables), the formula $φ_{\sf univ}$ has one free variable, and each formula $φ_{\sf R}$ has as many free variables as the arity of ${\sf R}.$
The free variables in the above formulas are first-order variables.
Given a $τ$-structure $\mathfrak{A}$ such that $\mathfrak{A}\models φ_{\sf dom},$ the output of the interpretation is the $σ$-structure $\mathfrak{B},$ where
\begin{itemize}
\item $V(\mathfrak{B})=\{a∈ V(\mathfrak{A})\mid \mathfrak{A}\models φ_{\sf univ}(a)\}$ and
\item for every ${\sf R}∈ σ$ of arity $r ≥ 1,$ ${\sf R}^{\mathfrak{B}} = \{ (a_1, \ldots, a_r) ∈ {V(\mathfrak{B})}^r \mid \mathfrak{A}\models φ_{\sf R} (a_1,\ldots, a_r)\}.$
\end{itemize}
If $\mathfrak{A}\not\models φ_{\sf dom},$ then the output of the interpretation is not defined.
Intuitively, the formula $φ_{\sf dom}$ specifies the {\sl domain} of the interpretation, by ``filtering out'' all structures that do not satisfy it.
Also, the formula $φ_{\sf univ}$ defines the {\sl universe} of the structure $\mathfrak{B},$ while the formulas $φ_{\sf R}$ allow us to ``interpret'' the relation symbols in $σ.$
\end{itemize}

A relation ${\cal I}$ between $τ$-structures and $σ$-structures is called an {\em \MSOL-transduction with input vocabulary $τ$ and output vocabulary $σ$} if there exists a \red{$k∈\mathbb{N}_{≥ 1}$} such that ${\cal I} = {\cal R}_k \circ \ldots \circ {\cal R}_1,$ where, for every $i∈ [k],$ ${\cal R}_i$ is a copying/coloring/interpreting between $τ_i$-structures and $σ_i$-structures, $τ_1 = τ,$ and $σ_k = σ.$

The reason why we call the above relations \MSOL-transductions is based on the fact that the formulas
we use in the definition of interpretation are formulas in $\MSOL[τ].$
We can define {\em \FOL-transductions} analogously,
by demanding that these formulas are $\FOL$-formulas.
{Notice that since every \FOL-formula is also an \MSOL-formula, an \FOL-transduction is also an \MSOL-transduction.}


\myskip\paragraph{Expressing operations as transductions.}
We now prove that all operations defined in~\autoref{sec_ope_str} are \MSOL-transductions.

\begin{lemma}
\labels{@essencialmente}
Let $τ$ be a vocabulary, ${\sf X}\notinτ$ be a unary relation symbol, and ${\sf E}\notinτ$ be a binary relation symbol.
\begin{itemize}
\item ${\sf ind}_{\sf X}$ and ${\sf rm}_{\sf X}$ are \MSOL-transductions from  $(τ\cup\{{\sf X}\})$-structures to $τ$-structures.

\item ${\sf cl}_{\sf X}$ is a \MSOL-transduction from  $(τ\cup\{{\sf X}\})$-structures to $(τ\cup\{{\sf E},{\sf X}\})$-structures.
\item ${\sf star}_{\sf X}$ is a \MSOL-transduction from  $(τ\cup\{{\sf X}\})$-structures to $(τ\cup\{{\sf X}\})$-structures.
\item The function that maps a $τ$-structure to its Gaifman graph is a \MSOL-transduction from $τ$-structures to $\{{\sf E}\}$-structures.
\end{itemize}
\end{lemma}

\begin{proof}
To see why ${\sf ind}_{\sf X}$ is a \MSOL-transduction
from $(τ\cup\{{\sf X}\})$-structures to $τ$-structures, observe that, for every
$(τ\cup\{{\sf X}\})$-structure $\mathfrak{A},$ using the interpretation where $φ_{\sf dom}$ is always true,
$φ_{\sf univ}({\sf x}) = {\sf x}∈ {\sf X},$
and for every ${\sf R}∈ τ$ of arity $r≥ 1,$ $φ_{\sf R}({\sf x}_1, \ldots, {\sf x}_r) = (({\sf x}_1, \ldots, {\sf x}_r)∈ {\sf R}),$
we get the $τ$-structure $\mathfrak{A}[{\sf X}^{\mathfrak{A}}]\upharpoonright{τ}.$

We next argue why ${\sf cl}_{\sf X}$ is a \MSOL-transduction
from $(τ\cup\{{\sf X}\})$-structures to $(τ\cup\{{\sf E},{\sf X}\})$-structures.
Let $\mathfrak{A}$ be a $(τ\cup\{{\sf X}\})$-structure. To obtain
$\mathfrak{B} = {\sf cl}_{\sf X}(\mathfrak{A}),$ we use interpretation, where
\begin{itemize}
\item $φ_{\sf dom},$ $φ_{\sf univ}({\sf x})$ are always true,
\item  for every ${\sf R}∈ (τ\cup\{{\sf X}\})$ of arity $r≥ 1,$ $φ_{\sf R}({\sf x}_1, \ldots, {\sf x}_r) =(({\sf x}_1, \ldots, {\sf x}_r)∈ {\sf R}),$
and
\item $φ_{\sf E} ({\sf x},{\sf y}) = \exists {\sf z}∈ {\sf X}\ \bigvee_{{\sf R},{\sf Q}∈ τ} \exists {\sf a}_1, \ldots, {\sf a}_{{\sf ar}({\sf R})} \ \exists {\sf b}_1,\ldots, {\sf b}_{{\sf ar}(Q)} \Big(\big(({\sf a}_1,\ldots, {\sf a}_{{\sf ar}({\sf R})})∈ {\sf R}\ \wedge ({\sf b}_1,\ldots, {\sf b}_{{\sf ar}(Q)})∈ Q\big)\to \big(\{{\sf x},{\sf z}\}\subseteq \{{\sf a}_1,\ldots, {\sf a}_{{\sf ar}({\sf R})}\}
\wedge \{{\sf y},{\sf z}\}\subseteq \{{\sf b}_1,\ldots, {\sf b}_{{\sf ar}(Q)}\} \big) \Big).$
\end{itemize}

We will prove that ${\sf star}_{\sf X}$ is a \MSOL-transduction from $(τ\cup\{{\sf X}\})$-structures to $(τ\cup\{{\sf X}\})$-structures.
Let $\mathfrak{A}$ be a $(τ\cup\{{\sf X}\})$-structure. To obtain $\mathfrak{B} = {\sf star}_{\sf X}(\mathfrak{A}),$
we first use coloring and add a new unary predicate ${\sf U}$ in $\mathfrak{A}$ and guess an interpretation $U$ of ${\sf U}$ in $V(\mathfrak{A}),$ which corresponds to a choice of representatives, one for every $C∈ {\sf cc}(G_{\mathfrak{A}},{\sf X}^{\mathfrak{A}}).$
We call $\mathfrak{A}'$ this new $(τ\cup\{{\sf X},{\sf U}\})$-structure.
Then, we use interpretation to transform $\mathfrak{A}'$ to $\mathfrak{B},$ by setting $φ_{\sf dom}$ to be always true, $φ_{\sf univ}({\sf x}) = ({\sf x}∈ {\sf X}\vee {\sf x}∈ {\sf U}),$ $φ_{{\sf X}}({\sf x}) = ({\sf x}∈ {\sf U}),$ and, for every ${\sf R}∈ τ$ of arity $r≥ 1,$
$$φ_{{\sf R}}({\sf x}_1, \ldots, {\sf x}_r) = \exists {\sf a}_1, \ldots, {\sf a}_r\Big(({\sf a}_1,\ldots, {\sf a}_r)∈ {\sf R}\ \wedge \bigwedge_{i∈[r]}\big( ({\sf x}_i = {\sf a}_i \wedge {\sf a}_i ∈ {\sf X})\vee ({\sf x}_i = {\sf U} \wedge {\sf a}_i ∈ {\sf U})\big)\Big).$$


To see how to obtain the Gaifman graph of a $τ$-structure through a \MSOL-transduction, consider the interpretation where $φ_{\sf dom}$ and $φ_{\sf univ}({\sf x})$ are always true and $φ_{\sf E} (x,y)$ is the formula that checks whether there is an ${\sf R}∈ τ$ such that $x,y$ belong simultaneously to some tuple in ${\sf R}^{\mathfrak{A}}.$
\end{proof}


The following result allows us to translate a question in one structure to an ``equivalent'' question in another structure through \MSOL-transductions. It is known as the {\em Backwards Translation Theorem}~\cite[Theorem 1.40]{CourcelleE12grap} (see also~\cite[Lemma B.1]{BojanczykP16defi}).
We state it for sentences, i.e., formulas without free variables.

\begin{proposition}
\labels{@imposibilitada}
Let ${\cal L}$ be either $\MSOL$ or $\FOL$ and let $τ$ and $σ$ be vocabularies without constant symbols.
Let ${\cal I}$ be an ${\cal L}$-transduction with input vocabulary $τ$ and output vocabulary $σ.$
If $φ$ is a sentence in ${\cal L}[σ],$
then there is a sentence $ψ∈ {\cal L}[τ]$ such that for every $σ$-structure $\mathfrak{B},$ if $\mathfrak{B}\in{\cal I}(\mathfrak{A})$ for some $τ$-structure $\mathfrak{A}$, it holds that
$$\mathfrak{A}\models ψ \iff \mathfrak{B}\models φ.$$
\end{proposition}

{
We now state the following result. Intuitively, it says that in the case of structures whose Gaifman graphs have bounded Hadwiger number, one can transduce the original structure from its Gaifman graph. This was proved in a more general setting in~\cite[Lemma 3.1]{BonnetNOST21twin} for the case where the Gaifman graphs have bounded star chromatic number, a property satisfied in classes of bounded expansion such as classes of bounded Hadwiger number.

\begin{proposition}
\label{@transducing_str_from_gr}
Let $τ$ be a vocabulary without constant symbols, let ${\sf E}\notin τ$ be a binary relation symbol, let $c\in\mathbb{N}$, and let ${\cal C}\subseteq \mathbb{STR}[τ]$.
There is an \FOL-transduction ${\cal I}$ from ${\sf E}$-structures to $τ$-structures
such that if all graphs in $\{G_{\mathfrak{A}}\mid \mathfrak{A}\in {\cal C}\}$ have Hadwiger number at most $c$, then, if $G=G_{\mathfrak{A}}$ for some $\mathfrak{A}\in {\cal C}$, it holds that ${\cal I}(G) = \mathfrak{A}$.
\end{proposition}

At this point, we should comment that, in~\autoref{sec_ourlogic}, we prove that the Gaifman graph of every structure that is a model of a formula in $Θ$ has bounded Hadwiger number (\autoref{@disillusioned}).
Therefore, due to~\autoref{@transducing_str_from_gr}, we can transduce every structure that is a model of a formula in $Θ$ from its Gaifman graph.
This, in turn, together with~\autoref{@imposibilitada}, the fact that $Θ\subseteq\MSOL,$ and the observation that any \FOL-transduction is also an \MSOL-transduction, indicates that
the problem of model-checking for $Θ$ in general structures is essentially not more general than in graphs.}\medskip

Combining~\autoref{@essencialmente} and~\autoref{@imposibilitada}, we get the following result.
\begin{corollary}\labels{@passablement}
Let $τ$ be a vocabulary without constant symbols, ${\sf X}\notin τ$ be a unary relation
symbol, and ${\sf E}\notin τ$ be a binary relation symbol.
\begin{itemize}
\item For every sentence
$ψ∈\MSOL[τ],$
there is a sentence $ψ|_{{\sf ind}_{\sf X}}∈ \MSOL[τ\cup\{{\sf X}\}]$
such that for every $(τ\cup\{{\sf X}\})$-structure $\mathfrak{A},$ it holds that
$\mathfrak{A}\models ψ|_{{\sf ind}_{\sf X}}\iff  {\sf ind}_{\sf X}(\mathfrak{A})\models ψ,$

\item for every sentence $ψ∈ \MSOL[τ],$
there is a sentence $ψ|_{{\sf rm}_{\sf X}}∈ \MSOL[τ\cup\{{\sf X}\}]$
such that for every $(τ\cup\{{\sf X}\})$-structure $\mathfrak{A},$ it holds that
$\mathfrak{A}\models ψ|_{{\sf rm}_{\sf X}}\iff  {\sf rm}_{\sf X}(\mathfrak{A})\models ψ,$

\item for every sentence $ψ∈ \MSOL[τ\cup\{{\sf E},{\sf X}\}],$
there is a sentence $ψ|_{{\sf cl}_{\sf X}}∈ \MSOL[τ\cup\{{\sf X}\}]$
such that for every $(τ\cup\{{\sf X}\})$-structure $\mathfrak{A},$ it holds that
$\mathfrak{A}\models ψ|_{{\sf cl}_{\sf X}} \iff  {\sf cl}_{\sf X}(\mathfrak{A})\models ψ,$

\item for every sentence $ψ∈ \MSOL[τ],$
there is a sentence $ψ|_{{\sf star}_{\sf X}}∈ \MSOL[τ\cup\{{\sf X}\}]$
such that for every $(τ\cup\{{\sf X}\})$-structure $\mathfrak{A},$ it holds that
$\mathfrak{A}\models ψ|_{{\sf star}_{\sf X}} \iff  {\sf star}_{\sf X}(\mathfrak{A})\models ψ,$ and

\item for every sentence $ψ∈ \MSOL[\{{\sf E}\}],$
there is a sentence $ψ|_{\sf gf}∈ \MSOL[τ]$
such that for every $τ$-structure $\mathfrak{A},$ it holds that
$\mathfrak{A}\models ψ|_{\sf gf} \iff  G_\mathfrak{A}\models ψ.$

\end{itemize}
\end{corollary}

Later in the paper, we will apply the transduction ${\sf ind}_{\sf X}$ to structures with constants (i.e., structures whose vocabulary contains constant symbols). In this case,~\autoref{@passablement} cannot be applied to obtain a sentence $ψ|_{{\sf ind}_{\sf X}}∈ \MSOL[τ\cup\{{\sf X}\}]$ from a sentence  $ψ∈\MSOL[τ].$ However, we can directly set $ψ|_{{\sf ind}_{\sf X}}$ to be the sentence obtained from $ψ$ after replacing, for each first-order variable, every occurrence of ``$\exists/\forall\ {\sf x}$'' with ``$\exists/\forall\ {\sf x}∈ {\sf X}^{\mathfrak{A}}$'' and, for each set variable $Y,$ every occurrence of ``$\exists/\forall\ {\sf Y}$'' with ``$\exists/\forall\ {\sf Y} \subseteq {\sf X}^{\mathfrak{A}}$''.
\bigskip

We define the operation $\triangleright$ as follows.
Given $β∈\MSOL[τ\cup\{{\sf X}\}]$ and $γ∈ \MSOL[τ],$ we define
\begin{eqnarray}
β\triangleright γ:=\exists {\sf X}\ β|_{{\sf star}_{\sf X}}\wedge γ|_{{\sf rm}_{\sf X}}.\label{def_triangle_op}
\end{eqnarray}


As a byproduct of~\autoref{@passablement}, we get the following.

\begin{corollary}
\labels{@confiscating}
If $β∈\MSOL[τ\cup\{{\sf X}\}]$ and $γ∈ \MSOL[τ],$ then
$β\triangleright γ∈\MSOL[τ].$
\end{corollary}	
	
\myskip\subsection{Classes of formulas}\labels{sec_form}
In this subsection we define several classes of formulas that will be used in the definition of our logic $Θ$ in~\autoref{sec_ourlogic}.

\myskip\paragraph{The class \NTMC.}
We say that a class ${\cal C}$ of structures is {\em minor-closed} if the graph class $\{G_{\mathfrak{A}} \mid \mathfrak{A}∈ {\cal C}\}$ is minor-closed.

The following lemma asserts that minor-exclusion can be expressed in \MSOL.
Its proof is a direct implication of~\cite[Corollary 1.14]{CourcelleE12grap} (see also~\cite[Appendix D]{KimLPRRSS16line} for an explicit \MSOL-formula for the case of topological minors).
\begin{lemma}\labels{@favoleggiare}
Let $G$ be a graph and ${\cal F}$ be a finite family of graphs.
There is a \MSOL-sentence $μ_{\cal F}$ that is evaluated on graphs such that $G\models μ_{\cal F}\iff G∈ \excl({\cal F}).$
\end{lemma}

Combining~\autoref{@passablement} and~\autoref{@favoleggiare}, we get the following:
\begin{corollary}\labels{@desabrocharle}
For every $τ$-structure $\mathfrak{A}$ and every finite family ${\cal F}$ of graphs, there is a sentence $μ_{\cal F}|_{\sf gf}∈ \MSOL[τ]$ such that $\mathfrak{A}\models μ_{\cal F}|_{\sf gf}\iff G_{\mathfrak{A}}∈ \excl({\cal F}).$
\end{corollary}

Let $τ$ be a vocabulary and $φ∈ \MSOL[τ].$ We say that ${\rm Mod}(φ)$ is {\em trivial} if it contains  all $τ$-structures.
We denote by $\NTMC[τ]$ the class of all sentences $μ$ in $\MSOL[τ]$ such that
${\rm Mod}(μ)$ is proper minor-closed.
Notice that for every sentence $μ∈ \NTMC[τ],$ because of  Robertson-Seymour's theorem~\cite{RobertsonS04GMXX} and based on~\autoref{@desabrocharle}, there is some positive integer $c_μ$ such that $\{G_{\mathfrak{A}}\mid \mathfrak{A}∈{\rm Mod}(μ)\}\subseteq \excl(K_{c_{μ}}).$

\myskip\paragraph{\MSOL-sentence of bounded treewidth.}
Let $τ$ be a vocabulary without constant symbols and let ${\sf X}\notin τ$ be a unary relation symbol.
We say that  a sentence $β∈\MSOL[τ\cup\{{\sf X}\}]$  has
{\em bounded treewidth}  if there is a constant $c_{β}$ such that
all $(τ\cup\{{\sf X}\})$-structures in ${\rm Mod}(β_{\sf cl}),$ that is, in the set $\{{\sf cl}_{\sf X}(\mathfrak{A},{\sf X})\mid (\mathfrak{A},{\sf X})∈ {\rm Mod}(β)\},$
have treewidth at most $c_{β}.$ 
We denote by \MSOL$^\tw[τ\cup\{{\sf X}\}]$ the set of all
sentences in $\MSOL[τ\cup\{{\sf X}\}]$ that have bounded treewidth.

\myskip\paragraph{Connected component closure of a formula.}
We first observe that, given a $τ$-structure $\mathfrak{A}$ and a set $C\subseteq V(\mathfrak{A}),$
the fact that $C$ is the vertex set of a connected component of $G_{\mathfrak{A}}$ can be expressed in \MSOL.

\begin{observation}\labels{@extinguishes}
Let $τ$ be a vocabulary and ${\sf X}\notin τ$ be a unary relation symbol.
There is a sentence $κ_{\sf X}∈ \MSOL[τ\cup\{{\sf X}\}]$ such that for every $(\mathfrak{A},C)∈ \mathbb{STR}[τ\cup\{{\sf X}\}]$ (where ${\sf X}$ is interpreted as $C$),
it holds that $(\mathfrak{A},C)\models κ_{\sf X} \iff C$~is the vertex set of a connected component of $G_\mathfrak{A}.$
\end{observation}

We consider the sentence $κ_{\sf X}∈ \MSOL[τ\cup\{{\sf X}\}]$ as in~\autoref{@extinguishes}.
For every formula $φ∈\MSOL[τ],$ we define the formula $φ^{({\sf c})}$ as $$φ^{({\sf c})}=\forall {\sf X}\ κ_{\sf X} \to φ|_{{\sf ind}_{\sf X}}.$$
Recall that $φ|_{{\sf ind}_{\sf X}}$ is a formula in $\MSOL[τ\cup \{{\sf X}\}]$ and note that $φ^{({\sf c})}∈\MSOL[τ].$
Alternatively, for every $τ$-structure $\mathfrak{A},$
$$\mathfrak{A}\models φ^{({\sf c})}\iff \mbox{for every $C∈ {\sf cc}(G_\mathfrak{A},\emptyset)$ it holds that~} \mathfrak{A}[C]\models φ.$$

Given a set ${\cal L}\subseteq\MSOL[τ],$ we define $${\cal L}^{({\sf c})}= {\cal L}\cup\{φ^{({\sf c})}\mid φ ∈ {\cal L}\}.$$

\myskip\paragraph{Boolean combination of formulas.}
Given a set of formulas ${\cal L}\subseteq\MSOL[τ],$ for some vocabulary $τ,$  we define $\bool({\cal L})$ as the set of all positive Boolean combinations of formulas in ${\cal L},$ i.e., all formulas constructed from formulas in ${\cal L}$ using the Boolean connectives $\vee$ and $\wedge.$

\myskip\subsection{Our compound logic $Θ$}\labels{sec_ourlogic}
We are now in position to define our compound logic $Θ$ for general structures. First, we define the intermediate logic $\bar{Θ}.$

\myskip\paragraph{The intermediate logic $\bar{Θ}.$}
Let $τ$ be a vocabulary.
Recall that, given a sentence $β∈\MSOL[τ\cup\{{\sf X}\}]$ and a sentence $γ∈ \MSOL[τ],$
$β\triangleright γ$ is defined as the sentence
$\exists {\sf X}\ β|_{{\sf star}_{\sf X}}\wedge γ|_{{\sf rm}_{\sf X}}$ (see~\autoref{def_triangle_op}).
We now recursively define, for every $i∈ \mathbb{N},$ the subclass $\bar{Θ}_{i}[τ]$  of $\MSOL[τ]$ so that

%

\begin{eqnarray*}
\bar{Θ}_{0}[τ] & = & \{σ\wedge μ\mid  σ∈ \FOL[τ]\mbox{~and~} μ∈ \NTMC[τ]\} \mbox{~and} \\
\mbox{for $i≥ 1,$ } \bar{Θ}_{i}[τ] & = & \{β\triangleright γ \mid  β∈ \MSOL^\tw[τ\cup\{{\sf X}\}]  \mbox{~and~} γ∈  {\bar{Θ}_{i-1}[τ]}^{({\sf c})}\}
\end{eqnarray*}
On a semantical level, a sentence $θ∈ \MSOL[τ]$ belongs to $\bar{Θ}_{i}[τ],i>0,$ if and only if
there exists a
$β∈ \MSOL^\tw[τ\cup\{{\sf X}\}]$ and  a $γ∈ {\bar{Θ}_{i-1}[τ]}^{({\sf c})}$
such that, for every $τ$-structure $\mathfrak{A},$
$$\mbox{there exists some set } X\subseteq V(\mathfrak{A})\mbox{~such that~} {\sf star}_{\sf X}(\mathfrak{A},X)\models β\mbox{~and~}\mathfrak{A}\setminus X\models γ.$$
	
We stress that, since $β∈ \MSOL^\tw[τ\cup\{{\sf X}\}],$ the fact that $ {\sf star}_{\sf X}(\mathfrak{A},X)\models β$ implies that there is a constant $c_β∈\mathbb{N}_{≥ 1},$ such that the structure ${\sf cl}_{\sf X}({\sf star}_{\sf X}(\mathfrak{A},X))$ has treewidth at most $c_β.$
See~\autoref{fig_transfor} for the Gaifman graph of a structure $(\mathfrak{A},X)$ and the Gaifman graph of the structure ${\sf cl}_{\sf X}({\sf star}_{\sf X}(\mathfrak{A},X)).$

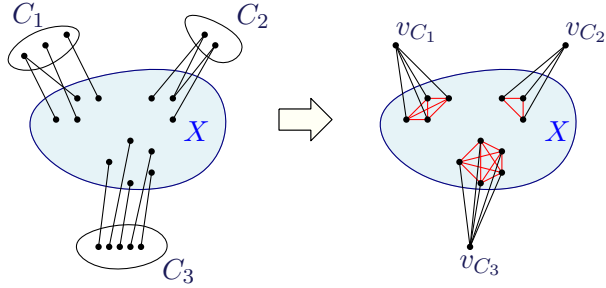
\begin{figure}[ht]
\centering
\tikzstyle{ipe stylesheet} = [
  ipe import,
  even odd rule,
  line join=round,
  line cap=butt,
  ipe pen normal/.style={line width=0.4},
  ipe pen heavier/.style={line width=0.8},
  ipe pen fat/.style={line width=1.2},
  ipe pen ultrafat/.style={line width=2},
  ipe pen normal,
  ipe mark normal/.style={ipe mark scale=3},
  ipe mark large/.style={ipe mark scale=5},
  ipe mark small/.style={ipe mark scale=2},
  ipe mark tiny/.style={ipe mark scale=1.1},
  ipe mark normal,
  /pgf/arrow keys/.cd,
  ipe arrow normal/.style={scale=7},
  ipe arrow large/.style={scale=10},
  ipe arrow small/.style={scale=5},
  ipe arrow tiny/.style={scale=3},
  ipe arrow normal,
  /tikz/.cd,
  ipe arrows, 
  <->/.tip = ipe normal,
  ipe dash normal/.style={dash pattern=},
  ipe dash dotted/.style={dash pattern=on 1bp off 3bp},
  ipe dash dashed/.style={dash pattern=on 4bp off 4bp},
  ipe dash dash dotted/.style={dash pattern=on 4bp off 2bp on 1bp off 2bp},
  ipe dash dash dot dotted/.style={dash pattern=on 4bp off 2bp on 1bp off 2bp on 1bp off 2bp},
  ipe dash normal,
  ipe node/.append style={font=\normalsize},
  ipe stretch normal/.style={ipe node stretch=1},
  ipe stretch normal,
  ipe opacity 10/.style={opacity=0.1},
  ipe opacity 30/.style={opacity=0.3},
  ipe opacity 50/.style={opacity=0.5},
  ipe opacity 75/.style={opacity=0.75},
  ipe opacity opaque/.style={opacity=1},
  ipe opacity opaque,
]
\definecolor{red}{rgb}{1,0,0}
\definecolor{blue}{rgb}{0,0,1}
\definecolor{green}{rgb}{0,1,0}
\definecolor{yellow}{rgb}{1,1,0}
\definecolor{orange}{rgb}{1,0.647,0}
\definecolor{gold}{rgb}{1,0.843,0}
\definecolor{purple}{rgb}{0.627,0.125,0.941}
\definecolor{gray}{rgb}{0.745,0.745,0.745}
\definecolor{brown}{rgb}{0.647,0.165,0.165}
\definecolor{navy}{rgb}{0,0,0.502}
\definecolor{pink}{rgb}{1,0.753,0.796}
\definecolor{seagreen}{rgb}{0.18,0.545,0.341}
\definecolor{turquoise}{rgb}{0.251,0.878,0.816}
\definecolor{violet}{rgb}{0.933,0.51,0.933}
\definecolor{darkblue}{rgb}{0,0,0.545}
\definecolor{darkcyan}{rgb}{0,0.545,0.545}
\definecolor{darkgray}{rgb}{0.663,0.663,0.663}
\definecolor{darkgreen}{rgb}{0,0.392,0}
\definecolor{darkmagenta}{rgb}{0.545,0,0.545}
\definecolor{darkorange}{rgb}{1,0.549,0}
\definecolor{darkred}{rgb}{0.545,0,0}
\definecolor{lightblue}{rgb}{0.678,0.847,0.902}
\definecolor{lightcyan}{rgb}{0.878,1,1}
\definecolor{lightgray}{rgb}{0.827,0.827,0.827}
\definecolor{lightgreen}{rgb}{0.565,0.933,0.565}
\definecolor{lightyellow}{rgb}{1,1,0.878}
\definecolor{black}{rgb}{0,0,0}
\definecolor{white}{rgb}{1,1,1}
\scalebox{1}{
\begin{tikzpicture}[ipe stylesheet]
  \filldraw[draw=navy, fill=lightblue, ipe opacity 30]
    (214.6667, 741.3333)
     .. controls (212, 733.3333) and (220, 722.6667) .. (234, 717.3333)
     .. controls (248, 712) and (268, 712) .. (278.6667, 720)
     .. controls (289.3333, 728) and (290.6667, 744) .. (283.3333, 752)
     .. controls (276, 760) and (260, 760) .. (245.3333, 757.3333)
     .. controls (230.6667, 754.6667) and (217.3333, 749.3333) .. cycle;
  \draw[draw=navy, ipe opacity opaque]
    (214.6667, 741.3333)
     .. controls (212, 733.3333) and (220, 722.6667) .. (234, 717.3333)
     .. controls (248, 712) and (268, 712) .. (278.6667, 720)
     .. controls (289.3333, 728) and (290.6667, 744) .. (283.3333, 752)
     .. controls (276, 760) and (260, 760) .. (245.3333, 757.3333)
     .. controls (230.6667, 754.6667) and (217.3333, 749.3333) .. cycle;
  \draw[red]
    (224, 740)
     -- (232, 748);
  \draw[red]
    (224, 740)
     -- (232, 740)
     -- (232, 740);
  \draw[red]
    (232, 748)
     -- (240, 748);
  \draw[red]
    (240, 748)
     -- (232, 740);
  \draw[red]
    (232, 748)
     -- (232, 740);
  \draw[red]
    (224, 740)
     -- (240, 748)
     -- (240, 748);
  \draw[red]
    (260, 748)
     -- (268, 748);
  \draw[red]
    (268, 748)
     -- (268, 740);
  \draw[red]
    (268, 740)
     -- (260, 748);
  \draw[red]
    (244, 724)
     -- (252, 732);
  \draw[red]
    (252, 732)
     -- (260, 728);
  \draw[red]
    (260, 728)
     -- (260, 720);
  \draw[red]
    (260, 720)
     -- (252, 716);
  \draw[red]
    (252, 716)
     -- (244, 724);
  \draw[red]
    (244, 724)
     -- (260, 728);
  \draw[red]
    (260, 728)
     -- (252, 716);
  \draw[red]
    (252, 716)
     -- (252, 732);
  \draw[red]
    (252, 732)
     -- (260, 720);
  \draw[red]
    (260, 720)
     -- (244, 724);
  \filldraw[draw=navy, fill=lightblue, ipe opacity 30]
    (82.6667, 741.3333)
     .. controls (80, 733.3333) and (88, 722.6667) .. (102, 717.3333)
     .. controls (116, 712) and (136, 712) .. (146.6667, 720)
     .. controls (157.3333, 728) and (158.6667, 744) .. (151.3333, 752)
     .. controls (144, 760) and (128, 760) .. (113.3333, 757.3333)
     .. controls (98.6667, 754.6667) and (85.3333, 749.3333) .. cycle;
     \draw[draw=navy, ipe opacity opaque]
    (82.6667, 741.3333)
     .. controls (80, 733.3333) and (88, 722.6667) .. (102, 717.3333)
     .. controls (116, 712) and (136, 712) .. (146.6667, 720)
     .. controls (157.3333, 728) and (158.6667, 744) .. (151.3333, 752)
     .. controls (144, 760) and (128, 760) .. (113.3333, 757.3333)
     .. controls (98.6667, 754.6667) and (85.3333, 749.3333) .. cycle;
  \draw
    (74, 762)
     .. controls (74.6667, 758.6667) and (81.3333, 757.3333) .. (88, 760)
     .. controls (94.6667, 762.6667) and (101.3333, 769.3333) .. (100.6667, 772.6667)
     .. controls (100, 776) and (92, 776) .. (85.3333, 773.3333)
     .. controls (78.6667, 770.6667) and (73.3333, 765.3333) .. cycle;
  \draw
    (140.6667, 776)
     .. controls (141.3333, 772) and (146.6667, 764) .. (152, 762)
     .. controls (157.3333, 760) and (162.6667, 764) .. (162, 768)
     .. controls (161.3333, 772) and (154.6667, 776) .. (149.3333, 778)
     .. controls (144, 780) and (140, 780) .. cycle;
  \draw
    (100.6667, 694.6667)
     .. controls (97.3333, 690.6667) and (102.6667, 685.3333) .. (110.6667, 684)
     .. controls (118.6667, 682.6667) and (129.3333, 685.3333) .. (132.6667, 689.3333)
     .. controls (136, 693.3333) and (132, 698.6667) .. (124, 700)
     .. controls (116, 701.3333) and (104, 698.6667) .. cycle;
  \draw
    (80, 764)
     -- (92, 740);
  \draw
    (80, 764)
     -- (100, 748);
  \draw
    (96, 772)
     -- (108, 748);
  \draw
    (88, 768)
     -- (100, 740);
  \draw
    (148, 772)
     -- (128, 748);
  \draw
    (152, 768)
     -- (136, 740);
  \draw
    (136, 748)
     -- (152, 768);
  \draw
    (148, 772)
     -- (136, 748);
  \draw
    (108, 692)
     -- (112, 724);
  \draw
    (116, 692)
     -- (120, 716);
  \draw
    (120, 732)
     -- (112, 692);
  \draw
    (124, 692)
     -- (128, 720);
  \draw
    (120, 692)
     -- (128, 728);
  \pic[ipe mark small]
     at (108, 692) {ipe disk};
  \pic[ipe mark small]
     at (112, 692) {ipe disk};
  \pic[ipe mark small]
     at (116, 692) {ipe disk};
  \pic[ipe mark small]
     at (120, 692) {ipe disk};
  \pic[ipe mark small]
     at (124, 692) {ipe disk};
  \pic[ipe mark small]
     at (80, 764) {ipe disk};
  \pic[ipe mark small]
     at (88, 768) {ipe disk};
  \pic[ipe mark small]
     at (96, 772) {ipe disk};
  \pic[ipe mark small]
     at (148, 772) {ipe disk};
  \pic[ipe mark small]
     at (152, 768) {ipe disk};
  \pic[ipe mark small]
     at (92, 740) {ipe disk};
  \pic[ipe mark small]
     at (100, 748) {ipe disk};
  \pic[ipe mark small]
     at (100, 740) {ipe disk};
  \pic[ipe mark small]
     at (108, 748) {ipe disk};
  \pic[ipe mark small]
     at (128, 748) {ipe disk};
  \pic[ipe mark small]
     at (136, 748) {ipe disk};
  \pic[ipe mark small]
     at (136, 740) {ipe disk};
  \pic[ipe mark small]
     at (128, 728) {ipe disk};
  \pic[ipe mark small]
     at (128, 720) {ipe disk};
  \pic[ipe mark small]
     at (120, 716) {ipe disk};
  \pic[ipe mark small]
     at (120, 732) {ipe disk};
  \pic[ipe mark small]
     at (112, 724) {ipe disk};
  \draw
    (220, 768)
     -- (224, 740);
  \draw
    (220, 768)
     -- (232, 748);
  \draw
    (220, 768)
     -- (240, 748);
  \draw
    (220, 768)
     -- (232, 740);
  \draw
    (248, 692)
     -- (252, 716);
  \pic[ipe mark small]
     at (248, 692) {ipe disk};
  \pic[ipe mark small]
     at (220, 768) {ipe disk};
  \pic[ipe mark small]
     at (224, 740) {ipe disk};
  \pic[ipe mark small]
     at (232, 748) {ipe disk};
  \pic[ipe mark small]
     at (232, 740) {ipe disk};
  \pic[ipe mark small]
     at (240, 748) {ipe disk};
  \pic[ipe mark small]
     at (260, 748) {ipe disk};
  \pic[ipe mark small]
     at (268, 748) {ipe disk};
  \pic[ipe mark small]
     at (268, 740) {ipe disk};
  \pic[ipe mark small]
     at (260, 728) {ipe disk};
  \pic[ipe mark small]
     at (260, 720) {ipe disk};
  \pic[ipe mark small]
     at (252, 716) {ipe disk};
  \pic[ipe mark small]
     at (252, 732) {ipe disk};
  \pic[ipe mark small]
     at (244, 724) {ipe disk};
  \pic[ipe mark small]
     at (284, 768) {ipe disk};
  \draw
    (260, 748)
     -- (284, 768);
  \draw
    (268, 748)
     -- (284, 768)
     -- (284, 768);
  \draw
    (268, 740)
     -- (284, 768);
  \draw
    (244, 724)
     -- (248, 692);
  \draw
    (252, 732)
     -- (248, 692);
  \draw
    (260, 728)
     -- (248, 692);
  \draw
    (260, 720)
     -- (248, 692);
  \node[ipe node, text=blue]
     at (140, 732) {$\blue{X}$};
  \node[ipe node, text=blue]
     at (276, 732) {$\blue{X}$};
  \filldraw[fill=lightyellow, ipe opacity 30]
    (176, 744)
     -- (188, 744)
     -- (188, 748)
     -- (196, 740)
     -- (188, 732)
     -- (188, 736)
     -- (176, 736)
     -- (176, 744);
      \draw[ipe opacity opaque]
    (176, 744)
     -- (188, 744)
     -- (188, 748)
     -- (196, 740)
     -- (188, 732)
     -- (188, 736)
     -- (176, 736)
     -- (176, 744);
  \node[ipe node]
     at (220, 772) {$v_{C_1}$};
  \node[ipe node]
     at (284, 772) {$v_{C_2}$};
  \node[ipe node]
     at (244, 684) {$v_{C_3}$};
  \node[ipe node]
     at (75.429, 776.479) {$C_1$};
  \node[ipe node]
     at (160, 776) {$C_2$};
  \node[ipe node]
     at (132, 680) {$C_3$};
\end{tikzpicture}
}
\caption{Left: The Gaifman graph of a structure $(\mathfrak{A},X),$ where $C_1, C_2,$ and $C_3$ are the vertex sets of the connected components of $G_{\mathfrak{A}}\setminus X.$
Right: The Gaifman graph of the structure ${\sf cl}_{\sf X}({\sf star}_{\sf X}(\mathfrak{A},X)).$}\labels{fig_transfor}
\end{figure}

Notice that  $\bar{Θ}_0[τ]\subseteq \bar{Θ}_{1}[τ]\subseteq \bar{Θ}_{2}[τ]\subseteq \cdots.$
We also set   $$\bar{Θ}[τ]=\bigcup_{i∈ \mathbb{N}}\bar{Θ}_{i}[τ].$$

\myskip\paragraph{The compound logic ${Θ}.$}
Let $τ$ be a vocabulary.
We now recursively define, for every $i∈ \mathbb{N},$ the subclass $Θ_{i}[τ]$  of $\MSOL[τ]$ so that
\begin{eqnarray*}
Θ_{0}[τ] & = & \{σ\wedge μ\mid  σ∈ \FOL[τ]\mbox{~and~} μ∈ \NTMC[τ]\} \mbox{~and} \\
\mbox{for $i≥ 1,$ } Θ_{i}[τ] & = & \{β\triangleright γ \mid  β∈ \MSOL^\tw[τ\cup\{{\sf X}\}]  \mbox{~and~} γ∈ \bool(Θ_{i-1}[τ]^{({\sf c})})\}.
\end{eqnarray*}

Notice that  $Θ_0\subseteq Θ_{1}\subseteq Θ_{2}\subseteq \cdots.$
We also set   $$Θ=\bigcup_{i∈ \mathbb{N}}Θ_{i}.$$

We stress that the difference between $\bar{Θ}$ and $Θ$ is that in the former we ask that $γ∈ {\bar{Θ}_{i-1}[τ]}^{({\sf c})},$ while in the latter we ask that $γ∈ \bool(Θ_{i-1}[τ]^{({\sf c})}).$
Observe that $\bar{Θ}\subseteq Θ.$

For every sentence $θ∈{Θ}[τ],$ we call every sentence $σ∈\FOL[τ]$ used in the definition of $θ$ a {\em \FOL-target sentence} of $θ$ and every sentence $μ∈\NTMC[τ]$ used the definition of $θ$ an {\em \NTMC-target sentence} of $θ.$
Also, every sentence $β∈\MSOL^\tw[τ\cup\{{\sf X}\}]$ used in the definition of $θ$ is called a {\em modulator sentence} of $θ.$\medskip

We are now in position to state our main result {using structures instead of graphs}.
Recall that in the introduction we mentioned the version of this theorem on graphs, that is when $τ=\{{\sf E}\}$ (\autoref{@decendientes1}).

\begin{theorem}
\label{@decendientes}
For every  vocabulary $τ$ and every $θ∈ Θ[τ],$
there exists an algorithm that, given a $\mathfrak{A}∈\mathbb{STR}[τ]$, outputs whether $\mathfrak{A}\models θ$
in   time $\O_{|θ|}(n^2)$.
\end{theorem}

As we mentioned in the introduction, we may also define the extension $\tilde{Θ}$ of $Θ$, by setting
$\tilde{Θ}_{0}[τ]=\FOL[τ]$ as the base case. The following is an easy corollary of \autoref{@decendientes}, that is the version of \autoref{@pertrechando1} stated on structures.
\begin{theorem}
\label{@pertrechando}
For every  vocabulary $τ$ and every $\tilde{θ}∈ \tilde{Θ}[τ],$
there exists an algorithm that, given a $\mathfrak{A}∈\mathbb{STR}[τ]$, outputs whether $\mathfrak{A}\models \tilde{θ}$
in   time $\O_{|θ|,\hw(G_\mathfrak{A})}(n^2)$.
\end{theorem}

To see this how  \autoref{@pertrechando} follows from \autoref{@decendientes} let $μ∈\MSOL[\{{\sf E}\}]$
express minor exclusion of $K_{\hw(G)}$ and enhance all target \FOL-sentences of $\tilde{θ}∈\tilde{Θ}$
by adding on them the conjunction with the {\sf EM}-target sentence $μ$. As this creates some new sentence  $θ∈ Θ$ where
$\mathfrak{A}\models \tilde{θ}\iff \mathfrak{A}\models {θ}$, \autoref{@pertrechando} follows as a consequence of \autoref{@decendientes}.

\myskip\paragraph{Quantifying the sentences in $Θ$.}
If $θ∈ Θ,$ then we define the {\em height} of $θ,$
denoted by ${\sf height}(θ),$ as the minimum $i$ such that $θ∈ Θ_{i}.$
We define the {\em treewidth} of $θ,$ denoted by $\tw(θ),$
as the maximum $c_{β}$ over all modulator sentences $β$ of $θ.$
Also we define the {\em Hadwiger number} of $θ,$ denoted by $\hw(θ),$
as the maximum $c_{μ}$ over all \NTMC-target sentences $μ$ of $θ.$
\medskip

We prove that for every model of a sentence in $Θ,$ its Gaifman graph excludes some ``big enough'' clique as a minor.
\begin{lemma}\labels{@disillusioned}
Let $τ$ be a vocabulary.
For every $θ∈ Θ[τ]$ and every $τ$-structure $\mathfrak{A}∈{\rm Mod}(θ),$ it holds that $G_\mathfrak{A}∈\excl(\{K_{c}\}),$ where $c=\hw(θ)+(\tw(θ)+1)\cdot {\sf height}(θ).$
\end{lemma}
\begin{proof}
Let $θ∈Θ[τ]$ of height $h$ and let $\mathfrak{A}$ be a $τ$-structure such that $\mathfrak{A}
∈{\rm Mod}(θ).$
We will prove that $G_{\mathfrak{A}}∈ \excl(\{K_c\}),$ where  $c=\hw(θ)+(\tw(θ)+1)\cdot h,$
by induction on $h.$

In the case where $h= 0,$ by definition there exist
a sentence $σ∈ \FOL[τ]$ and a sentence $μ∈ \NTMC[τ]$ such that
$θ=σ\wedge μ.$
The fact that $\mathfrak{A}∈ {\rm Mod}(θ)$
implies that $\mathfrak{A}\models σ\wedge μ$ and therefore, by~\autoref{@desabrocharle}, $G_{\mathfrak{A}}∈ \excl (\{K_{c_μ}\}).$
Also, by the assumption that $θ∈Θ_0,$ we have that $\hw(θ) = c_μ.$

Suppose now that $h≥ 1$ and assume that the lemma holds for all sentences of smaller height.
In this case, there is a sentence $β∈ \MSOL^\tw[τ\cup \{{\sf X}\}],$ an $\ell∈\mathbb{N}_{≥ 1},$ some $k_1,\ldots, k_\ell∈ \mathbb{N}_{≥ 1},$ and for every $i∈[\ell]$ and for every $j∈[k_i]$ there is a sentence $θ_{i,j}∈Θ_{h-1}[τ]^{({\sf c})}$ such that
$$θ = \exists {\sf X}\ β|_{{\sf star}_{\sf X}}\wedge \Big((θ_{1,1}\wedge \ldots \wedge θ_{1,k_1})\vee \ldots \vee (θ_{\ell,1}\wedge\ldots\wedge θ_{\ell, k_\ell})\Big)|_{{\sf rm}_{\sf X}}.$$
Suppose now that, given that $\mathfrak{A}\models θ$ and $c = \hw(θ)+(\tw(θ)+1)\cdot h,$ we have that $K_c\preceq_{\sf m}G_{\mathfrak{A}}.$
Let $M$ be a minor-model of $K_c$ in $G_{\mathfrak{A}},$ i.e., a subgraph of $G_{\mathfrak{A}}$ that can be transformed to $K_c$ after a series of edge contractions.
We call {\em bag} of $M$ the vertex set of every connected subgraph of $M$ whose edges are contracted in order to obtain a vertex of $K_c.$
The fact that $\mathfrak{A}\models θ$ implies that
there is a set $X\subseteq V(\mathfrak{A})$
such that
\begin{itemize}
\item ${{\sf star}_{\sf X}}(\mathfrak{A},X)\models β$ and
\item ${{\sf rm}_{\sf X}}(\mathfrak{A},X)\models (θ_{1,1}\wedge \ldots \wedge θ_{1,k_1})\vee \ldots \vee (θ_{\ell,1}\wedge\ldots\wedge θ_{\ell, k_\ell}).$
\end{itemize}
Since ${{\sf star}_{\sf X}}(\mathfrak{A},X)\models β,$ we know that the treewidth of ${\sf cl}_{\sf X}({{\sf star}_{\sf X}}(\mathfrak{A},X))$ is at most $c_β.$
Observe that this implies that $S$ intersects at most $c_β+1$ bags of $M.$
Let $M^\star$ to be the graph induced by the union of all bags in $M$ that $X$ does not intersect.
Note that $M^\star$ is a minor-model of $K_{c-(c_β+1)}.$

We now set $\mathfrak{B}: ={\sf rm}_{\sf X}(\mathfrak{A},X).$
The fact that $\mathfrak{B}\models (θ_{1,1}\wedge \ldots \wedge θ_{1,k_1})\vee \ldots \vee (θ_{\ell,1}\wedge\ldots\wedge θ_{\ell, k_\ell})$ implies that
there is an $i∈[\ell]$ such that
$\mathfrak{B}\models θ_{i,1}\wedge \ldots \wedge θ_{i,k_i}.$
Observe also that $M^\star$ is a subgraph of $G_{\mathfrak{B}}$ and recall that $M^\star$ is a minor-model of $K_{c- (c_β+1)}.$
Let  $t:= \max\{\tw(θ_{i,1}),\ldots, \tw(θ_{i,k_i})\},$ let $j∈[k_i]$ such that
$\tw(θ_{i,j}) = t,$ and let
$m:= \max\{\hw(θ_{i,1}),\ldots, \hw(θ_{i,k_i})\}.$
Since $\tw(θ) ≥ \max\{t, c_β\}$ and
$\hw(θ)≥ m,$ it holds that
$c-(c_β+1) = \hw(θ)+(\tw(θ)+1)\cdot h  - (c_β +1) ≥  m+(\tw(θ_{i,j})+1)\cdot (h-1)≥ \hw(θ_{i,j}) +(\tw(θ_{i,j})+1)\cdot (h-1).$
We set $q:=\hw(θ_{i,j}) +(\tw(θ_{i,j})+1)\cdot (h-1)$ and observe that since $K_{c-(c_β +1)}\preceq_{\sf m} G_{\mathfrak{B}}$ and $c-(c_β +1)≥ q,$ then $K_q\preceq_{\sf m} G_{\mathfrak{B}}.$
But  $θ_{i,j}∈ Θ_{h-1}^{({\sf c})}$ and by induction hypothesis, we have that $G_{\mathfrak{B}}∈ \excl(\{K_q\}),$ a contradiction.
\end{proof}

\myskip\section{An annotated version of the problem}
\label{sec_equivalent_version}
In this section we aim to define an enhanced version of every $θ∈ Θ$.
This is done in two steps.

In~\autoref{@vorstellende}, we focus on ``neutralizing'' a tuple ${\bf a}$ of elements of a structure $\mathfrak{A}$ and transforming a question on $\mathfrak{A}$ to a question on the structure obtained after ``neutralizing'' ${\bf a}$ (\autoref{@escrostonades}).
We will apply this tool under the existence of an apex set and a flat wall in the Gaifman graph of our structure, in order to ``neutralize'' the apex set and
ask the final \FOL-question of our sentence in a ``flattened'' structure, where apices can no longer ``bring close'' any distant parts of the wall. This transformation of the problem will allow the application of the ``locality-based'' strategy that uses Gaifman's locality theorem.

In~\autoref{sec_equivalent-versions_first-floor} we define an enhanced version of the problem, by replacing, in a given $θ∈Θ[τ],$
each \FOL-target sentence $σ$ of $θ$ with the sentence obtained from $σ$ after
(i) ``projecting'' it with respect to a set ${\bf c}$ of constant symbols
(using the definitions in~\autoref{@vorstellende}),
(ii) taking a Gaifman equivalent sentence of the obtained sentence,
and (iii) requiring that the ``scattered'' variables of the basic local sentences of the Gaifman sentence belong to an annotated set $R.$
We prove that the initial sentence $θ$ and the obtained sentence, denoted by $θ_{{\sf R},{\bf c}},$
are ``equivalent'' for any choice of ${\bf a}$ interpreting ${\bf c}$ and when ${\sf R}$ is interpreted as the whole universe of the given structure (\autoref{obs_addingR}).
Our algorithms will work with the sentence $θ_{{\sf R},{\bf c}}.$

\myskip\subsection{Dealing with apices}\labels{@vorstellende}
In this subsection
we introduce all necessary tools to handle the (possible) apices in the Gaifman graph of the input structure.
As we mentioned in the overview (see~\autoref{sec_overview}),
apices are an obstacle to the locality arguments needed for the part of the proof that concerns \FOL.
To be able to work in a ``flat'' graph, without the presence of the apices that possibly connect ``distant'' parts of the graph, we introduce an {\sl apex-projection} of our structure and the corresponding {\sl apex-projection} of a given \FOL-sentence.
This construction is an extension to general structures of a trick in~\cite{FlumG01fixe} (that deals with  graphs)
and gives an equivalent sentence (see~\autoref{@escrostonades}).
Finally, we express this transformation in terms of \FOL-transductions (see~\autoref{@disadvantages}) and by~\autoref{@imposibilitada},
we obtain a ``backwards translation'' of the latter question to a questions in structures of the form $(\mathfrak{A}, {\bf a})$ (see~\autoref{@recognitions}).

\myskip\paragraph{Apex-tuples of structures.}
Let $τ$ be a vocabulary, let $\mathfrak{A}$ be a $τ$-structure, and let $l∈ \mathbb{N}.$
A tuple ${\bf a}=(a_{1},\ldots, a_{l})$ where each $a_{i}$ is either an element of $V(\mathfrak{A})$ or {$\varnothing$},
is called an {\em apex-tuple} of $\mathfrak{A}$ of size $l.$
We use $V({\bf a})$ for the set containing the {non-$\varnothing$}
elements in ${\bf a}.$
Also, if $S\subseteq V(\mathfrak{A}),$ we define ${\bf  a}\cap S=(a_{1}',\ldots,a_{l}')$
so that if $a_{i}∈ S,$ then $a_i'=a_i,$  and otherwise  $a_i'=\varnothing.$
We also define ${\bf  a}\setminus S={\bf  a}\cap  (V(\mathfrak{A})\setminus S).$
Given an apex-tuple ${\bf a}$ of $\mathfrak{A}$ of size $l,$ we denote by ${\bf a}^\star$ the tuple $(a_1^\star, \ldots, a_{|V({\bf a})|}^\star),$ where for each $i∈[|V({\bf a})|],$ $a_i^\star$ is the $i$-th {non-$\varnothing$}
element of ${\bf a}$ (intuitively, ${\bf a}^\star$ can be seen as the substring of ${\bf a}$ obtained from ${\bf a}$ after removing  every occurrence of $\varnothing$).

\myskip\paragraph{Constant-projections of vocabularies.}
Let $τ$ be a vocabulary, let $l∈ \mathbb{N},$ and let ${\bf c} = \{{\sf c}_1,\ldots, {\sf c}_l\}$ be a collection of $l$ constant symbols that are not contained in $τ.$
We define the {\em constant-projection $τ^{\langle \bf c\rangle}$ of $(τ\cup{\bf c})$} to be the vocabulary obtained from $(τ\cup{\bf c})$ as follows:
For each ${\sf R}∈ τ$ of arity $r≥ 2,$ we consider a collection
${\cal R}_{\sf R} = \{{\sf R}_r, {\sf R}_{r-1}, \ldots, {\sf R}_{1}\}$ of relation symbols,
where ${\sf R}_i$ has arity $i,$
a collection ${\cal R}_{\sf R}^{({\sf ap})} = \{{\sf R}_r^{({\sf ap})},{\sf R}_{r-1}^{({\sf ap})}, \ldots, {\sf R}_1^{({\sf ap})}\}$ of relation symbols,
where ${\sf R}_i^{({\sf ap})}$ has arity $i,$ and a collection ${\cal Y}_{\sf R} = \{Y_{\sf R}^{(1)}, \ldots, Y_{\sf R}^{(l)}\}$ of unary relation symbols.
We set $$τ^{\langle \bf c\rangle}:= {\bf c}\cup\cupall\{{\cal R}_{\sf R}\cup {\cal R}^{({\sf ap})}_{\sf R}\cup {\cal Y}_{\sf R}\mid {\sf R}∈ τ\text{ and has arity at least two}\}\cup \{{\sf R}\mid {\sf R}∈ τ \text{ and has arity one}\}.$$

\myskip\paragraph{Projecting a structure with respect to an apex-tuple.}
Let $τ$ be a vocabulary, let $l∈ \mathbb{N},$ and let ${\bf c} = \{{\sf c}_1,\ldots, {\sf c}_l\}$ be a collection of $l$ constant symbols that are not contained in $τ.$
Let also  $τ^{\langle \bf c\rangle}$  be the constant-projection of $(τ\cup{\bf c}).$
Given a $(τ\cup{\bf c})$-structure $(\mathfrak{A}, {\bf a}),$ where ${\bf a}=(a_1, \ldots, a_l)$ is an apex-tuple of $\mathfrak{A}$ of size $l$ and, for every $i∈ [l]$ ,${\sf c}_i^{(\mathfrak{A},{\bf a})} = a_i,$
we define the structure ${\sf ap}_{\bf c}(\mathfrak{A},{\bf a})$ to be the $τ^{\langle \bf c\rangle}$-structure obtained as follows:
\begin{itemize}
\item $V({\sf ap}_{\bf c}(\mathfrak{A},{\bf a})) = V(\mathfrak{A}),$
\item for every $i∈ [l]$ ${\sf c}_i^{{\sf ap}_{\bf c}(\mathfrak{A},{\bf a})} = a_i,$
\item every ${\sf R}∈ τ$ of arity one is interpreted in ${{\sf ap}_{\bf c}(\mathfrak{A},{\bf a})}$ as in ${\mathfrak{A}},$
\item for every ${\sf R}∈ τ$ of arity $r≥ 2$ and for every $j∈ [r],$
${\sf R}_{j}$ is interpreted in ${\sf ap}_{\bf c}(\mathfrak{A},{\bf a})$ as the set
$${\{{\bf x}∈ {V({\sf ap}_{\bf c}(\mathfrak{A},{\bf a}))}^{j}\mid \exists {\bf y}∈ {\sf R}^{\mathfrak{A}} \mbox{ such that }({\bf y}\setminus V({\bf a}))^\star = {\bf x} \},}$$
\item for every ${\sf R}∈ τ$ of arity $r≥ 2$ and for every $j∈ [r],$
${\sf R}_j^{({\sf ap})}$ is interpreted in ${\sf ap}_{\bf c}(\mathfrak{A},{\bf a})$ as the set
$$\{{\bf x}∈ {V({\sf ap}_{\bf c}(\mathfrak{A},{\bf a}))}^{j}\mid \exists {\bf y}∈ {\sf R}^{\mathfrak{A}} \mbox{ such that }({\bf y}\cap V({\bf a}))^\star = {\bf x} \},\text{ and}$$
\item for every ${\sf R}∈ τ$ of arity $r≥ 2$ and for every $i∈ [l],$  $Y_{\sf R}^{(i)}$ is interpreted in ${\sf ap}_{\bf c}(\mathfrak{A},{\bf a})$ as the set $$\{v∈ V(\mathfrak{A}) \mid v\notin V({\bf a}) \mbox{ and }\exists {\bf x}∈ {\sf R}^{\mathfrak{A}}\mbox{ such that } a_i, v∈ V({\bf x})\}.$$
\end{itemize}
Notice that if $a_i = \varnothing,$ $Y_{\sf R}^{(i)}$ is interpreted in ${\sf ap}_{\bf c}(\mathfrak{A},{\bf a})$ as the empty set.

Intuitively, given a structure $(\mathfrak{A},{\bf a}),$ ${\sf ap}_{\bf c}(\mathfrak{A},{\bf a})$ is obtained from $(\mathfrak{A},{\bf a})$ after ``coloring'' with the color $Y_{\sf R}^{(i)}$ every element of $V(\mathfrak{A})\setminus V({\bf a})$ that is related (in ${\sf R}$) to $a_i$ (or, adjacent to $a_i,$ in the case of graphs) and after ``restricting'' every relation on $\mathfrak{A}$ to the set $V(\mathfrak{A})\setminus V({\bf a}).$ For the latter, we correspond each ${\sf R}∈ τ$ of arity $r≥ 2$ to a collection ${\sf R}_{1}, \ldots, {\sf R}_r$ of relation symbols that are interpreted as the ``restricted'' tuples in ${\sf R}^{\mathfrak{A}}.$
In other words, for every $i∈[r],$ ${\sf R}_i$ is interpreted as the set of all $i$-tuples of elements of $V(\mathfrak{A})$ that are obtained from an $r$-tuple in ${\sf R}^{\mathfrak{A}}$ after removing from it  all elements in $V({\bf a})$ (that are eventually $(l-i)$-many). Similarly, we correspond each ${\sf R}∈ τ$ of arity $r≥ 2$ to a collection ${\sf R}_{1}^{({\sf ap})}, \ldots, {\sf R}_r^{({\sf ap})}$ of relation symbols that are interpreted as the restriction of ${\sf R}^{\mathfrak{A}}$ to the set $V({\bf a}).$

It is crucial to observe that the Gaifman graph of ${\sf ap}_{\bf c}(\mathfrak{A},{\bf a})$ is a subgraph of $G_{\mathfrak{A}}.$
In fact, $G_{{\sf ap}_{\bf c}(\mathfrak{A},{\bf a})}$ is obtained from $G_{\mathfrak{A}}$ after removing every edge that is between a vertex in $V({\bf a})$ and a vertex in $V(\mathfrak{A})\setminus V({\bf a}).$ This removal permits us
to deal with ``flat structures'' that are amenable to the application of Gaifman's Theorem.

\myskip\paragraph{Apex-projected sentences.}
{Let ${\bf x}$ be a tuple of variables of size $r.$
For every $j∈[r],$ we define ${\cal A}_j({\bf x})$ to be the set of all possible partitions of ${\bf x}$ to two subtuples ${\bf z},{\bf w}$ of size $j$ and $r-j,$ respectively.
}
Let $τ$ be a vocabulary, let $l∈ \mathbb{N},$ and let ${\bf c} = \{{\sf c}_1,\ldots, {\sf c}_l\}$ be a collection of $l$ constant symbols that are not contained in $τ.$
For every sentence $σ∈\FOL[τ],$ we define its {\em $l$-apex-projected sentence} $σ^l$ to be the sentence obtained from $σ$ by replacing, for every ${\sf R}∈ τ$ of arity $r≥ 2,$ every term ${\bf x}∈ {\sf R},$ where ${\bf x}$ is a tuple of size $r,$ by
\begin{eqnarray*}
\bigvee_{j∈ [r]} \bigvee_{({\bf z},{\bf w})∈ {\cal A}_j({\bf x})}
\Bigg({\bf z}∈ {\sf R}_j \wedge {\bf w}∈ {\sf R}_{r-j}^{({\sf ap})}
\wedge   \bigvee_{{\bf t}∈ [l]^{(r-j)}} \Big(\bigwedge_{i∈ [r-j]} \big({\sf w}_i  = {\sf c}_{t_i} \wedge \bigwedge_{{\sf z}∈ {\bf z}} {\sf z}∈ Y_{\sf R}^{(t_i)}\big) \Big)\Bigg).
\end{eqnarray*}

To get some intuition of the meaning of the above sentence,
notice that the part ``$\bigvee_{j∈ [r]} \bigvee_{({\bf z},{\bf w})∈ {\cal A}_j({\bf x})}$''
corresponds to some ``guessing'' of a subtuple ${\bf z}$ of ${\bf x}$ of size $j.$
The tuple ${\bf z}$ corresponds to the tuple $({\bf x}\setminus V({\bf a}))^\star,$
while the tuple ${\bf w}$ corresponds to the tuple $({\bf x}\cap V({\bf a}))^\star.$
Therefore, in ``${\bf z}∈ {\sf R}_{j}$'', we ask that ${\bf z}$ belongs to the ``projection'' of ${\sf R}$ away from ${\bf a},$
while in ``${\bf w}∈ {\sf R}_{r-j}^{({\sf ap})}$'' we ask that ${\bf w}$ belongs to the ``projection'' of ${\sf R}$ inside ${\bf a}.$
We then guess a tuple of apices in $V({\bf a})$ (this corresponds to ``$\bigvee_{{\bf t}∈ [l]^{(r-j)}}$'').
Having these guessed apices,
for each one of them (``$\bigwedge_{i∈ [r-j]}$'') we ask that the order of the elements in ${\bf w}$ coincides with the order of the guessed apices in ${\bf c}$ (``${\sf w}_i = {\sf c}_{t_i}$'')
and
in ``$\bigwedge_{{\sf z}∈ {\bf z}} {\sf z}∈ Y_{\sf R}^{(t_i)}$'' we ask that all the elements in ${\bf z}$ are colored with the color corresponding to each guessed apex.
\smallskip

The definition of the $l$-apex-projected sentence $σ^l$ together with the above discussion imply the following lemma, which can be seen as a generalization to general structures
of~\cite[Lemma 26]{FlumG01fixe} that deals with graphs.

\begin{lemma}
\labels{@escrostonades}
Let $τ$ be a vocabulary, let $l∈ \mathbb{N},$ and let ${\bf c}$ be a collection of $l$ constant symbols.
For every $σ∈ \FOL[τ],$ every $τ$-structure $\mathfrak{A},$ and every apex-tuple  ${\bf a}$ of $\mathfrak{A}$ of size $l,$ it holds that $\mathfrak{A}\models σ \iff {\sf ap}_{\bf c}(\mathfrak{A},{\bf a})\models σ^l$ (where ${\bf c}$ is interpreted as ${\bf a}$).
\end{lemma}

%

\myskip\paragraph{Backwards translating an apex-projected sentence.}
We now aim to prove that
given a vocabulary $τ,$ an $l∈\mathbb{N},$ a collection ${\bf c}$ of $l$ constant symbols, and a sentence $σ∈\FOL[τ],$
we can find a sentence $σ'∈ \FOL[τ\cup {\bf c}]$ such that
for every $τ$-structure $\mathfrak{A}$ and every apex-tuple ${\bf a}$ of $\mathfrak{A}$ of size $l,$
$(\mathfrak{A},{\bf a})\models σ'\iff {\sf ap}_{\bf c}(\mathfrak{A},{\bf a})\models σ^l.$
For this reason, we first prove that the function ${\sf ap}_{\bf c}$ is an \FOL-transduction
and we then use~\autoref{@imposibilitada} to obtain the desired sentence $σ'$ (see~\autoref{@recognitions}).
{We stress that, in~\autoref{sec_trans}, we avoided to define transductions as relations between structures of vocabularies with constant symbols, for the sake of simplicity.
In our current case, we slightly abuse the definition of transductions and allow constant symbols, since
the function ${\sf ap}_{\bf c}$ leaves the interpretation of ${\bf c}$ intact and therfore we can safely extend the definition of transduction and the statement of~\autoref{@imposibilitada} to capture this case.
We refer the reader to~\cite[Section 7.1.2]{CourcelleE12grap} for a discussion on transductions between structures with constants.}

\begin{lemma}
\labels{@disadvantages}
Let $τ$ be a vocabulary, let $l∈ \mathbb{N},$ let ${\bf c}$ be a collection of $l$ constant symbols, and let $τ^{\langle \bf c\rangle}$ be the constant-projection of $τ\cup{\bf c}.$
The function that maps every $(τ\cup{\bf c})$-structure $(\mathfrak{A}, {\bf a})$ to the $τ^{\langle \bf c\rangle}$-structure ${\sf ap}_{\bf c}(\mathfrak{A},{\bf a})$ is an \FOL-transduction.
{Moreover, there is an \FOL-transduction from $τ^{\langle \bf c\rangle}$ to $τ\cup{\bf c}$ that maps  ${\sf ap}_{\bf c}(\mathfrak{A},{\bf a})$ to  $(\mathfrak{A}, {\bf a})$, if $G_{\mathfrak{A}}$ has bounded Hadwiger number.}
\end{lemma}
\begin{proof}
Let $(\mathfrak{A}, {\bf a})$ be a $(τ\cup{\bf c})$-structure.
We will describe an interpretation ${\cal I}$ with input vocabulary $τ\cup{\bf c}$ and output vocabulary $τ^{\langle \bf c\rangle},$ such that ${\cal I}(\mathfrak{A}, {\bf a}) ={\sf ap}_{\bf c}(\mathfrak{A},{\bf a}).$
To define this interpretation, we have to provide the formulas
$\{φ_{\sf dom}, φ_{\sf univ}\} \cup \{φ_{\sf R}\}_{{\sf R}∈ τ^{\langle \bf c\rangle}}.$
The formulas that we give are formulas of $\FOL[τ\cup{\bf c}],$ and therefore ${\cal I}$ is an $\FOL$-transduction.

First, we define $φ_{\sf dom}$ and $φ_{\sf univ}({\sf x})$ to be two formulas in $\FOL[τ\cup{\bf c}]$ that are always true.
Then, it remains to define, for each relation symbol ${\sf R}∈ τ^{\langle \bf c\rangle},$
a formula $φ_{\sf R}$ (the interpretation of the constants remains unchanged).

Recall that  $τ^{\langle \bf c\rangle}=  {\bf c}\cup\cupall\{{\cal R}_{\sf R}\cup {\cal R}^{({\sf ap})}_{\sf R}\cup {\cal Y}_{\sf R}\mid {\sf R}∈ τ\text{ and has arity $r≥ 2$}\}\cup \{{\sf R}\mid {\sf R}∈ τ \text{ and has arity one}\}$ and for every ${\sf R}∈ τ$ of arity $r≥ 2,$ ${\cal R}_{\sf R} = \{{\sf R}_r, {\sf R}_{r-1}, \ldots, {\sf R}_{1}\}$ is a collection of relation symbols,
where ${\sf R}_i$ has arity $i,$
${\cal R}_{\sf R}^{({\sf ap})} = \{{\sf R}_r^{({\sf ap})},{\sf R}_{r-1}^{({\sf ap})}, \ldots, {\sf R}_1^{({\sf ap})}\}$ is a collection of relation symbols,
where ${\sf R}_i^{({\sf ap})}$ has arity $i,$ and ${\cal Y}_{\sf R} = \{Y_{\sf R}^{(1)}, \ldots, Y_{\sf R}^{(l)}\}$ is a collection of unary relation symbols.

First, for each ${\sf R}∈ τ$ of arity one, we set $φ_{\sf R}({\sf x}) := ({\sf x}∈ {\sf R}).$
Let now ${\sf R}∈ τ$ of arity $r≥ 2.$
For every $j∈[r],$ we define
$φ_{{\sf R}_j}({\sf x}_1,\ldots, {\sf x}_j)$ to be the formula that checks whether there exist ${\sf y}_1, \ldots, {\sf y}_r$ such that $({\sf y}_1,\ldots,{\sf y}_r)∈ {\sf R}$ and whether there is a set $I\subseteq [r]$ of size $j$ such that
for every $i\notin I,$ ${\sf y}_i$ is equal to some ${\sf c}∈ {\bf c}$ and if $I = \{i_1, \ldots, i_j\}$ then for every $k∈[j],$ ${\sf x}_k = {\sf y}_{i_k}.$
More formally,
\begin{eqnarray*}
φ_{{\sf R}_j}({\sf x}_1,\ldots, {\sf x}_j)& :=  &\exists {\sf y}_1, \ldots, {\sf y}_r (({\sf y}_1,\ldots,{\sf y}_r)∈ {\sf R})\wedge\\
& &\bigvee_{\substack{\text{all subsets}\\ I = \{i_1, \ldots, i_j\}\\ \text{of $[r]$ of size $j$}}}
\Bigg(\Big(i\notin I \to (\exists {\sf c}∈{\bf c}\  {\sf y}_i = {\sf c})\Big)\wedge
\bigwedge_{k∈[j]} {\sf x}_i = {\sf y}_{i_k}\Bigg).
\end{eqnarray*}
Also, for every $j∈[r],$ we define
$φ_{{\sf R}_j^{({\sf ap})}}({\sf x}_1,\ldots, {\sf x}_j)$ to be the formula that
checks whether there exist ${\sf y}_1, \ldots, {\sf y}_r$
and whether there is a set $I\subseteq[r]$ of size $j$ such that for every $i\notin I,$
${\sf y}_i\notin{\bf c}$ and
if $I = \{i_1, \ldots, i_j\}$ then for every $k∈[j],$ ${\sf x}_k = {\sf y}_{i_k}.$
More formally,
\begin{eqnarray*}
φ_{{\sf R}_j^{({\sf ap})}}({\sf x}_1,\ldots, {\sf x}_j)& :=  &\exists {\sf y}_1, \ldots, {\sf y}_r (({\sf y}_1,\ldots,{\sf y}_r)∈ {\sf R})\wedge\\
& &\bigvee_{\substack{\text{all subsets}\\ I = \{i_1, \ldots, i_j\}\\ \text{of $[r]$ of size $j$}}}
\Bigg(\Big(i\notin I \to \neg(\exists {\sf c}∈{\bf c}\  {\sf y}_i = {\sf c})\Big)\wedge
\bigwedge_{k∈[j]} ({\sf y}_{i_k}∈ {\bf c}\wedge {\sf x}_i = {\sf y}_{i_k})\Bigg).
\end{eqnarray*}
Finally, for every $i∈ [l],$ we define
$$
φ_{Y_{\sf R}^{(i)}}({\sf x}) = (\forall {\sf c}∈ {\bf c}\ {\sf x}\neq {\sf c}) \wedge \exists {\sf y}_1, \ldots, {\sf y}_r \Big(({\sf y}_1, \ldots, {\sf y}_r)∈ {\sf R} \wedge \bigvee_{j,k∈ [r], j\neq k} ({\sf c}_i = {\sf y}_i \wedge {\sf x} = {\sf y}_k)\Big).
$$
It is easy to see that all above formulas are in $\FOL[τ\cup{\bf c}]$ and that ${\cal I}(\mathfrak{A},{\bf a}) = {\sf ap}_{\bf c}(\mathfrak{A},{\bf a}).$

{
The existence of a \FOL-transduction from $τ^{\langle \bf c\rangle}$ to $τ\cup{\bf c}$ that maps  ${\sf ap}_{\bf c}(\mathfrak{A},{\bf a})$ to  $(\mathfrak{A}, {\bf a})$, if $G_{\mathfrak{A}}$ has bounded Hadwiger number, follows from~\autoref{@transducing_str_from_gr}.}
\end{proof}

Combining~\autoref{@imposibilitada} and~\autoref{@disadvantages}, we get the following:
\begin{corollary}\labels{@recognitions}
Let $τ$ be a vocabulary, let $l∈ \mathbb{N},$ let ${\bf c}$ be a collection of $l$ constant symbols, and let $τ^{\langle \bf c\rangle}$ be the constant-projection of $τ\cup{\bf c}.$
For every sentence $φ∈ \FOL[τ^{\langle \bf c\rangle}],$ there exists a sentence $φ|_{{\sf ap}_{\bf c}}∈ \FOL[τ\cup{\bf c}]$ such that for every
$τ^{\langle \bf c\rangle}$-structure $\mathfrak{B}$, if $\mathfrak{B}={\sf ap}_{\bf c}(\mathfrak{A},{\bf a})$ for some
$(τ\cup{\bf c})$-structure $(\mathfrak{A}, {\bf a}),$ it holds that $(\mathfrak{A}, {\bf a})\models φ|_{{\sf ap}_{\bf c}} \iff {\sf ap}_{\bf c}(\mathfrak{A},{\bf a})\models φ.$
\end{corollary}

Concluding this subsection, we present~\autoref{@productrices} that summarizes the notations introduced above for the different kinds of formulas that we consider.
\begin{table}[H]
\centering
 \begin{tabular}{|| c | c | c||}
 \hline
 Formulas & Relation with $σ$ & Supporting results\\ [0.5ex]
 \hline\hline
$\breve{σ}$ & \mbox{Equivalent Gaifman sentence of a sentence $σ∈\FOL[τ]$} &~\autoref{lem_gaifman} \\
 \hline
 $σ|_{\sf f}$ & Given a transduction ${\sf f},$ $\mathfrak{A}\models σ|_{\sf f} \iff {\sf f}(\mathfrak{A})\models σ$ &~\autoref{@imposibilitada}\\
 \hline
 $σ^l$ & Formula obtained after ``projecting'' w.r.t. a tuple ${\bf c}$ of size $l$ &~\autoref{@escrostonades} \\
 \hline
\end{tabular}\caption{List of notations used on formulas, with their respective meaning and the results indicating their relation to an initial formula $σ.$}\labels{@productrices}
\end{table}

\myskip\subsection{Introducing an annotation}
\labels{sec_equivalent-versions_first-floor}

In this subsection we present a way to ``slightly modify'' our sentences in order to construct an enhanced version of every sentence in $Θ.$
Based on the results of~\autoref{@vorstellende}, we first consider for each \FOL-target sentence of our given sentence $θ∈Θ$ its $l$-apex-projected sentence $σ^l.$
For each one of them, we take an equivalent Gaifman sentence of it.
Finally, we add an additional unary relation symbol ${\sf R}$ to our vocabulary and we ask that the interpretation of the ``scattered'' variables of each Gaifman sentence are made ``inside'' the interpretation of ${\sf R}$ in our structure.
This idea is borrowed from~\cite{FominGST20analgo} but here, on the top of it,  we also incorporate the ``apex-projection'' in order to be able to apply locality arguments inside a ``flat'' graph.

\myskip\paragraph{Restricting the domain of variables.}
Let $τ$ be a vocabulary, let $l∈\mathbb{N},$ let ${\bf c}$ be a collection of $l$ constant symbols,
and let ${\sf R}\notin τ$ be a unary relation symbol.
We now describe how to define an {\em enhanced version} $θ_{{\sf R},{\bf c}}$ of a sentence $θ∈ Θ[τ].$

For every \FOL-target sentence $σ$ of $θ$ we do the following:
we consider the $l$-apex-projected sentence $σ^l∈ \FOL[τ^{\langle \bf c\rangle}]$ and we denote it by $ζ.$
By~\autoref{lem_gaifman}, there is a Gaifman sentence $\breve{ζ}∈\FOL[τ^{\langle \bf c\rangle}]$ that is equivalent to $ζ.$
Since $\breve{ζ}$ is a Gaifman sentence,
there exist $p∈ \mathbb{N}_{≥ 1},$ $r_1,\ldots, r_p, \ell_1, \ldots, \ell_p∈ \mathbb{N}_{≥ 1},$ and a collection of sentences $ζ_1,\ldots, ζ_p∈ \FOL[τ^{\langle \bf c\rangle}]$ such that $\breve{ζ}$ is a Boolean combination of $ζ_1 ,\ldots, ζ_p$ and, for every $h∈[p],$ every $ζ_h$ is a basic local sentence with parameters $\ell_h$ and $r_h,$ i.e.,
\begin{eqnarray*}
ζ_h = \exists {\sf x}_{1}\ldots\exists {\sf x}_{\ell_{h}}\big(\bigwedge_{1≤ i<j≤ \ell_{h}} d({\sf x}_{i}, {\sf x}_{j})> 2 r_{h}\wedge  \bigwedge_{i∈ [\ell_{h}]}ψ_{h}({\sf x}_{i})\big),
\end{eqnarray*}
where $ψ_h$ is an $r_h$-local formula  in $\FOL[τ^{\langle \bf c\rangle}]$ with one free variable.

Given a Gaifman sentence $\breve{ζ}∈\FOL[τ^{\langle \bf c\rangle}]$ as above that is a Boolean combination of sentences $ζ_1, \ldots, ζ_p∈ \FOL[τ],$ we define the sentence $\breve{ζ}_{\sf R}$ to be the sentence in $\FOL[τ^{\langle \bf c\rangle} \cup\{{\sf R}\}]$ that is the same Boolean combination of sentences $\tilde{ζ}_1, \ldots, \tilde{ζ}_p∈ \FOL[τ^{\langle \bf c\rangle}\cup\{{\sf R}\}]$ such that, for every $h∈[p],$
\begin{eqnarray*}
\tilde{ζ}_{h}=\exists {\sf x}_{1}\ldots\exists {\sf x}_{\ell_{h}}\big(\bigwedge_{i∈ [\ell_{h}]}{\sf x}_{i}∈ {\sf R}\wedge \bigwedge_{1≤ i<j≤ \ell_{h}} d({\sf x}_{i}, {\sf x}_{j})> 2r_{h}\wedge \bigwedge_{i∈ [\ell_{h}]}ψ_{h}({\sf x}_{i})\big).
\end{eqnarray*}

We define an {\em enhanced version} $θ_{{\sf R},{\bf c}}$ of $θ$  to be a sentence obtained
from $θ$ after replacing each \FOL-target sentence $σ$ of $θ$  with $\breve{ζ}_{\sf R}|_{{\sf ap}_{\bf c}},$ where $ζ=σ^l.$
Note that since $\breve{ζ}_{\sf R}∈\FOL[τ^{\langle \bf c\rangle}\cup\{{\sf R}\}],$ it holds that $\breve{ζ}_{\sf R}|_{{\sf ap}_{\bf c}}∈ \FOL[τ\cup \{{\sf R}\}\cup{\bf c}],$ which in turn implies that $θ_{{\sf R},{\bf c}}∈ \MSOL[τ\cup \{{\sf R}\}\cup{\bf c}].$
We also stress that, because of Gaifman's theorem (\autoref{lem_gaifman}),
for every sentence $ζ,$ there may exist {\sl many} different Gaifman sentences that are equivalent to $ζ.$
Due to this fact, a sentence $θ∈ Θ[τ]$ can have {\sl many} enhanced versions.
However, all the enhanced versions of $θ$ are equivalent.
{On the other hand, the proof of Gaifman's theorem implies that there is
{one} effectively computable Gaifman sentence that is equivalent to the given sentence $ζ$.}
\medskip

We now prove the equivalence between $θ$ and an enhanced version $θ_{{\sf R},{\bf c}}$ of $θ.$\smallskip

\begin{lemma}\labels{obs_addingR}
Let $τ$ be a vocabulary, ${\sf R}\notin τ$ be a unary relation symbol, and ${\bf c}$ be a collection of $l$ constant symbols, where $l∈\mathbb{N}_{≥ 1}.$
Also, let $θ∈Θ[τ]$ and let $θ_{{\sf R},{\bf c}}$ be an enhanced version of $θ.$
For every $τ$-structure $\mathfrak{A}$ and for every apex-tuple ${\bf a}$ of $\mathfrak{A}$ of size $l$, it holds that $\mathfrak{A}\models θ \iff (\mathfrak{A},V(\mathfrak{A}), {\bf a})\models θ_{{\sf R},{\bf c}},$ where ${\sf R}$ is interpreted as $V(\mathfrak{A})$ and ${\bf c}$ is interpreted as ${\bf a}.$
\end{lemma}

\begin{proof}
Let $σ$ be an \FOL-target sentence of $θ$ and let $ζ=σ^l.$
By~\autoref{@escrostonades},
for every $τ$-structure $\mathfrak{B}$ and every apex-tuple ${\bf a}$ of $\mathfrak{B}$ of size $l$, it holds that
 $\mathfrak{B}\models σ \iff {\sf ap}_{\bf c}(\mathfrak{B},{\bf a})\models ζ,$ where ${\bf c}$ is interpreted as ${\bf a}.$
 Also, observe that
 since ${\sf R}$ is a unary relation symbol and by the definition of the function ${\sf ap}_{\bf c},$
the structures $({\sf ap}_{\bf c}(\mathfrak{B},{\bf a}), V({\sf ap}_{\bf c}(\mathfrak{B},{\bf a})))$ and ${\sf ap}_{\bf c}(\mathfrak{B},V(\mathfrak{B}),{\bf a})$ are the same.
This implies that
${\sf ap}_{\bf c}(\mathfrak{B},{\bf a})\models ζ \iff {\sf ap}_{\bf c}(\mathfrak{B},V(\mathfrak{B}),{\bf a})\models \breve{ζ}_{\sf R},$ where ${\sf R}$ is interpreted as $V(\mathfrak{B}).$
Thus, by~\autoref{@recognitions}, $\mathfrak{B}\models σ \iff (\mathfrak{B},V(\mathfrak{B}),{\bf a})\models \breve{ζ}_{\sf R}|_{{\sf ap}_{\bf c}}.$
\end{proof}

Observe that, for every \FOL-target sentence $σ$ of $θ,$
by~\autoref{@escrostonades},
for every $τ$-structure $\mathfrak{A}$
and for every two apex-tuples ${\bf a}_1, {\bf a}_2$ of $\mathfrak{A}$ of size $l$,
it holds that
${\sf ap}_{\bf c}(\mathfrak{A},{\bf a}_1)\models σ^l\iff {\sf ap}_{\bf c}(\mathfrak{A},{\bf a}_2)\modelsσ^l.$
Therefore, we can prove the following:

\begin{lemma}\labels{lem_no_matter_which_apex}
Let $τ$ be a vocabulary, ${\sf R}\notin τ$ be a unary relation symbol, and ${\bf c}$ be a collection of $l$ constant symbols, where $l∈\mathbb{N}_{≥ 1}.$
Also, let $θ∈Θ[τ]$ and let $θ_{{\sf R},{\bf c}}$ be an enhanced version of $θ.$
For every $τ$-structure $\mathfrak{A},$ for every $R\subseteq V(\mathfrak{A}),$ and for every two apex-tuples ${\bf a}_1, {\bf a}_2$ of $\mathfrak{A}$ of size $l$, it holds that $(\mathfrak{A},R, {\bf a}_1)\models θ_{{\sf R},{\bf c}} \iff (\mathfrak{A},R, {\bf a}_2)\models θ_{{\sf R},{\bf c}}.$
\end{lemma}

In~\autoref{@connaissances}, we present all formulas needed  to define $θ_{{\sf R},{\bf c}}.$
\begin{table}[H]
\centering
\bgroup
\def\arraystretch{1.2}
 \begin{tabular}{|| c | c||}
 \hline
 Formulas & Meaning\\
 \hline\hline
 $σ$ & a \FOL-target sentence of $θ$\\
 \hline
 $ζ$ & the $l$-apex-projected sentence $σ^l$ of $σ$\\
\hline
$\breve{ζ}$ & a Gaifman sentence equivalent to $ζ$\\
\hline
 $ψ_h$ & $r$-local formulas of the basic local sentences of $\breve{ζ}$\\
 \hline
$\breve{ζ}_{\sf R}$ & the Gaifman sentence $\breve{ζ}$ after adding ${\sf R}$ (whose model is of the form ${\sf ap}_{\bf c} (\mathfrak{A},{\sf R})$)\\
\hline
$\breve{ζ}_{\sf R} |_{{\sf ap}_{\bf c}}$ & the ``backwards translation'' of $\breve{ζ}_{\sf R}$ to structures without ``projecting'' ${\bf c}$\\
\hline
\raisebox{-2.5mm}{$θ_{{\sf R},{\bf c}}$} & the sentence obtained from $θ$ after replacing\\[-2.5mm]
& every \FOL-target sentence $σ$ of $θ$ with the respective $\breve{ζ}_{\sf R} |_{{\sf ap}_{\bf c}}$\\
\hline
\end{tabular}\caption{List of formulas to define an enhanced version of a sentence $θ∈ Θ.$}\labels{@connaissances}
\egroup
\end{table}

\myskip\section{Preliminary tools}
\labels{sec_main_tools}
In this section we present a series of preliminary results required for our algorithm and its proof of correctness.

Our first tool, presented in~\autoref{sec_courc}, deals with boundaried structures (a generalization of boundaried graphs).
Given a sentence $φ∈ \MSOL[τ],$ we define an equivalence relation on boundaried structures with respect to the (partial) satisfaction of $φ.$ A variant of Courcelle's theorem (\autoref{cou_more}) indicates that there is a finite set of sentences that are evaluated on boundaried structures and are ``representatives'' of the equivalence classes defined by the above equivalence relation.
These ``representatives'' will help us to ``finitize'' the way  a sentence is partially  satisfied (or not) in a boundaried part of our structure.

In~\autoref{sec_bidimensionality} we define the notion of {\sl bidimensionality} of a vertex set $X$ of a graph $G$ with respect to a flatness pair (by flatness pair, here, we mean a flat wall $W$ together with a tuple $\mathfrak{R}$ that certifies its flatness, as defined in \cite{SauST21amor}; see~\autoref{label_exceptionalness} for a formal definition).
This concept is a measure of  the ``dispersion'' of $X$ inside the ``bidimensional territories'' of a flatness pair. We present two results on these notions, namely \autoref{@congregation} and \autoref{lemma_many_apices}. These two results will be crucial for our algorithm and its correctness.

In~\autoref{sec_privi} we define the notion of privileged component, that will express the part of our input structure that, under the presence of a ``big enough'' flatness pair, contains the ``bulk'' of the wall of this flatness pair.


\myskip\subsection{A variant of Courcelle's theorem}\labels{sec_courc}
In this subsection we aim to present a variant of Courcelle's theorem (\autoref{cou_more}).
We start with some definitions on {\sl boundaried structures}.

\myskip\paragraph{Boundaried structures.}
Given a vocabulary $τ$ and a non-negative integer $\ell,$ an {\em $\ell$-boundaried $τ$-structure} is a tuple $(\mathfrak{A},x_1, \ldots, x_\ell),$ also denoted by $(\mathfrak{A}, {\bf x}),$ where $\mathfrak{A}$ is a $τ$-structure and $x_i∈ V(\mathfrak{A}),$ $i∈[\ell].$
A {\em boundaried $τ$-structure}  is an {$\ell$-boundaried $τ$-structure}, for some $\ell∈ \mathbb{N}.$
Given a vocabulary $τ,$ we denote by ${\cal B}_τ$ the set of all boundaried $τ$-structures
and, given an $\ell∈ \mathbb{N},$ we denote by ${\cal B}_τ^{(\ell)}$ the set of all $\ell$-boundaried $τ$-structures.
We treat  \MSOL-sentences evaluated on $\ell$-boundaried $τ$-structures, as sentences in $\MSOL[τ\cup\{{\sf b}_1,\ldots, {\sf b}_\ell\}],$
where ${\sf b}_1, \ldots, {\sf b}_\ell$ are constant symbols not contained in $τ.$

Let $\ell∈ \mathbb{N}.$
We say that two $\ell$-boundaried $τ$-structures  $(\mathfrak{A},{\bf x}),(\mathfrak{B}, {\bf y})∈ {\cal B}_τ^{(\ell)}$
are {\em compatible} if there is a function that maps $x_i$ to $y_i,$ for every $i∈ [\ell],$ that is an isomorphism from $\mathfrak{A}[V({\bf x})]$ to $\mathfrak{B}[V({\bf y})].$
Given two compatible $\ell$-boundaried $τ$-structures $(\mathfrak{A},{\bf x})$ and $(\mathfrak{B}, {\bf y}),$
we define $(\mathfrak{A},{\bf x})\oplus (\mathfrak{B}, {\bf y})$ as the $τ$-structure obtained
if we take the disjoint union of $\mathfrak{A}$ and $\mathfrak{B}$
and, for every $i ∈ [\ell],$ we identify the elements $x_i$ and $y_i.$

Let $τ$ be a vocabulary and let $φ∈ \MSOL[τ].$
We say that two $\ell$-boundaried $τ$-structures $(\mathfrak{A},{\bf x}),(\mathfrak{B}, {\bf y})∈ {\cal B}_τ^{(\ell)}$
are {\em $(φ,\ell)$-equivalent}, and we denote it by $(\mathfrak{A},{\bf x}) \equiv_{φ,\ell} (\mathfrak{B}, {\bf y}),$ if they are compatible and $$\forall \ (\mathfrak{C},{\bf z})∈ {\cal B}_τ^{(\ell)}, \  (\mathfrak{C},{\bf z})\oplus(\mathfrak{A},{\bf x})  \models φ \iff (\mathfrak{C},{\bf z})\oplus (\mathfrak{B}, {\bf y}) \models φ.$$

Note that $\equiv_{φ,\ell}$ is an equivalence relation on ${\cal B}_τ^{(\ell)}.$
\bigskip

{
The following result is a variant of Courcelle's theorem~\cite{Courcelle90them,Courcelle92,Courcelle97}. It essentially says that the dynamic programming tables constructed by the proof of  Courcelle's theorem are
also definable in \MSOL. This fact is implicit in the proof of Courcelle's theorem. For instance, it can easily be derived from the proof of~\cite[Lemma 3.2]{BodlaenderFLPST16meta}.}
\begin{proposition}[Courcelle]
\labels{cou_more}
There is a function $f:\mathbb{N}^3 \to \mathbb{N}$ such that for every vocabulary $τ,$  every $φ∈ \MSOL[τ],$ and every $\ell∈ \mathbb{N},$ it holds that  $|{\cal B}_τ^{(\ell)} /_{\equiv_{φ,\ell}}|≤ f(|φ|,\ell,|τ|).$
\end{proposition}
An alternative way to see~\autoref{cou_more} is to say that, for every vocabulary $τ,$ every $φ∈\MSOL[τ],$ and every $\ell∈\mathbb{N},$ there is a collection ${\sf rep}_τ^{(\ell)}(φ) = \{φ_1, \ldots, φ_{m}\}$ of sentences {on $\ell$-boundaried $τ$-structures} (i.e., sentences in  $\MSOL[τ\cup\{{\sf b}_1,\ldots, {\sf b}_\ell\}]$)
where $m≤ f(|φ|,\ell,|τ|)$ and such that
\begin{itemize}
\item for every $(\mathfrak{A},{\bf x})∈ {\cal B}_τ^{(\ell)}$ there exists {\sl exactly one} $i∈[m]$ such that $(\mathfrak{A},{\bf x})\models φ_{i}$ and
\item for every compatible $(\mathfrak{A},{\bf x}),(\mathfrak{B}, {\bf y})∈ {\cal B}_τ^{(\ell)}$ and every $i∈ [m],$ if $(\mathfrak{A},{\bf x})\models φ_i$ and $(\mathfrak{B}, {\bf y})\models φ_i,$ then $(\mathfrak{A},{\bf x}) \equiv_{φ,\ell} (\mathfrak{B}, {\bf y}).$
\end{itemize}

{The elements of ${\sf rep}_τ^{(\ell)}(φ)$ are called {\em types} and can be seen
as an \MSOL-definable encoding of the tables of the dynamic programming  generated by Courcelle's theorem.
This representation of $φ,$ in what concerns boundary structures, provides
an abstract representation that does not depend on the ``internal part'' of a  boundary graph and will be  used as a key ingredient of the encodings in Sections~\ref{sec_first_floor}, \ref{the_second_level}, and \ref{sec_final_combo}.
}

\myskip\subsection{Bidimensionality of sets in flatness pairs}
\label{sec_bidimensionality}

\myskip\paragraph{Bidimensionality of a vertex set with respect to a flatness pair.}
See~\autoref{sec_canonical} for the definition of a $(W,\mathfrak{R})$-canonical partition of a graph, for some flatness pair $(W,\mathfrak{R}).$
Let $G$ be a graph and let $(W,\mathfrak{R})$ be a flatness pair of $G.$
For every set $X\subseteq V(G)$ and every $(W,\mathfrak{R})$-canonical partition $\tilde{\cal Q}$ of $G,$
we define the {\em $\tilde{\cal Q}$-bidimensionality of $X$}
to be the number of internal bags of $\tilde{\cal Q}$ that
contain a vertex of $X.$
For every set $X\subseteq V(G),$ we define the {\em bidimensionality of $X$ with respect to $(W,\mathfrak{R})$}, and we denote it by ${\sf bid}_{(W,\mathfrak{R})}(X),$
to be
the maximum $\tilde{\cal Q}$-bidimensionality of $X$ over all $(W,\mathfrak{R})$-canonical partitions $\tilde{\cal Q}$ of $G.$

At this point we wish to notice, even we do not use it in this paper, that  the choice of the canonical partition does not affect substantially the bidimensionality of a set.
\begin{observation}\labels{obse_bidi}
Let $k∈\mathbb{N},$ let $G$ be a graph, let $(W,\mathfrak{R})$ be a flatness pair of $G,$ and let
$\tilde{\cal Q}$ be a
$(W,\mathfrak{R})$-canonical partition  of $G.$
If the $\tilde{\cal Q}$-bidimensionality of $X$ is $k,$ then ${\sf bid}_{(W,\mathfrak{R})}(X)≤ 6k.$
\end{observation}

\myskip\paragraph{Brambles.}
Let $G$ be a graph. Two sets $V_1,V_2\subseteq V(G)$ are said to {\em touch} if they have a vertex in common or there is an edge $\{v_1,v_2\}∈ E(G)$ with $v_1∈ V_1$ and $v_2∈ V_2.$
A set ${\cal B}$ of pairwise touching vertex sets of $V(G)$ that induce connected subgraphs of $G$ is called a {\em bramble} of $G.$
The {\em order} of a bramble ${\cal B}$ is the minimum size of a vertex set that intersects every element of ${\cal B}.$

The following relation between treewidth and a maximum order bramble is proved in~\cite{SeymourT93graph} (see also~\cite[Theorem 5]{BellenbaumD02twosh}).
\begin{proposition}\labels{@almohadillas}
Let $k$ be a non-negative integer, let $G$ be a graph.
The treewidth of $G$ is at most $k$ if and only if every bramble of $G$ has order at most $k+1.$
\end{proposition}

We will prove the following key result:

\begin{lemma}\labels{@congregation}
Let $l ,q∈\mathbb{N},$
let $\mathfrak{A}$ be a $τ$-structure, let ${\bf a}=(a_{1},\ldots,a_l)$ be an apex-tuple of
$G_{\mathfrak{A}},$ and let $(W,\mathfrak{R})$ be a
flatness pair of $G_\mathfrak{A}\setminus V({\bf a}).$
For every set $X\subseteq V(\mathfrak{A}),$ if
${\sf cl}_{\sf X}({\sf star}_{\sf X}(\mathfrak{A},X))$ has treewidth at most $q,$ then
${\sf bid}_{(W,\mathfrak{R})}(X\setminus V({\bf a}))≤ (q+1)^2.$
\end{lemma}

\begin{proof}
Let $X\subseteq V(\mathfrak{A})$ such that ${\sf cl}_{\sf X}({\sf star}_{\sf X}(\mathfrak{A},X))$ has treewidth at most $q.$
Also, let $\tilde{\cal Q}$ be a $(W,\mathfrak{R})$-canonical partition of $G_{\mathfrak{A}}\setminus V({\bf a}).$
We will show that $X$ intersects at most $(q+1)^2$ internal bags of $\tilde{\cal Q}.$

Let $h$ be the height of $(W,\mathfrak{R}).$
Also, let $H$ be the Gaifman graph of ${\sf cl}_{\sf X}({\sf star}_{\sf X}(\mathfrak{A},S))$ and keep in mind that $\tw(H)≤ q.$
For every $i∈[h],$ let $P_i$ be the union of the vertex sets of all internal bags of $\tilde{\cal Q}$ that intersect the $i$-th horizontal path of $W,$ i.e., $P_i := \bigcup_{j∈[2,h-1]} V(Q^{(i,j)}).$
Also, let $L_i$ be the union of the vertex sets of all internal bags of $\tilde{\cal Q}$ that intersect the $i$-th vertical path of $W,$ i.e., $L_i:= \bigcup_{j∈[2,h-1]} V(Q^{(j,i)}).$
We also define $T^{(i,j)}:=P_i\cup L_j, i,j∈[h].$ 
We let $T_X^{(i,j)}:=T^{(i,j)}\cap X.$
We now consider the collection $${\cal D} = \{H[T_X^{(i,j)}]\mid i,j∈[h] \mbox{ and }T_X^{(i,j)}\neq \emptyset\}.$$

We will prove that ${\cal D}$ is a bramble of $H.$
For this, we have to prove that ${\cal D}$ consists of pairwise touching connected subgraphs of $H.$
Recall that ${\sf compass}_{\mathfrak{R}}(W)$ is connected
and notice that
if $v,u∈ X$ and there is a path $P$ in ${\sf compass}_{\mathfrak{R}}(W)$ connecting $v$ and $u$ such that no internal vertex of $P$ is in $X,$ then $\{v,u\}∈ E(H).$
This implies that every $D∈ {\cal D}$ is connected and every two $D_1,D_2∈ {\cal D}$ are touching, thus ${\cal D}$ is a bramble.

By~\autoref{@almohadillas}, we have that $\tw(H)≤ q$ implies that ${\cal D}$ has order at most $q+1.$
This, in turn, implies that $X$ intersects at most $(q+1)^2$
internal bags of $\tilde{\cal Q}.$
\end{proof}

The next result intuitively states that given a flat wall and some apices, we can find another flat wall inside the first one such that the set of apices that are adjacent to the compass of the new flat wall has ``big enough'' bidimensionality to the flat wall, i.e., is adjacent to ``many enough'' internal bags of every canonical partition of the graph defined by the latter flat wall.
We refer the reader to~\autoref{label_exceptionalness} for the definition of the tilt of a wall inside a flatness pair.
%
%
\begin{lemma}\labels{lemma_many_apices}
There is a function $\newfun{@lacedaemonians}:\mathbb{N}^3\to \mathbb{N}$
and an algorithm that receives as an input two integers $l,d∈ \mathbb{N},$ an odd integer $r≥ 3,$ a graph $G,$ a set $A\subseteq V(G)$ of size at most $l,$
and a flatness pair $(W,\mathfrak{R})$ of $G\setminus A$ of height $\funref{@lacedaemonians}(r,l,d),$ and outputs, in time ${\cal O}_{r,l,d} (n),$ an set $A'\subseteq A$ and a flatness pair $(\tilde{W}, \tilde{\mathfrak{R}})$ of
$G\setminus A'$
of height at least $r$ that is a $W'$-tilt of some subwall $W'$ of $W$ and
for every $a∈ A',$
${\sf bid}_{(\tilde{W}, \tilde{\mathfrak{R}})}(N_{G}(a)\setminus A')≥ d.$
\end{lemma}
\begin{proof}
Let $l,d∈ \mathbb{N}$ and let an odd integer $r≥ 3.$
We define the function
$\funref{@lacedaemonians}:\mathbb{N}^3\to\mathbb{N}$ so that,
for every $x,z∈\mathbb{N},$ $\funref{@lacedaemonians}(x,0,z) := x,$ while, for $y≥ 1,$ we set
$\funref{@lacedaemonians}(x,y,z) :={\sf odd}(\lceil\sqrt{z+1}\rceil)\cdot (\funref{@lacedaemonians}(x,y-1,z)+2).$

Let $G$ be a graph, let $A\subseteq V(G)$ of size at most $l,$ and let $(W,\mathfrak{R})$ be a flatness pair of $G\setminus A$ of height $\funref{@lacedaemonians}(r,l,d).$
We will prove the lemma by induction on $l.$
In the case that $l = 0,$ $W$ has height $\funref{@lacedaemonians}(r,0,d)=r$ and $A= \emptyset,$ so the lemma holds trivially for $(W,\mathfrak{R}).$
Suppose now that $l≥ 1$ and that the lemma holds for smaller values of $l.$
We set  $q:=\funref{@lacedaemonians}(x,y-1,z).$
Let $\tilde{\cal Q}$ be a $(W,\mathfrak{R})$-canonical partition of $G\setminus A.$
If every vertex in $A$ is adjacent in $G$ to at least $d$ internal bags of $\tilde{Q},$
then the algorithm outputs $A$ and $(W,\mathfrak{R}).$
Otherwise, there is a vertex $a∈ A$ that is adjacent, in $G,$ to less than $d$ internal bags of $\tilde{\cal Q}.$
In this case, we consider a collection ${\cal W} = \{W_1, \ldots, W_{d+1}\}$ of $d+1$ subwalls of $W$ of height $q$ such that, for every $i,j∈[d+1], i\neq j,$ if $(\tilde{W}_i,\tilde{\mathfrak{R}}_i)$ and $(\tilde{W}_j, \tilde{\mathfrak{R}}_j)$ are some $W_i$-tilt and $W_j$-tilt of $(W,\mathfrak{R})$ respectively, then $V(\cupall{\sf influence}_{\tilde{\mathfrak{R}}_i}(\tilde{W}_i))$ and $V(\cupall{\sf influence}_{\tilde{\mathfrak{R}}_j}(\tilde{W}_j))$ are disjoint. The existence of this collection is guaranteed by the fact that $\funref{@lacedaemonians}(r,l,d) ≥ \lceil\sqrt{d+1}\rceil\cdot (q+2)$ and it can be found in time ${\cal O}_{r,l,d} (n).$
Now notice that since $a$  is adjacent, in $G,$ to less than $d$ internal bags of $\tilde{\cal Q},$ then there is a wall $W_i, i∈[d+1]$ in ${\cal W}$ such that $a$ is adjacent, in $G,$ to no internal bag of any $(\tilde{W}_i,\tilde{\mathfrak{R}}_i)$-canonical partition of $G\setminus A.$
From the induction hypothesis, we have that we can compute, in time ${\cal O}_{r,l,d} (n),$ a $W''$-tilt $(\breve{W},\breve{\mathfrak{R}})$ of $(W,\mathfrak{R}),$ for some $W''$ that is a subwall of $W_i$ (and therefore of $W$), that has height at least $r$ and set $A'\subseteq A\setminus \{a\}$ of $G$ of size $l'<l$ such that every vertex in $A'$ is adjacent, in $G,$ to at least $d$ internal bags of every $(\breve{W},\breve{\mathfrak{R}})$-canonical partition of $G\setminus A'.$
\end{proof}

The following result is the core result of \cite{SauST21kapiI} (which, in turn, is based on the results of~\cite{BasteST20acom} and~\cite{GolovachST20hitti}).
We refer the reader to~\autoref{label_exceptionalness} for a definition of a flatness pair.
The framework of flatness pairs was recently introduced in~\cite{SauST21amor} to deal with some technical issues in the proof of the Flat Wall theorem in~\cite{KawarabayashiTW18anew} (see~\autoref{label_exceptionalness} and~\autoref{sec_homogeneouswalls} for the definitions of a regular and a homogeneous flatness pair, respectively).
We use $c_{\cal F}$ to denote $|{\cal F}|,$ i.e, the biggest size of a graph in ${\cal F}.$

 \begin{proposition}\labels{icalp_irrelevancy}
There exist two functions $\newfun{@denominations}:\mathbb{N}^4\to\mathbb{N}$ and $\newfun{@surreptitiously}:\mathbb{N}^2\to \mathbb{N}$ such that for every finite collection ${\cal F}$ of graphs,
if  $r,l,k∈ \mathbb{N},$ $G$ is a graph,
$A$ is a subset of $V(G)$ of size at most $l,$  $(W,\mathfrak{R})$ is a regular flatness pair of $G\setminus A$ of height $\funref{@denominations}(c_{\cal F},k,r,l)$
that is $\funref{@surreptitiously}(c_{\cal F}, l)$-homogeneous with respect to $2^{A},$ $V_{K}$
is the vertex set of the compass of the
central $r$-subwall of $W,$
then for every $X\subseteq V(G)$ such that ${\sf bid}_{(W,\mathfrak{R})}(X\setminus A)≤ k$ it holds that
$G\setminus X∈ \excl({\cal F})\iff G\setminus(X\setminus V_K)∈ \excl({\cal F}).$
\end{proposition}

\myskip\subsection{Privileged components}
\labels{sec_privi}
In this subsection we define pseudogrids and we introduce the notion of a {\sl privileged connected component}
of a graph with respect to a pseudogrid.
A pseudogrid is a collection of equally many ``vertical'' and ``horizontal'' paths that intersect in a ``grid-like'' way.
The size of a pseudogrid is the number of vertical/horizontal paths it contains.
Intuitively, a vertex set $C$ of a graph is privileged with respect to a set $X$ and a pseudogrid ${\bf W}_q$  if $C$ is the vertex set of a connected component of $G\setminus X$ that contains at least one vertical and one horizontal path (see formal definitions below).
We notice that, given a set $X$ and a pseudogrid ${\bf W}_q$ that is ``big enough'',
there is at most one vertex set $C$ that is privileged with respect to $X$ and ${\bf W}_q$ (see~\autoref{@vuitcentistes}).
After this intuitive introduction, we present the formal definitions of the notions presented above.
\smallskip

\begin{figure}[ht]
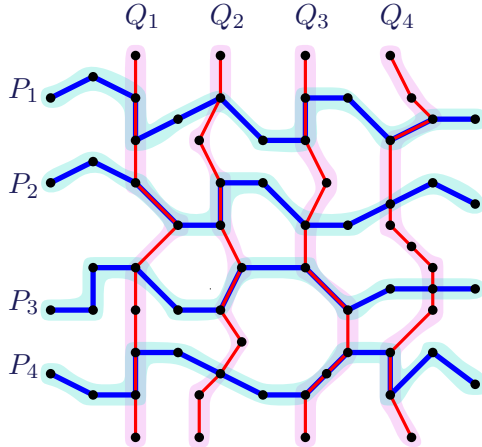

\centering
\tikzstyle{ipe stylesheet} = [
  ipe import,
  even odd rule,
  line join=round,
  line cap=butt,
  ipe pen normal/.style={line width=0.4},
  ipe pen heavier/.style={line width=0.8},
  ipe pen fat/.style={line width=1.2},
  ipe pen ultrafat/.style={line width=2},
  ipe pen normal,
  ipe mark normal/.style={ipe mark scale=3},
  ipe mark large/.style={ipe mark scale=5},
  ipe mark small/.style={ipe mark scale=2},
  ipe mark tiny/.style={ipe mark scale=1.1},
  ipe mark normal,
  /pgf/arrow keys/.cd,
  ipe arrow normal/.style={scale=7},
  ipe arrow large/.style={scale=10},
  ipe arrow small/.style={scale=5},
  ipe arrow tiny/.style={scale=3},
  ipe arrow normal,
  /tikz/.cd,
  ipe arrows, 
  <->/.tip = ipe normal,
  ipe dash normal/.style={dash pattern=},
  ipe dash dotted/.style={dash pattern=on 1bp off 3bp},
  ipe dash dashed/.style={dash pattern=on 4bp off 4bp},
  ipe dash dash dotted/.style={dash pattern=on 4bp off 2bp on 1bp off 2bp},
  ipe dash dash dot dotted/.style={dash pattern=on 4bp off 2bp on 1bp off 2bp on 1bp off 2bp},
  ipe dash normal,
  ipe node/.append style={font=\normalsize},
  ipe stretch normal/.style={ipe node stretch=1},
  ipe stretch normal,
  ipe opacity 10/.style={opacity=0.1},
  ipe opacity 30/.style={opacity=0.3},
  ipe opacity 50/.style={opacity=0.5},
  ipe opacity 75/.style={opacity=0.75},
  ipe opacity opaque/.style={opacity=1},
  ipe opacity opaque,
]
\definecolor{red}{rgb}{1,0,0}
\definecolor{blue}{rgb}{0,0,1}
\definecolor{green}{rgb}{0,1,0}
\definecolor{yellow}{rgb}{1,1,0}
\definecolor{orange}{rgb}{1,0.647,0}
\definecolor{gold}{rgb}{1,0.843,0}
\definecolor{purple}{rgb}{0.627,0.125,0.941}
\definecolor{gray}{rgb}{0.745,0.745,0.745}
\definecolor{brown}{rgb}{0.647,0.165,0.165}
\definecolor{navy}{rgb}{0,0,0.502}
\definecolor{pink}{rgb}{1,0.753,0.796}
\definecolor{seagreen}{rgb}{0.18,0.545,0.341}
\definecolor{turquoise}{rgb}{0.251,0.878,0.816}
\definecolor{violet}{rgb}{0.933,0.51,0.933}
\definecolor{darkblue}{rgb}{0,0,0.545}
\definecolor{darkcyan}{rgb}{0,0.545,0.545}
\definecolor{darkgray}{rgb}{0.663,0.663,0.663}
\definecolor{darkgreen}{rgb}{0,0.392,0}
\definecolor{darkmagenta}{rgb}{0.545,0,0.545}
\definecolor{darkorange}{rgb}{1,0.549,0}
\definecolor{darkred}{rgb}{0.545,0,0}
\definecolor{lightblue}{rgb}{0.678,0.847,0.902}
\definecolor{lightcyan}{rgb}{0.878,1,1}
\definecolor{lightgray}{rgb}{0.827,0.827,0.827}
\definecolor{lightgreen}{rgb}{0.565,0.933,0.565}
\definecolor{lightyellow}{rgb}{1,1,0.878}
\definecolor{black}{rgb}{0,0,0}
\definecolor{white}{rgb}{1,1,1}

\caption{An example of a $4$-pseudogrid ${\bf W}_4 = (P_1,\ldots, P_4, Q_1, \ldots, Q_4).$}
\labels{fig_pseudo}
\end{figure}

\myskip\paragraph{Respecting a collection of vertex sets.}
Let $G$ be a graph, let ${q}∈ \mathbb{N},$ and let ${\bf W}_{{q}}=(P_{1},\ldots,P_{{q}},Q_{1},\ldots,Q_{{q}})∈  (2^{V(G)})^{2q}.$  We say that  a vertex set $C\subseteq V(G)$ {\em respects} ${\bf W}_{{q}}$ if $C$  contains at least one element of  $\{P_{1},\ldots,P_{{q}}\}$ and  at least one element of  $\{Q_{1},\ldots,Q_{{q}}\}.$
Given an $X\subseteq V(G),$ we define $${\sf pr}(G,{\bf W}_{{q}},X)=\{C∈{\sf cc}(G, X)\mid \mbox{~$C$ respects ${\bf W}_{{q}}$}\}.$$
We call every set $C∈ {\sf pr}(G,{\bf W}_{{q}},X)$ a \emph{privileged component of $G$ with respect to ${\bf W}_q$ and $X$}.

\myskip\paragraph{Pseudogrids.}
We now give the definition of a pseudogrid, as given in~\cite[Definition 6]{KunOPS21polyn}.
Let $G$ be a graph, let ${q}∈ \mathbb{N},$ and let ${\bf W}_{{q}}=(P_{1},\ldots,P_{{q}},Q_{1},\ldots,Q_{{q}})∈ (2^{V(G)})^{2q}.$
We say that ${\bf W}_{{q}}$ is a {\em $q$-pseudogrid} of $G$ if ${\cal P}= (P_1, \ldots, P_q)$ and ${\cal Q} = (Q_1, \ldots, Q_q)$ are two sequences of vertex-disjoint paths and, for every $i∈[q],$
the path $P_i$ is a concatenation of paths $P_{i,0}, P_{i,1}^{Q}, P_{i,1}, P_{i,2}^{Q}, P_{i,2},\ldots, P_{i,q}^{Q}, P_{i,q}$ in this order such that, for every $j∈ [q],$ each path $P_{i,j}^{Q}$ is a non-empty subpath of $Q_j$ (possibly consisting of a single vertex) and, for every $j∈ [0,q],$  every path $P_{i,j}$ does not contain any edge nor internal vertex of any path $Q_j$ ($P_{i,0}$ and $P_{i,q}$ are allowed to be paths of length 0), and the symmetric conditions hold with the roles of ${\cal P}$ and ${\cal Q}$ swapped (see~\autoref{fig_pseudo}).
We refer to the paths in ${\cal P}$ (resp. ${\cal Q}$) as the {\em horizontal paths} (resp. {\em vertical paths}) of ${\bf W}_q.$

%

The next observation indicates that for every graph $G,$ every $q$-pseudogrid ${\bf W}_q,$ and every set $X\subseteq V(G),$ there is at most one set in ${\sf cc}(G,X)$ that respects ${\bf W}_q.$

\begin{observation}
\labels{@vuitcentistes}
If ${\bf W}_q$ is a  ${q}$-pseudogrid of a graph $G$ and $X$ is a subset of $V(G),$ then ${\sf pr}(G,{\bf W}_{{q}},X)$ is either a singleton or the empty set.
\end{observation}

To see why~\autoref{@vuitcentistes} holds, first notice that a $q$-pseudogrid of a graph $G$ induces
a connected subgraph of $G.$
Also notice that, if for a set $X\subseteq \cupall {\bf W}_q$ it holds that $|{\sf pr}(G,{\bf W}_{{q}},X)|≥ 2,$ then there are at least two distinct sets $C_1, C_2∈{\sf cc}(G, X)$ that respect ${\bf W}_q.$
This implies that there is a $P_i$ (resp. $P_j$) and a $Q_i$ (resp. $Q_j$) in ${\bf W}_q,$ such that $P_i, Q_i\subseteq C_1$ (resp. $P_j,Q_j\subseteq C_2$).
But $P_i\cup P_j \cup Q_i\cup Q_j$ induces a connected subgraph of $G\setminus X,$ a contradiction to the fact that $C_1$ and $C_2$ are distinct elements of ${\sf cc}(G, X).$
\medskip

\myskip\section{The algorithm}
\labels{sec_scheme}

In this section we aim to present the general  scheme of our algorithm for~\autoref{@decendientes}.
In~\autoref{sec_reduce_instance},
we present the main subroutine of our algorithm that
reduces the annotated set $R$ and the universe of the structure
under the presence of a flatness pair of ``big enough'' height in our structure, which is a certificate
that the treewidth of the structure is ``big enough'' (\autoref{lemma_irrele_flat}). The proof of \autoref{lemma_irrele_flat} is strongly based on~\autoref{@desmembramientos}, whose proof  is the main technical part of this paper and
 is postponed to Sections~\ref{sec_first_floor},~\ref{the_second_level}, and~\ref{sec_final_combo}.
A brief explanation of the proof idea is given in~\autoref{sec_what_3_sec}.
Assuming the claimed algorithm of~\autoref{lemma_irrele_flat}, in~\autoref{sec_gen_algo} we show how to use this subroutine in order to design an algorithm for~\autoref{@decendientes} and we provide the proof of the latter.


\myskip\subsection{Reducing the instance}
\labels{sec_reduce_instance}
As we mention in the overview of the proof presented in \autoref{sec_overview}, we use the {\sl irrelevant vertex technique} to reduce the problem to instances of bounded treewidth.
This idea is materialized in the next lemma that provides an algorithm that, given an instance $(\mathfrak{A},R,{\bf a}),$ where ${\bf a}$ is an apex-tuple of $\mathfrak{A},$ and a regular flatness pair $(W,\mathfrak{R})$ of $G_{\mathfrak{A}}\setminus V({\bf a})$ of ``big enough'' height, such that ${\sf compass}_{\mathfrak{R}}(W)$ has bounded treewidth, outputs an instance $(\mathfrak{A}',R',{\bf a})$ such that $V(\mathfrak{A}')\subsetneq V(\mathfrak{A}),$ $R'\subsetneq R,$ and $(\mathfrak{A}, R, {\bf a})\models θ_{{\sf R},{\bf c}}\iff (\mathfrak{A}', R',{\bf a})\models θ_{{\sf R},{\bf c}}.$

\begin{lemma}\labels{lemma_irrele_flat}
Let $τ$ be a vocabulary, ${\sf R}\notin τ$ be a unary relation symbol, and ${\bf c}$ be a collection of $l$ constant symbols, where $l∈\mathbb{N}_{≥ 1}.$
There is a function $\newfun{@ausgeschlossen}:\mathbb{N}^3 \to \mathbb{N}$ and an algorithm that receives as an input
\begin{itemize}
\item   an enhanced version $θ_{{\sf R},{\bf c}}$ of a sentence $θ∈ Θ[τ],$
\item  a $t∈\mathbb{N}$ and an odd integer $g≥ 3,$
\item  a $τ$-structure $\mathfrak{A},$   a set $R\subseteq V(\mathfrak{A}),$ and   an apex-tuple ${\bf a}$ of $\mathfrak{A}$ of size $l,$ and
\item
a regular flatness pair $(W,\mathfrak{R})$ of $G_{\mathfrak{A}}\setminus V({\bf a})$ of height $\funref{@ausgeschlossen}(|θ|,l, g)$
such that ${\sf compass}_{\mathfrak{R}}(W)$ has treewidth at most $t,$
\end{itemize}
and
outputs,
in time ${\cal O}_{|θ_{{\sf R},{\bf c}}|, l,g,t} (n),$ a
set $Y\subseteq V(\mathfrak{A})\setminus V({\bf a})$
and a flatness
pair $(\tilde{W}',\tilde{\mathfrak{R}}')$ of $G\setminus V({\bf a})$ that is a $W'$-tilt of some subwall $W'$ of $W$ of height $g$ such that
 $V({\sf compass}_{\tilde{\mathfrak{R}}'}(\tilde{W}')) \subseteq Y$
and $(\mathfrak{A},R, {\bf a})\models θ_{{\sf R},{\bf c}} \iff (\mathfrak{A}\setminus V({\sf compass}_{\tilde{\mathfrak{R}}'}(\tilde{W}')), R\setminus Y, {\bf a})\models θ_{{\sf R},{\bf c}}.$
\end{lemma}

To prove~\autoref{lemma_irrele_flat}, we aim to reduce the annotated set $R$ and to characterize some non-annotated vertices as ``irrelevant'' to the existence of a solution to the problem, which allows us to reduce our problem to ``simpler'' equivalent instances.
Since our problem has three basic elements\footnote{Throughout the reminder of the article, we use consistently this color coding using \blue{blue}/\green{green}/\red{red} to easily identify the three parts of our problem.}
\begin{itemize}
\item[\blue{1)}] the satisfaction of $β$ in the modulator sets,

\item[\green{2)}] the satisfaction of a \FOL-sentence in the remaining ``terminal part'' of the structure, and
\item[\red{3)}] the exclusion of a minor from the Gaifman graph of the remaining ``terminal part'' of the structure,
\end{itemize}
our ``irrelevancy'' arguments also decompose into three parts.

Concerning the ``irrelevancy'' for minor-exclusion (Item $\red{3)}$), we use
\autoref{icalp_irrelevancy}  in order to
obtain a flatness pair whose compass is ``irrelevant'' with respect to any set of ``small enough'' bidimensionality.
After finding this flatness pair, our attention is focused on  items $\blue{1)}$ and $\green{2)}.$

\begin{lemma}\labels{@desmembramientos}
Let $τ$ be a vocabulary, ${\sf R}\notin τ$ be a unary relation symbol, and ${\bf c}$ be a collection of $l$ constant symbols, where $l∈\mathbb{N}_{≥ 1}.$
There are three functions $\newfun{@occidentales}:\mathbb{N}^5 \to \mathbb{N},$ $\newfun{@riguardavano}:\mathbb{N}^3\to \mathbb{N},$  and $\newfun{@carthaginoise}:\mathbb{N}^2\to \mathbb{N},$
and an algorithm that receives as an input
\begin{enumerate}
\item a sentence $θ∈ Θ$ and an enhanced version $θ_{{\sf R},{\bf c}}$ of $θ,$
\item two integers $l,z∈ \mathbb{N},$ an odd integer $g≥ 3,$
\item a $τ$-structure $\mathfrak{A},$  a set $R\subseteq V(\mathfrak{A}),$ and a
tuple ${\bf a}=(a_{1},\ldots,a_l),$ where $V({\bf a})\subseteq V(\mathfrak{A}),$
\item a flatness pair $(W,\mathfrak{R})$ 
of $G_{\mathfrak{A}}\setminus V({\bf a})$ of height at least $\funref{@occidentales}(\hw(θ),\tw(θ),c,l,g),$ where $c$ is the maximum size of a \FOL-target sentence of $θ,$ such that
\begin{itemize}
\item ${\sf compass}_{\mathfrak{R}}(W)$ has treewidth at most $z$ and
\item for every $a∈ V({\bf a}),$ ${\sf bid}_{(W,\mathfrak{R})}(N_{G_{\mathfrak{A}}}(a)\setminus V({\bf a}))≥ \funref{@riguardavano}(|θ|,\tw(θ),g),$ and
\end{itemize}

\item a vertex set $D\subseteq V(\mathfrak{A})$ such that $V({\sf compass}_{\mathfrak{R}} (W))\subseteq D$ and for every
$X\subseteq V(G)$ such that ${\sf bid}_{(W,\mathfrak{R})}(X\setminus V({\bf a}))≤ \funref{@carthaginoise}(|θ|,\tw(θ))$ it holds that
$G_{\mathfrak{A}}\setminus X∈ \excl(\{K_{\hw(θ)}\})\iff G_{\mathfrak{A}}\setminus(X\setminus D)∈ \excl(\{K_{\hw(θ)}\}),$
\end{enumerate}
and outputs, in time ${\cal O}_{|θ|,l,z,j'}(n),$
a set $Y\subseteq V(\mathfrak{A})\setminus V({\bf a})$
and a flatness
pair $(\tilde{W}',\tilde{\mathfrak{R}}')$ of $G\setminus V({\bf a})$ that is a $W'$-tilt of some subwall $W'$ of $W$ of height $g$ such that
 $V({\sf compass}_{\tilde{\mathfrak{R}}'}(\tilde{W}')) \subseteq Y$
 and $(\mathfrak{A},R, {\bf a})\models θ_{{\sf R},{\bf c}} \iff (\mathfrak{A}\setminus V({\sf compass}_{\tilde{\mathfrak{R}}'}(\tilde{W}')), R\setminus Y, {\bf a})\models θ_{{\sf R},{\bf c}}.$
\end{lemma}

The proof of~\autoref{@desmembramientos} is based on the algorithm ${\tt Find\_Equiv\_FlatPairs},$
 presented in~\autoref{@inhumainement} (also informally sketched in~\autoref{sec_what_3_sec}).
First, in~\autoref{sec_proof_correctness}, we prove the correctness of this algorithm for sentences in $\bar{Θ}_1.$
Then, in the end of~\autoref{@enthaltenden} we provide a sketch of the proof for sentences in $\bar{Θ}.$
The full proof for sentences in $\bar{Θ}$ can be found in~\autoref{second_level_more}.
Finally, in~\autoref{sec_final_combo}, we explain how to extend the proof in~\autoref{second_level_more} so  to capture the full generality of $Θ.$

\medskip

We now provide the proof of~\autoref{lemma_irrele_flat}, assuming the correctness of \autoref{@desmembramientos}. See the down-right green rectangle of \autoref{fig_general_sch_alg} for a summary of the main ideas and supporting results of the proof of \autoref{lemma_irrele_flat}.

\begin{proof}[Proof of~\autoref{lemma_irrele_flat}]
Let $c$ be the maximum size of a \FOL-target sentence of $θ.$
We set
\begin{align*}
λ : = & \funref{@surreptitiously}(\hw(θ), l),\\
d:= & \funref{@carthaginoise}(|θ|,\tw(θ)),\\
m:= & \funref{@riguardavano}(|θ|,\tw(θ)),\\
p : = & \funref{@occidentales}(\hw(θ),\tw(θ),c,l, j),\\
z : = & \funref{@lacedaemonians}(p,l,m),\\
r  : = & \funref{@denominations}(\hw(θ),d,z,l),\\
w : = & \funref{@poderosamente}(r,l,\ell), \text{~and}\\
 \funref{@ausgeschlossen}(|θ|, j)  : = & w.
\end{align*}


We first run the algorithm of~\autoref{label_highlighting} with input $w,l,λ,$ $G_{\mathfrak{A}},$ ${\bf a},$ and $(W,\mathfrak{R}).$
It outputs, in time ${\cal O}_{|θ_{{\sf R},{\bf c}}|} (n),$ a flatness pair $(\breve{W},\breve{\frR})$ of $G_{\mathfrak{A}}\setminus V({\bf a})$  of height $r$ that is $λ$-homogeneous with respect to $2^{V({\bf a})}$ and
is a $\hat{W}$-tilt of $(W,\frR)$ for some subwall $\hat{W}$ of $W.$
The wall $\breve{W}$ virtually corresponds to the second wall of~\autoref{figure_walls_zoom}.
We set $D$ to be the vertex set of the compass of the central $z$-subwall $\overline{W}$ of $\breve{W}.$
The wall $\overline{W}$ corresponds to the inner part of the third wall of~\autoref{figure_walls_zoom} (that is also the fourth wall of the same figure).
Notice that $\overline{W}$ is also a subwall of $W.$
By~\autoref{icalp_irrelevancy},  for every
$X\subseteq V(G)$ such that ${\sf bid}_{(\breve{W}, \breve{\mathfrak{R}})}(X\setminus V({\bf a}))≤ d$ it holds that
$$G_{\mathfrak{A}}\setminus X∈ \excl(\{K_{\hw(θ)}\})\iff G_{\mathfrak{A}}\setminus(X\setminus D)∈ \excl(\{K_{\hw(θ)}\}).$$
Then, by applying the algorithm of~\autoref{lemma_many_apices} for $l,m,r,$ $G_{\mathfrak{A}},$ $V({\bf a}),$ and a $\overline{W}$-tilt of $(W,\mathfrak{R}),$ we can find, in time ${\cal O}_{|θ_{{\sf R},{\bf c}}|} (n),$ an apex-tuple ${\bf a}'$ of $G_{\mathfrak{A}}$ of size $l'≤ l$ and a
flatness pair $(W^\bullet, \mathfrak{R}^\bullet)$ of $G_{\mathfrak{A}}\setminus V({\bf a}')$
of height $r$ that is a $W^\star$-tilt of a subwall $W^\star$ of $\overline{W}$
such that for every $a∈ V({\bf a}'),$ ${\sf bid}_{(W^\bullet,\mathfrak{R}^\bullet)}(N_{G_{\mathfrak{A}}}(a)\setminus V({\bf a}'))≥ m.$
Note that  $(W^\bullet, \mathfrak{R}^\bullet)$ is also a $W^*$-tilt of a subwall $W^*$ of $W.$
The wall $W^\bullet$ corresponds to the selected wall inside the fourth wall of~\autoref{figure_walls_zoom}.
By applying~\autoref{@desmembramientos}, we can find,  in time ${\cal O}_{|θ_{{\sf R},{\bf c}}|} (n),$
a set $Y\subseteq V(\mathfrak{A})\setminus V({\bf a}')$
and a flatness
pair $(\tilde{W}',\tilde{\mathfrak{R}}')$ of $G\setminus V({\bf a}')$ that is a $W'$-tilt of some subwall $W'$ of $W$ of height $j$ such that
 $V({\sf compass}_{\tilde{\mathfrak{R}}'}(\tilde{W}')) \subseteq Y$
and $(\mathfrak{A},R, {\bf a})\models θ_{{\sf R},{\bf c}} \iff (\mathfrak{A}\setminus V({\sf compass}_{\tilde{\mathfrak{R}}'}(\tilde{W}')), R\setminus Y, {\bf a})\models θ_{{\sf R},{\bf c}}.$
\end{proof}

Notice that in the case of a sentence $θ∈Θ[τ]$ whose \FOL-target sentences are always true, the analogue of~\autoref{obs_addingR} is the following:

\begin{lemma}\labels{lemm_nominor}
Let $τ$ be a vocabulary, ${\sf R}\notin τ$ be a unary relation symbol, and ${\bf c}$ be a collection of $l$ constant symbols, where $l∈\mathbb{N}_{≥ 1}.$
Also, let $θ∈Θ[τ]$ whose \FOL-target sentences are always true and let $θ_{{\sf R},{\bf c}}$ be an enhanced version of $θ.$
For every $τ$-structure $\mathfrak{A},$ for every $R\subseteq V(\mathfrak{A}),$ and for every apex-tuple ${\bf a}$ of $\mathfrak{A}$ of size $l$, it holds that $\mathfrak{A}\models θ \iff (\mathfrak{A},R, {\bf a})\models θ_{{\sf R},{\bf c}},$ where ${\bf c}$ is interpreted as ${\bf a}.$
\end{lemma}

Therefore, combining~\autoref{lemma_irrele_flat} and~\autoref{lemm_nominor},
we get the following corollary:

\begin{corollary}\labels{corr_withoutR_irrele_flat}
Let $τ$ be a vocabulary.
There is a function $\newfun{@ausausgeschlossen}:\mathbb{N}^3 \to \mathbb{N}$ and an algorithm that receives as an input a $t,l∈\mathbb{N},$ a sentence $θ∈ Θ[τ]$ whose \FOL-target sentences are always true, a $τ$-structure $\mathfrak{A},$
an apex-tuple ${\bf a}$ of $\mathfrak{A}$ of size $l,$ and
a regular flatness pair $(W,\mathfrak{R})$ of $G_{\mathfrak{A}}\setminus V({\bf a})$ of height $\funref{@ausausgeschlossen}(|θ|,l, g)$
such that ${\sf compass}_{\mathfrak{R}}(W)$ has treewidth at most $t,$
and
outputs,
in time ${\cal O}_{|θ|, l,g,t} (n),$ a
flatness
pair $(\tilde{W}',\tilde{\mathfrak{R}}')$ of $G\setminus V({\bf a})$ that is a $W'$-tilt of some subwall $W'$ of $W$ of height $g$ such that
 $\mathfrak{A}\models θ \iff \mathfrak{A}\setminus V({\sf compass}_{\tilde{\mathfrak{R}}'}(\tilde{W}'))\models θ.$
\end{corollary}

\autoref{corr_withoutR_irrele_flat} has some special consequences on the constructibility of  Robertson-Seymour's theorem that will be discussed in \autoref{@insuperables}.

\myskip\subsection{The algorithm of~\autoref{@decendientes}}
\labels{sec_gen_algo}

We are now ready to present the proof of~\autoref{@decendientes} ({assuming the correctness} of \autoref{@desmembramientos} and therefore of~\autoref{lemma_irrele_flat} as well).

\begin{proof}[Proof of~\autoref{@decendientes}]
Given a sentence $θ∈ Θ[τ],$ we set
\begin{eqnarray*}
c&:=&\hw(θ)+(\tw(θ)+1)\cdot {\sf height}(θ),\\
l&:=&\funref{@carlovingios}(c)\mbox{ where $\funref{@carlovingios}$ is the function of~\autoref{label_proletarians}}, and\\
r&:= &\funref{@ausgeschlossen}(|θ|,l,3).
\end{eqnarray*}
Our algorithm consists of four steps,
which are summarized in \autoref{fig_general_sch_alg}, along with the supporting results:
\medskip

\noindent{\bf Step 1}:
Consider an enhanced version $θ_{{\sf R},{\bf c}}$ of $θ.$
Consider an arbitrary apex-tuple ${\bf a}_0$ of $\mathfrak{A}$ of size $l$.
By~\autoref{obs_addingR}, we have that
$\mathfrak{A}\models θ \iff (\mathfrak{A},V(\mathfrak{A}), {\bf a}_0)\models θ_{{\sf R},{\bf c}},$ where ${\sf R}$ is interpreted as $V(\mathfrak{A})$ and ${\bf c}$ is interpreted as ${\bf a}_0.$
We set $R_0:= V(\mathfrak{A})$ and we proceed to Step 2.
\medskip

\noindent{\bf Step 2}:
Run the algorithm of~\autoref{label_proletarians} for $G_{\mathfrak{A}},$ $r,$ and $c.$
This algorithm outputs, in linear time, either
a
report that $K_{c}\preceq_{\sf m} G_{\mathfrak{A}},$
or a tree decomposition of $G_{\mathfrak{A}}$ of width at most $\funref{@interference}(c)\cdot r,$ or
a set $A\subseteq V(G),$  where $|A|≤ l,$ a regular flatness pair $(W,\mathfrak{R})$ of $G\setminus A$ of height $r,$
and a tree decomposition of ${\sf compass}_{\mathfrak{R}}(W)$ of width at most $\funref{@interference}(c)\cdot r.$
In the first possible output,
by~\autoref{@disillusioned}, we can safely report that $\mathfrak{A}\notin{\rm Mod}(θ).$
In the second possible output,
 i.e., a tree decomposition of $G_{\mathfrak{A}}$ of width at most $\funref{@interference}(c)\cdot r,$
 proceed to Step 4.
In the third possible output,
proceed to Step 3.
\medskip

\noindent{\bf Step 3}:
We first consider an ordering $a_1, \ldots, a_{l}$ of the vertices in $A,$ and set ${\bf a} = (a_1, \ldots, a_l).$
By~\autoref{lem_no_matter_which_apex}, we have that
$(\mathfrak{A},R_0, {\bf a}_0)\models θ_{{\sf R},{\bf c}} \iff (\mathfrak{A},R_0, {\bf a})\models θ_{{\sf R},{\bf c}}.$
We run the algorithm of~\autoref{lemma_irrele_flat} for $θ_{{\sf R},{\bf c}},$ $\mathfrak{A},$ $R_0,$ ${\bf a},$ and $(W,\mathfrak{R}),$ and we obtain, in linear time, a
set $X\subseteq V(\mathfrak{A})\setminus V({\bf a})$
and a flatness
pair $(\tilde{W}',\tilde{\mathfrak{R}}')$ of $G\setminus V({\bf a})$ that is a $W'$-tilt of some subwall $W'$ of $W$ of height $3$ such that
 $V({\sf compass}_{\tilde{\mathfrak{R}}'}(\tilde{W}')) \subseteq X$
and $(\mathfrak{A},R_0, {\bf a})\models θ_{{\sf R},{\bf c}} \iff (\mathfrak{A}\setminus V({\sf compass}_{\tilde{\mathfrak{R}}'}(\tilde{W}')), R_0\setminus X, {\bf a})\models θ_{{\sf R},{\bf c}}.$
Then, we set $\mathfrak{A}:=\mathfrak{A}\setminus V({\sf compass}_{\tilde{\mathfrak{R}}'}(\tilde{W}')),$ ${\bf a}_0:={\bf a},$ $R_0:= R_0\setminus X,$ and we run again Step 2.
\medskip

\noindent{\bf Step 4}:
Given a tree decomposition of $G_{\mathfrak{A}}$ of width at most $\funref{@interference}(c)\cdot r,$ and since
$θ_{{\sf R},{\bf c}}∈\MSOL[τ\cup\{{\sf R}\}\cup{\bf c}],$ we decide whether $(\mathfrak{A},R_0,{\bf a}_0)\models θ_{{\sf R},{\bf c}}$ in linear time by using Courcelle's theorem.
\medskip

Observe that the second and the third step of the algorithm are executed in linear time and they can be repeated no more than a linear number of times.
Therefore, the overall algorithm runs in quadratic time, as claimed.
\end{proof}

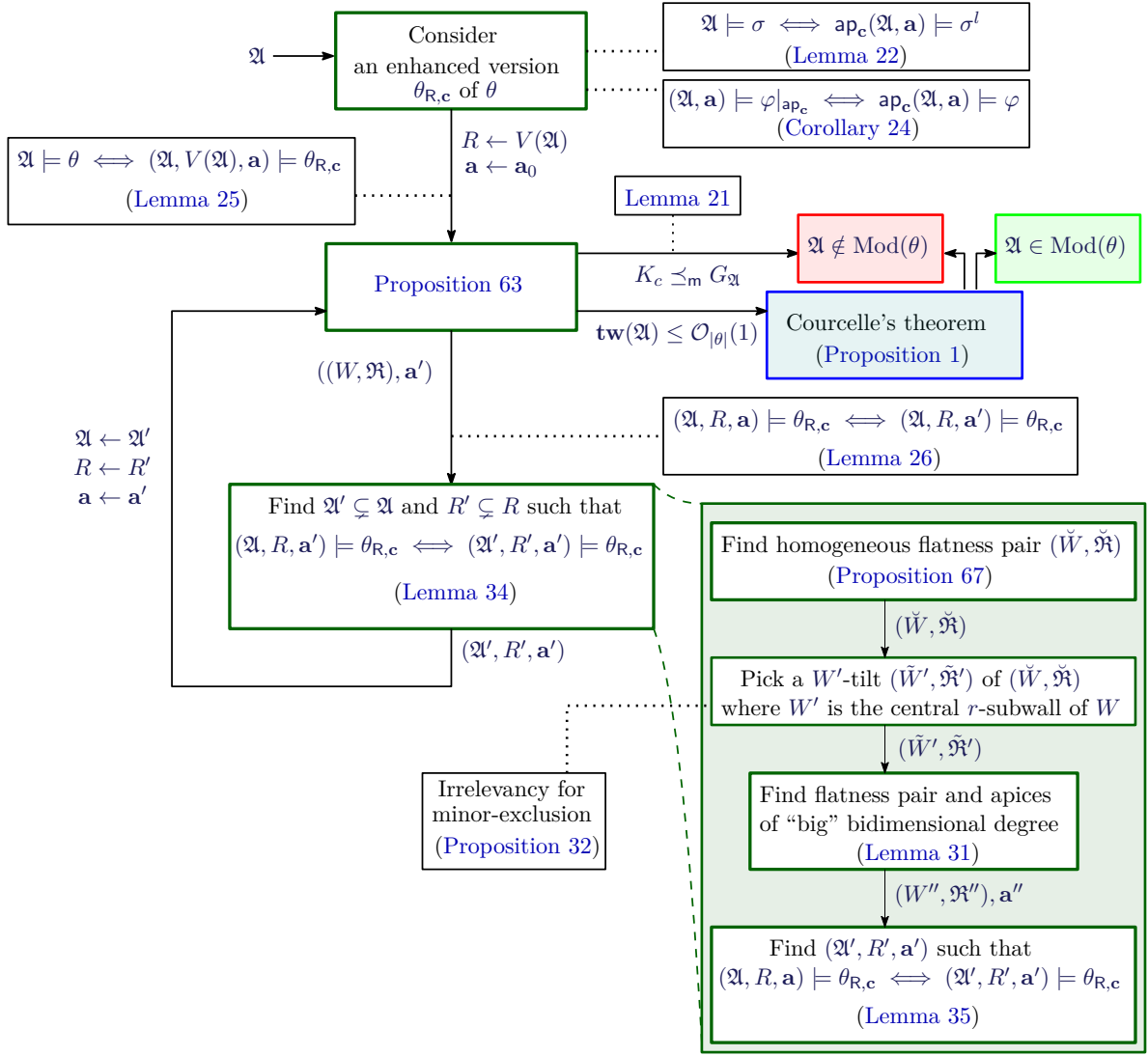
\begin{figure}[ht]
\!\!\!\!\!\!\!\!\!\!\!\!\!\!\!\!\!\!\!\!\!\!\!\!\!\!\!\!\!\!\!\!\!\!\!\!\!\!\!\!\!\!\!\!\!\!\!\!\!\!\!\!\!\!\scalebox{1.04}{%
\tikzstyle{ipe stylesheet} = [
  ipe import,
  even odd rule,
  line join=round,
  line cap=butt,
  ipe pen normal/.style={line width=0.4},
  ipe pen heavier/.style={line width=0.8},
  ipe pen fat/.style={line width=1.2},
  ipe pen ultrafat/.style={line width=1.5},
  ipe pen normal,
  ipe mark normal/.style={ipe mark scale=3},
  ipe mark large/.style={ipe mark scale=5},
  ipe mark small/.style={ipe mark scale=2},
  ipe mark tiny/.style={ipe mark scale=1.1},
  ipe mark normal,
  /pgf/arrow keys/.cd,
  ipe arrow normal/.style={scale=7},
  ipe arrow large/.style={scale=10},
  ipe arrow small/.style={scale=5},
  ipe arrow tiny/.style={scale=3},
  ipe arrow normal,
  /tikz/.cd,
  ipe arrows, 
  <->/.tip = ipe normal,
  ipe dash normal/.style={dash pattern=},
  ipe dash dotted/.style={dash pattern=on 1bp off 3bp},
  ipe dash dashed/.style={dash pattern=on 4bp off 4bp},
  ipe dash dash dotted/.style={dash pattern=on 4bp off 2bp on 1bp off 2bp},
  ipe dash dash dot dotted/.style={dash pattern=on 4bp off 2bp on 1bp off 2bp on 1bp off 2bp},
  ipe dash normal,
  ipe node/.append style={font=\normalsize},
  ipe stretch normal/.style={ipe node stretch=1},
  ipe stretch normal,
  ipe opacity 10/.style={opacity=0.1},
  ipe opacity 30/.style={opacity=0.3},
  ipe opacity 50/.style={opacity=0.5},
  ipe opacity 75/.style={opacity=0.75},
  ipe opacity opaque/.style={opacity=1},
  ipe opacity opaque,
]
\definecolor{red}{rgb}{1,0,0}
\definecolor{blue}{rgb}{0,0,1}
\definecolor{green}{rgb}{0,1,0}
\definecolor{yellow}{rgb}{1,1,0}
\definecolor{orange}{rgb}{1,0.647,0}
\definecolor{gold}{rgb}{1,0.843,0}
\definecolor{purple}{rgb}{0.627,0.125,0.941}
\definecolor{gray}{rgb}{0.745,0.745,0.745}
\definecolor{brown}{rgb}{0.647,0.165,0.165}
\definecolor{navy}{rgb}{0,0,0.502}
\definecolor{pink}{rgb}{1,0.753,0.796}
\definecolor{seagreen}{rgb}{0.18,0.545,0.341}
\definecolor{turquoise}{rgb}{0.251,0.878,0.816}
\definecolor{violet}{rgb}{0.933,0.51,0.933}
\definecolor{darkblue}{rgb}{0,0,0.545}
\definecolor{darkcyan}{rgb}{0,0.545,0.545}
\definecolor{darkgray}{rgb}{0.663,0.663,0.663}
\definecolor{darkgreen}{rgb}{0,0.392,0}
\definecolor{darkmagenta}{rgb}{0.545,0,0.545}
\definecolor{darkorange}{rgb}{1,0.549,0}
\definecolor{darkred}{rgb}{0.545,0,0}
\definecolor{lightblue}{rgb}{0.678,0.847,0.902}
\definecolor{lightcyan}{rgb}{0.878,1,1}
\definecolor{lightgray}{rgb}{0.827,0.827,0.827}
\definecolor{lightgreen}{rgb}{0.565,0.933,0.565}
\definecolor{lightyellow}{rgb}{1,1,0.878}
\definecolor{black}{rgb}{0,0,0}
\definecolor{white}{rgb}{1,1,1}
\scalebox{0.8}{
\begin{tikzpicture}[ipe stylesheet,scale=1.13]
 \draw[ipe pen heavier, -{ipe pointed[ipe arrow small]}]
    (150, 666)
     -- (175, 666);
  \node[ipe node]
     at (140, 663) {$\frak{A}$};
  \draw[blue, ipe pen fat]
 (355, 568) rectangle (461, 532);
   \filldraw[darkgreen, ipe pen ultrafat, ipe opacity 10]
    (328, 480) rectangle (512, 252);
  \filldraw[draw=darkgreen, ipe pen ultrafat, fill=white]
    (332, 304) rectangle (508, 256);
  \filldraw[draw=darkgreen, ipe pen ultrafat, fill=white]
    (348, 368) rectangle (484, 328);
  \filldraw[draw=darkgreen, ipe pen ultrafat, fill=white]
    (332, 472) rectangle (508, 440);
  \node[ipe node]
     at (296, 604) {\autoref{@disillusioned}};
  \node[ipe node]
     at (88.586, 603.346) {(\autoref{obs_addingR})};
  \node[ipe node]
     at (228, 628) {$R\leftarrow V(\frak{A}) $};
  \node[ipe node]
     at (229.72, 617.511) {${\bf a} \leftarrow {\bf a}_0$};
  \node[ipe node]
     at (192, 568) {\autoref{label_proletarians}};
  \node[ipe node]
     at (284, 548) {${\bf tw}(\frak{A})\leq {\cal O}_{|\theta|}(1)$};
  \node[ipe node]
     at (363, 552) {Courcelle's theorem};
  \node[ipe node]
     at (375, 538.736) {(\autoref{@originallypublishedin})};
  \node[ipe node]
     at (316, 512) {$(\frak{A}, R,{\bf a})\models \theta_{{\sf R},{\bf c}}
\iff
(\frak{A}, R,{\bf a}')\models \theta_{{\sf R},{\bf c}}$};
  \node[ipe node]
     at (376, 496) {(\autoref{lem_no_matter_which_apex})};
  \node[ipe node]
     at (200, 440) {(\autoref{lemma_irrele_flat})};
  \node[ipe node]
     at (228, 416) {$(\frak{A}',R',{\bf a}')$};
  \node[ipe node]
     at (68, 504) {$\frak{A}\leftarrow\frak{A}'$};
  \node[ipe node]
     at (67.196, 491.84) {$R\leftarrow R'$};
  \node[ipe node]
     at (69.188, 480.044) {${\bf a}\leftarrow {\bf a}'$};
  \node[ipe node]
     at (380, 447) {(\autoref{label_highlighting})};
  \node[ipe node]
     at (216.925, 334.813) {(\autoref{icalp_irrelevancy})};
  \node[ipe node]
     at (392, 332) {(\autoref{lemma_many_apices})};
  \node[ipe node]
     at (392, 264) {(\autoref{@desmembramientos})};
  \draw[ipe pen heavier, -{ipe pointed[ipe arrow small]}]
    (276, 584)
     -- (368, 584);
  \node[ipe node]
     at (300, 572) {$K_c\preceq_{\sf m} G_{\frak{A}}$};
  \filldraw[red, ipe opacity 10]
    (368, 600) rectangle (428, 572);
  \filldraw[green, ipe opacity 10]
    (450, 600) rectangle (510, 572);
  \draw[green, ipe pen fat]
	(450, 600) rectangle (510, 572);
\node[ipe node]
     at (454, 584) {$\frak{A}\in{\rm Mod}(\theta)$};
       \draw[ipe pen heavier, -{ipe pointed[ipe arrow small]}]
    (442, 569)
    -- (442,584)
     -- (450, 584);
       \draw[ipe pen heavier, -{ipe pointed[ipe arrow small]}]
      (437, 569)
    -- (437,584)
     -- (429, 584);
  \node[ipe node]
     at (372, 584) {$\frak{A}\notin{\rm Mod}(\theta)$};
  \draw[ipe pen heavier]
    (292, 616) rectangle (344, 600);
  \draw[ipe pen heavier, ipe dash dotted]
    (316, 600)
     -- (316, 584);
  \draw[darkgreen, ipe pen ultrafat]
    (176, 684) rectangle (280, 644);
  \draw[ipe pen heavier, -{ipe pointed[ipe arrow small]}]
    (224, 644)
     -- (224.0319, 588.9887);
  \draw[darkgreen, ipe pen ultrafat]
    (172, 588) rectangle (276, 552);
  \draw[ipe pen heavier]
    (40, 632) rectangle (184, 596);
  \draw[ipe pen fat, ipe dash dotted]
    (184, 608)
     -- (224, 608);
  \draw[ipe pen heavier, -{ipe pointed[ipe arrow small]}]
    (276, 560)
     -- (355, 560);
  \filldraw[draw=blue, ipe pen fat, fill=darkcyan, ipe opacity 10]
    (355, 568) rectangle (461, 532);
  \draw[ipe pen heavier]
    (312, 524) rectangle (484, 492);
  \draw[ipe pen fat, ipe dash dotted]
    (224, 508)
     -- (312, 508);
  \draw[ipe pen heavier, -{ipe pointed[ipe arrow small]}]
    (224, 552)
     -- (224, 488);
  \draw[darkgreen, ipe pen ultrafat]
    (132, 488) rectangle (308, 428);
  \draw[ipe pen heavier, -{ipe pointed[ipe arrow small]}]
    (224, 428)
     -- (224, 404)
     -- (108, 404)
     -- (108, 560)
     -- (172, 560);
  \filldraw[draw=darkgreen, ipe pen fat, fill=white]
    (332, 416) rectangle (508, 388);
  \node[ipe node]
     at (344, 404) {Pick a $W'$-tilt $(\tilde{W}',\tilde{\frak{R}}')$ of $(\breve{W},\breve{\frak{R}})$};
  \node[ipe node]
     at (336, 392) {where $W'$ is the central $r$-subwall of $W$};
  \node[ipe node]
     at (336, 460) {Find homogeneous flatness pair $(\breve{W},\breve{\frak{R}})$};
  \draw[ipe pen heavier]
    (212, 368) rectangle (288, 328);
  \draw[ipe pen fat, ipe dash dotted]
    (272, 368)
     -- (272, 396)
     -- (332, 396);
  \node[ipe node]
     at (352, 356) {Find flatness pair and apices};
  \node[ipe node]
     at (352, 344) {of ``big'' bidimensional degree};
  \node[ipe node]
     at (356, 292) {Find $(\frak{A}',R',{\bf a}')$ such that};
  \node[ipe node]
     at (334.766, 280) {$(\frak{A}, R,{\bf a})\models \theta_{{\sf R},{\bf c}}
\iff
(\frak{A}', R',{\bf a}')\models \theta_{{\sf R},{\bf c}}$};
  \draw[-{ipe pointed[ipe arrow small]}]
    (404, 440)
     -- (404, 416);
  \draw[-{ipe pointed[ipe arrow small]}]
    (404, 388)
     -- (404, 368);
  \draw[-{ipe pointed[ipe arrow small]}]
    (404, 328)
     -- (404, 304);
  \node[ipe node]
     at (408, 426.023) {$(\breve{W}, \breve{\frak{R}})$};
  \node[ipe node]
     at (409.402, 374.997) {$(\tilde{W}', \tilde{\frak{R}}')$};
  \node[ipe node]
     at (408.02, 314.538) {$(W'',\frak{R}''), {\bf a}''$};
  \node[ipe node]
     at (206.132, 671.568) {Consider};
  \node[ipe node]
     at (44, 620) {$\frak{A}\models \theta
\iff
(\frak{A}, V(\frak{A}),{\bf a})\models \theta_{{\sf R},{\bf c}}$};
  \node[ipe node]
     at (208.379, 648.831) {$\theta_{{\sf R},{\bf c}}$ of $\theta$};
  \node[ipe node]
     at (183.585, 658.92) { an enhanced version};
  \draw[red, ipe pen fat]
    (368.0001, 599.9996) rectangle (428.0001, 571.9996);
  \node[ipe node]
     at (216, 348) {minor-exclusion};
  \node[ipe node]
     at (218.498, 358.45) {Irrelevancy for};
  \draw[darkgreen, ipe pen fat]
    (327.9996, 480.0005) rectangle (511.9996, 252.0005);
  \node[ipe node]
     at (168, 532) {$((W,\frak{R}), {\bf a}')$};
  \node[ipe node]
     at (148, 476) {Find $\frak{A}'\subsetneq \frak{A}$ and $R'\subsetneq R$ such that};
  \node[ipe node]
     at (134, 460) {$(\frak{A}, R,{\bf a}')\models \theta_{{\sf R},{\bf c}}\iff
(\frak{A}', R',{\bf a}')\models \theta_{{\sf R},{\bf c}}$};
  \draw[darkgreen, ipe pen heavier, ipe dash dashed]
    (308, 428)
     .. controls (316, 406.6667) and (322.6667, 349.3333) .. (328, 256);
  \draw[darkgreen, ipe pen heavier, ipe dash dashed]
    (308, 488)
     .. controls (313.3333, 482.6667) and (320, 480) .. (328, 480);
  \draw[ipe pen heavier]
    (312, 688) rectangle (464, 660);
  \draw[ipe pen heavier]
    (312, 656) rectangle (464, 628);
  \node[ipe node]
     at (327.84, 676.287) {$\frak{A}\models \sigma \iff {\sf ap}_{\bf c}(\frak{A},{\bf a})\models \sigma^l$};
  \node[ipe node]
     at (314, 645.222) {$(\frak{A}, {\bf a})\models \varphi|_{{\sf ap}_{\bf c}} \iff {\sf ap}_{\bf c}(\frak{A},{\bf a})\models \varphi$};
  \node[ipe node]
     at (362.057, 663.121) {(\autoref{@escrostonades})};
  \node[ipe node]
     at (358.839, 633.156) {(\autoref{@recognitions})};
  \draw[ipe pen fat, ipe dash dotted]
    (280, 668)
     -- (312, 668);
  \draw[ipe pen fat, ipe dash dotted]
    (280, 652)
     -- (312, 652);
\end{tikzpicture}}
}
\caption{{The flow of the algorithm in the proof of~\autoref{@decendientes} along with the supporting results.}}
\labels{fig_general_sch_alg}
\end{figure}

\myskip\subsection{Sketch of proof of \autoref{@desmembramientos}}
\label{sec_what_3_sec}
In the next three sections,
we aim to provide a proof
for~\autoref{@desmembramientos}. In this subsection we give a brief description of the main ideas of this proof.

\myskip\paragraph{Dealing with $\bar{Θ}_1.$}
In~\autoref{sec_first_floor} we prove~\autoref{@desmembramientos} for sentences in $\bar{Θ}_1[τ].$
A sentence $θ∈\bar{Θ}_1[τ]$ can be written as $β\trianglerightγ,$ where $β∈\MSOL^\tw[τ\cup\{{\sf X}\}]$ and $γ = σ\wedge μ$ or $γ=(σ\wedge μ)^{({\sf c})},$ for some $σ∈\FOL[τ]$ and $μ∈\NTMC[τ].$
We associate each sentence $θ∈\bar{Θ}_1[τ]$ with a $\circ/\bullet$-flag, that is the symbol $\circ$ if $γ=(σ\wedge μ)^{({\sf c})}$, and the symbol $\bullet$ if $γ = σ\wedge μ.$
From this viewpoint,
depending on the $\circ/\bullet$-flag of $w,$
the target sentences $σ$ and $μ$ are asked either in the substructure of $\mathfrak{A}$ induced by the vertex set of each connected component of $G_{\mathfrak{A}}\setminus X,$ or in the whole structure $\mathfrak{A}\setminus X.$

Assume the existence of a ``big enough'' flatness pair $(W,\mathfrak{R})$ in $G_{\mathfrak{A}}$
and a ``big enough'' pseudogrid ${\bf W}_q$ defined by some vertical and horizontal paths of $W.$
Note that since $β∈\MSOL^\tw[τ\cup\{{\sf X}\}],$
${\sf cl}_{\sf X}({\sf star}_{\sf X}(\mathfrak{A},X))$ has bounded treewidth
and therefore, by~\autoref{@congregation}, $X$ has ``small'' bidimensionality with respect to $(W,\mathfrak{R}).$
Therefore, the removal of $X$ from $G_{\mathfrak{A}}$ leaves a ``big bulk'' of the wall in some connected component of $G_{\mathfrak{A}}\setminus X,$ i.e., there is a $\breve{C}∈{\sf cc}(G_{\mathfrak{A}},X)$ that is privileged with respect to ${\bf W}_q$ and $X.$
Depending on the $\circ/\bullet$-flag of $θ,$
we define the $w$-privileged set $C$ of $G_{\mathfrak{A}}$ with respect to ${\bf W}_q$ and $X$ to be either $\breve{C}$ or the whole set $V(\mathfrak{A})\setminus X.$

The above allows us to define a sentence $\tilde{θ}_q$ obtained from $θ_{{\sf R},{\bf c}}$ after splitting it into three parts (see~\autoref{@reconquistasen}).
The first part of  $\tilde{θ}_q$ is a new sentence $θ^{\sf out}_q$ that contains the modulator sentence $β$ and the target sentence $\breve{ζ}_{\sf R}|_{{\sf ap}_{\bf c}}\wedge μ$ asked in every connected component that is not in the $w$-privileged set $C.$
Note that the latter question appears only when $C=\breve{C}.$
The other two parts of  $\tilde{θ}_q$ are $\breve{ζ}_{\sf R}|_{{\sf ap}_{\bf c}}$
and the \NTMC-target sentence $μ,$ asked in the $w$-privileged set $C.$
The obtained sentence $\tilde{θ}_q$ is called the {\sl split version} of $θ_{{\sf R},{\bf c}}.$
Under the presence of a ``big enough'' pseudogrid, $\tilde{θ}_q$ and $θ_{{\sf R},{\bf c}}$ are proven to be equivalent (\autoref{lemma_equiva}).

Therefore, to reduce a structure $(\mathfrak{A},R,{\bf a})$ to an equivalent one with respect to the satisfaction of $θ_{{\sf R},{\bf c}},$
assuming the existence of a ``big enough'' wall in the input structure,
we will use the sentence $\tilde{θ}_q$ that ``separates'' the questions to the non-privileged and the privileged part of the structure.

The irrelevancy for the ``minor-exclusion'' part of $\tilde{θ}_q$ is guaranteed by assumption 5 of~\autoref{@desmembramientos}.
To deal with the sentences $θ^{\sf out}_q$ and $\breve{ζ}_{\sf R}|_{{\sf ap}_{\bf c}},$ we define the out-signature (\autoref{sec_out-sig_first-floor}) and the in-signature (\autoref{sec_in-sig_first-floor}) of a flatness pair, respectively, and the combination of these two constitutes the characteristic of a flatness pair.
This characteristic is an ``encoding'' of the partial satisfaction of $θ^{\sf out}_q$ and $\breve{ζ}_{\sf R}|_{{\sf ap}_{\bf c}}$ inside the flatness pair, and it is worth noting that it is \MSOL-definable.
After defining this characteristic, we use the following algorithm, that is formally presented in~\autoref{@inhumainement}.

\myskip\paragraph{The algorithm ${\tt Find\_Equiv\_FlatPairs}.$}
\begin{itemize}
\item Compute a packing of $z$ subwalls $W_1,\ldots, W_z$ of $W,$ where $z$ is some ``big enough'' integer depending on the sentence $θ,$ such that the compasses of all $W_i$-tilts of $(W,\mathfrak{R})$ are pairwise disjoint (this packing of walls virtually corresponds to the packing of walls inside the fifth wall of~\autoref{figure_walls_zoom}).
\item Compute a $W_i$-tilt of $(W,\mathfrak{R})$ for each $i∈[z].$ These define a collection $\tilde{\cal W}$ of $z$ flatness pairs.
\item For each of the flatness pairs in $\tilde{\cal W},$ compute its characteristic.
\item Output a collection $\tilde{\cal W}'$ of at least $m$ flatness pairs that have all the same characteristic and,
for some $(W_0,\mathfrak{R}_0)∈ \tilde{\cal W}'$ (that virtually corresponds to the sixth wall in~\autoref{figure_walls_zoom}),
the set $Y:=V({\sf compass}_{\breve{\mathfrak{R}}'}(\breve{W}')),$ where
 $(\breve{W}',\breve{\mathfrak{R}}')$ is a $\breve{W}$-tilt
of $(W,\mathfrak{R})$ and $\breve{W}$ is the central $j'$-subwall of ${W}_0,$
and
a $W^\bullet$-tilt
$(\tilde{W}',\tilde{\mathfrak{R}}')$
of $(W,\mathfrak{R}),$
where $W^\bullet$ is be the central $g$-subwall of ${W}_0$
(in the sixth wall of~\autoref{figure_walls_zoom}, $Y$ corresponds to the light blue area and $\breve{W}'$ to the innermost part of the wall).
\end{itemize}

After detecting $Y$ and $(\tilde{W}',\tilde{\mathfrak{R}}'),$ what remains is to prove that
$(\mathfrak{A},R, {\bf a})\models θ_{{\sf R},{\bf c}} \iff (\mathfrak{A}\setminus V({\sf compass}_{\tilde{\mathfrak{R}}'}(\tilde{W}')), R\setminus Y, {\bf a})\models θ_{{\sf R},{\bf c}}.$
The proof of the above is presented in~\autoref{sec_proof_correctness} and is split into three parts, corresponding to \autoref{claim_1}, \autoref{claim_2}, and \autoref{claim_3} (see \autoref{figu_levels_1and2}).

\begin{figure}[ht]
\centering
\medskip
\tikzstyle{ipe stylesheet} = [
  ipe import,
  even odd rule,
  line join=round,
  line cap=butt,
  ipe pen normal/.style={line width=0.4},
  ipe pen heavier/.style={line width=0.8},
  ipe pen fat/.style={line width=1.2},
  ipe pen ultrafat/.style={line width=2},
  ipe pen normal,
  ipe mark normal/.style={ipe mark scale=3},
  ipe mark large/.style={ipe mark scale=5},
  ipe mark small/.style={ipe mark scale=2},
  ipe mark tiny/.style={ipe mark scale=1.1},
  ipe mark normal,
  /pgf/arrow keys/.cd,
  ipe arrow normal/.style={scale=7},
  ipe arrow large/.style={scale=10},
  ipe arrow small/.style={scale=5},
  ipe arrow tiny/.style={scale=3},
  ipe arrow normal,
  /tikz/.cd,
  ipe arrows, 
  <->/.tip = ipe normal,
  ipe dash normal/.style={dash pattern=},
  ipe dash dotted/.style={dash pattern=on 1bp off 3bp},
  ipe dash dashed/.style={dash pattern=on 4bp off 4bp},
  ipe dash dash dotted/.style={dash pattern=on 4bp off 2bp on 1bp off 2bp},
  ipe dash dash dot dotted/.style={dash pattern=on 4bp off 2bp on 1bp off 2bp on 1bp off 2bp},
  ipe dash normal,
  ipe node/.append style={font=\normalsize},
  ipe stretch normal/.style={ipe node stretch=1},
  ipe stretch normal,
  ipe opacity 10/.style={opacity=0.1},
  ipe opacity 30/.style={opacity=0.3},
  ipe opacity 50/.style={opacity=0.5},
  ipe opacity 75/.style={opacity=0.75},
  ipe opacity opaque/.style={opacity=1},
  ipe opacity opaque,
]
\definecolor{red}{rgb}{1,0,0}
\definecolor{blue}{rgb}{0,0,1}
\definecolor{green}{rgb}{0,1,0}
\definecolor{yellow}{rgb}{1,1,0}
\definecolor{orange}{rgb}{1,0.647,0}
\definecolor{gold}{rgb}{1,0.843,0}
\definecolor{purple}{rgb}{0.627,0.125,0.941}
\definecolor{gray}{rgb}{0.745,0.745,0.745}
\definecolor{brown}{rgb}{0.647,0.165,0.165}
\definecolor{navy}{rgb}{0,0,0.502}
\definecolor{pink}{rgb}{1,0.753,0.796}
\definecolor{seagreen}{rgb}{0.18,0.545,0.341}
\definecolor{turquoise}{rgb}{0.251,0.878,0.816}
\definecolor{violet}{rgb}{0.933,0.51,0.933}
\definecolor{darkblue}{rgb}{0,0,0.545}
\definecolor{darkcyan}{rgb}{0,0.545,0.545}
\definecolor{darkgray}{rgb}{0.663,0.663,0.663}
\definecolor{darkgreen}{rgb}{0,0.392,0}
\definecolor{darkmagenta}{rgb}{0.545,0,0.545}
\definecolor{darkorange}{rgb}{1,0.549,0}
\definecolor{darkred}{rgb}{0.545,0,0}
\definecolor{lightblue}{rgb}{0.678,0.847,0.902}
\definecolor{lightcyan}{rgb}{0.878,1,1}
\definecolor{lightgray}{rgb}{0.827,0.827,0.827}
\definecolor{lightgreen}{rgb}{0.565,0.933,0.565}
\definecolor{lightyellow}{rgb}{1,1,0.878}
\definecolor{black}{rgb}{0,0,0}
\definecolor{white}{rgb}{1,1,1}
\begin{tikzpicture}[ipe stylesheet]
  \filldraw[draw=darkgreen, ipe pen heavier, fill=lightgreen, ipe opacity 50]
    (156.0689, 717.4239) rectangle (188.0689, 697.4239);
  \filldraw[blue, ipe pen heavier, ipe opacity 30]
    (156.069, 741.424) rectangle (188.069, 721.424);
  \draw[-{ipe pointed[ipe arrow small]}]
    (51.7679, 708.103)
     -- (116, 708);
  \node[ipe node]
     at (120, 704) {$\tilde{\theta}_q$};
  \node[ipe node]
     at (34.141, 704.311) {$\theta_{{\sf R},{\bf c}}$};
  \draw[-{ipe fptarc[ipe arrow small]}]
    (132, 708)
     -- (156, 708);
  \draw[-{ipe fptarc[ipe arrow small]}]
    (132, 708)
     -- (155.12, 728.96);
  \draw[-{ipe fptarc[ipe arrow small]}]
    (132, 708)
     -- (154.495, 683.072);
  \filldraw[red, ipe pen heavier, ipe opacity 30]
    (156.0689, 693.4239) rectangle (188.0689, 673.4239);
  \filldraw[draw=green, ipe pen heavier, fill=lightgreen, ipe opacity 30]
    (156.1342, 717.438) rectangle (188.1342, 697.438);
  \node[ipe node]
     at (168.069, 681.424) {$\mu$};
  \node[ipe node]
     at (160.728, 704.492) {$\breve{\zeta}_{\sf R}|_{{\sf ap}_{\bf c}}$};
  \node[ipe node]
     at (164.639, 729.424) {$\theta_q^{\sf out}$};
  \draw[ipe dash dashed]
    (188, 732)
     -- (232, 732);
  \node[ipe node, text=blue]
     at (234.522, 729.779) {{\sf out-sig}};
  \draw[ipe dash dashed]
    (188, 708)
     -- (236, 708);
  \node[ipe node, text=green, ipe opacity 50]
     at (237.248, 705.503) {{\sf in-sig}};
  \node[ipe node, text=darkgreen, ipe opacity 50]
     at (237.31, 705.449) {{\sf in-sig}};
  \draw[ipe dash dashed]
    (264.6, 731.994)
     -- (312, 732);
  \draw[ipe dash dashed]
    (264, 708)
     -- (312, 708);
  \node[ipe node]
     at (313.47, 729.296) {\autoref{claim_1}/\autoref{claim_4}};
  \node[ipe node]
     at (218.591, 686.01) {Assumption 5};
  \node[ipe node]
     at (220.474, 674.264) {of~\autoref{@desmembramientos}};
  \draw[ipe dash dashed]
    (188, 684)
     -- (312, 684);
  \draw[blue, ipe pen heavier, ipe opacity 50]
    (311.598, 741.501)
     -- (311.598, 721.501)
     -- (397.0547, 721.61)
     -- (397.0547, 741.675)
     -- cycle;
  \node[ipe node]
     at (313.543, 704.899) {\autoref{claim_2}/\autoref{claim_5}};
  \draw[green, ipe pen heavier, ipe opacity 50]
    (311.6707, 717.1044)
     -- (311.6707, 697.1044)
     -- (397.0547, 697.2134)
     -- (397.0547, 717.2784)
     -- cycle;
  \node[ipe node]
     at (313.655, 681.26) {\autoref{claim_3}/\autoref{claim_6}};
  \draw[red, ipe pen heavier, ipe opacity 75]
    (311.783, 693.4652)
     -- (311.783, 673.4652)
     -- (397.167, 673.5742)
     -- (397.071, 693.6392)
     -- cycle;
  \draw[darkgreen, ipe pen heavier, ipe opacity 50]
    (311.671, 717.1045)
     -- (311.671, 697.1045)
     -- (397.055, 697.2135)
     -- (397.055, 717.2785)
     -- cycle;
  \node[ipe node]
     at (58.335, 710.999) {\autoref{lemma_equiva}};
\end{tikzpicture}
\medskip
\caption{The structure of the proof of~\autoref{@desmembramientos} for sentences in $\bar{Θ}_1[τ]$ and  in $\bar{Θ}[τ].$ Claims~\ref{claim_1},~\ref{claim_2}, and~\ref{claim_3} correspond to the fragment $\bar{Θ}_1$, while Claims~\ref{claim_4},~\ref{claim_5}, and~\ref{claim_6} correspond to the fragment $\bar{Θ}$.}
\labels{figu_levels_1and2}
\end{figure}
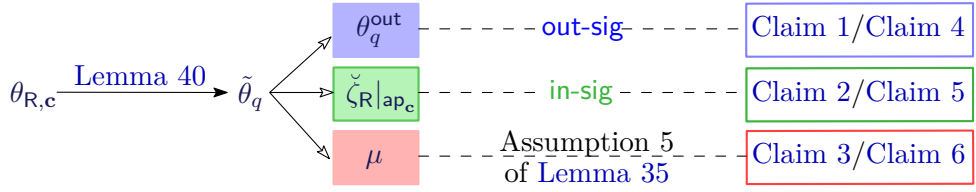

\myskip\paragraph{Dealing with $\bar{Θ}.$}
In~\autoref{the_second_level} we extend the above idea to sentences in $\bar{Θ}.$
We associate each sentence $θ∈ \bar{Θ}$ to a $\circ/\bullet$-scenario $w,$ that is a string of alphabet $\{\circ,\bullet\}$ that encodes whether each recursive question in the definition of $θ$ is asked in the whole remaining structure or in each of its connected components (see~\autoref{sec_scenarios}).
In this sense,
the notion of $w$-privileged set defined for sentences in $\bar{Θ}_1$ is generalized to the notion of $w$-privileged {\sl sequence}, that is the sequence of recursively obtained $w$-privileged sets as dictated by the $\circ/\bullet$-scenario $w.$
Also, for every $θ∈ \bar{Θ},$ in~\autoref{@arrangements} we define analogously the split version $\tilde{θ}_q$ of $θ_{{\sf R},{\bf c}}$
and prove that under the presence of
a ``big enough'' pseudogrid, $\tilde{θ}_q$ and $θ_{{\sf R},{\bf c}}$
are equivalent (\autoref{@interruptions}).

Also, in the case where $θ∈\bar{Θ}[τ],$ the out-signature and the in-signature of the extended compass of a flatness pair are defined analogously: we encode the partial satisfaction of $θ^{\sf out}_q$ for each recursive level  of the definition of $θ$ and the satisfaction of $\breve{ζ}_{\sf R}|_{{\sf ap}_{\bf c}}$ in the privileged connected component of $G_{\mathfrak{A}}$ with respect to a ``big enough'' pseudogrid ${\bf W}_q$ and
the union of all modulators $X_i$ (see~\autoref{@enthaltenden}).

In order to prove~\autoref{@desmembramientos} for sentences in $\bar{Θ}[τ],$ we use the algorithm ${\tt Find\_Equiv\_FlatPairs}$ for the ``more general'' characteristic mentioned above.
The proof of correctness of the algorithm ${\tt Find\_Equiv\_FlatPairs}$
for sentences in $\bar{Θ}[τ]$ is
again split into three main claims, namely
\autoref{claim_4},~\autoref{claim_5},~and~\autoref{claim_6} (see \autoref{figu_levels_1and2}).
Since the proof is very similar to the one for sentences in $\bar{Θ}_1[τ]$ presented in~\autoref{sec_proof_correctness},
we provide a sketch of proof in the end of~\autoref{@enthaltenden}, and a complete proof in~\autoref{second_level_more}.

\myskip\paragraph{Dealing with $Θ.$}
In~\autoref{sec_final_combo} we describe how to further modify the signatures in order to capture the full generality of $Θ.$
We start by changing the given sentence $θ∈Θ[τ]$ to a sentence $θ'$ that is equivalent to $θ$ but, instead of allowing arbitrary positive Boolean combinations, we allow only disjunctions of conjunctions.
This new sentence is equivalent to $θ$ (\autoref{@inexpressibly}).
This ``simplification'' allows us to do a two-step approach,  by first allowing only conjunctions (see~\autoref{@determinadas})  and then also allowing disjunctions of conjunctions (see~\autoref{@insinuations}).
In both steps, we perform a trick to express these more general sentences as
a finite combination of sentences in $\bar{Θ}$ and we build out-signatures and in-signatures based on this ``reduction''.
In~\autoref{sec_final_combo} we define the analogue of split version of $θ_{{\sf R},{\bf c}}$ when allowing conjunctions or even disjunctions of conjunctions, and the corresponding characteristics of extended compasses of flatness pairs.
To prove~\autoref{@desmembramientos} for sentences in $Θ[τ],$ we again use the algorithm ${\tt Find\_Equiv\_FlatPairs}$ and, in the end of~\autoref{sec_final_combo}, we describe how to prove its correctness for sentences in $Θ[τ].$

\myskip\section{Dealing with \texorpdfstring{$\bar{\Theta}_1$}{bar-Theta-1}}
\labels{sec_first_floor}

Let $τ$ be a vocabulary.
In this section we aim to present the proof of~\autoref{@desmembramientos} for a sentence $θ∈\bar{Θ}_1[τ].$
We fix ${\sf X}$ to be a second-order variable.
Let $σ∈ \FOL[τ]$ and $μ∈ \NTMC[τ]$ be the target sentences of $θ$ and $β∈ \MSOL^\tw[τ\cup \{{\sf X}\}]$ be the modulator sentence of $θ.$
By definition,
\begin{eqnarray}
θ  & = & β\triangleright γ ,\labels{@technologies}
\end{eqnarray}
where $γ = (σ\wedge μ)^{({\sf c})}$ or $γ  = σ\wedge μ.$

In other words, $θ$ asks whether, given a $τ$-structure $\mathfrak{A},$
there exists a set $X\subseteq V(\mathfrak{A})$ such that
 \begin{itemize}
 \item ${\sf star}_{\sf X}(\mathfrak{A},X)\models β$  \mbox{~and~}
 \item depending whether $γ = (σ\wedge μ)^{({\sf c})}$ or $γ  = σ\wedge μ$:
 \begin{itemize}
 \item either for every connected component $C$ of $G_\mathfrak{A}\setminus X,$ it holds that
 $\mathfrak{A}[C]\models σ\wedge μ,$ or,
 \item ${\sf rm}_{\sf X} (\mathfrak{A},X)\models σ\wedge μ.$
\end{itemize}
\end{itemize}
\bigskip

Before proceeding to the proof of~\autoref{@desmembramientos} for a sentence $θ∈\bar{Θ}_1[τ],$ we gather all formulas used in the definition of a $θ∈\bar{Θ}_1[τ]$ in~\autoref{@individuality}.

\begin{table}[H]
\centering
 \begin{tabular}{|| c | c||}
 \hline
 Formulas & Meaning\\ [0.5ex]
 \hline\hline
$β$ & modulator sentence expressing an \MSOL-property on bounded treewidth graphs\\
\hline
 $σ$ & target \FOL-sentence in (each connected component of) the ``remaining'' structure\\
 \hline
 $μ$ & target \NTMC-sentence in (each connected component of) the ``remaining'' structure\\
\hline
$γ$ & $(σ\wedge μ)^{({\sf c})}$ or $σ\wedge μ$\\
\hline
$θ$ & $\exists {\sf X}\ β|_{{\sf star}_{\sf X}}\wedge γ|_{{\sf rm}_{\sf X}}$ \\
\hline
\end{tabular}\caption{List of formulas introduced in the definition of a $θ∈\bar{Θ}_1[τ].$}\labels{@individuality}
\end{table}

We now associate each sentence $θ∈\bar{Θ}_1 [τ]$ with a $\circ/\bullet$-flag $w,$ that encodes whether $γ=(σ\wedge μ)^{({\sf c})}$ or
$γ=(σ\wedge μ)^{({\sf c})}.$
According to this, we define the $w$-privileged set of the Gaifman graph of the given structure to be the either the privileged component or the whole remaining graph,
depending on whether $w=\circ$ or $w=\bullet.$

\myskip\paragraph{The $\circ/\bullet$-flag.}
Let $θ∈\bar{Θ}_1[τ]$ and $σ, μ$ be its target sentences.
We define the {\em $\circ/\bullet$-flag} $w$ of $θ$ to be either the symbol ``$\circ$'' if $γ = (σ\wedge μ)^{({\sf c})},$ or the symbol ``$\bullet$'' if $γ= σ\wedge μ.$
Every sentence $θ∈\bar{Θ}_1[τ]$ is associated with its $\circ/\bullet$-flag.

\myskip\paragraph{Choosing the privileged set given by a $\circ/\bullet$-flag.}
Let $G$ be a graph, let $q∈ \mathbb{N},$ let ${\bf W}_q∈ (2^{V(G)})^{2q},$ and let $X$ be a subset of $V(G).$
Also, let $w∈\{\circ,\bullet\}.$

A set $C$ is {\em $w$-privileged set of $G$ with respect to ${\bf W}_q$ and $X$}, if
$$C =
\begin{cases}
\text{an element in }{\sf pr}(G, {\bf W}_q, X), & \text{if } w_i = \circ \text{ and }{\sf pr}(G, {\bf W}_q, X)\neq \emptyset,\\
\emptyset, & \text{if } w_i = \circ \text{ and }{\sf pr}(G, {\bf W}_q, X)= \emptyset, \text{ and}\\
V(G)\setminus X, & \text{if } w_i = \bullet.
\end{cases}
$$

By~\autoref{@vuitcentistes},
if ${\bf W}_q$ is a  ${q}$-pseudogrid of a graph $G$ and $X$ is a subset of $V(G),$ then ${\sf pr}(G,{\bf W}_{{q}},X)$ is either a singleton or the empty set.
Therefore, we get the following:\smallskip

\begin{observation}
\labels{@preconditions}
If ${\bf W}_q$ is a  ${q}$-pseudogrid of a graph $G,$ $X$ is a subset of $V(G),$ and $w∈\{\circ/\bullet\},$ then there is exactly one $w$-privileged set of $G$ with respect to ${\bf W}_q$ and $X.$
\end{observation}

It is trivial to see that the notion of a $q$-pseudogrid and the notion of a $w$-privileged set can be expressed in $\MSOL.$

\begin{lemma}\labels{@pejroanlmaaragall}
Let $q∈ \mathbb{N},$ let $τ$ be a vocabulary, let ${\bf Q}\cup\{{\sf X},{\sf C}\}$
be a set of $2q+2$ unary relation symbols that are not contained in $τ,$ and let $w∈\{\circ,\bullet\}.$
There is a sentence $η_{w\text{-}{\sf pr}_{{\sf X},{\sf C}}}∈ \MSOL[τ\cup{\bf Q}\cup\{{\sf X},{\sf C}\}]$
such that for every $τ$-structure $\mathfrak{A},$
every ${\bf W}_q∈ {(2^{V(\mathfrak{A})})}^{2q},$ and every $X,C\subseteq V(\mathfrak{A}),$
$(\mathfrak{A},{\bf W}_{q},X,C) \models η_{w\text{-}{\sf pr}_{{\sf X},{\sf C}}}$ (where ${\sf X}$ is interpreted as $X$ and
${\sf C}$ is interpreted as $C$) if and only if ${\bf W}_q$ is a $q$-pseudogrid
of $G_{\mathfrak{A}}$ and $C$ is the $w$-privileged set of $G_{\mathfrak{A}}$ with respect to ${\bf W}_q$ and $X.$
\end{lemma}

The rest of this section is structured as follows:
In~\autoref{@reconquistasen} we define the {\sl split sentence} $\tilde{θ}_q$ of $θ,$ that is a sentence equivalent to $θ$ that separates the questions of $θ$ that concern the non-privileged parts of the structure, and the questions that correspond to the privileged one.
Then, in~\autoref{@verwechslungen}, we define the {\sl extended compass} of a flatness pair, that
is a tuple that contains all necessary information around a flatness pair.
Then, in~\autoref{sec_out-sig_first-floor} and~\autoref{sec_in-sig_first-floor}, we define the
{\sl out-signature} and the {\sl in-signature} of
the extended compass of a flatness pair
that encodes how a partial solution (partial assignment of vertices to the variables)
satisfies the ``non-privileged'' and the ``privileged'' part of $\tilde{θ}_q,$ respectively.
Finally, in~\autoref{@inhumainement}, we present the algorithm
${\tt Fins\_Equiv\_FlatPairs}$
and, in~\autoref{sec_proof_correctness},
we prove that this algorithm correctly returns the claimed output of~\autoref{@desmembramientos} for the particular case of sentences in $\bar{Θ}_1[τ].$

\myskip\subsection{The sentences \texorpdfstring{$θ^{\sf out}_{{q}}$}{theta-out-q} and  \texorpdfstring{$\tilde{θ}_{{q}}$}{tilde-theta-q}}\labels{@reconquistasen}
In this subsection we aim to define two sentences $θ^{\sf out}_{{q}}$ and  $\tilde{θ}_{{q}}$ that will allow us
to ``break'' $θ$ into two questions:
one that concerns the ``privileged'' part and one that concerns the ``non-privileged'' part of the graph.

Let $q,l∈ \mathbb{N},$ let $τ$ be a vocabulary, let ${\bf c}$ be a collection of $l$ constant symbols not contained in $τ,$ and let ${\bf Q}\cup\{{\sf R},{\sf X}\}$ be a set of $2q+2$ unary relation symbols not contained in $τ.$
Let $θ∈\bar{Θ}_1[τ].$
Also, let $w$ be the $\circ/\bullet$-flag of $θ,$
let $β∈ \MSOL^\tw[τ\cup\{X\}]$ be the modulator sentence of $θ$ and
$σ∈\FOL[τ]$ and $μ∈\NTMC[τ]$ be the \FOL-target and \NTMC-target sentences of $θ,$ respectively.
Also, let $ζ=σ^l$ be the $l$-apex-projection of $σ$ and $\breve{ζ}$ be a Gaifman sentence that is equivalent to $ζ.$
Let $θ_{{\sf R},{\bf c}}$ be the enhanced version of $θ$
obtained from $θ$ after replacing $σ$with $\breve{ζ}_{\sf R}|_{{\sf ap}_{\bf c}}.$

\myskip\paragraph{The sentence $θ^{\sf out}_{{q}}.$}
We define the sentence $θ^{\sf out}_{{q}}∈ \MSOL[τ\cup{\bf Q}\cup\{{\sf R},{\sf X}\}\cup{\bf c}]$ such that for every $τ$-structure $\mathfrak{A},$ every ${\bf W}_q∈ {(2^{V(\mathfrak{A})})}^{2q},$
every apex-tuple ${\bf a}$ of $\mathfrak{A}$ of size $l,$ and every $X,R\subseteq V(\mathfrak{A}),$ $(\mathfrak{A},R,{\bf W}_{{q}},{\bf a},X)\models θ^{\sf out}_{{q}}$ if and only if the following conditions are satisfied:
\begin{itemize}
\item $(\mathfrak{A},X)\models β|_{{\sf star}_{\sf X}}$ and
\item ${\bf W}_q$ is a $q$-pseudogrid of $G_\mathfrak{A}$ and
\item if $C$ is the $w$-privileged set of $G_{\mathfrak{A}}$ with respect to ${\bf W}_q$ and $X,$ then for every $C'∈ {\sf cc}(G_{\mathfrak{A}},S)$ that is not a subset of $C$ it holds that $(\mathfrak{A},R, {\bf a})[C']\models \breve{ζ}_{\sf R}|_{{\sf ap}_{\bf c}}\wedge μ.$
\end{itemize}
Note that $θ^{\sf out}_{{q}}∈ \MSOL[τ\cup{\bf Q}\cup\{{\sf R},{\sf X}\}\cup{\bf c}]$ since $β|_{{\sf star}_{\sf X}}∈ \MSOL[τ\cup\{{\sf X}\}]$ and $\breve{ζ}_{\sf R}|_{{\sf ap}_{\bf c}}∈ \FOL[τ\cup\{{\sf R}\}\cup{\bf c}].$
{We stress that, in the definition of $θ^{\sf out}_{{q}}$ we ask that
${\bf W}_q$ is a $q$-pseudogrid of $G_\mathfrak{A}$ since we need to guarantee this fact in order to ask the third item, the question $\breve{ζ}_{\sf R}|_{{\sf ap}_{\bf c}}\wedge μ$ for every $C'∈ {\sf cc}(G_{\mathfrak{A}},S)$ that is not subset of the $w$-privileged set (to define the $w$-privileged set we need ${\bf W}_q$ to be a pseudogrid).}

Intuitively, if $w=\bullet,$ the sentence $θ^{\sf out}_q$
is the conjunction of $β|_{{\sf star}_{\sf X}}$ and a sentence that asks that ${\bf W}_q$ is a $q$-pseudogrid, while, if $w=\circ,$ $θ^{\sf out}_q$ is the conjunction of $β|_{{\sf star}_{\sf X}},$ a sentence that asks that ${\bf W}_q$ is a $q$-pseudogrid, and a sentence that asks that
$\breve{ζ}_{\sf R}\wedge μ$ is satisfied in every ``non-privileged'' connected component.
In other words,
$θ^{\sf out}_q$
does everything that $θ_{{\sf R},{\bf c}}$ does, except from the $w$-privileged set $C,$ and also asks that ${\bf W}_q$ is a $q$-pseudogrid.
%

\myskip\paragraph{The sentence $\tilde{θ}_{{q}}.$}
We now define the sentence $\tilde{θ}_{{q}}∈\MSOL[τ\cup{\bf Q}\cup\{{\sf R}\}\cup {\bf c}]$ such that

\begin{equation}
\tilde{θ}_{{q}}  = \exists {\sf X}\    \blue{\fbox{$θ^{\sf out}_{{q}}$}}\  \wedge
\green{\fbox{$\exists {\sf C} \ η_{w\text{-}{\sf pr}_{{\sf X},{\sf C}}}\ \wedge \breve{ζ}_{\sf R} |_{{\sf ap}_{\bf c}}|_{{\sf ind}_{\sf C}}$}}\ \wedge
 \red{\fbox{$\exists {\sf C} \ η_{w\text{-}{\sf pr}_{{\sf X},{\sf C}}}\ \wedge  μ|_{{\sf ind}_{\sf C}}$}}
\labels{@desembarazaron}
\end{equation}
(recall that $θ^{\sf out}_{q}∈ \MSOL[τ\cup{\bf Q}\cup\{{\sf R},{\sf X}\}\cup{\bf c}],$ $η_{w\text{-}{\sf pr}_{{\sf X},{\sf C}}}∈ \MSOL[τ\cup {\bf Q}\cup\{{\sf X},{\sf C}\}],$ $μ∈ \MSOL[τ],$ and $\breve{ζ}_{\sf R} |_{{\sf ap}_{\bf c}}∈ \FOL[τ\cup \{{\sf R}\}\cup{\bf c}]$).

Alternatively,  if $\mathfrak{A}$ is a $τ$-structure, $R\subseteq V(\mathfrak{A}),$
${\bf W}_{{q}}∈ {(2^{V(\mathfrak{A})})}^{2q},$ and ${\bf a}$ is an apex-tuple of $\mathfrak{A}$ of size $l,$
then $(\mathfrak{A},R,{\bf W}_{{q}}, {\bf a})\models \tilde{θ}_{{q}} \iff \exists X\subseteq V(\mathfrak{A})$ such that

\begin{itemize}
\item  $\blue{\fbox{$(\mathfrak{A},R,{\bf W}_{{q}},X)\models θ^{\sf out}_{{q}}$}},$
\item \green{\fbox{\black{$\exists C\subseteq V(\mathfrak{A})$ that is $w$-privileged with respect to ${\bf W}_q$ and $X$ and $(\mathfrak{A}, R, {\bf a})[C]\models \breve{ζ}_{\sf R} |_{{\sf ap}_{\bf c}},$}}} and
\item \red{\fbox{\black{$\exists C\subseteq V(\mathfrak{A})$ that is $w$-privileged with respect to ${\bf W}_q$ and $X$ and $\mathfrak{A}[C] \models μ$}}}.
\end{itemize}

We stress that in the above sentence, the part ``$\exists {\sf C} \ η_{w\text{-}{\sf pr}_{{\sf X},{\sf C}}}$'' appears twice, but essentially refers to the same set since, by~\autoref{@vuitcentistes}, ${\sf pr}(G_\mathfrak{A},{\bf W}_{{q}},X)$ is either a singleton or the empty set.

Intuitively, if the Gaifman graph of $\mathfrak{A}$ contains a ``big enough'' wall, we can ``separate'' the questions that concern
$X$ and every non-privileged connected component of $G\setminus X$ (expressed by $θ^{\sf out}_{{q}}$) and
the question $\breve{ζ}_{\sf R} |_{{\sf ap}_{\bf c}}\wedge μ$ that concerns the $w$-privileged set of $G$
with respect to ${\bf W}_q$ and $X.$

We call $\tilde{θ}_q$ the {\em split version} of $θ_{{\sf R},{\bf c}}.$

\begin{lemma}\labels{lemma_equiva}
Let $τ$ be a vocabulary, ${\sf R}\notin τ$ be a unary relation symbol, and ${\bf c}$ be a collection of $l$ constant symbols, where $l∈\mathbb{N}_{≥ 1}.$
Let $θ∈ \bar{Θ}_1[τ],$ let ${q}=(\tw(θ)+1)^2+1,$ let $θ_{{\sf R},{\bf c}}$ be an enhanced version of $θ,$ and let $\tilde{θ}_q$ be the split version of $θ_{{\sf R},{\bf c}}.$
If $\mathfrak{A}$ is a $τ$-structure, $R\subseteq V(\mathfrak{A}),$
${\bf W}_{{q}}$ is a $q$-pseudogrid of $G_{\mathfrak{A}},$
and ${\bf a}$ is an apex-tuple of $\mathfrak{A}$ of size $l,$
then $(\mathfrak{A},R, {\bf a})\models θ_{{\sf R},{\bf c}}\iff (\mathfrak{A},R,{\bf W}_{{q}}, {\bf a})\models \tilde{θ}_{{q}}.$
\end{lemma}

\begin{proof}
We will prove the lemma for the case where the $\circ/\bullet$-flag $w$ of $θ$ is $\circ,$ i.e., $γ=(σ\wedge μ)^{({\sf c})},$ since the
case where $w=\bullet$ is trivial.
Let $\mathfrak{A}$ be a $τ$-structure, let $R\subseteq V(\mathfrak{A}),$ let ${\bf W}_{{q}}$ be a $q$-pseudogrid of $G_{\mathfrak{A}},$ and let ${\bf a}$ be an apex-tuple of $\mathfrak{A}$ of size $l.$
We first prove that if $(\mathfrak{A},R, {\bf a})\models θ_{{\sf R},{\bf c}},$ then $(\mathfrak{A},R,{\bf W}_{{q}}, {\bf a})\models \tilde{θ}_{{q}}.$
Suppose that $(\mathfrak{A},R, {\bf a})\models θ_{{\sf R},{\bf c}}.$
By definition, this implies that there is a set $X\subseteq V(\mathfrak{A})$ such that $(\mathfrak{A},X)\models β_{{\sf star}_{\sf X}}$ and for every $C∈ {\sf cc}(G_{\mathfrak{A}},X),$ it holds that $(\mathfrak{A},R,{\bf a})[C]\models \breve{ζ}_{\sf R} |_{{\sf ap}_{\bf c}} \wedge μ.$
Since $(\mathfrak{A},X)\models β_{{\sf star}_{\sf X}}$ and $β∈\MSOL^\tw [τ\cup\{{\sf X}\}],$ we have that ${\sf cl}_{\sf X} ({\sf star}_{\sf X}(\mathfrak{A},X))$ has treewidth at most $\tw(θ).$
Therefore, by~\autoref{@congregation}, $X$ intersects at most $(\tw(θ)+1)^2$ bags of every $(W,\mathfrak{R})$-canonical partition of $G_{\mathfrak{A}}\setminus V({\bf a}).$
Since ${\bf W}_q$ is a $q$-pseudogrid, where $q=(\tw(θ)+1)^2+1,$
there is a $C∈ {\sf pr}(G_{\mathfrak{A}}, {\bf W}_q, X)$ and by~\autoref{@vuitcentistes} $\{C\} = {\sf pr}(G_{\mathfrak{A}}, {\bf W}_q, X).$
Thus, we have that $(\mathfrak{A}, R,{\bf W}_q,{\bf a},X)\models θ_q^{\sf out}$ and $(\mathfrak{A},R,{\bf a})[C]\models  \breve{ζ}_{\sf R} |_{{\sf ap}_{\bf c}} \wedge μ.$
This implies that  $(\mathfrak{A},R,{\bf W}_{{q}}, {\bf a})\models \tilde{θ}_{{q}}.$
To prove the inverse implication, note that if $(\mathfrak{A},R,{\bf W}_{{q}}, {\bf a})\models \tilde{θ}_{{q}},$ then there is a
set $X\subseteq V(\mathfrak{A})$ such that $(\mathfrak{A}, R,{\bf W}_q, {\bf a},X)\models θ_q^{\sf out}$ and there is a $C∈{\sf pr}(G_{\mathfrak{A}},{\bf W}_q, X)$ such that $(\mathfrak{A}, R,{\bf a})[C]\models  \breve{ζ}_{\sf R} |_{{\sf ap}_{\bf c}}  \wedge μ.$
Thus, by the definition of $θ_q^{\sf out},$ we deduce that $(\mathfrak{A},R, {\bf a})\models θ_{{\sf R},{\bf c}}.$
\end{proof}

By~\autoref{lemma_equiva}, we also get the following:
\begin{corollary}\labels{corrollaa_pseudo}
Let $τ$ be a vocabulary, ${\sf R}\notin τ$ be a unary relation symbol, and ${\bf c}$ be a collection of $l$ constant symbols, where $l∈\mathbb{N}_{≥ 1}.$
Let $θ∈ \bar{Θ}_1[τ],$ let ${q}=(\tw(θ)+1)^2+1,$ let $θ_{{\sf R},{\bf c}}$ be an enhanced version of $θ,$ and let $\tilde{θ}_q$ be the split version of $θ_{{\sf R},{\bf c}}.$
If $\mathfrak{A}$ is a $τ$-structure, $R\subseteq V(\mathfrak{A}),$
${\bf W}_{{q}}^{(1)}, {\bf W}_q^{(2)}$ are two $q$-pseudogrids of $G_{\mathfrak{A}},$
and ${\bf a}$ is an apex-tuple of $\mathfrak{A}$ of size $l,$
then $(\mathfrak{A},R,{\bf W}_{{q}}^{(1)}, {\bf a})\models \tilde{θ}_{{q}}\iff (\mathfrak{A},R,{\bf W}_{{q}}^{(2)}, {\bf a})\models \tilde{θ}_{{q}}.$
\end{corollary}

See~\autoref{@administraciones} for the list of formulas introduced in the last two subsections.
\begin{table}[H]
\centering
\bgroup
\def\arraystretch{1.2}
 \begin{tabular}{|| c | c||}
 \hline
 Formulas & Meaning\\ [0.5ex]
 \hline\hline
$η_{w\text{-}{\sf pr}_{{\sf X},{\sf C}}}$ & $w$-privileged set \\
\hline
$θ^{\sf out}_q$& the question $β$ in $X$ and  questions $ \breve{ζ}_{\sf R} |_{{\sf ap}_{\bf c}}\wedge μ$ in ``non-privileged'' components\\
\hline
$\tilde{θ}_q$  & sentence equivalent to $θ_{{\sf R}, {\bf c}},$ separating question $ \breve{ζ}_{\sf R} |_{{\sf ap}_{\bf c}}\wedge μ$ on the $w$-privileged set \\

\hline
\end{tabular}
\egroup
\caption{List of formulas used for the definition of $\tilde{θ}_q.$}\labels{@administraciones}
\end{table}

\myskip\subsection{Extended compasses of flatness pairs}
\labels{@verwechslungen}
\myskip\paragraph{Extended compasses of flatness pairs.}
Let $l,r∈\mathbb{N}$ and let $j,q∈\mathbb{N}_{≥ 3}$ be two integers, such that $2r+j≥ q.$
Let $G$ be a graph, let ${\bf a}=(a_{1},\ldots,a_l)$ be an apex-tuple  of $G,$ and let $(W,\mathfrak{R})$ be a flatness pair of $G\setminus V({\bf a})$ of height $2r+j$.
For every subwall $W'$ of $W,$ we denote by ${\sf influence}_\mathfrak{R} (W')$ the set of the flaps of the flat wall $W$ that either contain an edge of the perimeter of $W'$ or are ``embedded'' inside the disk ``cropped'' by the perimeter of $W'.$ Intuitively, ${\sf influence}_\mathfrak{R} (W')$ contains all flaps ``captured'' by the wall $W'.$ See~\autoref{@consideracions} for a formal definition of the above notions.
The graph ${\sf compass}_{\mathfrak{R}}(W)$ is always assumed to be connected.
We set $K:= {\sf compass}_{\mathfrak{R}}(W)$ and $K^{\bf a}:=G[V({\bf a})\cup V(K)].$
Also, for every $i∈[r],$ let $I^{(i)} = V(\cupall {\sf influence}_\mathfrak{R} (W^{(2i+j)}))$ and let ${\bf I}=(I^{(1)},\ldots,I^{(r)}).$

Let $\mathfrak{A}$ be a $τ$-structure, let $G_{\mathfrak{A}}$ be its Gaifman graph, let ${\bf a}=(a_{1},\ldots,a_l)$ be an apex-tuple  of $G_{\mathfrak{A}},$ and let $(W,\mathfrak{R})$ be a flatness pair of $G_{\mathfrak{A}}\setminus V({\bf a})$ of height $2r+j.$
Also, let ${\bf W}_{{q}}$ be the $q$-pseudogrid defined by the horizontal and vertical paths of the central $q$-subwall of $W.$
We call the tuple $\mathfrak{K}=(\mathfrak{A}[V(K^{\bf a})],{\bf a}, {\bf I}, {\bf W}_{{q}})$  {\em the extended compass
of the flatness pair $(W,\mathfrak{R})$ of $G_{\mathfrak{A}}\setminus V({\bf a})$}.
Given a $Z\subseteq V(K),$ we define $\partial_{\mathfrak{K}}(Z)$ to be the set $\partial_{K}(Z).$
Also,  if $L\subseteq [l],$ then $V_L ({\bf a})$ contains all non-$\varnothing$ elements in ${\bf a}$ indexed by $L.$

Intuitively,
$\mathfrak{K}=(\mathfrak{A}[V(K^{\bf a})],{\bf a}, {\bf I}, {\bf W}_{{q}})$
contains all the ``useful information'' around the flatness pair $(W,\mathfrak{A}).$
The  structure $\mathfrak{A}[V(K^{\bf a})]$ induced by the union of the $\mathfrak{R}$-compass $K$ of $W,$
the apices $V({\bf a}),$ the homocentric zones of influence of
the layers of $W$ (away from its $j$-central part),
and the $q$-pseudogrid ${\bf W}$ of its ``$q$-central'' part.

\myskip\subsection{Out-signature}
\labels{sec_out-sig_first-floor}

In this subsection we aim to ``encode'' all necessary information that concerns the satisfiability of the sentence $θ^{\sf out}_q$ in the ``non-privileged'' part of the input structure.
To do this, given an extended compass of a flatness pair, we define a certain boundaried structure and we will use the finite set of representatives of $θ^{\sf out}_q$ given by~\autoref{cou_more}, to ``associate'' this boundaried structure with a representative.
Since parts of the ``non-privileged'' vertex set of the graph may lie outside the extended compass of the considered flatness pair,
we may have to ``extend'' the boundary of our structure in order to capture this information.
This is achieved by ``guessing'' all ways an (abstract) graph with a bounded number of vertices can ``extend'' the boundary.
\medskip

Before presenting some additional definitions,
we first set up the sentences and the constants in which we will build the out-signature.
Let $τ$ be a vocabulary, ${\sf R}\notin τ$ be a unary relation symbol, and ${\bf c}$ be a collection of $l$ constant symbols, where $l∈\mathbb{N}_{≥ 1}.$
Let $θ∈ \bar{Θ}_1[τ],$ let ${q}:=(\tw(θ)+1)^2+1,$ let $θ_{{\sf R},{\bf c}}$ be an enhanced version of $θ,$ and let $\tilde{θ}_q$ be the split version of $θ_{{\sf R},{\bf c}}.$
Recall that
 $$\tilde{θ}_{{q}}=\exists {\sf X}\ \bigg(θ^{\sf out}_{q}\ \wedge
\exists {\sf C} \Big( η_{w\text{-}{\sf pr}_{{\sf X},{\sf C}}}\ \wedge \breve{ζ}_{\sf R} |_{\sf ap_{\bf c}}|_{{\sf ind}_{\sf C}}\Big) \wedge \exists {\sf C}\Big(η_{w\text{-}{\sf pr}_{{\sf X},{\sf C}}}\ \wedge  μ|_{{\sf ind}_{\sf C}} \Big) \bigg)$$
and $θ^{\sf out}_{q}∈ \MSOL[τ \cup{\bf Q}\cup\{{\sf R},{\sf X}\}\cup{\bf c}],$ $\breve{ζ}_{\sf R}|_{\sf ap_{\bf c}}∈ \FOL[τ\cup\{{\sf R}\}\cup {\bf c}],$ and $μ∈ \NTMC[τ].$
\medskip

For the rest of this subsection, keep in mind that $q=(\tw(θ)+1)^2+1.$
Let $j',r∈ \mathbb{N}.$
Also, we set
\begin{eqnarray*}
j & := &  {\sf odd}(\max\{q/2,j'\}) \mbox{~and~}\\
w & :=& (r+2)\cdot q.\\
\end{eqnarray*}

To give an intuition for the above, let us explain what each of the above represents:
First, $j'$ is the size of the wall that we want to declare irrelevant in the proof of~\autoref{@desmembramientos}.
We set $q=(\tw(θ)+1)^2+1$ in order to have a $q$-pseudogrid ${\bf W}_{q}$ that defines a unique privileged component.
We will consider a wall $W$ of height $2w+j$ that, apart from its $j$-central part, contains $q$ ``annulus buffers'' of thickness $(r+2).$
We stress that while, for now, $r$ is a given constant, in~\autoref{sec_in-sig_first-floor}, it will be a particular constant depending on the parameters of the basic local sentences in the definition of $\breve{ζ}_{\sf R}$).
As a modulator $X$ cannot affect more than $(\tw(θ)+1)^2 =q-1$ of these $q$ ``annulus buffers''   (\autoref{@congregation}), one of them will not be affected by the solution, and therefore it will be altogether in its privileged component.
\medskip

\myskip\paragraph{Guessing an extension of a vertex set.}
Let $G$ be a graph.
Given a set of vertices $S\subseteq V(G)$ and an $\ell∈\mathbb{N},$ we define the collection of graphs ${\cal F}^{S}_\ell,$ such that $F∈ {\cal F}^{S}_\ell$  if and only if there exists a graph $F'$ on $\max\{\ell-|S|,0\}$ vertices and a set $E$ of edges each with one endpoint in $S$ and the other endpoint in $V(F'),$ such that $F = F'\cup (S\cup V(F'),E)$ (see~\autoref{figure_extension} for an example). We stress that, for every $F∈ {\cal F}^{S}_\ell,$ $F[S]$ is an edgeless graph.

\begin{figure}[ht]
\centering
\tikzstyle{ipe stylesheet} = [
  ipe import,
  even odd rule,
  line join=round,
  line cap=butt,
  ipe pen normal/.style={line width=0.4},
  ipe pen heavier/.style={line width=0.8},
  ipe pen fat/.style={line width=1.2},
  ipe pen ultrafat/.style={line width=2},
  ipe pen normal,
  ipe mark normal/.style={ipe mark scale=3},
  ipe mark large/.style={ipe mark scale=5},
  ipe mark small/.style={ipe mark scale=2},
  ipe mark tiny/.style={ipe mark scale=1.1},
  ipe mark normal,
  /pgf/arrow keys/.cd,
  ipe arrow normal/.style={scale=7},
  ipe arrow large/.style={scale=10},
  ipe arrow small/.style={scale=5},
  ipe arrow tiny/.style={scale=3},
  ipe arrow normal,
  /tikz/.cd,
  ipe arrows, 
  <->/.tip = ipe normal,
  ipe dash normal/.style={dash pattern=},
  ipe dash dotted/.style={dash pattern=on 1bp off 3bp},
  ipe dash dashed/.style={dash pattern=on 4bp off 4bp},
  ipe dash dash dotted/.style={dash pattern=on 4bp off 2bp on 1bp off 2bp},
  ipe dash dash dot dotted/.style={dash pattern=on 4bp off 2bp on 1bp off 2bp on 1bp off 2bp},
  ipe dash normal,
  ipe node/.append style={font=\normalsize},
  ipe stretch normal/.style={ipe node stretch=1},
  ipe stretch normal,
  ipe opacity 10/.style={opacity=0.1},
  ipe opacity 30/.style={opacity=0.3},
  ipe opacity 50/.style={opacity=0.5},
  ipe opacity 75/.style={opacity=0.75},
  ipe opacity opaque/.style={opacity=1},
  ipe opacity opaque,
]
\definecolor{black}{rgb}{0,0,0}
\definecolor{white}{rgb}{1,1,1}
\definecolor{red}{rgb}{1,0,0}
\definecolor{blue}{rgb}{0,0,1}
\definecolor{green}{rgb}{0,1,0}
\definecolor{yellow}{rgb}{1,1,0}
\definecolor{orange}{rgb}{1,0.647,0}
\definecolor{gold}{rgb}{1,0.843,0}
\definecolor{purple}{rgb}{0.627,0.125,0.941}
\definecolor{gray}{rgb}{0.745,0.745,0.745}
\definecolor{brown}{rgb}{0.647,0.165,0.165}
\definecolor{navy}{rgb}{0,0,0.502}
\definecolor{pink}{rgb}{1,0.753,0.796}
\definecolor{seagreen}{rgb}{0.18,0.545,0.341}
\definecolor{turquoise}{rgb}{0.251,0.878,0.816}
\definecolor{violet}{rgb}{0.933,0.51,0.933}
\definecolor{darkblue}{rgb}{0,0,0.545}
\definecolor{darkcyan}{rgb}{0,0.545,0.545}
\definecolor{darkgray}{rgb}{0.663,0.663,0.663}
\definecolor{darkgreen}{rgb}{0,0.392,0}
\definecolor{darkmagenta}{rgb}{0.545,0,0.545}
\definecolor{darkorange}{rgb}{1,0.549,0}
\definecolor{darkred}{rgb}{0.545,0,0}
\definecolor{lightblue}{rgb}{0.678,0.847,0.902}
\definecolor{lightcyan}{rgb}{0.878,1,1}
\definecolor{lightgray}{rgb}{0.827,0.827,0.827}
\definecolor{lightgreen}{rgb}{0.565,0.933,0.565}
\definecolor{lightyellow}{rgb}{1,1,0.878}
\definecolor{aliceblue}{rgb}{0.941,0.973,1}
\definecolor{antiquewhite}{rgb}{0.98,0.922,0.843}
\definecolor{antiquewhite1}{rgb}{1,0.937,0.859}
\definecolor{antiquewhite2}{rgb}{0.933,0.875,0.8}
\definecolor{antiquewhite3}{rgb}{0.804,0.753,0.69}
\definecolor{antiquewhite4}{rgb}{0.545,0.514,0.471}
\definecolor{aquamarine}{rgb}{0.498,1,0.831}
\definecolor{aquamarine1}{rgb}{0.498,1,0.831}
\definecolor{aquamarine2}{rgb}{0.463,0.933,0.776}
\definecolor{aquamarine3}{rgb}{0.4,0.804,0.667}
\definecolor{aquamarine4}{rgb}{0.271,0.545,0.455}
\definecolor{azure}{rgb}{0.941,1,1}
\definecolor{azure1}{rgb}{0.941,1,1}
\definecolor{azure2}{rgb}{0.878,0.933,0.933}
\definecolor{azure3}{rgb}{0.757,0.804,0.804}
\definecolor{azure4}{rgb}{0.514,0.545,0.545}
\definecolor{beige}{rgb}{0.961,0.961,0.863}
\definecolor{bisque}{rgb}{1,0.894,0.769}
\definecolor{bisque1}{rgb}{1,0.894,0.769}
\definecolor{bisque2}{rgb}{0.933,0.835,0.718}
\definecolor{bisque3}{rgb}{0.804,0.718,0.62}
\definecolor{bisque4}{rgb}{0.545,0.49,0.42}
\definecolor{blanchedalmond}{rgb}{1,0.922,0.804}
\definecolor{blue1}{rgb}{0,0,1}
\definecolor{blue2}{rgb}{0,0,0.933}
\definecolor{blue3}{rgb}{0,0,0.804}
\definecolor{blue4}{rgb}{0,0,0.545}
\definecolor{blueviolet}{rgb}{0.541,0.169,0.886}
\definecolor{brown1}{rgb}{1,0.251,0.251}
\definecolor{brown2}{rgb}{0.933,0.231,0.231}
\definecolor{brown3}{rgb}{0.804,0.2,0.2}
\definecolor{brown4}{rgb}{0.545,0.137,0.137}
\definecolor{burlywood}{rgb}{0.871,0.722,0.529}
\definecolor{burlywood1}{rgb}{1,0.827,0.608}
\definecolor{burlywood2}{rgb}{0.933,0.773,0.569}
\definecolor{burlywood3}{rgb}{0.804,0.667,0.49}
\definecolor{burlywood4}{rgb}{0.545,0.451,0.333}
\definecolor{cadetblue}{rgb}{0.373,0.62,0.627}
\definecolor{cadetblue1}{rgb}{0.596,0.961,1}
\definecolor{cadetblue2}{rgb}{0.557,0.898,0.933}
\definecolor{cadetblue3}{rgb}{0.478,0.773,0.804}
\definecolor{cadetblue4}{rgb}{0.325,0.525,0.545}
\definecolor{chartreuse}{rgb}{0.498,1,0}
\definecolor{chartreuse1}{rgb}{0.498,1,0}
\definecolor{chartreuse2}{rgb}{0.463,0.933,0}
\definecolor{chartreuse3}{rgb}{0.4,0.804,0}
\definecolor{chartreuse4}{rgb}{0.271,0.545,0}
\definecolor{chocolate}{rgb}{0.824,0.412,0.118}
\definecolor{chocolate1}{rgb}{1,0.498,0.141}
\definecolor{chocolate2}{rgb}{0.933,0.463,0.129}
\definecolor{chocolate3}{rgb}{0.804,0.4,0.114}
\definecolor{chocolate4}{rgb}{0.545,0.271,0.075}
\definecolor{coral}{rgb}{1,0.498,0.314}
\definecolor{coral1}{rgb}{1,0.447,0.337}
\definecolor{coral2}{rgb}{0.933,0.416,0.314}
\definecolor{coral3}{rgb}{0.804,0.357,0.271}
\definecolor{coral4}{rgb}{0.545,0.243,0.184}
\definecolor{cornflowerblue}{rgb}{0.392,0.584,0.929}
\definecolor{cornsilk}{rgb}{1,0.973,0.863}
\definecolor{cornsilk1}{rgb}{1,0.973,0.863}
\definecolor{cornsilk2}{rgb}{0.933,0.91,0.804}
\definecolor{cornsilk3}{rgb}{0.804,0.784,0.694}
\definecolor{cornsilk4}{rgb}{0.545,0.533,0.471}
\definecolor{cyan}{rgb}{0,1,1}
\definecolor{cyan1}{rgb}{0,1,1}
\definecolor{cyan2}{rgb}{0,0.933,0.933}
\definecolor{cyan3}{rgb}{0,0.804,0.804}
\definecolor{cyan4}{rgb}{0,0.545,0.545}
\definecolor{darkgoldenrod}{rgb}{0.722,0.525,0.043}
\definecolor{darkgoldenrod1}{rgb}{1,0.725,0.059}
\definecolor{darkgoldenrod2}{rgb}{0.933,0.678,0.055}
\definecolor{darkgoldenrod3}{rgb}{0.804,0.584,0.047}
\definecolor{darkgoldenrod4}{rgb}{0.545,0.396,0.031}
\definecolor{darkgrey}{rgb}{0.663,0.663,0.663}
\definecolor{darkkhaki}{rgb}{0.741,0.718,0.42}
\definecolor{darkolivegreen}{rgb}{0.333,0.42,0.184}
\definecolor{darkolivegreen1}{rgb}{0.792,1,0.439}
\definecolor{darkolivegreen2}{rgb}{0.737,0.933,0.408}
\definecolor{darkolivegreen3}{rgb}{0.635,0.804,0.353}
\definecolor{darkolivegreen4}{rgb}{0.431,0.545,0.239}
\definecolor{darkorange1}{rgb}{1,0.498,0}
\definecolor{darkorange2}{rgb}{0.933,0.463,0}
\definecolor{darkorange3}{rgb}{0.804,0.4,0}
\definecolor{darkorange4}{rgb}{0.545,0.271,0}
\definecolor{darkorchid}{rgb}{0.6,0.196,0.8}
\definecolor{darkorchid1}{rgb}{0.749,0.243,1}
\definecolor{darkorchid2}{rgb}{0.698,0.227,0.933}
\definecolor{darkorchid3}{rgb}{0.604,0.196,0.804}
\definecolor{darkorchid4}{rgb}{0.408,0.133,0.545}
\definecolor{darksalmon}{rgb}{0.914,0.588,0.478}
\definecolor{darkseagreen}{rgb}{0.561,0.737,0.561}
\definecolor{darkseagreen1}{rgb}{0.757,1,0.757}
\definecolor{darkseagreen2}{rgb}{0.706,0.933,0.706}
\definecolor{darkseagreen3}{rgb}{0.608,0.804,0.608}
\definecolor{darkseagreen4}{rgb}{0.412,0.545,0.412}
\definecolor{darkslateblue}{rgb}{0.282,0.239,0.545}
\definecolor{darkslategray}{rgb}{0.184,0.31,0.31}
\definecolor{darkslategray1}{rgb}{0.592,1,1}
\definecolor{darkslategray2}{rgb}{0.553,0.933,0.933}
\definecolor{darkslategray3}{rgb}{0.475,0.804,0.804}
\definecolor{darkslategray4}{rgb}{0.322,0.545,0.545}
\definecolor{darkslategrey}{rgb}{0.184,0.31,0.31}
\definecolor{darkturquoise}{rgb}{0,0.808,0.82}
\definecolor{darkviolet}{rgb}{0.58,0,0.827}
\definecolor{deeppink}{rgb}{1,0.078,0.576}
\definecolor{deeppink1}{rgb}{1,0.078,0.576}
\definecolor{deeppink2}{rgb}{0.933,0.071,0.537}
\definecolor{deeppink3}{rgb}{0.804,0.063,0.463}
\definecolor{deeppink4}{rgb}{0.545,0.039,0.314}
\definecolor{deepskyblue}{rgb}{0,0.749,1}
\definecolor{deepskyblue1}{rgb}{0,0.749,1}
\definecolor{deepskyblue2}{rgb}{0,0.698,0.933}
\definecolor{deepskyblue3}{rgb}{0,0.604,0.804}
\definecolor{deepskyblue4}{rgb}{0,0.408,0.545}
\definecolor{dimgray}{rgb}{0.412,0.412,0.412}
\definecolor{dimgrey}{rgb}{0.412,0.412,0.412}
\definecolor{dodgerblue}{rgb}{0.118,0.565,1}
\definecolor{dodgerblue1}{rgb}{0.118,0.565,1}
\definecolor{dodgerblue2}{rgb}{0.11,0.525,0.933}
\definecolor{dodgerblue3}{rgb}{0.094,0.455,0.804}
\definecolor{dodgerblue4}{rgb}{0.063,0.306,0.545}
\definecolor{firebrick}{rgb}{0.698,0.133,0.133}
\definecolor{firebrick1}{rgb}{1,0.188,0.188}
\definecolor{firebrick2}{rgb}{0.933,0.173,0.173}
\definecolor{firebrick3}{rgb}{0.804,0.149,0.149}
\definecolor{firebrick4}{rgb}{0.545,0.102,0.102}
\definecolor{floralwhite}{rgb}{1,0.98,0.941}
\definecolor{forestgreen}{rgb}{0.133,0.545,0.133}
\definecolor{gainsboro}{rgb}{0.863,0.863,0.863}
\definecolor{ghostwhite}{rgb}{0.973,0.973,1}
\definecolor{gold1}{rgb}{1,0.843,0}
\definecolor{gold2}{rgb}{0.933,0.788,0}
\definecolor{gold3}{rgb}{0.804,0.678,0}
\definecolor{gold4}{rgb}{0.545,0.459,0}
\definecolor{goldenrod}{rgb}{0.855,0.647,0.125}
\definecolor{goldenrod1}{rgb}{1,0.757,0.145}
\definecolor{goldenrod2}{rgb}{0.933,0.706,0.133}
\definecolor{goldenrod3}{rgb}{0.804,0.608,0.114}
\definecolor{goldenrod4}{rgb}{0.545,0.412,0.078}
\definecolor{gray0}{rgb}{0,0,0}
\definecolor{gray1}{rgb}{0.012,0.012,0.012}
\definecolor{gray10}{rgb}{0.102,0.102,0.102}
\definecolor{gray100}{rgb}{1,1,1}
\definecolor{gray11}{rgb}{0.11,0.11,0.11}
\definecolor{gray12}{rgb}{0.122,0.122,0.122}
\definecolor{gray13}{rgb}{0.129,0.129,0.129}
\definecolor{gray14}{rgb}{0.141,0.141,0.141}
\definecolor{gray15}{rgb}{0.149,0.149,0.149}
\definecolor{gray16}{rgb}{0.161,0.161,0.161}
\definecolor{gray17}{rgb}{0.169,0.169,0.169}
\definecolor{gray18}{rgb}{0.18,0.18,0.18}
\definecolor{gray19}{rgb}{0.188,0.188,0.188}
\definecolor{gray2}{rgb}{0.02,0.02,0.02}
\definecolor{gray20}{rgb}{0.2,0.2,0.2}
\definecolor{gray21}{rgb}{0.212,0.212,0.212}
\definecolor{gray22}{rgb}{0.22,0.22,0.22}
\definecolor{gray23}{rgb}{0.231,0.231,0.231}
\definecolor{gray24}{rgb}{0.239,0.239,0.239}
\definecolor{gray25}{rgb}{0.251,0.251,0.251}
\definecolor{gray26}{rgb}{0.259,0.259,0.259}
\definecolor{gray27}{rgb}{0.271,0.271,0.271}
\definecolor{gray28}{rgb}{0.278,0.278,0.278}
\definecolor{gray29}{rgb}{0.29,0.29,0.29}
\definecolor{gray3}{rgb}{0.031,0.031,0.031}
\definecolor{gray30}{rgb}{0.302,0.302,0.302}
\definecolor{gray31}{rgb}{0.31,0.31,0.31}
\definecolor{gray32}{rgb}{0.322,0.322,0.322}
\definecolor{gray33}{rgb}{0.329,0.329,0.329}
\definecolor{gray34}{rgb}{0.341,0.341,0.341}
\definecolor{gray35}{rgb}{0.349,0.349,0.349}
\definecolor{gray36}{rgb}{0.361,0.361,0.361}
\definecolor{gray37}{rgb}{0.369,0.369,0.369}
\definecolor{gray38}{rgb}{0.38,0.38,0.38}
\definecolor{gray39}{rgb}{0.388,0.388,0.388}
\definecolor{gray4}{rgb}{0.039,0.039,0.039}
\definecolor{gray40}{rgb}{0.4,0.4,0.4}
\definecolor{gray41}{rgb}{0.412,0.412,0.412}
\definecolor{gray42}{rgb}{0.42,0.42,0.42}
\definecolor{gray43}{rgb}{0.431,0.431,0.431}
\definecolor{gray44}{rgb}{0.439,0.439,0.439}
\definecolor{gray45}{rgb}{0.451,0.451,0.451}
\definecolor{gray46}{rgb}{0.459,0.459,0.459}
\definecolor{gray47}{rgb}{0.471,0.471,0.471}
\definecolor{gray48}{rgb}{0.478,0.478,0.478}
\definecolor{gray49}{rgb}{0.49,0.49,0.49}
\definecolor{gray5}{rgb}{0.051,0.051,0.051}
\definecolor{gray50}{rgb}{0.498,0.498,0.498}
\definecolor{gray51}{rgb}{0.51,0.51,0.51}
\definecolor{gray52}{rgb}{0.522,0.522,0.522}
\definecolor{gray53}{rgb}{0.529,0.529,0.529}
\definecolor{gray54}{rgb}{0.541,0.541,0.541}
\definecolor{gray55}{rgb}{0.549,0.549,0.549}
\definecolor{gray56}{rgb}{0.561,0.561,0.561}
\definecolor{gray57}{rgb}{0.569,0.569,0.569}
\definecolor{gray58}{rgb}{0.58,0.58,0.58}
\definecolor{gray59}{rgb}{0.588,0.588,0.588}
\definecolor{gray6}{rgb}{0.059,0.059,0.059}
\definecolor{gray60}{rgb}{0.6,0.6,0.6}
\definecolor{gray61}{rgb}{0.612,0.612,0.612}
\definecolor{gray62}{rgb}{0.62,0.62,0.62}
\definecolor{gray63}{rgb}{0.631,0.631,0.631}
\definecolor{gray64}{rgb}{0.639,0.639,0.639}
\definecolor{gray65}{rgb}{0.651,0.651,0.651}
\definecolor{gray66}{rgb}{0.659,0.659,0.659}
\definecolor{gray67}{rgb}{0.671,0.671,0.671}
\definecolor{gray68}{rgb}{0.678,0.678,0.678}
\definecolor{gray69}{rgb}{0.69,0.69,0.69}
\definecolor{gray7}{rgb}{0.071,0.071,0.071}
\definecolor{gray70}{rgb}{0.702,0.702,0.702}
\definecolor{gray71}{rgb}{0.71,0.71,0.71}
\definecolor{gray72}{rgb}{0.722,0.722,0.722}
\definecolor{gray73}{rgb}{0.729,0.729,0.729}
\definecolor{gray74}{rgb}{0.741,0.741,0.741}
\definecolor{gray75}{rgb}{0.749,0.749,0.749}
\definecolor{gray76}{rgb}{0.761,0.761,0.761}
\definecolor{gray77}{rgb}{0.769,0.769,0.769}
\definecolor{gray78}{rgb}{0.78,0.78,0.78}
\definecolor{gray79}{rgb}{0.788,0.788,0.788}
\definecolor{gray8}{rgb}{0.078,0.078,0.078}
\definecolor{gray80}{rgb}{0.8,0.8,0.8}
\definecolor{gray81}{rgb}{0.812,0.812,0.812}
\definecolor{gray82}{rgb}{0.82,0.82,0.82}
\definecolor{gray83}{rgb}{0.831,0.831,0.831}
\definecolor{gray84}{rgb}{0.839,0.839,0.839}
\definecolor{gray85}{rgb}{0.851,0.851,0.851}
\definecolor{gray86}{rgb}{0.859,0.859,0.859}
\definecolor{gray87}{rgb}{0.871,0.871,0.871}
\definecolor{gray88}{rgb}{0.878,0.878,0.878}
\definecolor{gray89}{rgb}{0.89,0.89,0.89}
\definecolor{gray9}{rgb}{0.09,0.09,0.09}
\definecolor{gray90}{rgb}{0.898,0.898,0.898}
\definecolor{gray91}{rgb}{0.91,0.91,0.91}
\definecolor{gray92}{rgb}{0.922,0.922,0.922}
\definecolor{gray93}{rgb}{0.929,0.929,0.929}
\definecolor{gray94}{rgb}{0.941,0.941,0.941}
\definecolor{gray95}{rgb}{0.949,0.949,0.949}
\definecolor{gray96}{rgb}{0.961,0.961,0.961}
\definecolor{gray97}{rgb}{0.969,0.969,0.969}
\definecolor{gray98}{rgb}{0.98,0.98,0.98}
\definecolor{gray99}{rgb}{0.988,0.988,0.988}
\definecolor{green1}{rgb}{0,1,0}
\definecolor{green2}{rgb}{0,0.933,0}
\definecolor{green3}{rgb}{0,0.804,0}
\definecolor{green4}{rgb}{0,0.545,0}
\definecolor{greenyellow}{rgb}{0.678,1,0.184}
\definecolor{grey}{rgb}{0.745,0.745,0.745}
\definecolor{grey0}{rgb}{0,0,0}
\definecolor{grey1}{rgb}{0.012,0.012,0.012}
\definecolor{grey10}{rgb}{0.102,0.102,0.102}
\definecolor{grey100}{rgb}{1,1,1}
\definecolor{grey11}{rgb}{0.11,0.11,0.11}
\definecolor{grey12}{rgb}{0.122,0.122,0.122}
\definecolor{grey13}{rgb}{0.129,0.129,0.129}
\definecolor{grey14}{rgb}{0.141,0.141,0.141}
\definecolor{grey15}{rgb}{0.149,0.149,0.149}
\definecolor{grey16}{rgb}{0.161,0.161,0.161}
\definecolor{grey17}{rgb}{0.169,0.169,0.169}
\definecolor{grey18}{rgb}{0.18,0.18,0.18}
\definecolor{grey19}{rgb}{0.188,0.188,0.188}
\definecolor{grey2}{rgb}{0.02,0.02,0.02}
\definecolor{grey20}{rgb}{0.2,0.2,0.2}
\definecolor{grey21}{rgb}{0.212,0.212,0.212}
\definecolor{grey22}{rgb}{0.22,0.22,0.22}
\definecolor{grey23}{rgb}{0.231,0.231,0.231}
\definecolor{grey24}{rgb}{0.239,0.239,0.239}
\definecolor{grey25}{rgb}{0.251,0.251,0.251}
\definecolor{grey26}{rgb}{0.259,0.259,0.259}
\definecolor{grey27}{rgb}{0.271,0.271,0.271}
\definecolor{grey28}{rgb}{0.278,0.278,0.278}
\definecolor{grey29}{rgb}{0.29,0.29,0.29}
\definecolor{grey3}{rgb}{0.031,0.031,0.031}
\definecolor{grey30}{rgb}{0.302,0.302,0.302}
\definecolor{grey31}{rgb}{0.31,0.31,0.31}
\definecolor{grey32}{rgb}{0.322,0.322,0.322}
\definecolor{grey33}{rgb}{0.329,0.329,0.329}
\definecolor{grey34}{rgb}{0.341,0.341,0.341}
\definecolor{grey35}{rgb}{0.349,0.349,0.349}
\definecolor{grey36}{rgb}{0.361,0.361,0.361}
\definecolor{grey37}{rgb}{0.369,0.369,0.369}
\definecolor{grey38}{rgb}{0.38,0.38,0.38}
\definecolor{grey39}{rgb}{0.388,0.388,0.388}
\definecolor{grey4}{rgb}{0.039,0.039,0.039}
\definecolor{grey40}{rgb}{0.4,0.4,0.4}
\definecolor{grey41}{rgb}{0.412,0.412,0.412}
\definecolor{grey42}{rgb}{0.42,0.42,0.42}
\definecolor{grey43}{rgb}{0.431,0.431,0.431}
\definecolor{grey44}{rgb}{0.439,0.439,0.439}
\definecolor{grey45}{rgb}{0.451,0.451,0.451}
\definecolor{grey46}{rgb}{0.459,0.459,0.459}
\definecolor{grey47}{rgb}{0.471,0.471,0.471}
\definecolor{grey48}{rgb}{0.478,0.478,0.478}
\definecolor{grey49}{rgb}{0.49,0.49,0.49}
\definecolor{grey5}{rgb}{0.051,0.051,0.051}
\definecolor{grey50}{rgb}{0.498,0.498,0.498}
\definecolor{grey51}{rgb}{0.51,0.51,0.51}
\definecolor{grey52}{rgb}{0.522,0.522,0.522}
\definecolor{grey53}{rgb}{0.529,0.529,0.529}
\definecolor{grey54}{rgb}{0.541,0.541,0.541}
\definecolor{grey55}{rgb}{0.549,0.549,0.549}
\definecolor{grey56}{rgb}{0.561,0.561,0.561}
\definecolor{grey57}{rgb}{0.569,0.569,0.569}
\definecolor{grey58}{rgb}{0.58,0.58,0.58}
\definecolor{grey59}{rgb}{0.588,0.588,0.588}
\definecolor{grey6}{rgb}{0.059,0.059,0.059}
\definecolor{grey60}{rgb}{0.6,0.6,0.6}
\definecolor{grey61}{rgb}{0.612,0.612,0.612}
\definecolor{grey62}{rgb}{0.62,0.62,0.62}
\definecolor{grey63}{rgb}{0.631,0.631,0.631}
\definecolor{grey64}{rgb}{0.639,0.639,0.639}
\definecolor{grey65}{rgb}{0.651,0.651,0.651}
\definecolor{grey66}{rgb}{0.659,0.659,0.659}
\definecolor{grey67}{rgb}{0.671,0.671,0.671}
\definecolor{grey68}{rgb}{0.678,0.678,0.678}
\definecolor{grey69}{rgb}{0.69,0.69,0.69}
\definecolor{grey7}{rgb}{0.071,0.071,0.071}
\definecolor{grey70}{rgb}{0.702,0.702,0.702}
\definecolor{grey71}{rgb}{0.71,0.71,0.71}
\definecolor{grey72}{rgb}{0.722,0.722,0.722}
\definecolor{grey73}{rgb}{0.729,0.729,0.729}
\definecolor{grey74}{rgb}{0.741,0.741,0.741}
\definecolor{grey75}{rgb}{0.749,0.749,0.749}
\definecolor{grey76}{rgb}{0.761,0.761,0.761}
\definecolor{grey77}{rgb}{0.769,0.769,0.769}
\definecolor{grey78}{rgb}{0.78,0.78,0.78}
\definecolor{grey79}{rgb}{0.788,0.788,0.788}
\definecolor{grey8}{rgb}{0.078,0.078,0.078}
\definecolor{grey80}{rgb}{0.8,0.8,0.8}
\definecolor{grey81}{rgb}{0.812,0.812,0.812}
\definecolor{grey82}{rgb}{0.82,0.82,0.82}
\definecolor{grey83}{rgb}{0.831,0.831,0.831}
\definecolor{grey84}{rgb}{0.839,0.839,0.839}
\definecolor{grey85}{rgb}{0.851,0.851,0.851}
\definecolor{grey86}{rgb}{0.859,0.859,0.859}
\definecolor{grey87}{rgb}{0.871,0.871,0.871}
\definecolor{grey88}{rgb}{0.878,0.878,0.878}
\definecolor{grey89}{rgb}{0.89,0.89,0.89}
\definecolor{grey9}{rgb}{0.09,0.09,0.09}
\definecolor{grey90}{rgb}{0.898,0.898,0.898}
\definecolor{grey91}{rgb}{0.91,0.91,0.91}
\definecolor{grey92}{rgb}{0.922,0.922,0.922}
\definecolor{grey93}{rgb}{0.929,0.929,0.929}
\definecolor{grey94}{rgb}{0.941,0.941,0.941}
\definecolor{grey95}{rgb}{0.949,0.949,0.949}
\definecolor{grey96}{rgb}{0.961,0.961,0.961}
\definecolor{grey97}{rgb}{0.969,0.969,0.969}
\definecolor{grey98}{rgb}{0.98,0.98,0.98}
\definecolor{grey99}{rgb}{0.988,0.988,0.988}
\definecolor{honeydew}{rgb}{0.941,1,0.941}
\definecolor{honeydew1}{rgb}{0.941,1,0.941}
\definecolor{honeydew2}{rgb}{0.878,0.933,0.878}
\definecolor{honeydew3}{rgb}{0.757,0.804,0.757}
\definecolor{honeydew4}{rgb}{0.514,0.545,0.514}
\definecolor{hotpink}{rgb}{1,0.412,0.706}
\definecolor{hotpink1}{rgb}{1,0.431,0.706}
\definecolor{hotpink2}{rgb}{0.933,0.416,0.655}
\definecolor{hotpink3}{rgb}{0.804,0.376,0.565}
\definecolor{hotpink4}{rgb}{0.545,0.227,0.384}
\definecolor{indianred}{rgb}{0.804,0.361,0.361}
\definecolor{indianred1}{rgb}{1,0.416,0.416}
\definecolor{indianred2}{rgb}{0.933,0.388,0.388}
\definecolor{indianred3}{rgb}{0.804,0.333,0.333}
\definecolor{indianred4}{rgb}{0.545,0.227,0.227}
\definecolor{ivory}{rgb}{1,1,0.941}
\definecolor{ivory1}{rgb}{1,1,0.941}
\definecolor{ivory2}{rgb}{0.933,0.933,0.878}
\definecolor{ivory3}{rgb}{0.804,0.804,0.757}
\definecolor{ivory4}{rgb}{0.545,0.545,0.514}
\definecolor{khaki}{rgb}{0.941,0.902,0.549}
\definecolor{khaki1}{rgb}{1,0.965,0.561}
\definecolor{khaki2}{rgb}{0.933,0.902,0.522}
\definecolor{khaki3}{rgb}{0.804,0.776,0.451}
\definecolor{khaki4}{rgb}{0.545,0.525,0.306}
\definecolor{lavender}{rgb}{0.902,0.902,0.98}
\definecolor{lavenderblush}{rgb}{1,0.941,0.961}
\definecolor{lavenderblush1}{rgb}{1,0.941,0.961}
\definecolor{lavenderblush2}{rgb}{0.933,0.878,0.898}
\definecolor{lavenderblush3}{rgb}{0.804,0.757,0.773}
\definecolor{lavenderblush4}{rgb}{0.545,0.514,0.525}
\definecolor{lawngreen}{rgb}{0.486,0.988,0}
\definecolor{lemonchiffon}{rgb}{1,0.98,0.804}
\definecolor{lemonchiffon1}{rgb}{1,0.98,0.804}
\definecolor{lemonchiffon2}{rgb}{0.933,0.914,0.749}
\definecolor{lemonchiffon3}{rgb}{0.804,0.788,0.647}
\definecolor{lemonchiffon4}{rgb}{0.545,0.537,0.439}
\definecolor{lightblue1}{rgb}{0.749,0.937,1}
\definecolor{lightblue2}{rgb}{0.698,0.875,0.933}
\definecolor{lightblue3}{rgb}{0.604,0.753,0.804}
\definecolor{lightblue4}{rgb}{0.408,0.514,0.545}
\definecolor{lightcoral}{rgb}{0.941,0.502,0.502}
\definecolor{lightcyan1}{rgb}{0.878,1,1}
\definecolor{lightcyan2}{rgb}{0.82,0.933,0.933}
\definecolor{lightcyan3}{rgb}{0.706,0.804,0.804}
\definecolor{lightcyan4}{rgb}{0.478,0.545,0.545}
\definecolor{lightgoldenrod}{rgb}{0.933,0.867,0.51}
\definecolor{lightgoldenrod1}{rgb}{1,0.925,0.545}
\definecolor{lightgoldenrod2}{rgb}{0.933,0.863,0.51}
\definecolor{lightgoldenrod3}{rgb}{0.804,0.745,0.439}
\definecolor{lightgoldenrod4}{rgb}{0.545,0.506,0.298}
\definecolor{lightgoldenrodyellow}{rgb}{0.98,0.98,0.824}
\definecolor{lightgrey}{rgb}{0.827,0.827,0.827}
\definecolor{lightpink}{rgb}{1,0.714,0.757}
\definecolor{lightpink1}{rgb}{1,0.682,0.725}
\definecolor{lightpink2}{rgb}{0.933,0.635,0.678}
\definecolor{lightpink3}{rgb}{0.804,0.549,0.584}
\definecolor{lightpink4}{rgb}{0.545,0.373,0.396}
\definecolor{lightsalmon}{rgb}{1,0.627,0.478}
\definecolor{lightsalmon1}{rgb}{1,0.627,0.478}
\definecolor{lightsalmon2}{rgb}{0.933,0.584,0.447}
\definecolor{lightsalmon3}{rgb}{0.804,0.506,0.384}
\definecolor{lightsalmon4}{rgb}{0.545,0.341,0.259}
\definecolor{lightseagreen}{rgb}{0.125,0.698,0.667}
\definecolor{lightskyblue}{rgb}{0.529,0.808,0.98}
\definecolor{lightskyblue1}{rgb}{0.69,0.886,1}
\definecolor{lightskyblue2}{rgb}{0.643,0.827,0.933}
\definecolor{lightskyblue3}{rgb}{0.553,0.714,0.804}
\definecolor{lightskyblue4}{rgb}{0.376,0.482,0.545}
\definecolor{lightslateblue}{rgb}{0.518,0.439,1}
\definecolor{lightslategray}{rgb}{0.467,0.533,0.6}
\definecolor{lightslategrey}{rgb}{0.467,0.533,0.6}
\definecolor{lightsteelblue}{rgb}{0.69,0.769,0.871}
\definecolor{lightsteelblue1}{rgb}{0.792,0.882,1}
\definecolor{lightsteelblue2}{rgb}{0.737,0.824,0.933}
\definecolor{lightsteelblue3}{rgb}{0.635,0.71,0.804}
\definecolor{lightsteelblue4}{rgb}{0.431,0.482,0.545}
\definecolor{lightyellow1}{rgb}{1,1,0.878}
\definecolor{lightyellow2}{rgb}{0.933,0.933,0.82}
\definecolor{lightyellow3}{rgb}{0.804,0.804,0.706}
\definecolor{lightyellow4}{rgb}{0.545,0.545,0.478}
\definecolor{limegreen}{rgb}{0.196,0.804,0.196}
\definecolor{linen}{rgb}{0.98,0.941,0.902}
\definecolor{magenta}{rgb}{1,0,1}
\definecolor{magenta1}{rgb}{1,0,1}
\definecolor{magenta2}{rgb}{0.933,0,0.933}
\definecolor{magenta3}{rgb}{0.804,0,0.804}
\definecolor{magenta4}{rgb}{0.545,0,0.545}
\definecolor{maroon}{rgb}{0.69,0.188,0.376}
\definecolor{maroon1}{rgb}{1,0.204,0.702}
\definecolor{maroon2}{rgb}{0.933,0.188,0.655}
\definecolor{maroon3}{rgb}{0.804,0.161,0.565}
\definecolor{maroon4}{rgb}{0.545,0.11,0.384}
\definecolor{mediumaquamarine}{rgb}{0.4,0.804,0.667}
\definecolor{mediumblue}{rgb}{0,0,0.804}
\definecolor{mediumorchid}{rgb}{0.729,0.333,0.827}
\definecolor{mediumorchid1}{rgb}{0.878,0.4,1}
\definecolor{mediumorchid2}{rgb}{0.82,0.373,0.933}
\definecolor{mediumorchid3}{rgb}{0.706,0.322,0.804}
\definecolor{mediumorchid4}{rgb}{0.478,0.216,0.545}
\definecolor{mediumpurple}{rgb}{0.576,0.439,0.859}
\definecolor{mediumpurple1}{rgb}{0.671,0.51,1}
\definecolor{mediumpurple2}{rgb}{0.624,0.475,0.933}
\definecolor{mediumpurple3}{rgb}{0.537,0.408,0.804}
\definecolor{mediumpurple4}{rgb}{0.365,0.278,0.545}
\definecolor{mediumseagreen}{rgb}{0.235,0.702,0.443}
\definecolor{mediumslateblue}{rgb}{0.482,0.408,0.933}
\definecolor{mediumspringgreen}{rgb}{0,0.98,0.604}
\definecolor{mediumturquoise}{rgb}{0.282,0.82,0.8}
\definecolor{mediumvioletred}{rgb}{0.78,0.082,0.522}
\definecolor{midnightblue}{rgb}{0.098,0.098,0.439}
\definecolor{mintcream}{rgb}{0.961,1,0.98}
\definecolor{mistyrose}{rgb}{1,0.894,0.882}
\definecolor{mistyrose1}{rgb}{1,0.894,0.882}
\definecolor{mistyrose2}{rgb}{0.933,0.835,0.824}
\definecolor{mistyrose3}{rgb}{0.804,0.718,0.71}
\definecolor{mistyrose4}{rgb}{0.545,0.49,0.482}
\definecolor{moccasin}{rgb}{1,0.894,0.71}
\definecolor{navajowhite}{rgb}{1,0.871,0.678}
\definecolor{navajowhite1}{rgb}{1,0.871,0.678}
\definecolor{navajowhite2}{rgb}{0.933,0.812,0.631}
\definecolor{navajowhite3}{rgb}{0.804,0.702,0.545}
\definecolor{navajowhite4}{rgb}{0.545,0.475,0.369}
\definecolor{navyblue}{rgb}{0,0,0.502}
\definecolor{oldlace}{rgb}{0.992,0.961,0.902}
\definecolor{olivedrab}{rgb}{0.42,0.557,0.137}
\definecolor{olivedrab1}{rgb}{0.753,1,0.243}
\definecolor{olivedrab2}{rgb}{0.702,0.933,0.227}
\definecolor{olivedrab3}{rgb}{0.604,0.804,0.196}
\definecolor{olivedrab4}{rgb}{0.412,0.545,0.133}
\definecolor{orange1}{rgb}{1,0.647,0}
\definecolor{orange2}{rgb}{0.933,0.604,0}
\definecolor{orange3}{rgb}{0.804,0.522,0}
\definecolor{orange4}{rgb}{0.545,0.353,0}
\definecolor{orangered}{rgb}{1,0.271,0}
\definecolor{orangered1}{rgb}{1,0.271,0}
\definecolor{orangered2}{rgb}{0.933,0.251,0}
\definecolor{orangered3}{rgb}{0.804,0.216,0}
\definecolor{orangered4}{rgb}{0.545,0.145,0}
\definecolor{orchid}{rgb}{0.855,0.439,0.839}
\definecolor{orchid1}{rgb}{1,0.514,0.98}
\definecolor{orchid2}{rgb}{0.933,0.478,0.914}
\definecolor{orchid3}{rgb}{0.804,0.412,0.788}
\definecolor{orchid4}{rgb}{0.545,0.278,0.537}
\definecolor{palegoldenrod}{rgb}{0.933,0.91,0.667}
\definecolor{palegreen}{rgb}{0.596,0.984,0.596}
\definecolor{palegreen1}{rgb}{0.604,1,0.604}
\definecolor{palegreen2}{rgb}{0.565,0.933,0.565}
\definecolor{palegreen3}{rgb}{0.486,0.804,0.486}
\definecolor{palegreen4}{rgb}{0.329,0.545,0.329}
\definecolor{paleturquoise}{rgb}{0.686,0.933,0.933}
\definecolor{paleturquoise1}{rgb}{0.733,1,1}
\definecolor{paleturquoise2}{rgb}{0.682,0.933,0.933}
\definecolor{paleturquoise3}{rgb}{0.588,0.804,0.804}
\definecolor{paleturquoise4}{rgb}{0.4,0.545,0.545}
\definecolor{palevioletred}{rgb}{0.859,0.439,0.576}
\definecolor{palevioletred1}{rgb}{1,0.51,0.671}
\definecolor{palevioletred2}{rgb}{0.933,0.475,0.624}
\definecolor{palevioletred3}{rgb}{0.804,0.408,0.537}
\definecolor{palevioletred4}{rgb}{0.545,0.278,0.365}
\definecolor{papayawhip}{rgb}{1,0.937,0.835}
\definecolor{peachpuff}{rgb}{1,0.855,0.725}
\definecolor{peachpuff1}{rgb}{1,0.855,0.725}
\definecolor{peachpuff2}{rgb}{0.933,0.796,0.678}
\definecolor{peachpuff3}{rgb}{0.804,0.686,0.584}
\definecolor{peachpuff4}{rgb}{0.545,0.467,0.396}
\definecolor{peru}{rgb}{0.804,0.522,0.247}
\definecolor{pink1}{rgb}{1,0.71,0.773}
\definecolor{pink2}{rgb}{0.933,0.663,0.722}
\definecolor{pink3}{rgb}{0.804,0.569,0.62}
\definecolor{pink4}{rgb}{0.545,0.388,0.424}
\definecolor{plum}{rgb}{0.867,0.627,0.867}
\definecolor{plum1}{rgb}{1,0.733,1}
\definecolor{plum2}{rgb}{0.933,0.682,0.933}
\definecolor{plum3}{rgb}{0.804,0.588,0.804}
\definecolor{plum4}{rgb}{0.545,0.4,0.545}
\definecolor{powderblue}{rgb}{0.69,0.878,0.902}
\definecolor{purple1}{rgb}{0.608,0.188,1}
\definecolor{purple2}{rgb}{0.569,0.173,0.933}
\definecolor{purple3}{rgb}{0.49,0.149,0.804}
\definecolor{purple4}{rgb}{0.333,0.102,0.545}
\definecolor{red1}{rgb}{1,0,0}
\definecolor{red2}{rgb}{0.933,0,0}
\definecolor{red3}{rgb}{0.804,0,0}
\definecolor{red4}{rgb}{0.545,0,0}
\definecolor{rosybrown}{rgb}{0.737,0.561,0.561}
\definecolor{rosybrown1}{rgb}{1,0.757,0.757}
\definecolor{rosybrown2}{rgb}{0.933,0.706,0.706}
\definecolor{rosybrown3}{rgb}{0.804,0.608,0.608}
\definecolor{rosybrown4}{rgb}{0.545,0.412,0.412}
\definecolor{royalblue}{rgb}{0.255,0.412,0.882}
\definecolor{royalblue1}{rgb}{0.282,0.463,1}
\definecolor{royalblue2}{rgb}{0.263,0.431,0.933}
\definecolor{royalblue3}{rgb}{0.227,0.373,0.804}
\definecolor{royalblue4}{rgb}{0.153,0.251,0.545}
\definecolor{saddlebrown}{rgb}{0.545,0.271,0.075}
\definecolor{salmon}{rgb}{0.98,0.502,0.447}
\definecolor{salmon1}{rgb}{1,0.549,0.412}
\definecolor{salmon2}{rgb}{0.933,0.51,0.384}
\definecolor{salmon3}{rgb}{0.804,0.439,0.329}
\definecolor{salmon4}{rgb}{0.545,0.298,0.224}
\definecolor{sandybrown}{rgb}{0.957,0.643,0.376}
\definecolor{seagreen1}{rgb}{0.329,1,0.624}
\definecolor{seagreen2}{rgb}{0.306,0.933,0.58}
\definecolor{seagreen3}{rgb}{0.263,0.804,0.502}
\definecolor{seagreen4}{rgb}{0.18,0.545,0.341}
\definecolor{seashell}{rgb}{1,0.961,0.933}
\definecolor{seashell1}{rgb}{1,0.961,0.933}
\definecolor{seashell2}{rgb}{0.933,0.898,0.871}
\definecolor{seashell3}{rgb}{0.804,0.773,0.749}
\definecolor{seashell4}{rgb}{0.545,0.525,0.51}
\definecolor{sienna}{rgb}{0.627,0.322,0.176}
\definecolor{sienna1}{rgb}{1,0.51,0.278}
\definecolor{sienna2}{rgb}{0.933,0.475,0.259}
\definecolor{sienna3}{rgb}{0.804,0.408,0.224}
\definecolor{sienna4}{rgb}{0.545,0.278,0.149}
\definecolor{skyblue}{rgb}{0.529,0.808,0.922}
\definecolor{skyblue1}{rgb}{0.529,0.808,1}
\definecolor{skyblue2}{rgb}{0.494,0.753,0.933}
\definecolor{skyblue3}{rgb}{0.424,0.651,0.804}
\definecolor{skyblue4}{rgb}{0.29,0.439,0.545}
\definecolor{slateblue}{rgb}{0.416,0.353,0.804}
\definecolor{slateblue1}{rgb}{0.514,0.435,1}
\definecolor{slateblue2}{rgb}{0.478,0.404,0.933}
\definecolor{slateblue3}{rgb}{0.412,0.349,0.804}
\definecolor{slateblue4}{rgb}{0.278,0.235,0.545}
\definecolor{slategray}{rgb}{0.439,0.502,0.565}
\definecolor{slategray1}{rgb}{0.776,0.886,1}
\definecolor{slategray2}{rgb}{0.725,0.827,0.933}
\definecolor{slategray3}{rgb}{0.624,0.714,0.804}
\definecolor{slategray4}{rgb}{0.424,0.482,0.545}
\definecolor{slategrey}{rgb}{0.439,0.502,0.565}
\definecolor{snow}{rgb}{1,0.98,0.98}
\definecolor{snow1}{rgb}{1,0.98,0.98}
\definecolor{snow2}{rgb}{0.933,0.914,0.914}
\definecolor{snow3}{rgb}{0.804,0.788,0.788}
\definecolor{snow4}{rgb}{0.545,0.537,0.537}
\definecolor{springgreen}{rgb}{0,1,0.498}
\definecolor{springgreen1}{rgb}{0,1,0.498}
\definecolor{springgreen2}{rgb}{0,0.933,0.463}
\definecolor{springgreen3}{rgb}{0,0.804,0.4}
\definecolor{springgreen4}{rgb}{0,0.545,0.271}
\definecolor{steelblue}{rgb}{0.275,0.51,0.706}
\definecolor{steelblue1}{rgb}{0.388,0.722,1}
\definecolor{steelblue2}{rgb}{0.361,0.675,0.933}
\definecolor{steelblue3}{rgb}{0.31,0.58,0.804}
\definecolor{steelblue4}{rgb}{0.212,0.392,0.545}
\definecolor{tan}{rgb}{0.824,0.706,0.549}
\definecolor{tan1}{rgb}{1,0.647,0.31}
\definecolor{tan2}{rgb}{0.933,0.604,0.286}
\definecolor{tan3}{rgb}{0.804,0.522,0.247}
\definecolor{tan4}{rgb}{0.545,0.353,0.169}
\definecolor{thistle}{rgb}{0.847,0.749,0.847}
\definecolor{thistle1}{rgb}{1,0.882,1}
\definecolor{thistle2}{rgb}{0.933,0.824,0.933}
\definecolor{thistle3}{rgb}{0.804,0.71,0.804}
\definecolor{thistle4}{rgb}{0.545,0.482,0.545}
\definecolor{tomato}{rgb}{1,0.388,0.278}
\definecolor{tomato1}{rgb}{1,0.388,0.278}
\definecolor{tomato2}{rgb}{0.933,0.361,0.259}
\definecolor{tomato3}{rgb}{0.804,0.31,0.224}
\definecolor{tomato4}{rgb}{0.545,0.212,0.149}
\definecolor{turquoise1}{rgb}{0,0.961,1}
\definecolor{turquoise2}{rgb}{0,0.898,0.933}
\definecolor{turquoise3}{rgb}{0,0.773,0.804}
\definecolor{turquoise4}{rgb}{0,0.525,0.545}
\definecolor{violetred}{rgb}{0.816,0.125,0.565}
\definecolor{violetred1}{rgb}{1,0.243,0.588}
\definecolor{violetred2}{rgb}{0.933,0.227,0.549}
\definecolor{violetred3}{rgb}{0.804,0.196,0.471}
\definecolor{violetred4}{rgb}{0.545,0.133,0.322}
\definecolor{wheat}{rgb}{0.961,0.871,0.702}
\definecolor{wheat1}{rgb}{1,0.906,0.729}
\definecolor{wheat2}{rgb}{0.933,0.847,0.682}
\definecolor{wheat3}{rgb}{0.804,0.729,0.588}
\definecolor{wheat4}{rgb}{0.545,0.494,0.4}
\definecolor{whitesmoke}{rgb}{0.961,0.961,0.961}
\definecolor{yellow1}{rgb}{1,1,0}
\definecolor{yellow2}{rgb}{0.933,0.933,0}
\definecolor{yellow3}{rgb}{0.804,0.804,0}
\definecolor{yellow4}{rgb}{0.545,0.545,0}
\definecolor{yellowgreen}{rgb}{0.604,0.804,0.196}
\begin{tikzpicture}[ipe stylesheet]
  \draw
    (132, 736)
     -- (144, 736);
  \draw
    (144, 736)
     -- (152, 716);
  \draw[gold, ipe pen fat, ipe opacity 50]
    (220, 756)
     -- (228, 732);
  \draw[gold, ipe pen fat, ipe opacity 50]
    (208, 712)
     -- (228, 732);
  \draw[gold, ipe pen fat, ipe opacity 50]
    (228, 732)
     -- (248, 752);
  \draw[gold, ipe pen fat, ipe opacity 50]
    (220, 756)
     -- (248, 752)
     -- (248, 752);
  \draw[gold, ipe pen fat, ipe opacity 50]
    (228, 732)
     -- (248, 708);
  \draw[gold, ipe pen fat, ipe opacity 50]
    (248, 708)
     -- (248, 752);
  \draw[darkorange, ipe pen heavier, ipe opacity 50]
    (220, 756)
     -- (144, 752);
  \draw[darkorange, ipe pen heavier, ipe opacity 50]
    (228, 732)
     -- (160, 740);
  \draw[darkorange, ipe pen heavier, ipe opacity 50]
    (152, 716)
     -- (228, 732);
  \draw[darkorange, ipe pen heavier, ipe opacity 50]
    (128, 712)
     -- (208, 712);
  \draw[darkorange, ipe pen heavier, ipe opacity 50]
    (208, 712)
     -- (160, 740);
  \draw[darkorange, ipe pen heavier, ipe opacity 50]
    (152, 716)
     -- (208, 712);
  \draw[darkorange, ipe pen heavier, ipe opacity 50]
    (160, 740)
     -- (248, 752);
  \draw
    (144, 752)
     -- (132, 736)
     -- (132, 736)
     -- (132, 736);
  \draw
    (144, 752)
     -- (160, 740);
  \draw
    (128, 712)
     -- (152, 716);
  \draw
    (152, 716)
     -- (160, 740);
  \draw
    (128, 712)
     -- (160, 740);
  \node[ipe node, text=red]
     at (140, 692) {$S$};
  \draw
    (144, 752)
     -- (144, 736);
  \node[ipe node]
     at (232, 692) {$F'$};
  \node[ipe node, text=darkorange]
     at (184, 692) {$E$};
  \pic[fill=darkgoldenrod2]
     at (248, 752) {ipe fdisk};
  \pic[fill=darkgoldenrod2]
     at (248, 708) {ipe fdisk};
  \pic[fill=darkgoldenrod2]
     at (208, 712) {ipe fdisk};
  \pic[fill=darkgoldenrod2]
     at (228, 732) {ipe fdisk};
  \pic[fill=darkgoldenrod2]
     at (220, 756) {ipe fdisk};
  \pic[red]
     at (144, 752) {ipe disk};
  \pic[red]
     at (132, 736) {ipe disk};
  \pic[red]
     at (152, 716) {ipe disk};
  \pic[red]
     at (128, 712) {ipe disk};
  \pic[red]
     at (160, 740) {ipe disk};
  \pic[red]
     at (144, 736) {ipe disk};
\end{tikzpicture}
\caption{A set $S$ of $6$ vertices of a graph, a graph $F'$ on $5$ vertices and a set $E$ of edges between vertices of $S$ and of $V(F').$ The graph $(F'\cup (S\cup V(F'),E)$ belongs to ${\cal F}^{S}_{11}.$}\labels{figure_extension}
\end{figure}
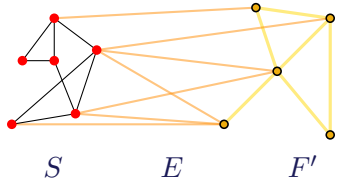
Notice that if $\ell ≤ |S|$ then ${\cal F}^{S}_\ell$ contains only the graph with vertex set $S$ and no edges.

\myskip\paragraph{Towards constructing a boundaried structure.}
Let $q,j',j,r,w,$ and $l$ as above.
Let $\mathfrak{A}$ be a $τ$-structure, let $G_{\mathfrak{A}}$ be its Gaifman graph, let ${\bf a}=(a_{1},\ldots,a_l)$ be an apex-tuple  of $G_{\mathfrak{A}},$ let $(W,\mathfrak{R})$ be a flatness pair of $G_{\mathfrak{A}}\setminus V({\bf a})$ of height $2w+j,$ and let ${\bf W}_{{q}}$ be the $q$-pseudogrid defined by the horizontal and vertical paths of the central ${q}$-subwall of $W.$
Also, let
$\mathfrak{K}=(\mathfrak{A}[V(K^{\bf a})],{\bf a}, {\bf I}, {\bf W}_{{q}})$  be the extended compass
of the flatness pair $(W,\mathfrak{R})$ of $G_{\mathfrak{A}}\setminus V({\bf a})$ and let $\ell∈[0,\tw(θ)-1].$
Given a $d∈[r,w],$ an $L\subseteq [l],$ a vertex set $Z\subseteq I^{(d-r+1)},$ and a graph $F∈ {\cal F}^{V_L ({\bf a})}_{\ell},$
we define the graph $K^{(d,Z,L,F)}$ as the one obtained from ${K^{\bf a}[I^{(d)}\cup V_L ({\bf a})]\cup F}$ by making every vertex in $V(F\setminus V_L ({\bf a}))$ adjacent to an arbitrarily chosen vertex in  $I^{(d)}\setminus Z.$
We also define the structure $\mathfrak{A}^{(d,Z,L,F)}$ to be the one obtained from $\mathfrak{A}[I^{(d)}\cup V_L ({\bf a})]$ after adding $|V(F)|$ new elements to its universe and a binary relation symbol that is interpreted as pairs of elements (corresponding to the additional edges of $F$ and every edge between a vertex in  $V(F\setminus V_L ({\bf a}))$ and a vertex in  $I^{(d)}\setminus Z$).
Notice that if $G$ is the Gaifman graph of $\mathfrak{A},$ then $K^{(d,Z,L,F)}$ is the Gaifman graph of  $\mathfrak{A}^{(d,Z,L,F)}.$
For simplicity, we will assume that $\mathfrak{A}^{(d,Z,L,F)}$ is also
a $τ$-structure.

See~\autoref{@incomportable} for a visualization of $K^{(d,Z,L,F)}.$
Intuitively, it contains the influence of the $d$-layer $I^{(d)},$ the apices that we guess that will belong to the modulator $X,$
the set $Z$ that is the part of $I^{(d-r)}$ that will not belong to the privileged component after the removal of $X,$
and the
part $F'$ of $F$ that corresponds to $X_{\rm out},$ i.e., the portion of the modulator $X$ that will not be part of $I^{(d)}.$
The graph $F$ in~\autoref{@incomportable} is the graph containing all vertices $V_L ({\bf a})$ and the ``extra'' guessed part $F'$ together with the extra edges from $V(F')$ to $V_L ({\bf a}).$
Let us explain the motivation behind adding these extra edges:
The reason we consider the structure $\mathfrak{A}^{(d,Z,L,F)}$ is to ``focus'' inside $I^{(d)}$ and temporarily ``forget'' what happens outside $I^{(d)}.$
However, we need to keep record of the fact that $I^{(d)}$ is in the same connected component as $V(F').$ This is why we add the extra edges.

\begin{figure}[ht]
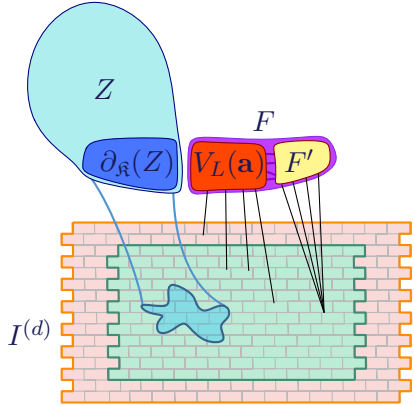

	\begin{center}
\tikzstyle{ipe stylesheet} = [
  ipe import,
  even odd rule,
  line join=round,
  line cap=butt,
  ipe pen normal/.style={line width=0.4},
  ipe pen heavier/.style={line width=0.8},
  ipe pen fat/.style={line width=1.2},
  ipe pen ultrafat/.style={line width=2},
  ipe pen normal,
  ipe mark normal/.style={ipe mark scale=3},
  ipe mark large/.style={ipe mark scale=5},
  ipe mark small/.style={ipe mark scale=2},
  ipe mark tiny/.style={ipe mark scale=1.1},
  ipe mark normal,
  /pgf/arrow keys/.cd,
  ipe arrow normal/.style={scale=7},
  ipe arrow large/.style={scale=10},
  ipe arrow small/.style={scale=5},
  ipe arrow tiny/.style={scale=3},
  ipe arrow normal,
  /tikz/.cd,
  ipe arrows, 
  <->/.tip = ipe normal,
  ipe dash normal/.style={dash pattern=},
  ipe dash dotted/.style={dash pattern=on 1bp off 3bp},
  ipe dash dashed/.style={dash pattern=on 4bp off 4bp},
  ipe dash dash dotted/.style={dash pattern=on 4bp off 2bp on 1bp off 2bp},
  ipe dash dash dot dotted/.style={dash pattern=on 4bp off 2bp on 1bp off 2bp on 1bp off 2bp},
  ipe dash normal,
  ipe node/.append style={font=\normalsize},
  ipe stretch normal/.style={ipe node stretch=1},
  ipe stretch normal,
  ipe opacity 10/.style={opacity=0.1},
  ipe opacity 30/.style={opacity=0.3},
  ipe opacity 50/.style={opacity=0.5},
  ipe opacity 75/.style={opacity=0.75},
  ipe opacity opaque/.style={opacity=1},
  ipe opacity opaque,
]
\definecolor{black}{rgb}{0,0,0}
\definecolor{white}{rgb}{1,1,1}
\definecolor{red}{rgb}{1,0,0}
\definecolor{blue}{rgb}{0,0,1}
\definecolor{green}{rgb}{0,1,0}
\definecolor{yellow}{rgb}{1,1,0}
\definecolor{orange}{rgb}{1,0.647,0}
\definecolor{gold}{rgb}{1,0.843,0}
\definecolor{purple}{rgb}{0.627,0.125,0.941}
\definecolor{gray}{rgb}{0.745,0.745,0.745}
\definecolor{brown}{rgb}{0.647,0.165,0.165}
\definecolor{navy}{rgb}{0,0,0.502}
\definecolor{pink}{rgb}{1,0.753,0.796}
\definecolor{seagreen}{rgb}{0.18,0.545,0.341}
\definecolor{turquoise}{rgb}{0.251,0.878,0.816}
\definecolor{violet}{rgb}{0.933,0.51,0.933}
\definecolor{darkblue}{rgb}{0,0,0.545}
\definecolor{darkcyan}{rgb}{0,0.545,0.545}
\definecolor{darkgray}{rgb}{0.663,0.663,0.663}
\definecolor{darkgreen}{rgb}{0,0.392,0}
\definecolor{darkmagenta}{rgb}{0.545,0,0.545}
\definecolor{darkorange}{rgb}{1,0.549,0}
\definecolor{darkred}{rgb}{0.545,0,0}
\definecolor{lightblue}{rgb}{0.678,0.847,0.902}
\definecolor{lightcyan}{rgb}{0.878,1,1}
\definecolor{lightgray}{rgb}{0.827,0.827,0.827}
\definecolor{lightgreen}{rgb}{0.565,0.933,0.565}
\definecolor{lightyellow}{rgb}{1,1,0.878}
\definecolor{aliceblue}{rgb}{0.941,0.973,1}
\definecolor{antiquewhite}{rgb}{0.98,0.922,0.843}
\definecolor{antiquewhite1}{rgb}{1,0.937,0.859}
\definecolor{antiquewhite2}{rgb}{0.933,0.875,0.8}
\definecolor{antiquewhite3}{rgb}{0.804,0.753,0.69}
\definecolor{antiquewhite4}{rgb}{0.545,0.514,0.471}
\definecolor{aquamarine}{rgb}{0.498,1,0.831}
\definecolor{aquamarine1}{rgb}{0.498,1,0.831}
\definecolor{aquamarine2}{rgb}{0.463,0.933,0.776}
\definecolor{aquamarine3}{rgb}{0.4,0.804,0.667}
\definecolor{aquamarine4}{rgb}{0.271,0.545,0.455}
\definecolor{azure}{rgb}{0.941,1,1}
\definecolor{azure1}{rgb}{0.941,1,1}
\definecolor{azure2}{rgb}{0.878,0.933,0.933}
\definecolor{azure3}{rgb}{0.757,0.804,0.804}
\definecolor{azure4}{rgb}{0.514,0.545,0.545}
\definecolor{beige}{rgb}{0.961,0.961,0.863}
\definecolor{bisque}{rgb}{1,0.894,0.769}
\definecolor{bisque1}{rgb}{1,0.894,0.769}
\definecolor{bisque2}{rgb}{0.933,0.835,0.718}
\definecolor{bisque3}{rgb}{0.804,0.718,0.62}
\definecolor{bisque4}{rgb}{0.545,0.49,0.42}
\definecolor{blanchedalmond}{rgb}{1,0.922,0.804}
\definecolor{blue1}{rgb}{0,0,1}
\definecolor{blue2}{rgb}{0,0,0.933}
\definecolor{blue3}{rgb}{0,0,0.804}
\definecolor{blue4}{rgb}{0,0,0.545}
\definecolor{blueviolet}{rgb}{0.541,0.169,0.886}
\definecolor{brown1}{rgb}{1,0.251,0.251}
\definecolor{brown2}{rgb}{0.933,0.231,0.231}
\definecolor{brown3}{rgb}{0.804,0.2,0.2}
\definecolor{brown4}{rgb}{0.545,0.137,0.137}
\definecolor{burlywood}{rgb}{0.871,0.722,0.529}
\definecolor{burlywood1}{rgb}{1,0.827,0.608}
\definecolor{burlywood2}{rgb}{0.933,0.773,0.569}
\definecolor{burlywood3}{rgb}{0.804,0.667,0.49}
\definecolor{burlywood4}{rgb}{0.545,0.451,0.333}
\definecolor{cadetblue}{rgb}{0.373,0.62,0.627}
\definecolor{cadetblue1}{rgb}{0.596,0.961,1}
\definecolor{cadetblue2}{rgb}{0.557,0.898,0.933}
\definecolor{cadetblue3}{rgb}{0.478,0.773,0.804}
\definecolor{cadetblue4}{rgb}{0.325,0.525,0.545}
\definecolor{chartreuse}{rgb}{0.498,1,0}
\definecolor{chartreuse1}{rgb}{0.498,1,0}
\definecolor{chartreuse2}{rgb}{0.463,0.933,0}
\definecolor{chartreuse3}{rgb}{0.4,0.804,0}
\definecolor{chartreuse4}{rgb}{0.271,0.545,0}
\definecolor{chocolate}{rgb}{0.824,0.412,0.118}
\definecolor{chocolate1}{rgb}{1,0.498,0.141}
\definecolor{chocolate2}{rgb}{0.933,0.463,0.129}
\definecolor{chocolate3}{rgb}{0.804,0.4,0.114}
\definecolor{chocolate4}{rgb}{0.545,0.271,0.075}
\definecolor{coral}{rgb}{1,0.498,0.314}
\definecolor{coral1}{rgb}{1,0.447,0.337}
\definecolor{coral2}{rgb}{0.933,0.416,0.314}
\definecolor{coral3}{rgb}{0.804,0.357,0.271}
\definecolor{coral4}{rgb}{0.545,0.243,0.184}
\definecolor{cornflowerblue}{rgb}{0.392,0.584,0.929}
\definecolor{cornsilk}{rgb}{1,0.973,0.863}
\definecolor{cornsilk1}{rgb}{1,0.973,0.863}
\definecolor{cornsilk2}{rgb}{0.933,0.91,0.804}
\definecolor{cornsilk3}{rgb}{0.804,0.784,0.694}
\definecolor{cornsilk4}{rgb}{0.545,0.533,0.471}
\definecolor{cyan}{rgb}{0,1,1}
\definecolor{cyan1}{rgb}{0,1,1}
\definecolor{cyan2}{rgb}{0,0.933,0.933}
\definecolor{cyan3}{rgb}{0,0.804,0.804}
\definecolor{cyan4}{rgb}{0,0.545,0.545}
\definecolor{darkgoldenrod}{rgb}{0.722,0.525,0.043}
\definecolor{darkgoldenrod1}{rgb}{1,0.725,0.059}
\definecolor{darkgoldenrod2}{rgb}{0.933,0.678,0.055}
\definecolor{darkgoldenrod3}{rgb}{0.804,0.584,0.047}
\definecolor{darkgoldenrod4}{rgb}{0.545,0.396,0.031}
\definecolor{darkgrey}{rgb}{0.663,0.663,0.663}
\definecolor{darkkhaki}{rgb}{0.741,0.718,0.42}
\definecolor{darkolivegreen}{rgb}{0.333,0.42,0.184}
\definecolor{darkolivegreen1}{rgb}{0.792,1,0.439}
\definecolor{darkolivegreen2}{rgb}{0.737,0.933,0.408}
\definecolor{darkolivegreen3}{rgb}{0.635,0.804,0.353}
\definecolor{darkolivegreen4}{rgb}{0.431,0.545,0.239}
\definecolor{darkorange1}{rgb}{1,0.498,0}
\definecolor{darkorange2}{rgb}{0.933,0.463,0}
\definecolor{darkorange3}{rgb}{0.804,0.4,0}
\definecolor{darkorange4}{rgb}{0.545,0.271,0}
\definecolor{darkorchid}{rgb}{0.6,0.196,0.8}
\definecolor{darkorchid1}{rgb}{0.749,0.243,1}
\definecolor{darkorchid2}{rgb}{0.698,0.227,0.933}
\definecolor{darkorchid3}{rgb}{0.604,0.196,0.804}
\definecolor{darkorchid4}{rgb}{0.408,0.133,0.545}
\definecolor{darksalmon}{rgb}{0.914,0.588,0.478}
\definecolor{darkseagreen}{rgb}{0.561,0.737,0.561}
\definecolor{darkseagreen1}{rgb}{0.757,1,0.757}
\definecolor{darkseagreen2}{rgb}{0.706,0.933,0.706}
\definecolor{darkseagreen3}{rgb}{0.608,0.804,0.608}
\definecolor{darkseagreen4}{rgb}{0.412,0.545,0.412}
\definecolor{darkslateblue}{rgb}{0.282,0.239,0.545}
\definecolor{darkslategray}{rgb}{0.184,0.31,0.31}
\definecolor{darkslategray1}{rgb}{0.592,1,1}
\definecolor{darkslategray2}{rgb}{0.553,0.933,0.933}
\definecolor{darkslategray3}{rgb}{0.475,0.804,0.804}
\definecolor{darkslategray4}{rgb}{0.322,0.545,0.545}
\definecolor{darkslategrey}{rgb}{0.184,0.31,0.31}
\definecolor{darkturquoise}{rgb}{0,0.808,0.82}
\definecolor{darkviolet}{rgb}{0.58,0,0.827}
\definecolor{deeppink}{rgb}{1,0.078,0.576}
\definecolor{deeppink1}{rgb}{1,0.078,0.576}
\definecolor{deeppink2}{rgb}{0.933,0.071,0.537}
\definecolor{deeppink3}{rgb}{0.804,0.063,0.463}
\definecolor{deeppink4}{rgb}{0.545,0.039,0.314}
\definecolor{deepskyblue}{rgb}{0,0.749,1}
\definecolor{deepskyblue1}{rgb}{0,0.749,1}
\definecolor{deepskyblue2}{rgb}{0,0.698,0.933}
\definecolor{deepskyblue3}{rgb}{0,0.604,0.804}
\definecolor{deepskyblue4}{rgb}{0,0.408,0.545}
\definecolor{dimgray}{rgb}{0.412,0.412,0.412}
\definecolor{dimgrey}{rgb}{0.412,0.412,0.412}
\definecolor{dodgerblue}{rgb}{0.118,0.565,1}
\definecolor{dodgerblue1}{rgb}{0.118,0.565,1}
\definecolor{dodgerblue2}{rgb}{0.11,0.525,0.933}
\definecolor{dodgerblue3}{rgb}{0.094,0.455,0.804}
\definecolor{dodgerblue4}{rgb}{0.063,0.306,0.545}
\definecolor{firebrick}{rgb}{0.698,0.133,0.133}
\definecolor{firebrick1}{rgb}{1,0.188,0.188}
\definecolor{firebrick2}{rgb}{0.933,0.173,0.173}
\definecolor{firebrick3}{rgb}{0.804,0.149,0.149}
\definecolor{firebrick4}{rgb}{0.545,0.102,0.102}
\definecolor{floralwhite}{rgb}{1,0.98,0.941}
\definecolor{forestgreen}{rgb}{0.133,0.545,0.133}
\definecolor{gainsboro}{rgb}{0.863,0.863,0.863}
\definecolor{ghostwhite}{rgb}{0.973,0.973,1}
\definecolor{gold1}{rgb}{1,0.843,0}
\definecolor{gold2}{rgb}{0.933,0.788,0}
\definecolor{gold3}{rgb}{0.804,0.678,0}
\definecolor{gold4}{rgb}{0.545,0.459,0}
\definecolor{goldenrod}{rgb}{0.855,0.647,0.125}
\definecolor{goldenrod1}{rgb}{1,0.757,0.145}
\definecolor{goldenrod2}{rgb}{0.933,0.706,0.133}
\definecolor{goldenrod3}{rgb}{0.804,0.608,0.114}
\definecolor{goldenrod4}{rgb}{0.545,0.412,0.078}
\definecolor{gray0}{rgb}{0,0,0}
\definecolor{gray1}{rgb}{0.012,0.012,0.012}
\definecolor{gray10}{rgb}{0.102,0.102,0.102}
\definecolor{gray100}{rgb}{1,1,1}
\definecolor{gray11}{rgb}{0.11,0.11,0.11}
\definecolor{gray12}{rgb}{0.122,0.122,0.122}
\definecolor{gray13}{rgb}{0.129,0.129,0.129}
\definecolor{gray14}{rgb}{0.141,0.141,0.141}
\definecolor{gray15}{rgb}{0.149,0.149,0.149}
\definecolor{gray16}{rgb}{0.161,0.161,0.161}
\definecolor{gray17}{rgb}{0.169,0.169,0.169}
\definecolor{gray18}{rgb}{0.18,0.18,0.18}
\definecolor{gray19}{rgb}{0.188,0.188,0.188}
\definecolor{gray2}{rgb}{0.02,0.02,0.02}
\definecolor{gray20}{rgb}{0.2,0.2,0.2}
\definecolor{gray21}{rgb}{0.212,0.212,0.212}
\definecolor{gray22}{rgb}{0.22,0.22,0.22}
\definecolor{gray23}{rgb}{0.231,0.231,0.231}
\definecolor{gray24}{rgb}{0.239,0.239,0.239}
\definecolor{gray25}{rgb}{0.251,0.251,0.251}
\definecolor{gray26}{rgb}{0.259,0.259,0.259}
\definecolor{gray27}{rgb}{0.271,0.271,0.271}
\definecolor{gray28}{rgb}{0.278,0.278,0.278}
\definecolor{gray29}{rgb}{0.29,0.29,0.29}
\definecolor{gray3}{rgb}{0.031,0.031,0.031}
\definecolor{gray30}{rgb}{0.302,0.302,0.302}
\definecolor{gray31}{rgb}{0.31,0.31,0.31}
\definecolor{gray32}{rgb}{0.322,0.322,0.322}
\definecolor{gray33}{rgb}{0.329,0.329,0.329}
\definecolor{gray34}{rgb}{0.341,0.341,0.341}
\definecolor{gray35}{rgb}{0.349,0.349,0.349}
\definecolor{gray36}{rgb}{0.361,0.361,0.361}
\definecolor{gray37}{rgb}{0.369,0.369,0.369}
\definecolor{gray38}{rgb}{0.38,0.38,0.38}
\definecolor{gray39}{rgb}{0.388,0.388,0.388}
\definecolor{gray4}{rgb}{0.039,0.039,0.039}
\definecolor{gray40}{rgb}{0.4,0.4,0.4}
\definecolor{gray41}{rgb}{0.412,0.412,0.412}
\definecolor{gray42}{rgb}{0.42,0.42,0.42}
\definecolor{gray43}{rgb}{0.431,0.431,0.431}
\definecolor{gray44}{rgb}{0.439,0.439,0.439}
\definecolor{gray45}{rgb}{0.451,0.451,0.451}
\definecolor{gray46}{rgb}{0.459,0.459,0.459}
\definecolor{gray47}{rgb}{0.471,0.471,0.471}
\definecolor{gray48}{rgb}{0.478,0.478,0.478}
\definecolor{gray49}{rgb}{0.49,0.49,0.49}
\definecolor{gray5}{rgb}{0.051,0.051,0.051}
\definecolor{gray50}{rgb}{0.498,0.498,0.498}
\definecolor{gray51}{rgb}{0.51,0.51,0.51}
\definecolor{gray52}{rgb}{0.522,0.522,0.522}
\definecolor{gray53}{rgb}{0.529,0.529,0.529}
\definecolor{gray54}{rgb}{0.541,0.541,0.541}
\definecolor{gray55}{rgb}{0.549,0.549,0.549}
\definecolor{gray56}{rgb}{0.561,0.561,0.561}
\definecolor{gray57}{rgb}{0.569,0.569,0.569}
\definecolor{gray58}{rgb}{0.58,0.58,0.58}
\definecolor{gray59}{rgb}{0.588,0.588,0.588}
\definecolor{gray6}{rgb}{0.059,0.059,0.059}
\definecolor{gray60}{rgb}{0.6,0.6,0.6}
\definecolor{gray61}{rgb}{0.612,0.612,0.612}
\definecolor{gray62}{rgb}{0.62,0.62,0.62}
\definecolor{gray63}{rgb}{0.631,0.631,0.631}
\definecolor{gray64}{rgb}{0.639,0.639,0.639}
\definecolor{gray65}{rgb}{0.651,0.651,0.651}
\definecolor{gray66}{rgb}{0.659,0.659,0.659}
\definecolor{gray67}{rgb}{0.671,0.671,0.671}
\definecolor{gray68}{rgb}{0.678,0.678,0.678}
\definecolor{gray69}{rgb}{0.69,0.69,0.69}
\definecolor{gray7}{rgb}{0.071,0.071,0.071}
\definecolor{gray70}{rgb}{0.702,0.702,0.702}
\definecolor{gray71}{rgb}{0.71,0.71,0.71}
\definecolor{gray72}{rgb}{0.722,0.722,0.722}
\definecolor{gray73}{rgb}{0.729,0.729,0.729}
\definecolor{gray74}{rgb}{0.741,0.741,0.741}
\definecolor{gray75}{rgb}{0.749,0.749,0.749}
\definecolor{gray76}{rgb}{0.761,0.761,0.761}
\definecolor{gray77}{rgb}{0.769,0.769,0.769}
\definecolor{gray78}{rgb}{0.78,0.78,0.78}
\definecolor{gray79}{rgb}{0.788,0.788,0.788}
\definecolor{gray8}{rgb}{0.078,0.078,0.078}
\definecolor{gray80}{rgb}{0.8,0.8,0.8}
\definecolor{gray81}{rgb}{0.812,0.812,0.812}
\definecolor{gray82}{rgb}{0.82,0.82,0.82}
\definecolor{gray83}{rgb}{0.831,0.831,0.831}
\definecolor{gray84}{rgb}{0.839,0.839,0.839}
\definecolor{gray85}{rgb}{0.851,0.851,0.851}
\definecolor{gray86}{rgb}{0.859,0.859,0.859}
\definecolor{gray87}{rgb}{0.871,0.871,0.871}
\definecolor{gray88}{rgb}{0.878,0.878,0.878}
\definecolor{gray89}{rgb}{0.89,0.89,0.89}
\definecolor{gray9}{rgb}{0.09,0.09,0.09}
\definecolor{gray90}{rgb}{0.898,0.898,0.898}
\definecolor{gray91}{rgb}{0.91,0.91,0.91}
\definecolor{gray92}{rgb}{0.922,0.922,0.922}
\definecolor{gray93}{rgb}{0.929,0.929,0.929}
\definecolor{gray94}{rgb}{0.941,0.941,0.941}
\definecolor{gray95}{rgb}{0.949,0.949,0.949}
\definecolor{gray96}{rgb}{0.961,0.961,0.961}
\definecolor{gray97}{rgb}{0.969,0.969,0.969}
\definecolor{gray98}{rgb}{0.98,0.98,0.98}
\definecolor{gray99}{rgb}{0.988,0.988,0.988}
\definecolor{green1}{rgb}{0,1,0}
\definecolor{green2}{rgb}{0,0.933,0}
\definecolor{green3}{rgb}{0,0.804,0}
\definecolor{green4}{rgb}{0,0.545,0}
\definecolor{greenyellow}{rgb}{0.678,1,0.184}
\definecolor{grey}{rgb}{0.745,0.745,0.745}
\definecolor{grey0}{rgb}{0,0,0}
\definecolor{grey1}{rgb}{0.012,0.012,0.012}
\definecolor{grey10}{rgb}{0.102,0.102,0.102}
\definecolor{grey100}{rgb}{1,1,1}
\definecolor{grey11}{rgb}{0.11,0.11,0.11}
\definecolor{grey12}{rgb}{0.122,0.122,0.122}
\definecolor{grey13}{rgb}{0.129,0.129,0.129}
\definecolor{grey14}{rgb}{0.141,0.141,0.141}
\definecolor{grey15}{rgb}{0.149,0.149,0.149}
\definecolor{grey16}{rgb}{0.161,0.161,0.161}
\definecolor{grey17}{rgb}{0.169,0.169,0.169}
\definecolor{grey18}{rgb}{0.18,0.18,0.18}
\definecolor{grey19}{rgb}{0.188,0.188,0.188}
\definecolor{grey2}{rgb}{0.02,0.02,0.02}
\definecolor{grey20}{rgb}{0.2,0.2,0.2}
\definecolor{grey21}{rgb}{0.212,0.212,0.212}
\definecolor{grey22}{rgb}{0.22,0.22,0.22}
\definecolor{grey23}{rgb}{0.231,0.231,0.231}
\definecolor{grey24}{rgb}{0.239,0.239,0.239}
\definecolor{grey25}{rgb}{0.251,0.251,0.251}
\definecolor{grey26}{rgb}{0.259,0.259,0.259}
\definecolor{grey27}{rgb}{0.271,0.271,0.271}
\definecolor{grey28}{rgb}{0.278,0.278,0.278}
\definecolor{grey29}{rgb}{0.29,0.29,0.29}
\definecolor{grey3}{rgb}{0.031,0.031,0.031}
\definecolor{grey30}{rgb}{0.302,0.302,0.302}
\definecolor{grey31}{rgb}{0.31,0.31,0.31}
\definecolor{grey32}{rgb}{0.322,0.322,0.322}
\definecolor{grey33}{rgb}{0.329,0.329,0.329}
\definecolor{grey34}{rgb}{0.341,0.341,0.341}
\definecolor{grey35}{rgb}{0.349,0.349,0.349}
\definecolor{grey36}{rgb}{0.361,0.361,0.361}
\definecolor{grey37}{rgb}{0.369,0.369,0.369}
\definecolor{grey38}{rgb}{0.38,0.38,0.38}
\definecolor{grey39}{rgb}{0.388,0.388,0.388}
\definecolor{grey4}{rgb}{0.039,0.039,0.039}
\definecolor{grey40}{rgb}{0.4,0.4,0.4}
\definecolor{grey41}{rgb}{0.412,0.412,0.412}
\definecolor{grey42}{rgb}{0.42,0.42,0.42}
\definecolor{grey43}{rgb}{0.431,0.431,0.431}
\definecolor{grey44}{rgb}{0.439,0.439,0.439}
\definecolor{grey45}{rgb}{0.451,0.451,0.451}
\definecolor{grey46}{rgb}{0.459,0.459,0.459}
\definecolor{grey47}{rgb}{0.471,0.471,0.471}
\definecolor{grey48}{rgb}{0.478,0.478,0.478}
\definecolor{grey49}{rgb}{0.49,0.49,0.49}
\definecolor{grey5}{rgb}{0.051,0.051,0.051}
\definecolor{grey50}{rgb}{0.498,0.498,0.498}
\definecolor{grey51}{rgb}{0.51,0.51,0.51}
\definecolor{grey52}{rgb}{0.522,0.522,0.522}
\definecolor{grey53}{rgb}{0.529,0.529,0.529}
\definecolor{grey54}{rgb}{0.541,0.541,0.541}
\definecolor{grey55}{rgb}{0.549,0.549,0.549}
\definecolor{grey56}{rgb}{0.561,0.561,0.561}
\definecolor{grey57}{rgb}{0.569,0.569,0.569}
\definecolor{grey58}{rgb}{0.58,0.58,0.58}
\definecolor{grey59}{rgb}{0.588,0.588,0.588}
\definecolor{grey6}{rgb}{0.059,0.059,0.059}
\definecolor{grey60}{rgb}{0.6,0.6,0.6}
\definecolor{grey61}{rgb}{0.612,0.612,0.612}
\definecolor{grey62}{rgb}{0.62,0.62,0.62}
\definecolor{grey63}{rgb}{0.631,0.631,0.631}
\definecolor{grey64}{rgb}{0.639,0.639,0.639}
\definecolor{grey65}{rgb}{0.651,0.651,0.651}
\definecolor{grey66}{rgb}{0.659,0.659,0.659}
\definecolor{grey67}{rgb}{0.671,0.671,0.671}
\definecolor{grey68}{rgb}{0.678,0.678,0.678}
\definecolor{grey69}{rgb}{0.69,0.69,0.69}
\definecolor{grey7}{rgb}{0.071,0.071,0.071}
\definecolor{grey70}{rgb}{0.702,0.702,0.702}
\definecolor{grey71}{rgb}{0.71,0.71,0.71}
\definecolor{grey72}{rgb}{0.722,0.722,0.722}
\definecolor{grey73}{rgb}{0.729,0.729,0.729}
\definecolor{grey74}{rgb}{0.741,0.741,0.741}
\definecolor{grey75}{rgb}{0.749,0.749,0.749}
\definecolor{grey76}{rgb}{0.761,0.761,0.761}
\definecolor{grey77}{rgb}{0.769,0.769,0.769}
\definecolor{grey78}{rgb}{0.78,0.78,0.78}
\definecolor{grey79}{rgb}{0.788,0.788,0.788}
\definecolor{grey8}{rgb}{0.078,0.078,0.078}
\definecolor{grey80}{rgb}{0.8,0.8,0.8}
\definecolor{grey81}{rgb}{0.812,0.812,0.812}
\definecolor{grey82}{rgb}{0.82,0.82,0.82}
\definecolor{grey83}{rgb}{0.831,0.831,0.831}
\definecolor{grey84}{rgb}{0.839,0.839,0.839}
\definecolor{grey85}{rgb}{0.851,0.851,0.851}
\definecolor{grey86}{rgb}{0.859,0.859,0.859}
\definecolor{grey87}{rgb}{0.871,0.871,0.871}
\definecolor{grey88}{rgb}{0.878,0.878,0.878}
\definecolor{grey89}{rgb}{0.89,0.89,0.89}
\definecolor{grey9}{rgb}{0.09,0.09,0.09}
\definecolor{grey90}{rgb}{0.898,0.898,0.898}
\definecolor{grey91}{rgb}{0.91,0.91,0.91}
\definecolor{grey92}{rgb}{0.922,0.922,0.922}
\definecolor{grey93}{rgb}{0.929,0.929,0.929}
\definecolor{grey94}{rgb}{0.941,0.941,0.941}
\definecolor{grey95}{rgb}{0.949,0.949,0.949}
\definecolor{grey96}{rgb}{0.961,0.961,0.961}
\definecolor{grey97}{rgb}{0.969,0.969,0.969}
\definecolor{grey98}{rgb}{0.98,0.98,0.98}
\definecolor{grey99}{rgb}{0.988,0.988,0.988}
\definecolor{honeydew}{rgb}{0.941,1,0.941}
\definecolor{honeydew1}{rgb}{0.941,1,0.941}
\definecolor{honeydew2}{rgb}{0.878,0.933,0.878}
\definecolor{honeydew3}{rgb}{0.757,0.804,0.757}
\definecolor{honeydew4}{rgb}{0.514,0.545,0.514}
\definecolor{hotpink}{rgb}{1,0.412,0.706}
\definecolor{hotpink1}{rgb}{1,0.431,0.706}
\definecolor{hotpink2}{rgb}{0.933,0.416,0.655}
\definecolor{hotpink3}{rgb}{0.804,0.376,0.565}
\definecolor{hotpink4}{rgb}{0.545,0.227,0.384}
\definecolor{indianred}{rgb}{0.804,0.361,0.361}
\definecolor{indianred1}{rgb}{1,0.416,0.416}
\definecolor{indianred2}{rgb}{0.933,0.388,0.388}
\definecolor{indianred3}{rgb}{0.804,0.333,0.333}
\definecolor{indianred4}{rgb}{0.545,0.227,0.227}
\definecolor{ivory}{rgb}{1,1,0.941}
\definecolor{ivory1}{rgb}{1,1,0.941}
\definecolor{ivory2}{rgb}{0.933,0.933,0.878}
\definecolor{ivory3}{rgb}{0.804,0.804,0.757}
\definecolor{ivory4}{rgb}{0.545,0.545,0.514}
\definecolor{khaki}{rgb}{0.941,0.902,0.549}
\definecolor{khaki1}{rgb}{1,0.965,0.561}
\definecolor{khaki2}{rgb}{0.933,0.902,0.522}
\definecolor{khaki3}{rgb}{0.804,0.776,0.451}
\definecolor{khaki4}{rgb}{0.545,0.525,0.306}
\definecolor{lavender}{rgb}{0.902,0.902,0.98}
\definecolor{lavenderblush}{rgb}{1,0.941,0.961}
\definecolor{lavenderblush1}{rgb}{1,0.941,0.961}
\definecolor{lavenderblush2}{rgb}{0.933,0.878,0.898}
\definecolor{lavenderblush3}{rgb}{0.804,0.757,0.773}
\definecolor{lavenderblush4}{rgb}{0.545,0.514,0.525}
\definecolor{lawngreen}{rgb}{0.486,0.988,0}
\definecolor{lemonchiffon}{rgb}{1,0.98,0.804}
\definecolor{lemonchiffon1}{rgb}{1,0.98,0.804}
\definecolor{lemonchiffon2}{rgb}{0.933,0.914,0.749}
\definecolor{lemonchiffon3}{rgb}{0.804,0.788,0.647}
\definecolor{lemonchiffon4}{rgb}{0.545,0.537,0.439}
\definecolor{lightblue1}{rgb}{0.749,0.937,1}
\definecolor{lightblue2}{rgb}{0.698,0.875,0.933}
\definecolor{lightblue3}{rgb}{0.604,0.753,0.804}
\definecolor{lightblue4}{rgb}{0.408,0.514,0.545}
\definecolor{lightcoral}{rgb}{0.941,0.502,0.502}
\definecolor{lightcyan1}{rgb}{0.878,1,1}
\definecolor{lightcyan2}{rgb}{0.82,0.933,0.933}
\definecolor{lightcyan3}{rgb}{0.706,0.804,0.804}
\definecolor{lightcyan4}{rgb}{0.478,0.545,0.545}
\definecolor{lightgoldenrod}{rgb}{0.933,0.867,0.51}
\definecolor{lightgoldenrod1}{rgb}{1,0.925,0.545}
\definecolor{lightgoldenrod2}{rgb}{0.933,0.863,0.51}
\definecolor{lightgoldenrod3}{rgb}{0.804,0.745,0.439}
\definecolor{lightgoldenrod4}{rgb}{0.545,0.506,0.298}
\definecolor{lightgoldenrodyellow}{rgb}{0.98,0.98,0.824}
\definecolor{lightgrey}{rgb}{0.827,0.827,0.827}
\definecolor{lightpink}{rgb}{1,0.714,0.757}
\definecolor{lightpink1}{rgb}{1,0.682,0.725}
\definecolor{lightpink2}{rgb}{0.933,0.635,0.678}
\definecolor{lightpink3}{rgb}{0.804,0.549,0.584}
\definecolor{lightpink4}{rgb}{0.545,0.373,0.396}
\definecolor{lightsalmon}{rgb}{1,0.627,0.478}
\definecolor{lightsalmon1}{rgb}{1,0.627,0.478}
\definecolor{lightsalmon2}{rgb}{0.933,0.584,0.447}
\definecolor{lightsalmon3}{rgb}{0.804,0.506,0.384}
\definecolor{lightsalmon4}{rgb}{0.545,0.341,0.259}
\definecolor{lightseagreen}{rgb}{0.125,0.698,0.667}
\definecolor{lightskyblue}{rgb}{0.529,0.808,0.98}
\definecolor{lightskyblue1}{rgb}{0.69,0.886,1}
\definecolor{lightskyblue2}{rgb}{0.643,0.827,0.933}
\definecolor{lightskyblue3}{rgb}{0.553,0.714,0.804}
\definecolor{lightskyblue4}{rgb}{0.376,0.482,0.545}
\definecolor{lightslateblue}{rgb}{0.518,0.439,1}
\definecolor{lightslategray}{rgb}{0.467,0.533,0.6}
\definecolor{lightslategrey}{rgb}{0.467,0.533,0.6}
\definecolor{lightsteelblue}{rgb}{0.69,0.769,0.871}
\definecolor{lightsteelblue1}{rgb}{0.792,0.882,1}
\definecolor{lightsteelblue2}{rgb}{0.737,0.824,0.933}
\definecolor{lightsteelblue3}{rgb}{0.635,0.71,0.804}
\definecolor{lightsteelblue4}{rgb}{0.431,0.482,0.545}
\definecolor{lightyellow1}{rgb}{1,1,0.878}
\definecolor{lightyellow2}{rgb}{0.933,0.933,0.82}
\definecolor{lightyellow3}{rgb}{0.804,0.804,0.706}
\definecolor{lightyellow4}{rgb}{0.545,0.545,0.478}
\definecolor{limegreen}{rgb}{0.196,0.804,0.196}
\definecolor{linen}{rgb}{0.98,0.941,0.902}
\definecolor{magenta}{rgb}{1,0,1}
\definecolor{magenta1}{rgb}{1,0,1}
\definecolor{magenta2}{rgb}{0.933,0,0.933}
\definecolor{magenta3}{rgb}{0.804,0,0.804}
\definecolor{magenta4}{rgb}{0.545,0,0.545}
\definecolor{maroon}{rgb}{0.69,0.188,0.376}
\definecolor{maroon1}{rgb}{1,0.204,0.702}
\definecolor{maroon2}{rgb}{0.933,0.188,0.655}
\definecolor{maroon3}{rgb}{0.804,0.161,0.565}
\definecolor{maroon4}{rgb}{0.545,0.11,0.384}
\definecolor{mediumaquamarine}{rgb}{0.4,0.804,0.667}
\definecolor{mediumblue}{rgb}{0,0,0.804}
\definecolor{mediumorchid}{rgb}{0.729,0.333,0.827}
\definecolor{mediumorchid1}{rgb}{0.878,0.4,1}
\definecolor{mediumorchid2}{rgb}{0.82,0.373,0.933}
\definecolor{mediumorchid3}{rgb}{0.706,0.322,0.804}
\definecolor{mediumorchid4}{rgb}{0.478,0.216,0.545}
\definecolor{mediumpurple}{rgb}{0.576,0.439,0.859}
\definecolor{mediumpurple1}{rgb}{0.671,0.51,1}
\definecolor{mediumpurple2}{rgb}{0.624,0.475,0.933}
\definecolor{mediumpurple3}{rgb}{0.537,0.408,0.804}
\definecolor{mediumpurple4}{rgb}{0.365,0.278,0.545}
\definecolor{mediumseagreen}{rgb}{0.235,0.702,0.443}
\definecolor{mediumslateblue}{rgb}{0.482,0.408,0.933}
\definecolor{mediumspringgreen}{rgb}{0,0.98,0.604}
\definecolor{mediumturquoise}{rgb}{0.282,0.82,0.8}
\definecolor{mediumvioletred}{rgb}{0.78,0.082,0.522}
\definecolor{midnightblue}{rgb}{0.098,0.098,0.439}
\definecolor{mintcream}{rgb}{0.961,1,0.98}
\definecolor{mistyrose}{rgb}{1,0.894,0.882}
\definecolor{mistyrose1}{rgb}{1,0.894,0.882}
\definecolor{mistyrose2}{rgb}{0.933,0.835,0.824}
\definecolor{mistyrose3}{rgb}{0.804,0.718,0.71}
\definecolor{mistyrose4}{rgb}{0.545,0.49,0.482}
\definecolor{moccasin}{rgb}{1,0.894,0.71}
\definecolor{navajowhite}{rgb}{1,0.871,0.678}
\definecolor{navajowhite1}{rgb}{1,0.871,0.678}
\definecolor{navajowhite2}{rgb}{0.933,0.812,0.631}
\definecolor{navajowhite3}{rgb}{0.804,0.702,0.545}
\definecolor{navajowhite4}{rgb}{0.545,0.475,0.369}
\definecolor{navyblue}{rgb}{0,0,0.502}
\definecolor{oldlace}{rgb}{0.992,0.961,0.902}
\definecolor{olivedrab}{rgb}{0.42,0.557,0.137}
\definecolor{olivedrab1}{rgb}{0.753,1,0.243}
\definecolor{olivedrab2}{rgb}{0.702,0.933,0.227}
\definecolor{olivedrab3}{rgb}{0.604,0.804,0.196}
\definecolor{olivedrab4}{rgb}{0.412,0.545,0.133}
\definecolor{orange1}{rgb}{1,0.647,0}
\definecolor{orange2}{rgb}{0.933,0.604,0}
\definecolor{orange3}{rgb}{0.804,0.522,0}
\definecolor{orange4}{rgb}{0.545,0.353,0}
\definecolor{orangered}{rgb}{1,0.271,0}
\definecolor{orangered1}{rgb}{1,0.271,0}
\definecolor{orangered2}{rgb}{0.933,0.251,0}
\definecolor{orangered3}{rgb}{0.804,0.216,0}
\definecolor{orangered4}{rgb}{0.545,0.145,0}
\definecolor{orchid}{rgb}{0.855,0.439,0.839}
\definecolor{orchid1}{rgb}{1,0.514,0.98}
\definecolor{orchid2}{rgb}{0.933,0.478,0.914}
\definecolor{orchid3}{rgb}{0.804,0.412,0.788}
\definecolor{orchid4}{rgb}{0.545,0.278,0.537}
\definecolor{palegoldenrod}{rgb}{0.933,0.91,0.667}
\definecolor{palegreen}{rgb}{0.596,0.984,0.596}
\definecolor{palegreen1}{rgb}{0.604,1,0.604}
\definecolor{palegreen2}{rgb}{0.565,0.933,0.565}
\definecolor{palegreen3}{rgb}{0.486,0.804,0.486}
\definecolor{palegreen4}{rgb}{0.329,0.545,0.329}
\definecolor{paleturquoise}{rgb}{0.686,0.933,0.933}
\definecolor{paleturquoise1}{rgb}{0.733,1,1}
\definecolor{paleturquoise2}{rgb}{0.682,0.933,0.933}
\definecolor{paleturquoise3}{rgb}{0.588,0.804,0.804}
\definecolor{paleturquoise4}{rgb}{0.4,0.545,0.545}
\definecolor{palevioletred}{rgb}{0.859,0.439,0.576}
\definecolor{palevioletred1}{rgb}{1,0.51,0.671}
\definecolor{palevioletred2}{rgb}{0.933,0.475,0.624}
\definecolor{palevioletred3}{rgb}{0.804,0.408,0.537}
\definecolor{palevioletred4}{rgb}{0.545,0.278,0.365}
\definecolor{papayawhip}{rgb}{1,0.937,0.835}
\definecolor{peachpuff}{rgb}{1,0.855,0.725}
\definecolor{peachpuff1}{rgb}{1,0.855,0.725}
\definecolor{peachpuff2}{rgb}{0.933,0.796,0.678}
\definecolor{peachpuff3}{rgb}{0.804,0.686,0.584}
\definecolor{peachpuff4}{rgb}{0.545,0.467,0.396}
\definecolor{peru}{rgb}{0.804,0.522,0.247}
\definecolor{pink1}{rgb}{1,0.71,0.773}
\definecolor{pink2}{rgb}{0.933,0.663,0.722}
\definecolor{pink3}{rgb}{0.804,0.569,0.62}
\definecolor{pink4}{rgb}{0.545,0.388,0.424}
\definecolor{plum}{rgb}{0.867,0.627,0.867}
\definecolor{plum1}{rgb}{1,0.733,1}
\definecolor{plum2}{rgb}{0.933,0.682,0.933}
\definecolor{plum3}{rgb}{0.804,0.588,0.804}
\definecolor{plum4}{rgb}{0.545,0.4,0.545}
\definecolor{powderblue}{rgb}{0.69,0.878,0.902}
\definecolor{purple1}{rgb}{0.608,0.188,1}
\definecolor{purple2}{rgb}{0.569,0.173,0.933}
\definecolor{purple3}{rgb}{0.49,0.149,0.804}
\definecolor{purple4}{rgb}{0.333,0.102,0.545}
\definecolor{red1}{rgb}{1,0,0}
\definecolor{red2}{rgb}{0.933,0,0}
\definecolor{red3}{rgb}{0.804,0,0}
\definecolor{red4}{rgb}{0.545,0,0}
\definecolor{rosybrown}{rgb}{0.737,0.561,0.561}
\definecolor{rosybrown1}{rgb}{1,0.757,0.757}
\definecolor{rosybrown2}{rgb}{0.933,0.706,0.706}
\definecolor{rosybrown3}{rgb}{0.804,0.608,0.608}
\definecolor{rosybrown4}{rgb}{0.545,0.412,0.412}
\definecolor{royalblue}{rgb}{0.255,0.412,0.882}
\definecolor{royalblue1}{rgb}{0.282,0.463,1}
\definecolor{royalblue2}{rgb}{0.263,0.431,0.933}
\definecolor{royalblue3}{rgb}{0.227,0.373,0.804}
\definecolor{royalblue4}{rgb}{0.153,0.251,0.545}
\definecolor{saddlebrown}{rgb}{0.545,0.271,0.075}
\definecolor{salmon}{rgb}{0.98,0.502,0.447}
\definecolor{salmon1}{rgb}{1,0.549,0.412}
\definecolor{salmon2}{rgb}{0.933,0.51,0.384}
\definecolor{salmon3}{rgb}{0.804,0.439,0.329}
\definecolor{salmon4}{rgb}{0.545,0.298,0.224}
\definecolor{sandybrown}{rgb}{0.957,0.643,0.376}
\definecolor{seagreen1}{rgb}{0.329,1,0.624}
\definecolor{seagreen2}{rgb}{0.306,0.933,0.58}
\definecolor{seagreen3}{rgb}{0.263,0.804,0.502}
\definecolor{seagreen4}{rgb}{0.18,0.545,0.341}
\definecolor{seashell}{rgb}{1,0.961,0.933}
\definecolor{seashell1}{rgb}{1,0.961,0.933}
\definecolor{seashell2}{rgb}{0.933,0.898,0.871}
\definecolor{seashell3}{rgb}{0.804,0.773,0.749}
\definecolor{seashell4}{rgb}{0.545,0.525,0.51}
\definecolor{sienna}{rgb}{0.627,0.322,0.176}
\definecolor{sienna1}{rgb}{1,0.51,0.278}
\definecolor{sienna2}{rgb}{0.933,0.475,0.259}
\definecolor{sienna3}{rgb}{0.804,0.408,0.224}
\definecolor{sienna4}{rgb}{0.545,0.278,0.149}
\definecolor{skyblue}{rgb}{0.529,0.808,0.922}
\definecolor{skyblue1}{rgb}{0.529,0.808,1}
\definecolor{skyblue2}{rgb}{0.494,0.753,0.933}
\definecolor{skyblue3}{rgb}{0.424,0.651,0.804}
\definecolor{skyblue4}{rgb}{0.29,0.439,0.545}
\definecolor{slateblue}{rgb}{0.416,0.353,0.804}
\definecolor{slateblue1}{rgb}{0.514,0.435,1}
\definecolor{slateblue2}{rgb}{0.478,0.404,0.933}
\definecolor{slateblue3}{rgb}{0.412,0.349,0.804}
\definecolor{slateblue4}{rgb}{0.278,0.235,0.545}
\definecolor{slategray}{rgb}{0.439,0.502,0.565}
\definecolor{slategray1}{rgb}{0.776,0.886,1}
\definecolor{slategray2}{rgb}{0.725,0.827,0.933}
\definecolor{slategray3}{rgb}{0.624,0.714,0.804}
\definecolor{slategray4}{rgb}{0.424,0.482,0.545}
\definecolor{slategrey}{rgb}{0.439,0.502,0.565}
\definecolor{snow}{rgb}{1,0.98,0.98}
\definecolor{snow1}{rgb}{1,0.98,0.98}
\definecolor{snow2}{rgb}{0.933,0.914,0.914}
\definecolor{snow3}{rgb}{0.804,0.788,0.788}
\definecolor{snow4}{rgb}{0.545,0.537,0.537}
\definecolor{springgreen}{rgb}{0,1,0.498}
\definecolor{springgreen1}{rgb}{0,1,0.498}
\definecolor{springgreen2}{rgb}{0,0.933,0.463}
\definecolor{springgreen3}{rgb}{0,0.804,0.4}
\definecolor{springgreen4}{rgb}{0,0.545,0.271}
\definecolor{steelblue}{rgb}{0.275,0.51,0.706}
\definecolor{steelblue1}{rgb}{0.388,0.722,1}
\definecolor{steelblue2}{rgb}{0.361,0.675,0.933}
\definecolor{steelblue3}{rgb}{0.31,0.58,0.804}
\definecolor{steelblue4}{rgb}{0.212,0.392,0.545}
\definecolor{tan}{rgb}{0.824,0.706,0.549}
\definecolor{tan1}{rgb}{1,0.647,0.31}
\definecolor{tan2}{rgb}{0.933,0.604,0.286}
\definecolor{tan3}{rgb}{0.804,0.522,0.247}
\definecolor{tan4}{rgb}{0.545,0.353,0.169}
\definecolor{thistle}{rgb}{0.847,0.749,0.847}
\definecolor{thistle1}{rgb}{1,0.882,1}
\definecolor{thistle2}{rgb}{0.933,0.824,0.933}
\definecolor{thistle3}{rgb}{0.804,0.71,0.804}
\definecolor{thistle4}{rgb}{0.545,0.482,0.545}
\definecolor{tomato}{rgb}{1,0.388,0.278}
\definecolor{tomato1}{rgb}{1,0.388,0.278}
\definecolor{tomato2}{rgb}{0.933,0.361,0.259}
\definecolor{tomato3}{rgb}{0.804,0.31,0.224}
\definecolor{tomato4}{rgb}{0.545,0.212,0.149}
\definecolor{turquoise1}{rgb}{0,0.961,1}
\definecolor{turquoise2}{rgb}{0,0.898,0.933}
\definecolor{turquoise3}{rgb}{0,0.773,0.804}
\definecolor{turquoise4}{rgb}{0,0.525,0.545}
\definecolor{violetred}{rgb}{0.816,0.125,0.565}
\definecolor{violetred1}{rgb}{1,0.243,0.588}
\definecolor{violetred2}{rgb}{0.933,0.227,0.549}
\definecolor{violetred3}{rgb}{0.804,0.196,0.471}
\definecolor{violetred4}{rgb}{0.545,0.133,0.322}
\definecolor{wheat}{rgb}{0.961,0.871,0.702}
\definecolor{wheat1}{rgb}{1,0.906,0.729}
\definecolor{wheat2}{rgb}{0.933,0.847,0.682}
\definecolor{wheat3}{rgb}{0.804,0.729,0.588}
\definecolor{wheat4}{rgb}{0.545,0.494,0.4}
\definecolor{whitesmoke}{rgb}{0.961,0.961,0.961}
\definecolor{yellow1}{rgb}{1,1,0}
\definecolor{yellow2}{rgb}{0.933,0.933,0}
\definecolor{yellow3}{rgb}{0.804,0.804,0}
\definecolor{yellow4}{rgb}{0.545,0.545,0}
\definecolor{yellowgreen}{rgb}{0.604,0.804,0.196}

	\end{center}
	\caption{The graph $K^{(d,Z,L,F)}.$}
	\labels{@incomportable}
\end{figure}

\myskip\paragraph{Strongly isomorphic graphs.}
Let $G$ be a graph.
A {\em nice 3-partition} of $G$ is an ordered partition ${\cal V} = (V_1,V_2, V_3)$ of $V(G)$ such that $(V_1\cup V_2, V_2 \cup V_3)$ is a separation of $G$ (see~\autoref{figure_partition} for an example).

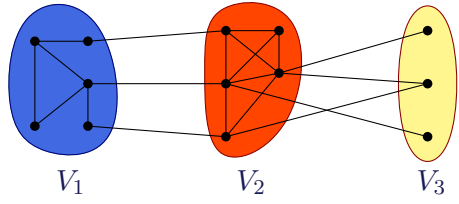
\begin{figure}[ht]
\centering
\tikzstyle{ipe stylesheet} = [
  ipe import,
  even odd rule,
  line join=round,
  line cap=butt,
  ipe pen normal/.style={line width=0.4},
  ipe pen heavier/.style={line width=0.8},
  ipe pen fat/.style={line width=1.2},
  ipe pen ultrafat/.style={line width=2},
  ipe pen normal,
  ipe mark normal/.style={ipe mark scale=3},
  ipe mark large/.style={ipe mark scale=5},
  ipe mark small/.style={ipe mark scale=2},
  ipe mark tiny/.style={ipe mark scale=1.1},
  ipe mark normal,
  /pgf/arrow keys/.cd,
  ipe arrow normal/.style={scale=7},
  ipe arrow large/.style={scale=10},
  ipe arrow small/.style={scale=5},
  ipe arrow tiny/.style={scale=3},
  ipe arrow normal,
  /tikz/.cd,
  ipe arrows, 
  <->/.tip = ipe normal,
  ipe dash normal/.style={dash pattern=},
  ipe dash dotted/.style={dash pattern=on 1bp off 3bp},
  ipe dash dashed/.style={dash pattern=on 4bp off 4bp},
  ipe dash dash dotted/.style={dash pattern=on 4bp off 2bp on 1bp off 2bp},
  ipe dash dash dot dotted/.style={dash pattern=on 4bp off 2bp on 1bp off 2bp on 1bp off 2bp},
  ipe dash normal,
  ipe node/.append style={font=\normalsize},
  ipe stretch normal/.style={ipe node stretch=1},
  ipe stretch normal,
  ipe opacity 10/.style={opacity=0.1},
  ipe opacity 30/.style={opacity=0.3},
  ipe opacity 50/.style={opacity=0.5},
  ipe opacity 75/.style={opacity=0.75},
  ipe opacity opaque/.style={opacity=1},
  ipe opacity opaque,
]
\definecolor{black}{rgb}{0,0,0}
\definecolor{white}{rgb}{1,1,1}
\definecolor{red}{rgb}{1,0,0}
\definecolor{blue}{rgb}{0,0,1}
\definecolor{green}{rgb}{0,1,0}
\definecolor{yellow}{rgb}{1,1,0}
\definecolor{orange}{rgb}{1,0.647,0}
\definecolor{gold}{rgb}{1,0.843,0}
\definecolor{purple}{rgb}{0.627,0.125,0.941}
\definecolor{gray}{rgb}{0.745,0.745,0.745}
\definecolor{brown}{rgb}{0.647,0.165,0.165}
\definecolor{navy}{rgb}{0,0,0.502}
\definecolor{pink}{rgb}{1,0.753,0.796}
\definecolor{seagreen}{rgb}{0.18,0.545,0.341}
\definecolor{turquoise}{rgb}{0.251,0.878,0.816}
\definecolor{violet}{rgb}{0.933,0.51,0.933}
\definecolor{darkblue}{rgb}{0,0,0.545}
\definecolor{darkcyan}{rgb}{0,0.545,0.545}
\definecolor{darkgray}{rgb}{0.663,0.663,0.663}
\definecolor{darkgreen}{rgb}{0,0.392,0}
\definecolor{darkmagenta}{rgb}{0.545,0,0.545}
\definecolor{darkorange}{rgb}{1,0.549,0}
\definecolor{darkred}{rgb}{0.545,0,0}
\definecolor{lightblue}{rgb}{0.678,0.847,0.902}
\definecolor{lightcyan}{rgb}{0.878,1,1}
\definecolor{lightgray}{rgb}{0.827,0.827,0.827}
\definecolor{lightgreen}{rgb}{0.565,0.933,0.565}
\definecolor{lightyellow}{rgb}{1,1,0.878}
\definecolor{aliceblue}{rgb}{0.941,0.973,1}
\definecolor{antiquewhite}{rgb}{0.98,0.922,0.843}
\definecolor{antiquewhite1}{rgb}{1,0.937,0.859}
\definecolor{antiquewhite2}{rgb}{0.933,0.875,0.8}
\definecolor{antiquewhite3}{rgb}{0.804,0.753,0.69}
\definecolor{antiquewhite4}{rgb}{0.545,0.514,0.471}
\definecolor{aquamarine}{rgb}{0.498,1,0.831}
\definecolor{aquamarine1}{rgb}{0.498,1,0.831}
\definecolor{aquamarine2}{rgb}{0.463,0.933,0.776}
\definecolor{aquamarine3}{rgb}{0.4,0.804,0.667}
\definecolor{aquamarine4}{rgb}{0.271,0.545,0.455}
\definecolor{azure}{rgb}{0.941,1,1}
\definecolor{azure1}{rgb}{0.941,1,1}
\definecolor{azure2}{rgb}{0.878,0.933,0.933}
\definecolor{azure3}{rgb}{0.757,0.804,0.804}
\definecolor{azure4}{rgb}{0.514,0.545,0.545}
\definecolor{beige}{rgb}{0.961,0.961,0.863}
\definecolor{bisque}{rgb}{1,0.894,0.769}
\definecolor{bisque1}{rgb}{1,0.894,0.769}
\definecolor{bisque2}{rgb}{0.933,0.835,0.718}
\definecolor{bisque3}{rgb}{0.804,0.718,0.62}
\definecolor{bisque4}{rgb}{0.545,0.49,0.42}
\definecolor{blanchedalmond}{rgb}{1,0.922,0.804}
\definecolor{blue1}{rgb}{0,0,1}
\definecolor{blue2}{rgb}{0,0,0.933}
\definecolor{blue3}{rgb}{0,0,0.804}
\definecolor{blue4}{rgb}{0,0,0.545}
\definecolor{blueviolet}{rgb}{0.541,0.169,0.886}
\definecolor{brown1}{rgb}{1,0.251,0.251}
\definecolor{brown2}{rgb}{0.933,0.231,0.231}
\definecolor{brown3}{rgb}{0.804,0.2,0.2}
\definecolor{brown4}{rgb}{0.545,0.137,0.137}
\definecolor{burlywood}{rgb}{0.871,0.722,0.529}
\definecolor{burlywood1}{rgb}{1,0.827,0.608}
\definecolor{burlywood2}{rgb}{0.933,0.773,0.569}
\definecolor{burlywood3}{rgb}{0.804,0.667,0.49}
\definecolor{burlywood4}{rgb}{0.545,0.451,0.333}
\definecolor{cadetblue}{rgb}{0.373,0.62,0.627}
\definecolor{cadetblue1}{rgb}{0.596,0.961,1}
\definecolor{cadetblue2}{rgb}{0.557,0.898,0.933}
\definecolor{cadetblue3}{rgb}{0.478,0.773,0.804}
\definecolor{cadetblue4}{rgb}{0.325,0.525,0.545}
\definecolor{chartreuse}{rgb}{0.498,1,0}
\definecolor{chartreuse1}{rgb}{0.498,1,0}
\definecolor{chartreuse2}{rgb}{0.463,0.933,0}
\definecolor{chartreuse3}{rgb}{0.4,0.804,0}
\definecolor{chartreuse4}{rgb}{0.271,0.545,0}
\definecolor{chocolate}{rgb}{0.824,0.412,0.118}
\definecolor{chocolate1}{rgb}{1,0.498,0.141}
\definecolor{chocolate2}{rgb}{0.933,0.463,0.129}
\definecolor{chocolate3}{rgb}{0.804,0.4,0.114}
\definecolor{chocolate4}{rgb}{0.545,0.271,0.075}
\definecolor{coral}{rgb}{1,0.498,0.314}
\definecolor{coral1}{rgb}{1,0.447,0.337}
\definecolor{coral2}{rgb}{0.933,0.416,0.314}
\definecolor{coral3}{rgb}{0.804,0.357,0.271}
\definecolor{coral4}{rgb}{0.545,0.243,0.184}
\definecolor{cornflowerblue}{rgb}{0.392,0.584,0.929}
\definecolor{cornsilk}{rgb}{1,0.973,0.863}
\definecolor{cornsilk1}{rgb}{1,0.973,0.863}
\definecolor{cornsilk2}{rgb}{0.933,0.91,0.804}
\definecolor{cornsilk3}{rgb}{0.804,0.784,0.694}
\definecolor{cornsilk4}{rgb}{0.545,0.533,0.471}
\definecolor{cyan}{rgb}{0,1,1}
\definecolor{cyan1}{rgb}{0,1,1}
\definecolor{cyan2}{rgb}{0,0.933,0.933}
\definecolor{cyan3}{rgb}{0,0.804,0.804}
\definecolor{cyan4}{rgb}{0,0.545,0.545}
\definecolor{darkgoldenrod}{rgb}{0.722,0.525,0.043}
\definecolor{darkgoldenrod1}{rgb}{1,0.725,0.059}
\definecolor{darkgoldenrod2}{rgb}{0.933,0.678,0.055}
\definecolor{darkgoldenrod3}{rgb}{0.804,0.584,0.047}
\definecolor{darkgoldenrod4}{rgb}{0.545,0.396,0.031}
\definecolor{darkgrey}{rgb}{0.663,0.663,0.663}
\definecolor{darkkhaki}{rgb}{0.741,0.718,0.42}
\definecolor{darkolivegreen}{rgb}{0.333,0.42,0.184}
\definecolor{darkolivegreen1}{rgb}{0.792,1,0.439}
\definecolor{darkolivegreen2}{rgb}{0.737,0.933,0.408}
\definecolor{darkolivegreen3}{rgb}{0.635,0.804,0.353}
\definecolor{darkolivegreen4}{rgb}{0.431,0.545,0.239}
\definecolor{darkorange1}{rgb}{1,0.498,0}
\definecolor{darkorange2}{rgb}{0.933,0.463,0}
\definecolor{darkorange3}{rgb}{0.804,0.4,0}
\definecolor{darkorange4}{rgb}{0.545,0.271,0}
\definecolor{darkorchid}{rgb}{0.6,0.196,0.8}
\definecolor{darkorchid1}{rgb}{0.749,0.243,1}
\definecolor{darkorchid2}{rgb}{0.698,0.227,0.933}
\definecolor{darkorchid3}{rgb}{0.604,0.196,0.804}
\definecolor{darkorchid4}{rgb}{0.408,0.133,0.545}
\definecolor{darksalmon}{rgb}{0.914,0.588,0.478}
\definecolor{darkseagreen}{rgb}{0.561,0.737,0.561}
\definecolor{darkseagreen1}{rgb}{0.757,1,0.757}
\definecolor{darkseagreen2}{rgb}{0.706,0.933,0.706}
\definecolor{darkseagreen3}{rgb}{0.608,0.804,0.608}
\definecolor{darkseagreen4}{rgb}{0.412,0.545,0.412}
\definecolor{darkslateblue}{rgb}{0.282,0.239,0.545}
\definecolor{darkslategray}{rgb}{0.184,0.31,0.31}
\definecolor{darkslategray1}{rgb}{0.592,1,1}
\definecolor{darkslategray2}{rgb}{0.553,0.933,0.933}
\definecolor{darkslategray3}{rgb}{0.475,0.804,0.804}
\definecolor{darkslategray4}{rgb}{0.322,0.545,0.545}
\definecolor{darkslategrey}{rgb}{0.184,0.31,0.31}
\definecolor{darkturquoise}{rgb}{0,0.808,0.82}
\definecolor{darkviolet}{rgb}{0.58,0,0.827}
\definecolor{deeppink}{rgb}{1,0.078,0.576}
\definecolor{deeppink1}{rgb}{1,0.078,0.576}
\definecolor{deeppink2}{rgb}{0.933,0.071,0.537}
\definecolor{deeppink3}{rgb}{0.804,0.063,0.463}
\definecolor{deeppink4}{rgb}{0.545,0.039,0.314}
\definecolor{deepskyblue}{rgb}{0,0.749,1}
\definecolor{deepskyblue1}{rgb}{0,0.749,1}
\definecolor{deepskyblue2}{rgb}{0,0.698,0.933}
\definecolor{deepskyblue3}{rgb}{0,0.604,0.804}
\definecolor{deepskyblue4}{rgb}{0,0.408,0.545}
\definecolor{dimgray}{rgb}{0.412,0.412,0.412}
\definecolor{dimgrey}{rgb}{0.412,0.412,0.412}
\definecolor{dodgerblue}{rgb}{0.118,0.565,1}
\definecolor{dodgerblue1}{rgb}{0.118,0.565,1}
\definecolor{dodgerblue2}{rgb}{0.11,0.525,0.933}
\definecolor{dodgerblue3}{rgb}{0.094,0.455,0.804}
\definecolor{dodgerblue4}{rgb}{0.063,0.306,0.545}
\definecolor{firebrick}{rgb}{0.698,0.133,0.133}
\definecolor{firebrick1}{rgb}{1,0.188,0.188}
\definecolor{firebrick2}{rgb}{0.933,0.173,0.173}
\definecolor{firebrick3}{rgb}{0.804,0.149,0.149}
\definecolor{firebrick4}{rgb}{0.545,0.102,0.102}
\definecolor{floralwhite}{rgb}{1,0.98,0.941}
\definecolor{forestgreen}{rgb}{0.133,0.545,0.133}
\definecolor{gainsboro}{rgb}{0.863,0.863,0.863}
\definecolor{ghostwhite}{rgb}{0.973,0.973,1}
\definecolor{gold1}{rgb}{1,0.843,0}
\definecolor{gold2}{rgb}{0.933,0.788,0}
\definecolor{gold3}{rgb}{0.804,0.678,0}
\definecolor{gold4}{rgb}{0.545,0.459,0}
\definecolor{goldenrod}{rgb}{0.855,0.647,0.125}
\definecolor{goldenrod1}{rgb}{1,0.757,0.145}
\definecolor{goldenrod2}{rgb}{0.933,0.706,0.133}
\definecolor{goldenrod3}{rgb}{0.804,0.608,0.114}
\definecolor{goldenrod4}{rgb}{0.545,0.412,0.078}
\definecolor{gray0}{rgb}{0,0,0}
\definecolor{gray1}{rgb}{0.012,0.012,0.012}
\definecolor{gray10}{rgb}{0.102,0.102,0.102}
\definecolor{gray100}{rgb}{1,1,1}
\definecolor{gray11}{rgb}{0.11,0.11,0.11}
\definecolor{gray12}{rgb}{0.122,0.122,0.122}
\definecolor{gray13}{rgb}{0.129,0.129,0.129}
\definecolor{gray14}{rgb}{0.141,0.141,0.141}
\definecolor{gray15}{rgb}{0.149,0.149,0.149}
\definecolor{gray16}{rgb}{0.161,0.161,0.161}
\definecolor{gray17}{rgb}{0.169,0.169,0.169}
\definecolor{gray18}{rgb}{0.18,0.18,0.18}
\definecolor{gray19}{rgb}{0.188,0.188,0.188}
\definecolor{gray2}{rgb}{0.02,0.02,0.02}
\definecolor{gray20}{rgb}{0.2,0.2,0.2}
\definecolor{gray21}{rgb}{0.212,0.212,0.212}
\definecolor{gray22}{rgb}{0.22,0.22,0.22}
\definecolor{gray23}{rgb}{0.231,0.231,0.231}
\definecolor{gray24}{rgb}{0.239,0.239,0.239}
\definecolor{gray25}{rgb}{0.251,0.251,0.251}
\definecolor{gray26}{rgb}{0.259,0.259,0.259}
\definecolor{gray27}{rgb}{0.271,0.271,0.271}
\definecolor{gray28}{rgb}{0.278,0.278,0.278}
\definecolor{gray29}{rgb}{0.29,0.29,0.29}
\definecolor{gray3}{rgb}{0.031,0.031,0.031}
\definecolor{gray30}{rgb}{0.302,0.302,0.302}
\definecolor{gray31}{rgb}{0.31,0.31,0.31}
\definecolor{gray32}{rgb}{0.322,0.322,0.322}
\definecolor{gray33}{rgb}{0.329,0.329,0.329}
\definecolor{gray34}{rgb}{0.341,0.341,0.341}
\definecolor{gray35}{rgb}{0.349,0.349,0.349}
\definecolor{gray36}{rgb}{0.361,0.361,0.361}
\definecolor{gray37}{rgb}{0.369,0.369,0.369}
\definecolor{gray38}{rgb}{0.38,0.38,0.38}
\definecolor{gray39}{rgb}{0.388,0.388,0.388}
\definecolor{gray4}{rgb}{0.039,0.039,0.039}
\definecolor{gray40}{rgb}{0.4,0.4,0.4}
\definecolor{gray41}{rgb}{0.412,0.412,0.412}
\definecolor{gray42}{rgb}{0.42,0.42,0.42}
\definecolor{gray43}{rgb}{0.431,0.431,0.431}
\definecolor{gray44}{rgb}{0.439,0.439,0.439}
\definecolor{gray45}{rgb}{0.451,0.451,0.451}
\definecolor{gray46}{rgb}{0.459,0.459,0.459}
\definecolor{gray47}{rgb}{0.471,0.471,0.471}
\definecolor{gray48}{rgb}{0.478,0.478,0.478}
\definecolor{gray49}{rgb}{0.49,0.49,0.49}
\definecolor{gray5}{rgb}{0.051,0.051,0.051}
\definecolor{gray50}{rgb}{0.498,0.498,0.498}
\definecolor{gray51}{rgb}{0.51,0.51,0.51}
\definecolor{gray52}{rgb}{0.522,0.522,0.522}
\definecolor{gray53}{rgb}{0.529,0.529,0.529}
\definecolor{gray54}{rgb}{0.541,0.541,0.541}
\definecolor{gray55}{rgb}{0.549,0.549,0.549}
\definecolor{gray56}{rgb}{0.561,0.561,0.561}
\definecolor{gray57}{rgb}{0.569,0.569,0.569}
\definecolor{gray58}{rgb}{0.58,0.58,0.58}
\definecolor{gray59}{rgb}{0.588,0.588,0.588}
\definecolor{gray6}{rgb}{0.059,0.059,0.059}
\definecolor{gray60}{rgb}{0.6,0.6,0.6}
\definecolor{gray61}{rgb}{0.612,0.612,0.612}
\definecolor{gray62}{rgb}{0.62,0.62,0.62}
\definecolor{gray63}{rgb}{0.631,0.631,0.631}
\definecolor{gray64}{rgb}{0.639,0.639,0.639}
\definecolor{gray65}{rgb}{0.651,0.651,0.651}
\definecolor{gray66}{rgb}{0.659,0.659,0.659}
\definecolor{gray67}{rgb}{0.671,0.671,0.671}
\definecolor{gray68}{rgb}{0.678,0.678,0.678}
\definecolor{gray69}{rgb}{0.69,0.69,0.69}
\definecolor{gray7}{rgb}{0.071,0.071,0.071}
\definecolor{gray70}{rgb}{0.702,0.702,0.702}
\definecolor{gray71}{rgb}{0.71,0.71,0.71}
\definecolor{gray72}{rgb}{0.722,0.722,0.722}
\definecolor{gray73}{rgb}{0.729,0.729,0.729}
\definecolor{gray74}{rgb}{0.741,0.741,0.741}
\definecolor{gray75}{rgb}{0.749,0.749,0.749}
\definecolor{gray76}{rgb}{0.761,0.761,0.761}
\definecolor{gray77}{rgb}{0.769,0.769,0.769}
\definecolor{gray78}{rgb}{0.78,0.78,0.78}
\definecolor{gray79}{rgb}{0.788,0.788,0.788}
\definecolor{gray8}{rgb}{0.078,0.078,0.078}
\definecolor{gray80}{rgb}{0.8,0.8,0.8}
\definecolor{gray81}{rgb}{0.812,0.812,0.812}
\definecolor{gray82}{rgb}{0.82,0.82,0.82}
\definecolor{gray83}{rgb}{0.831,0.831,0.831}
\definecolor{gray84}{rgb}{0.839,0.839,0.839}
\definecolor{gray85}{rgb}{0.851,0.851,0.851}
\definecolor{gray86}{rgb}{0.859,0.859,0.859}
\definecolor{gray87}{rgb}{0.871,0.871,0.871}
\definecolor{gray88}{rgb}{0.878,0.878,0.878}
\definecolor{gray89}{rgb}{0.89,0.89,0.89}
\definecolor{gray9}{rgb}{0.09,0.09,0.09}
\definecolor{gray90}{rgb}{0.898,0.898,0.898}
\definecolor{gray91}{rgb}{0.91,0.91,0.91}
\definecolor{gray92}{rgb}{0.922,0.922,0.922}
\definecolor{gray93}{rgb}{0.929,0.929,0.929}
\definecolor{gray94}{rgb}{0.941,0.941,0.941}
\definecolor{gray95}{rgb}{0.949,0.949,0.949}
\definecolor{gray96}{rgb}{0.961,0.961,0.961}
\definecolor{gray97}{rgb}{0.969,0.969,0.969}
\definecolor{gray98}{rgb}{0.98,0.98,0.98}
\definecolor{gray99}{rgb}{0.988,0.988,0.988}
\definecolor{green1}{rgb}{0,1,0}
\definecolor{green2}{rgb}{0,0.933,0}
\definecolor{green3}{rgb}{0,0.804,0}
\definecolor{green4}{rgb}{0,0.545,0}
\definecolor{greenyellow}{rgb}{0.678,1,0.184}
\definecolor{grey}{rgb}{0.745,0.745,0.745}
\definecolor{grey0}{rgb}{0,0,0}
\definecolor{grey1}{rgb}{0.012,0.012,0.012}
\definecolor{grey10}{rgb}{0.102,0.102,0.102}
\definecolor{grey100}{rgb}{1,1,1}
\definecolor{grey11}{rgb}{0.11,0.11,0.11}
\definecolor{grey12}{rgb}{0.122,0.122,0.122}
\definecolor{grey13}{rgb}{0.129,0.129,0.129}
\definecolor{grey14}{rgb}{0.141,0.141,0.141}
\definecolor{grey15}{rgb}{0.149,0.149,0.149}
\definecolor{grey16}{rgb}{0.161,0.161,0.161}
\definecolor{grey17}{rgb}{0.169,0.169,0.169}
\definecolor{grey18}{rgb}{0.18,0.18,0.18}
\definecolor{grey19}{rgb}{0.188,0.188,0.188}
\definecolor{grey2}{rgb}{0.02,0.02,0.02}
\definecolor{grey20}{rgb}{0.2,0.2,0.2}
\definecolor{grey21}{rgb}{0.212,0.212,0.212}
\definecolor{grey22}{rgb}{0.22,0.22,0.22}
\definecolor{grey23}{rgb}{0.231,0.231,0.231}
\definecolor{grey24}{rgb}{0.239,0.239,0.239}
\definecolor{grey25}{rgb}{0.251,0.251,0.251}
\definecolor{grey26}{rgb}{0.259,0.259,0.259}
\definecolor{grey27}{rgb}{0.271,0.271,0.271}
\definecolor{grey28}{rgb}{0.278,0.278,0.278}
\definecolor{grey29}{rgb}{0.29,0.29,0.29}
\definecolor{grey3}{rgb}{0.031,0.031,0.031}
\definecolor{grey30}{rgb}{0.302,0.302,0.302}
\definecolor{grey31}{rgb}{0.31,0.31,0.31}
\definecolor{grey32}{rgb}{0.322,0.322,0.322}
\definecolor{grey33}{rgb}{0.329,0.329,0.329}
\definecolor{grey34}{rgb}{0.341,0.341,0.341}
\definecolor{grey35}{rgb}{0.349,0.349,0.349}
\definecolor{grey36}{rgb}{0.361,0.361,0.361}
\definecolor{grey37}{rgb}{0.369,0.369,0.369}
\definecolor{grey38}{rgb}{0.38,0.38,0.38}
\definecolor{grey39}{rgb}{0.388,0.388,0.388}
\definecolor{grey4}{rgb}{0.039,0.039,0.039}
\definecolor{grey40}{rgb}{0.4,0.4,0.4}
\definecolor{grey41}{rgb}{0.412,0.412,0.412}
\definecolor{grey42}{rgb}{0.42,0.42,0.42}
\definecolor{grey43}{rgb}{0.431,0.431,0.431}
\definecolor{grey44}{rgb}{0.439,0.439,0.439}
\definecolor{grey45}{rgb}{0.451,0.451,0.451}
\definecolor{grey46}{rgb}{0.459,0.459,0.459}
\definecolor{grey47}{rgb}{0.471,0.471,0.471}
\definecolor{grey48}{rgb}{0.478,0.478,0.478}
\definecolor{grey49}{rgb}{0.49,0.49,0.49}
\definecolor{grey5}{rgb}{0.051,0.051,0.051}
\definecolor{grey50}{rgb}{0.498,0.498,0.498}
\definecolor{grey51}{rgb}{0.51,0.51,0.51}
\definecolor{grey52}{rgb}{0.522,0.522,0.522}
\definecolor{grey53}{rgb}{0.529,0.529,0.529}
\definecolor{grey54}{rgb}{0.541,0.541,0.541}
\definecolor{grey55}{rgb}{0.549,0.549,0.549}
\definecolor{grey56}{rgb}{0.561,0.561,0.561}
\definecolor{grey57}{rgb}{0.569,0.569,0.569}
\definecolor{grey58}{rgb}{0.58,0.58,0.58}
\definecolor{grey59}{rgb}{0.588,0.588,0.588}
\definecolor{grey6}{rgb}{0.059,0.059,0.059}
\definecolor{grey60}{rgb}{0.6,0.6,0.6}
\definecolor{grey61}{rgb}{0.612,0.612,0.612}
\definecolor{grey62}{rgb}{0.62,0.62,0.62}
\definecolor{grey63}{rgb}{0.631,0.631,0.631}
\definecolor{grey64}{rgb}{0.639,0.639,0.639}
\definecolor{grey65}{rgb}{0.651,0.651,0.651}
\definecolor{grey66}{rgb}{0.659,0.659,0.659}
\definecolor{grey67}{rgb}{0.671,0.671,0.671}
\definecolor{grey68}{rgb}{0.678,0.678,0.678}
\definecolor{grey69}{rgb}{0.69,0.69,0.69}
\definecolor{grey7}{rgb}{0.071,0.071,0.071}
\definecolor{grey70}{rgb}{0.702,0.702,0.702}
\definecolor{grey71}{rgb}{0.71,0.71,0.71}
\definecolor{grey72}{rgb}{0.722,0.722,0.722}
\definecolor{grey73}{rgb}{0.729,0.729,0.729}
\definecolor{grey74}{rgb}{0.741,0.741,0.741}
\definecolor{grey75}{rgb}{0.749,0.749,0.749}
\definecolor{grey76}{rgb}{0.761,0.761,0.761}
\definecolor{grey77}{rgb}{0.769,0.769,0.769}
\definecolor{grey78}{rgb}{0.78,0.78,0.78}
\definecolor{grey79}{rgb}{0.788,0.788,0.788}
\definecolor{grey8}{rgb}{0.078,0.078,0.078}
\definecolor{grey80}{rgb}{0.8,0.8,0.8}
\definecolor{grey81}{rgb}{0.812,0.812,0.812}
\definecolor{grey82}{rgb}{0.82,0.82,0.82}
\definecolor{grey83}{rgb}{0.831,0.831,0.831}
\definecolor{grey84}{rgb}{0.839,0.839,0.839}
\definecolor{grey85}{rgb}{0.851,0.851,0.851}
\definecolor{grey86}{rgb}{0.859,0.859,0.859}
\definecolor{grey87}{rgb}{0.871,0.871,0.871}
\definecolor{grey88}{rgb}{0.878,0.878,0.878}
\definecolor{grey89}{rgb}{0.89,0.89,0.89}
\definecolor{grey9}{rgb}{0.09,0.09,0.09}
\definecolor{grey90}{rgb}{0.898,0.898,0.898}
\definecolor{grey91}{rgb}{0.91,0.91,0.91}
\definecolor{grey92}{rgb}{0.922,0.922,0.922}
\definecolor{grey93}{rgb}{0.929,0.929,0.929}
\definecolor{grey94}{rgb}{0.941,0.941,0.941}
\definecolor{grey95}{rgb}{0.949,0.949,0.949}
\definecolor{grey96}{rgb}{0.961,0.961,0.961}
\definecolor{grey97}{rgb}{0.969,0.969,0.969}
\definecolor{grey98}{rgb}{0.98,0.98,0.98}
\definecolor{grey99}{rgb}{0.988,0.988,0.988}
\definecolor{honeydew}{rgb}{0.941,1,0.941}
\definecolor{honeydew1}{rgb}{0.941,1,0.941}
\definecolor{honeydew2}{rgb}{0.878,0.933,0.878}
\definecolor{honeydew3}{rgb}{0.757,0.804,0.757}
\definecolor{honeydew4}{rgb}{0.514,0.545,0.514}
\definecolor{hotpink}{rgb}{1,0.412,0.706}
\definecolor{hotpink1}{rgb}{1,0.431,0.706}
\definecolor{hotpink2}{rgb}{0.933,0.416,0.655}
\definecolor{hotpink3}{rgb}{0.804,0.376,0.565}
\definecolor{hotpink4}{rgb}{0.545,0.227,0.384}
\definecolor{indianred}{rgb}{0.804,0.361,0.361}
\definecolor{indianred1}{rgb}{1,0.416,0.416}
\definecolor{indianred2}{rgb}{0.933,0.388,0.388}
\definecolor{indianred3}{rgb}{0.804,0.333,0.333}
\definecolor{indianred4}{rgb}{0.545,0.227,0.227}
\definecolor{ivory}{rgb}{1,1,0.941}
\definecolor{ivory1}{rgb}{1,1,0.941}
\definecolor{ivory2}{rgb}{0.933,0.933,0.878}
\definecolor{ivory3}{rgb}{0.804,0.804,0.757}
\definecolor{ivory4}{rgb}{0.545,0.545,0.514}
\definecolor{khaki}{rgb}{0.941,0.902,0.549}
\definecolor{khaki1}{rgb}{1,0.965,0.561}
\definecolor{khaki2}{rgb}{0.933,0.902,0.522}
\definecolor{khaki3}{rgb}{0.804,0.776,0.451}
\definecolor{khaki4}{rgb}{0.545,0.525,0.306}
\definecolor{lavender}{rgb}{0.902,0.902,0.98}
\definecolor{lavenderblush}{rgb}{1,0.941,0.961}
\definecolor{lavenderblush1}{rgb}{1,0.941,0.961}
\definecolor{lavenderblush2}{rgb}{0.933,0.878,0.898}
\definecolor{lavenderblush3}{rgb}{0.804,0.757,0.773}
\definecolor{lavenderblush4}{rgb}{0.545,0.514,0.525}
\definecolor{lawngreen}{rgb}{0.486,0.988,0}
\definecolor{lemonchiffon}{rgb}{1,0.98,0.804}
\definecolor{lemonchiffon1}{rgb}{1,0.98,0.804}
\definecolor{lemonchiffon2}{rgb}{0.933,0.914,0.749}
\definecolor{lemonchiffon3}{rgb}{0.804,0.788,0.647}
\definecolor{lemonchiffon4}{rgb}{0.545,0.537,0.439}
\definecolor{lightblue1}{rgb}{0.749,0.937,1}
\definecolor{lightblue2}{rgb}{0.698,0.875,0.933}
\definecolor{lightblue3}{rgb}{0.604,0.753,0.804}
\definecolor{lightblue4}{rgb}{0.408,0.514,0.545}
\definecolor{lightcoral}{rgb}{0.941,0.502,0.502}
\definecolor{lightcyan1}{rgb}{0.878,1,1}
\definecolor{lightcyan2}{rgb}{0.82,0.933,0.933}
\definecolor{lightcyan3}{rgb}{0.706,0.804,0.804}
\definecolor{lightcyan4}{rgb}{0.478,0.545,0.545}
\definecolor{lightgoldenrod}{rgb}{0.933,0.867,0.51}
\definecolor{lightgoldenrod1}{rgb}{1,0.925,0.545}
\definecolor{lightgoldenrod2}{rgb}{0.933,0.863,0.51}
\definecolor{lightgoldenrod3}{rgb}{0.804,0.745,0.439}
\definecolor{lightgoldenrod4}{rgb}{0.545,0.506,0.298}
\definecolor{lightgoldenrodyellow}{rgb}{0.98,0.98,0.824}
\definecolor{lightgrey}{rgb}{0.827,0.827,0.827}
\definecolor{lightpink}{rgb}{1,0.714,0.757}
\definecolor{lightpink1}{rgb}{1,0.682,0.725}
\definecolor{lightpink2}{rgb}{0.933,0.635,0.678}
\definecolor{lightpink3}{rgb}{0.804,0.549,0.584}
\definecolor{lightpink4}{rgb}{0.545,0.373,0.396}
\definecolor{lightsalmon}{rgb}{1,0.627,0.478}
\definecolor{lightsalmon1}{rgb}{1,0.627,0.478}
\definecolor{lightsalmon2}{rgb}{0.933,0.584,0.447}
\definecolor{lightsalmon3}{rgb}{0.804,0.506,0.384}
\definecolor{lightsalmon4}{rgb}{0.545,0.341,0.259}
\definecolor{lightseagreen}{rgb}{0.125,0.698,0.667}
\definecolor{lightskyblue}{rgb}{0.529,0.808,0.98}
\definecolor{lightskyblue1}{rgb}{0.69,0.886,1}
\definecolor{lightskyblue2}{rgb}{0.643,0.827,0.933}
\definecolor{lightskyblue3}{rgb}{0.553,0.714,0.804}
\definecolor{lightskyblue4}{rgb}{0.376,0.482,0.545}
\definecolor{lightslateblue}{rgb}{0.518,0.439,1}
\definecolor{lightslategray}{rgb}{0.467,0.533,0.6}
\definecolor{lightslategrey}{rgb}{0.467,0.533,0.6}
\definecolor{lightsteelblue}{rgb}{0.69,0.769,0.871}
\definecolor{lightsteelblue1}{rgb}{0.792,0.882,1}
\definecolor{lightsteelblue2}{rgb}{0.737,0.824,0.933}
\definecolor{lightsteelblue3}{rgb}{0.635,0.71,0.804}
\definecolor{lightsteelblue4}{rgb}{0.431,0.482,0.545}
\definecolor{lightyellow1}{rgb}{1,1,0.878}
\definecolor{lightyellow2}{rgb}{0.933,0.933,0.82}
\definecolor{lightyellow3}{rgb}{0.804,0.804,0.706}
\definecolor{lightyellow4}{rgb}{0.545,0.545,0.478}
\definecolor{limegreen}{rgb}{0.196,0.804,0.196}
\definecolor{linen}{rgb}{0.98,0.941,0.902}
\definecolor{magenta}{rgb}{1,0,1}
\definecolor{magenta1}{rgb}{1,0,1}
\definecolor{magenta2}{rgb}{0.933,0,0.933}
\definecolor{magenta3}{rgb}{0.804,0,0.804}
\definecolor{magenta4}{rgb}{0.545,0,0.545}
\definecolor{maroon}{rgb}{0.69,0.188,0.376}
\definecolor{maroon1}{rgb}{1,0.204,0.702}
\definecolor{maroon2}{rgb}{0.933,0.188,0.655}
\definecolor{maroon3}{rgb}{0.804,0.161,0.565}
\definecolor{maroon4}{rgb}{0.545,0.11,0.384}
\definecolor{mediumaquamarine}{rgb}{0.4,0.804,0.667}
\definecolor{mediumblue}{rgb}{0,0,0.804}
\definecolor{mediumorchid}{rgb}{0.729,0.333,0.827}
\definecolor{mediumorchid1}{rgb}{0.878,0.4,1}
\definecolor{mediumorchid2}{rgb}{0.82,0.373,0.933}
\definecolor{mediumorchid3}{rgb}{0.706,0.322,0.804}
\definecolor{mediumorchid4}{rgb}{0.478,0.216,0.545}
\definecolor{mediumpurple}{rgb}{0.576,0.439,0.859}
\definecolor{mediumpurple1}{rgb}{0.671,0.51,1}
\definecolor{mediumpurple2}{rgb}{0.624,0.475,0.933}
\definecolor{mediumpurple3}{rgb}{0.537,0.408,0.804}
\definecolor{mediumpurple4}{rgb}{0.365,0.278,0.545}
\definecolor{mediumseagreen}{rgb}{0.235,0.702,0.443}
\definecolor{mediumslateblue}{rgb}{0.482,0.408,0.933}
\definecolor{mediumspringgreen}{rgb}{0,0.98,0.604}
\definecolor{mediumturquoise}{rgb}{0.282,0.82,0.8}
\definecolor{mediumvioletred}{rgb}{0.78,0.082,0.522}
\definecolor{midnightblue}{rgb}{0.098,0.098,0.439}
\definecolor{mintcream}{rgb}{0.961,1,0.98}
\definecolor{mistyrose}{rgb}{1,0.894,0.882}
\definecolor{mistyrose1}{rgb}{1,0.894,0.882}
\definecolor{mistyrose2}{rgb}{0.933,0.835,0.824}
\definecolor{mistyrose3}{rgb}{0.804,0.718,0.71}
\definecolor{mistyrose4}{rgb}{0.545,0.49,0.482}
\definecolor{moccasin}{rgb}{1,0.894,0.71}
\definecolor{navajowhite}{rgb}{1,0.871,0.678}
\definecolor{navajowhite1}{rgb}{1,0.871,0.678}
\definecolor{navajowhite2}{rgb}{0.933,0.812,0.631}
\definecolor{navajowhite3}{rgb}{0.804,0.702,0.545}
\definecolor{navajowhite4}{rgb}{0.545,0.475,0.369}
\definecolor{navyblue}{rgb}{0,0,0.502}
\definecolor{oldlace}{rgb}{0.992,0.961,0.902}
\definecolor{olivedrab}{rgb}{0.42,0.557,0.137}
\definecolor{olivedrab1}{rgb}{0.753,1,0.243}
\definecolor{olivedrab2}{rgb}{0.702,0.933,0.227}
\definecolor{olivedrab3}{rgb}{0.604,0.804,0.196}
\definecolor{olivedrab4}{rgb}{0.412,0.545,0.133}
\definecolor{orange1}{rgb}{1,0.647,0}
\definecolor{orange2}{rgb}{0.933,0.604,0}
\definecolor{orange3}{rgb}{0.804,0.522,0}
\definecolor{orange4}{rgb}{0.545,0.353,0}
\definecolor{orangered}{rgb}{1,0.271,0}
\definecolor{orangered1}{rgb}{1,0.271,0}
\definecolor{orangered2}{rgb}{0.933,0.251,0}
\definecolor{orangered3}{rgb}{0.804,0.216,0}
\definecolor{orangered4}{rgb}{0.545,0.145,0}
\definecolor{orchid}{rgb}{0.855,0.439,0.839}
\definecolor{orchid1}{rgb}{1,0.514,0.98}
\definecolor{orchid2}{rgb}{0.933,0.478,0.914}
\definecolor{orchid3}{rgb}{0.804,0.412,0.788}
\definecolor{orchid4}{rgb}{0.545,0.278,0.537}
\definecolor{palegoldenrod}{rgb}{0.933,0.91,0.667}
\definecolor{palegreen}{rgb}{0.596,0.984,0.596}
\definecolor{palegreen1}{rgb}{0.604,1,0.604}
\definecolor{palegreen2}{rgb}{0.565,0.933,0.565}
\definecolor{palegreen3}{rgb}{0.486,0.804,0.486}
\definecolor{palegreen4}{rgb}{0.329,0.545,0.329}
\definecolor{paleturquoise}{rgb}{0.686,0.933,0.933}
\definecolor{paleturquoise1}{rgb}{0.733,1,1}
\definecolor{paleturquoise2}{rgb}{0.682,0.933,0.933}
\definecolor{paleturquoise3}{rgb}{0.588,0.804,0.804}
\definecolor{paleturquoise4}{rgb}{0.4,0.545,0.545}
\definecolor{palevioletred}{rgb}{0.859,0.439,0.576}
\definecolor{palevioletred1}{rgb}{1,0.51,0.671}
\definecolor{palevioletred2}{rgb}{0.933,0.475,0.624}
\definecolor{palevioletred3}{rgb}{0.804,0.408,0.537}
\definecolor{palevioletred4}{rgb}{0.545,0.278,0.365}
\definecolor{papayawhip}{rgb}{1,0.937,0.835}
\definecolor{peachpuff}{rgb}{1,0.855,0.725}
\definecolor{peachpuff1}{rgb}{1,0.855,0.725}
\definecolor{peachpuff2}{rgb}{0.933,0.796,0.678}
\definecolor{peachpuff3}{rgb}{0.804,0.686,0.584}
\definecolor{peachpuff4}{rgb}{0.545,0.467,0.396}
\definecolor{peru}{rgb}{0.804,0.522,0.247}
\definecolor{pink1}{rgb}{1,0.71,0.773}
\definecolor{pink2}{rgb}{0.933,0.663,0.722}
\definecolor{pink3}{rgb}{0.804,0.569,0.62}
\definecolor{pink4}{rgb}{0.545,0.388,0.424}
\definecolor{plum}{rgb}{0.867,0.627,0.867}
\definecolor{plum1}{rgb}{1,0.733,1}
\definecolor{plum2}{rgb}{0.933,0.682,0.933}
\definecolor{plum3}{rgb}{0.804,0.588,0.804}
\definecolor{plum4}{rgb}{0.545,0.4,0.545}
\definecolor{powderblue}{rgb}{0.69,0.878,0.902}
\definecolor{purple1}{rgb}{0.608,0.188,1}
\definecolor{purple2}{rgb}{0.569,0.173,0.933}
\definecolor{purple3}{rgb}{0.49,0.149,0.804}
\definecolor{purple4}{rgb}{0.333,0.102,0.545}
\definecolor{red1}{rgb}{1,0,0}
\definecolor{red2}{rgb}{0.933,0,0}
\definecolor{red3}{rgb}{0.804,0,0}
\definecolor{red4}{rgb}{0.545,0,0}
\definecolor{rosybrown}{rgb}{0.737,0.561,0.561}
\definecolor{rosybrown1}{rgb}{1,0.757,0.757}
\definecolor{rosybrown2}{rgb}{0.933,0.706,0.706}
\definecolor{rosybrown3}{rgb}{0.804,0.608,0.608}
\definecolor{rosybrown4}{rgb}{0.545,0.412,0.412}
\definecolor{royalblue}{rgb}{0.255,0.412,0.882}
\definecolor{royalblue1}{rgb}{0.282,0.463,1}
\definecolor{royalblue2}{rgb}{0.263,0.431,0.933}
\definecolor{royalblue3}{rgb}{0.227,0.373,0.804}
\definecolor{royalblue4}{rgb}{0.153,0.251,0.545}
\definecolor{saddlebrown}{rgb}{0.545,0.271,0.075}
\definecolor{salmon}{rgb}{0.98,0.502,0.447}
\definecolor{salmon1}{rgb}{1,0.549,0.412}
\definecolor{salmon2}{rgb}{0.933,0.51,0.384}
\definecolor{salmon3}{rgb}{0.804,0.439,0.329}
\definecolor{salmon4}{rgb}{0.545,0.298,0.224}
\definecolor{sandybrown}{rgb}{0.957,0.643,0.376}
\definecolor{seagreen1}{rgb}{0.329,1,0.624}
\definecolor{seagreen2}{rgb}{0.306,0.933,0.58}
\definecolor{seagreen3}{rgb}{0.263,0.804,0.502}
\definecolor{seagreen4}{rgb}{0.18,0.545,0.341}
\definecolor{seashell}{rgb}{1,0.961,0.933}
\definecolor{seashell1}{rgb}{1,0.961,0.933}
\definecolor{seashell2}{rgb}{0.933,0.898,0.871}
\definecolor{seashell3}{rgb}{0.804,0.773,0.749}
\definecolor{seashell4}{rgb}{0.545,0.525,0.51}
\definecolor{sienna}{rgb}{0.627,0.322,0.176}
\definecolor{sienna1}{rgb}{1,0.51,0.278}
\definecolor{sienna2}{rgb}{0.933,0.475,0.259}
\definecolor{sienna3}{rgb}{0.804,0.408,0.224}
\definecolor{sienna4}{rgb}{0.545,0.278,0.149}
\definecolor{skyblue}{rgb}{0.529,0.808,0.922}
\definecolor{skyblue1}{rgb}{0.529,0.808,1}
\definecolor{skyblue2}{rgb}{0.494,0.753,0.933}
\definecolor{skyblue3}{rgb}{0.424,0.651,0.804}
\definecolor{skyblue4}{rgb}{0.29,0.439,0.545}
\definecolor{slateblue}{rgb}{0.416,0.353,0.804}
\definecolor{slateblue1}{rgb}{0.514,0.435,1}
\definecolor{slateblue2}{rgb}{0.478,0.404,0.933}
\definecolor{slateblue3}{rgb}{0.412,0.349,0.804}
\definecolor{slateblue4}{rgb}{0.278,0.235,0.545}
\definecolor{slategray}{rgb}{0.439,0.502,0.565}
\definecolor{slategray1}{rgb}{0.776,0.886,1}
\definecolor{slategray2}{rgb}{0.725,0.827,0.933}
\definecolor{slategray3}{rgb}{0.624,0.714,0.804}
\definecolor{slategray4}{rgb}{0.424,0.482,0.545}
\definecolor{slategrey}{rgb}{0.439,0.502,0.565}
\definecolor{snow}{rgb}{1,0.98,0.98}
\definecolor{snow1}{rgb}{1,0.98,0.98}
\definecolor{snow2}{rgb}{0.933,0.914,0.914}
\definecolor{snow3}{rgb}{0.804,0.788,0.788}
\definecolor{snow4}{rgb}{0.545,0.537,0.537}
\definecolor{springgreen}{rgb}{0,1,0.498}
\definecolor{springgreen1}{rgb}{0,1,0.498}
\definecolor{springgreen2}{rgb}{0,0.933,0.463}
\definecolor{springgreen3}{rgb}{0,0.804,0.4}
\definecolor{springgreen4}{rgb}{0,0.545,0.271}
\definecolor{steelblue}{rgb}{0.275,0.51,0.706}
\definecolor{steelblue1}{rgb}{0.388,0.722,1}
\definecolor{steelblue2}{rgb}{0.361,0.675,0.933}
\definecolor{steelblue3}{rgb}{0.31,0.58,0.804}
\definecolor{steelblue4}{rgb}{0.212,0.392,0.545}
\definecolor{tan}{rgb}{0.824,0.706,0.549}
\definecolor{tan1}{rgb}{1,0.647,0.31}
\definecolor{tan2}{rgb}{0.933,0.604,0.286}
\definecolor{tan3}{rgb}{0.804,0.522,0.247}
\definecolor{tan4}{rgb}{0.545,0.353,0.169}
\definecolor{thistle}{rgb}{0.847,0.749,0.847}
\definecolor{thistle1}{rgb}{1,0.882,1}
\definecolor{thistle2}{rgb}{0.933,0.824,0.933}
\definecolor{thistle3}{rgb}{0.804,0.71,0.804}
\definecolor{thistle4}{rgb}{0.545,0.482,0.545}
\definecolor{tomato}{rgb}{1,0.388,0.278}
\definecolor{tomato1}{rgb}{1,0.388,0.278}
\definecolor{tomato2}{rgb}{0.933,0.361,0.259}
\definecolor{tomato3}{rgb}{0.804,0.31,0.224}
\definecolor{tomato4}{rgb}{0.545,0.212,0.149}
\definecolor{turquoise1}{rgb}{0,0.961,1}
\definecolor{turquoise2}{rgb}{0,0.898,0.933}
\definecolor{turquoise3}{rgb}{0,0.773,0.804}
\definecolor{turquoise4}{rgb}{0,0.525,0.545}
\definecolor{violetred}{rgb}{0.816,0.125,0.565}
\definecolor{violetred1}{rgb}{1,0.243,0.588}
\definecolor{violetred2}{rgb}{0.933,0.227,0.549}
\definecolor{violetred3}{rgb}{0.804,0.196,0.471}
\definecolor{violetred4}{rgb}{0.545,0.133,0.322}
\definecolor{wheat}{rgb}{0.961,0.871,0.702}
\definecolor{wheat1}{rgb}{1,0.906,0.729}
\definecolor{wheat2}{rgb}{0.933,0.847,0.682}
\definecolor{wheat3}{rgb}{0.804,0.729,0.588}
\definecolor{wheat4}{rgb}{0.545,0.494,0.4}
\definecolor{whitesmoke}{rgb}{0.961,0.961,0.961}
\definecolor{yellow1}{rgb}{1,1,0}
\definecolor{yellow2}{rgb}{0.933,0.933,0}
\definecolor{yellow3}{rgb}{0.804,0.804,0}
\definecolor{yellow4}{rgb}{0.545,0.545,0}
\definecolor{yellowgreen}{rgb}{0.604,0.804,0.196}
\begin{tikzpicture}[ipe stylesheet]
  \filldraw[seagreen, ipe opacity 30]
    (94.6667, 712.6667)
     .. controls (96, 702.6667) and (100, 697.3333) .. (108, 694.6667)
     .. controls (116, 692) and (128, 692) .. (132.6667, 702)
     .. controls (137.3333, 712) and (134.6667, 732) .. (127.3333, 742)
     .. controls (120, 752) and (108, 752) .. (101.3333, 744.6667)
     .. controls (94.6667, 737.3333) and (93.3333, 722.6667) .. cycle;
  \filldraw[draw=darkgreen, fill=green, ipe opacity 30]
    (94.6667, 712.6667)
     .. controls (96, 702.6667) and (100, 697.3333) .. (108, 694.6667)
     .. controls (116, 692) and (128, 692) .. (132.6667, 702)
     .. controls (137.3333, 712) and (134.6667, 732) .. (127.3333, 742)
     .. controls (120, 752) and (108, 752) .. (101.3333, 744.6667)
     .. controls (94.6667, 737.3333) and (93.3333, 722.6667) .. cycle;
  \filldraw[draw=navy, fill=royalblue]
    (94.6667, 712.6667)
     .. controls (96, 702.6667) and (100, 697.3333) .. (108, 694.6667)
     .. controls (116, 692) and (128, 692) .. (132.6667, 702)
     .. controls (137.3333, 712) and (134.6667, 732) .. (127.3333, 742)
     .. controls (120, 752) and (108, 752) .. (101.3333, 744.6667)
     .. controls (94.6667, 737.3333) and (93.3333, 722.6667) .. cycle;
  \filldraw[draw=darkred, fill=orangered]
    (168, 724.6667)
     .. controls (168, 713.3333) and (168, 698.6667) .. (174, 694)
     .. controls (180, 689.3333) and (192, 694.6667) .. (198.6667, 706)
     .. controls (205.3333, 717.3333) and (206.6667, 734.6667) .. (200.6667, 743.3333)
     .. controls (194.6667, 752) and (181.3333, 752) .. (174.6667, 748)
     .. controls (168, 744) and (168, 736) .. cycle;
  \filldraw[draw=darkred, fill=khaki1]
    (244, 741.3333)
     .. controls (240, 730.6667) and (240, 709.3333) .. (244, 698.6667)
     .. controls (248, 688) and (256, 688) .. (260, 698.6667)
     .. controls (264, 709.3333) and (264, 730.6667) .. (260, 741.3333)
     .. controls (256, 752) and (248, 752) .. cycle;
  \draw
    (124, 736)
     -- (176, 740);
  \draw
    (124, 720)
     -- (176, 720);
  \draw
    (124, 704)
     -- (176, 700);
  \draw
    (124, 736)
     -- (104, 736);
  \draw
    (104, 736)
     -- (124, 720);
  \draw
    (124, 720)
     -- (104, 704);
  \draw
    (104, 704)
     -- (104, 736);
  \draw
    (124, 704)
     -- (124, 720);
  \draw
    (176, 700)
     -- (196, 724);
  \draw
    (196, 724)
     -- (176, 720);
  \draw
    (176, 740)
     -- (196, 740);
  \draw
    (176, 740)
     -- (196, 724);
  \draw
    (176, 720)
     -- (252, 700);
  \draw
    (196, 724)
     -- (252, 720);
  \draw
    (176, 700)
     -- (252, 720);
  \draw
    (252, 740)
     -- (196, 724);
  \pic
     at (104, 736) {ipe disk};
  \pic
     at (124, 736) {ipe disk};
  \pic
     at (124, 720) {ipe disk};
  \pic
     at (124, 704) {ipe disk};
  \pic
     at (104, 704) {ipe disk};
  \pic
     at (176, 720) {ipe disk};
  \pic
     at (196, 740) {ipe disk};
  \pic
     at (176, 740) {ipe disk};
  \pic
     at (196, 724) {ipe disk};
  \pic
     at (176, 700) {ipe disk};
  \pic
     at (252, 740) {ipe disk};
  \pic
     at (252, 720) {ipe disk};
  \pic
     at (252, 700) {ipe disk};
  \node[ipe node]
     at (112, 680) {$V_1$};
  \node[ipe node]
     at (180, 680) {$V_2$};
  \node[ipe node]
     at (248, 680) {$V_3$};
  \draw
    (196, 740)
     -- (196, 724);
  \draw
    (176, 740)
     -- (176, 720);
  \draw
    (176, 720)
     -- (176, 700);
  \draw
    (176, 720)
     -- (196, 740);
\end{tikzpicture}

\caption{An example of a nice $3$-partition $(V_1, V_2, V_3)$ of a graph.}
\labels{figure_partition}
\end{figure}

For every $\ell∈\mathbb{N},$ let
\begin{eqnarray*}
{\cal H}^{(\ell)} = \{ (H,{\cal V}) \mid H\mbox{~is a graph on $\ell$ vertices and ${\cal V}$ is a nice 3-partition of $H$}\}.
\end{eqnarray*}
Let $G$ and $H$ be two graphs and ${\cal V} = (V_1, V_2, V_3)$ and ${\cal U} = (U_1, U_2, U_3)$
be nice 3-partitions of $G$ and $H,$ respectively.
We say that $G$ is {\em strongly isomorphic to $H$ with respect to $({\cal V}, {\cal U})$}, if $G$ is isomorphic to $H,$ $G[V_1\cup V_2]$ is isomorphic to $H[U_1\cup U_2],$ $G[V_2\cup V_3]$ is isomorphic to $H[U_2\cup U_3],$ and these two last isomorphisms are identical when restricted to $V_2.$

Let us exlain why we introduce strongly isomorphic graphs.
Having defined the structure $\mathfrak{A}^{(d,Z,L,F)},$ we aim to define a boundaried structure that we will associate with a representative of $θ^{\sf out}_q.$
The boundary of our boundaried structure will be the set $\partial_{\mathfrak{K}}(Z)\cup V(F).$ By definition, the graph induced by this set has an obvious nice $3$-partition (since there is no edge between $F'$ and $\partial_{\mathfrak{K}} (Z)$).
The information we want to store is not just a boundary but the ``inner-structure'' of this boundary, which is mirrored by the nice 3-partition. We demand this ``stronger'' notion of isomorphism to be able to find another boundaried structure that corresponds to the same representative of $θ^{\sf out}_q$ and still its boundary is ``nicely 3-partitioned'' in the same way as the boundary of the initial boundaried structure, since (as we will see later in the course of the proof) the set $V(F)$ remains ``invariant'' no matter of which flatness pair the extended compass we consider - thus, our isomorphism needs to keep $V(F)$ ``intact''.

\myskip\paragraph{The out-signature of an extended compass.}
We now define the out-signature of an extended compass.
We encode all possible sets $F∈{\cal F}^{V_L ({\bf a})}_{i - |\partial_{\mathfrak{K}}(Z)|},$ where $i∈[0,\tw(θ)-1],$ and all representatives $\bar{φ}$ of $θ^{\sf out}_q$ such that when extending $\partial_{\mathfrak{K}}(Z)\cup V_L ({\bf a})$ to $\partial_{\mathfrak{K}}(Z)\cup V(F),$
the boundaried structure obtained from
$\mathfrak{A}^{(d,Z,L,F)}$
after considering $\partial_{\mathfrak{K}}(Z)\cup V(F)$
as its boundary,
satisfies $\bar{φ}.$

We set $τ':=τ\cup{\bf Q}\cup\{{\sf R},{\sf X}\}\cup{\bf c}$ and, for every $\ell∈ [0,\tw(θ)-1],$
following~\autoref{cou_more},
we consider the collection ${\sf rep}_{τ'}^{(\ell)}(θ^{\sf out}_q)$ of sentences on $\ell$-boundaried $τ'$-structures that are ``representatives'' of the sentence $θ^{\sf out}_q$ (that is a sentence in $\MSOL[τ']$).

We set $$\blue{{\sf SIG}_{\sf out}}:=\{({\bf H},\bar{φ})\mid \exists \ell∈[0,\tw(θ)-1] \mbox{\rm ~such that ${\bf H}∈ {\cal H}^{(\ell)}$~and~}\bar{φ}∈ {{\sf rep}^{(\ell)}_{τ'}(θ^{\sf out}_q)}\}.$$

Let $\mathfrak{K}=(\mathfrak{A}[V(K^{\bf a})],{\bf a}, {\bf I}, {\bf W}_{{q}})$ be  the extended compass of a flatness pair $(W,\mathfrak{R})$ of $G\setminus V({\bf a})$ of height $2w+j,$  $R\subseteq V(K^{\bf a}),$ $d∈[r,w],$ $L\subseteq [l],$ and  $Z\subseteq I^{(d-r+1)}.$

We define

\begin{figure}[ht]
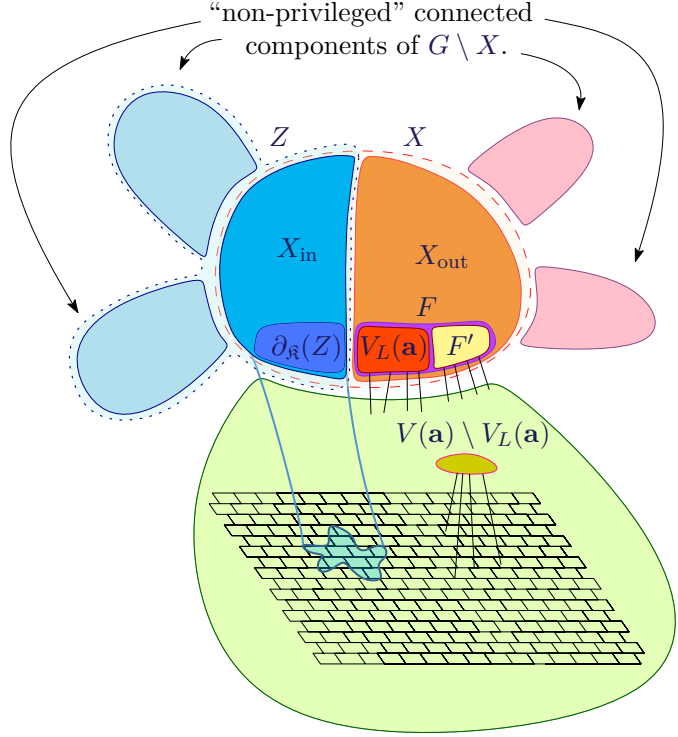

	\begin{center}
\tikzstyle{ipe stylesheet} = [
  ipe import,
  even odd rule,
  line join=round,
  line cap=butt,
  ipe pen normal/.style={line width=0.4},
  ipe pen heavier/.style={line width=0.8},
  ipe pen fat/.style={line width=1.2},
  ipe pen ultrafat/.style={line width=2},
  ipe pen normal,
  ipe mark normal/.style={ipe mark scale=3},
  ipe mark large/.style={ipe mark scale=5},
  ipe mark small/.style={ipe mark scale=2},
  ipe mark tiny/.style={ipe mark scale=1.1},
  ipe mark normal,
  /pgf/arrow keys/.cd,
  ipe arrow normal/.style={scale=7},
  ipe arrow large/.style={scale=10},
  ipe arrow small/.style={scale=5},
  ipe arrow tiny/.style={scale=3},
  ipe arrow normal,
  /tikz/.cd,
  ipe arrows, 
  <->/.tip = ipe normal,
  ipe dash normal/.style={dash pattern=},
  ipe dash dotted/.style={dash pattern=on 1bp off 3bp},
  ipe dash dashed/.style={dash pattern=on 4bp off 4bp},
  ipe dash dash dotted/.style={dash pattern=on 4bp off 2bp on 1bp off 2bp},
  ipe dash dash dot dotted/.style={dash pattern=on 4bp off 2bp on 1bp off 2bp on 1bp off 2bp},
  ipe dash normal,
  ipe node/.append style={font=\normalsize},
  ipe stretch normal/.style={ipe node stretch=1},
  ipe stretch normal,
  ipe opacity 10/.style={opacity=0.1},
  ipe opacity 30/.style={opacity=0.3},
  ipe opacity 50/.style={opacity=0.5},
  ipe opacity 75/.style={opacity=0.75},
  ipe opacity opaque/.style={opacity=1},
  ipe opacity opaque,
]
\definecolor{black}{rgb}{0,0,0}
\definecolor{white}{rgb}{1,1,1}
\definecolor{red}{rgb}{1,0,0}
\definecolor{blue}{rgb}{0,0,1}
\definecolor{green}{rgb}{0,1,0}
\definecolor{yellow}{rgb}{1,1,0}
\definecolor{orange}{rgb}{1,0.647,0}
\definecolor{gold}{rgb}{1,0.843,0}
\definecolor{purple}{rgb}{0.627,0.125,0.941}
\definecolor{gray}{rgb}{0.745,0.745,0.745}
\definecolor{brown}{rgb}{0.647,0.165,0.165}
\definecolor{navy}{rgb}{0,0,0.502}
\definecolor{pink}{rgb}{1,0.753,0.796}
\definecolor{seagreen}{rgb}{0.18,0.545,0.341}
\definecolor{turquoise}{rgb}{0.251,0.878,0.816}
\definecolor{violet}{rgb}{0.933,0.51,0.933}
\definecolor{darkblue}{rgb}{0,0,0.545}
\definecolor{darkcyan}{rgb}{0,0.545,0.545}
\definecolor{darkgray}{rgb}{0.663,0.663,0.663}
\definecolor{darkgreen}{rgb}{0,0.392,0}
\definecolor{darkmagenta}{rgb}{0.545,0,0.545}
\definecolor{darkorange}{rgb}{1,0.549,0}
\definecolor{darkred}{rgb}{0.545,0,0}
\definecolor{lightblue}{rgb}{0.678,0.847,0.902}
\definecolor{lightcyan}{rgb}{0.878,1,1}
\definecolor{lightgray}{rgb}{0.827,0.827,0.827}
\definecolor{lightgreen}{rgb}{0.565,0.933,0.565}
\definecolor{lightyellow}{rgb}{1,1,0.878}
\definecolor{aliceblue}{rgb}{0.941,0.973,1}
\definecolor{antiquewhite}{rgb}{0.98,0.922,0.843}
\definecolor{antiquewhite1}{rgb}{1,0.937,0.859}
\definecolor{antiquewhite2}{rgb}{0.933,0.875,0.8}
\definecolor{antiquewhite3}{rgb}{0.804,0.753,0.69}
\definecolor{antiquewhite4}{rgb}{0.545,0.514,0.471}
\definecolor{aquamarine}{rgb}{0.498,1,0.831}
\definecolor{aquamarine1}{rgb}{0.498,1,0.831}
\definecolor{aquamarine2}{rgb}{0.463,0.933,0.776}
\definecolor{aquamarine3}{rgb}{0.4,0.804,0.667}
\definecolor{aquamarine4}{rgb}{0.271,0.545,0.455}
\definecolor{azure}{rgb}{0.941,1,1}
\definecolor{azure1}{rgb}{0.941,1,1}
\definecolor{azure2}{rgb}{0.878,0.933,0.933}
\definecolor{azure3}{rgb}{0.757,0.804,0.804}
\definecolor{azure4}{rgb}{0.514,0.545,0.545}
\definecolor{beige}{rgb}{0.961,0.961,0.863}
\definecolor{bisque}{rgb}{1,0.894,0.769}
\definecolor{bisque1}{rgb}{1,0.894,0.769}
\definecolor{bisque2}{rgb}{0.933,0.835,0.718}
\definecolor{bisque3}{rgb}{0.804,0.718,0.62}
\definecolor{bisque4}{rgb}{0.545,0.49,0.42}
\definecolor{blanchedalmond}{rgb}{1,0.922,0.804}
\definecolor{blue1}{rgb}{0,0,1}
\definecolor{blue2}{rgb}{0,0,0.933}
\definecolor{blue3}{rgb}{0,0,0.804}
\definecolor{blue4}{rgb}{0,0,0.545}
\definecolor{blueviolet}{rgb}{0.541,0.169,0.886}
\definecolor{brown1}{rgb}{1,0.251,0.251}
\definecolor{brown2}{rgb}{0.933,0.231,0.231}
\definecolor{brown3}{rgb}{0.804,0.2,0.2}
\definecolor{brown4}{rgb}{0.545,0.137,0.137}
\definecolor{burlywood}{rgb}{0.871,0.722,0.529}
\definecolor{burlywood1}{rgb}{1,0.827,0.608}
\definecolor{burlywood2}{rgb}{0.933,0.773,0.569}
\definecolor{burlywood3}{rgb}{0.804,0.667,0.49}
\definecolor{burlywood4}{rgb}{0.545,0.451,0.333}
\definecolor{cadetblue}{rgb}{0.373,0.62,0.627}
\definecolor{cadetblue1}{rgb}{0.596,0.961,1}
\definecolor{cadetblue2}{rgb}{0.557,0.898,0.933}
\definecolor{cadetblue3}{rgb}{0.478,0.773,0.804}
\definecolor{cadetblue4}{rgb}{0.325,0.525,0.545}
\definecolor{chartreuse}{rgb}{0.498,1,0}
\definecolor{chartreuse1}{rgb}{0.498,1,0}
\definecolor{chartreuse2}{rgb}{0.463,0.933,0}
\definecolor{chartreuse3}{rgb}{0.4,0.804,0}
\definecolor{chartreuse4}{rgb}{0.271,0.545,0}
\definecolor{chocolate}{rgb}{0.824,0.412,0.118}
\definecolor{chocolate1}{rgb}{1,0.498,0.141}
\definecolor{chocolate2}{rgb}{0.933,0.463,0.129}
\definecolor{chocolate3}{rgb}{0.804,0.4,0.114}
\definecolor{chocolate4}{rgb}{0.545,0.271,0.075}
\definecolor{coral}{rgb}{1,0.498,0.314}
\definecolor{coral1}{rgb}{1,0.447,0.337}
\definecolor{coral2}{rgb}{0.933,0.416,0.314}
\definecolor{coral3}{rgb}{0.804,0.357,0.271}
\definecolor{coral4}{rgb}{0.545,0.243,0.184}
\definecolor{cornflowerblue}{rgb}{0.392,0.584,0.929}
\definecolor{cornsilk}{rgb}{1,0.973,0.863}
\definecolor{cornsilk1}{rgb}{1,0.973,0.863}
\definecolor{cornsilk2}{rgb}{0.933,0.91,0.804}
\definecolor{cornsilk3}{rgb}{0.804,0.784,0.694}
\definecolor{cornsilk4}{rgb}{0.545,0.533,0.471}
\definecolor{cyan}{rgb}{0,1,1}
\definecolor{cyan1}{rgb}{0,1,1}
\definecolor{cyan2}{rgb}{0,0.933,0.933}
\definecolor{cyan3}{rgb}{0,0.804,0.804}
\definecolor{cyan4}{rgb}{0,0.545,0.545}
\definecolor{darkgoldenrod}{rgb}{0.722,0.525,0.043}
\definecolor{darkgoldenrod1}{rgb}{1,0.725,0.059}
\definecolor{darkgoldenrod2}{rgb}{0.933,0.678,0.055}
\definecolor{darkgoldenrod3}{rgb}{0.804,0.584,0.047}
\definecolor{darkgoldenrod4}{rgb}{0.545,0.396,0.031}
\definecolor{darkgrey}{rgb}{0.663,0.663,0.663}
\definecolor{darkkhaki}{rgb}{0.741,0.718,0.42}
\definecolor{darkolivegreen}{rgb}{0.333,0.42,0.184}
\definecolor{darkolivegreen1}{rgb}{0.792,1,0.439}
\definecolor{darkolivegreen2}{rgb}{0.737,0.933,0.408}
\definecolor{darkolivegreen3}{rgb}{0.635,0.804,0.353}
\definecolor{darkolivegreen4}{rgb}{0.431,0.545,0.239}
\definecolor{darkorange1}{rgb}{1,0.498,0}
\definecolor{darkorange2}{rgb}{0.933,0.463,0}
\definecolor{darkorange3}{rgb}{0.804,0.4,0}
\definecolor{darkorange4}{rgb}{0.545,0.271,0}
\definecolor{darkorchid}{rgb}{0.6,0.196,0.8}
\definecolor{darkorchid1}{rgb}{0.749,0.243,1}
\definecolor{darkorchid2}{rgb}{0.698,0.227,0.933}
\definecolor{darkorchid3}{rgb}{0.604,0.196,0.804}
\definecolor{darkorchid4}{rgb}{0.408,0.133,0.545}
\definecolor{darksalmon}{rgb}{0.914,0.588,0.478}
\definecolor{darkseagreen}{rgb}{0.561,0.737,0.561}
\definecolor{darkseagreen1}{rgb}{0.757,1,0.757}
\definecolor{darkseagreen2}{rgb}{0.706,0.933,0.706}
\definecolor{darkseagreen3}{rgb}{0.608,0.804,0.608}
\definecolor{darkseagreen4}{rgb}{0.412,0.545,0.412}
\definecolor{darkslateblue}{rgb}{0.282,0.239,0.545}
\definecolor{darkslategray}{rgb}{0.184,0.31,0.31}
\definecolor{darkslategray1}{rgb}{0.592,1,1}
\definecolor{darkslategray2}{rgb}{0.553,0.933,0.933}
\definecolor{darkslategray3}{rgb}{0.475,0.804,0.804}
\definecolor{darkslategray4}{rgb}{0.322,0.545,0.545}
\definecolor{darkslategrey}{rgb}{0.184,0.31,0.31}
\definecolor{darkturquoise}{rgb}{0,0.808,0.82}
\definecolor{darkviolet}{rgb}{0.58,0,0.827}
\definecolor{deeppink}{rgb}{1,0.078,0.576}
\definecolor{deeppink1}{rgb}{1,0.078,0.576}
\definecolor{deeppink2}{rgb}{0.933,0.071,0.537}
\definecolor{deeppink3}{rgb}{0.804,0.063,0.463}
\definecolor{deeppink4}{rgb}{0.545,0.039,0.314}
\definecolor{deepskyblue}{rgb}{0,0.749,1}
\definecolor{deepskyblue1}{rgb}{0,0.749,1}
\definecolor{deepskyblue2}{rgb}{0,0.698,0.933}
\definecolor{deepskyblue3}{rgb}{0,0.604,0.804}
\definecolor{deepskyblue4}{rgb}{0,0.408,0.545}
\definecolor{dimgray}{rgb}{0.412,0.412,0.412}
\definecolor{dimgrey}{rgb}{0.412,0.412,0.412}
\definecolor{dodgerblue}{rgb}{0.118,0.565,1}
\definecolor{dodgerblue1}{rgb}{0.118,0.565,1}
\definecolor{dodgerblue2}{rgb}{0.11,0.525,0.933}
\definecolor{dodgerblue3}{rgb}{0.094,0.455,0.804}
\definecolor{dodgerblue4}{rgb}{0.063,0.306,0.545}
\definecolor{firebrick}{rgb}{0.698,0.133,0.133}
\definecolor{firebrick1}{rgb}{1,0.188,0.188}
\definecolor{firebrick2}{rgb}{0.933,0.173,0.173}
\definecolor{firebrick3}{rgb}{0.804,0.149,0.149}
\definecolor{firebrick4}{rgb}{0.545,0.102,0.102}
\definecolor{floralwhite}{rgb}{1,0.98,0.941}
\definecolor{forestgreen}{rgb}{0.133,0.545,0.133}
\definecolor{gainsboro}{rgb}{0.863,0.863,0.863}
\definecolor{ghostwhite}{rgb}{0.973,0.973,1}
\definecolor{gold1}{rgb}{1,0.843,0}
\definecolor{gold2}{rgb}{0.933,0.788,0}
\definecolor{gold3}{rgb}{0.804,0.678,0}
\definecolor{gold4}{rgb}{0.545,0.459,0}
\definecolor{goldenrod}{rgb}{0.855,0.647,0.125}
\definecolor{goldenrod1}{rgb}{1,0.757,0.145}
\definecolor{goldenrod2}{rgb}{0.933,0.706,0.133}
\definecolor{goldenrod3}{rgb}{0.804,0.608,0.114}
\definecolor{goldenrod4}{rgb}{0.545,0.412,0.078}
\definecolor{gray0}{rgb}{0,0,0}
\definecolor{gray1}{rgb}{0.012,0.012,0.012}
\definecolor{gray10}{rgb}{0.102,0.102,0.102}
\definecolor{gray100}{rgb}{1,1,1}
\definecolor{gray11}{rgb}{0.11,0.11,0.11}
\definecolor{gray12}{rgb}{0.122,0.122,0.122}
\definecolor{gray13}{rgb}{0.129,0.129,0.129}
\definecolor{gray14}{rgb}{0.141,0.141,0.141}
\definecolor{gray15}{rgb}{0.149,0.149,0.149}
\definecolor{gray16}{rgb}{0.161,0.161,0.161}
\definecolor{gray17}{rgb}{0.169,0.169,0.169}
\definecolor{gray18}{rgb}{0.18,0.18,0.18}
\definecolor{gray19}{rgb}{0.188,0.188,0.188}
\definecolor{gray2}{rgb}{0.02,0.02,0.02}
\definecolor{gray20}{rgb}{0.2,0.2,0.2}
\definecolor{gray21}{rgb}{0.212,0.212,0.212}
\definecolor{gray22}{rgb}{0.22,0.22,0.22}
\definecolor{gray23}{rgb}{0.231,0.231,0.231}
\definecolor{gray24}{rgb}{0.239,0.239,0.239}
\definecolor{gray25}{rgb}{0.251,0.251,0.251}
\definecolor{gray26}{rgb}{0.259,0.259,0.259}
\definecolor{gray27}{rgb}{0.271,0.271,0.271}
\definecolor{gray28}{rgb}{0.278,0.278,0.278}
\definecolor{gray29}{rgb}{0.29,0.29,0.29}
\definecolor{gray3}{rgb}{0.031,0.031,0.031}
\definecolor{gray30}{rgb}{0.302,0.302,0.302}
\definecolor{gray31}{rgb}{0.31,0.31,0.31}
\definecolor{gray32}{rgb}{0.322,0.322,0.322}
\definecolor{gray33}{rgb}{0.329,0.329,0.329}
\definecolor{gray34}{rgb}{0.341,0.341,0.341}
\definecolor{gray35}{rgb}{0.349,0.349,0.349}
\definecolor{gray36}{rgb}{0.361,0.361,0.361}
\definecolor{gray37}{rgb}{0.369,0.369,0.369}
\definecolor{gray38}{rgb}{0.38,0.38,0.38}
\definecolor{gray39}{rgb}{0.388,0.388,0.388}
\definecolor{gray4}{rgb}{0.039,0.039,0.039}
\definecolor{gray40}{rgb}{0.4,0.4,0.4}
\definecolor{gray41}{rgb}{0.412,0.412,0.412}
\definecolor{gray42}{rgb}{0.42,0.42,0.42}
\definecolor{gray43}{rgb}{0.431,0.431,0.431}
\definecolor{gray44}{rgb}{0.439,0.439,0.439}
\definecolor{gray45}{rgb}{0.451,0.451,0.451}
\definecolor{gray46}{rgb}{0.459,0.459,0.459}
\definecolor{gray47}{rgb}{0.471,0.471,0.471}
\definecolor{gray48}{rgb}{0.478,0.478,0.478}
\definecolor{gray49}{rgb}{0.49,0.49,0.49}
\definecolor{gray5}{rgb}{0.051,0.051,0.051}
\definecolor{gray50}{rgb}{0.498,0.498,0.498}
\definecolor{gray51}{rgb}{0.51,0.51,0.51}
\definecolor{gray52}{rgb}{0.522,0.522,0.522}
\definecolor{gray53}{rgb}{0.529,0.529,0.529}
\definecolor{gray54}{rgb}{0.541,0.541,0.541}
\definecolor{gray55}{rgb}{0.549,0.549,0.549}
\definecolor{gray56}{rgb}{0.561,0.561,0.561}
\definecolor{gray57}{rgb}{0.569,0.569,0.569}
\definecolor{gray58}{rgb}{0.58,0.58,0.58}
\definecolor{gray59}{rgb}{0.588,0.588,0.588}
\definecolor{gray6}{rgb}{0.059,0.059,0.059}
\definecolor{gray60}{rgb}{0.6,0.6,0.6}
\definecolor{gray61}{rgb}{0.612,0.612,0.612}
\definecolor{gray62}{rgb}{0.62,0.62,0.62}
\definecolor{gray63}{rgb}{0.631,0.631,0.631}
\definecolor{gray64}{rgb}{0.639,0.639,0.639}
\definecolor{gray65}{rgb}{0.651,0.651,0.651}
\definecolor{gray66}{rgb}{0.659,0.659,0.659}
\definecolor{gray67}{rgb}{0.671,0.671,0.671}
\definecolor{gray68}{rgb}{0.678,0.678,0.678}
\definecolor{gray69}{rgb}{0.69,0.69,0.69}
\definecolor{gray7}{rgb}{0.071,0.071,0.071}
\definecolor{gray70}{rgb}{0.702,0.702,0.702}
\definecolor{gray71}{rgb}{0.71,0.71,0.71}
\definecolor{gray72}{rgb}{0.722,0.722,0.722}
\definecolor{gray73}{rgb}{0.729,0.729,0.729}
\definecolor{gray74}{rgb}{0.741,0.741,0.741}
\definecolor{gray75}{rgb}{0.749,0.749,0.749}
\definecolor{gray76}{rgb}{0.761,0.761,0.761}
\definecolor{gray77}{rgb}{0.769,0.769,0.769}
\definecolor{gray78}{rgb}{0.78,0.78,0.78}
\definecolor{gray79}{rgb}{0.788,0.788,0.788}
\definecolor{gray8}{rgb}{0.078,0.078,0.078}
\definecolor{gray80}{rgb}{0.8,0.8,0.8}
\definecolor{gray81}{rgb}{0.812,0.812,0.812}
\definecolor{gray82}{rgb}{0.82,0.82,0.82}
\definecolor{gray83}{rgb}{0.831,0.831,0.831}
\definecolor{gray84}{rgb}{0.839,0.839,0.839}
\definecolor{gray85}{rgb}{0.851,0.851,0.851}
\definecolor{gray86}{rgb}{0.859,0.859,0.859}
\definecolor{gray87}{rgb}{0.871,0.871,0.871}
\definecolor{gray88}{rgb}{0.878,0.878,0.878}
\definecolor{gray89}{rgb}{0.89,0.89,0.89}
\definecolor{gray9}{rgb}{0.09,0.09,0.09}
\definecolor{gray90}{rgb}{0.898,0.898,0.898}
\definecolor{gray91}{rgb}{0.91,0.91,0.91}
\definecolor{gray92}{rgb}{0.922,0.922,0.922}
\definecolor{gray93}{rgb}{0.929,0.929,0.929}
\definecolor{gray94}{rgb}{0.941,0.941,0.941}
\definecolor{gray95}{rgb}{0.949,0.949,0.949}
\definecolor{gray96}{rgb}{0.961,0.961,0.961}
\definecolor{gray97}{rgb}{0.969,0.969,0.969}
\definecolor{gray98}{rgb}{0.98,0.98,0.98}
\definecolor{gray99}{rgb}{0.988,0.988,0.988}
\definecolor{green1}{rgb}{0,1,0}
\definecolor{green2}{rgb}{0,0.933,0}
\definecolor{green3}{rgb}{0,0.804,0}
\definecolor{green4}{rgb}{0,0.545,0}
\definecolor{greenyellow}{rgb}{0.678,1,0.184}
\definecolor{grey}{rgb}{0.745,0.745,0.745}
\definecolor{grey0}{rgb}{0,0,0}
\definecolor{grey1}{rgb}{0.012,0.012,0.012}
\definecolor{grey10}{rgb}{0.102,0.102,0.102}
\definecolor{grey100}{rgb}{1,1,1}
\definecolor{grey11}{rgb}{0.11,0.11,0.11}
\definecolor{grey12}{rgb}{0.122,0.122,0.122}
\definecolor{grey13}{rgb}{0.129,0.129,0.129}
\definecolor{grey14}{rgb}{0.141,0.141,0.141}
\definecolor{grey15}{rgb}{0.149,0.149,0.149}
\definecolor{grey16}{rgb}{0.161,0.161,0.161}
\definecolor{grey17}{rgb}{0.169,0.169,0.169}
\definecolor{grey18}{rgb}{0.18,0.18,0.18}
\definecolor{grey19}{rgb}{0.188,0.188,0.188}
\definecolor{grey2}{rgb}{0.02,0.02,0.02}
\definecolor{grey20}{rgb}{0.2,0.2,0.2}
\definecolor{grey21}{rgb}{0.212,0.212,0.212}
\definecolor{grey22}{rgb}{0.22,0.22,0.22}
\definecolor{grey23}{rgb}{0.231,0.231,0.231}
\definecolor{grey24}{rgb}{0.239,0.239,0.239}
\definecolor{grey25}{rgb}{0.251,0.251,0.251}
\definecolor{grey26}{rgb}{0.259,0.259,0.259}
\definecolor{grey27}{rgb}{0.271,0.271,0.271}
\definecolor{grey28}{rgb}{0.278,0.278,0.278}
\definecolor{grey29}{rgb}{0.29,0.29,0.29}
\definecolor{grey3}{rgb}{0.031,0.031,0.031}
\definecolor{grey30}{rgb}{0.302,0.302,0.302}
\definecolor{grey31}{rgb}{0.31,0.31,0.31}
\definecolor{grey32}{rgb}{0.322,0.322,0.322}
\definecolor{grey33}{rgb}{0.329,0.329,0.329}
\definecolor{grey34}{rgb}{0.341,0.341,0.341}
\definecolor{grey35}{rgb}{0.349,0.349,0.349}
\definecolor{grey36}{rgb}{0.361,0.361,0.361}
\definecolor{grey37}{rgb}{0.369,0.369,0.369}
\definecolor{grey38}{rgb}{0.38,0.38,0.38}
\definecolor{grey39}{rgb}{0.388,0.388,0.388}
\definecolor{grey4}{rgb}{0.039,0.039,0.039}
\definecolor{grey40}{rgb}{0.4,0.4,0.4}
\definecolor{grey41}{rgb}{0.412,0.412,0.412}
\definecolor{grey42}{rgb}{0.42,0.42,0.42}
\definecolor{grey43}{rgb}{0.431,0.431,0.431}
\definecolor{grey44}{rgb}{0.439,0.439,0.439}
\definecolor{grey45}{rgb}{0.451,0.451,0.451}
\definecolor{grey46}{rgb}{0.459,0.459,0.459}
\definecolor{grey47}{rgb}{0.471,0.471,0.471}
\definecolor{grey48}{rgb}{0.478,0.478,0.478}
\definecolor{grey49}{rgb}{0.49,0.49,0.49}
\definecolor{grey5}{rgb}{0.051,0.051,0.051}
\definecolor{grey50}{rgb}{0.498,0.498,0.498}
\definecolor{grey51}{rgb}{0.51,0.51,0.51}
\definecolor{grey52}{rgb}{0.522,0.522,0.522}
\definecolor{grey53}{rgb}{0.529,0.529,0.529}
\definecolor{grey54}{rgb}{0.541,0.541,0.541}
\definecolor{grey55}{rgb}{0.549,0.549,0.549}
\definecolor{grey56}{rgb}{0.561,0.561,0.561}
\definecolor{grey57}{rgb}{0.569,0.569,0.569}
\definecolor{grey58}{rgb}{0.58,0.58,0.58}
\definecolor{grey59}{rgb}{0.588,0.588,0.588}
\definecolor{grey6}{rgb}{0.059,0.059,0.059}
\definecolor{grey60}{rgb}{0.6,0.6,0.6}
\definecolor{grey61}{rgb}{0.612,0.612,0.612}
\definecolor{grey62}{rgb}{0.62,0.62,0.62}
\definecolor{grey63}{rgb}{0.631,0.631,0.631}
\definecolor{grey64}{rgb}{0.639,0.639,0.639}
\definecolor{grey65}{rgb}{0.651,0.651,0.651}
\definecolor{grey66}{rgb}{0.659,0.659,0.659}
\definecolor{grey67}{rgb}{0.671,0.671,0.671}
\definecolor{grey68}{rgb}{0.678,0.678,0.678}
\definecolor{grey69}{rgb}{0.69,0.69,0.69}
\definecolor{grey7}{rgb}{0.071,0.071,0.071}
\definecolor{grey70}{rgb}{0.702,0.702,0.702}
\definecolor{grey71}{rgb}{0.71,0.71,0.71}
\definecolor{grey72}{rgb}{0.722,0.722,0.722}
\definecolor{grey73}{rgb}{0.729,0.729,0.729}
\definecolor{grey74}{rgb}{0.741,0.741,0.741}
\definecolor{grey75}{rgb}{0.749,0.749,0.749}
\definecolor{grey76}{rgb}{0.761,0.761,0.761}
\definecolor{grey77}{rgb}{0.769,0.769,0.769}
\definecolor{grey78}{rgb}{0.78,0.78,0.78}
\definecolor{grey79}{rgb}{0.788,0.788,0.788}
\definecolor{grey8}{rgb}{0.078,0.078,0.078}
\definecolor{grey80}{rgb}{0.8,0.8,0.8}
\definecolor{grey81}{rgb}{0.812,0.812,0.812}
\definecolor{grey82}{rgb}{0.82,0.82,0.82}
\definecolor{grey83}{rgb}{0.831,0.831,0.831}
\definecolor{grey84}{rgb}{0.839,0.839,0.839}
\definecolor{grey85}{rgb}{0.851,0.851,0.851}
\definecolor{grey86}{rgb}{0.859,0.859,0.859}
\definecolor{grey87}{rgb}{0.871,0.871,0.871}
\definecolor{grey88}{rgb}{0.878,0.878,0.878}
\definecolor{grey89}{rgb}{0.89,0.89,0.89}
\definecolor{grey9}{rgb}{0.09,0.09,0.09}
\definecolor{grey90}{rgb}{0.898,0.898,0.898}
\definecolor{grey91}{rgb}{0.91,0.91,0.91}
\definecolor{grey92}{rgb}{0.922,0.922,0.922}
\definecolor{grey93}{rgb}{0.929,0.929,0.929}
\definecolor{grey94}{rgb}{0.941,0.941,0.941}
\definecolor{grey95}{rgb}{0.949,0.949,0.949}
\definecolor{grey96}{rgb}{0.961,0.961,0.961}
\definecolor{grey97}{rgb}{0.969,0.969,0.969}
\definecolor{grey98}{rgb}{0.98,0.98,0.98}
\definecolor{grey99}{rgb}{0.988,0.988,0.988}
\definecolor{honeydew}{rgb}{0.941,1,0.941}
\definecolor{honeydew1}{rgb}{0.941,1,0.941}
\definecolor{honeydew2}{rgb}{0.878,0.933,0.878}
\definecolor{honeydew3}{rgb}{0.757,0.804,0.757}
\definecolor{honeydew4}{rgb}{0.514,0.545,0.514}
\definecolor{hotpink}{rgb}{1,0.412,0.706}
\definecolor{hotpink1}{rgb}{1,0.431,0.706}
\definecolor{hotpink2}{rgb}{0.933,0.416,0.655}
\definecolor{hotpink3}{rgb}{0.804,0.376,0.565}
\definecolor{hotpink4}{rgb}{0.545,0.227,0.384}
\definecolor{indianred}{rgb}{0.804,0.361,0.361}
\definecolor{indianred1}{rgb}{1,0.416,0.416}
\definecolor{indianred2}{rgb}{0.933,0.388,0.388}
\definecolor{indianred3}{rgb}{0.804,0.333,0.333}
\definecolor{indianred4}{rgb}{0.545,0.227,0.227}
\definecolor{ivory}{rgb}{1,1,0.941}
\definecolor{ivory1}{rgb}{1,1,0.941}
\definecolor{ivory2}{rgb}{0.933,0.933,0.878}
\definecolor{ivory3}{rgb}{0.804,0.804,0.757}
\definecolor{ivory4}{rgb}{0.545,0.545,0.514}
\definecolor{khaki}{rgb}{0.941,0.902,0.549}
\definecolor{khaki1}{rgb}{1,0.965,0.561}
\definecolor{khaki2}{rgb}{0.933,0.902,0.522}
\definecolor{khaki3}{rgb}{0.804,0.776,0.451}
\definecolor{khaki4}{rgb}{0.545,0.525,0.306}
\definecolor{lavender}{rgb}{0.902,0.902,0.98}
\definecolor{lavenderblush}{rgb}{1,0.941,0.961}
\definecolor{lavenderblush1}{rgb}{1,0.941,0.961}
\definecolor{lavenderblush2}{rgb}{0.933,0.878,0.898}
\definecolor{lavenderblush3}{rgb}{0.804,0.757,0.773}
\definecolor{lavenderblush4}{rgb}{0.545,0.514,0.525}
\definecolor{lawngreen}{rgb}{0.486,0.988,0}
\definecolor{lemonchiffon}{rgb}{1,0.98,0.804}
\definecolor{lemonchiffon1}{rgb}{1,0.98,0.804}
\definecolor{lemonchiffon2}{rgb}{0.933,0.914,0.749}
\definecolor{lemonchiffon3}{rgb}{0.804,0.788,0.647}
\definecolor{lemonchiffon4}{rgb}{0.545,0.537,0.439}
\definecolor{lightblue1}{rgb}{0.749,0.937,1}
\definecolor{lightblue2}{rgb}{0.698,0.875,0.933}
\definecolor{lightblue3}{rgb}{0.604,0.753,0.804}
\definecolor{lightblue4}{rgb}{0.408,0.514,0.545}
\definecolor{lightcoral}{rgb}{0.941,0.502,0.502}
\definecolor{lightcyan1}{rgb}{0.878,1,1}
\definecolor{lightcyan2}{rgb}{0.82,0.933,0.933}
\definecolor{lightcyan3}{rgb}{0.706,0.804,0.804}
\definecolor{lightcyan4}{rgb}{0.478,0.545,0.545}
\definecolor{lightgoldenrod}{rgb}{0.933,0.867,0.51}
\definecolor{lightgoldenrod1}{rgb}{1,0.925,0.545}
\definecolor{lightgoldenrod2}{rgb}{0.933,0.863,0.51}
\definecolor{lightgoldenrod3}{rgb}{0.804,0.745,0.439}
\definecolor{lightgoldenrod4}{rgb}{0.545,0.506,0.298}
\definecolor{lightgoldenrodyellow}{rgb}{0.98,0.98,0.824}
\definecolor{lightgrey}{rgb}{0.827,0.827,0.827}
\definecolor{lightpink}{rgb}{1,0.714,0.757}
\definecolor{lightpink1}{rgb}{1,0.682,0.725}
\definecolor{lightpink2}{rgb}{0.933,0.635,0.678}
\definecolor{lightpink3}{rgb}{0.804,0.549,0.584}
\definecolor{lightpink4}{rgb}{0.545,0.373,0.396}
\definecolor{lightsalmon}{rgb}{1,0.627,0.478}
\definecolor{lightsalmon1}{rgb}{1,0.627,0.478}
\definecolor{lightsalmon2}{rgb}{0.933,0.584,0.447}
\definecolor{lightsalmon3}{rgb}{0.804,0.506,0.384}
\definecolor{lightsalmon4}{rgb}{0.545,0.341,0.259}
\definecolor{lightseagreen}{rgb}{0.125,0.698,0.667}
\definecolor{lightskyblue}{rgb}{0.529,0.808,0.98}
\definecolor{lightskyblue1}{rgb}{0.69,0.886,1}
\definecolor{lightskyblue2}{rgb}{0.643,0.827,0.933}
\definecolor{lightskyblue3}{rgb}{0.553,0.714,0.804}
\definecolor{lightskyblue4}{rgb}{0.376,0.482,0.545}
\definecolor{lightslateblue}{rgb}{0.518,0.439,1}
\definecolor{lightslategray}{rgb}{0.467,0.533,0.6}
\definecolor{lightslategrey}{rgb}{0.467,0.533,0.6}
\definecolor{lightsteelblue}{rgb}{0.69,0.769,0.871}
\definecolor{lightsteelblue1}{rgb}{0.792,0.882,1}
\definecolor{lightsteelblue2}{rgb}{0.737,0.824,0.933}
\definecolor{lightsteelblue3}{rgb}{0.635,0.71,0.804}
\definecolor{lightsteelblue4}{rgb}{0.431,0.482,0.545}
\definecolor{lightyellow1}{rgb}{1,1,0.878}
\definecolor{lightyellow2}{rgb}{0.933,0.933,0.82}
\definecolor{lightyellow3}{rgb}{0.804,0.804,0.706}
\definecolor{lightyellow4}{rgb}{0.545,0.545,0.478}
\definecolor{limegreen}{rgb}{0.196,0.804,0.196}
\definecolor{linen}{rgb}{0.98,0.941,0.902}
\definecolor{magenta}{rgb}{1,0,1}
\definecolor{magenta1}{rgb}{1,0,1}
\definecolor{magenta2}{rgb}{0.933,0,0.933}
\definecolor{magenta3}{rgb}{0.804,0,0.804}
\definecolor{magenta4}{rgb}{0.545,0,0.545}
\definecolor{maroon}{rgb}{0.69,0.188,0.376}
\definecolor{maroon1}{rgb}{1,0.204,0.702}
\definecolor{maroon2}{rgb}{0.933,0.188,0.655}
\definecolor{maroon3}{rgb}{0.804,0.161,0.565}
\definecolor{maroon4}{rgb}{0.545,0.11,0.384}
\definecolor{mediumaquamarine}{rgb}{0.4,0.804,0.667}
\definecolor{mediumblue}{rgb}{0,0,0.804}
\definecolor{mediumorchid}{rgb}{0.729,0.333,0.827}
\definecolor{mediumorchid1}{rgb}{0.878,0.4,1}
\definecolor{mediumorchid2}{rgb}{0.82,0.373,0.933}
\definecolor{mediumorchid3}{rgb}{0.706,0.322,0.804}
\definecolor{mediumorchid4}{rgb}{0.478,0.216,0.545}
\definecolor{mediumpurple}{rgb}{0.576,0.439,0.859}
\definecolor{mediumpurple1}{rgb}{0.671,0.51,1}
\definecolor{mediumpurple2}{rgb}{0.624,0.475,0.933}
\definecolor{mediumpurple3}{rgb}{0.537,0.408,0.804}
\definecolor{mediumpurple4}{rgb}{0.365,0.278,0.545}
\definecolor{mediumseagreen}{rgb}{0.235,0.702,0.443}
\definecolor{mediumslateblue}{rgb}{0.482,0.408,0.933}
\definecolor{mediumspringgreen}{rgb}{0,0.98,0.604}
\definecolor{mediumturquoise}{rgb}{0.282,0.82,0.8}
\definecolor{mediumvioletred}{rgb}{0.78,0.082,0.522}
\definecolor{midnightblue}{rgb}{0.098,0.098,0.439}
\definecolor{mintcream}{rgb}{0.961,1,0.98}
\definecolor{mistyrose}{rgb}{1,0.894,0.882}
\definecolor{mistyrose1}{rgb}{1,0.894,0.882}
\definecolor{mistyrose2}{rgb}{0.933,0.835,0.824}
\definecolor{mistyrose3}{rgb}{0.804,0.718,0.71}
\definecolor{mistyrose4}{rgb}{0.545,0.49,0.482}
\definecolor{moccasin}{rgb}{1,0.894,0.71}
\definecolor{navajowhite}{rgb}{1,0.871,0.678}
\definecolor{navajowhite1}{rgb}{1,0.871,0.678}
\definecolor{navajowhite2}{rgb}{0.933,0.812,0.631}
\definecolor{navajowhite3}{rgb}{0.804,0.702,0.545}
\definecolor{navajowhite4}{rgb}{0.545,0.475,0.369}
\definecolor{navyblue}{rgb}{0,0,0.502}
\definecolor{oldlace}{rgb}{0.992,0.961,0.902}
\definecolor{olivedrab}{rgb}{0.42,0.557,0.137}
\definecolor{olivedrab1}{rgb}{0.753,1,0.243}
\definecolor{olivedrab2}{rgb}{0.702,0.933,0.227}
\definecolor{olivedrab3}{rgb}{0.604,0.804,0.196}
\definecolor{olivedrab4}{rgb}{0.412,0.545,0.133}
\definecolor{orange1}{rgb}{1,0.647,0}
\definecolor{orange2}{rgb}{0.933,0.604,0}
\definecolor{orange3}{rgb}{0.804,0.522,0}
\definecolor{orange4}{rgb}{0.545,0.353,0}
\definecolor{orangered}{rgb}{1,0.271,0}
\definecolor{orangered1}{rgb}{1,0.271,0}
\definecolor{orangered2}{rgb}{0.933,0.251,0}
\definecolor{orangered3}{rgb}{0.804,0.216,0}
\definecolor{orangered4}{rgb}{0.545,0.145,0}
\definecolor{orchid}{rgb}{0.855,0.439,0.839}
\definecolor{orchid1}{rgb}{1,0.514,0.98}
\definecolor{orchid2}{rgb}{0.933,0.478,0.914}
\definecolor{orchid3}{rgb}{0.804,0.412,0.788}
\definecolor{orchid4}{rgb}{0.545,0.278,0.537}
\definecolor{palegoldenrod}{rgb}{0.933,0.91,0.667}
\definecolor{palegreen}{rgb}{0.596,0.984,0.596}
\definecolor{palegreen1}{rgb}{0.604,1,0.604}
\definecolor{palegreen2}{rgb}{0.565,0.933,0.565}
\definecolor{palegreen3}{rgb}{0.486,0.804,0.486}
\definecolor{palegreen4}{rgb}{0.329,0.545,0.329}
\definecolor{paleturquoise}{rgb}{0.686,0.933,0.933}
\definecolor{paleturquoise1}{rgb}{0.733,1,1}
\definecolor{paleturquoise2}{rgb}{0.682,0.933,0.933}
\definecolor{paleturquoise3}{rgb}{0.588,0.804,0.804}
\definecolor{paleturquoise4}{rgb}{0.4,0.545,0.545}
\definecolor{palevioletred}{rgb}{0.859,0.439,0.576}
\definecolor{palevioletred1}{rgb}{1,0.51,0.671}
\definecolor{palevioletred2}{rgb}{0.933,0.475,0.624}
\definecolor{palevioletred3}{rgb}{0.804,0.408,0.537}
\definecolor{palevioletred4}{rgb}{0.545,0.278,0.365}
\definecolor{papayawhip}{rgb}{1,0.937,0.835}
\definecolor{peachpuff}{rgb}{1,0.855,0.725}
\definecolor{peachpuff1}{rgb}{1,0.855,0.725}
\definecolor{peachpuff2}{rgb}{0.933,0.796,0.678}
\definecolor{peachpuff3}{rgb}{0.804,0.686,0.584}
\definecolor{peachpuff4}{rgb}{0.545,0.467,0.396}
\definecolor{peru}{rgb}{0.804,0.522,0.247}
\definecolor{pink1}{rgb}{1,0.71,0.773}
\definecolor{pink2}{rgb}{0.933,0.663,0.722}
\definecolor{pink3}{rgb}{0.804,0.569,0.62}
\definecolor{pink4}{rgb}{0.545,0.388,0.424}
\definecolor{plum}{rgb}{0.867,0.627,0.867}
\definecolor{plum1}{rgb}{1,0.733,1}
\definecolor{plum2}{rgb}{0.933,0.682,0.933}
\definecolor{plum3}{rgb}{0.804,0.588,0.804}
\definecolor{plum4}{rgb}{0.545,0.4,0.545}
\definecolor{powderblue}{rgb}{0.69,0.878,0.902}
\definecolor{purple1}{rgb}{0.608,0.188,1}
\definecolor{purple2}{rgb}{0.569,0.173,0.933}
\definecolor{purple3}{rgb}{0.49,0.149,0.804}
\definecolor{purple4}{rgb}{0.333,0.102,0.545}
\definecolor{red1}{rgb}{1,0,0}
\definecolor{red2}{rgb}{0.933,0,0}
\definecolor{red3}{rgb}{0.804,0,0}
\definecolor{red4}{rgb}{0.545,0,0}
\definecolor{rosybrown}{rgb}{0.737,0.561,0.561}
\definecolor{rosybrown1}{rgb}{1,0.757,0.757}
\definecolor{rosybrown2}{rgb}{0.933,0.706,0.706}
\definecolor{rosybrown3}{rgb}{0.804,0.608,0.608}
\definecolor{rosybrown4}{rgb}{0.545,0.412,0.412}
\definecolor{royalblue}{rgb}{0.255,0.412,0.882}
\definecolor{royalblue1}{rgb}{0.282,0.463,1}
\definecolor{royalblue2}{rgb}{0.263,0.431,0.933}
\definecolor{royalblue3}{rgb}{0.227,0.373,0.804}
\definecolor{royalblue4}{rgb}{0.153,0.251,0.545}
\definecolor{saddlebrown}{rgb}{0.545,0.271,0.075}
\definecolor{salmon}{rgb}{0.98,0.502,0.447}
\definecolor{salmon1}{rgb}{1,0.549,0.412}
\definecolor{salmon2}{rgb}{0.933,0.51,0.384}
\definecolor{salmon3}{rgb}{0.804,0.439,0.329}
\definecolor{salmon4}{rgb}{0.545,0.298,0.224}
\definecolor{sandybrown}{rgb}{0.957,0.643,0.376}
\definecolor{seagreen1}{rgb}{0.329,1,0.624}
\definecolor{seagreen2}{rgb}{0.306,0.933,0.58}
\definecolor{seagreen3}{rgb}{0.263,0.804,0.502}
\definecolor{seagreen4}{rgb}{0.18,0.545,0.341}
\definecolor{seashell}{rgb}{1,0.961,0.933}
\definecolor{seashell1}{rgb}{1,0.961,0.933}
\definecolor{seashell2}{rgb}{0.933,0.898,0.871}
\definecolor{seashell3}{rgb}{0.804,0.773,0.749}
\definecolor{seashell4}{rgb}{0.545,0.525,0.51}
\definecolor{sienna}{rgb}{0.627,0.322,0.176}
\definecolor{sienna1}{rgb}{1,0.51,0.278}
\definecolor{sienna2}{rgb}{0.933,0.475,0.259}
\definecolor{sienna3}{rgb}{0.804,0.408,0.224}
\definecolor{sienna4}{rgb}{0.545,0.278,0.149}
\definecolor{skyblue}{rgb}{0.529,0.808,0.922}
\definecolor{skyblue1}{rgb}{0.529,0.808,1}
\definecolor{skyblue2}{rgb}{0.494,0.753,0.933}
\definecolor{skyblue3}{rgb}{0.424,0.651,0.804}
\definecolor{skyblue4}{rgb}{0.29,0.439,0.545}
\definecolor{slateblue}{rgb}{0.416,0.353,0.804}
\definecolor{slateblue1}{rgb}{0.514,0.435,1}
\definecolor{slateblue2}{rgb}{0.478,0.404,0.933}
\definecolor{slateblue3}{rgb}{0.412,0.349,0.804}
\definecolor{slateblue4}{rgb}{0.278,0.235,0.545}
\definecolor{slategray}{rgb}{0.439,0.502,0.565}
\definecolor{slategray1}{rgb}{0.776,0.886,1}
\definecolor{slategray2}{rgb}{0.725,0.827,0.933}
\definecolor{slategray3}{rgb}{0.624,0.714,0.804}
\definecolor{slategray4}{rgb}{0.424,0.482,0.545}
\definecolor{slategrey}{rgb}{0.439,0.502,0.565}
\definecolor{snow}{rgb}{1,0.98,0.98}
\definecolor{snow1}{rgb}{1,0.98,0.98}
\definecolor{snow2}{rgb}{0.933,0.914,0.914}
\definecolor{snow3}{rgb}{0.804,0.788,0.788}
\definecolor{snow4}{rgb}{0.545,0.537,0.537}
\definecolor{springgreen}{rgb}{0,1,0.498}
\definecolor{springgreen1}{rgb}{0,1,0.498}
\definecolor{springgreen2}{rgb}{0,0.933,0.463}
\definecolor{springgreen3}{rgb}{0,0.804,0.4}
\definecolor{springgreen4}{rgb}{0,0.545,0.271}
\definecolor{steelblue}{rgb}{0.275,0.51,0.706}
\definecolor{steelblue1}{rgb}{0.388,0.722,1}
\definecolor{steelblue2}{rgb}{0.361,0.675,0.933}
\definecolor{steelblue3}{rgb}{0.31,0.58,0.804}
\definecolor{steelblue4}{rgb}{0.212,0.392,0.545}
\definecolor{tan}{rgb}{0.824,0.706,0.549}
\definecolor{tan1}{rgb}{1,0.647,0.31}
\definecolor{tan2}{rgb}{0.933,0.604,0.286}
\definecolor{tan3}{rgb}{0.804,0.522,0.247}
\definecolor{tan4}{rgb}{0.545,0.353,0.169}
\definecolor{thistle}{rgb}{0.847,0.749,0.847}
\definecolor{thistle1}{rgb}{1,0.882,1}
\definecolor{thistle2}{rgb}{0.933,0.824,0.933}
\definecolor{thistle3}{rgb}{0.804,0.71,0.804}
\definecolor{thistle4}{rgb}{0.545,0.482,0.545}
\definecolor{tomato}{rgb}{1,0.388,0.278}
\definecolor{tomato1}{rgb}{1,0.388,0.278}
\definecolor{tomato2}{rgb}{0.933,0.361,0.259}
\definecolor{tomato3}{rgb}{0.804,0.31,0.224}
\definecolor{tomato4}{rgb}{0.545,0.212,0.149}
\definecolor{turquoise1}{rgb}{0,0.961,1}
\definecolor{turquoise2}{rgb}{0,0.898,0.933}
\definecolor{turquoise3}{rgb}{0,0.773,0.804}
\definecolor{turquoise4}{rgb}{0,0.525,0.545}
\definecolor{violetred}{rgb}{0.816,0.125,0.565}
\definecolor{violetred1}{rgb}{1,0.243,0.588}
\definecolor{violetred2}{rgb}{0.933,0.227,0.549}
\definecolor{violetred3}{rgb}{0.804,0.196,0.471}
\definecolor{violetred4}{rgb}{0.545,0.133,0.322}
\definecolor{wheat}{rgb}{0.961,0.871,0.702}
\definecolor{wheat1}{rgb}{1,0.906,0.729}
\definecolor{wheat2}{rgb}{0.933,0.847,0.682}
\definecolor{wheat3}{rgb}{0.804,0.729,0.588}
\definecolor{wheat4}{rgb}{0.545,0.494,0.4}
\definecolor{whitesmoke}{rgb}{0.961,0.961,0.961}
\definecolor{yellow1}{rgb}{1,1,0}
\definecolor{yellow2}{rgb}{0.933,0.933,0}
\definecolor{yellow3}{rgb}{0.804,0.804,0}
\definecolor{yellow4}{rgb}{0.545,0.545,0}
\definecolor{yellowgreen}{rgb}{0.604,0.804,0.196}
\scalebox{0.95}{

}
	\end{center}
	\caption{The set $X$ and the connected components of $G\setminus X.$}
	\labels{@governorship}
\end{figure}

\begin{eqnarray}
\labels{eq_out-sig_first-floor}\blue{{\sf out}\text{-}{\sf sig}}(\mathfrak{K},R,d,L,Z)=\{({\bf H},\bar{φ})∈ \blue{{\sf SIG}_{\sf out}}& \mid
 & \exists\  F∈ {\cal F}^{V_L ({\bf a})}_{|V(H)| - |\partial_{\mathfrak{K}}(Z)|}, \mbox{ such that if ${\bf H} = (H,{\cal U})$}\\
 \notag & &~~~~\mbox{and ${\cal V} = (\partial_{\mathfrak{K}}(Z),V_L ({\bf a}), V(F)\setminus V_L ({\bf a})),$ then}\\
\notag & &~~~~\mbox{${\cal V}$ is a nice 3-partition of $K^{\bf a}[\partial_{\mathfrak{K}}(Z)\cup V_L ({\bf a})]\cup F$}\\
\notag & &~~~~\mbox{and $K^{\bf a}[\partial_{\mathfrak{K}}(Z)\cup V_L ({\bf a})]\cup F$ is strongly isomorphic}\\
\notag & &~~~~\mbox{to $H$ with respect to $({\cal V}, {\cal U}),$}\\
\notag & & \mbox{$\exists$ an ordering ${\bf b}$ of $\partial_{\mathfrak{K}} (Z)\cup V(F),$ and}\\
\notag & & \exists\ \tilde{X} \subseteq Z\cup  V(F)\mbox{~such that~} \partial_{\mathfrak{K}} (Z)\cup V(F)\subseteq \tilde{X} \mbox{~and}\\
\notag & &~~~~\mbox{if $R'=(Z\setminus  \partial_{\mathfrak{K}}(Z))\cap R,$ then}\\
\notag & &~~~~\big(\mathfrak{A}^{(d,Z,L,F)},R',{\bf W}_{q}, \varnothing^l,\tilde{X},{\bf b}\big)\models \bar{φ}\}.
\end{eqnarray}

Intuitively,
for each ${\bf H}∈{\cal H}^{(\ell)},$ where $\ell∈[0,\tw(θ)-1]$ and  ${\bf H}$ is a graph $H$ together with a nice $3$-partition,
and each $\bar{φ}∈ {\sf rep}^{(\ell)}_{τ'}(θ^{\sf out}_q),$
we are asked to guess three
objects:
a graph $F∈ {\cal F}^{V_L ({\bf a})}_{|V(H)| - |\partial_{\mathfrak{K}}(Z)|},$
a function $ρ$ and
a set $X.$
The guessed additional part $F'$ of $F$
represents the boundary of $X_{\rm out}$ that is the portion of the modulator that will be away from $I^{(d)}.$
The set $\partial_{\mathfrak{K}} (Z)\cup V(F)$
is the boundary  of the boundaried structure $\big(\mathfrak{A}^{(d,Z,L,F)},R',{\bf W}_{q}, \varnothing^l,X_{\rm in}, {\bf b}\big).$
This $|\partial_{\mathfrak{K}} (Z)\cup V(F)|$-boundaried structure should
be a model of $\bar{φ}$ and its boundary
(that is the union of $F$ and $K^{\bf a}[\partial_{\mathfrak{K}}(Z)\cup V_L ({\bf a})]$) should be isomorphic to $H.$
The ordering ${\bf b}$ of the boundary is guessed.
In $Z,$ we will guess
the portion $\tilde{X}$ of the solution that will be part of $I^{(d)}$ (see~\autoref{@governorship} and~\autoref{@accomplishes} for the situation of these sets inside the $d$-layer).

Also, keep in mind that, since $θ^{\sf out}_q$ is a sentence in $\MSOL[τ\cup{\bf Q}\cup\{{\sf R},{\sf X}\}\cup{\bf c}]$ and $\bar{φ}∈ {{\sf rep}^{(\ell)}_{τ'}(θ^{\sf out}_q)},$
we have that $\bar{φ}∈\MSOL[τ\cup{\bf Q}\cup\{{\sf R},{\sf X}\}\cup{\bf c}\cup\{{\sf b}_1,\ldots, {\sf b}_\ell\}],$
where ${\sf b}_1,\ldots, {\sf b}_\ell$ are constant symbols different than the symbols in ${\bf c}.$
When asking whether $\big(\mathfrak{A}^{(d,Z,L,F)},R',{\bf W}_{q}, \varnothing^l,\tilde{X},{\bf b}\big)\models\bar{φ},$ we interpret ${\bf c}$ by $\varnothing^l$ and ${\sf b}_1,\ldots, {\sf b}_\ell$ by ${\bf b}.$
{We stress that, in the out-signature of every extended compass, we always interpret ${\bf c}$ by $\varnothing^l.$
This choice will be supported by the assumption of~\autoref{@desmembramientos}, that the neighborhood of the apices of the input flatness pair $(W,\mathfrak{R})$ is asked to have ``big enough'' bidimensionality with respect to $(W,\mathfrak{R}),$
and therefore, as we will prove in~\autoref{sec_proof_correctness},
every apex vertex should be either contained in $\partial_{\mathfrak{K}}(X)$ or inside the privileged component.}

\begin{figure}[ht]
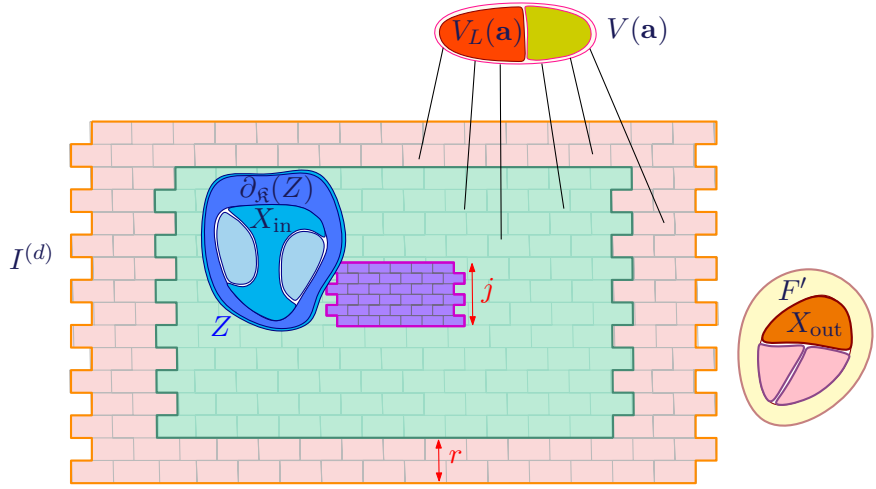

	\begin{center}
\tikzstyle{ipe stylesheet} = [
  ipe import,
  even odd rule,
  line join=round,
  line cap=butt,
  ipe pen normal/.style={line width=0.4},
  ipe pen heavier/.style={line width=0.8},
  ipe pen fat/.style={line width=1.2},
  ipe pen ultrafat/.style={line width=2},
  ipe pen normal,
  ipe mark normal/.style={ipe mark scale=3},
  ipe mark large/.style={ipe mark scale=5},
  ipe mark small/.style={ipe mark scale=2},
  ipe mark tiny/.style={ipe mark scale=1.1},
  ipe mark normal,
  /pgf/arrow keys/.cd,
  ipe arrow normal/.style={scale=7},
  ipe arrow large/.style={scale=10},
  ipe arrow small/.style={scale=5},
  ipe arrow tiny/.style={scale=3},
  ipe arrow normal,
  /tikz/.cd,
  ipe arrows, 
  <->/.tip = ipe normal,
  ipe dash normal/.style={dash pattern=},
  ipe dash dotted/.style={dash pattern=on 1bp off 3bp},
  ipe dash dashed/.style={dash pattern=on 4bp off 4bp},
  ipe dash dash dotted/.style={dash pattern=on 4bp off 2bp on 1bp off 2bp},
  ipe dash dash dot dotted/.style={dash pattern=on 4bp off 2bp on 1bp off 2bp on 1bp off 2bp},
  ipe dash normal,
  ipe node/.append style={font=\normalsize},
  ipe stretch normal/.style={ipe node stretch=1},
  ipe stretch normal,
  ipe opacity 10/.style={opacity=0.1},
  ipe opacity 30/.style={opacity=0.3},
  ipe opacity 50/.style={opacity=0.5},
  ipe opacity 75/.style={opacity=0.75},
  ipe opacity opaque/.style={opacity=1},
  ipe opacity opaque,
]
\definecolor{black}{rgb}{0,0,0}
\definecolor{white}{rgb}{1,1,1}
\definecolor{red}{rgb}{1,0,0}
\definecolor{blue}{rgb}{0,0,1}
\definecolor{green}{rgb}{0,1,0}
\definecolor{yellow}{rgb}{1,1,0}
\definecolor{orange}{rgb}{1,0.647,0}
\definecolor{gold}{rgb}{1,0.843,0}
\definecolor{purple}{rgb}{0.627,0.125,0.941}
\definecolor{gray}{rgb}{0.745,0.745,0.745}
\definecolor{brown}{rgb}{0.647,0.165,0.165}
\definecolor{navy}{rgb}{0,0,0.502}
\definecolor{pink}{rgb}{1,0.753,0.796}
\definecolor{seagreen}{rgb}{0.18,0.545,0.341}
\definecolor{turquoise}{rgb}{0.251,0.878,0.816}
\definecolor{violet}{rgb}{0.933,0.51,0.933}
\definecolor{darkblue}{rgb}{0,0,0.545}
\definecolor{darkcyan}{rgb}{0,0.545,0.545}
\definecolor{darkgray}{rgb}{0.663,0.663,0.663}
\definecolor{darkgreen}{rgb}{0,0.392,0}
\definecolor{darkmagenta}{rgb}{0.545,0,0.545}
\definecolor{darkorange}{rgb}{1,0.549,0}
\definecolor{darkred}{rgb}{0.545,0,0}
\definecolor{lightblue}{rgb}{0.678,0.847,0.902}
\definecolor{lightcyan}{rgb}{0.878,1,1}
\definecolor{lightgray}{rgb}{0.827,0.827,0.827}
\definecolor{lightgreen}{rgb}{0.565,0.933,0.565}
\definecolor{lightyellow}{rgb}{1,1,0.878}
\definecolor{aliceblue}{rgb}{0.941,0.973,1}
\definecolor{antiquewhite}{rgb}{0.98,0.922,0.843}
\definecolor{antiquewhite1}{rgb}{1,0.937,0.859}
\definecolor{antiquewhite2}{rgb}{0.933,0.875,0.8}
\definecolor{antiquewhite3}{rgb}{0.804,0.753,0.69}
\definecolor{antiquewhite4}{rgb}{0.545,0.514,0.471}
\definecolor{aquamarine}{rgb}{0.498,1,0.831}
\definecolor{aquamarine1}{rgb}{0.498,1,0.831}
\definecolor{aquamarine2}{rgb}{0.463,0.933,0.776}
\definecolor{aquamarine3}{rgb}{0.4,0.804,0.667}
\definecolor{aquamarine4}{rgb}{0.271,0.545,0.455}
\definecolor{azure}{rgb}{0.941,1,1}
\definecolor{azure1}{rgb}{0.941,1,1}
\definecolor{azure2}{rgb}{0.878,0.933,0.933}
\definecolor{azure3}{rgb}{0.757,0.804,0.804}
\definecolor{azure4}{rgb}{0.514,0.545,0.545}
\definecolor{beige}{rgb}{0.961,0.961,0.863}
\definecolor{bisque}{rgb}{1,0.894,0.769}
\definecolor{bisque1}{rgb}{1,0.894,0.769}
\definecolor{bisque2}{rgb}{0.933,0.835,0.718}
\definecolor{bisque3}{rgb}{0.804,0.718,0.62}
\definecolor{bisque4}{rgb}{0.545,0.49,0.42}
\definecolor{blanchedalmond}{rgb}{1,0.922,0.804}
\definecolor{blue1}{rgb}{0,0,1}
\definecolor{blue2}{rgb}{0,0,0.933}
\definecolor{blue3}{rgb}{0,0,0.804}
\definecolor{blue4}{rgb}{0,0,0.545}
\definecolor{blueviolet}{rgb}{0.541,0.169,0.886}
\definecolor{brown1}{rgb}{1,0.251,0.251}
\definecolor{brown2}{rgb}{0.933,0.231,0.231}
\definecolor{brown3}{rgb}{0.804,0.2,0.2}
\definecolor{brown4}{rgb}{0.545,0.137,0.137}
\definecolor{burlywood}{rgb}{0.871,0.722,0.529}
\definecolor{burlywood1}{rgb}{1,0.827,0.608}
\definecolor{burlywood2}{rgb}{0.933,0.773,0.569}
\definecolor{burlywood3}{rgb}{0.804,0.667,0.49}
\definecolor{burlywood4}{rgb}{0.545,0.451,0.333}
\definecolor{cadetblue}{rgb}{0.373,0.62,0.627}
\definecolor{cadetblue1}{rgb}{0.596,0.961,1}
\definecolor{cadetblue2}{rgb}{0.557,0.898,0.933}
\definecolor{cadetblue3}{rgb}{0.478,0.773,0.804}
\definecolor{cadetblue4}{rgb}{0.325,0.525,0.545}
\definecolor{chartreuse}{rgb}{0.498,1,0}
\definecolor{chartreuse1}{rgb}{0.498,1,0}
\definecolor{chartreuse2}{rgb}{0.463,0.933,0}
\definecolor{chartreuse3}{rgb}{0.4,0.804,0}
\definecolor{chartreuse4}{rgb}{0.271,0.545,0}
\definecolor{chocolate}{rgb}{0.824,0.412,0.118}
\definecolor{chocolate1}{rgb}{1,0.498,0.141}
\definecolor{chocolate2}{rgb}{0.933,0.463,0.129}
\definecolor{chocolate3}{rgb}{0.804,0.4,0.114}
\definecolor{chocolate4}{rgb}{0.545,0.271,0.075}
\definecolor{coral}{rgb}{1,0.498,0.314}
\definecolor{coral1}{rgb}{1,0.447,0.337}
\definecolor{coral2}{rgb}{0.933,0.416,0.314}
\definecolor{coral3}{rgb}{0.804,0.357,0.271}
\definecolor{coral4}{rgb}{0.545,0.243,0.184}
\definecolor{cornflowerblue}{rgb}{0.392,0.584,0.929}
\definecolor{cornsilk}{rgb}{1,0.973,0.863}
\definecolor{cornsilk1}{rgb}{1,0.973,0.863}
\definecolor{cornsilk2}{rgb}{0.933,0.91,0.804}
\definecolor{cornsilk3}{rgb}{0.804,0.784,0.694}
\definecolor{cornsilk4}{rgb}{0.545,0.533,0.471}
\definecolor{cyan}{rgb}{0,1,1}
\definecolor{cyan1}{rgb}{0,1,1}
\definecolor{cyan2}{rgb}{0,0.933,0.933}
\definecolor{cyan3}{rgb}{0,0.804,0.804}
\definecolor{cyan4}{rgb}{0,0.545,0.545}
\definecolor{darkgoldenrod}{rgb}{0.722,0.525,0.043}
\definecolor{darkgoldenrod1}{rgb}{1,0.725,0.059}
\definecolor{darkgoldenrod2}{rgb}{0.933,0.678,0.055}
\definecolor{darkgoldenrod3}{rgb}{0.804,0.584,0.047}
\definecolor{darkgoldenrod4}{rgb}{0.545,0.396,0.031}
\definecolor{darkgrey}{rgb}{0.663,0.663,0.663}
\definecolor{darkkhaki}{rgb}{0.741,0.718,0.42}
\definecolor{darkolivegreen}{rgb}{0.333,0.42,0.184}
\definecolor{darkolivegreen1}{rgb}{0.792,1,0.439}
\definecolor{darkolivegreen2}{rgb}{0.737,0.933,0.408}
\definecolor{darkolivegreen3}{rgb}{0.635,0.804,0.353}
\definecolor{darkolivegreen4}{rgb}{0.431,0.545,0.239}
\definecolor{darkorange1}{rgb}{1,0.498,0}
\definecolor{darkorange2}{rgb}{0.933,0.463,0}
\definecolor{darkorange3}{rgb}{0.804,0.4,0}
\definecolor{darkorange4}{rgb}{0.545,0.271,0}
\definecolor{darkorchid}{rgb}{0.6,0.196,0.8}
\definecolor{darkorchid1}{rgb}{0.749,0.243,1}
\definecolor{darkorchid2}{rgb}{0.698,0.227,0.933}
\definecolor{darkorchid3}{rgb}{0.604,0.196,0.804}
\definecolor{darkorchid4}{rgb}{0.408,0.133,0.545}
\definecolor{darksalmon}{rgb}{0.914,0.588,0.478}
\definecolor{darkseagreen}{rgb}{0.561,0.737,0.561}
\definecolor{darkseagreen1}{rgb}{0.757,1,0.757}
\definecolor{darkseagreen2}{rgb}{0.706,0.933,0.706}
\definecolor{darkseagreen3}{rgb}{0.608,0.804,0.608}
\definecolor{darkseagreen4}{rgb}{0.412,0.545,0.412}
\definecolor{darkslateblue}{rgb}{0.282,0.239,0.545}
\definecolor{darkslategray}{rgb}{0.184,0.31,0.31}
\definecolor{darkslategray1}{rgb}{0.592,1,1}
\definecolor{darkslategray2}{rgb}{0.553,0.933,0.933}
\definecolor{darkslategray3}{rgb}{0.475,0.804,0.804}
\definecolor{darkslategray4}{rgb}{0.322,0.545,0.545}
\definecolor{darkslategrey}{rgb}{0.184,0.31,0.31}
\definecolor{darkturquoise}{rgb}{0,0.808,0.82}
\definecolor{darkviolet}{rgb}{0.58,0,0.827}
\definecolor{deeppink}{rgb}{1,0.078,0.576}
\definecolor{deeppink1}{rgb}{1,0.078,0.576}
\definecolor{deeppink2}{rgb}{0.933,0.071,0.537}
\definecolor{deeppink3}{rgb}{0.804,0.063,0.463}
\definecolor{deeppink4}{rgb}{0.545,0.039,0.314}
\definecolor{deepskyblue}{rgb}{0,0.749,1}
\definecolor{deepskyblue1}{rgb}{0,0.749,1}
\definecolor{deepskyblue2}{rgb}{0,0.698,0.933}
\definecolor{deepskyblue3}{rgb}{0,0.604,0.804}
\definecolor{deepskyblue4}{rgb}{0,0.408,0.545}
\definecolor{dimgray}{rgb}{0.412,0.412,0.412}
\definecolor{dimgrey}{rgb}{0.412,0.412,0.412}
\definecolor{dodgerblue}{rgb}{0.118,0.565,1}
\definecolor{dodgerblue1}{rgb}{0.118,0.565,1}
\definecolor{dodgerblue2}{rgb}{0.11,0.525,0.933}
\definecolor{dodgerblue3}{rgb}{0.094,0.455,0.804}
\definecolor{dodgerblue4}{rgb}{0.063,0.306,0.545}
\definecolor{firebrick}{rgb}{0.698,0.133,0.133}
\definecolor{firebrick1}{rgb}{1,0.188,0.188}
\definecolor{firebrick2}{rgb}{0.933,0.173,0.173}
\definecolor{firebrick3}{rgb}{0.804,0.149,0.149}
\definecolor{firebrick4}{rgb}{0.545,0.102,0.102}
\definecolor{floralwhite}{rgb}{1,0.98,0.941}
\definecolor{forestgreen}{rgb}{0.133,0.545,0.133}
\definecolor{gainsboro}{rgb}{0.863,0.863,0.863}
\definecolor{ghostwhite}{rgb}{0.973,0.973,1}
\definecolor{gold1}{rgb}{1,0.843,0}
\definecolor{gold2}{rgb}{0.933,0.788,0}
\definecolor{gold3}{rgb}{0.804,0.678,0}
\definecolor{gold4}{rgb}{0.545,0.459,0}
\definecolor{goldenrod}{rgb}{0.855,0.647,0.125}
\definecolor{goldenrod1}{rgb}{1,0.757,0.145}
\definecolor{goldenrod2}{rgb}{0.933,0.706,0.133}
\definecolor{goldenrod3}{rgb}{0.804,0.608,0.114}
\definecolor{goldenrod4}{rgb}{0.545,0.412,0.078}
\definecolor{gray0}{rgb}{0,0,0}
\definecolor{gray1}{rgb}{0.012,0.012,0.012}
\definecolor{gray10}{rgb}{0.102,0.102,0.102}
\definecolor{gray100}{rgb}{1,1,1}
\definecolor{gray11}{rgb}{0.11,0.11,0.11}
\definecolor{gray12}{rgb}{0.122,0.122,0.122}
\definecolor{gray13}{rgb}{0.129,0.129,0.129}
\definecolor{gray14}{rgb}{0.141,0.141,0.141}
\definecolor{gray15}{rgb}{0.149,0.149,0.149}
\definecolor{gray16}{rgb}{0.161,0.161,0.161}
\definecolor{gray17}{rgb}{0.169,0.169,0.169}
\definecolor{gray18}{rgb}{0.18,0.18,0.18}
\definecolor{gray19}{rgb}{0.188,0.188,0.188}
\definecolor{gray2}{rgb}{0.02,0.02,0.02}
\definecolor{gray20}{rgb}{0.2,0.2,0.2}
\definecolor{gray21}{rgb}{0.212,0.212,0.212}
\definecolor{gray22}{rgb}{0.22,0.22,0.22}
\definecolor{gray23}{rgb}{0.231,0.231,0.231}
\definecolor{gray24}{rgb}{0.239,0.239,0.239}
\definecolor{gray25}{rgb}{0.251,0.251,0.251}
\definecolor{gray26}{rgb}{0.259,0.259,0.259}
\definecolor{gray27}{rgb}{0.271,0.271,0.271}
\definecolor{gray28}{rgb}{0.278,0.278,0.278}
\definecolor{gray29}{rgb}{0.29,0.29,0.29}
\definecolor{gray3}{rgb}{0.031,0.031,0.031}
\definecolor{gray30}{rgb}{0.302,0.302,0.302}
\definecolor{gray31}{rgb}{0.31,0.31,0.31}
\definecolor{gray32}{rgb}{0.322,0.322,0.322}
\definecolor{gray33}{rgb}{0.329,0.329,0.329}
\definecolor{gray34}{rgb}{0.341,0.341,0.341}
\definecolor{gray35}{rgb}{0.349,0.349,0.349}
\definecolor{gray36}{rgb}{0.361,0.361,0.361}
\definecolor{gray37}{rgb}{0.369,0.369,0.369}
\definecolor{gray38}{rgb}{0.38,0.38,0.38}
\definecolor{gray39}{rgb}{0.388,0.388,0.388}
\definecolor{gray4}{rgb}{0.039,0.039,0.039}
\definecolor{gray40}{rgb}{0.4,0.4,0.4}
\definecolor{gray41}{rgb}{0.412,0.412,0.412}
\definecolor{gray42}{rgb}{0.42,0.42,0.42}
\definecolor{gray43}{rgb}{0.431,0.431,0.431}
\definecolor{gray44}{rgb}{0.439,0.439,0.439}
\definecolor{gray45}{rgb}{0.451,0.451,0.451}
\definecolor{gray46}{rgb}{0.459,0.459,0.459}
\definecolor{gray47}{rgb}{0.471,0.471,0.471}
\definecolor{gray48}{rgb}{0.478,0.478,0.478}
\definecolor{gray49}{rgb}{0.49,0.49,0.49}
\definecolor{gray5}{rgb}{0.051,0.051,0.051}
\definecolor{gray50}{rgb}{0.498,0.498,0.498}
\definecolor{gray51}{rgb}{0.51,0.51,0.51}
\definecolor{gray52}{rgb}{0.522,0.522,0.522}
\definecolor{gray53}{rgb}{0.529,0.529,0.529}
\definecolor{gray54}{rgb}{0.541,0.541,0.541}
\definecolor{gray55}{rgb}{0.549,0.549,0.549}
\definecolor{gray56}{rgb}{0.561,0.561,0.561}
\definecolor{gray57}{rgb}{0.569,0.569,0.569}
\definecolor{gray58}{rgb}{0.58,0.58,0.58}
\definecolor{gray59}{rgb}{0.588,0.588,0.588}
\definecolor{gray6}{rgb}{0.059,0.059,0.059}
\definecolor{gray60}{rgb}{0.6,0.6,0.6}
\definecolor{gray61}{rgb}{0.612,0.612,0.612}
\definecolor{gray62}{rgb}{0.62,0.62,0.62}
\definecolor{gray63}{rgb}{0.631,0.631,0.631}
\definecolor{gray64}{rgb}{0.639,0.639,0.639}
\definecolor{gray65}{rgb}{0.651,0.651,0.651}
\definecolor{gray66}{rgb}{0.659,0.659,0.659}
\definecolor{gray67}{rgb}{0.671,0.671,0.671}
\definecolor{gray68}{rgb}{0.678,0.678,0.678}
\definecolor{gray69}{rgb}{0.69,0.69,0.69}
\definecolor{gray7}{rgb}{0.071,0.071,0.071}
\definecolor{gray70}{rgb}{0.702,0.702,0.702}
\definecolor{gray71}{rgb}{0.71,0.71,0.71}
\definecolor{gray72}{rgb}{0.722,0.722,0.722}
\definecolor{gray73}{rgb}{0.729,0.729,0.729}
\definecolor{gray74}{rgb}{0.741,0.741,0.741}
\definecolor{gray75}{rgb}{0.749,0.749,0.749}
\definecolor{gray76}{rgb}{0.761,0.761,0.761}
\definecolor{gray77}{rgb}{0.769,0.769,0.769}
\definecolor{gray78}{rgb}{0.78,0.78,0.78}
\definecolor{gray79}{rgb}{0.788,0.788,0.788}
\definecolor{gray8}{rgb}{0.078,0.078,0.078}
\definecolor{gray80}{rgb}{0.8,0.8,0.8}
\definecolor{gray81}{rgb}{0.812,0.812,0.812}
\definecolor{gray82}{rgb}{0.82,0.82,0.82}
\definecolor{gray83}{rgb}{0.831,0.831,0.831}
\definecolor{gray84}{rgb}{0.839,0.839,0.839}
\definecolor{gray85}{rgb}{0.851,0.851,0.851}
\definecolor{gray86}{rgb}{0.859,0.859,0.859}
\definecolor{gray87}{rgb}{0.871,0.871,0.871}
\definecolor{gray88}{rgb}{0.878,0.878,0.878}
\definecolor{gray89}{rgb}{0.89,0.89,0.89}
\definecolor{gray9}{rgb}{0.09,0.09,0.09}
\definecolor{gray90}{rgb}{0.898,0.898,0.898}
\definecolor{gray91}{rgb}{0.91,0.91,0.91}
\definecolor{gray92}{rgb}{0.922,0.922,0.922}
\definecolor{gray93}{rgb}{0.929,0.929,0.929}
\definecolor{gray94}{rgb}{0.941,0.941,0.941}
\definecolor{gray95}{rgb}{0.949,0.949,0.949}
\definecolor{gray96}{rgb}{0.961,0.961,0.961}
\definecolor{gray97}{rgb}{0.969,0.969,0.969}
\definecolor{gray98}{rgb}{0.98,0.98,0.98}
\definecolor{gray99}{rgb}{0.988,0.988,0.988}
\definecolor{green1}{rgb}{0,1,0}
\definecolor{green2}{rgb}{0,0.933,0}
\definecolor{green3}{rgb}{0,0.804,0}
\definecolor{green4}{rgb}{0,0.545,0}
\definecolor{greenyellow}{rgb}{0.678,1,0.184}
\definecolor{grey}{rgb}{0.745,0.745,0.745}
\definecolor{grey0}{rgb}{0,0,0}
\definecolor{grey1}{rgb}{0.012,0.012,0.012}
\definecolor{grey10}{rgb}{0.102,0.102,0.102}
\definecolor{grey100}{rgb}{1,1,1}
\definecolor{grey11}{rgb}{0.11,0.11,0.11}
\definecolor{grey12}{rgb}{0.122,0.122,0.122}
\definecolor{grey13}{rgb}{0.129,0.129,0.129}
\definecolor{grey14}{rgb}{0.141,0.141,0.141}
\definecolor{grey15}{rgb}{0.149,0.149,0.149}
\definecolor{grey16}{rgb}{0.161,0.161,0.161}
\definecolor{grey17}{rgb}{0.169,0.169,0.169}
\definecolor{grey18}{rgb}{0.18,0.18,0.18}
\definecolor{grey19}{rgb}{0.188,0.188,0.188}
\definecolor{grey2}{rgb}{0.02,0.02,0.02}
\definecolor{grey20}{rgb}{0.2,0.2,0.2}
\definecolor{grey21}{rgb}{0.212,0.212,0.212}
\definecolor{grey22}{rgb}{0.22,0.22,0.22}
\definecolor{grey23}{rgb}{0.231,0.231,0.231}
\definecolor{grey24}{rgb}{0.239,0.239,0.239}
\definecolor{grey25}{rgb}{0.251,0.251,0.251}
\definecolor{grey26}{rgb}{0.259,0.259,0.259}
\definecolor{grey27}{rgb}{0.271,0.271,0.271}
\definecolor{grey28}{rgb}{0.278,0.278,0.278}
\definecolor{grey29}{rgb}{0.29,0.29,0.29}
\definecolor{grey3}{rgb}{0.031,0.031,0.031}
\definecolor{grey30}{rgb}{0.302,0.302,0.302}
\definecolor{grey31}{rgb}{0.31,0.31,0.31}
\definecolor{grey32}{rgb}{0.322,0.322,0.322}
\definecolor{grey33}{rgb}{0.329,0.329,0.329}
\definecolor{grey34}{rgb}{0.341,0.341,0.341}
\definecolor{grey35}{rgb}{0.349,0.349,0.349}
\definecolor{grey36}{rgb}{0.361,0.361,0.361}
\definecolor{grey37}{rgb}{0.369,0.369,0.369}
\definecolor{grey38}{rgb}{0.38,0.38,0.38}
\definecolor{grey39}{rgb}{0.388,0.388,0.388}
\definecolor{grey4}{rgb}{0.039,0.039,0.039}
\definecolor{grey40}{rgb}{0.4,0.4,0.4}
\definecolor{grey41}{rgb}{0.412,0.412,0.412}
\definecolor{grey42}{rgb}{0.42,0.42,0.42}
\definecolor{grey43}{rgb}{0.431,0.431,0.431}
\definecolor{grey44}{rgb}{0.439,0.439,0.439}
\definecolor{grey45}{rgb}{0.451,0.451,0.451}
\definecolor{grey46}{rgb}{0.459,0.459,0.459}
\definecolor{grey47}{rgb}{0.471,0.471,0.471}
\definecolor{grey48}{rgb}{0.478,0.478,0.478}
\definecolor{grey49}{rgb}{0.49,0.49,0.49}
\definecolor{grey5}{rgb}{0.051,0.051,0.051}
\definecolor{grey50}{rgb}{0.498,0.498,0.498}
\definecolor{grey51}{rgb}{0.51,0.51,0.51}
\definecolor{grey52}{rgb}{0.522,0.522,0.522}
\definecolor{grey53}{rgb}{0.529,0.529,0.529}
\definecolor{grey54}{rgb}{0.541,0.541,0.541}
\definecolor{grey55}{rgb}{0.549,0.549,0.549}
\definecolor{grey56}{rgb}{0.561,0.561,0.561}
\definecolor{grey57}{rgb}{0.569,0.569,0.569}
\definecolor{grey58}{rgb}{0.58,0.58,0.58}
\definecolor{grey59}{rgb}{0.588,0.588,0.588}
\definecolor{grey6}{rgb}{0.059,0.059,0.059}
\definecolor{grey60}{rgb}{0.6,0.6,0.6}
\definecolor{grey61}{rgb}{0.612,0.612,0.612}
\definecolor{grey62}{rgb}{0.62,0.62,0.62}
\definecolor{grey63}{rgb}{0.631,0.631,0.631}
\definecolor{grey64}{rgb}{0.639,0.639,0.639}
\definecolor{grey65}{rgb}{0.651,0.651,0.651}
\definecolor{grey66}{rgb}{0.659,0.659,0.659}
\definecolor{grey67}{rgb}{0.671,0.671,0.671}
\definecolor{grey68}{rgb}{0.678,0.678,0.678}
\definecolor{grey69}{rgb}{0.69,0.69,0.69}
\definecolor{grey7}{rgb}{0.071,0.071,0.071}
\definecolor{grey70}{rgb}{0.702,0.702,0.702}
\definecolor{grey71}{rgb}{0.71,0.71,0.71}
\definecolor{grey72}{rgb}{0.722,0.722,0.722}
\definecolor{grey73}{rgb}{0.729,0.729,0.729}
\definecolor{grey74}{rgb}{0.741,0.741,0.741}
\definecolor{grey75}{rgb}{0.749,0.749,0.749}
\definecolor{grey76}{rgb}{0.761,0.761,0.761}
\definecolor{grey77}{rgb}{0.769,0.769,0.769}
\definecolor{grey78}{rgb}{0.78,0.78,0.78}
\definecolor{grey79}{rgb}{0.788,0.788,0.788}
\definecolor{grey8}{rgb}{0.078,0.078,0.078}
\definecolor{grey80}{rgb}{0.8,0.8,0.8}
\definecolor{grey81}{rgb}{0.812,0.812,0.812}
\definecolor{grey82}{rgb}{0.82,0.82,0.82}
\definecolor{grey83}{rgb}{0.831,0.831,0.831}
\definecolor{grey84}{rgb}{0.839,0.839,0.839}
\definecolor{grey85}{rgb}{0.851,0.851,0.851}
\definecolor{grey86}{rgb}{0.859,0.859,0.859}
\definecolor{grey87}{rgb}{0.871,0.871,0.871}
\definecolor{grey88}{rgb}{0.878,0.878,0.878}
\definecolor{grey89}{rgb}{0.89,0.89,0.89}
\definecolor{grey9}{rgb}{0.09,0.09,0.09}
\definecolor{grey90}{rgb}{0.898,0.898,0.898}
\definecolor{grey91}{rgb}{0.91,0.91,0.91}
\definecolor{grey92}{rgb}{0.922,0.922,0.922}
\definecolor{grey93}{rgb}{0.929,0.929,0.929}
\definecolor{grey94}{rgb}{0.941,0.941,0.941}
\definecolor{grey95}{rgb}{0.949,0.949,0.949}
\definecolor{grey96}{rgb}{0.961,0.961,0.961}
\definecolor{grey97}{rgb}{0.969,0.969,0.969}
\definecolor{grey98}{rgb}{0.98,0.98,0.98}
\definecolor{grey99}{rgb}{0.988,0.988,0.988}
\definecolor{honeydew}{rgb}{0.941,1,0.941}
\definecolor{honeydew1}{rgb}{0.941,1,0.941}
\definecolor{honeydew2}{rgb}{0.878,0.933,0.878}
\definecolor{honeydew3}{rgb}{0.757,0.804,0.757}
\definecolor{honeydew4}{rgb}{0.514,0.545,0.514}
\definecolor{hotpink}{rgb}{1,0.412,0.706}
\definecolor{hotpink1}{rgb}{1,0.431,0.706}
\definecolor{hotpink2}{rgb}{0.933,0.416,0.655}
\definecolor{hotpink3}{rgb}{0.804,0.376,0.565}
\definecolor{hotpink4}{rgb}{0.545,0.227,0.384}
\definecolor{indianred}{rgb}{0.804,0.361,0.361}
\definecolor{indianred1}{rgb}{1,0.416,0.416}
\definecolor{indianred2}{rgb}{0.933,0.388,0.388}
\definecolor{indianred3}{rgb}{0.804,0.333,0.333}
\definecolor{indianred4}{rgb}{0.545,0.227,0.227}
\definecolor{ivory}{rgb}{1,1,0.941}
\definecolor{ivory1}{rgb}{1,1,0.941}
\definecolor{ivory2}{rgb}{0.933,0.933,0.878}
\definecolor{ivory3}{rgb}{0.804,0.804,0.757}
\definecolor{ivory4}{rgb}{0.545,0.545,0.514}
\definecolor{khaki}{rgb}{0.941,0.902,0.549}
\definecolor{khaki1}{rgb}{1,0.965,0.561}
\definecolor{khaki2}{rgb}{0.933,0.902,0.522}
\definecolor{khaki3}{rgb}{0.804,0.776,0.451}
\definecolor{khaki4}{rgb}{0.545,0.525,0.306}
\definecolor{lavender}{rgb}{0.902,0.902,0.98}
\definecolor{lavenderblush}{rgb}{1,0.941,0.961}
\definecolor{lavenderblush1}{rgb}{1,0.941,0.961}
\definecolor{lavenderblush2}{rgb}{0.933,0.878,0.898}
\definecolor{lavenderblush3}{rgb}{0.804,0.757,0.773}
\definecolor{lavenderblush4}{rgb}{0.545,0.514,0.525}
\definecolor{lawngreen}{rgb}{0.486,0.988,0}
\definecolor{lemonchiffon}{rgb}{1,0.98,0.804}
\definecolor{lemonchiffon1}{rgb}{1,0.98,0.804}
\definecolor{lemonchiffon2}{rgb}{0.933,0.914,0.749}
\definecolor{lemonchiffon3}{rgb}{0.804,0.788,0.647}
\definecolor{lemonchiffon4}{rgb}{0.545,0.537,0.439}
\definecolor{lightblue1}{rgb}{0.749,0.937,1}
\definecolor{lightblue2}{rgb}{0.698,0.875,0.933}
\definecolor{lightblue3}{rgb}{0.604,0.753,0.804}
\definecolor{lightblue4}{rgb}{0.408,0.514,0.545}
\definecolor{lightcoral}{rgb}{0.941,0.502,0.502}
\definecolor{lightcyan1}{rgb}{0.878,1,1}
\definecolor{lightcyan2}{rgb}{0.82,0.933,0.933}
\definecolor{lightcyan3}{rgb}{0.706,0.804,0.804}
\definecolor{lightcyan4}{rgb}{0.478,0.545,0.545}
\definecolor{lightgoldenrod}{rgb}{0.933,0.867,0.51}
\definecolor{lightgoldenrod1}{rgb}{1,0.925,0.545}
\definecolor{lightgoldenrod2}{rgb}{0.933,0.863,0.51}
\definecolor{lightgoldenrod3}{rgb}{0.804,0.745,0.439}
\definecolor{lightgoldenrod4}{rgb}{0.545,0.506,0.298}
\definecolor{lightgoldenrodyellow}{rgb}{0.98,0.98,0.824}
\definecolor{lightgrey}{rgb}{0.827,0.827,0.827}
\definecolor{lightpink}{rgb}{1,0.714,0.757}
\definecolor{lightpink1}{rgb}{1,0.682,0.725}
\definecolor{lightpink2}{rgb}{0.933,0.635,0.678}
\definecolor{lightpink3}{rgb}{0.804,0.549,0.584}
\definecolor{lightpink4}{rgb}{0.545,0.373,0.396}
\definecolor{lightsalmon}{rgb}{1,0.627,0.478}
\definecolor{lightsalmon1}{rgb}{1,0.627,0.478}
\definecolor{lightsalmon2}{rgb}{0.933,0.584,0.447}
\definecolor{lightsalmon3}{rgb}{0.804,0.506,0.384}
\definecolor{lightsalmon4}{rgb}{0.545,0.341,0.259}
\definecolor{lightseagreen}{rgb}{0.125,0.698,0.667}
\definecolor{lightskyblue}{rgb}{0.529,0.808,0.98}
\definecolor{lightskyblue1}{rgb}{0.69,0.886,1}
\definecolor{lightskyblue2}{rgb}{0.643,0.827,0.933}
\definecolor{lightskyblue3}{rgb}{0.553,0.714,0.804}
\definecolor{lightskyblue4}{rgb}{0.376,0.482,0.545}
\definecolor{lightslateblue}{rgb}{0.518,0.439,1}
\definecolor{lightslategray}{rgb}{0.467,0.533,0.6}
\definecolor{lightslategrey}{rgb}{0.467,0.533,0.6}
\definecolor{lightsteelblue}{rgb}{0.69,0.769,0.871}
\definecolor{lightsteelblue1}{rgb}{0.792,0.882,1}
\definecolor{lightsteelblue2}{rgb}{0.737,0.824,0.933}
\definecolor{lightsteelblue3}{rgb}{0.635,0.71,0.804}
\definecolor{lightsteelblue4}{rgb}{0.431,0.482,0.545}
\definecolor{lightyellow1}{rgb}{1,1,0.878}
\definecolor{lightyellow2}{rgb}{0.933,0.933,0.82}
\definecolor{lightyellow3}{rgb}{0.804,0.804,0.706}
\definecolor{lightyellow4}{rgb}{0.545,0.545,0.478}
\definecolor{limegreen}{rgb}{0.196,0.804,0.196}
\definecolor{linen}{rgb}{0.98,0.941,0.902}
\definecolor{magenta}{rgb}{1,0,1}
\definecolor{magenta1}{rgb}{1,0,1}
\definecolor{magenta2}{rgb}{0.933,0,0.933}
\definecolor{magenta3}{rgb}{0.804,0,0.804}
\definecolor{magenta4}{rgb}{0.545,0,0.545}
\definecolor{maroon}{rgb}{0.69,0.188,0.376}
\definecolor{maroon1}{rgb}{1,0.204,0.702}
\definecolor{maroon2}{rgb}{0.933,0.188,0.655}
\definecolor{maroon3}{rgb}{0.804,0.161,0.565}
\definecolor{maroon4}{rgb}{0.545,0.11,0.384}
\definecolor{mediumaquamarine}{rgb}{0.4,0.804,0.667}
\definecolor{mediumblue}{rgb}{0,0,0.804}
\definecolor{mediumorchid}{rgb}{0.729,0.333,0.827}
\definecolor{mediumorchid1}{rgb}{0.878,0.4,1}
\definecolor{mediumorchid2}{rgb}{0.82,0.373,0.933}
\definecolor{mediumorchid3}{rgb}{0.706,0.322,0.804}
\definecolor{mediumorchid4}{rgb}{0.478,0.216,0.545}
\definecolor{mediumpurple}{rgb}{0.576,0.439,0.859}
\definecolor{mediumpurple1}{rgb}{0.671,0.51,1}
\definecolor{mediumpurple2}{rgb}{0.624,0.475,0.933}
\definecolor{mediumpurple3}{rgb}{0.537,0.408,0.804}
\definecolor{mediumpurple4}{rgb}{0.365,0.278,0.545}
\definecolor{mediumseagreen}{rgb}{0.235,0.702,0.443}
\definecolor{mediumslateblue}{rgb}{0.482,0.408,0.933}
\definecolor{mediumspringgreen}{rgb}{0,0.98,0.604}
\definecolor{mediumturquoise}{rgb}{0.282,0.82,0.8}
\definecolor{mediumvioletred}{rgb}{0.78,0.082,0.522}
\definecolor{midnightblue}{rgb}{0.098,0.098,0.439}
\definecolor{mintcream}{rgb}{0.961,1,0.98}
\definecolor{mistyrose}{rgb}{1,0.894,0.882}
\definecolor{mistyrose1}{rgb}{1,0.894,0.882}
\definecolor{mistyrose2}{rgb}{0.933,0.835,0.824}
\definecolor{mistyrose3}{rgb}{0.804,0.718,0.71}
\definecolor{mistyrose4}{rgb}{0.545,0.49,0.482}
\definecolor{moccasin}{rgb}{1,0.894,0.71}
\definecolor{navajowhite}{rgb}{1,0.871,0.678}
\definecolor{navajowhite1}{rgb}{1,0.871,0.678}
\definecolor{navajowhite2}{rgb}{0.933,0.812,0.631}
\definecolor{navajowhite3}{rgb}{0.804,0.702,0.545}
\definecolor{navajowhite4}{rgb}{0.545,0.475,0.369}
\definecolor{navyblue}{rgb}{0,0,0.502}
\definecolor{oldlace}{rgb}{0.992,0.961,0.902}
\definecolor{olivedrab}{rgb}{0.42,0.557,0.137}
\definecolor{olivedrab1}{rgb}{0.753,1,0.243}
\definecolor{olivedrab2}{rgb}{0.702,0.933,0.227}
\definecolor{olivedrab3}{rgb}{0.604,0.804,0.196}
\definecolor{olivedrab4}{rgb}{0.412,0.545,0.133}
\definecolor{orange1}{rgb}{1,0.647,0}
\definecolor{orange2}{rgb}{0.933,0.604,0}
\definecolor{orange3}{rgb}{0.804,0.522,0}
\definecolor{orange4}{rgb}{0.545,0.353,0}
\definecolor{orangered}{rgb}{1,0.271,0}
\definecolor{orangered1}{rgb}{1,0.271,0}
\definecolor{orangered2}{rgb}{0.933,0.251,0}
\definecolor{orangered3}{rgb}{0.804,0.216,0}
\definecolor{orangered4}{rgb}{0.545,0.145,0}
\definecolor{orchid}{rgb}{0.855,0.439,0.839}
\definecolor{orchid1}{rgb}{1,0.514,0.98}
\definecolor{orchid2}{rgb}{0.933,0.478,0.914}
\definecolor{orchid3}{rgb}{0.804,0.412,0.788}
\definecolor{orchid4}{rgb}{0.545,0.278,0.537}
\definecolor{palegoldenrod}{rgb}{0.933,0.91,0.667}
\definecolor{palegreen}{rgb}{0.596,0.984,0.596}
\definecolor{palegreen1}{rgb}{0.604,1,0.604}
\definecolor{palegreen2}{rgb}{0.565,0.933,0.565}
\definecolor{palegreen3}{rgb}{0.486,0.804,0.486}
\definecolor{palegreen4}{rgb}{0.329,0.545,0.329}
\definecolor{paleturquoise}{rgb}{0.686,0.933,0.933}
\definecolor{paleturquoise1}{rgb}{0.733,1,1}
\definecolor{paleturquoise2}{rgb}{0.682,0.933,0.933}
\definecolor{paleturquoise3}{rgb}{0.588,0.804,0.804}
\definecolor{paleturquoise4}{rgb}{0.4,0.545,0.545}
\definecolor{palevioletred}{rgb}{0.859,0.439,0.576}
\definecolor{palevioletred1}{rgb}{1,0.51,0.671}
\definecolor{palevioletred2}{rgb}{0.933,0.475,0.624}
\definecolor{palevioletred3}{rgb}{0.804,0.408,0.537}
\definecolor{palevioletred4}{rgb}{0.545,0.278,0.365}
\definecolor{papayawhip}{rgb}{1,0.937,0.835}
\definecolor{peachpuff}{rgb}{1,0.855,0.725}
\definecolor{peachpuff1}{rgb}{1,0.855,0.725}
\definecolor{peachpuff2}{rgb}{0.933,0.796,0.678}
\definecolor{peachpuff3}{rgb}{0.804,0.686,0.584}
\definecolor{peachpuff4}{rgb}{0.545,0.467,0.396}
\definecolor{peru}{rgb}{0.804,0.522,0.247}
\definecolor{pink1}{rgb}{1,0.71,0.773}
\definecolor{pink2}{rgb}{0.933,0.663,0.722}
\definecolor{pink3}{rgb}{0.804,0.569,0.62}
\definecolor{pink4}{rgb}{0.545,0.388,0.424}
\definecolor{plum}{rgb}{0.867,0.627,0.867}
\definecolor{plum1}{rgb}{1,0.733,1}
\definecolor{plum2}{rgb}{0.933,0.682,0.933}
\definecolor{plum3}{rgb}{0.804,0.588,0.804}
\definecolor{plum4}{rgb}{0.545,0.4,0.545}
\definecolor{powderblue}{rgb}{0.69,0.878,0.902}
\definecolor{purple1}{rgb}{0.608,0.188,1}
\definecolor{purple2}{rgb}{0.569,0.173,0.933}
\definecolor{purple3}{rgb}{0.49,0.149,0.804}
\definecolor{purple4}{rgb}{0.333,0.102,0.545}
\definecolor{red1}{rgb}{1,0,0}
\definecolor{red2}{rgb}{0.933,0,0}
\definecolor{red3}{rgb}{0.804,0,0}
\definecolor{red4}{rgb}{0.545,0,0}
\definecolor{rosybrown}{rgb}{0.737,0.561,0.561}
\definecolor{rosybrown1}{rgb}{1,0.757,0.757}
\definecolor{rosybrown2}{rgb}{0.933,0.706,0.706}
\definecolor{rosybrown3}{rgb}{0.804,0.608,0.608}
\definecolor{rosybrown4}{rgb}{0.545,0.412,0.412}
\definecolor{royalblue}{rgb}{0.255,0.412,0.882}
\definecolor{royalblue1}{rgb}{0.282,0.463,1}
\definecolor{royalblue2}{rgb}{0.263,0.431,0.933}
\definecolor{royalblue3}{rgb}{0.227,0.373,0.804}
\definecolor{royalblue4}{rgb}{0.153,0.251,0.545}
\definecolor{saddlebrown}{rgb}{0.545,0.271,0.075}
\definecolor{salmon}{rgb}{0.98,0.502,0.447}
\definecolor{salmon1}{rgb}{1,0.549,0.412}
\definecolor{salmon2}{rgb}{0.933,0.51,0.384}
\definecolor{salmon3}{rgb}{0.804,0.439,0.329}
\definecolor{salmon4}{rgb}{0.545,0.298,0.224}
\definecolor{sandybrown}{rgb}{0.957,0.643,0.376}
\definecolor{seagreen1}{rgb}{0.329,1,0.624}
\definecolor{seagreen2}{rgb}{0.306,0.933,0.58}
\definecolor{seagreen3}{rgb}{0.263,0.804,0.502}
\definecolor{seagreen4}{rgb}{0.18,0.545,0.341}
\definecolor{seashell}{rgb}{1,0.961,0.933}
\definecolor{seashell1}{rgb}{1,0.961,0.933}
\definecolor{seashell2}{rgb}{0.933,0.898,0.871}
\definecolor{seashell3}{rgb}{0.804,0.773,0.749}
\definecolor{seashell4}{rgb}{0.545,0.525,0.51}
\definecolor{sienna}{rgb}{0.627,0.322,0.176}
\definecolor{sienna1}{rgb}{1,0.51,0.278}
\definecolor{sienna2}{rgb}{0.933,0.475,0.259}
\definecolor{sienna3}{rgb}{0.804,0.408,0.224}
\definecolor{sienna4}{rgb}{0.545,0.278,0.149}
\definecolor{skyblue}{rgb}{0.529,0.808,0.922}
\definecolor{skyblue1}{rgb}{0.529,0.808,1}
\definecolor{skyblue2}{rgb}{0.494,0.753,0.933}
\definecolor{skyblue3}{rgb}{0.424,0.651,0.804}
\definecolor{skyblue4}{rgb}{0.29,0.439,0.545}
\definecolor{slateblue}{rgb}{0.416,0.353,0.804}
\definecolor{slateblue1}{rgb}{0.514,0.435,1}
\definecolor{slateblue2}{rgb}{0.478,0.404,0.933}
\definecolor{slateblue3}{rgb}{0.412,0.349,0.804}
\definecolor{slateblue4}{rgb}{0.278,0.235,0.545}
\definecolor{slategray}{rgb}{0.439,0.502,0.565}
\definecolor{slategray1}{rgb}{0.776,0.886,1}
\definecolor{slategray2}{rgb}{0.725,0.827,0.933}
\definecolor{slategray3}{rgb}{0.624,0.714,0.804}
\definecolor{slategray4}{rgb}{0.424,0.482,0.545}
\definecolor{slategrey}{rgb}{0.439,0.502,0.565}
\definecolor{snow}{rgb}{1,0.98,0.98}
\definecolor{snow1}{rgb}{1,0.98,0.98}
\definecolor{snow2}{rgb}{0.933,0.914,0.914}
\definecolor{snow3}{rgb}{0.804,0.788,0.788}
\definecolor{snow4}{rgb}{0.545,0.537,0.537}
\definecolor{springgreen}{rgb}{0,1,0.498}
\definecolor{springgreen1}{rgb}{0,1,0.498}
\definecolor{springgreen2}{rgb}{0,0.933,0.463}
\definecolor{springgreen3}{rgb}{0,0.804,0.4}
\definecolor{springgreen4}{rgb}{0,0.545,0.271}
\definecolor{steelblue}{rgb}{0.275,0.51,0.706}
\definecolor{steelblue1}{rgb}{0.388,0.722,1}
\definecolor{steelblue2}{rgb}{0.361,0.675,0.933}
\definecolor{steelblue3}{rgb}{0.31,0.58,0.804}
\definecolor{steelblue4}{rgb}{0.212,0.392,0.545}
\definecolor{tan}{rgb}{0.824,0.706,0.549}
\definecolor{tan1}{rgb}{1,0.647,0.31}
\definecolor{tan2}{rgb}{0.933,0.604,0.286}
\definecolor{tan3}{rgb}{0.804,0.522,0.247}
\definecolor{tan4}{rgb}{0.545,0.353,0.169}
\definecolor{thistle}{rgb}{0.847,0.749,0.847}
\definecolor{thistle1}{rgb}{1,0.882,1}
\definecolor{thistle2}{rgb}{0.933,0.824,0.933}
\definecolor{thistle3}{rgb}{0.804,0.71,0.804}
\definecolor{thistle4}{rgb}{0.545,0.482,0.545}
\definecolor{tomato}{rgb}{1,0.388,0.278}
\definecolor{tomato1}{rgb}{1,0.388,0.278}
\definecolor{tomato2}{rgb}{0.933,0.361,0.259}
\definecolor{tomato3}{rgb}{0.804,0.31,0.224}
\definecolor{tomato4}{rgb}{0.545,0.212,0.149}
\definecolor{turquoise1}{rgb}{0,0.961,1}
\definecolor{turquoise2}{rgb}{0,0.898,0.933}
\definecolor{turquoise3}{rgb}{0,0.773,0.804}
\definecolor{turquoise4}{rgb}{0,0.525,0.545}
\definecolor{violetred}{rgb}{0.816,0.125,0.565}
\definecolor{violetred1}{rgb}{1,0.243,0.588}
\definecolor{violetred2}{rgb}{0.933,0.227,0.549}
\definecolor{violetred3}{rgb}{0.804,0.196,0.471}
\definecolor{violetred4}{rgb}{0.545,0.133,0.322}
\definecolor{wheat}{rgb}{0.961,0.871,0.702}
\definecolor{wheat1}{rgb}{1,0.906,0.729}
\definecolor{wheat2}{rgb}{0.933,0.847,0.682}
\definecolor{wheat3}{rgb}{0.804,0.729,0.588}
\definecolor{wheat4}{rgb}{0.545,0.494,0.4}
\definecolor{whitesmoke}{rgb}{0.961,0.961,0.961}
\definecolor{yellow1}{rgb}{1,1,0}
\definecolor{yellow2}{rgb}{0.933,0.933,0}
\definecolor{yellow3}{rgb}{0.804,0.804,0}
\definecolor{yellow4}{rgb}{0.545,0.545,0}
\definecolor{yellowgreen}{rgb}{0.604,0.804,0.196}

	\end{center}
	\caption{An example of a set $I^{(d)}$ inside the extended compass of a flatness pair of a given graph and the position of $Z,{X}_{\rm in},X_{\rm out}, V_L ({\bf a}),$ and $F'.$}
	\labels{@accomplishes}
\end{figure}

In the proof of~\autoref{@desmembramientos}, we will find two extended compasses $(\mathfrak{K},R),$ $(\mathfrak{K}',R')$ with the same $\blue{{\sf out}\text{-}{\sf sig}}$ for a particular choice of $d$ and $L$ and some choices $Z$ and $Z',$ respectively.
In the proof, $Z$ and $Z'$
will be exchanged as well as the $\tilde{X}$'s inside those $Z$ and $Z'.$
Here it is important to notice that the graph $F$ is always the same (for both $(\mathfrak{K},R)$ and $(\mathfrak{K}',R')$)
and constitutes the fictitious ``invariant'' part of the graph, that is not affected during this exchange.
See~\autoref{@governorship} for the great picture –– what is $F$ will not be exchanged, while $\partial_{\mathfrak{K}} (Z)$ will be substituted by the isomorphic $\partial_{\mathfrak{K}'} (Z')$ (see also~\autoref{figure_boundariedgraph5}).

\myskip\subsection{In-signature}
\labels{sec_in-sig_first-floor}

Let $\tilde{θ}_q$ be the split version of $θ_{{\sf R},{\bf c}},$ as in the previous subsection.
Recall that $\breve{ζ}_{\sf R}∈\FOL[τ^{\langle{\bf c}\rangle}\cup\{{\sf R}\}].$
there exist $p∈ \mathbb{N}_{≥ 1},$ $\ell_1, \ldots, \ell_p, r_1, \ldots, r_p∈ \mathbb{N}_{≥ 1},$ and sentences $\tilde{ζ}_1, \ldots, \tilde{ζ}_p∈ \FOL[τ^{\langle{\bf c}\rangle}\cup\{{\sf R}\}]$ such that $\breve{ζ}_{\sf R}$ is a Boolean combination of $\tilde{ζ}_1, \ldots, \tilde{ζ}_p$ and for every $h∈ [p],$ $\tilde{ζ}_h$ is a basic local sentence with parameters $\ell_h$ and $r_h,$ i.e.,
\begin{eqnarray*}
\tilde{ζ}_h = \exists {\sf x}_{1}\ldots\exists {\sf x}_{\ell_{h}}\big(\bigwedge_{i∈ [\ell_{h}]}{\sf x}_{i}∈ {\sf R}\wedge \bigwedge_{1≤ i<j≤ \ell_{h}} d({\sf x}_{i}, {\sf x}_{j})> 2r_{h}\wedge \bigwedge_{i∈ [\ell_{h}]}ψ_{h}({\sf x}_{i})\big),
\end{eqnarray*}
where $ψ_h$ is an $r_h$-local formula  in $\FOL[τ^{\langle{\bf c}\rangle}]$ with one free variable.
Keep in mind that, since $\breve{ζ}_{\sf R}∈ \FOL[τ^{\langle{\bf c}\rangle}\cup\{{\sf R}\}],$ distances are measured in the Gaifman graph of $τ^{\langle{\bf c}\rangle}$-structures.
\medskip

We set $\hat{r} := \max_{h∈[p]}\{r_h\},$ $\hat{\ell} :=  \max_{h∈ [p]}\{\ell_h\},$ and
$r :=2\cdot (\hat{\ell}+ 3)\cdot \hat{r}.$
As in the previous subsection,
we keep in mind that $q=(\tw(θ)+1)^2+1,$ we let $j'∈ \mathbb{N},$ and we set
\begin{eqnarray*}
j & := &  {\sf odd}(\max\{q/2,j'\}) \mbox{~and~}\\
w & :=& (r+2)\cdot q.\\
\end{eqnarray*}
The reason that $r$ is set to be equal to $2\cdot (\hat{\ell}+ 3)\cdot \hat{r}$ will be clear in the proof of~\autoref{@desmembramientos}
and is based on an idea already present in~\cite{FominGST20analgo}.
\medskip

\myskip\paragraph{Scattered sets in structures.}
Let $\mathfrak{A}$ be a $τ$-structure and let $X\subseteq V(\mathfrak{A}).$
We say that $X$ is {\em $(\ell,r)$-scattered in $\mathfrak{A}$}, if $|X| = \ell$ and for every two distinct vertices in $X,$ their distance in the Gaifman graph $G_{\mathfrak{A}}$ is more than $2r,$ i.e., for every $a,b∈ X, a\neq b,$ it holds that
$d^{\mathfrak{A}}(a,b)>2r.$

\myskip\paragraph{The in-signature of an extended compass.}
We now define the in-signature of an extended compass.
In this, using the approach of~\cite{FominGST20analgo},
we encode all (partial) sets of variables, one set for each basic local sentence of the Gaifman sentence $\breve{ζ}_{\sf R},$ such that these variables are lying inside an ``inner part'' of the compass, they are scattered in this inner part, and they satisfy the local formulas $ψ_i.$
These arguments are always applied in some $τ^{\langle \bf c\rangle}$-structure of the form ${\sf ap}_{\bf c}(\mathfrak{A}, {\bf a}).$
We define
$$\green{{\sf SIG}_{\sf in}} = 2^{[\ell_1]}\times\cdots\times 2^{[\ell_{p}]}\times[w].$$

Let $\mathfrak{K}=(\mathfrak{A}[V(K^{\bf a})],K^{\bf a},{\bf a}, {\bf I}, {\bf W}_{{q}})$ be  an extended compass of the flatness pair $(W,\mathfrak{R})$ of $G\setminus V({\bf a})$ of height $2w+j,$  $R\subseteq V(K^{\bf a}),$ $d∈[r,w],$  $L\subseteq [l],$ and $Z\subseteq I^{(d-r+1)}.$
We set
\begin{eqnarray}
\labels{eq_in-sig_first-floor}\green{{\sf in\mbox{-}sig}}(\mathfrak{K},R,d,L,Z) & := & \{(Y_{1},\ldots,Y_{p},t)∈ \green{{\sf SIG}_{\sf in}}\mid t≤ d\mbox{~and~} \exists\  C∈ {\sf pr}(K^{\bf a}[I^{(d)}],{\bf W}_q, Z) \mbox{~and}\\
\notag &  &~~~~~~~~~~~~~~~~~~~~~~~~~~~~~~~~ \exists \ (\tilde{X}_{1},\ldots,\tilde{X}_p)\mbox{~such that~} \forall h∈[p]\  \\
\notag & & ~~~~~~~~~~~~~~~~~~~~~~~~~~~~~~~~~~\tilde{X}_{h}=\{x_{i}^{h}\mid i∈ Y_{h}\},\\
\notag & & ~~~~~~~~~~~~~~~~~~~~~~~~~~~~~~~~~~\tilde{X}_{h}\subseteq (I^{(t-\hat{r}+1)}\setminus Z)\cap R,\text{ and}\\
\notag & &~~~~~~~~~~~~~~~~~~~~~~~~~~~~~~~~~~\text{if~}{\bf a}'={\bf a}\setminus V_L ({\bf a}), \text{ then}\\
\notag &  &~~~~~~~~~~~~~~~~~~~~~~~~~~~~~~~~~~\tilde{X}_h \text{~is~} (|Y_{h}|, r_{h})\text{-scattered in } {\sf ap}_{\bf c}(\mathfrak{A}, {\bf a}')[I^{(t)}\setminus Z]\\
\notag & &~~~~~~~~~~~~~~~~~~~~~~~~~~~~~~~~~~\text{and }{\sf ap}_{\bf c}(\mathfrak{A}, {\bf a'})[C]\models \bigwedge_{x∈ \tilde{X}_h} ψ_{h}(x)\}.
\end{eqnarray}

Here, $C$ represents the ``privileged'' part of the graph $K^{\bf a}[I^{(d)}]$ obtained after the removal of the whole $Z$ (that is the set $X_{\rm in}$ together with all the non-privileged components of $G\setminus X$ that are connected with $X_{\rm in}$).
Notice that this is a more restricted part of the privileged component of $G\setminus X$ but it is also ``flat''.
Then, we guess how the scattered sets of each of the basic local sentences of the Gaifman sentence can intersect this graph (a buffer that ``crops'' the area that contains the vertices that intersect an inner-area of $C$ corresponds to $t,$ and the numbers of the selected vertices correspond to the sets $Y_1, \ldots, Y_p$) and how these variables satisfy the $r_i$-local formulas $ψ_i.$
The ``scatteredness'' and the satisfaction of $ψ_i$ are evaluated on the structure after ``projecting'' with respect to the apices.

We finally define
$${\sf CHAR} = {[r,w]} \times 2^{[l]}\times 2^{\blue{{\sf SIG}_{\sf out}}} \times 2^{\green{{\sf SIG}_{\sf in}}}
$$
and
\begin{eqnarray}
\labels{eq_char_first-floor}θ\text{-}{\sf char}(\mathfrak{K},R) & = &\{(d,L,\blue{{\sf sig}_{\sf out}},\green{{\sf sig}_{\sf in}})∈ {\sf CHAR} \mid
\exists\ Z\subseteq I^{(d - r+1)}\mbox{~such that},\\
\notag & &\hspace{6cm}  \blue{{\sf out}\text{-}{\sf sig}}(\mathfrak{K},R,d,L,Z)= \blue{{\sf sig}_{\sf out}},  \mbox{~and~}\\
\notag & &\hspace{6cm} \green{{\sf in\mbox{-}sig}}(\mathfrak{K},R,d,L,Z)= \green{{\sf sig}_{\sf in}}\}.
\end{eqnarray}

Observe that $|{\sf CHAR}| = {\cal O}_{|θ|,l,q,j'}(1).$

\myskip\subsection{An algorithm for finding equivalent flatness pairs}
\labels{@inhumainement}
In this subsection we present an algorithm ${\tt Find\_Equiv\_FlatPairs}$ that will serve as the algorithm for~\autoref{@desmembramientos}.
Given the inputs in~\autoref{@desmembramientos},
the algorithm ${\tt Find\_Equiv\_FlatPairs}$ will return, in linear time,
a set $Y\subseteq V(\mathfrak{A})\setminus V({\bf a})$ and a flatness pair $(W',\mathfrak{R}')$ of $G_{\mathfrak{A}}\setminus V({\bf a})$ of height $g$
that is a $W''$-tilt of $(W,\mathfrak{R})$ for a subwall $W''$ of $W,$ where $\mathfrak{A}$ is the input $τ$-structure, ${\bf a}$ is an apex-tuple of $\mathfrak{A},$ and $(W,\mathfrak{R})$ is a flatness pair of $G_{\mathfrak{A}}\setminus V({\bf a})$ of ``sufficiently big'' height,
with the property that  $(\mathfrak{A},R, {\bf a})\models θ_{{\sf R},{\bf c}}\iff (\mathfrak{A}\setminus V({\sf compass}_{\mathfrak{R}'}(W')),R\setminus Y, {\bf a})\models θ_{{\sf R},{\bf c}}.$
The proof of correctness of the algorithm ${\tt Find\_Equiv\_FlatPairs}$ will prove~\autoref{@desmembramientos} and it is in~\autoref{sec_proof_correctness}.

Before presenting the the algorithm ${\tt Find\_Equiv\_FlatPairs},$ we present~\autoref{@transvaluation} that summarizes all different formulas that we consider up to this point, with their corresponding meanings.
\begin{table}[H]
\centering
\bgroup
\def\arraystretch{1.25}
 \begin{tabular}{|| c | c||}
 \hline
 Formulas & Meaning\\ [0.5ex]
 \hline\hline
$β$ & modulator sentence expressing \MSOL-property on bounded treewidth structures\\
\hline
 $σ$ & \FOL-target sentence\\
 \hline
 $μ$ & \NTMC-target sentence\\
\hline
$γ$ & $(σ\wedge μ)^{({\sf c})}$ or $σ\wedgeμ$\\
\hline
$θ$ & $ \exists {\sf X}\ β|_{{\sf star}_{\sf X}}\wedge γ|_{{\sf rm}_{\sf X}}$ \\
\hline
 $ζ$ & the $l$-apex-projected sentence $σ^l$ of $σ$\\
\hline
$\breve{ζ}$ & a Gaifman sentence equivalent to $ζ$\\
\hline
 $ψ_h$ & $r$-local formulas of the basic local sentences of $\breve{ζ}$\\
 \hline
$\breve{ζ}_{\sf R}$ & the Gaifman sentence $\breve{ζ}$ after adding ${\sf R}$ (whose model is of the form ${\sf ap}_{\bf c} (\mathfrak{A},R)$)\\
\hline
$\breve{ζ}_{\sf R} |_{{\sf ap}_{\bf c}}$ & the ``backwards translation'' of $\breve{ζ}_{\sf R}$ to structures without ``projecting'' ${\bf c}$\\
\hline
\raisebox{-2.5mm}{$θ_{{\sf R},{\bf c}}$} & the sentence obtained from $θ$ after replacing\\[-2.5mm]
& every \FOL-target sentence $σ$ of $θ$ with the respective $\breve{ζ}_{\sf R} |_{{\sf ap}_{\bf c}}$\\
\hline
$η_{w\text{-}{\sf pr}_{{\sf X},{\sf Y}}}$ & privileged connected component \\
\hline
$θ^{\sf out}_q$& the question $β$ in $X$ and questions $\breve{ζ}_{\sf R} |_{{\sf ap}_{\bf c}}\wedge μ$ in ``non-privileged'' components\\
\hline
$\tilde{θ}_q$  & sentence equivalent to $θ_{{\sf R},{\bf c}},$ separating question $\breve{ζ}_{\sf R} |_{{\sf ap}_{\bf c}}\wedge μ$ on the $w$-privileged set\\
\hline
$\bar{φ}$ & a representative of $\tilde{θ}_q$ given by  Courcelle's theorem\\
\hline
$ψ_h$ & $r$-local formulas of the basic local sentences of the Gaifman sentence $\breve{ζ}$\\
 \hline
\end{tabular}\caption{List of formulas used in the proof of~\autoref{@desmembramientos} for sentences in  $\bar{Θ}_1[τ]$ with their respective meanings.}\labels{@transvaluation}
\egroup
\end{table}

\myskip\paragraph{The algorithm ${\tt Find\_Equiv\_FlatPairs}.$}
The algorithm has four steps.
First, recall that
there exist $p∈ \mathbb{N}_{≥ 1},$ $\ell_1, \ldots, \ell_p, r_1, \ldots, r_p∈ \mathbb{N}_{≥ 1},$ and sentences $\tilde{ζ}_1, \ldots, \tilde{ζ}_p∈ \FOL[τ^{\langle{\bf c}\rangle}\cup\{{\sf R}\}]$ such that $\breve{ζ}_{\sf R}$ is a Boolean combination of $\tilde{ζ}_1, \ldots, \tilde{ζ}_p$ and for every $h∈ [p],$ $\tilde{ζ}_h$ is a basic local sentence with parameters $\ell_h$ and $r_h,$ i.e.,
\begin{eqnarray*}
\tilde{ζ}_h = \exists {\sf x}_{1}\ldots\exists {\sf x}_{\ell_{h}}\big(\bigwedge_{i∈ [\ell_{h}]}{\sf x}_{i}∈ {\sf R}\wedge \bigwedge_{1≤ i<j≤ \ell_{h}} d({\sf x}_{i}, {\sf x}_{j})> 2r_{h}\wedge \bigwedge_{i∈ [\ell_{h}]}ψ_{h}({\sf x}_{i})\big),
\end{eqnarray*}
where $ψ_h$ is an $r_h$-local formula  in $\FOL[τ^{\langle {\bf c}\rangle}]$ with one free variable.

Let $\hat{r}:= \max_{h∈[p]}\{r_h\}$ and $\hat{\ell}:=\max_{h∈ [p]}\{\ell_h\}.$
We set $c$ to be the size of the \FOL-target sentence $σ$ of $θ,$
\begin{align*}
	q &: = (\tw(θ)+1)^2+1,\\
	\funref{@riguardavano}(|θ|,\tw(θ),g) & := \max\{q,(g+1)^2+1\},\\
	\funref{@carthaginoise}(|θ|,\tw(θ)) & := q-1,\\
	j' & := g+2\hat{r} +2,\\
	j & :=   {\sf odd}(\max\{q/2,j'\}),\\
	r &: = 2\cdot (\hat{\ell}+ 3)\cdot \hat{r}, \\
	w & := (r+2)\cdot q,\\
	m & := 2^{|{\sf CHAR}|} \cdot q\cdot (\hat{\ell} +3),\text{ and}\\
	\funref{@occidentales}(\hw(θ),\tw(θ),c,l, g)& := \lceil (2w+j)\cdot \sqrt{m}\rceil. \\
\end{align*}

\noindent{\bf Step 1}:
We first find a ``packing'' of subwalls of $W,$ i.e., a collection ${\cal W}$ of $m$ $(2w+j)$-subwalls of $W$ such that their influences are pairwise disjoint.
This collection exists because $W$ has height at least $\funref{@occidentales}(\hw(θ),\tw(θ),|σ|,l, j') = \lceil (2w+j)\cdot \sqrt{m}\rceil$
and because, due to~\autoref{label_grundfalscher}, for every distinct $W_i, W_j∈ {\cal W},$ there are no cells of $\mathfrak{R}$ that are both $W_i$-perimetric and $W_j$-perimetric.
Observe that the collection ${\cal W}$ can be computed in linear time.
\medskip

\noindent{\bf Step 2}:
Then, for every wall $W_i ∈ {\cal W},$ we compute a $W_i$-tilt of $(W,\mathfrak{R}),$ which we denote by $(\tilde{W}_i, \tilde{\mathfrak{R}}_i),$ and we consider the collection $\tilde{\cal W} := \{(\tilde{W}_i, \tilde{\mathfrak{R}}_i)\mid W_i ∈ {\cal W}\}$ of $m$ flatness pairs of $G_{\mathfrak{A}}\setminus V({\bf a})$ of height $2w+j.$
Note that $\tilde{\cal W}$ can be computed in time ${\cal O}(n),$ due to~\autoref{label_proporcionada}.
\medskip

\noindent{\bf Step 3}:
For every $i∈[m],$ let
$K_i := {\sf compass}_{\tilde{\mathfrak{R}}_i}(\tilde{W}_i),$
$K_i^{{\bf a}} := G_{\mathfrak{A}}[V({\bf a})\cup V(K_i)],$ and
${\bf W}_q^{(i)}$ be the $q$-pseudogrid defined by the horizontal and the vertical paths of the central $q$-subwall of $\tilde{W}_i.$
Also, for every $d∈ [w],$ let $I_i^{(d)} : = V(\cupall {\sf influence}_{\tilde{\mathfrak{R}}_i} (W_i^{(2d+j')}))$ and let ${\bf I}_i:=(I_i^{(1)},\ldots,I_i^{(w)}).$
Let $\mathfrak{K}_i := (\mathfrak{A}[V(K_i^{\bf a})],{\bf a}, {\bf I}_i, {\bf W}^{(i)}_{{q}})$ be the extended compass of $(\tilde{W}_i, \tilde{\mathfrak{R}}_i)$ in $G_{\mathfrak{A}}\setminus V({\bf a}),$ $R_i := R\cap V(K_i^{\bf a}),$ and observe that for every $i,j∈[m],$ $R_i \cap R_j = R\cap V({\bf a}).$
After defining the above collection $\{(\mathfrak{K}_1, R_1),\ldots, (\mathfrak{K}_m,R_m)\}$ of extended compasses of flatness pairs of $G_{\mathfrak{A}}\setminus V({\bf a}),$ we compute their characteristics:
Since, by the hypothesis of the lemma, $K_i, i∈[m]$ has treewidth at most $z,$ by Courcelle's theorem (\autoref{@originallypublishedin}), $θ\text{-}{\sf char}(\mathfrak{K}_i,R_i)$ can be computed in time ${\cal O}_{|θ|} (n).$
We say that two flatness pairs $(\tilde{W}_i, \tilde{\mathfrak{R}}_i), (\tilde{W}_j, \tilde{\mathfrak{R}}_j)∈ \tilde{\cal W}$ are {\em $θ$-equivalent} if $θ\text{-}{\sf char}(\mathfrak{K}_i,R_i)= θ\text{-}{\sf char}(\mathfrak{K}_j,R_j).$
\medskip

\noindent{\bf Step 4}:
Since $m = 2^{|{\sf CHAR}|} \cdot q\cdot (\hat{\ell} +3)$ and for every $i∈ [m],$ $θ\text{-}{\sf char}(\mathfrak{K}_i, R_i)\subseteq {\sf CHAR},$
we can find a collection $\tilde{\cal W}'\subseteq \tilde{\cal W}$ of pairwise
 $θ$-equivalent flatness pairs
such that
$|\tilde{\cal W}'| = q\cdot (\hat{\ell} +3).$
Without loss of generality, we assume that $(\tilde{W}_{1}, \tilde{\mathfrak{R}}_{1})∈ \tilde{\cal W}'.$
We set $\breve{W}$ to be the central $j'$-subwall of $\tilde{W}_1,$ $W^\bullet$ to be the central $g$-subwall of $\tilde{W}_1,$ and keep in mind that $j'=g+2\hat{r}+2.$
Note that $\breve{W}$ (resp. $W^\bullet$) is also the central $j'$-subwall (resp. $g$-subwall) of $W_1$ and, therefore, it is a subwall of $W$ of height $j'$ (resp. $g$).
Again, using~\autoref{label_proporcionada}, we compute, in time ${\cal O}(n),$
a $\breve{W}$-tilt
$(\breve{W}',\breve{\mathfrak{R}}')$
of $(W,\mathfrak{R})$
and
a $W^\bullet$-tilt
$(\tilde{W}',\tilde{\mathfrak{R}}')$
of $(W,\mathfrak{R}).$
We set $Y:=V({\sf compass}(\breve{W}',\breve{\mathfrak{R}}')).$
We output the set $Y$ and the flatness pair
$(\tilde{W}',\tilde{\mathfrak{R}}').$
\medskip

Observe that the overall algorithm runs in linear time.

\myskip\subsection{Proof of correctness of the algorithm for sentences in $\bar{Θ}_1$}
\labels{sec_proof_correctness}

In order to complete the proof of~\autoref{@desmembramientos} for a sentence $θ ∈ \bar{Θ}_1[τ],$
we have to prove that
$(\mathfrak{A},R, {\bf a})\models θ_{{\sf R},{\bf c}}\iff (\mathfrak{A}\setminus V({\sf compass}_{\tilde{\mathfrak{R}}'}(\tilde{W}')),R\setminus Y, {\bf a})\models θ_{{\sf R},{\bf c}}.$

For sake of  simplicity, we use $G$ to denote the Gaifman graph $G_{\mathfrak{A}}$ of $\mathfrak{A}.$

\myskip\paragraph{Observations on the collection $\tilde{\cal W}'.$}
Recall that, for every two $(\tilde{W}_i, \tilde{\mathfrak{R}}_i),(\tilde{W}_j, \tilde{\mathfrak{R}}_j)∈ \tilde{\cal W}',$
$(\tilde{W}_i, \tilde{\mathfrak{R}}_i)$ (resp. $(\tilde{W}_j, \tilde{\mathfrak{R}}_j)$) is a
$W_i$-tilt (resp. $W_j$-tilt) of $(W,\mathfrak{R}),$
where $W_i, W_j∈ {\cal W}$ and $V(\cupall{\sf influence}_{\mathfrak{R}} (W_i)) \cap V(\cupall {\sf influence}_{\mathfrak{R}} (W_j)) = \emptyset.$
This implies that
 $V({\sf compass}_{\tilde{\mathfrak{R}}_i}(\tilde{W}_i)) \cap V({\sf compass}_{\tilde{\mathfrak{R}}_j} (\tilde{W}_j)) = \emptyset.$
 Moreover, observe that
 if $\tilde{\cal Q}$ is a
 $(W, \mathfrak{R})$-canonical partition of $G\setminus V({\bf a}),$
 then no internal bag of $\tilde{\cal Q}$ intersects both $V(\cupall {\sf influence}_{\mathfrak{R}} (W_i))$ and $V(\cupall {\sf influence}_{\mathfrak{R}} (W_j)),$ for every $i,j∈[m], i\neq j.$

\myskip\paragraph{Shifting to the split version of $θ_{{\sf R},{\bf c}}.$}
Suppose that $(\mathfrak{A},R, {\bf a})\models θ_{{\sf R},{\bf c}}.$
Recall that
${\bf W}_q^{(1)}$ is the $q$-pseudogrid defined by the horizontal and the vertical paths of the central $q$-subwall of $\tilde{W}_1.$
Due to~\autoref{lemma_equiva}, we have that $(\mathfrak{A},R,{\bf W}_{{q}}^{(1)}, {\bf a})\models \tilde{θ}_{{q}},$
i.e.,
there is a set $X\subseteq V(\mathfrak{A})$ such that
$(\mathfrak{A},R,{\bf W}_{{q}},{\bf a},X)\models θ^{\sf out}_q$ and there is a set $C\subseteq V(\mathfrak{A})$ that is a $w$-privileged set of
$\mathfrak{A}$ with respect to ${\bf W}_q^{(1)}$ and $X$
and it holds that
$\mathfrak{A}[C]\models μ$
and $(\mathfrak{A},R, {\bf a})[C]\models\breve{ζ}_{\sf R} |_{\sf ap_{\bf c}}.$
Notice that, due to~\autoref{@preconditions}, $C$ is unique.

\myskip\paragraph{Finding a $θ$-equivalent extended compass that is disjoint from $X.$}
Recall that $\tilde{\cal W}'$ is a collection of $q\cdot (\hat{\ell}+3)$ flatness pairs of $G\setminus V({\bf a})$ of height $2w+j$ that are $θ$-equivalent to $(\tilde{W}_1, \tilde{\mathfrak{R}}_1).$
The fact that ${\sf star}(\mathfrak{A},X)\models β$ and $β∈ \MSOL^\tw[τ\cup\{{\sf X}\}]$ implies that ${\sf cl}_{\sf X}({\sf star}_{\sf X} (\mathfrak{A},X))$ has treewidth at most $\tw(θ).$
Therefore, by~\autoref{@congregation}, $X$ intersects at most $(\tw(θ)+1)^2 = q-1$ internal bags of every $(W, \mathfrak{R})$-canonical partition of $G\setminus V({\bf a}).$
This, together with the fact that  $|\tilde{\cal W}'| = q\cdot (\hat{\ell} +3)$
and that,
 if $\tilde{\cal Q}$ is a
 $(W, \mathfrak{R})$-canonical partition of $G\setminus V({\bf a}),$
 then no internal bag of $\tilde{\cal Q}$ intersects both $V(\cupall {\sf influence}_{\mathfrak{R}} (W_i))$ and $V(\cupall {\sf influence}_{\mathfrak{R}} (W_j)),$ for every $i,j∈[m], i\neq j,$
 implies that there is a collection $\tilde{\cal W}''\subseteq \tilde{\cal W}'$ of size $\hat{\ell} +2$ such that $(\tilde{W}_{1}, \tilde{\mathfrak{R}}_{1})\notin \tilde{\cal W}'',$ every flatness pair in $\tilde{\cal W}''$ is $θ$-equivalent to  $(\tilde{W}_{1}, \tilde{\mathfrak{R}}_{1}),$ and the vertex set of its influence is disjoint from $X.$
Assume, without loss of generality, that $(\tilde{W}_{2}, \tilde{\mathfrak{R}}_{2})∈ \tilde{\cal W}'',$ which implies that $θ\text{-}{\sf char}(\mathfrak{K}_1,R_1)= θ\text{-}{\sf char}(\mathfrak{K}_2,R_2)$ and $I_2^{(w)}\cap X = \emptyset.$

\myskip\paragraph{Every modulator leaves an intact buffer.}
We fix $\tilde{\cal Q}$ to be a
$(\tilde{W}_1,\tilde{\mathfrak{R}}_1)$-canonical partition of $G\setminus V({\bf a}).$
By~\autoref{@congregation},
$X$ intersects at most $q-1$ bags of $\tilde{\cal Q}.$
This implies that, given that $\tilde{W}_1$ has height $2w+j$ and $w = (r+2)\cdot q,$
there is an $i∈ [q]$ such that $X\cap (I_1^{(i\cdot r-1)}\setminus I_1^{(i\cdot r-r)}) = \emptyset.$
Let $X_{\rm in}= X\cap I_1^{(i\cdot r-r)}$ and $X_{\rm out} = X\setminus I_1^{(i\cdot r-1)}$ (see~\autoref{figure_xinxout} for a visualization of an example).
We set $d := i\cdot r-1.$

\begin{figure}[ht]
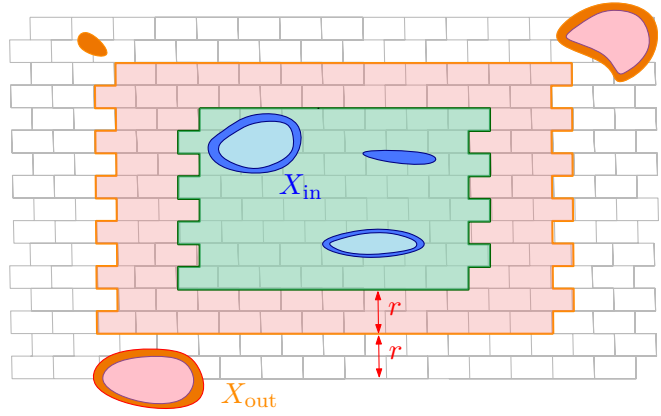

\centering
\tikzstyle{ipe stylesheet} = [
  ipe import,
  even odd rule,
  line join=round,
  line cap=butt,
  ipe pen normal/.style={line width=0.4},
  ipe pen heavier/.style={line width=0.8},
  ipe pen fat/.style={line width=1.2},
  ipe pen ultrafat/.style={line width=2},
  ipe pen normal,
  ipe mark normal/.style={ipe mark scale=3},
  ipe mark large/.style={ipe mark scale=5},
  ipe mark small/.style={ipe mark scale=2},
  ipe mark tiny/.style={ipe mark scale=1.1},
  ipe mark normal,
  /pgf/arrow keys/.cd,
  ipe arrow normal/.style={scale=7},
  ipe arrow large/.style={scale=10},
  ipe arrow small/.style={scale=5},
  ipe arrow tiny/.style={scale=3},
  ipe arrow normal,
  /tikz/.cd,
  ipe arrows, 
  <->/.tip = ipe normal,
  ipe dash normal/.style={dash pattern=},
  ipe dash dotted/.style={dash pattern=on 1bp off 3bp},
  ipe dash dashed/.style={dash pattern=on 4bp off 4bp},
  ipe dash dash dotted/.style={dash pattern=on 4bp off 2bp on 1bp off 2bp},
  ipe dash dash dot dotted/.style={dash pattern=on 4bp off 2bp on 1bp off 2bp on 1bp off 2bp},
  ipe dash normal,
  ipe node/.append style={font=\normalsize},
  ipe stretch normal/.style={ipe node stretch=1},
  ipe stretch normal,
  ipe opacity 10/.style={opacity=0.1},
  ipe opacity 30/.style={opacity=0.3},
  ipe opacity 50/.style={opacity=0.5},
  ipe opacity 75/.style={opacity=0.75},
  ipe opacity opaque/.style={opacity=1},
  ipe opacity opaque,
]
\definecolor{black}{rgb}{0,0,0}
\definecolor{white}{rgb}{1,1,1}
\definecolor{red}{rgb}{1,0,0}
\definecolor{blue}{rgb}{0,0,1}
\definecolor{green}{rgb}{0,1,0}
\definecolor{yellow}{rgb}{1,1,0}
\definecolor{orange}{rgb}{1,0.647,0}
\definecolor{gold}{rgb}{1,0.843,0}
\definecolor{purple}{rgb}{0.627,0.125,0.941}
\definecolor{gray}{rgb}{0.745,0.745,0.745}
\definecolor{brown}{rgb}{0.647,0.165,0.165}
\definecolor{navy}{rgb}{0,0,0.502}
\definecolor{pink}{rgb}{1,0.753,0.796}
\definecolor{seagreen}{rgb}{0.18,0.545,0.341}
\definecolor{turquoise}{rgb}{0.251,0.878,0.816}
\definecolor{violet}{rgb}{0.933,0.51,0.933}
\definecolor{darkblue}{rgb}{0,0,0.545}
\definecolor{darkcyan}{rgb}{0,0.545,0.545}
\definecolor{darkgray}{rgb}{0.663,0.663,0.663}
\definecolor{darkgreen}{rgb}{0,0.392,0}
\definecolor{darkmagenta}{rgb}{0.545,0,0.545}
\definecolor{darkorange}{rgb}{1,0.549,0}
\definecolor{darkred}{rgb}{0.545,0,0}
\definecolor{lightblue}{rgb}{0.678,0.847,0.902}
\definecolor{lightcyan}{rgb}{0.878,1,1}
\definecolor{lightgray}{rgb}{0.827,0.827,0.827}
\definecolor{lightgreen}{rgb}{0.565,0.933,0.565}
\definecolor{lightyellow}{rgb}{1,1,0.878}
\definecolor{lightcoral}{rgb}{0.941,0.502,0.502}
\definecolor{gray73}{rgb}{0.729,0.729,0.729}
\definecolor{aquamarine}{rgb}{0.498,1,0.831}
\definecolor{aquamarine4}{rgb}{0.271,0.545,0.455}
\definecolor{deepskyblue2}{rgb}{0,0.698,0.933}
\definecolor{magenta3}{rgb}{0.804,0,0.804}
\definecolor{mediumpurple1}{rgb}{0.671,0.51,1}
\definecolor{gray37}{rgb}{0.369,0.369,0.369}
\definecolor{royalblue1}{rgb}{0.282,0.463,1}
\definecolor{lightskyblue2}{rgb}{0.643,0.827,0.933}
\definecolor{aliceblue}{rgb}{0.941,0.973,1}
\definecolor{antiquewhite}{rgb}{0.98,0.922,0.843}
\definecolor{antiquewhite1}{rgb}{1,0.937,0.859}
\definecolor{antiquewhite2}{rgb}{0.933,0.875,0.8}
\definecolor{antiquewhite3}{rgb}{0.804,0.753,0.69}
\definecolor{antiquewhite4}{rgb}{0.545,0.514,0.471}
\definecolor{aquamarine1}{rgb}{0.498,1,0.831}
\definecolor{aquamarine2}{rgb}{0.463,0.933,0.776}
\definecolor{aquamarine3}{rgb}{0.4,0.804,0.667}
\definecolor{azure}{rgb}{0.941,1,1}
\definecolor{azure1}{rgb}{0.941,1,1}
\definecolor{azure2}{rgb}{0.878,0.933,0.933}
\definecolor{azure3}{rgb}{0.757,0.804,0.804}
\definecolor{azure4}{rgb}{0.514,0.545,0.545}
\definecolor{beige}{rgb}{0.961,0.961,0.863}
\definecolor{bisque}{rgb}{1,0.894,0.769}
\definecolor{bisque1}{rgb}{1,0.894,0.769}
\definecolor{bisque2}{rgb}{0.933,0.835,0.718}
\definecolor{bisque3}{rgb}{0.804,0.718,0.62}
\definecolor{bisque4}{rgb}{0.545,0.49,0.42}
\definecolor{blanchedalmond}{rgb}{1,0.922,0.804}
\definecolor{blue1}{rgb}{0,0,1}
\definecolor{blue2}{rgb}{0,0,0.933}
\definecolor{blue3}{rgb}{0,0,0.804}
\definecolor{blue4}{rgb}{0,0,0.545}
\definecolor{blueviolet}{rgb}{0.541,0.169,0.886}
\definecolor{brown1}{rgb}{1,0.251,0.251}
\definecolor{brown2}{rgb}{0.933,0.231,0.231}
\definecolor{brown3}{rgb}{0.804,0.2,0.2}
\definecolor{brown4}{rgb}{0.545,0.137,0.137}
\definecolor{burlywood}{rgb}{0.871,0.722,0.529}
\definecolor{burlywood1}{rgb}{1,0.827,0.608}
\definecolor{burlywood2}{rgb}{0.933,0.773,0.569}
\definecolor{burlywood3}{rgb}{0.804,0.667,0.49}
\definecolor{burlywood4}{rgb}{0.545,0.451,0.333}
\definecolor{cadetblue}{rgb}{0.373,0.62,0.627}
\definecolor{cadetblue1}{rgb}{0.596,0.961,1}
\definecolor{cadetblue2}{rgb}{0.557,0.898,0.933}
\definecolor{cadetblue3}{rgb}{0.478,0.773,0.804}
\definecolor{cadetblue4}{rgb}{0.325,0.525,0.545}
\definecolor{chartreuse}{rgb}{0.498,1,0}
\definecolor{chartreuse1}{rgb}{0.498,1,0}
\definecolor{chartreuse2}{rgb}{0.463,0.933,0}
\definecolor{chartreuse3}{rgb}{0.4,0.804,0}
\definecolor{chartreuse4}{rgb}{0.271,0.545,0}
\definecolor{chocolate}{rgb}{0.824,0.412,0.118}
\definecolor{chocolate1}{rgb}{1,0.498,0.141}
\definecolor{chocolate2}{rgb}{0.933,0.463,0.129}
\definecolor{chocolate3}{rgb}{0.804,0.4,0.114}
\definecolor{chocolate4}{rgb}{0.545,0.271,0.075}
\definecolor{coral}{rgb}{1,0.498,0.314}
\definecolor{coral1}{rgb}{1,0.447,0.337}
\definecolor{coral2}{rgb}{0.933,0.416,0.314}
\definecolor{coral3}{rgb}{0.804,0.357,0.271}
\definecolor{coral4}{rgb}{0.545,0.243,0.184}
\definecolor{cornflowerblue}{rgb}{0.392,0.584,0.929}
\definecolor{cornsilk}{rgb}{1,0.973,0.863}
\definecolor{cornsilk1}{rgb}{1,0.973,0.863}
\definecolor{cornsilk2}{rgb}{0.933,0.91,0.804}
\definecolor{cornsilk3}{rgb}{0.804,0.784,0.694}
\definecolor{cornsilk4}{rgb}{0.545,0.533,0.471}
\definecolor{cyan}{rgb}{0,1,1}
\definecolor{cyan1}{rgb}{0,1,1}
\definecolor{cyan2}{rgb}{0,0.933,0.933}
\definecolor{cyan3}{rgb}{0,0.804,0.804}
\definecolor{cyan4}{rgb}{0,0.545,0.545}
\definecolor{darkgoldenrod}{rgb}{0.722,0.525,0.043}
\definecolor{darkgoldenrod1}{rgb}{1,0.725,0.059}
\definecolor{darkgoldenrod2}{rgb}{0.933,0.678,0.055}
\definecolor{darkgoldenrod3}{rgb}{0.804,0.584,0.047}
\definecolor{darkgoldenrod4}{rgb}{0.545,0.396,0.031}
\definecolor{darkgrey}{rgb}{0.663,0.663,0.663}
\definecolor{darkkhaki}{rgb}{0.741,0.718,0.42}
\definecolor{darkolivegreen}{rgb}{0.333,0.42,0.184}
\definecolor{darkolivegreen1}{rgb}{0.792,1,0.439}
\definecolor{darkolivegreen2}{rgb}{0.737,0.933,0.408}
\definecolor{darkolivegreen3}{rgb}{0.635,0.804,0.353}
\definecolor{darkolivegreen4}{rgb}{0.431,0.545,0.239}
\definecolor{darkorange1}{rgb}{1,0.498,0}
\definecolor{darkorange2}{rgb}{0.933,0.463,0}
\definecolor{darkorange3}{rgb}{0.804,0.4,0}
\definecolor{darkorange4}{rgb}{0.545,0.271,0}
\definecolor{darkorchid}{rgb}{0.6,0.196,0.8}
\definecolor{darkorchid1}{rgb}{0.749,0.243,1}
\definecolor{darkorchid2}{rgb}{0.698,0.227,0.933}
\definecolor{darkorchid3}{rgb}{0.604,0.196,0.804}
\definecolor{darkorchid4}{rgb}{0.408,0.133,0.545}
\definecolor{darksalmon}{rgb}{0.914,0.588,0.478}
\definecolor{darkseagreen}{rgb}{0.561,0.737,0.561}
\definecolor{darkseagreen1}{rgb}{0.757,1,0.757}
\definecolor{darkseagreen2}{rgb}{0.706,0.933,0.706}
\definecolor{darkseagreen3}{rgb}{0.608,0.804,0.608}
\definecolor{darkseagreen4}{rgb}{0.412,0.545,0.412}
\definecolor{darkslateblue}{rgb}{0.282,0.239,0.545}
\definecolor{darkslategray}{rgb}{0.184,0.31,0.31}
\definecolor{darkslategray1}{rgb}{0.592,1,1}
\definecolor{darkslategray2}{rgb}{0.553,0.933,0.933}
\definecolor{darkslategray3}{rgb}{0.475,0.804,0.804}
\definecolor{darkslategray4}{rgb}{0.322,0.545,0.545}
\definecolor{darkslategrey}{rgb}{0.184,0.31,0.31}
\definecolor{darkturquoise}{rgb}{0,0.808,0.82}
\definecolor{darkviolet}{rgb}{0.58,0,0.827}
\definecolor{deeppink}{rgb}{1,0.078,0.576}
\definecolor{deeppink1}{rgb}{1,0.078,0.576}
\definecolor{deeppink2}{rgb}{0.933,0.071,0.537}
\definecolor{deeppink3}{rgb}{0.804,0.063,0.463}
\definecolor{deeppink4}{rgb}{0.545,0.039,0.314}
\definecolor{deepskyblue}{rgb}{0,0.749,1}
\definecolor{deepskyblue1}{rgb}{0,0.749,1}
\definecolor{deepskyblue3}{rgb}{0,0.604,0.804}
\definecolor{deepskyblue4}{rgb}{0,0.408,0.545}
\definecolor{dimgray}{rgb}{0.412,0.412,0.412}
\definecolor{dimgrey}{rgb}{0.412,0.412,0.412}
\definecolor{dodgerblue}{rgb}{0.118,0.565,1}
\definecolor{dodgerblue1}{rgb}{0.118,0.565,1}
\definecolor{dodgerblue2}{rgb}{0.11,0.525,0.933}
\definecolor{dodgerblue3}{rgb}{0.094,0.455,0.804}
\definecolor{dodgerblue4}{rgb}{0.063,0.306,0.545}
\definecolor{firebrick}{rgb}{0.698,0.133,0.133}
\definecolor{firebrick1}{rgb}{1,0.188,0.188}
\definecolor{firebrick2}{rgb}{0.933,0.173,0.173}
\definecolor{firebrick3}{rgb}{0.804,0.149,0.149}
\definecolor{firebrick4}{rgb}{0.545,0.102,0.102}
\definecolor{floralwhite}{rgb}{1,0.98,0.941}
\definecolor{forestgreen}{rgb}{0.133,0.545,0.133}
\definecolor{gainsboro}{rgb}{0.863,0.863,0.863}
\definecolor{ghostwhite}{rgb}{0.973,0.973,1}
\definecolor{gold1}{rgb}{1,0.843,0}
\definecolor{gold2}{rgb}{0.933,0.788,0}
\definecolor{gold3}{rgb}{0.804,0.678,0}
\definecolor{gold4}{rgb}{0.545,0.459,0}
\definecolor{goldenrod}{rgb}{0.855,0.647,0.125}
\definecolor{goldenrod1}{rgb}{1,0.757,0.145}
\definecolor{goldenrod2}{rgb}{0.933,0.706,0.133}
\definecolor{goldenrod3}{rgb}{0.804,0.608,0.114}
\definecolor{goldenrod4}{rgb}{0.545,0.412,0.078}
\definecolor{gray0}{rgb}{0,0,0}
\definecolor{gray1}{rgb}{0.012,0.012,0.012}
\definecolor{gray10}{rgb}{0.102,0.102,0.102}
\definecolor{gray100}{rgb}{1,1,1}
\definecolor{gray11}{rgb}{0.11,0.11,0.11}
\definecolor{gray12}{rgb}{0.122,0.122,0.122}
\definecolor{gray13}{rgb}{0.129,0.129,0.129}
\definecolor{gray14}{rgb}{0.141,0.141,0.141}
\definecolor{gray15}{rgb}{0.149,0.149,0.149}
\definecolor{gray16}{rgb}{0.161,0.161,0.161}
\definecolor{gray17}{rgb}{0.169,0.169,0.169}
\definecolor{gray18}{rgb}{0.18,0.18,0.18}
\definecolor{gray19}{rgb}{0.188,0.188,0.188}
\definecolor{gray2}{rgb}{0.02,0.02,0.02}
\definecolor{gray20}{rgb}{0.2,0.2,0.2}
\definecolor{gray21}{rgb}{0.212,0.212,0.212}
\definecolor{gray22}{rgb}{0.22,0.22,0.22}
\definecolor{gray23}{rgb}{0.231,0.231,0.231}
\definecolor{gray24}{rgb}{0.239,0.239,0.239}
\definecolor{gray25}{rgb}{0.251,0.251,0.251}
\definecolor{gray26}{rgb}{0.259,0.259,0.259}
\definecolor{gray27}{rgb}{0.271,0.271,0.271}
\definecolor{gray28}{rgb}{0.278,0.278,0.278}
\definecolor{gray29}{rgb}{0.29,0.29,0.29}
\definecolor{gray3}{rgb}{0.031,0.031,0.031}
\definecolor{gray30}{rgb}{0.302,0.302,0.302}
\definecolor{gray31}{rgb}{0.31,0.31,0.31}
\definecolor{gray32}{rgb}{0.322,0.322,0.322}
\definecolor{gray33}{rgb}{0.329,0.329,0.329}
\definecolor{gray34}{rgb}{0.341,0.341,0.341}
\definecolor{gray35}{rgb}{0.349,0.349,0.349}
\definecolor{gray36}{rgb}{0.361,0.361,0.361}
\definecolor{gray38}{rgb}{0.38,0.38,0.38}
\definecolor{gray39}{rgb}{0.388,0.388,0.388}
\definecolor{gray4}{rgb}{0.039,0.039,0.039}
\definecolor{gray40}{rgb}{0.4,0.4,0.4}
\definecolor{gray41}{rgb}{0.412,0.412,0.412}
\definecolor{gray42}{rgb}{0.42,0.42,0.42}
\definecolor{gray43}{rgb}{0.431,0.431,0.431}
\definecolor{gray44}{rgb}{0.439,0.439,0.439}
\definecolor{gray45}{rgb}{0.451,0.451,0.451}
\definecolor{gray46}{rgb}{0.459,0.459,0.459}
\definecolor{gray47}{rgb}{0.471,0.471,0.471}
\definecolor{gray48}{rgb}{0.478,0.478,0.478}
\definecolor{gray49}{rgb}{0.49,0.49,0.49}
\definecolor{gray5}{rgb}{0.051,0.051,0.051}
\definecolor{gray50}{rgb}{0.498,0.498,0.498}
\definecolor{gray51}{rgb}{0.51,0.51,0.51}
\definecolor{gray52}{rgb}{0.522,0.522,0.522}
\definecolor{gray53}{rgb}{0.529,0.529,0.529}
\definecolor{gray54}{rgb}{0.541,0.541,0.541}
\definecolor{gray55}{rgb}{0.549,0.549,0.549}
\definecolor{gray56}{rgb}{0.561,0.561,0.561}
\definecolor{gray57}{rgb}{0.569,0.569,0.569}
\definecolor{gray58}{rgb}{0.58,0.58,0.58}
\definecolor{gray59}{rgb}{0.588,0.588,0.588}
\definecolor{gray6}{rgb}{0.059,0.059,0.059}
\definecolor{gray60}{rgb}{0.6,0.6,0.6}
\definecolor{gray61}{rgb}{0.612,0.612,0.612}
\definecolor{gray62}{rgb}{0.62,0.62,0.62}
\definecolor{gray63}{rgb}{0.631,0.631,0.631}
\definecolor{gray64}{rgb}{0.639,0.639,0.639}
\definecolor{gray65}{rgb}{0.651,0.651,0.651}
\definecolor{gray66}{rgb}{0.659,0.659,0.659}
\definecolor{gray67}{rgb}{0.671,0.671,0.671}
\definecolor{gray68}{rgb}{0.678,0.678,0.678}
\definecolor{gray69}{rgb}{0.69,0.69,0.69}
\definecolor{gray7}{rgb}{0.071,0.071,0.071}
\definecolor{gray70}{rgb}{0.702,0.702,0.702}
\definecolor{gray71}{rgb}{0.71,0.71,0.71}
\definecolor{gray72}{rgb}{0.722,0.722,0.722}
\definecolor{gray74}{rgb}{0.741,0.741,0.741}
\definecolor{gray75}{rgb}{0.749,0.749,0.749}
\definecolor{gray76}{rgb}{0.761,0.761,0.761}
\definecolor{gray77}{rgb}{0.769,0.769,0.769}
\definecolor{gray78}{rgb}{0.78,0.78,0.78}
\definecolor{gray79}{rgb}{0.788,0.788,0.788}
\definecolor{gray8}{rgb}{0.078,0.078,0.078}
\definecolor{gray80}{rgb}{0.8,0.8,0.8}
\definecolor{gray81}{rgb}{0.812,0.812,0.812}
\definecolor{gray82}{rgb}{0.82,0.82,0.82}
\definecolor{gray83}{rgb}{0.831,0.831,0.831}
\definecolor{gray84}{rgb}{0.839,0.839,0.839}
\definecolor{gray85}{rgb}{0.851,0.851,0.851}
\definecolor{gray86}{rgb}{0.859,0.859,0.859}
\definecolor{gray87}{rgb}{0.871,0.871,0.871}
\definecolor{gray88}{rgb}{0.878,0.878,0.878}
\definecolor{gray89}{rgb}{0.89,0.89,0.89}
\definecolor{gray9}{rgb}{0.09,0.09,0.09}
\definecolor{gray90}{rgb}{0.898,0.898,0.898}
\definecolor{gray91}{rgb}{0.91,0.91,0.91}
\definecolor{gray92}{rgb}{0.922,0.922,0.922}
\definecolor{gray93}{rgb}{0.929,0.929,0.929}
\definecolor{gray94}{rgb}{0.941,0.941,0.941}
\definecolor{gray95}{rgb}{0.949,0.949,0.949}
\definecolor{gray96}{rgb}{0.961,0.961,0.961}
\definecolor{gray97}{rgb}{0.969,0.969,0.969}
\definecolor{gray98}{rgb}{0.98,0.98,0.98}
\definecolor{gray99}{rgb}{0.988,0.988,0.988}
\definecolor{green1}{rgb}{0,1,0}
\definecolor{green2}{rgb}{0,0.933,0}
\definecolor{green3}{rgb}{0,0.804,0}
\definecolor{green4}{rgb}{0,0.545,0}
\definecolor{greenyellow}{rgb}{0.678,1,0.184}
\definecolor{grey}{rgb}{0.745,0.745,0.745}
\definecolor{grey0}{rgb}{0,0,0}
\definecolor{grey1}{rgb}{0.012,0.012,0.012}
\definecolor{grey10}{rgb}{0.102,0.102,0.102}
\definecolor{grey100}{rgb}{1,1,1}
\definecolor{grey11}{rgb}{0.11,0.11,0.11}
\definecolor{grey12}{rgb}{0.122,0.122,0.122}
\definecolor{grey13}{rgb}{0.129,0.129,0.129}
\definecolor{grey14}{rgb}{0.141,0.141,0.141}
\definecolor{grey15}{rgb}{0.149,0.149,0.149}
\definecolor{grey16}{rgb}{0.161,0.161,0.161}
\definecolor{grey17}{rgb}{0.169,0.169,0.169}
\definecolor{grey18}{rgb}{0.18,0.18,0.18}
\definecolor{grey19}{rgb}{0.188,0.188,0.188}
\definecolor{grey2}{rgb}{0.02,0.02,0.02}
\definecolor{grey20}{rgb}{0.2,0.2,0.2}
\definecolor{grey21}{rgb}{0.212,0.212,0.212}
\definecolor{grey22}{rgb}{0.22,0.22,0.22}
\definecolor{grey23}{rgb}{0.231,0.231,0.231}
\definecolor{grey24}{rgb}{0.239,0.239,0.239}
\definecolor{grey25}{rgb}{0.251,0.251,0.251}
\definecolor{grey26}{rgb}{0.259,0.259,0.259}
\definecolor{grey27}{rgb}{0.271,0.271,0.271}
\definecolor{grey28}{rgb}{0.278,0.278,0.278}
\definecolor{grey29}{rgb}{0.29,0.29,0.29}
\definecolor{grey3}{rgb}{0.031,0.031,0.031}
\definecolor{grey30}{rgb}{0.302,0.302,0.302}
\definecolor{grey31}{rgb}{0.31,0.31,0.31}
\definecolor{grey32}{rgb}{0.322,0.322,0.322}
\definecolor{grey33}{rgb}{0.329,0.329,0.329}
\definecolor{grey34}{rgb}{0.341,0.341,0.341}
\definecolor{grey35}{rgb}{0.349,0.349,0.349}
\definecolor{grey36}{rgb}{0.361,0.361,0.361}
\definecolor{grey37}{rgb}{0.369,0.369,0.369}
\definecolor{grey38}{rgb}{0.38,0.38,0.38}
\definecolor{grey39}{rgb}{0.388,0.388,0.388}
\definecolor{grey4}{rgb}{0.039,0.039,0.039}
\definecolor{grey40}{rgb}{0.4,0.4,0.4}
\definecolor{grey41}{rgb}{0.412,0.412,0.412}
\definecolor{grey42}{rgb}{0.42,0.42,0.42}
\definecolor{grey43}{rgb}{0.431,0.431,0.431}
\definecolor{grey44}{rgb}{0.439,0.439,0.439}
\definecolor{grey45}{rgb}{0.451,0.451,0.451}
\definecolor{grey46}{rgb}{0.459,0.459,0.459}
\definecolor{grey47}{rgb}{0.471,0.471,0.471}
\definecolor{grey48}{rgb}{0.478,0.478,0.478}
\definecolor{grey49}{rgb}{0.49,0.49,0.49}
\definecolor{grey5}{rgb}{0.051,0.051,0.051}
\definecolor{grey50}{rgb}{0.498,0.498,0.498}
\definecolor{grey51}{rgb}{0.51,0.51,0.51}
\definecolor{grey52}{rgb}{0.522,0.522,0.522}
\definecolor{grey53}{rgb}{0.529,0.529,0.529}
\definecolor{grey54}{rgb}{0.541,0.541,0.541}
\definecolor{grey55}{rgb}{0.549,0.549,0.549}
\definecolor{grey56}{rgb}{0.561,0.561,0.561}
\definecolor{grey57}{rgb}{0.569,0.569,0.569}
\definecolor{grey58}{rgb}{0.58,0.58,0.58}
\definecolor{grey59}{rgb}{0.588,0.588,0.588}
\definecolor{grey6}{rgb}{0.059,0.059,0.059}
\definecolor{grey60}{rgb}{0.6,0.6,0.6}
\definecolor{grey61}{rgb}{0.612,0.612,0.612}
\definecolor{grey62}{rgb}{0.62,0.62,0.62}
\definecolor{grey63}{rgb}{0.631,0.631,0.631}
\definecolor{grey64}{rgb}{0.639,0.639,0.639}
\definecolor{grey65}{rgb}{0.651,0.651,0.651}
\definecolor{grey66}{rgb}{0.659,0.659,0.659}
\definecolor{grey67}{rgb}{0.671,0.671,0.671}
\definecolor{grey68}{rgb}{0.678,0.678,0.678}
\definecolor{grey69}{rgb}{0.69,0.69,0.69}
\definecolor{grey7}{rgb}{0.071,0.071,0.071}
\definecolor{grey70}{rgb}{0.702,0.702,0.702}
\definecolor{grey71}{rgb}{0.71,0.71,0.71}
\definecolor{grey72}{rgb}{0.722,0.722,0.722}
\definecolor{grey73}{rgb}{0.729,0.729,0.729}
\definecolor{grey74}{rgb}{0.741,0.741,0.741}
\definecolor{grey75}{rgb}{0.749,0.749,0.749}
\definecolor{grey76}{rgb}{0.761,0.761,0.761}
\definecolor{grey77}{rgb}{0.769,0.769,0.769}
\definecolor{grey78}{rgb}{0.78,0.78,0.78}
\definecolor{grey79}{rgb}{0.788,0.788,0.788}
\definecolor{grey8}{rgb}{0.078,0.078,0.078}
\definecolor{grey80}{rgb}{0.8,0.8,0.8}
\definecolor{grey81}{rgb}{0.812,0.812,0.812}
\definecolor{grey82}{rgb}{0.82,0.82,0.82}
\definecolor{grey83}{rgb}{0.831,0.831,0.831}
\definecolor{grey84}{rgb}{0.839,0.839,0.839}
\definecolor{grey85}{rgb}{0.851,0.851,0.851}
\definecolor{grey86}{rgb}{0.859,0.859,0.859}
\definecolor{grey87}{rgb}{0.871,0.871,0.871}
\definecolor{grey88}{rgb}{0.878,0.878,0.878}
\definecolor{grey89}{rgb}{0.89,0.89,0.89}
\definecolor{grey9}{rgb}{0.09,0.09,0.09}
\definecolor{grey90}{rgb}{0.898,0.898,0.898}
\definecolor{grey91}{rgb}{0.91,0.91,0.91}
\definecolor{grey92}{rgb}{0.922,0.922,0.922}
\definecolor{grey93}{rgb}{0.929,0.929,0.929}
\definecolor{grey94}{rgb}{0.941,0.941,0.941}
\definecolor{grey95}{rgb}{0.949,0.949,0.949}
\definecolor{grey96}{rgb}{0.961,0.961,0.961}
\definecolor{grey97}{rgb}{0.969,0.969,0.969}
\definecolor{grey98}{rgb}{0.98,0.98,0.98}
\definecolor{grey99}{rgb}{0.988,0.988,0.988}
\definecolor{honeydew}{rgb}{0.941,1,0.941}
\definecolor{honeydew1}{rgb}{0.941,1,0.941}
\definecolor{honeydew2}{rgb}{0.878,0.933,0.878}
\definecolor{honeydew3}{rgb}{0.757,0.804,0.757}
\definecolor{honeydew4}{rgb}{0.514,0.545,0.514}
\definecolor{hotpink}{rgb}{1,0.412,0.706}
\definecolor{hotpink1}{rgb}{1,0.431,0.706}
\definecolor{hotpink2}{rgb}{0.933,0.416,0.655}
\definecolor{hotpink3}{rgb}{0.804,0.376,0.565}
\definecolor{hotpink4}{rgb}{0.545,0.227,0.384}
\definecolor{indianred}{rgb}{0.804,0.361,0.361}
\definecolor{indianred1}{rgb}{1,0.416,0.416}
\definecolor{indianred2}{rgb}{0.933,0.388,0.388}
\definecolor{indianred3}{rgb}{0.804,0.333,0.333}
\definecolor{indianred4}{rgb}{0.545,0.227,0.227}
\definecolor{ivory}{rgb}{1,1,0.941}
\definecolor{ivory1}{rgb}{1,1,0.941}
\definecolor{ivory2}{rgb}{0.933,0.933,0.878}
\definecolor{ivory3}{rgb}{0.804,0.804,0.757}
\definecolor{ivory4}{rgb}{0.545,0.545,0.514}
\definecolor{khaki}{rgb}{0.941,0.902,0.549}
\definecolor{khaki1}{rgb}{1,0.965,0.561}
\definecolor{khaki2}{rgb}{0.933,0.902,0.522}
\definecolor{khaki3}{rgb}{0.804,0.776,0.451}
\definecolor{khaki4}{rgb}{0.545,0.525,0.306}
\definecolor{lavender}{rgb}{0.902,0.902,0.98}
\definecolor{lavenderblush}{rgb}{1,0.941,0.961}
\definecolor{lavenderblush1}{rgb}{1,0.941,0.961}
\definecolor{lavenderblush2}{rgb}{0.933,0.878,0.898}
\definecolor{lavenderblush3}{rgb}{0.804,0.757,0.773}
\definecolor{lavenderblush4}{rgb}{0.545,0.514,0.525}
\definecolor{lawngreen}{rgb}{0.486,0.988,0}
\definecolor{lemonchiffon}{rgb}{1,0.98,0.804}
\definecolor{lemonchiffon1}{rgb}{1,0.98,0.804}
\definecolor{lemonchiffon2}{rgb}{0.933,0.914,0.749}
\definecolor{lemonchiffon3}{rgb}{0.804,0.788,0.647}
\definecolor{lemonchiffon4}{rgb}{0.545,0.537,0.439}
\definecolor{lightblue1}{rgb}{0.749,0.937,1}
\definecolor{lightblue2}{rgb}{0.698,0.875,0.933}
\definecolor{lightblue3}{rgb}{0.604,0.753,0.804}
\definecolor{lightblue4}{rgb}{0.408,0.514,0.545}
\definecolor{lightcyan1}{rgb}{0.878,1,1}
\definecolor{lightcyan2}{rgb}{0.82,0.933,0.933}
\definecolor{lightcyan3}{rgb}{0.706,0.804,0.804}
\definecolor{lightcyan4}{rgb}{0.478,0.545,0.545}
\definecolor{lightgoldenrod}{rgb}{0.933,0.867,0.51}
\definecolor{lightgoldenrod1}{rgb}{1,0.925,0.545}
\definecolor{lightgoldenrod2}{rgb}{0.933,0.863,0.51}
\definecolor{lightgoldenrod3}{rgb}{0.804,0.745,0.439}
\definecolor{lightgoldenrod4}{rgb}{0.545,0.506,0.298}
\definecolor{lightgoldenrodyellow}{rgb}{0.98,0.98,0.824}
\definecolor{lightgrey}{rgb}{0.827,0.827,0.827}
\definecolor{lightpink}{rgb}{1,0.714,0.757}
\definecolor{lightpink1}{rgb}{1,0.682,0.725}
\definecolor{lightpink2}{rgb}{0.933,0.635,0.678}
\definecolor{lightpink3}{rgb}{0.804,0.549,0.584}
\definecolor{lightpink4}{rgb}{0.545,0.373,0.396}
\definecolor{lightsalmon}{rgb}{1,0.627,0.478}
\definecolor{lightsalmon1}{rgb}{1,0.627,0.478}
\definecolor{lightsalmon2}{rgb}{0.933,0.584,0.447}
\definecolor{lightsalmon3}{rgb}{0.804,0.506,0.384}
\definecolor{lightsalmon4}{rgb}{0.545,0.341,0.259}
\definecolor{lightseagreen}{rgb}{0.125,0.698,0.667}
\definecolor{lightskyblue}{rgb}{0.529,0.808,0.98}
\definecolor{lightskyblue1}{rgb}{0.69,0.886,1}
\definecolor{lightskyblue3}{rgb}{0.553,0.714,0.804}
\definecolor{lightskyblue4}{rgb}{0.376,0.482,0.545}
\definecolor{lightslateblue}{rgb}{0.518,0.439,1}
\definecolor{lightslategray}{rgb}{0.467,0.533,0.6}
\definecolor{lightslategrey}{rgb}{0.467,0.533,0.6}
\definecolor{lightsteelblue}{rgb}{0.69,0.769,0.871}
\definecolor{lightsteelblue1}{rgb}{0.792,0.882,1}
\definecolor{lightsteelblue2}{rgb}{0.737,0.824,0.933}
\definecolor{lightsteelblue3}{rgb}{0.635,0.71,0.804}
\definecolor{lightsteelblue4}{rgb}{0.431,0.482,0.545}
\definecolor{lightyellow1}{rgb}{1,1,0.878}
\definecolor{lightyellow2}{rgb}{0.933,0.933,0.82}
\definecolor{lightyellow3}{rgb}{0.804,0.804,0.706}
\definecolor{lightyellow4}{rgb}{0.545,0.545,0.478}
\definecolor{limegreen}{rgb}{0.196,0.804,0.196}
\definecolor{linen}{rgb}{0.98,0.941,0.902}
\definecolor{magenta}{rgb}{1,0,1}
\definecolor{magenta1}{rgb}{1,0,1}
\definecolor{magenta2}{rgb}{0.933,0,0.933}
\definecolor{magenta4}{rgb}{0.545,0,0.545}
\definecolor{maroon}{rgb}{0.69,0.188,0.376}
\definecolor{maroon1}{rgb}{1,0.204,0.702}
\definecolor{maroon2}{rgb}{0.933,0.188,0.655}
\definecolor{maroon3}{rgb}{0.804,0.161,0.565}
\definecolor{maroon4}{rgb}{0.545,0.11,0.384}
\definecolor{mediumaquamarine}{rgb}{0.4,0.804,0.667}
\definecolor{mediumblue}{rgb}{0,0,0.804}
\definecolor{mediumorchid}{rgb}{0.729,0.333,0.827}
\definecolor{mediumorchid1}{rgb}{0.878,0.4,1}
\definecolor{mediumorchid2}{rgb}{0.82,0.373,0.933}
\definecolor{mediumorchid3}{rgb}{0.706,0.322,0.804}
\definecolor{mediumorchid4}{rgb}{0.478,0.216,0.545}
\definecolor{mediumpurple}{rgb}{0.576,0.439,0.859}
\definecolor{mediumpurple2}{rgb}{0.624,0.475,0.933}
\definecolor{mediumpurple3}{rgb}{0.537,0.408,0.804}
\definecolor{mediumpurple4}{rgb}{0.365,0.278,0.545}
\definecolor{mediumseagreen}{rgb}{0.235,0.702,0.443}
\definecolor{mediumslateblue}{rgb}{0.482,0.408,0.933}
\definecolor{mediumspringgreen}{rgb}{0,0.98,0.604}
\definecolor{mediumturquoise}{rgb}{0.282,0.82,0.8}
\definecolor{mediumvioletred}{rgb}{0.78,0.082,0.522}
\definecolor{midnightblue}{rgb}{0.098,0.098,0.439}
\definecolor{mintcream}{rgb}{0.961,1,0.98}
\definecolor{mistyrose}{rgb}{1,0.894,0.882}
\definecolor{mistyrose1}{rgb}{1,0.894,0.882}
\definecolor{mistyrose2}{rgb}{0.933,0.835,0.824}
\definecolor{mistyrose3}{rgb}{0.804,0.718,0.71}
\definecolor{mistyrose4}{rgb}{0.545,0.49,0.482}
\definecolor{moccasin}{rgb}{1,0.894,0.71}
\definecolor{navajowhite}{rgb}{1,0.871,0.678}
\definecolor{navajowhite1}{rgb}{1,0.871,0.678}
\definecolor{navajowhite2}{rgb}{0.933,0.812,0.631}
\definecolor{navajowhite3}{rgb}{0.804,0.702,0.545}
\definecolor{navajowhite4}{rgb}{0.545,0.475,0.369}
\definecolor{navyblue}{rgb}{0,0,0.502}
\definecolor{oldlace}{rgb}{0.992,0.961,0.902}
\definecolor{olivedrab}{rgb}{0.42,0.557,0.137}
\definecolor{olivedrab1}{rgb}{0.753,1,0.243}
\definecolor{olivedrab2}{rgb}{0.702,0.933,0.227}
\definecolor{olivedrab3}{rgb}{0.604,0.804,0.196}
\definecolor{olivedrab4}{rgb}{0.412,0.545,0.133}
\definecolor{orange1}{rgb}{1,0.647,0}
\definecolor{orange2}{rgb}{0.933,0.604,0}
\definecolor{orange3}{rgb}{0.804,0.522,0}
\definecolor{orange4}{rgb}{0.545,0.353,0}
\definecolor{orangered}{rgb}{1,0.271,0}
\definecolor{orangered1}{rgb}{1,0.271,0}
\definecolor{orangered2}{rgb}{0.933,0.251,0}
\definecolor{orangered3}{rgb}{0.804,0.216,0}
\definecolor{orangered4}{rgb}{0.545,0.145,0}
\definecolor{orchid}{rgb}{0.855,0.439,0.839}
\definecolor{orchid1}{rgb}{1,0.514,0.98}
\definecolor{orchid2}{rgb}{0.933,0.478,0.914}
\definecolor{orchid3}{rgb}{0.804,0.412,0.788}
\definecolor{orchid4}{rgb}{0.545,0.278,0.537}
\definecolor{palegoldenrod}{rgb}{0.933,0.91,0.667}
\definecolor{palegreen}{rgb}{0.596,0.984,0.596}
\definecolor{palegreen1}{rgb}{0.604,1,0.604}
\definecolor{palegreen2}{rgb}{0.565,0.933,0.565}
\definecolor{palegreen3}{rgb}{0.486,0.804,0.486}
\definecolor{palegreen4}{rgb}{0.329,0.545,0.329}
\definecolor{paleturquoise}{rgb}{0.686,0.933,0.933}
\definecolor{paleturquoise1}{rgb}{0.733,1,1}
\definecolor{paleturquoise2}{rgb}{0.682,0.933,0.933}
\definecolor{paleturquoise3}{rgb}{0.588,0.804,0.804}
\definecolor{paleturquoise4}{rgb}{0.4,0.545,0.545}
\definecolor{palevioletred}{rgb}{0.859,0.439,0.576}
\definecolor{palevioletred1}{rgb}{1,0.51,0.671}
\definecolor{palevioletred2}{rgb}{0.933,0.475,0.624}
\definecolor{palevioletred3}{rgb}{0.804,0.408,0.537}
\definecolor{palevioletred4}{rgb}{0.545,0.278,0.365}
\definecolor{papayawhip}{rgb}{1,0.937,0.835}
\definecolor{peachpuff}{rgb}{1,0.855,0.725}
\definecolor{peachpuff1}{rgb}{1,0.855,0.725}
\definecolor{peachpuff2}{rgb}{0.933,0.796,0.678}
\definecolor{peachpuff3}{rgb}{0.804,0.686,0.584}
\definecolor{peachpuff4}{rgb}{0.545,0.467,0.396}
\definecolor{peru}{rgb}{0.804,0.522,0.247}
\definecolor{pink1}{rgb}{1,0.71,0.773}
\definecolor{pink2}{rgb}{0.933,0.663,0.722}
\definecolor{pink3}{rgb}{0.804,0.569,0.62}
\definecolor{pink4}{rgb}{0.545,0.388,0.424}
\definecolor{plum}{rgb}{0.867,0.627,0.867}
\definecolor{plum1}{rgb}{1,0.733,1}
\definecolor{plum2}{rgb}{0.933,0.682,0.933}
\definecolor{plum3}{rgb}{0.804,0.588,0.804}
\definecolor{plum4}{rgb}{0.545,0.4,0.545}
\definecolor{powderblue}{rgb}{0.69,0.878,0.902}
\definecolor{purple1}{rgb}{0.608,0.188,1}
\definecolor{purple2}{rgb}{0.569,0.173,0.933}
\definecolor{purple3}{rgb}{0.49,0.149,0.804}
\definecolor{purple4}{rgb}{0.333,0.102,0.545}
\definecolor{red1}{rgb}{1,0,0}
\definecolor{red2}{rgb}{0.933,0,0}
\definecolor{red3}{rgb}{0.804,0,0}
\definecolor{red4}{rgb}{0.545,0,0}
\definecolor{rosybrown}{rgb}{0.737,0.561,0.561}
\definecolor{rosybrown1}{rgb}{1,0.757,0.757}
\definecolor{rosybrown2}{rgb}{0.933,0.706,0.706}
\definecolor{rosybrown3}{rgb}{0.804,0.608,0.608}
\definecolor{rosybrown4}{rgb}{0.545,0.412,0.412}
\definecolor{royalblue}{rgb}{0.255,0.412,0.882}
\definecolor{royalblue2}{rgb}{0.263,0.431,0.933}
\definecolor{royalblue3}{rgb}{0.227,0.373,0.804}
\definecolor{royalblue4}{rgb}{0.153,0.251,0.545}
\definecolor{saddlebrown}{rgb}{0.545,0.271,0.075}
\definecolor{salmon}{rgb}{0.98,0.502,0.447}
\definecolor{salmon1}{rgb}{1,0.549,0.412}
\definecolor{salmon2}{rgb}{0.933,0.51,0.384}
\definecolor{salmon3}{rgb}{0.804,0.439,0.329}
\definecolor{salmon4}{rgb}{0.545,0.298,0.224}
\definecolor{sandybrown}{rgb}{0.957,0.643,0.376}
\definecolor{seagreen1}{rgb}{0.329,1,0.624}
\definecolor{seagreen2}{rgb}{0.306,0.933,0.58}
\definecolor{seagreen3}{rgb}{0.263,0.804,0.502}
\definecolor{seagreen4}{rgb}{0.18,0.545,0.341}
\definecolor{seashell}{rgb}{1,0.961,0.933}
\definecolor{seashell1}{rgb}{1,0.961,0.933}
\definecolor{seashell2}{rgb}{0.933,0.898,0.871}
\definecolor{seashell3}{rgb}{0.804,0.773,0.749}
\definecolor{seashell4}{rgb}{0.545,0.525,0.51}
\definecolor{sienna}{rgb}{0.627,0.322,0.176}
\definecolor{sienna1}{rgb}{1,0.51,0.278}
\definecolor{sienna2}{rgb}{0.933,0.475,0.259}
\definecolor{sienna3}{rgb}{0.804,0.408,0.224}
\definecolor{sienna4}{rgb}{0.545,0.278,0.149}
\definecolor{skyblue}{rgb}{0.529,0.808,0.922}
\definecolor{skyblue1}{rgb}{0.529,0.808,1}
\definecolor{skyblue2}{rgb}{0.494,0.753,0.933}
\definecolor{skyblue3}{rgb}{0.424,0.651,0.804}
\definecolor{skyblue4}{rgb}{0.29,0.439,0.545}
\definecolor{slateblue}{rgb}{0.416,0.353,0.804}
\definecolor{slateblue1}{rgb}{0.514,0.435,1}
\definecolor{slateblue2}{rgb}{0.478,0.404,0.933}
\definecolor{slateblue3}{rgb}{0.412,0.349,0.804}
\definecolor{slateblue4}{rgb}{0.278,0.235,0.545}
\definecolor{slategray}{rgb}{0.439,0.502,0.565}
\definecolor{slategray1}{rgb}{0.776,0.886,1}
\definecolor{slategray2}{rgb}{0.725,0.827,0.933}
\definecolor{slategray3}{rgb}{0.624,0.714,0.804}
\definecolor{slategray4}{rgb}{0.424,0.482,0.545}
\definecolor{slategrey}{rgb}{0.439,0.502,0.565}
\definecolor{snow}{rgb}{1,0.98,0.98}
\definecolor{snow1}{rgb}{1,0.98,0.98}
\definecolor{snow2}{rgb}{0.933,0.914,0.914}
\definecolor{snow3}{rgb}{0.804,0.788,0.788}
\definecolor{snow4}{rgb}{0.545,0.537,0.537}
\definecolor{springgreen}{rgb}{0,1,0.498}
\definecolor{springgreen1}{rgb}{0,1,0.498}
\definecolor{springgreen2}{rgb}{0,0.933,0.463}
\definecolor{springgreen3}{rgb}{0,0.804,0.4}
\definecolor{springgreen4}{rgb}{0,0.545,0.271}
\definecolor{steelblue}{rgb}{0.275,0.51,0.706}
\definecolor{steelblue1}{rgb}{0.388,0.722,1}
\definecolor{steelblue2}{rgb}{0.361,0.675,0.933}
\definecolor{steelblue3}{rgb}{0.31,0.58,0.804}
\definecolor{steelblue4}{rgb}{0.212,0.392,0.545}
\definecolor{tan}{rgb}{0.824,0.706,0.549}
\definecolor{tan1}{rgb}{1,0.647,0.31}
\definecolor{tan2}{rgb}{0.933,0.604,0.286}
\definecolor{tan3}{rgb}{0.804,0.522,0.247}
\definecolor{tan4}{rgb}{0.545,0.353,0.169}
\definecolor{thistle}{rgb}{0.847,0.749,0.847}
\definecolor{thistle1}{rgb}{1,0.882,1}
\definecolor{thistle2}{rgb}{0.933,0.824,0.933}
\definecolor{thistle3}{rgb}{0.804,0.71,0.804}
\definecolor{thistle4}{rgb}{0.545,0.482,0.545}
\definecolor{tomato}{rgb}{1,0.388,0.278}
\definecolor{tomato1}{rgb}{1,0.388,0.278}
\definecolor{tomato2}{rgb}{0.933,0.361,0.259}
\definecolor{tomato3}{rgb}{0.804,0.31,0.224}
\definecolor{tomato4}{rgb}{0.545,0.212,0.149}
\definecolor{turquoise1}{rgb}{0,0.961,1}
\definecolor{turquoise2}{rgb}{0,0.898,0.933}
\definecolor{turquoise3}{rgb}{0,0.773,0.804}
\definecolor{turquoise4}{rgb}{0,0.525,0.545}
\definecolor{violetred}{rgb}{0.816,0.125,0.565}
\definecolor{violetred1}{rgb}{1,0.243,0.588}
\definecolor{violetred2}{rgb}{0.933,0.227,0.549}
\definecolor{violetred3}{rgb}{0.804,0.196,0.471}
\definecolor{violetred4}{rgb}{0.545,0.133,0.322}
\definecolor{wheat}{rgb}{0.961,0.871,0.702}
\definecolor{wheat1}{rgb}{1,0.906,0.729}
\definecolor{wheat2}{rgb}{0.933,0.847,0.682}
\definecolor{wheat3}{rgb}{0.804,0.729,0.588}
\definecolor{wheat4}{rgb}{0.545,0.494,0.4}
\definecolor{whitesmoke}{rgb}{0.961,0.961,0.961}
\definecolor{yellow1}{rgb}{1,1,0}
\definecolor{yellow2}{rgb}{0.933,0.933,0}
\definecolor{yellow3}{rgb}{0.804,0.804,0}
\definecolor{yellow4}{rgb}{0.545,0.545,0}
\definecolor{yellowgreen}{rgb}{0.604,0.804,0.196}

\caption{The wall $\tilde{W}_1$ together with some ``zones'' of $r$ consecutive layers.
The area bounded by the orange layer corresponds to the set $I_1^{(i\cdot r)},$ while the area bounded by the green layer corresponds to the set $I_1^{(i\cdot r-r+1)}.$
The sets $X_{\rm in}$ and $X_{\rm out}$ are depicted with blue and orange, repectively.
With light blue (resp. pink) we depict the ``non-privileged'' connected components of $G\setminus X$ that are adjacent to vertices of $X_{\rm in}$ (resp. $X_{\rm out}$).}
\labels{figure_xinxout}
\end{figure}

\myskip\paragraph{Picking the privileged component inside $G\setminus X.$}
Let $\breve{C}$ be the privileged connected component of $G$ with respect to ${\bf W}_q^{(1)}$ and $X.$
Let $Z = I_1^{(d-r+1)}\setminus \breve{C}$ and observe that $\partial_{\mathfrak{K}_1}(Z)\subseteq X_{\rm in}$ (in~\autoref{figure_xinxout}, $Z$ corresponds to the union of the set $X_{\rm in}$ and all connected components of $G\setminus X$ that are depicted in yellow).
We stress that, the target sentences are asked to be satisfied in $C$ but, depending on the $\circ/\bullet$-flag of $θ,$ $C$ is either equal to $\breve{C}$ or $V(\mathfrak{A})\setminus X.$
Note that $I_1^{(d-r+1)}\setminus \breve{C}$ is the union of $X_{\rm in}$ and of every $C∈{\sf cc}(G,X)$ that contains a vertex that is adjacent to a vertex of $X_{\rm in}.$
Therefore, if $w=\circ,$ for each $C∈{\sf cc}(G,X)$ that is a subset of $Z$ (that is equal to $I_1^{(d-r+1)}\setminus \breve{C}$), we ask that $(\mathfrak{A},R,{\bf a})[C]\models \breve{ζ}_{\sf R} |_{\sf ap_{\bf c}},$ while, if $w=\bullet,$ we ask that $(\mathfrak{A},R,{\bf a})\setminus X\models\breve{ζ}_{\sf R} |_{\sf ap_{\bf c}}$ and we keep in mind that $Z\setminus X\subseteq V(\mathfrak{A})\setminus X.$

\myskip\paragraph{All apices are adjacent to the privileged component.}
We also set $V_L({\bf a}) = X\cap V({\bf a})$ and $L$ be the set of indices of the vertices of ${\bf a}$ in $X.$
Observe that $V_L({\bf a}) \subseteq X_{\rm out}.$
We also claim that $V_L ({\bf a}) = V({\bf a}) \cap N_{G} (\breve{C}).$
More generally, we show that for every set $X'$ that
intersects at most $q-1$ bags of $\tilde{\cal Q},$
if  $C'$ is the privileged component of $G$ with respect to ${\bf W}_q^{(1)}$ and $X',$
then $V_L ({\bf a}) = V({\bf a}) \cap N_{G} (C').$

Suppose that a set $X'\subseteq V(\mathfrak{A})$ intersects at most $q-1$ bags of $\tilde{\cal Q}.$
By assumption,
every vertex in $V({\bf a})$ is adjacent, in $G,$ to at least $q$ internal bags of $\tilde{\cal Q}.$
Therefore, for every $a∈ V({\bf a}),$ there is an internal
bag $Q$ of $\tilde{\cal Q}$ such that $V(Q)\subseteq V(G\setminus X')$ and $a$ is adjacent, in $G,$ to a vertex in $V(Q).$
For every such $Q,$ since $C'$ is the privileged component of $G$ with respect to ${\bf W}_q^{(1)}$ and $X',$
it holds that $V(Q)\subseteq C'$ and therefore
every $a∈ V({\bf a})$ is adjacent, in $G,$ to a vertex in
$C'.$
This implies that every $a∈ V({\bf a})$ is either in $N_{G}(C')$ (that is a subset of $X$) or belongs to $C'.$
Therefore $V_L ({\bf a}) = V({\bf a}) \cap N_{G} (C').$
Hence, in the particular case of $\breve{C},$ we have that
$V_L ({\bf a}) = V({\bf a}) \cap N_{G} (\breve{C}).$

Note that, since for every $X'\subseteq V(\mathfrak{A})$ that
intersects at most $q-1$ bags of $\tilde{\cal Q}$ it holds that
$a∈ V({\bf a})$ is either in $N_{G}(C')$ (that is a subset of $X$) or belongs to $C',$ where $C'$ is the privileged component of $G$ with respect to ${\bf W}_q^{(1)}$ and $X',$ for every $C''∈ {\sf cc}(G,X')$ that is not privileged, it holds that $V({\bf a}\cap V(C'')=\emptyset.$
Therefore, whenever we assume that, given a $τ$-stracture $\mathfrak{A},$ an $R\subseteq V(\mathfrak{A}),$ a ${\bf W}_q∈ {(2^{V(\mathfrak{A})})}^{2q},$ an apex-tuple ${\bf a}$ of $\mathfrak{A}$ of size $l$, and a $X\subseteq V(\mathfrak{A}),$ $(\mathfrak{A},R,{\bf W}_{q},{\bf a}, X)\models θ^{\sf out}_q,$ we can replace ${\bf a}$ with $\varnothing^l$ without affecting the validity of $θ^{\sf out}_q.$
\bigskip

The fact that $θ\text{-}{\sf char}(\mathfrak{K}_1,R_1)= θ\text{-}{\sf char}(\mathfrak{K}_2,R_2)$ implies that there is a $Z'\subseteq I_2^{(d-r+1)},$ such that
\begin{itemize}
\item $\blue{{\sf out}\text{-}{\sf sig}}(\mathfrak{K}_1,R_1,d,L,Z)= \blue{{\sf out}\text{-}{\sf sig}}(\mathfrak{K}_2,R_2,d,L,Z')$ and
\item $\green{{\sf in\mbox{-}sig}}(\mathfrak{K}_1,R_1,d,L,Z)=  \green{{\sf in\mbox{-}sig}}(\mathfrak{K}_2,R_2,d,L,Z').$
\end{itemize}

We first prove the following:\medskip
\begin{theoone}\label{claim_1}
There is a set $X'\subseteq Z'$ such that $\partial_{\mathfrak{K}_2} (Z')\subseteq X'$ and for every $V\subseteq Y$
that is also a subset of $V(\cupall{\sf influence}_{\breve{\mathfrak{R}}'}(\overline{W}),$ where $\overline{W}$ is the central $(j'-2)$-subwall of $W_1,$ it holds that
$(\mathfrak{A},R,{\bf W}_{{q}}^{(1)},\varnothing^l, X)\models θ^{\sf out}_q \iff (\mathfrak{A}\setminus V,R\setminus Y,{\bf W}_q^{(1)},\varnothing^l, X_{\rm out}\cup X')\models θ^{\sf out}_q.$
\end{theoone}

\noindent{\em \blue{Proof of~\autoref{claim_1}}:}
The idea here is to build an $h$-boundaried $τ'$-structure (with respect to $W_1$) to fit the {\sf out}\text{-}{\sf sig}.
By~\autoref{cou_more}, this structure is associated with a sentence $\bar{φ}∈ {\sf rep}_{τ'}^{(h)} ( θ^{\sf out}_q).$
Next, we will consider another $h$-boundaried $τ'$-structure (with respect to $W_2$) that satisfies the same sentence, using the fact that $W_1$ and $W_2$ have the same {\sf out}\text{-}{\sf sig}.
The fact that both these boundaried structures, when ``completed'' from the other side by the same structure, give the same structure,
will imply that they are ``equivalent'' with respect to the satisfaction of $θ^{\sf out}_q.$

Let $h:= |N_{G} (\breve{C})|.$
Recall that ${\sf cl}_{\sf X}({\sf star}_{\sf X} (\mathfrak{A},X))$ has treewidth at most $\tw(θ).$
Since  the set $N_{G} (\breve{C})$ induces a $K_h$ on the Gaifman graph of ${\sf cl}_{\sf X}({\sf star}_{\sf X} (\mathfrak{A},X)),$ we have that $h∈[0, \tw(θ)-1].$

\myskip\paragraph{Defining the boundary of our boundaried structure.}
We set $F'$ to be the graph $G[(X_{\rm out}\setminus V_L ({\bf a}))\cap N_{G} (\breve{C})].$
In other words, $F'$ is the subgraph of $G$ induced by the vertices of $X_{\rm out}$ that are not apices and are adjacent to vertices in $\breve{C}$ (see~\autoref{@governorship} for an example).
Also, we set $F^\star$ to be the graph obtained from $G[V_L ({\bf a})\cup V(F')]$ after removing every edge that has both endpoints in $V_L ({\bf a}).$
Intuitively, we extend $F'$ to $F^\star$ by adding the vertices in $V_L ({\bf a})$ and the edges connecting vertices of  $V_L ({\bf a})$ and $V(F'),$ but not the edges that have both endpoints in $V_L ({\bf a}).$
This graph $F^\star$ will be later associated with a graph $F∈ {\cal F}_{h-|\partial_{\mathfrak{K}_1} (Z)|}^{V_L({\bf a})}.$

\myskip\paragraph{Separating $\mathfrak{A}$ into two boundaried structures.}
We now aim to ``break'' $\mathfrak{A}$ into two boundaried structures,
to be able to encode, using Courcelle's representatives, the ``partial satisfaction'' of $θ_q^{\rm out}$ inside
$\breve{C}$ and $Z.$
 Let $$\mathfrak{A}_{\rm out}^\star = \mathfrak{A}\setminus (\breve{C}\cup (Z\setminus \partial_{\mathfrak{K}_1} (Z)))\mbox{~~and~~}\mathfrak{A}^\star = \mathfrak{A}[\breve{C}\cup Z \cup V(F^\star)].$$
Keep in mind that $Z = I_1^{(d-r+1)}\setminus \breve{C}.$
See~\autoref{figure_boundariedgraph1} to get some intuition of the Gaifman graph of $\mathfrak{A}_{\rm out}^\star$ and $\mathfrak{A}^\star.$
Verbally, the structure $\mathfrak{A}_{\rm out}^\star$ is obtained from $\mathfrak{A}$ by removing from its universe the elements of $\breve{C}$ and the elements of $Z\setminus \partial_{\mathfrak{K}_1} (Z).$ In other words, apart from the vertices in $\partial_{\mathfrak{K}_1} (Z)\cup V(F^\star)$ (that are in the universe of both $\mathfrak{A}_{\rm out}^\star$ and $\mathfrak{A}^\star$),  the structure $\mathfrak{A}_{\rm out}^\star$ corresponds to the part of $\mathfrak{A}$ that is ``away'' from $\breve{C}$ and $Z,$ while the structure $\mathfrak{A}^\star$ corresponds to the part of $\mathfrak{A}$ induced by the union of $\breve{C},$ $Z$ and $V(F^\star).$ Keep in mind that $V(\mathfrak{A}_{\rm out}^\star) \cap V(\mathfrak{A}^\star) = \partial_{\mathfrak{K}_1} (Z)\cup V(F^\star).$
Next, we will define two boundaried structures corresponding to $\mathfrak{A}_{\rm out}^\star$ and $\mathfrak{A}^\star,$ whose boundary will be the set $\partial_{\mathfrak{K}_1} (Z)\cup V(F^\star).$

\begin{figure}[ht]
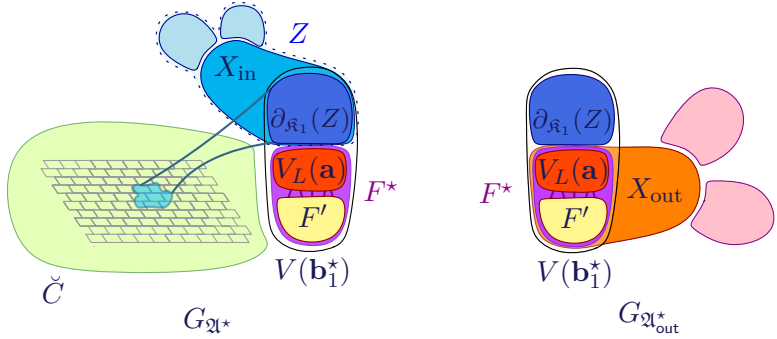

\centering
\tikzstyle{ipe stylesheet} = [
  ipe import,
  even odd rule,
  line join=round,
  line cap=butt,
  ipe pen normal/.style={line width=0.4},
  ipe pen heavier/.style={line width=0.8},
  ipe pen fat/.style={line width=1.2},
  ipe pen ultrafat/.style={line width=2},
  ipe pen normal,
  ipe mark normal/.style={ipe mark scale=3},
  ipe mark large/.style={ipe mark scale=5},
  ipe mark small/.style={ipe mark scale=2},
  ipe mark tiny/.style={ipe mark scale=1.1},
  ipe mark normal,
  /pgf/arrow keys/.cd,
  ipe arrow normal/.style={scale=7},
  ipe arrow large/.style={scale=10},
  ipe arrow small/.style={scale=5},
  ipe arrow tiny/.style={scale=3},
  ipe arrow normal,
  /tikz/.cd,
  ipe arrows, 
  <->/.tip = ipe normal,
  ipe dash normal/.style={dash pattern=},
  ipe dash dotted/.style={dash pattern=on 1bp off 3bp},
  ipe dash dashed/.style={dash pattern=on 4bp off 4bp},
  ipe dash dash dotted/.style={dash pattern=on 4bp off 2bp on 1bp off 2bp},
  ipe dash dash dot dotted/.style={dash pattern=on 4bp off 2bp on 1bp off 2bp on 1bp off 2bp},
  ipe dash normal,
  ipe node/.append style={font=\normalsize},
  ipe stretch normal/.style={ipe node stretch=1},
  ipe stretch normal,
  ipe opacity 10/.style={opacity=0.1},
  ipe opacity 30/.style={opacity=0.3},
  ipe opacity 50/.style={opacity=0.5},
  ipe opacity 75/.style={opacity=0.75},
  ipe opacity opaque/.style={opacity=1},
  ipe opacity opaque,
]
\definecolor{black}{rgb}{0,0,0}
\definecolor{white}{rgb}{1,1,1}
\definecolor{red}{rgb}{1,0,0}
\definecolor{blue}{rgb}{0,0,1}
\definecolor{green}{rgb}{0,1,0}
\definecolor{yellow}{rgb}{1,1,0}
\definecolor{orange}{rgb}{1,0.647,0}
\definecolor{gold}{rgb}{1,0.843,0}
\definecolor{purple}{rgb}{0.627,0.125,0.941}
\definecolor{gray}{rgb}{0.745,0.745,0.745}
\definecolor{brown}{rgb}{0.647,0.165,0.165}
\definecolor{navy}{rgb}{0,0,0.502}
\definecolor{pink}{rgb}{1,0.753,0.796}
\definecolor{seagreen}{rgb}{0.18,0.545,0.341}
\definecolor{turquoise}{rgb}{0.251,0.878,0.816}
\definecolor{violet}{rgb}{0.933,0.51,0.933}
\definecolor{darkblue}{rgb}{0,0,0.545}
\definecolor{darkcyan}{rgb}{0,0.545,0.545}
\definecolor{darkgray}{rgb}{0.663,0.663,0.663}
\definecolor{darkgreen}{rgb}{0,0.392,0}
\definecolor{darkmagenta}{rgb}{0.545,0,0.545}
\definecolor{darkorange}{rgb}{1,0.549,0}
\definecolor{darkred}{rgb}{0.545,0,0}
\definecolor{lightblue}{rgb}{0.678,0.847,0.902}
\definecolor{lightcyan}{rgb}{0.878,1,1}
\definecolor{lightgray}{rgb}{0.827,0.827,0.827}
\definecolor{lightgreen}{rgb}{0.565,0.933,0.565}
\definecolor{lightyellow}{rgb}{1,1,0.878}
\definecolor{aliceblue}{rgb}{0.941,0.973,1}
\definecolor{antiquewhite}{rgb}{0.98,0.922,0.843}
\definecolor{antiquewhite1}{rgb}{1,0.937,0.859}
\definecolor{antiquewhite2}{rgb}{0.933,0.875,0.8}
\definecolor{antiquewhite3}{rgb}{0.804,0.753,0.69}
\definecolor{antiquewhite4}{rgb}{0.545,0.514,0.471}
\definecolor{aquamarine}{rgb}{0.498,1,0.831}
\definecolor{aquamarine1}{rgb}{0.498,1,0.831}
\definecolor{aquamarine2}{rgb}{0.463,0.933,0.776}
\definecolor{aquamarine3}{rgb}{0.4,0.804,0.667}
\definecolor{aquamarine4}{rgb}{0.271,0.545,0.455}
\definecolor{azure}{rgb}{0.941,1,1}
\definecolor{azure1}{rgb}{0.941,1,1}
\definecolor{azure2}{rgb}{0.878,0.933,0.933}
\definecolor{azure3}{rgb}{0.757,0.804,0.804}
\definecolor{azure4}{rgb}{0.514,0.545,0.545}
\definecolor{beige}{rgb}{0.961,0.961,0.863}
\definecolor{bisque}{rgb}{1,0.894,0.769}
\definecolor{bisque1}{rgb}{1,0.894,0.769}
\definecolor{bisque2}{rgb}{0.933,0.835,0.718}
\definecolor{bisque3}{rgb}{0.804,0.718,0.62}
\definecolor{bisque4}{rgb}{0.545,0.49,0.42}
\definecolor{blanchedalmond}{rgb}{1,0.922,0.804}
\definecolor{blue1}{rgb}{0,0,1}
\definecolor{blue2}{rgb}{0,0,0.933}
\definecolor{blue3}{rgb}{0,0,0.804}
\definecolor{blue4}{rgb}{0,0,0.545}
\definecolor{blueviolet}{rgb}{0.541,0.169,0.886}
\definecolor{brown1}{rgb}{1,0.251,0.251}
\definecolor{brown2}{rgb}{0.933,0.231,0.231}
\definecolor{brown3}{rgb}{0.804,0.2,0.2}
\definecolor{brown4}{rgb}{0.545,0.137,0.137}
\definecolor{burlywood}{rgb}{0.871,0.722,0.529}
\definecolor{burlywood1}{rgb}{1,0.827,0.608}
\definecolor{burlywood2}{rgb}{0.933,0.773,0.569}
\definecolor{burlywood3}{rgb}{0.804,0.667,0.49}
\definecolor{burlywood4}{rgb}{0.545,0.451,0.333}
\definecolor{cadetblue}{rgb}{0.373,0.62,0.627}
\definecolor{cadetblue1}{rgb}{0.596,0.961,1}
\definecolor{cadetblue2}{rgb}{0.557,0.898,0.933}
\definecolor{cadetblue3}{rgb}{0.478,0.773,0.804}
\definecolor{cadetblue4}{rgb}{0.325,0.525,0.545}
\definecolor{chartreuse}{rgb}{0.498,1,0}
\definecolor{chartreuse1}{rgb}{0.498,1,0}
\definecolor{chartreuse2}{rgb}{0.463,0.933,0}
\definecolor{chartreuse3}{rgb}{0.4,0.804,0}
\definecolor{chartreuse4}{rgb}{0.271,0.545,0}
\definecolor{chocolate}{rgb}{0.824,0.412,0.118}
\definecolor{chocolate1}{rgb}{1,0.498,0.141}
\definecolor{chocolate2}{rgb}{0.933,0.463,0.129}
\definecolor{chocolate3}{rgb}{0.804,0.4,0.114}
\definecolor{chocolate4}{rgb}{0.545,0.271,0.075}
\definecolor{coral}{rgb}{1,0.498,0.314}
\definecolor{coral1}{rgb}{1,0.447,0.337}
\definecolor{coral2}{rgb}{0.933,0.416,0.314}
\definecolor{coral3}{rgb}{0.804,0.357,0.271}
\definecolor{coral4}{rgb}{0.545,0.243,0.184}
\definecolor{cornflowerblue}{rgb}{0.392,0.584,0.929}
\definecolor{cornsilk}{rgb}{1,0.973,0.863}
\definecolor{cornsilk1}{rgb}{1,0.973,0.863}
\definecolor{cornsilk2}{rgb}{0.933,0.91,0.804}
\definecolor{cornsilk3}{rgb}{0.804,0.784,0.694}
\definecolor{cornsilk4}{rgb}{0.545,0.533,0.471}
\definecolor{cyan}{rgb}{0,1,1}
\definecolor{cyan1}{rgb}{0,1,1}
\definecolor{cyan2}{rgb}{0,0.933,0.933}
\definecolor{cyan3}{rgb}{0,0.804,0.804}
\definecolor{cyan4}{rgb}{0,0.545,0.545}
\definecolor{darkgoldenrod}{rgb}{0.722,0.525,0.043}
\definecolor{darkgoldenrod1}{rgb}{1,0.725,0.059}
\definecolor{darkgoldenrod2}{rgb}{0.933,0.678,0.055}
\definecolor{darkgoldenrod3}{rgb}{0.804,0.584,0.047}
\definecolor{darkgoldenrod4}{rgb}{0.545,0.396,0.031}
\definecolor{darkgrey}{rgb}{0.663,0.663,0.663}
\definecolor{darkkhaki}{rgb}{0.741,0.718,0.42}
\definecolor{darkolivegreen}{rgb}{0.333,0.42,0.184}
\definecolor{darkolivegreen1}{rgb}{0.792,1,0.439}
\definecolor{darkolivegreen2}{rgb}{0.737,0.933,0.408}
\definecolor{darkolivegreen3}{rgb}{0.635,0.804,0.353}
\definecolor{darkolivegreen4}{rgb}{0.431,0.545,0.239}
\definecolor{darkorange1}{rgb}{1,0.498,0}
\definecolor{darkorange2}{rgb}{0.933,0.463,0}
\definecolor{darkorange3}{rgb}{0.804,0.4,0}
\definecolor{darkorange4}{rgb}{0.545,0.271,0}
\definecolor{darkorchid}{rgb}{0.6,0.196,0.8}
\definecolor{darkorchid1}{rgb}{0.749,0.243,1}
\definecolor{darkorchid2}{rgb}{0.698,0.227,0.933}
\definecolor{darkorchid3}{rgb}{0.604,0.196,0.804}
\definecolor{darkorchid4}{rgb}{0.408,0.133,0.545}
\definecolor{darksalmon}{rgb}{0.914,0.588,0.478}
\definecolor{darkseagreen}{rgb}{0.561,0.737,0.561}
\definecolor{darkseagreen1}{rgb}{0.757,1,0.757}
\definecolor{darkseagreen2}{rgb}{0.706,0.933,0.706}
\definecolor{darkseagreen3}{rgb}{0.608,0.804,0.608}
\definecolor{darkseagreen4}{rgb}{0.412,0.545,0.412}
\definecolor{darkslateblue}{rgb}{0.282,0.239,0.545}
\definecolor{darkslategray}{rgb}{0.184,0.31,0.31}
\definecolor{darkslategray1}{rgb}{0.592,1,1}
\definecolor{darkslategray2}{rgb}{0.553,0.933,0.933}
\definecolor{darkslategray3}{rgb}{0.475,0.804,0.804}
\definecolor{darkslategray4}{rgb}{0.322,0.545,0.545}
\definecolor{darkslategrey}{rgb}{0.184,0.31,0.31}
\definecolor{darkturquoise}{rgb}{0,0.808,0.82}
\definecolor{darkviolet}{rgb}{0.58,0,0.827}
\definecolor{deeppink}{rgb}{1,0.078,0.576}
\definecolor{deeppink1}{rgb}{1,0.078,0.576}
\definecolor{deeppink2}{rgb}{0.933,0.071,0.537}
\definecolor{deeppink3}{rgb}{0.804,0.063,0.463}
\definecolor{deeppink4}{rgb}{0.545,0.039,0.314}
\definecolor{deepskyblue}{rgb}{0,0.749,1}
\definecolor{deepskyblue1}{rgb}{0,0.749,1}
\definecolor{deepskyblue2}{rgb}{0,0.698,0.933}
\definecolor{deepskyblue3}{rgb}{0,0.604,0.804}
\definecolor{deepskyblue4}{rgb}{0,0.408,0.545}
\definecolor{dimgray}{rgb}{0.412,0.412,0.412}
\definecolor{dimgrey}{rgb}{0.412,0.412,0.412}
\definecolor{dodgerblue}{rgb}{0.118,0.565,1}
\definecolor{dodgerblue1}{rgb}{0.118,0.565,1}
\definecolor{dodgerblue2}{rgb}{0.11,0.525,0.933}
\definecolor{dodgerblue3}{rgb}{0.094,0.455,0.804}
\definecolor{dodgerblue4}{rgb}{0.063,0.306,0.545}
\definecolor{firebrick}{rgb}{0.698,0.133,0.133}
\definecolor{firebrick1}{rgb}{1,0.188,0.188}
\definecolor{firebrick2}{rgb}{0.933,0.173,0.173}
\definecolor{firebrick3}{rgb}{0.804,0.149,0.149}
\definecolor{firebrick4}{rgb}{0.545,0.102,0.102}
\definecolor{floralwhite}{rgb}{1,0.98,0.941}
\definecolor{forestgreen}{rgb}{0.133,0.545,0.133}
\definecolor{gainsboro}{rgb}{0.863,0.863,0.863}
\definecolor{ghostwhite}{rgb}{0.973,0.973,1}
\definecolor{gold1}{rgb}{1,0.843,0}
\definecolor{gold2}{rgb}{0.933,0.788,0}
\definecolor{gold3}{rgb}{0.804,0.678,0}
\definecolor{gold4}{rgb}{0.545,0.459,0}
\definecolor{goldenrod}{rgb}{0.855,0.647,0.125}
\definecolor{goldenrod1}{rgb}{1,0.757,0.145}
\definecolor{goldenrod2}{rgb}{0.933,0.706,0.133}
\definecolor{goldenrod3}{rgb}{0.804,0.608,0.114}
\definecolor{goldenrod4}{rgb}{0.545,0.412,0.078}
\definecolor{gray0}{rgb}{0,0,0}
\definecolor{gray1}{rgb}{0.012,0.012,0.012}
\definecolor{gray10}{rgb}{0.102,0.102,0.102}
\definecolor{gray100}{rgb}{1,1,1}
\definecolor{gray11}{rgb}{0.11,0.11,0.11}
\definecolor{gray12}{rgb}{0.122,0.122,0.122}
\definecolor{gray13}{rgb}{0.129,0.129,0.129}
\definecolor{gray14}{rgb}{0.141,0.141,0.141}
\definecolor{gray15}{rgb}{0.149,0.149,0.149}
\definecolor{gray16}{rgb}{0.161,0.161,0.161}
\definecolor{gray17}{rgb}{0.169,0.169,0.169}
\definecolor{gray18}{rgb}{0.18,0.18,0.18}
\definecolor{gray19}{rgb}{0.188,0.188,0.188}
\definecolor{gray2}{rgb}{0.02,0.02,0.02}
\definecolor{gray20}{rgb}{0.2,0.2,0.2}
\definecolor{gray21}{rgb}{0.212,0.212,0.212}
\definecolor{gray22}{rgb}{0.22,0.22,0.22}
\definecolor{gray23}{rgb}{0.231,0.231,0.231}
\definecolor{gray24}{rgb}{0.239,0.239,0.239}
\definecolor{gray25}{rgb}{0.251,0.251,0.251}
\definecolor{gray26}{rgb}{0.259,0.259,0.259}
\definecolor{gray27}{rgb}{0.271,0.271,0.271}
\definecolor{gray28}{rgb}{0.278,0.278,0.278}
\definecolor{gray29}{rgb}{0.29,0.29,0.29}
\definecolor{gray3}{rgb}{0.031,0.031,0.031}
\definecolor{gray30}{rgb}{0.302,0.302,0.302}
\definecolor{gray31}{rgb}{0.31,0.31,0.31}
\definecolor{gray32}{rgb}{0.322,0.322,0.322}
\definecolor{gray33}{rgb}{0.329,0.329,0.329}
\definecolor{gray34}{rgb}{0.341,0.341,0.341}
\definecolor{gray35}{rgb}{0.349,0.349,0.349}
\definecolor{gray36}{rgb}{0.361,0.361,0.361}
\definecolor{gray37}{rgb}{0.369,0.369,0.369}
\definecolor{gray38}{rgb}{0.38,0.38,0.38}
\definecolor{gray39}{rgb}{0.388,0.388,0.388}
\definecolor{gray4}{rgb}{0.039,0.039,0.039}
\definecolor{gray40}{rgb}{0.4,0.4,0.4}
\definecolor{gray41}{rgb}{0.412,0.412,0.412}
\definecolor{gray42}{rgb}{0.42,0.42,0.42}
\definecolor{gray43}{rgb}{0.431,0.431,0.431}
\definecolor{gray44}{rgb}{0.439,0.439,0.439}
\definecolor{gray45}{rgb}{0.451,0.451,0.451}
\definecolor{gray46}{rgb}{0.459,0.459,0.459}
\definecolor{gray47}{rgb}{0.471,0.471,0.471}
\definecolor{gray48}{rgb}{0.478,0.478,0.478}
\definecolor{gray49}{rgb}{0.49,0.49,0.49}
\definecolor{gray5}{rgb}{0.051,0.051,0.051}
\definecolor{gray50}{rgb}{0.498,0.498,0.498}
\definecolor{gray51}{rgb}{0.51,0.51,0.51}
\definecolor{gray52}{rgb}{0.522,0.522,0.522}
\definecolor{gray53}{rgb}{0.529,0.529,0.529}
\definecolor{gray54}{rgb}{0.541,0.541,0.541}
\definecolor{gray55}{rgb}{0.549,0.549,0.549}
\definecolor{gray56}{rgb}{0.561,0.561,0.561}
\definecolor{gray57}{rgb}{0.569,0.569,0.569}
\definecolor{gray58}{rgb}{0.58,0.58,0.58}
\definecolor{gray59}{rgb}{0.588,0.588,0.588}
\definecolor{gray6}{rgb}{0.059,0.059,0.059}
\definecolor{gray60}{rgb}{0.6,0.6,0.6}
\definecolor{gray61}{rgb}{0.612,0.612,0.612}
\definecolor{gray62}{rgb}{0.62,0.62,0.62}
\definecolor{gray63}{rgb}{0.631,0.631,0.631}
\definecolor{gray64}{rgb}{0.639,0.639,0.639}
\definecolor{gray65}{rgb}{0.651,0.651,0.651}
\definecolor{gray66}{rgb}{0.659,0.659,0.659}
\definecolor{gray67}{rgb}{0.671,0.671,0.671}
\definecolor{gray68}{rgb}{0.678,0.678,0.678}
\definecolor{gray69}{rgb}{0.69,0.69,0.69}
\definecolor{gray7}{rgb}{0.071,0.071,0.071}
\definecolor{gray70}{rgb}{0.702,0.702,0.702}
\definecolor{gray71}{rgb}{0.71,0.71,0.71}
\definecolor{gray72}{rgb}{0.722,0.722,0.722}
\definecolor{gray73}{rgb}{0.729,0.729,0.729}
\definecolor{gray74}{rgb}{0.741,0.741,0.741}
\definecolor{gray75}{rgb}{0.749,0.749,0.749}
\definecolor{gray76}{rgb}{0.761,0.761,0.761}
\definecolor{gray77}{rgb}{0.769,0.769,0.769}
\definecolor{gray78}{rgb}{0.78,0.78,0.78}
\definecolor{gray79}{rgb}{0.788,0.788,0.788}
\definecolor{gray8}{rgb}{0.078,0.078,0.078}
\definecolor{gray80}{rgb}{0.8,0.8,0.8}
\definecolor{gray81}{rgb}{0.812,0.812,0.812}
\definecolor{gray82}{rgb}{0.82,0.82,0.82}
\definecolor{gray83}{rgb}{0.831,0.831,0.831}
\definecolor{gray84}{rgb}{0.839,0.839,0.839}
\definecolor{gray85}{rgb}{0.851,0.851,0.851}
\definecolor{gray86}{rgb}{0.859,0.859,0.859}
\definecolor{gray87}{rgb}{0.871,0.871,0.871}
\definecolor{gray88}{rgb}{0.878,0.878,0.878}
\definecolor{gray89}{rgb}{0.89,0.89,0.89}
\definecolor{gray9}{rgb}{0.09,0.09,0.09}
\definecolor{gray90}{rgb}{0.898,0.898,0.898}
\definecolor{gray91}{rgb}{0.91,0.91,0.91}
\definecolor{gray92}{rgb}{0.922,0.922,0.922}
\definecolor{gray93}{rgb}{0.929,0.929,0.929}
\definecolor{gray94}{rgb}{0.941,0.941,0.941}
\definecolor{gray95}{rgb}{0.949,0.949,0.949}
\definecolor{gray96}{rgb}{0.961,0.961,0.961}
\definecolor{gray97}{rgb}{0.969,0.969,0.969}
\definecolor{gray98}{rgb}{0.98,0.98,0.98}
\definecolor{gray99}{rgb}{0.988,0.988,0.988}
\definecolor{green1}{rgb}{0,1,0}
\definecolor{green2}{rgb}{0,0.933,0}
\definecolor{green3}{rgb}{0,0.804,0}
\definecolor{green4}{rgb}{0,0.545,0}
\definecolor{greenyellow}{rgb}{0.678,1,0.184}
\definecolor{grey}{rgb}{0.745,0.745,0.745}
\definecolor{grey0}{rgb}{0,0,0}
\definecolor{grey1}{rgb}{0.012,0.012,0.012}
\definecolor{grey10}{rgb}{0.102,0.102,0.102}
\definecolor{grey100}{rgb}{1,1,1}
\definecolor{grey11}{rgb}{0.11,0.11,0.11}
\definecolor{grey12}{rgb}{0.122,0.122,0.122}
\definecolor{grey13}{rgb}{0.129,0.129,0.129}
\definecolor{grey14}{rgb}{0.141,0.141,0.141}
\definecolor{grey15}{rgb}{0.149,0.149,0.149}
\definecolor{grey16}{rgb}{0.161,0.161,0.161}
\definecolor{grey17}{rgb}{0.169,0.169,0.169}
\definecolor{grey18}{rgb}{0.18,0.18,0.18}
\definecolor{grey19}{rgb}{0.188,0.188,0.188}
\definecolor{grey2}{rgb}{0.02,0.02,0.02}
\definecolor{grey20}{rgb}{0.2,0.2,0.2}
\definecolor{grey21}{rgb}{0.212,0.212,0.212}
\definecolor{grey22}{rgb}{0.22,0.22,0.22}
\definecolor{grey23}{rgb}{0.231,0.231,0.231}
\definecolor{grey24}{rgb}{0.239,0.239,0.239}
\definecolor{grey25}{rgb}{0.251,0.251,0.251}
\definecolor{grey26}{rgb}{0.259,0.259,0.259}
\definecolor{grey27}{rgb}{0.271,0.271,0.271}
\definecolor{grey28}{rgb}{0.278,0.278,0.278}
\definecolor{grey29}{rgb}{0.29,0.29,0.29}
\definecolor{grey3}{rgb}{0.031,0.031,0.031}
\definecolor{grey30}{rgb}{0.302,0.302,0.302}
\definecolor{grey31}{rgb}{0.31,0.31,0.31}
\definecolor{grey32}{rgb}{0.322,0.322,0.322}
\definecolor{grey33}{rgb}{0.329,0.329,0.329}
\definecolor{grey34}{rgb}{0.341,0.341,0.341}
\definecolor{grey35}{rgb}{0.349,0.349,0.349}
\definecolor{grey36}{rgb}{0.361,0.361,0.361}
\definecolor{grey37}{rgb}{0.369,0.369,0.369}
\definecolor{grey38}{rgb}{0.38,0.38,0.38}
\definecolor{grey39}{rgb}{0.388,0.388,0.388}
\definecolor{grey4}{rgb}{0.039,0.039,0.039}
\definecolor{grey40}{rgb}{0.4,0.4,0.4}
\definecolor{grey41}{rgb}{0.412,0.412,0.412}
\definecolor{grey42}{rgb}{0.42,0.42,0.42}
\definecolor{grey43}{rgb}{0.431,0.431,0.431}
\definecolor{grey44}{rgb}{0.439,0.439,0.439}
\definecolor{grey45}{rgb}{0.451,0.451,0.451}
\definecolor{grey46}{rgb}{0.459,0.459,0.459}
\definecolor{grey47}{rgb}{0.471,0.471,0.471}
\definecolor{grey48}{rgb}{0.478,0.478,0.478}
\definecolor{grey49}{rgb}{0.49,0.49,0.49}
\definecolor{grey5}{rgb}{0.051,0.051,0.051}
\definecolor{grey50}{rgb}{0.498,0.498,0.498}
\definecolor{grey51}{rgb}{0.51,0.51,0.51}
\definecolor{grey52}{rgb}{0.522,0.522,0.522}
\definecolor{grey53}{rgb}{0.529,0.529,0.529}
\definecolor{grey54}{rgb}{0.541,0.541,0.541}
\definecolor{grey55}{rgb}{0.549,0.549,0.549}
\definecolor{grey56}{rgb}{0.561,0.561,0.561}
\definecolor{grey57}{rgb}{0.569,0.569,0.569}
\definecolor{grey58}{rgb}{0.58,0.58,0.58}
\definecolor{grey59}{rgb}{0.588,0.588,0.588}
\definecolor{grey6}{rgb}{0.059,0.059,0.059}
\definecolor{grey60}{rgb}{0.6,0.6,0.6}
\definecolor{grey61}{rgb}{0.612,0.612,0.612}
\definecolor{grey62}{rgb}{0.62,0.62,0.62}
\definecolor{grey63}{rgb}{0.631,0.631,0.631}
\definecolor{grey64}{rgb}{0.639,0.639,0.639}
\definecolor{grey65}{rgb}{0.651,0.651,0.651}
\definecolor{grey66}{rgb}{0.659,0.659,0.659}
\definecolor{grey67}{rgb}{0.671,0.671,0.671}
\definecolor{grey68}{rgb}{0.678,0.678,0.678}
\definecolor{grey69}{rgb}{0.69,0.69,0.69}
\definecolor{grey7}{rgb}{0.071,0.071,0.071}
\definecolor{grey70}{rgb}{0.702,0.702,0.702}
\definecolor{grey71}{rgb}{0.71,0.71,0.71}
\definecolor{grey72}{rgb}{0.722,0.722,0.722}
\definecolor{grey73}{rgb}{0.729,0.729,0.729}
\definecolor{grey74}{rgb}{0.741,0.741,0.741}
\definecolor{grey75}{rgb}{0.749,0.749,0.749}
\definecolor{grey76}{rgb}{0.761,0.761,0.761}
\definecolor{grey77}{rgb}{0.769,0.769,0.769}
\definecolor{grey78}{rgb}{0.78,0.78,0.78}
\definecolor{grey79}{rgb}{0.788,0.788,0.788}
\definecolor{grey8}{rgb}{0.078,0.078,0.078}
\definecolor{grey80}{rgb}{0.8,0.8,0.8}
\definecolor{grey81}{rgb}{0.812,0.812,0.812}
\definecolor{grey82}{rgb}{0.82,0.82,0.82}
\definecolor{grey83}{rgb}{0.831,0.831,0.831}
\definecolor{grey84}{rgb}{0.839,0.839,0.839}
\definecolor{grey85}{rgb}{0.851,0.851,0.851}
\definecolor{grey86}{rgb}{0.859,0.859,0.859}
\definecolor{grey87}{rgb}{0.871,0.871,0.871}
\definecolor{grey88}{rgb}{0.878,0.878,0.878}
\definecolor{grey89}{rgb}{0.89,0.89,0.89}
\definecolor{grey9}{rgb}{0.09,0.09,0.09}
\definecolor{grey90}{rgb}{0.898,0.898,0.898}
\definecolor{grey91}{rgb}{0.91,0.91,0.91}
\definecolor{grey92}{rgb}{0.922,0.922,0.922}
\definecolor{grey93}{rgb}{0.929,0.929,0.929}
\definecolor{grey94}{rgb}{0.941,0.941,0.941}
\definecolor{grey95}{rgb}{0.949,0.949,0.949}
\definecolor{grey96}{rgb}{0.961,0.961,0.961}
\definecolor{grey97}{rgb}{0.969,0.969,0.969}
\definecolor{grey98}{rgb}{0.98,0.98,0.98}
\definecolor{grey99}{rgb}{0.988,0.988,0.988}
\definecolor{honeydew}{rgb}{0.941,1,0.941}
\definecolor{honeydew1}{rgb}{0.941,1,0.941}
\definecolor{honeydew2}{rgb}{0.878,0.933,0.878}
\definecolor{honeydew3}{rgb}{0.757,0.804,0.757}
\definecolor{honeydew4}{rgb}{0.514,0.545,0.514}
\definecolor{hotpink}{rgb}{1,0.412,0.706}
\definecolor{hotpink1}{rgb}{1,0.431,0.706}
\definecolor{hotpink2}{rgb}{0.933,0.416,0.655}
\definecolor{hotpink3}{rgb}{0.804,0.376,0.565}
\definecolor{hotpink4}{rgb}{0.545,0.227,0.384}
\definecolor{indianred}{rgb}{0.804,0.361,0.361}
\definecolor{indianred1}{rgb}{1,0.416,0.416}
\definecolor{indianred2}{rgb}{0.933,0.388,0.388}
\definecolor{indianred3}{rgb}{0.804,0.333,0.333}
\definecolor{indianred4}{rgb}{0.545,0.227,0.227}
\definecolor{ivory}{rgb}{1,1,0.941}
\definecolor{ivory1}{rgb}{1,1,0.941}
\definecolor{ivory2}{rgb}{0.933,0.933,0.878}
\definecolor{ivory3}{rgb}{0.804,0.804,0.757}
\definecolor{ivory4}{rgb}{0.545,0.545,0.514}
\definecolor{khaki}{rgb}{0.941,0.902,0.549}
\definecolor{khaki1}{rgb}{1,0.965,0.561}
\definecolor{khaki2}{rgb}{0.933,0.902,0.522}
\definecolor{khaki3}{rgb}{0.804,0.776,0.451}
\definecolor{khaki4}{rgb}{0.545,0.525,0.306}
\definecolor{lavender}{rgb}{0.902,0.902,0.98}
\definecolor{lavenderblush}{rgb}{1,0.941,0.961}
\definecolor{lavenderblush1}{rgb}{1,0.941,0.961}
\definecolor{lavenderblush2}{rgb}{0.933,0.878,0.898}
\definecolor{lavenderblush3}{rgb}{0.804,0.757,0.773}
\definecolor{lavenderblush4}{rgb}{0.545,0.514,0.525}
\definecolor{lawngreen}{rgb}{0.486,0.988,0}
\definecolor{lemonchiffon}{rgb}{1,0.98,0.804}
\definecolor{lemonchiffon1}{rgb}{1,0.98,0.804}
\definecolor{lemonchiffon2}{rgb}{0.933,0.914,0.749}
\definecolor{lemonchiffon3}{rgb}{0.804,0.788,0.647}
\definecolor{lemonchiffon4}{rgb}{0.545,0.537,0.439}
\definecolor{lightblue1}{rgb}{0.749,0.937,1}
\definecolor{lightblue2}{rgb}{0.698,0.875,0.933}
\definecolor{lightblue3}{rgb}{0.604,0.753,0.804}
\definecolor{lightblue4}{rgb}{0.408,0.514,0.545}
\definecolor{lightcoral}{rgb}{0.941,0.502,0.502}
\definecolor{lightcyan1}{rgb}{0.878,1,1}
\definecolor{lightcyan2}{rgb}{0.82,0.933,0.933}
\definecolor{lightcyan3}{rgb}{0.706,0.804,0.804}
\definecolor{lightcyan4}{rgb}{0.478,0.545,0.545}
\definecolor{lightgoldenrod}{rgb}{0.933,0.867,0.51}
\definecolor{lightgoldenrod1}{rgb}{1,0.925,0.545}
\definecolor{lightgoldenrod2}{rgb}{0.933,0.863,0.51}
\definecolor{lightgoldenrod3}{rgb}{0.804,0.745,0.439}
\definecolor{lightgoldenrod4}{rgb}{0.545,0.506,0.298}
\definecolor{lightgoldenrodyellow}{rgb}{0.98,0.98,0.824}
\definecolor{lightgrey}{rgb}{0.827,0.827,0.827}
\definecolor{lightpink}{rgb}{1,0.714,0.757}
\definecolor{lightpink1}{rgb}{1,0.682,0.725}
\definecolor{lightpink2}{rgb}{0.933,0.635,0.678}
\definecolor{lightpink3}{rgb}{0.804,0.549,0.584}
\definecolor{lightpink4}{rgb}{0.545,0.373,0.396}
\definecolor{lightsalmon}{rgb}{1,0.627,0.478}
\definecolor{lightsalmon1}{rgb}{1,0.627,0.478}
\definecolor{lightsalmon2}{rgb}{0.933,0.584,0.447}
\definecolor{lightsalmon3}{rgb}{0.804,0.506,0.384}
\definecolor{lightsalmon4}{rgb}{0.545,0.341,0.259}
\definecolor{lightseagreen}{rgb}{0.125,0.698,0.667}
\definecolor{lightskyblue}{rgb}{0.529,0.808,0.98}
\definecolor{lightskyblue1}{rgb}{0.69,0.886,1}
\definecolor{lightskyblue2}{rgb}{0.643,0.827,0.933}
\definecolor{lightskyblue3}{rgb}{0.553,0.714,0.804}
\definecolor{lightskyblue4}{rgb}{0.376,0.482,0.545}
\definecolor{lightslateblue}{rgb}{0.518,0.439,1}
\definecolor{lightslategray}{rgb}{0.467,0.533,0.6}
\definecolor{lightslategrey}{rgb}{0.467,0.533,0.6}
\definecolor{lightsteelblue}{rgb}{0.69,0.769,0.871}
\definecolor{lightsteelblue1}{rgb}{0.792,0.882,1}
\definecolor{lightsteelblue2}{rgb}{0.737,0.824,0.933}
\definecolor{lightsteelblue3}{rgb}{0.635,0.71,0.804}
\definecolor{lightsteelblue4}{rgb}{0.431,0.482,0.545}
\definecolor{lightyellow1}{rgb}{1,1,0.878}
\definecolor{lightyellow2}{rgb}{0.933,0.933,0.82}
\definecolor{lightyellow3}{rgb}{0.804,0.804,0.706}
\definecolor{lightyellow4}{rgb}{0.545,0.545,0.478}
\definecolor{limegreen}{rgb}{0.196,0.804,0.196}
\definecolor{linen}{rgb}{0.98,0.941,0.902}
\definecolor{magenta}{rgb}{1,0,1}
\definecolor{magenta1}{rgb}{1,0,1}
\definecolor{magenta2}{rgb}{0.933,0,0.933}
\definecolor{magenta3}{rgb}{0.804,0,0.804}
\definecolor{magenta4}{rgb}{0.545,0,0.545}
\definecolor{maroon}{rgb}{0.69,0.188,0.376}
\definecolor{maroon1}{rgb}{1,0.204,0.702}
\definecolor{maroon2}{rgb}{0.933,0.188,0.655}
\definecolor{maroon3}{rgb}{0.804,0.161,0.565}
\definecolor{maroon4}{rgb}{0.545,0.11,0.384}
\definecolor{mediumaquamarine}{rgb}{0.4,0.804,0.667}
\definecolor{mediumblue}{rgb}{0,0,0.804}
\definecolor{mediumorchid}{rgb}{0.729,0.333,0.827}
\definecolor{mediumorchid1}{rgb}{0.878,0.4,1}
\definecolor{mediumorchid2}{rgb}{0.82,0.373,0.933}
\definecolor{mediumorchid3}{rgb}{0.706,0.322,0.804}
\definecolor{mediumorchid4}{rgb}{0.478,0.216,0.545}
\definecolor{mediumpurple}{rgb}{0.576,0.439,0.859}
\definecolor{mediumpurple1}{rgb}{0.671,0.51,1}
\definecolor{mediumpurple2}{rgb}{0.624,0.475,0.933}
\definecolor{mediumpurple3}{rgb}{0.537,0.408,0.804}
\definecolor{mediumpurple4}{rgb}{0.365,0.278,0.545}
\definecolor{mediumseagreen}{rgb}{0.235,0.702,0.443}
\definecolor{mediumslateblue}{rgb}{0.482,0.408,0.933}
\definecolor{mediumspringgreen}{rgb}{0,0.98,0.604}
\definecolor{mediumturquoise}{rgb}{0.282,0.82,0.8}
\definecolor{mediumvioletred}{rgb}{0.78,0.082,0.522}
\definecolor{midnightblue}{rgb}{0.098,0.098,0.439}
\definecolor{mintcream}{rgb}{0.961,1,0.98}
\definecolor{mistyrose}{rgb}{1,0.894,0.882}
\definecolor{mistyrose1}{rgb}{1,0.894,0.882}
\definecolor{mistyrose2}{rgb}{0.933,0.835,0.824}
\definecolor{mistyrose3}{rgb}{0.804,0.718,0.71}
\definecolor{mistyrose4}{rgb}{0.545,0.49,0.482}
\definecolor{moccasin}{rgb}{1,0.894,0.71}
\definecolor{navajowhite}{rgb}{1,0.871,0.678}
\definecolor{navajowhite1}{rgb}{1,0.871,0.678}
\definecolor{navajowhite2}{rgb}{0.933,0.812,0.631}
\definecolor{navajowhite3}{rgb}{0.804,0.702,0.545}
\definecolor{navajowhite4}{rgb}{0.545,0.475,0.369}
\definecolor{navyblue}{rgb}{0,0,0.502}
\definecolor{oldlace}{rgb}{0.992,0.961,0.902}
\definecolor{olivedrab}{rgb}{0.42,0.557,0.137}
\definecolor{olivedrab1}{rgb}{0.753,1,0.243}
\definecolor{olivedrab2}{rgb}{0.702,0.933,0.227}
\definecolor{olivedrab3}{rgb}{0.604,0.804,0.196}
\definecolor{olivedrab4}{rgb}{0.412,0.545,0.133}
\definecolor{orange1}{rgb}{1,0.647,0}
\definecolor{orange2}{rgb}{0.933,0.604,0}
\definecolor{orange3}{rgb}{0.804,0.522,0}
\definecolor{orange4}{rgb}{0.545,0.353,0}
\definecolor{orangered}{rgb}{1,0.271,0}
\definecolor{orangered1}{rgb}{1,0.271,0}
\definecolor{orangered2}{rgb}{0.933,0.251,0}
\definecolor{orangered3}{rgb}{0.804,0.216,0}
\definecolor{orangered4}{rgb}{0.545,0.145,0}
\definecolor{orchid}{rgb}{0.855,0.439,0.839}
\definecolor{orchid1}{rgb}{1,0.514,0.98}
\definecolor{orchid2}{rgb}{0.933,0.478,0.914}
\definecolor{orchid3}{rgb}{0.804,0.412,0.788}
\definecolor{orchid4}{rgb}{0.545,0.278,0.537}
\definecolor{palegoldenrod}{rgb}{0.933,0.91,0.667}
\definecolor{palegreen}{rgb}{0.596,0.984,0.596}
\definecolor{palegreen1}{rgb}{0.604,1,0.604}
\definecolor{palegreen2}{rgb}{0.565,0.933,0.565}
\definecolor{palegreen3}{rgb}{0.486,0.804,0.486}
\definecolor{palegreen4}{rgb}{0.329,0.545,0.329}
\definecolor{paleturquoise}{rgb}{0.686,0.933,0.933}
\definecolor{paleturquoise1}{rgb}{0.733,1,1}
\definecolor{paleturquoise2}{rgb}{0.682,0.933,0.933}
\definecolor{paleturquoise3}{rgb}{0.588,0.804,0.804}
\definecolor{paleturquoise4}{rgb}{0.4,0.545,0.545}
\definecolor{palevioletred}{rgb}{0.859,0.439,0.576}
\definecolor{palevioletred1}{rgb}{1,0.51,0.671}
\definecolor{palevioletred2}{rgb}{0.933,0.475,0.624}
\definecolor{palevioletred3}{rgb}{0.804,0.408,0.537}
\definecolor{palevioletred4}{rgb}{0.545,0.278,0.365}
\definecolor{papayawhip}{rgb}{1,0.937,0.835}
\definecolor{peachpuff}{rgb}{1,0.855,0.725}
\definecolor{peachpuff1}{rgb}{1,0.855,0.725}
\definecolor{peachpuff2}{rgb}{0.933,0.796,0.678}
\definecolor{peachpuff3}{rgb}{0.804,0.686,0.584}
\definecolor{peachpuff4}{rgb}{0.545,0.467,0.396}
\definecolor{peru}{rgb}{0.804,0.522,0.247}
\definecolor{pink1}{rgb}{1,0.71,0.773}
\definecolor{pink2}{rgb}{0.933,0.663,0.722}
\definecolor{pink3}{rgb}{0.804,0.569,0.62}
\definecolor{pink4}{rgb}{0.545,0.388,0.424}
\definecolor{plum}{rgb}{0.867,0.627,0.867}
\definecolor{plum1}{rgb}{1,0.733,1}
\definecolor{plum2}{rgb}{0.933,0.682,0.933}
\definecolor{plum3}{rgb}{0.804,0.588,0.804}
\definecolor{plum4}{rgb}{0.545,0.4,0.545}
\definecolor{powderblue}{rgb}{0.69,0.878,0.902}
\definecolor{purple1}{rgb}{0.608,0.188,1}
\definecolor{purple2}{rgb}{0.569,0.173,0.933}
\definecolor{purple3}{rgb}{0.49,0.149,0.804}
\definecolor{purple4}{rgb}{0.333,0.102,0.545}
\definecolor{red1}{rgb}{1,0,0}
\definecolor{red2}{rgb}{0.933,0,0}
\definecolor{red3}{rgb}{0.804,0,0}
\definecolor{red4}{rgb}{0.545,0,0}
\definecolor{rosybrown}{rgb}{0.737,0.561,0.561}
\definecolor{rosybrown1}{rgb}{1,0.757,0.757}
\definecolor{rosybrown2}{rgb}{0.933,0.706,0.706}
\definecolor{rosybrown3}{rgb}{0.804,0.608,0.608}
\definecolor{rosybrown4}{rgb}{0.545,0.412,0.412}
\definecolor{royalblue}{rgb}{0.255,0.412,0.882}
\definecolor{royalblue1}{rgb}{0.282,0.463,1}
\definecolor{royalblue2}{rgb}{0.263,0.431,0.933}
\definecolor{royalblue3}{rgb}{0.227,0.373,0.804}
\definecolor{royalblue4}{rgb}{0.153,0.251,0.545}
\definecolor{saddlebrown}{rgb}{0.545,0.271,0.075}
\definecolor{salmon}{rgb}{0.98,0.502,0.447}
\definecolor{salmon1}{rgb}{1,0.549,0.412}
\definecolor{salmon2}{rgb}{0.933,0.51,0.384}
\definecolor{salmon3}{rgb}{0.804,0.439,0.329}
\definecolor{salmon4}{rgb}{0.545,0.298,0.224}
\definecolor{sandybrown}{rgb}{0.957,0.643,0.376}
\definecolor{seagreen1}{rgb}{0.329,1,0.624}
\definecolor{seagreen2}{rgb}{0.306,0.933,0.58}
\definecolor{seagreen3}{rgb}{0.263,0.804,0.502}
\definecolor{seagreen4}{rgb}{0.18,0.545,0.341}
\definecolor{seashell}{rgb}{1,0.961,0.933}
\definecolor{seashell1}{rgb}{1,0.961,0.933}
\definecolor{seashell2}{rgb}{0.933,0.898,0.871}
\definecolor{seashell3}{rgb}{0.804,0.773,0.749}
\definecolor{seashell4}{rgb}{0.545,0.525,0.51}
\definecolor{sienna}{rgb}{0.627,0.322,0.176}
\definecolor{sienna1}{rgb}{1,0.51,0.278}
\definecolor{sienna2}{rgb}{0.933,0.475,0.259}
\definecolor{sienna3}{rgb}{0.804,0.408,0.224}
\definecolor{sienna4}{rgb}{0.545,0.278,0.149}
\definecolor{skyblue}{rgb}{0.529,0.808,0.922}
\definecolor{skyblue1}{rgb}{0.529,0.808,1}
\definecolor{skyblue2}{rgb}{0.494,0.753,0.933}
\definecolor{skyblue3}{rgb}{0.424,0.651,0.804}
\definecolor{skyblue4}{rgb}{0.29,0.439,0.545}
\definecolor{slateblue}{rgb}{0.416,0.353,0.804}
\definecolor{slateblue1}{rgb}{0.514,0.435,1}
\definecolor{slateblue2}{rgb}{0.478,0.404,0.933}
\definecolor{slateblue3}{rgb}{0.412,0.349,0.804}
\definecolor{slateblue4}{rgb}{0.278,0.235,0.545}
\definecolor{slategray}{rgb}{0.439,0.502,0.565}
\definecolor{slategray1}{rgb}{0.776,0.886,1}
\definecolor{slategray2}{rgb}{0.725,0.827,0.933}
\definecolor{slategray3}{rgb}{0.624,0.714,0.804}
\definecolor{slategray4}{rgb}{0.424,0.482,0.545}
\definecolor{slategrey}{rgb}{0.439,0.502,0.565}
\definecolor{snow}{rgb}{1,0.98,0.98}
\definecolor{snow1}{rgb}{1,0.98,0.98}
\definecolor{snow2}{rgb}{0.933,0.914,0.914}
\definecolor{snow3}{rgb}{0.804,0.788,0.788}
\definecolor{snow4}{rgb}{0.545,0.537,0.537}
\definecolor{springgreen}{rgb}{0,1,0.498}
\definecolor{springgreen1}{rgb}{0,1,0.498}
\definecolor{springgreen2}{rgb}{0,0.933,0.463}
\definecolor{springgreen3}{rgb}{0,0.804,0.4}
\definecolor{springgreen4}{rgb}{0,0.545,0.271}
\definecolor{steelblue}{rgb}{0.275,0.51,0.706}
\definecolor{steelblue1}{rgb}{0.388,0.722,1}
\definecolor{steelblue2}{rgb}{0.361,0.675,0.933}
\definecolor{steelblue3}{rgb}{0.31,0.58,0.804}
\definecolor{steelblue4}{rgb}{0.212,0.392,0.545}
\definecolor{tan}{rgb}{0.824,0.706,0.549}
\definecolor{tan1}{rgb}{1,0.647,0.31}
\definecolor{tan2}{rgb}{0.933,0.604,0.286}
\definecolor{tan3}{rgb}{0.804,0.522,0.247}
\definecolor{tan4}{rgb}{0.545,0.353,0.169}
\definecolor{thistle}{rgb}{0.847,0.749,0.847}
\definecolor{thistle1}{rgb}{1,0.882,1}
\definecolor{thistle2}{rgb}{0.933,0.824,0.933}
\definecolor{thistle3}{rgb}{0.804,0.71,0.804}
\definecolor{thistle4}{rgb}{0.545,0.482,0.545}
\definecolor{tomato}{rgb}{1,0.388,0.278}
\definecolor{tomato1}{rgb}{1,0.388,0.278}
\definecolor{tomato2}{rgb}{0.933,0.361,0.259}
\definecolor{tomato3}{rgb}{0.804,0.31,0.224}
\definecolor{tomato4}{rgb}{0.545,0.212,0.149}
\definecolor{turquoise1}{rgb}{0,0.961,1}
\definecolor{turquoise2}{rgb}{0,0.898,0.933}
\definecolor{turquoise3}{rgb}{0,0.773,0.804}
\definecolor{turquoise4}{rgb}{0,0.525,0.545}
\definecolor{violetred}{rgb}{0.816,0.125,0.565}
\definecolor{violetred1}{rgb}{1,0.243,0.588}
\definecolor{violetred2}{rgb}{0.933,0.227,0.549}
\definecolor{violetred3}{rgb}{0.804,0.196,0.471}
\definecolor{violetred4}{rgb}{0.545,0.133,0.322}
\definecolor{wheat}{rgb}{0.961,0.871,0.702}
\definecolor{wheat1}{rgb}{1,0.906,0.729}
\definecolor{wheat2}{rgb}{0.933,0.847,0.682}
\definecolor{wheat3}{rgb}{0.804,0.729,0.588}
\definecolor{wheat4}{rgb}{0.545,0.494,0.4}
\definecolor{whitesmoke}{rgb}{0.961,0.961,0.961}
\definecolor{yellow1}{rgb}{1,1,0}
\definecolor{yellow2}{rgb}{0.933,0.933,0}
\definecolor{yellow3}{rgb}{0.804,0.804,0}
\definecolor{yellow4}{rgb}{0.545,0.545,0}
\definecolor{yellowgreen}{rgb}{0.604,0.804,0.196}

\caption{The Gaifman graph of $\mathfrak{A}$ from~\autoref{@governorship} and~\autoref{@accomplishes}, ``separated'' into two parts. Left: The graph $G_{\mathfrak{A}^\star}$ with the set $V({\bf b}_1^\star)$ as boundary. Right: The graph $G_{\mathfrak{A}_{\sf out}^\star}$ with the set $V({\bf b}_1^\star)$ as boundary.}
\labels{figure_boundariedgraph1}
\end{figure}

\myskip\paragraph{An ordering on the (common) boundary of the two structures.}
We next claim that $\partial_{\mathfrak{K}_1} (Z)\cup V(F^\star) = N_G (\breve{C}),$ which directly implies that $|\partial_{\mathfrak{K}_1} (Z)\cup V(F^\star)| = h.$
To see why $\partial_{\mathfrak{K}_1} (Z)\cup V(F^\star) = N_G (\breve{C}),$ first observe that, since $\breve{C}∈{\sf cc}(G,X),$ it holds that $N_G(\breve{C}) \subseteq X$ and also notice that $X_{\rm in} \cap N_G (\breve{C}) = \partial_{\mathfrak{K}_1} (Z).$
Since $V_L ({\bf a}) =V({\bf a})\cap N_G (\breve{C})$ and
$V(F') = (X_{\rm out}\setminus V_L ({\bf a}) )\cap N_G (\breve{C}),$ we have that $N_G (\breve{C}) = \partial_{\mathfrak{K}_1} (Z)\cup V_L ({\bf a}) \cup V(F') = \partial_{\mathfrak{K}_1} (Z)\cup V(F^\star).$
Therefore, we can consider an ordering $v_1, \ldots, v_h$ of the vertices in $\partial_{\mathfrak{K}_1} (Z)\cup V(F^\star).$
Let ${\bf b}^\star_1 = (v_1, \ldots, v_h)$ and recall that $V(\mathfrak{A}_{\rm out}^\star) \cap V(\mathfrak{A}^\star) = \partial_{\mathfrak{K}_1} (Z)\cup V(F^\star).$
Now, consider the $h$-boundaried $τ$-structures $(\mathfrak{A}_{\rm out}^\star, {\bf b}_1^\star)$ and $(\mathfrak{A}^\star, {\bf b}_1^\star).$
Notice that $(\mathfrak{A}_{\rm out}^\star, {\bf b}_1^\star)$ and $(\mathfrak{A}^\star, {\bf b}_1^\star)$ are compatible
and that $(\mathfrak{A}_{\rm out}^\star, {\bf b}_1^\star)\oplus (\mathfrak{A}^\star, {\bf b}_1^\star)= \mathfrak{A}.$
To give a better understanding of the proof, we stress that our objective is
to find another ``solution'' $X'$ that is ``away'' from $Y,$ the vertex set of the influence of the central $j'$-subwall of $\tilde{W}_1.$ For this reason, we aim to use the equivalence of ${\sf out}\text{-}{\sf sig}$ to find another way to ``look'' at $\mathfrak{A}^\star,$ by ``changing'' its boundary and finding this other $X'.$

\myskip\paragraph{Adding $V(F^\star)$ to  $X_{\rm in}.$}
Let $\tilde{X}^\star_1 = X_{\rm in} \cup V(F^\star).$
Now,
the fact that
$\partial_{\mathfrak{K}_1} (Z) \subseteq X_{\rm in}$ implies that $V({\bf b}_1^\star)\subseteq \tilde{X}^\star_1.$
Also,
we set $R_1^{' \star}:=R\cap (\breve{C}\cup Z).$
Therefore, it holds that $(R\setminus (\breve{C}\cup Z)) \cup  R_1^{' \star} = R$ and, since $V(F^\star)\subseteq X_{\rm out},$  $X_{\rm out}\cup \tilde{X}^\star_1 = X.$

\myskip\paragraph{Separating $(\mathfrak{A},R,{\bf W}_q^{(1)}, \varnothing^l,X)$ into two boundaried structures.}
We consider the structure $(\mathfrak{A},R,{\bf W}_q^{(1)}, \varnothing^l, X).$
We choose to include the $q$-pseudogrid ${\bf W}_q^{(1)}$ in the aforementioned structure (and not another $q$-pseudogrid,
even if, due to~\autoref{corrollaa_pseudo}, they would yield equivalent instances)
in
order to be able to ``break'' $(\mathfrak{A},R,{\bf W}_q^{(1)}, \varnothing^l, X)$ into two $h$-boundaried structures corresponding to $\mathfrak{A}_{\rm out}$ and $\mathfrak{A}^\star,$ and $\mathfrak{A}^\star$ to contain $\cupall {\bf W}_q^{(1)},$ since $\cupall {\bf W}_q^{(1)}\subseteq \breve{C}\cup Z\subseteq V(\mathfrak{A}^\star).$
Thus, we have that
\begin{eqnarray}
(\mathfrak{A},R,{\bf W}_q^{(1)},\varnothing^l, X) = (\mathfrak{A}_{\rm out}^\star, R\setminus (\breve{C}\cup Z),\emptyset^{2q},\varnothing^l, X_{\rm out}, {\bf b}_1^\star ) \oplus (\mathfrak{A}^\star, R_1^{' \star},{\bf W}_{{q}}^{(1)},\varnothing^l,\tilde{X}^\star_1, {\bf b}_1^\star).\labels{@electronically}
\end{eqnarray}
Also,~\autoref{cou_more} implies that there is a $\bar{φ}∈ {\sf rep}_{τ'}^{(h)} ( θ^{\sf out}_q),$ where $τ' = τ \cup{\bf Q}\cup\{{\sf R},{\sf X}\}\cup{\bf c},$ such that
 $\big(\mathfrak{A}^\star,R_1^{'\star},{\bf W}_{{q}}^{(1)},\varnothing^l,\tilde{X}^\star_1, {\bf b}_1^\star\big)\models \bar{φ}.$

\myskip\paragraph{Shifting from $\mathfrak{A}^\star$ to $\mathfrak{A}^{(d,Z,L,F_1)}.$}
Now, consider a graph $F_1∈ {\cal F}_{h-|\partial_{\mathfrak{K}_1} (Z)|}^{V_L({\bf a})}$ that is isomorphic\footnote{In the rest of the proof of the claim, we will usually consider a subgraph of $G,$ or a structure with universe $V(G),$ and isomorphic graphs/structures of them, and the latter will be ``abstract'' graphs/structures. For example, here we consider an ``abstract'' graph $F_1$ that is isomorphic to the graph $F^\star$ that is a subgraph of $G.$ We will always use superscript ``$^\star$'' in order to denote the subgraphs/structures that are being given by the graph, while the lack of superscript reflects to the corresponding isomorphic ``abstract'' graphs/structures.} to $F^\star,$ via a bijection $ξ: V(F_1) \leftrightarrow V(F^\star)$ that maps every $a∈ V_L({\bf a})$ to itself.
Let $F_1':=F_1\setminus V_L({\bf a}).$
We set ${\cal V}_1 := (\partial_{\mathfrak{K}_1} (Z), V_L ({\bf a}), V(F_1'))$ and observe that ${\cal V}_1$ is a nice 3-partition of $K_1^{\bf a}[\partial_{\mathfrak{K}_1} (Z) \cup V_L ({\bf a})]\cup F_1.$
Also, observe that  the graph $V(K_1^{\bf a}[\partial_{\mathfrak{K}_1} (Z) \cup V_L ({\bf a})]\cup F_1)$ has $h$ vertices and therefore $(K_1^{\bf a}[\partial_{\mathfrak{K}_1} (Z) \cup V_L ({\bf a})]\cup F_1, {\cal V}_1)∈ {\cal H}^{(h)}.$
Let ${\bf H} := (K_1^{\bf a}[\partial_{\mathfrak{K}_1} (Z) \cup V_L ({\bf a})]\cup F_1,{\cal V}_1).$

\begin{figure}[ht]
\centering
\tikzstyle{ipe stylesheet} = [
  ipe import,
  even odd rule,
  line join=round,
  line cap=butt,
  ipe pen normal/.style={line width=0.4},
  ipe pen heavier/.style={line width=0.8},
  ipe pen fat/.style={line width=1.2},
  ipe pen ultrafat/.style={line width=2},
  ipe pen normal,
  ipe mark normal/.style={ipe mark scale=3},
  ipe mark large/.style={ipe mark scale=5},
  ipe mark small/.style={ipe mark scale=2},
  ipe mark tiny/.style={ipe mark scale=1.1},
  ipe mark normal,
  /pgf/arrow keys/.cd,
  ipe arrow normal/.style={scale=7},
  ipe arrow large/.style={scale=10},
  ipe arrow small/.style={scale=5},
  ipe arrow tiny/.style={scale=3},
  ipe arrow normal,
  /tikz/.cd,
  ipe arrows, 
  <->/.tip = ipe normal,
  ipe dash normal/.style={dash pattern=},
  ipe dash dotted/.style={dash pattern=on 1bp off 3bp},
  ipe dash dashed/.style={dash pattern=on 4bp off 4bp},
  ipe dash dash dotted/.style={dash pattern=on 4bp off 2bp on 1bp off 2bp},
  ipe dash dash dot dotted/.style={dash pattern=on 4bp off 2bp on 1bp off 2bp on 1bp off 2bp},
  ipe dash normal,
  ipe node/.append style={font=\normalsize},
  ipe stretch normal/.style={ipe node stretch=1},
  ipe stretch normal,
  ipe opacity 10/.style={opacity=0.1},
  ipe opacity 30/.style={opacity=0.3},
  ipe opacity 50/.style={opacity=0.5},
  ipe opacity 75/.style={opacity=0.75},
  ipe opacity opaque/.style={opacity=1},
  ipe opacity opaque,
]
\definecolor{black}{rgb}{0,0,0}
\definecolor{white}{rgb}{1,1,1}
\definecolor{red}{rgb}{1,0,0}
\definecolor{blue}{rgb}{0,0,1}
\definecolor{green}{rgb}{0,1,0}
\definecolor{yellow}{rgb}{1,1,0}
\definecolor{orange}{rgb}{1,0.647,0}
\definecolor{gold}{rgb}{1,0.843,0}
\definecolor{purple}{rgb}{0.627,0.125,0.941}
\definecolor{gray}{rgb}{0.745,0.745,0.745}
\definecolor{brown}{rgb}{0.647,0.165,0.165}
\definecolor{navy}{rgb}{0,0,0.502}
\definecolor{pink}{rgb}{1,0.753,0.796}
\definecolor{seagreen}{rgb}{0.18,0.545,0.341}
\definecolor{turquoise}{rgb}{0.251,0.878,0.816}
\definecolor{violet}{rgb}{0.933,0.51,0.933}
\definecolor{darkblue}{rgb}{0,0,0.545}
\definecolor{darkcyan}{rgb}{0,0.545,0.545}
\definecolor{darkgray}{rgb}{0.663,0.663,0.663}
\definecolor{darkgreen}{rgb}{0,0.392,0}
\definecolor{darkmagenta}{rgb}{0.545,0,0.545}
\definecolor{darkorange}{rgb}{1,0.549,0}
\definecolor{darkred}{rgb}{0.545,0,0}
\definecolor{lightblue}{rgb}{0.678,0.847,0.902}
\definecolor{lightcyan}{rgb}{0.878,1,1}
\definecolor{lightgray}{rgb}{0.827,0.827,0.827}
\definecolor{lightgreen}{rgb}{0.565,0.933,0.565}
\definecolor{lightyellow}{rgb}{1,1,0.878}
\definecolor{aliceblue}{rgb}{0.941,0.973,1}
\definecolor{antiquewhite}{rgb}{0.98,0.922,0.843}
\definecolor{antiquewhite1}{rgb}{1,0.937,0.859}
\definecolor{antiquewhite2}{rgb}{0.933,0.875,0.8}
\definecolor{antiquewhite3}{rgb}{0.804,0.753,0.69}
\definecolor{antiquewhite4}{rgb}{0.545,0.514,0.471}
\definecolor{aquamarine}{rgb}{0.498,1,0.831}
\definecolor{aquamarine1}{rgb}{0.498,1,0.831}
\definecolor{aquamarine2}{rgb}{0.463,0.933,0.776}
\definecolor{aquamarine3}{rgb}{0.4,0.804,0.667}
\definecolor{aquamarine4}{rgb}{0.271,0.545,0.455}
\definecolor{azure}{rgb}{0.941,1,1}
\definecolor{azure1}{rgb}{0.941,1,1}
\definecolor{azure2}{rgb}{0.878,0.933,0.933}
\definecolor{azure3}{rgb}{0.757,0.804,0.804}
\definecolor{azure4}{rgb}{0.514,0.545,0.545}
\definecolor{beige}{rgb}{0.961,0.961,0.863}
\definecolor{bisque}{rgb}{1,0.894,0.769}
\definecolor{bisque1}{rgb}{1,0.894,0.769}
\definecolor{bisque2}{rgb}{0.933,0.835,0.718}
\definecolor{bisque3}{rgb}{0.804,0.718,0.62}
\definecolor{bisque4}{rgb}{0.545,0.49,0.42}
\definecolor{blanchedalmond}{rgb}{1,0.922,0.804}
\definecolor{blue1}{rgb}{0,0,1}
\definecolor{blue2}{rgb}{0,0,0.933}
\definecolor{blue3}{rgb}{0,0,0.804}
\definecolor{blue4}{rgb}{0,0,0.545}
\definecolor{blueviolet}{rgb}{0.541,0.169,0.886}
\definecolor{brown1}{rgb}{1,0.251,0.251}
\definecolor{brown2}{rgb}{0.933,0.231,0.231}
\definecolor{brown3}{rgb}{0.804,0.2,0.2}
\definecolor{brown4}{rgb}{0.545,0.137,0.137}
\definecolor{burlywood}{rgb}{0.871,0.722,0.529}
\definecolor{burlywood1}{rgb}{1,0.827,0.608}
\definecolor{burlywood2}{rgb}{0.933,0.773,0.569}
\definecolor{burlywood3}{rgb}{0.804,0.667,0.49}
\definecolor{burlywood4}{rgb}{0.545,0.451,0.333}
\definecolor{cadetblue}{rgb}{0.373,0.62,0.627}
\definecolor{cadetblue1}{rgb}{0.596,0.961,1}
\definecolor{cadetblue2}{rgb}{0.557,0.898,0.933}
\definecolor{cadetblue3}{rgb}{0.478,0.773,0.804}
\definecolor{cadetblue4}{rgb}{0.325,0.525,0.545}
\definecolor{chartreuse}{rgb}{0.498,1,0}
\definecolor{chartreuse1}{rgb}{0.498,1,0}
\definecolor{chartreuse2}{rgb}{0.463,0.933,0}
\definecolor{chartreuse3}{rgb}{0.4,0.804,0}
\definecolor{chartreuse4}{rgb}{0.271,0.545,0}
\definecolor{chocolate}{rgb}{0.824,0.412,0.118}
\definecolor{chocolate1}{rgb}{1,0.498,0.141}
\definecolor{chocolate2}{rgb}{0.933,0.463,0.129}
\definecolor{chocolate3}{rgb}{0.804,0.4,0.114}
\definecolor{chocolate4}{rgb}{0.545,0.271,0.075}
\definecolor{coral}{rgb}{1,0.498,0.314}
\definecolor{coral1}{rgb}{1,0.447,0.337}
\definecolor{coral2}{rgb}{0.933,0.416,0.314}
\definecolor{coral3}{rgb}{0.804,0.357,0.271}
\definecolor{coral4}{rgb}{0.545,0.243,0.184}
\definecolor{cornflowerblue}{rgb}{0.392,0.584,0.929}
\definecolor{cornsilk}{rgb}{1,0.973,0.863}
\definecolor{cornsilk1}{rgb}{1,0.973,0.863}
\definecolor{cornsilk2}{rgb}{0.933,0.91,0.804}
\definecolor{cornsilk3}{rgb}{0.804,0.784,0.694}
\definecolor{cornsilk4}{rgb}{0.545,0.533,0.471}
\definecolor{cyan}{rgb}{0,1,1}
\definecolor{cyan1}{rgb}{0,1,1}
\definecolor{cyan2}{rgb}{0,0.933,0.933}
\definecolor{cyan3}{rgb}{0,0.804,0.804}
\definecolor{cyan4}{rgb}{0,0.545,0.545}
\definecolor{darkgoldenrod}{rgb}{0.722,0.525,0.043}
\definecolor{darkgoldenrod1}{rgb}{1,0.725,0.059}
\definecolor{darkgoldenrod2}{rgb}{0.933,0.678,0.055}
\definecolor{darkgoldenrod3}{rgb}{0.804,0.584,0.047}
\definecolor{darkgoldenrod4}{rgb}{0.545,0.396,0.031}
\definecolor{darkgrey}{rgb}{0.663,0.663,0.663}
\definecolor{darkkhaki}{rgb}{0.741,0.718,0.42}
\definecolor{darkolivegreen}{rgb}{0.333,0.42,0.184}
\definecolor{darkolivegreen1}{rgb}{0.792,1,0.439}
\definecolor{darkolivegreen2}{rgb}{0.737,0.933,0.408}
\definecolor{darkolivegreen3}{rgb}{0.635,0.804,0.353}
\definecolor{darkolivegreen4}{rgb}{0.431,0.545,0.239}
\definecolor{darkorange1}{rgb}{1,0.498,0}
\definecolor{darkorange2}{rgb}{0.933,0.463,0}
\definecolor{darkorange3}{rgb}{0.804,0.4,0}
\definecolor{darkorange4}{rgb}{0.545,0.271,0}
\definecolor{darkorchid}{rgb}{0.6,0.196,0.8}
\definecolor{darkorchid1}{rgb}{0.749,0.243,1}
\definecolor{darkorchid2}{rgb}{0.698,0.227,0.933}
\definecolor{darkorchid3}{rgb}{0.604,0.196,0.804}
\definecolor{darkorchid4}{rgb}{0.408,0.133,0.545}
\definecolor{darksalmon}{rgb}{0.914,0.588,0.478}
\definecolor{darkseagreen}{rgb}{0.561,0.737,0.561}
\definecolor{darkseagreen1}{rgb}{0.757,1,0.757}
\definecolor{darkseagreen2}{rgb}{0.706,0.933,0.706}
\definecolor{darkseagreen3}{rgb}{0.608,0.804,0.608}
\definecolor{darkseagreen4}{rgb}{0.412,0.545,0.412}
\definecolor{darkslateblue}{rgb}{0.282,0.239,0.545}
\definecolor{darkslategray}{rgb}{0.184,0.31,0.31}
\definecolor{darkslategray1}{rgb}{0.592,1,1}
\definecolor{darkslategray2}{rgb}{0.553,0.933,0.933}
\definecolor{darkslategray3}{rgb}{0.475,0.804,0.804}
\definecolor{darkslategray4}{rgb}{0.322,0.545,0.545}
\definecolor{darkslategrey}{rgb}{0.184,0.31,0.31}
\definecolor{darkturquoise}{rgb}{0,0.808,0.82}
\definecolor{darkviolet}{rgb}{0.58,0,0.827}
\definecolor{deeppink}{rgb}{1,0.078,0.576}
\definecolor{deeppink1}{rgb}{1,0.078,0.576}
\definecolor{deeppink2}{rgb}{0.933,0.071,0.537}
\definecolor{deeppink3}{rgb}{0.804,0.063,0.463}
\definecolor{deeppink4}{rgb}{0.545,0.039,0.314}
\definecolor{deepskyblue}{rgb}{0,0.749,1}
\definecolor{deepskyblue1}{rgb}{0,0.749,1}
\definecolor{deepskyblue2}{rgb}{0,0.698,0.933}
\definecolor{deepskyblue3}{rgb}{0,0.604,0.804}
\definecolor{deepskyblue4}{rgb}{0,0.408,0.545}
\definecolor{dimgray}{rgb}{0.412,0.412,0.412}
\definecolor{dimgrey}{rgb}{0.412,0.412,0.412}
\definecolor{dodgerblue}{rgb}{0.118,0.565,1}
\definecolor{dodgerblue1}{rgb}{0.118,0.565,1}
\definecolor{dodgerblue2}{rgb}{0.11,0.525,0.933}
\definecolor{dodgerblue3}{rgb}{0.094,0.455,0.804}
\definecolor{dodgerblue4}{rgb}{0.063,0.306,0.545}
\definecolor{firebrick}{rgb}{0.698,0.133,0.133}
\definecolor{firebrick1}{rgb}{1,0.188,0.188}
\definecolor{firebrick2}{rgb}{0.933,0.173,0.173}
\definecolor{firebrick3}{rgb}{0.804,0.149,0.149}
\definecolor{firebrick4}{rgb}{0.545,0.102,0.102}
\definecolor{floralwhite}{rgb}{1,0.98,0.941}
\definecolor{forestgreen}{rgb}{0.133,0.545,0.133}
\definecolor{gainsboro}{rgb}{0.863,0.863,0.863}
\definecolor{ghostwhite}{rgb}{0.973,0.973,1}
\definecolor{gold1}{rgb}{1,0.843,0}
\definecolor{gold2}{rgb}{0.933,0.788,0}
\definecolor{gold3}{rgb}{0.804,0.678,0}
\definecolor{gold4}{rgb}{0.545,0.459,0}
\definecolor{goldenrod}{rgb}{0.855,0.647,0.125}
\definecolor{goldenrod1}{rgb}{1,0.757,0.145}
\definecolor{goldenrod2}{rgb}{0.933,0.706,0.133}
\definecolor{goldenrod3}{rgb}{0.804,0.608,0.114}
\definecolor{goldenrod4}{rgb}{0.545,0.412,0.078}
\definecolor{gray0}{rgb}{0,0,0}
\definecolor{gray1}{rgb}{0.012,0.012,0.012}
\definecolor{gray10}{rgb}{0.102,0.102,0.102}
\definecolor{gray100}{rgb}{1,1,1}
\definecolor{gray11}{rgb}{0.11,0.11,0.11}
\definecolor{gray12}{rgb}{0.122,0.122,0.122}
\definecolor{gray13}{rgb}{0.129,0.129,0.129}
\definecolor{gray14}{rgb}{0.141,0.141,0.141}
\definecolor{gray15}{rgb}{0.149,0.149,0.149}
\definecolor{gray16}{rgb}{0.161,0.161,0.161}
\definecolor{gray17}{rgb}{0.169,0.169,0.169}
\definecolor{gray18}{rgb}{0.18,0.18,0.18}
\definecolor{gray19}{rgb}{0.188,0.188,0.188}
\definecolor{gray2}{rgb}{0.02,0.02,0.02}
\definecolor{gray20}{rgb}{0.2,0.2,0.2}
\definecolor{gray21}{rgb}{0.212,0.212,0.212}
\definecolor{gray22}{rgb}{0.22,0.22,0.22}
\definecolor{gray23}{rgb}{0.231,0.231,0.231}
\definecolor{gray24}{rgb}{0.239,0.239,0.239}
\definecolor{gray25}{rgb}{0.251,0.251,0.251}
\definecolor{gray26}{rgb}{0.259,0.259,0.259}
\definecolor{gray27}{rgb}{0.271,0.271,0.271}
\definecolor{gray28}{rgb}{0.278,0.278,0.278}
\definecolor{gray29}{rgb}{0.29,0.29,0.29}
\definecolor{gray3}{rgb}{0.031,0.031,0.031}
\definecolor{gray30}{rgb}{0.302,0.302,0.302}
\definecolor{gray31}{rgb}{0.31,0.31,0.31}
\definecolor{gray32}{rgb}{0.322,0.322,0.322}
\definecolor{gray33}{rgb}{0.329,0.329,0.329}
\definecolor{gray34}{rgb}{0.341,0.341,0.341}
\definecolor{gray35}{rgb}{0.349,0.349,0.349}
\definecolor{gray36}{rgb}{0.361,0.361,0.361}
\definecolor{gray37}{rgb}{0.369,0.369,0.369}
\definecolor{gray38}{rgb}{0.38,0.38,0.38}
\definecolor{gray39}{rgb}{0.388,0.388,0.388}
\definecolor{gray4}{rgb}{0.039,0.039,0.039}
\definecolor{gray40}{rgb}{0.4,0.4,0.4}
\definecolor{gray41}{rgb}{0.412,0.412,0.412}
\definecolor{gray42}{rgb}{0.42,0.42,0.42}
\definecolor{gray43}{rgb}{0.431,0.431,0.431}
\definecolor{gray44}{rgb}{0.439,0.439,0.439}
\definecolor{gray45}{rgb}{0.451,0.451,0.451}
\definecolor{gray46}{rgb}{0.459,0.459,0.459}
\definecolor{gray47}{rgb}{0.471,0.471,0.471}
\definecolor{gray48}{rgb}{0.478,0.478,0.478}
\definecolor{gray49}{rgb}{0.49,0.49,0.49}
\definecolor{gray5}{rgb}{0.051,0.051,0.051}
\definecolor{gray50}{rgb}{0.498,0.498,0.498}
\definecolor{gray51}{rgb}{0.51,0.51,0.51}
\definecolor{gray52}{rgb}{0.522,0.522,0.522}
\definecolor{gray53}{rgb}{0.529,0.529,0.529}
\definecolor{gray54}{rgb}{0.541,0.541,0.541}
\definecolor{gray55}{rgb}{0.549,0.549,0.549}
\definecolor{gray56}{rgb}{0.561,0.561,0.561}
\definecolor{gray57}{rgb}{0.569,0.569,0.569}
\definecolor{gray58}{rgb}{0.58,0.58,0.58}
\definecolor{gray59}{rgb}{0.588,0.588,0.588}
\definecolor{gray6}{rgb}{0.059,0.059,0.059}
\definecolor{gray60}{rgb}{0.6,0.6,0.6}
\definecolor{gray61}{rgb}{0.612,0.612,0.612}
\definecolor{gray62}{rgb}{0.62,0.62,0.62}
\definecolor{gray63}{rgb}{0.631,0.631,0.631}
\definecolor{gray64}{rgb}{0.639,0.639,0.639}
\definecolor{gray65}{rgb}{0.651,0.651,0.651}
\definecolor{gray66}{rgb}{0.659,0.659,0.659}
\definecolor{gray67}{rgb}{0.671,0.671,0.671}
\definecolor{gray68}{rgb}{0.678,0.678,0.678}
\definecolor{gray69}{rgb}{0.69,0.69,0.69}
\definecolor{gray7}{rgb}{0.071,0.071,0.071}
\definecolor{gray70}{rgb}{0.702,0.702,0.702}
\definecolor{gray71}{rgb}{0.71,0.71,0.71}
\definecolor{gray72}{rgb}{0.722,0.722,0.722}
\definecolor{gray73}{rgb}{0.729,0.729,0.729}
\definecolor{gray74}{rgb}{0.741,0.741,0.741}
\definecolor{gray75}{rgb}{0.749,0.749,0.749}
\definecolor{gray76}{rgb}{0.761,0.761,0.761}
\definecolor{gray77}{rgb}{0.769,0.769,0.769}
\definecolor{gray78}{rgb}{0.78,0.78,0.78}
\definecolor{gray79}{rgb}{0.788,0.788,0.788}
\definecolor{gray8}{rgb}{0.078,0.078,0.078}
\definecolor{gray80}{rgb}{0.8,0.8,0.8}
\definecolor{gray81}{rgb}{0.812,0.812,0.812}
\definecolor{gray82}{rgb}{0.82,0.82,0.82}
\definecolor{gray83}{rgb}{0.831,0.831,0.831}
\definecolor{gray84}{rgb}{0.839,0.839,0.839}
\definecolor{gray85}{rgb}{0.851,0.851,0.851}
\definecolor{gray86}{rgb}{0.859,0.859,0.859}
\definecolor{gray87}{rgb}{0.871,0.871,0.871}
\definecolor{gray88}{rgb}{0.878,0.878,0.878}
\definecolor{gray89}{rgb}{0.89,0.89,0.89}
\definecolor{gray9}{rgb}{0.09,0.09,0.09}
\definecolor{gray90}{rgb}{0.898,0.898,0.898}
\definecolor{gray91}{rgb}{0.91,0.91,0.91}
\definecolor{gray92}{rgb}{0.922,0.922,0.922}
\definecolor{gray93}{rgb}{0.929,0.929,0.929}
\definecolor{gray94}{rgb}{0.941,0.941,0.941}
\definecolor{gray95}{rgb}{0.949,0.949,0.949}
\definecolor{gray96}{rgb}{0.961,0.961,0.961}
\definecolor{gray97}{rgb}{0.969,0.969,0.969}
\definecolor{gray98}{rgb}{0.98,0.98,0.98}
\definecolor{gray99}{rgb}{0.988,0.988,0.988}
\definecolor{green1}{rgb}{0,1,0}
\definecolor{green2}{rgb}{0,0.933,0}
\definecolor{green3}{rgb}{0,0.804,0}
\definecolor{green4}{rgb}{0,0.545,0}
\definecolor{greenyellow}{rgb}{0.678,1,0.184}
\definecolor{grey}{rgb}{0.745,0.745,0.745}
\definecolor{grey0}{rgb}{0,0,0}
\definecolor{grey1}{rgb}{0.012,0.012,0.012}
\definecolor{grey10}{rgb}{0.102,0.102,0.102}
\definecolor{grey100}{rgb}{1,1,1}
\definecolor{grey11}{rgb}{0.11,0.11,0.11}
\definecolor{grey12}{rgb}{0.122,0.122,0.122}
\definecolor{grey13}{rgb}{0.129,0.129,0.129}
\definecolor{grey14}{rgb}{0.141,0.141,0.141}
\definecolor{grey15}{rgb}{0.149,0.149,0.149}
\definecolor{grey16}{rgb}{0.161,0.161,0.161}
\definecolor{grey17}{rgb}{0.169,0.169,0.169}
\definecolor{grey18}{rgb}{0.18,0.18,0.18}
\definecolor{grey19}{rgb}{0.188,0.188,0.188}
\definecolor{grey2}{rgb}{0.02,0.02,0.02}
\definecolor{grey20}{rgb}{0.2,0.2,0.2}
\definecolor{grey21}{rgb}{0.212,0.212,0.212}
\definecolor{grey22}{rgb}{0.22,0.22,0.22}
\definecolor{grey23}{rgb}{0.231,0.231,0.231}
\definecolor{grey24}{rgb}{0.239,0.239,0.239}
\definecolor{grey25}{rgb}{0.251,0.251,0.251}
\definecolor{grey26}{rgb}{0.259,0.259,0.259}
\definecolor{grey27}{rgb}{0.271,0.271,0.271}
\definecolor{grey28}{rgb}{0.278,0.278,0.278}
\definecolor{grey29}{rgb}{0.29,0.29,0.29}
\definecolor{grey3}{rgb}{0.031,0.031,0.031}
\definecolor{grey30}{rgb}{0.302,0.302,0.302}
\definecolor{grey31}{rgb}{0.31,0.31,0.31}
\definecolor{grey32}{rgb}{0.322,0.322,0.322}
\definecolor{grey33}{rgb}{0.329,0.329,0.329}
\definecolor{grey34}{rgb}{0.341,0.341,0.341}
\definecolor{grey35}{rgb}{0.349,0.349,0.349}
\definecolor{grey36}{rgb}{0.361,0.361,0.361}
\definecolor{grey37}{rgb}{0.369,0.369,0.369}
\definecolor{grey38}{rgb}{0.38,0.38,0.38}
\definecolor{grey39}{rgb}{0.388,0.388,0.388}
\definecolor{grey4}{rgb}{0.039,0.039,0.039}
\definecolor{grey40}{rgb}{0.4,0.4,0.4}
\definecolor{grey41}{rgb}{0.412,0.412,0.412}
\definecolor{grey42}{rgb}{0.42,0.42,0.42}
\definecolor{grey43}{rgb}{0.431,0.431,0.431}
\definecolor{grey44}{rgb}{0.439,0.439,0.439}
\definecolor{grey45}{rgb}{0.451,0.451,0.451}
\definecolor{grey46}{rgb}{0.459,0.459,0.459}
\definecolor{grey47}{rgb}{0.471,0.471,0.471}
\definecolor{grey48}{rgb}{0.478,0.478,0.478}
\definecolor{grey49}{rgb}{0.49,0.49,0.49}
\definecolor{grey5}{rgb}{0.051,0.051,0.051}
\definecolor{grey50}{rgb}{0.498,0.498,0.498}
\definecolor{grey51}{rgb}{0.51,0.51,0.51}
\definecolor{grey52}{rgb}{0.522,0.522,0.522}
\definecolor{grey53}{rgb}{0.529,0.529,0.529}
\definecolor{grey54}{rgb}{0.541,0.541,0.541}
\definecolor{grey55}{rgb}{0.549,0.549,0.549}
\definecolor{grey56}{rgb}{0.561,0.561,0.561}
\definecolor{grey57}{rgb}{0.569,0.569,0.569}
\definecolor{grey58}{rgb}{0.58,0.58,0.58}
\definecolor{grey59}{rgb}{0.588,0.588,0.588}
\definecolor{grey6}{rgb}{0.059,0.059,0.059}
\definecolor{grey60}{rgb}{0.6,0.6,0.6}
\definecolor{grey61}{rgb}{0.612,0.612,0.612}
\definecolor{grey62}{rgb}{0.62,0.62,0.62}
\definecolor{grey63}{rgb}{0.631,0.631,0.631}
\definecolor{grey64}{rgb}{0.639,0.639,0.639}
\definecolor{grey65}{rgb}{0.651,0.651,0.651}
\definecolor{grey66}{rgb}{0.659,0.659,0.659}
\definecolor{grey67}{rgb}{0.671,0.671,0.671}
\definecolor{grey68}{rgb}{0.678,0.678,0.678}
\definecolor{grey69}{rgb}{0.69,0.69,0.69}
\definecolor{grey7}{rgb}{0.071,0.071,0.071}
\definecolor{grey70}{rgb}{0.702,0.702,0.702}
\definecolor{grey71}{rgb}{0.71,0.71,0.71}
\definecolor{grey72}{rgb}{0.722,0.722,0.722}
\definecolor{grey73}{rgb}{0.729,0.729,0.729}
\definecolor{grey74}{rgb}{0.741,0.741,0.741}
\definecolor{grey75}{rgb}{0.749,0.749,0.749}
\definecolor{grey76}{rgb}{0.761,0.761,0.761}
\definecolor{grey77}{rgb}{0.769,0.769,0.769}
\definecolor{grey78}{rgb}{0.78,0.78,0.78}
\definecolor{grey79}{rgb}{0.788,0.788,0.788}
\definecolor{grey8}{rgb}{0.078,0.078,0.078}
\definecolor{grey80}{rgb}{0.8,0.8,0.8}
\definecolor{grey81}{rgb}{0.812,0.812,0.812}
\definecolor{grey82}{rgb}{0.82,0.82,0.82}
\definecolor{grey83}{rgb}{0.831,0.831,0.831}
\definecolor{grey84}{rgb}{0.839,0.839,0.839}
\definecolor{grey85}{rgb}{0.851,0.851,0.851}
\definecolor{grey86}{rgb}{0.859,0.859,0.859}
\definecolor{grey87}{rgb}{0.871,0.871,0.871}
\definecolor{grey88}{rgb}{0.878,0.878,0.878}
\definecolor{grey89}{rgb}{0.89,0.89,0.89}
\definecolor{grey9}{rgb}{0.09,0.09,0.09}
\definecolor{grey90}{rgb}{0.898,0.898,0.898}
\definecolor{grey91}{rgb}{0.91,0.91,0.91}
\definecolor{grey92}{rgb}{0.922,0.922,0.922}
\definecolor{grey93}{rgb}{0.929,0.929,0.929}
\definecolor{grey94}{rgb}{0.941,0.941,0.941}
\definecolor{grey95}{rgb}{0.949,0.949,0.949}
\definecolor{grey96}{rgb}{0.961,0.961,0.961}
\definecolor{grey97}{rgb}{0.969,0.969,0.969}
\definecolor{grey98}{rgb}{0.98,0.98,0.98}
\definecolor{grey99}{rgb}{0.988,0.988,0.988}
\definecolor{honeydew}{rgb}{0.941,1,0.941}
\definecolor{honeydew1}{rgb}{0.941,1,0.941}
\definecolor{honeydew2}{rgb}{0.878,0.933,0.878}
\definecolor{honeydew3}{rgb}{0.757,0.804,0.757}
\definecolor{honeydew4}{rgb}{0.514,0.545,0.514}
\definecolor{hotpink}{rgb}{1,0.412,0.706}
\definecolor{hotpink1}{rgb}{1,0.431,0.706}
\definecolor{hotpink2}{rgb}{0.933,0.416,0.655}
\definecolor{hotpink3}{rgb}{0.804,0.376,0.565}
\definecolor{hotpink4}{rgb}{0.545,0.227,0.384}
\definecolor{indianred}{rgb}{0.804,0.361,0.361}
\definecolor{indianred1}{rgb}{1,0.416,0.416}
\definecolor{indianred2}{rgb}{0.933,0.388,0.388}
\definecolor{indianred3}{rgb}{0.804,0.333,0.333}
\definecolor{indianred4}{rgb}{0.545,0.227,0.227}
\definecolor{ivory}{rgb}{1,1,0.941}
\definecolor{ivory1}{rgb}{1,1,0.941}
\definecolor{ivory2}{rgb}{0.933,0.933,0.878}
\definecolor{ivory3}{rgb}{0.804,0.804,0.757}
\definecolor{ivory4}{rgb}{0.545,0.545,0.514}
\definecolor{khaki}{rgb}{0.941,0.902,0.549}
\definecolor{khaki1}{rgb}{1,0.965,0.561}
\definecolor{khaki2}{rgb}{0.933,0.902,0.522}
\definecolor{khaki3}{rgb}{0.804,0.776,0.451}
\definecolor{khaki4}{rgb}{0.545,0.525,0.306}
\definecolor{lavender}{rgb}{0.902,0.902,0.98}
\definecolor{lavenderblush}{rgb}{1,0.941,0.961}
\definecolor{lavenderblush1}{rgb}{1,0.941,0.961}
\definecolor{lavenderblush2}{rgb}{0.933,0.878,0.898}
\definecolor{lavenderblush3}{rgb}{0.804,0.757,0.773}
\definecolor{lavenderblush4}{rgb}{0.545,0.514,0.525}
\definecolor{lawngreen}{rgb}{0.486,0.988,0}
\definecolor{lemonchiffon}{rgb}{1,0.98,0.804}
\definecolor{lemonchiffon1}{rgb}{1,0.98,0.804}
\definecolor{lemonchiffon2}{rgb}{0.933,0.914,0.749}
\definecolor{lemonchiffon3}{rgb}{0.804,0.788,0.647}
\definecolor{lemonchiffon4}{rgb}{0.545,0.537,0.439}
\definecolor{lightblue1}{rgb}{0.749,0.937,1}
\definecolor{lightblue2}{rgb}{0.698,0.875,0.933}
\definecolor{lightblue3}{rgb}{0.604,0.753,0.804}
\definecolor{lightblue4}{rgb}{0.408,0.514,0.545}
\definecolor{lightcoral}{rgb}{0.941,0.502,0.502}
\definecolor{lightcyan1}{rgb}{0.878,1,1}
\definecolor{lightcyan2}{rgb}{0.82,0.933,0.933}
\definecolor{lightcyan3}{rgb}{0.706,0.804,0.804}
\definecolor{lightcyan4}{rgb}{0.478,0.545,0.545}
\definecolor{lightgoldenrod}{rgb}{0.933,0.867,0.51}
\definecolor{lightgoldenrod1}{rgb}{1,0.925,0.545}
\definecolor{lightgoldenrod2}{rgb}{0.933,0.863,0.51}
\definecolor{lightgoldenrod3}{rgb}{0.804,0.745,0.439}
\definecolor{lightgoldenrod4}{rgb}{0.545,0.506,0.298}
\definecolor{lightgoldenrodyellow}{rgb}{0.98,0.98,0.824}
\definecolor{lightgrey}{rgb}{0.827,0.827,0.827}
\definecolor{lightpink}{rgb}{1,0.714,0.757}
\definecolor{lightpink1}{rgb}{1,0.682,0.725}
\definecolor{lightpink2}{rgb}{0.933,0.635,0.678}
\definecolor{lightpink3}{rgb}{0.804,0.549,0.584}
\definecolor{lightpink4}{rgb}{0.545,0.373,0.396}
\definecolor{lightsalmon}{rgb}{1,0.627,0.478}
\definecolor{lightsalmon1}{rgb}{1,0.627,0.478}
\definecolor{lightsalmon2}{rgb}{0.933,0.584,0.447}
\definecolor{lightsalmon3}{rgb}{0.804,0.506,0.384}
\definecolor{lightsalmon4}{rgb}{0.545,0.341,0.259}
\definecolor{lightseagreen}{rgb}{0.125,0.698,0.667}
\definecolor{lightskyblue}{rgb}{0.529,0.808,0.98}
\definecolor{lightskyblue1}{rgb}{0.69,0.886,1}
\definecolor{lightskyblue2}{rgb}{0.643,0.827,0.933}
\definecolor{lightskyblue3}{rgb}{0.553,0.714,0.804}
\definecolor{lightskyblue4}{rgb}{0.376,0.482,0.545}
\definecolor{lightslateblue}{rgb}{0.518,0.439,1}
\definecolor{lightslategray}{rgb}{0.467,0.533,0.6}
\definecolor{lightslategrey}{rgb}{0.467,0.533,0.6}
\definecolor{lightsteelblue}{rgb}{0.69,0.769,0.871}
\definecolor{lightsteelblue1}{rgb}{0.792,0.882,1}
\definecolor{lightsteelblue2}{rgb}{0.737,0.824,0.933}
\definecolor{lightsteelblue3}{rgb}{0.635,0.71,0.804}
\definecolor{lightsteelblue4}{rgb}{0.431,0.482,0.545}
\definecolor{lightyellow1}{rgb}{1,1,0.878}
\definecolor{lightyellow2}{rgb}{0.933,0.933,0.82}
\definecolor{lightyellow3}{rgb}{0.804,0.804,0.706}
\definecolor{lightyellow4}{rgb}{0.545,0.545,0.478}
\definecolor{limegreen}{rgb}{0.196,0.804,0.196}
\definecolor{linen}{rgb}{0.98,0.941,0.902}
\definecolor{magenta}{rgb}{1,0,1}
\definecolor{magenta1}{rgb}{1,0,1}
\definecolor{magenta2}{rgb}{0.933,0,0.933}
\definecolor{magenta3}{rgb}{0.804,0,0.804}
\definecolor{magenta4}{rgb}{0.545,0,0.545}
\definecolor{maroon}{rgb}{0.69,0.188,0.376}
\definecolor{maroon1}{rgb}{1,0.204,0.702}
\definecolor{maroon2}{rgb}{0.933,0.188,0.655}
\definecolor{maroon3}{rgb}{0.804,0.161,0.565}
\definecolor{maroon4}{rgb}{0.545,0.11,0.384}
\definecolor{mediumaquamarine}{rgb}{0.4,0.804,0.667}
\definecolor{mediumblue}{rgb}{0,0,0.804}
\definecolor{mediumorchid}{rgb}{0.729,0.333,0.827}
\definecolor{mediumorchid1}{rgb}{0.878,0.4,1}
\definecolor{mediumorchid2}{rgb}{0.82,0.373,0.933}
\definecolor{mediumorchid3}{rgb}{0.706,0.322,0.804}
\definecolor{mediumorchid4}{rgb}{0.478,0.216,0.545}
\definecolor{mediumpurple}{rgb}{0.576,0.439,0.859}
\definecolor{mediumpurple1}{rgb}{0.671,0.51,1}
\definecolor{mediumpurple2}{rgb}{0.624,0.475,0.933}
\definecolor{mediumpurple3}{rgb}{0.537,0.408,0.804}
\definecolor{mediumpurple4}{rgb}{0.365,0.278,0.545}
\definecolor{mediumseagreen}{rgb}{0.235,0.702,0.443}
\definecolor{mediumslateblue}{rgb}{0.482,0.408,0.933}
\definecolor{mediumspringgreen}{rgb}{0,0.98,0.604}
\definecolor{mediumturquoise}{rgb}{0.282,0.82,0.8}
\definecolor{mediumvioletred}{rgb}{0.78,0.082,0.522}
\definecolor{midnightblue}{rgb}{0.098,0.098,0.439}
\definecolor{mintcream}{rgb}{0.961,1,0.98}
\definecolor{mistyrose}{rgb}{1,0.894,0.882}
\definecolor{mistyrose1}{rgb}{1,0.894,0.882}
\definecolor{mistyrose2}{rgb}{0.933,0.835,0.824}
\definecolor{mistyrose3}{rgb}{0.804,0.718,0.71}
\definecolor{mistyrose4}{rgb}{0.545,0.49,0.482}
\definecolor{moccasin}{rgb}{1,0.894,0.71}
\definecolor{navajowhite}{rgb}{1,0.871,0.678}
\definecolor{navajowhite1}{rgb}{1,0.871,0.678}
\definecolor{navajowhite2}{rgb}{0.933,0.812,0.631}
\definecolor{navajowhite3}{rgb}{0.804,0.702,0.545}
\definecolor{navajowhite4}{rgb}{0.545,0.475,0.369}
\definecolor{navyblue}{rgb}{0,0,0.502}
\definecolor{oldlace}{rgb}{0.992,0.961,0.902}
\definecolor{olivedrab}{rgb}{0.42,0.557,0.137}
\definecolor{olivedrab1}{rgb}{0.753,1,0.243}
\definecolor{olivedrab2}{rgb}{0.702,0.933,0.227}
\definecolor{olivedrab3}{rgb}{0.604,0.804,0.196}
\definecolor{olivedrab4}{rgb}{0.412,0.545,0.133}
\definecolor{orange1}{rgb}{1,0.647,0}
\definecolor{orange2}{rgb}{0.933,0.604,0}
\definecolor{orange3}{rgb}{0.804,0.522,0}
\definecolor{orange4}{rgb}{0.545,0.353,0}
\definecolor{orangered}{rgb}{1,0.271,0}
\definecolor{orangered1}{rgb}{1,0.271,0}
\definecolor{orangered2}{rgb}{0.933,0.251,0}
\definecolor{orangered3}{rgb}{0.804,0.216,0}
\definecolor{orangered4}{rgb}{0.545,0.145,0}
\definecolor{orchid}{rgb}{0.855,0.439,0.839}
\definecolor{orchid1}{rgb}{1,0.514,0.98}
\definecolor{orchid2}{rgb}{0.933,0.478,0.914}
\definecolor{orchid3}{rgb}{0.804,0.412,0.788}
\definecolor{orchid4}{rgb}{0.545,0.278,0.537}
\definecolor{palegoldenrod}{rgb}{0.933,0.91,0.667}
\definecolor{palegreen}{rgb}{0.596,0.984,0.596}
\definecolor{palegreen1}{rgb}{0.604,1,0.604}
\definecolor{palegreen2}{rgb}{0.565,0.933,0.565}
\definecolor{palegreen3}{rgb}{0.486,0.804,0.486}
\definecolor{palegreen4}{rgb}{0.329,0.545,0.329}
\definecolor{paleturquoise}{rgb}{0.686,0.933,0.933}
\definecolor{paleturquoise1}{rgb}{0.733,1,1}
\definecolor{paleturquoise2}{rgb}{0.682,0.933,0.933}
\definecolor{paleturquoise3}{rgb}{0.588,0.804,0.804}
\definecolor{paleturquoise4}{rgb}{0.4,0.545,0.545}
\definecolor{palevioletred}{rgb}{0.859,0.439,0.576}
\definecolor{palevioletred1}{rgb}{1,0.51,0.671}
\definecolor{palevioletred2}{rgb}{0.933,0.475,0.624}
\definecolor{palevioletred3}{rgb}{0.804,0.408,0.537}
\definecolor{palevioletred4}{rgb}{0.545,0.278,0.365}
\definecolor{papayawhip}{rgb}{1,0.937,0.835}
\definecolor{peachpuff}{rgb}{1,0.855,0.725}
\definecolor{peachpuff1}{rgb}{1,0.855,0.725}
\definecolor{peachpuff2}{rgb}{0.933,0.796,0.678}
\definecolor{peachpuff3}{rgb}{0.804,0.686,0.584}
\definecolor{peachpuff4}{rgb}{0.545,0.467,0.396}
\definecolor{peru}{rgb}{0.804,0.522,0.247}
\definecolor{pink1}{rgb}{1,0.71,0.773}
\definecolor{pink2}{rgb}{0.933,0.663,0.722}
\definecolor{pink3}{rgb}{0.804,0.569,0.62}
\definecolor{pink4}{rgb}{0.545,0.388,0.424}
\definecolor{plum}{rgb}{0.867,0.627,0.867}
\definecolor{plum1}{rgb}{1,0.733,1}
\definecolor{plum2}{rgb}{0.933,0.682,0.933}
\definecolor{plum3}{rgb}{0.804,0.588,0.804}
\definecolor{plum4}{rgb}{0.545,0.4,0.545}
\definecolor{powderblue}{rgb}{0.69,0.878,0.902}
\definecolor{purple1}{rgb}{0.608,0.188,1}
\definecolor{purple2}{rgb}{0.569,0.173,0.933}
\definecolor{purple3}{rgb}{0.49,0.149,0.804}
\definecolor{purple4}{rgb}{0.333,0.102,0.545}
\definecolor{red1}{rgb}{1,0,0}
\definecolor{red2}{rgb}{0.933,0,0}
\definecolor{red3}{rgb}{0.804,0,0}
\definecolor{red4}{rgb}{0.545,0,0}
\definecolor{rosybrown}{rgb}{0.737,0.561,0.561}
\definecolor{rosybrown1}{rgb}{1,0.757,0.757}
\definecolor{rosybrown2}{rgb}{0.933,0.706,0.706}
\definecolor{rosybrown3}{rgb}{0.804,0.608,0.608}
\definecolor{rosybrown4}{rgb}{0.545,0.412,0.412}
\definecolor{royalblue}{rgb}{0.255,0.412,0.882}
\definecolor{royalblue1}{rgb}{0.282,0.463,1}
\definecolor{royalblue2}{rgb}{0.263,0.431,0.933}
\definecolor{royalblue3}{rgb}{0.227,0.373,0.804}
\definecolor{royalblue4}{rgb}{0.153,0.251,0.545}
\definecolor{saddlebrown}{rgb}{0.545,0.271,0.075}
\definecolor{salmon}{rgb}{0.98,0.502,0.447}
\definecolor{salmon1}{rgb}{1,0.549,0.412}
\definecolor{salmon2}{rgb}{0.933,0.51,0.384}
\definecolor{salmon3}{rgb}{0.804,0.439,0.329}
\definecolor{salmon4}{rgb}{0.545,0.298,0.224}
\definecolor{sandybrown}{rgb}{0.957,0.643,0.376}
\definecolor{seagreen1}{rgb}{0.329,1,0.624}
\definecolor{seagreen2}{rgb}{0.306,0.933,0.58}
\definecolor{seagreen3}{rgb}{0.263,0.804,0.502}
\definecolor{seagreen4}{rgb}{0.18,0.545,0.341}
\definecolor{seashell}{rgb}{1,0.961,0.933}
\definecolor{seashell1}{rgb}{1,0.961,0.933}
\definecolor{seashell2}{rgb}{0.933,0.898,0.871}
\definecolor{seashell3}{rgb}{0.804,0.773,0.749}
\definecolor{seashell4}{rgb}{0.545,0.525,0.51}
\definecolor{sienna}{rgb}{0.627,0.322,0.176}
\definecolor{sienna1}{rgb}{1,0.51,0.278}
\definecolor{sienna2}{rgb}{0.933,0.475,0.259}
\definecolor{sienna3}{rgb}{0.804,0.408,0.224}
\definecolor{sienna4}{rgb}{0.545,0.278,0.149}
\definecolor{skyblue}{rgb}{0.529,0.808,0.922}
\definecolor{skyblue1}{rgb}{0.529,0.808,1}
\definecolor{skyblue2}{rgb}{0.494,0.753,0.933}
\definecolor{skyblue3}{rgb}{0.424,0.651,0.804}
\definecolor{skyblue4}{rgb}{0.29,0.439,0.545}
\definecolor{slateblue}{rgb}{0.416,0.353,0.804}
\definecolor{slateblue1}{rgb}{0.514,0.435,1}
\definecolor{slateblue2}{rgb}{0.478,0.404,0.933}
\definecolor{slateblue3}{rgb}{0.412,0.349,0.804}
\definecolor{slateblue4}{rgb}{0.278,0.235,0.545}
\definecolor{slategray}{rgb}{0.439,0.502,0.565}
\definecolor{slategray1}{rgb}{0.776,0.886,1}
\definecolor{slategray2}{rgb}{0.725,0.827,0.933}
\definecolor{slategray3}{rgb}{0.624,0.714,0.804}
\definecolor{slategray4}{rgb}{0.424,0.482,0.545}
\definecolor{slategrey}{rgb}{0.439,0.502,0.565}
\definecolor{snow}{rgb}{1,0.98,0.98}
\definecolor{snow1}{rgb}{1,0.98,0.98}
\definecolor{snow2}{rgb}{0.933,0.914,0.914}
\definecolor{snow3}{rgb}{0.804,0.788,0.788}
\definecolor{snow4}{rgb}{0.545,0.537,0.537}
\definecolor{springgreen}{rgb}{0,1,0.498}
\definecolor{springgreen1}{rgb}{0,1,0.498}
\definecolor{springgreen2}{rgb}{0,0.933,0.463}
\definecolor{springgreen3}{rgb}{0,0.804,0.4}
\definecolor{springgreen4}{rgb}{0,0.545,0.271}
\definecolor{steelblue}{rgb}{0.275,0.51,0.706}
\definecolor{steelblue1}{rgb}{0.388,0.722,1}
\definecolor{steelblue2}{rgb}{0.361,0.675,0.933}
\definecolor{steelblue3}{rgb}{0.31,0.58,0.804}
\definecolor{steelblue4}{rgb}{0.212,0.392,0.545}
\definecolor{tan}{rgb}{0.824,0.706,0.549}
\definecolor{tan1}{rgb}{1,0.647,0.31}
\definecolor{tan2}{rgb}{0.933,0.604,0.286}
\definecolor{tan3}{rgb}{0.804,0.522,0.247}
\definecolor{tan4}{rgb}{0.545,0.353,0.169}
\definecolor{thistle}{rgb}{0.847,0.749,0.847}
\definecolor{thistle1}{rgb}{1,0.882,1}
\definecolor{thistle2}{rgb}{0.933,0.824,0.933}
\definecolor{thistle3}{rgb}{0.804,0.71,0.804}
\definecolor{thistle4}{rgb}{0.545,0.482,0.545}
\definecolor{tomato}{rgb}{1,0.388,0.278}
\definecolor{tomato1}{rgb}{1,0.388,0.278}
\definecolor{tomato2}{rgb}{0.933,0.361,0.259}
\definecolor{tomato3}{rgb}{0.804,0.31,0.224}
\definecolor{tomato4}{rgb}{0.545,0.212,0.149}
\definecolor{turquoise1}{rgb}{0,0.961,1}
\definecolor{turquoise2}{rgb}{0,0.898,0.933}
\definecolor{turquoise3}{rgb}{0,0.773,0.804}
\definecolor{turquoise4}{rgb}{0,0.525,0.545}
\definecolor{violetred}{rgb}{0.816,0.125,0.565}
\definecolor{violetred1}{rgb}{1,0.243,0.588}
\definecolor{violetred2}{rgb}{0.933,0.227,0.549}
\definecolor{violetred3}{rgb}{0.804,0.196,0.471}
\definecolor{violetred4}{rgb}{0.545,0.133,0.322}
\definecolor{wheat}{rgb}{0.961,0.871,0.702}
\definecolor{wheat1}{rgb}{1,0.906,0.729}
\definecolor{wheat2}{rgb}{0.933,0.847,0.682}
\definecolor{wheat3}{rgb}{0.804,0.729,0.588}
\definecolor{wheat4}{rgb}{0.545,0.494,0.4}
\definecolor{whitesmoke}{rgb}{0.961,0.961,0.961}
\definecolor{yellow1}{rgb}{1,1,0}
\definecolor{yellow2}{rgb}{0.933,0.933,0}
\definecolor{yellow3}{rgb}{0.804,0.804,0}
\definecolor{yellow4}{rgb}{0.545,0.545,0}
\definecolor{yellowgreen}{rgb}{0.604,0.804,0.196}

\caption{The Gaifman graph of $\mathfrak{A}_1.$}
\labels{figure_boundariedgraph2}
\end{figure}

\myskip\paragraph{A boundaried structure of bounded treewidth that satisfies $\bar{φ}.$}
Let ${\bf b}_1$ be the tuple obtained from ${\bf b}_1^\star$ after replacing, in ${\bf b}_1^\star,$ each vertex $v∈ V(F^\star)$ with the vertex $ξ^{-1}(v)∈ V(F_1).$
Also, let $\mathfrak{A}_1 = \mathfrak{A}^{(d,Z,L,F_1)}$ (see~\autoref{figure_boundariedgraph2} for a visualization of the Gaifman graph of $\mathfrak{A}^{(d,Z,L,F_1)}$).
Observe that $\cupall{\bf W}_q^{(1)}\subseteq V(\mathfrak{A}_1)$ and $R_1\cap I_1^{(d)}\subseteq V(\mathfrak{A}_1).$
We set  $\tilde{X}_1 := (\tilde{X}^\star_1\setminus V(F^\star))\cup V(F_1)$ and $R_1^{'} :=R_1\cap  (Z\setminus \partial_{\mathfrak{K}_1}(Z)).$
Observe that $\tilde{X}_1\subseteq V(\mathfrak{A}_1),$ $R_1^{'}\subseteq V(\mathfrak{A}_1),$ and  $R_1^{'}= R_1^{' \star} \setminus( \partial_{\mathfrak{K}_1}(Z)\cup \breve{C}).$
At this point, we stress that while $\tilde{X}_1$ is obtained from $\tilde{X}^\star_1$
after replacing $V(F^\star)$ with $V(F_1),$ $R_1^{'}$ is obtained from $R_1^{' \star}$ after removing all elements in $V(\mathfrak{A}^\star)$ that are in $\breve{C}$ and in $\partial_{\mathfrak{K}_1}(Z).$

We aim to show that $({\bf H}, \bar{φ})∈ {\sf out}\text{-}{\sf sig}(\mathfrak{K}_1,R_1,d,L,Z).$
To show this, by the definition of {\sf out}\text{-}{\sf sig} it remains to prove that $\big(\mathfrak{A}_1,R_1 ',{\bf W}_{{q}}^{(1)}, \varnothing^l,\tilde{X}_1, {\bf b}_1\big)\models \bar{φ}.$
To prove the latter, first notice that, since $F_1$ and $F^\star$ are isomorphic, we have that
$\mathfrak{A}_1[V({\bf b}_1)],$ $\mathfrak{A}^\star[V({\bf b}_1^\star)],$ and $\mathfrak{A}_{\rm out}^\star[V( {\bf b}_1^\star)]$ are (pairwise) isomorphic.
This implies that
 $(\mathfrak{A}_1, {\bf b}_1),$ $(\mathfrak{A}^\star, {\bf b}_1^\star),$ and $(\mathfrak{A}_{\rm out}^\star, {\bf b}_1^\star)$ are (pairwise) compatible.
We consider the $h$-boundaried $τ'$-structures
 $\big(\mathfrak{A}^\star, R_1^{' \star},{\bf W}_{{q}}^{(1)} ,\varnothing^l, \tilde{X}^\star_1, {\bf b}_1^\star\big)$
and
$\big(\mathfrak{A}_1,R_1 ',{\bf W}_{{q}}^{(1)},\varnothing^l,\tilde{X}_1, {\bf b}_1\big).$
These $h$-boundaried $τ'$-structures
are compatible.
We now show that they are also $(θ^{\sf out}_q,h)$-equivalent, which will imply that $\big(\mathfrak{A}^\star, R_1^{' \star},{\bf W}_{{q}}^{(1)},\varnothing^l, \tilde{X}^\star_1, {\bf b}_1^\star\big)\models \bar{φ}\iff\big(\mathfrak{A}_1,R_1 ',{\bf W}_{{q}}^{(1)},\varnothing^l,\tilde{X}_1, {\bf b}_1\big)\models \bar{φ}.$
\medskip

 \noindent{\em Subclaim:}
 $\big(\mathfrak{A}^\star, R_1^{' \star},{\bf W}_{{q}}^{(1)}, \varnothing^l,\tilde{X}^\star_1, {\bf b}_1^\star\big)$
 and
 $\big(\mathfrak{A}_1,R_1 ',{\bf W}_{{q}}^{(1)},\varnothing^l,\tilde{X}_1, {\bf b}_1\big)$
 are $(θ^{\sf out}_q,h)$-equivalent.
 \medskip

\noindent{\em Proof of Subclaim:}
Let $\mathfrak{C}^\circ$ be a $τ$-structure, $R^\circ\subseteq V(\mathfrak{C}^\circ),$
${\bf W}_q^{\circ}∈ (2^{V(\mathfrak{C}^\circ)})^{2q},$
${\bf a}^\circ$ be an apex-tuple of  $\mathfrak{C}^\circ$ of size $l,$
$X^\circ\subseteq V(\mathfrak{C}^\circ),$ and
${\bf b}^\circ$ be an apex-tuple of $\mathfrak{C}^\circ$ of size $h,$
such that
$(\mathfrak{C}^\circ,R^\circ,{\bf W}_q^{\circ},{\bf a}^\circ,X^\circ, {\bf b}^\circ)$
is an  $h$-boundaried $τ'$-structure
that is compatible with the $h$-boundaried $τ'$-structures $\big(\mathfrak{A}^\star, R_1^{' \star},{\bf W}_{{q}}^{(1)},\varnothing^l, \tilde{X}^\star_1, {\bf b}_1^\star\big)$
and $\big(\mathfrak{A}_1,R_1 ',{\bf W}_{{q}}^{(1)}, \varnothing^l, \tilde{X}_1, {\bf b}_1\big).$
Our goal is to show that
$(\mathfrak{C}^\circ, R^\circ, {\bf W}_q^\circ,{\bf a}^\circ, X^\circ, {\bf b}^\circ)\oplus \big(\mathfrak{A}^\star,R_1^{'\star},{\bf W}_{{q}}^{(1)},\varnothing^l, \tilde{X}^\star_1, {\bf b}_1^\star\big)\models θ^{\sf out}_q
\iff
(\mathfrak{C}^\circ, R^\circ, {\bf W}_q^\circ, {\bf a}^\circ, X^\circ, {\bf b}^\circ)\oplus \big( \mathfrak{A}_1,R_1 ',{\bf W}_{{q}}^{(1)}, \varnothing^l,\tilde{X}_1, {\bf b}_1^\star\big)\models θ^{\sf out}_q.$
We set
$\mathfrak{B}^\star := (\mathfrak{C}^\circ, {\bf b}^\circ) \oplus (\mathfrak{A}^\star, {\bf b}_1^\star)$
and
$\mathfrak{B} := (\mathfrak{C}^\circ, {\bf b}^\circ)\oplus(\mathfrak{A}_1, {\bf b}_1).$
Equivalently, it suffices to prove that
$$(\mathfrak{B}^\star, R^\circ \cup R_1^{'\star}, {\bf W}_q^\circ \cup {\bf W}_{{q}}^{(1)},{\bf a}^\circ, X^\circ \cup \tilde{X}^\star_1)\models θ^{\sf out}_q\iff (\mathfrak{B}, R^\circ \cup R_1^{'}, {\bf W}_q^\circ \cup{\bf W}_{{q}}^{(1)},{\bf a}^\circ, X^\circ \cup \tilde{X}_1)\models θ^{\sf out}_q.$$
%
%
%
By the definition of $ θ^{\sf out}_q$ (see~\autoref{@reconquistasen}), it will be enough to show that
\begin{itemize}
\item[(i).] $(\mathfrak{B}^\star,X^\circ \cup \tilde{X}^\star_1)\models β|_{{\sf star}_X}\iff (\mathfrak{B}, X^\circ \cup\tilde{X}_1)\models β|_{{\sf star}_{\sf X}},$
\item[(ii).] ${\bf W}_q^\circ \cup {\bf W}_{{q}}^{(1)}$ is a $q$-pseudogrid of $G_{\mathfrak{B}^\star}$ $\iff$ ${\bf W}_q^\circ \cup {\bf W}_{{q}}^{(1)}$ is a $q$-pseudogrid of $G_{\mathfrak{B}},$ and
\item[(iii).]
 for every $C'∈ {\sf cc}(\mathfrak{B}^\star,X^\circ \cup \tilde{X}^\star_1)$ that is not a subset of the $w$-privileged set of $G_{\mathfrak{B}^\star}$ with respect to ${\bf W}_q^\circ \cup {\bf W}_{{q}}^{(1)}$ and $X^\circ \cup \tilde{X}^\star_1,$ it holds that $(\mathfrak{B}^\star,R^\circ \cup R_1^{'\star}, {\bf a}^0)[C']\models  \breve{ζ}_R |_{\sf ap_{\bf c}}\wedge μ,$
%
%
if and only if
for every $C'∈ {\sf cc}(\mathfrak{B},X^\circ \cup\tilde{X}_1)$ that is not a subset of the $w$-privileged set of $G_{\mathfrak{B}}$ with respect to ${\bf W}_q^\circ \cup {\bf W}_{{q}}^{(1)}$ and $X^\circ \cup\tilde{X}_1,$
it holds that $(\mathfrak{B},R^\circ \cup R_1^{'}, {\bf a}^0)[C']\models  \breve{ζ}_{\sf R} |_{\sf ap_{\bf c}}\wedge μ.$
%
\end{itemize}
Note that, since $\cupall {\bf W}_q^{(1)} \subseteq V(\mathfrak{A}_1)\cap V(\mathfrak{A}_1^\star),$
$\cupall{\bf W}_q^\circ \cup {\bf W}_{{q}}^{(1)}\subseteq V(\mathfrak{B}^\star)\cap V(\mathfrak{B}).$
Thus, item~(ii) holds.

Let ${\cal C}^\star$ be the set of all
$C∈{\sf cc}(\mathfrak{B}^\star,X^\circ \cup \tilde{X}^\star_1)$
that are not subsets of the $w$-privileged set of $G_{\mathfrak{B}^\star}$ with respect to ${\bf W}_q^\circ \cup {\bf W}_{{q}}^{(1)}$ and $X^\circ \cup\tilde{X}_1^\star$
and let
 ${\cal C}$ be the set of all
$C∈{\sf cc}(\mathfrak{B},X^\circ \cup \tilde{X}_1)$
that are not subsets of the $w$-privileged set of $G_{\mathfrak{B}}$ with respect to ${\bf W}_q^\circ \cup {\bf W}_{{q}}^{(1)}$ and $X^\circ \cup\tilde{X}_1.$
Observe that every $C∈ {\cal C}^\star$
 is a subset of $Z\setminus \partial_{\mathfrak{K}_1}(Z).$
Therefore, since $\tilde{X}_1 = (\tilde{X}^\star_1\setminus V(F^\star))\cup V(F_1)$
and $G_{\mathfrak{A}^\star}[Z]$ is the same graph as $G_{\mathfrak{A}_1}[Z],$ it holds that
every $C∈ {\cal C}$ is also a subset of $Z\setminus \partial_{\mathfrak{K}_1}(Z)$ and, moreover,
${\cal C} = {\cal C}^\star.$
Also, observe that, since
$R_1^{'\star}  = R\cap (\breve{C}\cup Z)$
while
$R_1^{'} = R_1\cap  (Z\setminus \partial_{\mathfrak{K}_1}(Z)),$
we have that $R_1^{'\star}\cap (Z\setminus \partial_{\mathfrak{K}_1}(Z)) =
R_1^{'}\cap (Z\setminus \partial_{\mathfrak{K}_1}(Z)).$
Therefore, for every $C∈ {\cal C},$
$(\mathfrak{B}^\star, R^\circ \cup R_1^{'\star})[C] = (\mathfrak{B}, R^\circ \cup R_1^{'})[C].$
Thus, item (iii) above holds.

It remains to prove that
$$(\mathfrak{B}^\star,X^\circ \cup \tilde{X}^\star_1)\models β|_{{\sf star}_{\sf X}}\iff (\mathfrak{B}, X^\circ \cup\tilde{X}_1)\models β|_{{\sf star}_{\sf X}}.$$
To prove this, we will argue that the structure ${\sf star}_{\sf X}(\mathfrak{B},X^\circ \cup \tilde{X}_1)$ is isomorphic to the structure ${\sf star}_{\sf X}(\mathfrak{B}^\star,X^\circ \cup \tilde{X}^\star_1).$
To see why this holds, notice that
there is a bijection $ρ$ between the universes of ${\sf star}_{\sf X}(\mathfrak{B}^\star,X^\circ \cup \tilde{X}^\star_1)$ and of ${\sf star}_{\sf X}(\mathfrak{B},X^\circ \cup \tilde{X}_1),$ such that ${\bf x}∈ {\sf R}^{{\sf star}_X(\mathfrak{B}^\star,X^\circ \cup \tilde{X}^{\star}_1)}$ if and only if ${\bf x}∈ {\sf R}^{{\sf star}_{\sf X}(\mathfrak{B},X^\circ \cup \tilde{X}_1)}$ for all ${\sf R}$ of arity $r≥ 1$ and all ${\bf x}∈ V({\sf star}_{\sf X}(\mathfrak{B}^\star,X^\circ \cup \tilde{X}_1^\star)).$
This bijection is the identity function for every $x∈ X^\circ \cup \tilde{X}^\star_1$ and every $v∈ V({\sf star}_{\sf X}(\mathfrak{B}^\star,X^\circ \cup \tilde{X}^\star_1))$ that corresponds to a set in ${\sf cc}(\mathfrak{B}^\star,X^\circ \cup \tilde{X}^\star_1)\setminus {\sf pr}(G_{\mathfrak{B}^\star},{\bf W}_q^\circ \cup {\bf W}_{{q}}^{(1)},X^\circ \cup \tilde{X}^\star_1).$
The element $v∈ V({\sf star}_{\sf X}(\mathfrak{B}^\star,X^\circ \cup \tilde{X}^\star_1))$ that corresponds to the unique element in ${\sf pr}(G_{\mathfrak{B}^\star},{\bf W}_q^\circ \cup {\bf W}_{{q}}^{(1)},X^\circ \cup \tilde{X}^\star_1)$ is mapped to the element in $V({\sf star}(\mathfrak{B},X^\circ \cup \tilde{X}_1))$ that corresponds to the unique element in ${\sf pr}(G_{\mathfrak{B}},{\bf W}_q^\circ \cup {\bf W}_{{q}}^{(1)},X^\circ \cup\tilde{X}_1).$
To get some intuition why this bijection satisfies the above condition, notice that
if we ``contract'' the vertex set in ${\sf pr}(G_{\mathfrak{B}^\star},{\bf W}_q^\circ \cup {\bf W}_{{q}}^{(1)},X^\circ \cup \tilde{X}^\star_1)$ and the vertex set in ${\sf pr}(G_{\mathfrak{B}},{\bf W}_q^\circ \cup {\bf W}_{{q}}^{(1)},X^\circ \cup\tilde{X}_1)$
to two single vertices $v^\star$ and $v,$ respectively, then
the neighbors of $v^\star$ in $X^\circ \cup \tilde{X}^\star_1$ and the neighbors of $v$ in $X^\circ \cup\tilde{X}_1$ induce isomorphic graphs in $G_{\mathfrak{B}^\star}$ and $G_{\mathfrak{B}},$ respectively. The latter holds due to the fact that if a vertex of $X^\circ \cup \tilde{X}^\star_1$ is connected with a vertex in ${\sf pr}(G_{\mathfrak{B}^\star},{\bf W}_q^\circ \cup {\bf W}_{{q}}^{(1)},X^\circ \cup \tilde{X}^\star_1),$ then
the corresponding vertex in $X^\circ \cup\tilde{X}_1$ is also connected with a vertex in  ${\sf pr}(G_{\mathfrak{B}},{\bf W}_q^\circ \cup {\bf W}_{{q}}^{(1)},X^\circ \cup\tilde{X}_1).$

Therefore, $(\mathfrak{B}^\star, R^\circ \cup R_1^{'\star}, {\bf W}_{{q}}^{(1)}, {\bf a}^\circ,X^\circ \cup \tilde{X}^\star_1)\models θ^{\sf out}_q \iff (\mathfrak{B}, R^\circ \cup R_1^{'}, {\bf W}_{{q}}^{(1)},{\bf a}^\circ, X^\circ \cup \tilde{X}_1)\models θ^{\sf out}_q$ and the subclaim follows.\hfill$\diamond$
\bigskip

Since by the above subclaim,  $\big(\mathfrak{A}^\star, R_1^{' \star},{\bf W}_{{q}}^{(1)} ,\varnothing^l, \tilde{X}^\star_1, {\bf b}_1^\star\big)$
 and
 $\big(\mathfrak{A}_1,R_1 ',{\bf W}_{{q}}^{(1)}, \varnothing^l,\tilde{X}_1, {\bf b}_1\big)$
 are $(θ^{\sf out}_q,h)$-equivalent,
 it follows that
\begin{eqnarray}
\big(\mathfrak{A}^\star, R_1^{' \star},{\bf W}_{{q}}^{(1)} , \varnothing^l, \tilde{X}^\star_1, {\bf b}_1^\star\big)\models \bar{φ}\iff\big(\mathfrak{A}_1,R_1 ',{\bf W}_{{q}}^{(1)}, \varnothing^l, \tilde{X}_1, {\bf b}_1\big)\models \bar{φ}.\labels{@identificable}
\end{eqnarray}

Therefore, we conclude that $F_1, {\bf b}_1,$ and $\tilde{X}_1$
certify that $({\bf H}, \bar{φ})∈ {\sf out}\text{-}{\sf sig}(\mathfrak{K}_1,R_1,d,L,Z).$
Since ${\sf out}\text{-}{\sf sig}(\mathfrak{K}_1,R_1,d,L,Z)= {\sf out}\text{-}{\sf sig}(\mathfrak{K}_2,R_2,d,L,Z'),$ we have $({\bf H},\bar{φ})∈ {\sf out}\text{-}{\sf sig}(\mathfrak{K}_2,R_2,d,L,Z').$
Thus,
\begin{itemize}
\item[(a).] there is an $F_2∈ {\cal F}_{h-|\partial_{\mathfrak{K}_2} (Z')|}^{V_L({\bf a})}$ such that if ${\bf H} = (H, {\cal U})$ and
${\cal V}_2= (\partial_{\mathfrak{K}_2} (Z'),V_L ({\bf a}), V(F_2)\setminus V_L ({\bf a})),$
then ${\cal V}_2$ is a nice 3-partition of  $K_2^{\bf a}[\partial_{\mathfrak{K}_2} (Z') \cup V_L ({\bf a})]\cup F_2$  and $K_2^{\bf a}[\partial_{\mathfrak{K}_2} (Z') \cup V_L ({\bf a})]\cup F_2$ is strongly isomorphic to $H$ with respect to $({\cal V}_2, {\cal U})$ and
\item[(b).] there is an ordering ${\bf b}_2$ of $\partial_{\mathfrak{K}_2} (Z') \cup V(F_2)$
and an $\tilde{X}_2\subseteq Z'\cup V(F_2),$
such that $\partial_{\mathfrak{K}_2}(Z') \cup V(F_2)\subseteq \tilde{X}_2$ and,
if $R_2^{'} = (Z'\setminus \partial_{\mathfrak{K}_2}(Z'))\cap R_2,$
then
$\big(\mathfrak{A}_2^{(d,Z',L,F_2)},R_2 ',{\bf W}_{{q}}^{(2)},\varnothing^l, \tilde{X}_2, {\bf b}_2\big)\models \bar{φ}.$
\end{itemize}

Observe that,
since  $K_1^{\bf a}[\partial_{\mathfrak{K}_1} (Z) \cup V_L ({\bf a})]\cup F_1$
and $K_2^{\bf a}[\partial_{\mathfrak{K}_2} (Z') \cup V_L ({\bf a})]\cup F_2$
are strongly isomorphic to  $H$ with respect to
$({\cal V}_1, {\cal U})$ and $({\cal V}_2, {\cal U}),$ respectively,
we also have that  $K_1^{\bf a}[\partial_{\mathfrak{K}_1} (Z) \cup V_L ({\bf a})]\cup F_1$
is strongly isomorphic to
$K_2^{\bf a}[\partial_{\mathfrak{K}_2} (Z') \cup V_L ({\bf a})]\cup F_2$
with respect to $({\cal V}_1, {\cal V}_2).$
We now set
$F_2 ' = F_2\setminus V_L ({\bf a})$
and
$\mathfrak{A}_2 = \mathfrak{A}_2^{(d,Z',L,F_2)}$ (see~\autoref{figure_boundariedgraph3} for an example of its Gaifman graph).
\begin{figure}[ht]
\centering
\tikzstyle{ipe stylesheet} = [
  ipe import,
  even odd rule,
  line join=round,
  line cap=butt,
  ipe pen normal/.style={line width=0.4},
  ipe pen heavier/.style={line width=0.8},
  ipe pen fat/.style={line width=1.2},
  ipe pen ultrafat/.style={line width=2},
  ipe pen normal,
  ipe mark normal/.style={ipe mark scale=3},
  ipe mark large/.style={ipe mark scale=5},
  ipe mark small/.style={ipe mark scale=2},
  ipe mark tiny/.style={ipe mark scale=1.1},
  ipe mark normal,
  /pgf/arrow keys/.cd,
  ipe arrow normal/.style={scale=7},
  ipe arrow large/.style={scale=10},
  ipe arrow small/.style={scale=5},
  ipe arrow tiny/.style={scale=3},
  ipe arrow normal,
  /tikz/.cd,
  ipe arrows, 
  <->/.tip = ipe normal,
  ipe dash normal/.style={dash pattern=},
  ipe dash dotted/.style={dash pattern=on 1bp off 3bp},
  ipe dash dashed/.style={dash pattern=on 4bp off 4bp},
  ipe dash dash dotted/.style={dash pattern=on 4bp off 2bp on 1bp off 2bp},
  ipe dash dash dot dotted/.style={dash pattern=on 4bp off 2bp on 1bp off 2bp on 1bp off 2bp},
  ipe dash normal,
  ipe node/.append style={font=\normalsize},
  ipe stretch normal/.style={ipe node stretch=1},
  ipe stretch normal,
  ipe opacity 10/.style={opacity=0.1},
  ipe opacity 30/.style={opacity=0.3},
  ipe opacity 50/.style={opacity=0.5},
  ipe opacity 75/.style={opacity=0.75},
  ipe opacity opaque/.style={opacity=1},
  ipe opacity opaque,
]
\definecolor{black}{rgb}{0,0,0}
\definecolor{white}{rgb}{1,1,1}
\definecolor{red}{rgb}{1,0,0}
\definecolor{blue}{rgb}{0,0,1}
\definecolor{green}{rgb}{0,1,0}
\definecolor{yellow}{rgb}{1,1,0}
\definecolor{orange}{rgb}{1,0.647,0}
\definecolor{gold}{rgb}{1,0.843,0}
\definecolor{purple}{rgb}{0.627,0.125,0.941}
\definecolor{gray}{rgb}{0.745,0.745,0.745}
\definecolor{brown}{rgb}{0.647,0.165,0.165}
\definecolor{navy}{rgb}{0,0,0.502}
\definecolor{pink}{rgb}{1,0.753,0.796}
\definecolor{seagreen}{rgb}{0.18,0.545,0.341}
\definecolor{turquoise}{rgb}{0.251,0.878,0.816}
\definecolor{violet}{rgb}{0.933,0.51,0.933}
\definecolor{darkblue}{rgb}{0,0,0.545}
\definecolor{darkcyan}{rgb}{0,0.545,0.545}
\definecolor{darkgray}{rgb}{0.663,0.663,0.663}
\definecolor{darkgreen}{rgb}{0,0.392,0}
\definecolor{darkmagenta}{rgb}{0.545,0,0.545}
\definecolor{darkorange}{rgb}{1,0.549,0}
\definecolor{darkred}{rgb}{0.545,0,0}
\definecolor{lightblue}{rgb}{0.678,0.847,0.902}
\definecolor{lightcyan}{rgb}{0.878,1,1}
\definecolor{lightgray}{rgb}{0.827,0.827,0.827}
\definecolor{lightgreen}{rgb}{0.565,0.933,0.565}
\definecolor{lightyellow}{rgb}{1,1,0.878}
\definecolor{aliceblue}{rgb}{0.941,0.973,1}
\definecolor{antiquewhite}{rgb}{0.98,0.922,0.843}
\definecolor{antiquewhite1}{rgb}{1,0.937,0.859}
\definecolor{antiquewhite2}{rgb}{0.933,0.875,0.8}
\definecolor{antiquewhite3}{rgb}{0.804,0.753,0.69}
\definecolor{antiquewhite4}{rgb}{0.545,0.514,0.471}
\definecolor{aquamarine}{rgb}{0.498,1,0.831}
\definecolor{aquamarine1}{rgb}{0.498,1,0.831}
\definecolor{aquamarine2}{rgb}{0.463,0.933,0.776}
\definecolor{aquamarine3}{rgb}{0.4,0.804,0.667}
\definecolor{aquamarine4}{rgb}{0.271,0.545,0.455}
\definecolor{azure}{rgb}{0.941,1,1}
\definecolor{azure1}{rgb}{0.941,1,1}
\definecolor{azure2}{rgb}{0.878,0.933,0.933}
\definecolor{azure3}{rgb}{0.757,0.804,0.804}
\definecolor{azure4}{rgb}{0.514,0.545,0.545}
\definecolor{beige}{rgb}{0.961,0.961,0.863}
\definecolor{bisque}{rgb}{1,0.894,0.769}
\definecolor{bisque1}{rgb}{1,0.894,0.769}
\definecolor{bisque2}{rgb}{0.933,0.835,0.718}
\definecolor{bisque3}{rgb}{0.804,0.718,0.62}
\definecolor{bisque4}{rgb}{0.545,0.49,0.42}
\definecolor{blanchedalmond}{rgb}{1,0.922,0.804}
\definecolor{blue1}{rgb}{0,0,1}
\definecolor{blue2}{rgb}{0,0,0.933}
\definecolor{blue3}{rgb}{0,0,0.804}
\definecolor{blue4}{rgb}{0,0,0.545}
\definecolor{blueviolet}{rgb}{0.541,0.169,0.886}
\definecolor{brown1}{rgb}{1,0.251,0.251}
\definecolor{brown2}{rgb}{0.933,0.231,0.231}
\definecolor{brown3}{rgb}{0.804,0.2,0.2}
\definecolor{brown4}{rgb}{0.545,0.137,0.137}
\definecolor{burlywood}{rgb}{0.871,0.722,0.529}
\definecolor{burlywood1}{rgb}{1,0.827,0.608}
\definecolor{burlywood2}{rgb}{0.933,0.773,0.569}
\definecolor{burlywood3}{rgb}{0.804,0.667,0.49}
\definecolor{burlywood4}{rgb}{0.545,0.451,0.333}
\definecolor{cadetblue}{rgb}{0.373,0.62,0.627}
\definecolor{cadetblue1}{rgb}{0.596,0.961,1}
\definecolor{cadetblue2}{rgb}{0.557,0.898,0.933}
\definecolor{cadetblue3}{rgb}{0.478,0.773,0.804}
\definecolor{cadetblue4}{rgb}{0.325,0.525,0.545}
\definecolor{chartreuse}{rgb}{0.498,1,0}
\definecolor{chartreuse1}{rgb}{0.498,1,0}
\definecolor{chartreuse2}{rgb}{0.463,0.933,0}
\definecolor{chartreuse3}{rgb}{0.4,0.804,0}
\definecolor{chartreuse4}{rgb}{0.271,0.545,0}
\definecolor{chocolate}{rgb}{0.824,0.412,0.118}
\definecolor{chocolate1}{rgb}{1,0.498,0.141}
\definecolor{chocolate2}{rgb}{0.933,0.463,0.129}
\definecolor{chocolate3}{rgb}{0.804,0.4,0.114}
\definecolor{chocolate4}{rgb}{0.545,0.271,0.075}
\definecolor{coral}{rgb}{1,0.498,0.314}
\definecolor{coral1}{rgb}{1,0.447,0.337}
\definecolor{coral2}{rgb}{0.933,0.416,0.314}
\definecolor{coral3}{rgb}{0.804,0.357,0.271}
\definecolor{coral4}{rgb}{0.545,0.243,0.184}
\definecolor{cornflowerblue}{rgb}{0.392,0.584,0.929}
\definecolor{cornsilk}{rgb}{1,0.973,0.863}
\definecolor{cornsilk1}{rgb}{1,0.973,0.863}
\definecolor{cornsilk2}{rgb}{0.933,0.91,0.804}
\definecolor{cornsilk3}{rgb}{0.804,0.784,0.694}
\definecolor{cornsilk4}{rgb}{0.545,0.533,0.471}
\definecolor{cyan}{rgb}{0,1,1}
\definecolor{cyan1}{rgb}{0,1,1}
\definecolor{cyan2}{rgb}{0,0.933,0.933}
\definecolor{cyan3}{rgb}{0,0.804,0.804}
\definecolor{cyan4}{rgb}{0,0.545,0.545}
\definecolor{darkgoldenrod}{rgb}{0.722,0.525,0.043}
\definecolor{darkgoldenrod1}{rgb}{1,0.725,0.059}
\definecolor{darkgoldenrod2}{rgb}{0.933,0.678,0.055}
\definecolor{darkgoldenrod3}{rgb}{0.804,0.584,0.047}
\definecolor{darkgoldenrod4}{rgb}{0.545,0.396,0.031}
\definecolor{darkgrey}{rgb}{0.663,0.663,0.663}
\definecolor{darkkhaki}{rgb}{0.741,0.718,0.42}
\definecolor{darkolivegreen}{rgb}{0.333,0.42,0.184}
\definecolor{darkolivegreen1}{rgb}{0.792,1,0.439}
\definecolor{darkolivegreen2}{rgb}{0.737,0.933,0.408}
\definecolor{darkolivegreen3}{rgb}{0.635,0.804,0.353}
\definecolor{darkolivegreen4}{rgb}{0.431,0.545,0.239}
\definecolor{darkorange1}{rgb}{1,0.498,0}
\definecolor{darkorange2}{rgb}{0.933,0.463,0}
\definecolor{darkorange3}{rgb}{0.804,0.4,0}
\definecolor{darkorange4}{rgb}{0.545,0.271,0}
\definecolor{darkorchid}{rgb}{0.6,0.196,0.8}
\definecolor{darkorchid1}{rgb}{0.749,0.243,1}
\definecolor{darkorchid2}{rgb}{0.698,0.227,0.933}
\definecolor{darkorchid3}{rgb}{0.604,0.196,0.804}
\definecolor{darkorchid4}{rgb}{0.408,0.133,0.545}
\definecolor{darksalmon}{rgb}{0.914,0.588,0.478}
\definecolor{darkseagreen}{rgb}{0.561,0.737,0.561}
\definecolor{darkseagreen1}{rgb}{0.757,1,0.757}
\definecolor{darkseagreen2}{rgb}{0.706,0.933,0.706}
\definecolor{darkseagreen3}{rgb}{0.608,0.804,0.608}
\definecolor{darkseagreen4}{rgb}{0.412,0.545,0.412}
\definecolor{darkslateblue}{rgb}{0.282,0.239,0.545}
\definecolor{darkslategray}{rgb}{0.184,0.31,0.31}
\definecolor{darkslategray1}{rgb}{0.592,1,1}
\definecolor{darkslategray2}{rgb}{0.553,0.933,0.933}
\definecolor{darkslategray3}{rgb}{0.475,0.804,0.804}
\definecolor{darkslategray4}{rgb}{0.322,0.545,0.545}
\definecolor{darkslategrey}{rgb}{0.184,0.31,0.31}
\definecolor{darkturquoise}{rgb}{0,0.808,0.82}
\definecolor{darkviolet}{rgb}{0.58,0,0.827}
\definecolor{deeppink}{rgb}{1,0.078,0.576}
\definecolor{deeppink1}{rgb}{1,0.078,0.576}
\definecolor{deeppink2}{rgb}{0.933,0.071,0.537}
\definecolor{deeppink3}{rgb}{0.804,0.063,0.463}
\definecolor{deeppink4}{rgb}{0.545,0.039,0.314}
\definecolor{deepskyblue}{rgb}{0,0.749,1}
\definecolor{deepskyblue1}{rgb}{0,0.749,1}
\definecolor{deepskyblue2}{rgb}{0,0.698,0.933}
\definecolor{deepskyblue3}{rgb}{0,0.604,0.804}
\definecolor{deepskyblue4}{rgb}{0,0.408,0.545}
\definecolor{dimgray}{rgb}{0.412,0.412,0.412}
\definecolor{dimgrey}{rgb}{0.412,0.412,0.412}
\definecolor{dodgerblue}{rgb}{0.118,0.565,1}
\definecolor{dodgerblue1}{rgb}{0.118,0.565,1}
\definecolor{dodgerblue2}{rgb}{0.11,0.525,0.933}
\definecolor{dodgerblue3}{rgb}{0.094,0.455,0.804}
\definecolor{dodgerblue4}{rgb}{0.063,0.306,0.545}
\definecolor{firebrick}{rgb}{0.698,0.133,0.133}
\definecolor{firebrick1}{rgb}{1,0.188,0.188}
\definecolor{firebrick2}{rgb}{0.933,0.173,0.173}
\definecolor{firebrick3}{rgb}{0.804,0.149,0.149}
\definecolor{firebrick4}{rgb}{0.545,0.102,0.102}
\definecolor{floralwhite}{rgb}{1,0.98,0.941}
\definecolor{forestgreen}{rgb}{0.133,0.545,0.133}
\definecolor{gainsboro}{rgb}{0.863,0.863,0.863}
\definecolor{ghostwhite}{rgb}{0.973,0.973,1}
\definecolor{gold1}{rgb}{1,0.843,0}
\definecolor{gold2}{rgb}{0.933,0.788,0}
\definecolor{gold3}{rgb}{0.804,0.678,0}
\definecolor{gold4}{rgb}{0.545,0.459,0}
\definecolor{goldenrod}{rgb}{0.855,0.647,0.125}
\definecolor{goldenrod1}{rgb}{1,0.757,0.145}
\definecolor{goldenrod2}{rgb}{0.933,0.706,0.133}
\definecolor{goldenrod3}{rgb}{0.804,0.608,0.114}
\definecolor{goldenrod4}{rgb}{0.545,0.412,0.078}
\definecolor{gray0}{rgb}{0,0,0}
\definecolor{gray1}{rgb}{0.012,0.012,0.012}
\definecolor{gray10}{rgb}{0.102,0.102,0.102}
\definecolor{gray100}{rgb}{1,1,1}
\definecolor{gray11}{rgb}{0.11,0.11,0.11}
\definecolor{gray12}{rgb}{0.122,0.122,0.122}
\definecolor{gray13}{rgb}{0.129,0.129,0.129}
\definecolor{gray14}{rgb}{0.141,0.141,0.141}
\definecolor{gray15}{rgb}{0.149,0.149,0.149}
\definecolor{gray16}{rgb}{0.161,0.161,0.161}
\definecolor{gray17}{rgb}{0.169,0.169,0.169}
\definecolor{gray18}{rgb}{0.18,0.18,0.18}
\definecolor{gray19}{rgb}{0.188,0.188,0.188}
\definecolor{gray2}{rgb}{0.02,0.02,0.02}
\definecolor{gray20}{rgb}{0.2,0.2,0.2}
\definecolor{gray21}{rgb}{0.212,0.212,0.212}
\definecolor{gray22}{rgb}{0.22,0.22,0.22}
\definecolor{gray23}{rgb}{0.231,0.231,0.231}
\definecolor{gray24}{rgb}{0.239,0.239,0.239}
\definecolor{gray25}{rgb}{0.251,0.251,0.251}
\definecolor{gray26}{rgb}{0.259,0.259,0.259}
\definecolor{gray27}{rgb}{0.271,0.271,0.271}
\definecolor{gray28}{rgb}{0.278,0.278,0.278}
\definecolor{gray29}{rgb}{0.29,0.29,0.29}
\definecolor{gray3}{rgb}{0.031,0.031,0.031}
\definecolor{gray30}{rgb}{0.302,0.302,0.302}
\definecolor{gray31}{rgb}{0.31,0.31,0.31}
\definecolor{gray32}{rgb}{0.322,0.322,0.322}
\definecolor{gray33}{rgb}{0.329,0.329,0.329}
\definecolor{gray34}{rgb}{0.341,0.341,0.341}
\definecolor{gray35}{rgb}{0.349,0.349,0.349}
\definecolor{gray36}{rgb}{0.361,0.361,0.361}
\definecolor{gray37}{rgb}{0.369,0.369,0.369}
\definecolor{gray38}{rgb}{0.38,0.38,0.38}
\definecolor{gray39}{rgb}{0.388,0.388,0.388}
\definecolor{gray4}{rgb}{0.039,0.039,0.039}
\definecolor{gray40}{rgb}{0.4,0.4,0.4}
\definecolor{gray41}{rgb}{0.412,0.412,0.412}
\definecolor{gray42}{rgb}{0.42,0.42,0.42}
\definecolor{gray43}{rgb}{0.431,0.431,0.431}
\definecolor{gray44}{rgb}{0.439,0.439,0.439}
\definecolor{gray45}{rgb}{0.451,0.451,0.451}
\definecolor{gray46}{rgb}{0.459,0.459,0.459}
\definecolor{gray47}{rgb}{0.471,0.471,0.471}
\definecolor{gray48}{rgb}{0.478,0.478,0.478}
\definecolor{gray49}{rgb}{0.49,0.49,0.49}
\definecolor{gray5}{rgb}{0.051,0.051,0.051}
\definecolor{gray50}{rgb}{0.498,0.498,0.498}
\definecolor{gray51}{rgb}{0.51,0.51,0.51}
\definecolor{gray52}{rgb}{0.522,0.522,0.522}
\definecolor{gray53}{rgb}{0.529,0.529,0.529}
\definecolor{gray54}{rgb}{0.541,0.541,0.541}
\definecolor{gray55}{rgb}{0.549,0.549,0.549}
\definecolor{gray56}{rgb}{0.561,0.561,0.561}
\definecolor{gray57}{rgb}{0.569,0.569,0.569}
\definecolor{gray58}{rgb}{0.58,0.58,0.58}
\definecolor{gray59}{rgb}{0.588,0.588,0.588}
\definecolor{gray6}{rgb}{0.059,0.059,0.059}
\definecolor{gray60}{rgb}{0.6,0.6,0.6}
\definecolor{gray61}{rgb}{0.612,0.612,0.612}
\definecolor{gray62}{rgb}{0.62,0.62,0.62}
\definecolor{gray63}{rgb}{0.631,0.631,0.631}
\definecolor{gray64}{rgb}{0.639,0.639,0.639}
\definecolor{gray65}{rgb}{0.651,0.651,0.651}
\definecolor{gray66}{rgb}{0.659,0.659,0.659}
\definecolor{gray67}{rgb}{0.671,0.671,0.671}
\definecolor{gray68}{rgb}{0.678,0.678,0.678}
\definecolor{gray69}{rgb}{0.69,0.69,0.69}
\definecolor{gray7}{rgb}{0.071,0.071,0.071}
\definecolor{gray70}{rgb}{0.702,0.702,0.702}
\definecolor{gray71}{rgb}{0.71,0.71,0.71}
\definecolor{gray72}{rgb}{0.722,0.722,0.722}
\definecolor{gray73}{rgb}{0.729,0.729,0.729}
\definecolor{gray74}{rgb}{0.741,0.741,0.741}
\definecolor{gray75}{rgb}{0.749,0.749,0.749}
\definecolor{gray76}{rgb}{0.761,0.761,0.761}
\definecolor{gray77}{rgb}{0.769,0.769,0.769}
\definecolor{gray78}{rgb}{0.78,0.78,0.78}
\definecolor{gray79}{rgb}{0.788,0.788,0.788}
\definecolor{gray8}{rgb}{0.078,0.078,0.078}
\definecolor{gray80}{rgb}{0.8,0.8,0.8}
\definecolor{gray81}{rgb}{0.812,0.812,0.812}
\definecolor{gray82}{rgb}{0.82,0.82,0.82}
\definecolor{gray83}{rgb}{0.831,0.831,0.831}
\definecolor{gray84}{rgb}{0.839,0.839,0.839}
\definecolor{gray85}{rgb}{0.851,0.851,0.851}
\definecolor{gray86}{rgb}{0.859,0.859,0.859}
\definecolor{gray87}{rgb}{0.871,0.871,0.871}
\definecolor{gray88}{rgb}{0.878,0.878,0.878}
\definecolor{gray89}{rgb}{0.89,0.89,0.89}
\definecolor{gray9}{rgb}{0.09,0.09,0.09}
\definecolor{gray90}{rgb}{0.898,0.898,0.898}
\definecolor{gray91}{rgb}{0.91,0.91,0.91}
\definecolor{gray92}{rgb}{0.922,0.922,0.922}
\definecolor{gray93}{rgb}{0.929,0.929,0.929}
\definecolor{gray94}{rgb}{0.941,0.941,0.941}
\definecolor{gray95}{rgb}{0.949,0.949,0.949}
\definecolor{gray96}{rgb}{0.961,0.961,0.961}
\definecolor{gray97}{rgb}{0.969,0.969,0.969}
\definecolor{gray98}{rgb}{0.98,0.98,0.98}
\definecolor{gray99}{rgb}{0.988,0.988,0.988}
\definecolor{green1}{rgb}{0,1,0}
\definecolor{green2}{rgb}{0,0.933,0}
\definecolor{green3}{rgb}{0,0.804,0}
\definecolor{green4}{rgb}{0,0.545,0}
\definecolor{greenyellow}{rgb}{0.678,1,0.184}
\definecolor{grey}{rgb}{0.745,0.745,0.745}
\definecolor{grey0}{rgb}{0,0,0}
\definecolor{grey1}{rgb}{0.012,0.012,0.012}
\definecolor{grey10}{rgb}{0.102,0.102,0.102}
\definecolor{grey100}{rgb}{1,1,1}
\definecolor{grey11}{rgb}{0.11,0.11,0.11}
\definecolor{grey12}{rgb}{0.122,0.122,0.122}
\definecolor{grey13}{rgb}{0.129,0.129,0.129}
\definecolor{grey14}{rgb}{0.141,0.141,0.141}
\definecolor{grey15}{rgb}{0.149,0.149,0.149}
\definecolor{grey16}{rgb}{0.161,0.161,0.161}
\definecolor{grey17}{rgb}{0.169,0.169,0.169}
\definecolor{grey18}{rgb}{0.18,0.18,0.18}
\definecolor{grey19}{rgb}{0.188,0.188,0.188}
\definecolor{grey2}{rgb}{0.02,0.02,0.02}
\definecolor{grey20}{rgb}{0.2,0.2,0.2}
\definecolor{grey21}{rgb}{0.212,0.212,0.212}
\definecolor{grey22}{rgb}{0.22,0.22,0.22}
\definecolor{grey23}{rgb}{0.231,0.231,0.231}
\definecolor{grey24}{rgb}{0.239,0.239,0.239}
\definecolor{grey25}{rgb}{0.251,0.251,0.251}
\definecolor{grey26}{rgb}{0.259,0.259,0.259}
\definecolor{grey27}{rgb}{0.271,0.271,0.271}
\definecolor{grey28}{rgb}{0.278,0.278,0.278}
\definecolor{grey29}{rgb}{0.29,0.29,0.29}
\definecolor{grey3}{rgb}{0.031,0.031,0.031}
\definecolor{grey30}{rgb}{0.302,0.302,0.302}
\definecolor{grey31}{rgb}{0.31,0.31,0.31}
\definecolor{grey32}{rgb}{0.322,0.322,0.322}
\definecolor{grey33}{rgb}{0.329,0.329,0.329}
\definecolor{grey34}{rgb}{0.341,0.341,0.341}
\definecolor{grey35}{rgb}{0.349,0.349,0.349}
\definecolor{grey36}{rgb}{0.361,0.361,0.361}
\definecolor{grey37}{rgb}{0.369,0.369,0.369}
\definecolor{grey38}{rgb}{0.38,0.38,0.38}
\definecolor{grey39}{rgb}{0.388,0.388,0.388}
\definecolor{grey4}{rgb}{0.039,0.039,0.039}
\definecolor{grey40}{rgb}{0.4,0.4,0.4}
\definecolor{grey41}{rgb}{0.412,0.412,0.412}
\definecolor{grey42}{rgb}{0.42,0.42,0.42}
\definecolor{grey43}{rgb}{0.431,0.431,0.431}
\definecolor{grey44}{rgb}{0.439,0.439,0.439}
\definecolor{grey45}{rgb}{0.451,0.451,0.451}
\definecolor{grey46}{rgb}{0.459,0.459,0.459}
\definecolor{grey47}{rgb}{0.471,0.471,0.471}
\definecolor{grey48}{rgb}{0.478,0.478,0.478}
\definecolor{grey49}{rgb}{0.49,0.49,0.49}
\definecolor{grey5}{rgb}{0.051,0.051,0.051}
\definecolor{grey50}{rgb}{0.498,0.498,0.498}
\definecolor{grey51}{rgb}{0.51,0.51,0.51}
\definecolor{grey52}{rgb}{0.522,0.522,0.522}
\definecolor{grey53}{rgb}{0.529,0.529,0.529}
\definecolor{grey54}{rgb}{0.541,0.541,0.541}
\definecolor{grey55}{rgb}{0.549,0.549,0.549}
\definecolor{grey56}{rgb}{0.561,0.561,0.561}
\definecolor{grey57}{rgb}{0.569,0.569,0.569}
\definecolor{grey58}{rgb}{0.58,0.58,0.58}
\definecolor{grey59}{rgb}{0.588,0.588,0.588}
\definecolor{grey6}{rgb}{0.059,0.059,0.059}
\definecolor{grey60}{rgb}{0.6,0.6,0.6}
\definecolor{grey61}{rgb}{0.612,0.612,0.612}
\definecolor{grey62}{rgb}{0.62,0.62,0.62}
\definecolor{grey63}{rgb}{0.631,0.631,0.631}
\definecolor{grey64}{rgb}{0.639,0.639,0.639}
\definecolor{grey65}{rgb}{0.651,0.651,0.651}
\definecolor{grey66}{rgb}{0.659,0.659,0.659}
\definecolor{grey67}{rgb}{0.671,0.671,0.671}
\definecolor{grey68}{rgb}{0.678,0.678,0.678}
\definecolor{grey69}{rgb}{0.69,0.69,0.69}
\definecolor{grey7}{rgb}{0.071,0.071,0.071}
\definecolor{grey70}{rgb}{0.702,0.702,0.702}
\definecolor{grey71}{rgb}{0.71,0.71,0.71}
\definecolor{grey72}{rgb}{0.722,0.722,0.722}
\definecolor{grey73}{rgb}{0.729,0.729,0.729}
\definecolor{grey74}{rgb}{0.741,0.741,0.741}
\definecolor{grey75}{rgb}{0.749,0.749,0.749}
\definecolor{grey76}{rgb}{0.761,0.761,0.761}
\definecolor{grey77}{rgb}{0.769,0.769,0.769}
\definecolor{grey78}{rgb}{0.78,0.78,0.78}
\definecolor{grey79}{rgb}{0.788,0.788,0.788}
\definecolor{grey8}{rgb}{0.078,0.078,0.078}
\definecolor{grey80}{rgb}{0.8,0.8,0.8}
\definecolor{grey81}{rgb}{0.812,0.812,0.812}
\definecolor{grey82}{rgb}{0.82,0.82,0.82}
\definecolor{grey83}{rgb}{0.831,0.831,0.831}
\definecolor{grey84}{rgb}{0.839,0.839,0.839}
\definecolor{grey85}{rgb}{0.851,0.851,0.851}
\definecolor{grey86}{rgb}{0.859,0.859,0.859}
\definecolor{grey87}{rgb}{0.871,0.871,0.871}
\definecolor{grey88}{rgb}{0.878,0.878,0.878}
\definecolor{grey89}{rgb}{0.89,0.89,0.89}
\definecolor{grey9}{rgb}{0.09,0.09,0.09}
\definecolor{grey90}{rgb}{0.898,0.898,0.898}
\definecolor{grey91}{rgb}{0.91,0.91,0.91}
\definecolor{grey92}{rgb}{0.922,0.922,0.922}
\definecolor{grey93}{rgb}{0.929,0.929,0.929}
\definecolor{grey94}{rgb}{0.941,0.941,0.941}
\definecolor{grey95}{rgb}{0.949,0.949,0.949}
\definecolor{grey96}{rgb}{0.961,0.961,0.961}
\definecolor{grey97}{rgb}{0.969,0.969,0.969}
\definecolor{grey98}{rgb}{0.98,0.98,0.98}
\definecolor{grey99}{rgb}{0.988,0.988,0.988}
\definecolor{honeydew}{rgb}{0.941,1,0.941}
\definecolor{honeydew1}{rgb}{0.941,1,0.941}
\definecolor{honeydew2}{rgb}{0.878,0.933,0.878}
\definecolor{honeydew3}{rgb}{0.757,0.804,0.757}
\definecolor{honeydew4}{rgb}{0.514,0.545,0.514}
\definecolor{hotpink}{rgb}{1,0.412,0.706}
\definecolor{hotpink1}{rgb}{1,0.431,0.706}
\definecolor{hotpink2}{rgb}{0.933,0.416,0.655}
\definecolor{hotpink3}{rgb}{0.804,0.376,0.565}
\definecolor{hotpink4}{rgb}{0.545,0.227,0.384}
\definecolor{indianred}{rgb}{0.804,0.361,0.361}
\definecolor{indianred1}{rgb}{1,0.416,0.416}
\definecolor{indianred2}{rgb}{0.933,0.388,0.388}
\definecolor{indianred3}{rgb}{0.804,0.333,0.333}
\definecolor{indianred4}{rgb}{0.545,0.227,0.227}
\definecolor{ivory}{rgb}{1,1,0.941}
\definecolor{ivory1}{rgb}{1,1,0.941}
\definecolor{ivory2}{rgb}{0.933,0.933,0.878}
\definecolor{ivory3}{rgb}{0.804,0.804,0.757}
\definecolor{ivory4}{rgb}{0.545,0.545,0.514}
\definecolor{khaki}{rgb}{0.941,0.902,0.549}
\definecolor{khaki1}{rgb}{1,0.965,0.561}
\definecolor{khaki2}{rgb}{0.933,0.902,0.522}
\definecolor{khaki3}{rgb}{0.804,0.776,0.451}
\definecolor{khaki4}{rgb}{0.545,0.525,0.306}
\definecolor{lavender}{rgb}{0.902,0.902,0.98}
\definecolor{lavenderblush}{rgb}{1,0.941,0.961}
\definecolor{lavenderblush1}{rgb}{1,0.941,0.961}
\definecolor{lavenderblush2}{rgb}{0.933,0.878,0.898}
\definecolor{lavenderblush3}{rgb}{0.804,0.757,0.773}
\definecolor{lavenderblush4}{rgb}{0.545,0.514,0.525}
\definecolor{lawngreen}{rgb}{0.486,0.988,0}
\definecolor{lemonchiffon}{rgb}{1,0.98,0.804}
\definecolor{lemonchiffon1}{rgb}{1,0.98,0.804}
\definecolor{lemonchiffon2}{rgb}{0.933,0.914,0.749}
\definecolor{lemonchiffon3}{rgb}{0.804,0.788,0.647}
\definecolor{lemonchiffon4}{rgb}{0.545,0.537,0.439}
\definecolor{lightblue1}{rgb}{0.749,0.937,1}
\definecolor{lightblue2}{rgb}{0.698,0.875,0.933}
\definecolor{lightblue3}{rgb}{0.604,0.753,0.804}
\definecolor{lightblue4}{rgb}{0.408,0.514,0.545}
\definecolor{lightcoral}{rgb}{0.941,0.502,0.502}
\definecolor{lightcyan1}{rgb}{0.878,1,1}
\definecolor{lightcyan2}{rgb}{0.82,0.933,0.933}
\definecolor{lightcyan3}{rgb}{0.706,0.804,0.804}
\definecolor{lightcyan4}{rgb}{0.478,0.545,0.545}
\definecolor{lightgoldenrod}{rgb}{0.933,0.867,0.51}
\definecolor{lightgoldenrod1}{rgb}{1,0.925,0.545}
\definecolor{lightgoldenrod2}{rgb}{0.933,0.863,0.51}
\definecolor{lightgoldenrod3}{rgb}{0.804,0.745,0.439}
\definecolor{lightgoldenrod4}{rgb}{0.545,0.506,0.298}
\definecolor{lightgoldenrodyellow}{rgb}{0.98,0.98,0.824}
\definecolor{lightgrey}{rgb}{0.827,0.827,0.827}
\definecolor{lightpink}{rgb}{1,0.714,0.757}
\definecolor{lightpink1}{rgb}{1,0.682,0.725}
\definecolor{lightpink2}{rgb}{0.933,0.635,0.678}
\definecolor{lightpink3}{rgb}{0.804,0.549,0.584}
\definecolor{lightpink4}{rgb}{0.545,0.373,0.396}
\definecolor{lightsalmon}{rgb}{1,0.627,0.478}
\definecolor{lightsalmon1}{rgb}{1,0.627,0.478}
\definecolor{lightsalmon2}{rgb}{0.933,0.584,0.447}
\definecolor{lightsalmon3}{rgb}{0.804,0.506,0.384}
\definecolor{lightsalmon4}{rgb}{0.545,0.341,0.259}
\definecolor{lightseagreen}{rgb}{0.125,0.698,0.667}
\definecolor{lightskyblue}{rgb}{0.529,0.808,0.98}
\definecolor{lightskyblue1}{rgb}{0.69,0.886,1}
\definecolor{lightskyblue2}{rgb}{0.643,0.827,0.933}
\definecolor{lightskyblue3}{rgb}{0.553,0.714,0.804}
\definecolor{lightskyblue4}{rgb}{0.376,0.482,0.545}
\definecolor{lightslateblue}{rgb}{0.518,0.439,1}
\definecolor{lightslategray}{rgb}{0.467,0.533,0.6}
\definecolor{lightslategrey}{rgb}{0.467,0.533,0.6}
\definecolor{lightsteelblue}{rgb}{0.69,0.769,0.871}
\definecolor{lightsteelblue1}{rgb}{0.792,0.882,1}
\definecolor{lightsteelblue2}{rgb}{0.737,0.824,0.933}
\definecolor{lightsteelblue3}{rgb}{0.635,0.71,0.804}
\definecolor{lightsteelblue4}{rgb}{0.431,0.482,0.545}
\definecolor{lightyellow1}{rgb}{1,1,0.878}
\definecolor{lightyellow2}{rgb}{0.933,0.933,0.82}
\definecolor{lightyellow3}{rgb}{0.804,0.804,0.706}
\definecolor{lightyellow4}{rgb}{0.545,0.545,0.478}
\definecolor{limegreen}{rgb}{0.196,0.804,0.196}
\definecolor{linen}{rgb}{0.98,0.941,0.902}
\definecolor{magenta}{rgb}{1,0,1}
\definecolor{magenta1}{rgb}{1,0,1}
\definecolor{magenta2}{rgb}{0.933,0,0.933}
\definecolor{magenta3}{rgb}{0.804,0,0.804}
\definecolor{magenta4}{rgb}{0.545,0,0.545}
\definecolor{maroon}{rgb}{0.69,0.188,0.376}
\definecolor{maroon1}{rgb}{1,0.204,0.702}
\definecolor{maroon2}{rgb}{0.933,0.188,0.655}
\definecolor{maroon3}{rgb}{0.804,0.161,0.565}
\definecolor{maroon4}{rgb}{0.545,0.11,0.384}
\definecolor{mediumaquamarine}{rgb}{0.4,0.804,0.667}
\definecolor{mediumblue}{rgb}{0,0,0.804}
\definecolor{mediumorchid}{rgb}{0.729,0.333,0.827}
\definecolor{mediumorchid1}{rgb}{0.878,0.4,1}
\definecolor{mediumorchid2}{rgb}{0.82,0.373,0.933}
\definecolor{mediumorchid3}{rgb}{0.706,0.322,0.804}
\definecolor{mediumorchid4}{rgb}{0.478,0.216,0.545}
\definecolor{mediumpurple}{rgb}{0.576,0.439,0.859}
\definecolor{mediumpurple1}{rgb}{0.671,0.51,1}
\definecolor{mediumpurple2}{rgb}{0.624,0.475,0.933}
\definecolor{mediumpurple3}{rgb}{0.537,0.408,0.804}
\definecolor{mediumpurple4}{rgb}{0.365,0.278,0.545}
\definecolor{mediumseagreen}{rgb}{0.235,0.702,0.443}
\definecolor{mediumslateblue}{rgb}{0.482,0.408,0.933}
\definecolor{mediumspringgreen}{rgb}{0,0.98,0.604}
\definecolor{mediumturquoise}{rgb}{0.282,0.82,0.8}
\definecolor{mediumvioletred}{rgb}{0.78,0.082,0.522}
\definecolor{midnightblue}{rgb}{0.098,0.098,0.439}
\definecolor{mintcream}{rgb}{0.961,1,0.98}
\definecolor{mistyrose}{rgb}{1,0.894,0.882}
\definecolor{mistyrose1}{rgb}{1,0.894,0.882}
\definecolor{mistyrose2}{rgb}{0.933,0.835,0.824}
\definecolor{mistyrose3}{rgb}{0.804,0.718,0.71}
\definecolor{mistyrose4}{rgb}{0.545,0.49,0.482}
\definecolor{moccasin}{rgb}{1,0.894,0.71}
\definecolor{navajowhite}{rgb}{1,0.871,0.678}
\definecolor{navajowhite1}{rgb}{1,0.871,0.678}
\definecolor{navajowhite2}{rgb}{0.933,0.812,0.631}
\definecolor{navajowhite3}{rgb}{0.804,0.702,0.545}
\definecolor{navajowhite4}{rgb}{0.545,0.475,0.369}
\definecolor{navyblue}{rgb}{0,0,0.502}
\definecolor{oldlace}{rgb}{0.992,0.961,0.902}
\definecolor{olivedrab}{rgb}{0.42,0.557,0.137}
\definecolor{olivedrab1}{rgb}{0.753,1,0.243}
\definecolor{olivedrab2}{rgb}{0.702,0.933,0.227}
\definecolor{olivedrab3}{rgb}{0.604,0.804,0.196}
\definecolor{olivedrab4}{rgb}{0.412,0.545,0.133}
\definecolor{orange1}{rgb}{1,0.647,0}
\definecolor{orange2}{rgb}{0.933,0.604,0}
\definecolor{orange3}{rgb}{0.804,0.522,0}
\definecolor{orange4}{rgb}{0.545,0.353,0}
\definecolor{orangered}{rgb}{1,0.271,0}
\definecolor{orangered1}{rgb}{1,0.271,0}
\definecolor{orangered2}{rgb}{0.933,0.251,0}
\definecolor{orangered3}{rgb}{0.804,0.216,0}
\definecolor{orangered4}{rgb}{0.545,0.145,0}
\definecolor{orchid}{rgb}{0.855,0.439,0.839}
\definecolor{orchid1}{rgb}{1,0.514,0.98}
\definecolor{orchid2}{rgb}{0.933,0.478,0.914}
\definecolor{orchid3}{rgb}{0.804,0.412,0.788}
\definecolor{orchid4}{rgb}{0.545,0.278,0.537}
\definecolor{palegoldenrod}{rgb}{0.933,0.91,0.667}
\definecolor{palegreen}{rgb}{0.596,0.984,0.596}
\definecolor{palegreen1}{rgb}{0.604,1,0.604}
\definecolor{palegreen2}{rgb}{0.565,0.933,0.565}
\definecolor{palegreen3}{rgb}{0.486,0.804,0.486}
\definecolor{palegreen4}{rgb}{0.329,0.545,0.329}
\definecolor{paleturquoise}{rgb}{0.686,0.933,0.933}
\definecolor{paleturquoise1}{rgb}{0.733,1,1}
\definecolor{paleturquoise2}{rgb}{0.682,0.933,0.933}
\definecolor{paleturquoise3}{rgb}{0.588,0.804,0.804}
\definecolor{paleturquoise4}{rgb}{0.4,0.545,0.545}
\definecolor{palevioletred}{rgb}{0.859,0.439,0.576}
\definecolor{palevioletred1}{rgb}{1,0.51,0.671}
\definecolor{palevioletred2}{rgb}{0.933,0.475,0.624}
\definecolor{palevioletred3}{rgb}{0.804,0.408,0.537}
\definecolor{palevioletred4}{rgb}{0.545,0.278,0.365}
\definecolor{papayawhip}{rgb}{1,0.937,0.835}
\definecolor{peachpuff}{rgb}{1,0.855,0.725}
\definecolor{peachpuff1}{rgb}{1,0.855,0.725}
\definecolor{peachpuff2}{rgb}{0.933,0.796,0.678}
\definecolor{peachpuff3}{rgb}{0.804,0.686,0.584}
\definecolor{peachpuff4}{rgb}{0.545,0.467,0.396}
\definecolor{peru}{rgb}{0.804,0.522,0.247}
\definecolor{pink1}{rgb}{1,0.71,0.773}
\definecolor{pink2}{rgb}{0.933,0.663,0.722}
\definecolor{pink3}{rgb}{0.804,0.569,0.62}
\definecolor{pink4}{rgb}{0.545,0.388,0.424}
\definecolor{plum}{rgb}{0.867,0.627,0.867}
\definecolor{plum1}{rgb}{1,0.733,1}
\definecolor{plum2}{rgb}{0.933,0.682,0.933}
\definecolor{plum3}{rgb}{0.804,0.588,0.804}
\definecolor{plum4}{rgb}{0.545,0.4,0.545}
\definecolor{powderblue}{rgb}{0.69,0.878,0.902}
\definecolor{purple1}{rgb}{0.608,0.188,1}
\definecolor{purple2}{rgb}{0.569,0.173,0.933}
\definecolor{purple3}{rgb}{0.49,0.149,0.804}
\definecolor{purple4}{rgb}{0.333,0.102,0.545}
\definecolor{red1}{rgb}{1,0,0}
\definecolor{red2}{rgb}{0.933,0,0}
\definecolor{red3}{rgb}{0.804,0,0}
\definecolor{red4}{rgb}{0.545,0,0}
\definecolor{rosybrown}{rgb}{0.737,0.561,0.561}
\definecolor{rosybrown1}{rgb}{1,0.757,0.757}
\definecolor{rosybrown2}{rgb}{0.933,0.706,0.706}
\definecolor{rosybrown3}{rgb}{0.804,0.608,0.608}
\definecolor{rosybrown4}{rgb}{0.545,0.412,0.412}
\definecolor{royalblue}{rgb}{0.255,0.412,0.882}
\definecolor{royalblue1}{rgb}{0.282,0.463,1}
\definecolor{royalblue2}{rgb}{0.263,0.431,0.933}
\definecolor{royalblue3}{rgb}{0.227,0.373,0.804}
\definecolor{royalblue4}{rgb}{0.153,0.251,0.545}
\definecolor{saddlebrown}{rgb}{0.545,0.271,0.075}
\definecolor{salmon}{rgb}{0.98,0.502,0.447}
\definecolor{salmon1}{rgb}{1,0.549,0.412}
\definecolor{salmon2}{rgb}{0.933,0.51,0.384}
\definecolor{salmon3}{rgb}{0.804,0.439,0.329}
\definecolor{salmon4}{rgb}{0.545,0.298,0.224}
\definecolor{sandybrown}{rgb}{0.957,0.643,0.376}
\definecolor{seagreen1}{rgb}{0.329,1,0.624}
\definecolor{seagreen2}{rgb}{0.306,0.933,0.58}
\definecolor{seagreen3}{rgb}{0.263,0.804,0.502}
\definecolor{seagreen4}{rgb}{0.18,0.545,0.341}
\definecolor{seashell}{rgb}{1,0.961,0.933}
\definecolor{seashell1}{rgb}{1,0.961,0.933}
\definecolor{seashell2}{rgb}{0.933,0.898,0.871}
\definecolor{seashell3}{rgb}{0.804,0.773,0.749}
\definecolor{seashell4}{rgb}{0.545,0.525,0.51}
\definecolor{sienna}{rgb}{0.627,0.322,0.176}
\definecolor{sienna1}{rgb}{1,0.51,0.278}
\definecolor{sienna2}{rgb}{0.933,0.475,0.259}
\definecolor{sienna3}{rgb}{0.804,0.408,0.224}
\definecolor{sienna4}{rgb}{0.545,0.278,0.149}
\definecolor{skyblue}{rgb}{0.529,0.808,0.922}
\definecolor{skyblue1}{rgb}{0.529,0.808,1}
\definecolor{skyblue2}{rgb}{0.494,0.753,0.933}
\definecolor{skyblue3}{rgb}{0.424,0.651,0.804}
\definecolor{skyblue4}{rgb}{0.29,0.439,0.545}
\definecolor{slateblue}{rgb}{0.416,0.353,0.804}
\definecolor{slateblue1}{rgb}{0.514,0.435,1}
\definecolor{slateblue2}{rgb}{0.478,0.404,0.933}
\definecolor{slateblue3}{rgb}{0.412,0.349,0.804}
\definecolor{slateblue4}{rgb}{0.278,0.235,0.545}
\definecolor{slategray}{rgb}{0.439,0.502,0.565}
\definecolor{slategray1}{rgb}{0.776,0.886,1}
\definecolor{slategray2}{rgb}{0.725,0.827,0.933}
\definecolor{slategray3}{rgb}{0.624,0.714,0.804}
\definecolor{slategray4}{rgb}{0.424,0.482,0.545}
\definecolor{slategrey}{rgb}{0.439,0.502,0.565}
\definecolor{snow}{rgb}{1,0.98,0.98}
\definecolor{snow1}{rgb}{1,0.98,0.98}
\definecolor{snow2}{rgb}{0.933,0.914,0.914}
\definecolor{snow3}{rgb}{0.804,0.788,0.788}
\definecolor{snow4}{rgb}{0.545,0.537,0.537}
\definecolor{springgreen}{rgb}{0,1,0.498}
\definecolor{springgreen1}{rgb}{0,1,0.498}
\definecolor{springgreen2}{rgb}{0,0.933,0.463}
\definecolor{springgreen3}{rgb}{0,0.804,0.4}
\definecolor{springgreen4}{rgb}{0,0.545,0.271}
\definecolor{steelblue}{rgb}{0.275,0.51,0.706}
\definecolor{steelblue1}{rgb}{0.388,0.722,1}
\definecolor{steelblue2}{rgb}{0.361,0.675,0.933}
\definecolor{steelblue3}{rgb}{0.31,0.58,0.804}
\definecolor{steelblue4}{rgb}{0.212,0.392,0.545}
\definecolor{tan}{rgb}{0.824,0.706,0.549}
\definecolor{tan1}{rgb}{1,0.647,0.31}
\definecolor{tan2}{rgb}{0.933,0.604,0.286}
\definecolor{tan3}{rgb}{0.804,0.522,0.247}
\definecolor{tan4}{rgb}{0.545,0.353,0.169}
\definecolor{thistle}{rgb}{0.847,0.749,0.847}
\definecolor{thistle1}{rgb}{1,0.882,1}
\definecolor{thistle2}{rgb}{0.933,0.824,0.933}
\definecolor{thistle3}{rgb}{0.804,0.71,0.804}
\definecolor{thistle4}{rgb}{0.545,0.482,0.545}
\definecolor{tomato}{rgb}{1,0.388,0.278}
\definecolor{tomato1}{rgb}{1,0.388,0.278}
\definecolor{tomato2}{rgb}{0.933,0.361,0.259}
\definecolor{tomato3}{rgb}{0.804,0.31,0.224}
\definecolor{tomato4}{rgb}{0.545,0.212,0.149}
\definecolor{turquoise1}{rgb}{0,0.961,1}
\definecolor{turquoise2}{rgb}{0,0.898,0.933}
\definecolor{turquoise3}{rgb}{0,0.773,0.804}
\definecolor{turquoise4}{rgb}{0,0.525,0.545}
\definecolor{violetred}{rgb}{0.816,0.125,0.565}
\definecolor{violetred1}{rgb}{1,0.243,0.588}
\definecolor{violetred2}{rgb}{0.933,0.227,0.549}
\definecolor{violetred3}{rgb}{0.804,0.196,0.471}
\definecolor{violetred4}{rgb}{0.545,0.133,0.322}
\definecolor{wheat}{rgb}{0.961,0.871,0.702}
\definecolor{wheat1}{rgb}{1,0.906,0.729}
\definecolor{wheat2}{rgb}{0.933,0.847,0.682}
\definecolor{wheat3}{rgb}{0.804,0.729,0.588}
\definecolor{wheat4}{rgb}{0.545,0.494,0.4}
\definecolor{whitesmoke}{rgb}{0.961,0.961,0.961}
\definecolor{yellow1}{rgb}{1,1,0}
\definecolor{yellow2}{rgb}{0.933,0.933,0}
\definecolor{yellow3}{rgb}{0.804,0.804,0}
\definecolor{yellow4}{rgb}{0.545,0.545,0}
\definecolor{yellowgreen}{rgb}{0.604,0.804,0.196}
\scalebox{1.1}{

}
\caption{The Gaifman graph of ${\mathfrak{A}_2}.$}
\labels{figure_boundariedgraph3}
\end{figure}
Notice that the fact that
$K_1^{\bf a}[\partial_{\mathfrak{K}_1} (Z) \cup V_L ({\bf a})]\cup F_1$
is strongly isomorphic to
$K_2^{\bf a}[\partial_{\mathfrak{K}_2} (Z') \cup V_L ({\bf a})]\cup F_2$
with respect to $({\cal V}_1, {\cal V}_2)$
implies that the $h$-boundaried $τ'$-structures
$\big(\mathfrak{A}_2,R_2 ',{\bf W}_{{q}}^{(2)},\varnothing^l, \tilde{X}_2, {\bf b}_2\big)$
and $\big(\mathfrak{A}_1,R_1 ',{\bf W}_{{q}}^{(1)},\varnothing^l, \tilde{X}_1, {\bf b}_1\big)$
are compatible.
Thus, given that
 $\big(\mathfrak{A}_2,R_2 ',{\bf W}_{{q}}^{(2)},\varnothing^l, \tilde{X}_2, {\bf b}_2\big)\models \bar{φ},$ we have that  $\big(\mathfrak{A}_2,R_2 ',{\bf W}_{{q}}^{(2)},\varnothing^l, \tilde{X}_2, {\bf b}_2\big)$ and $\big(\mathfrak{A}_1,R_1 ',{\bf W}_{{q}}^{(1)},\varnothing^l, \tilde{X}_1, {\bf b}_1\big)$ are
 $(θ^{\sf out}_q,h)$-equivalent.
 Therefore,
 \begin{eqnarray}
\big(\mathfrak{A}_1,R_1 ',{\bf W}_{{q}}^{(1)},\varnothing^l, \tilde{X}_1, {\bf b}_1\big)\models \bar{φ}\iff
\big(\mathfrak{A}_2,R_2 ',{\bf W}_{{q}}^{(2)}, \varnothing^l, \tilde{X}_2, {\bf b}_2\big)\models \bar{φ}.\labels{@photographic}
\end{eqnarray}

At this point, to give some intuition, we underline that even if $(\mathfrak{A}_2,R_2 ',{\bf W}_{{q}}^{(2)}, \varnothing^l,\tilde{X}_2, {\bf b}_2)$ and $(\mathfrak{A}_1,R_1 ',{\bf W}_{{q}}^{(1)},\varnothing^l, \tilde{X}_1, {\bf b}_1)$ are
 $(θ^{\sf out}_q,h)$-equivalent, we did not yet provide
 a boundaried structure that is a {\sl substructure} of $(\mathfrak{A},R,{\bf W}_{{q}},\varnothing^l, X)$ and that
is $(θ^{\sf out}_q,h)$-equivalent to
$(\mathfrak{A}^\star,R_1^{'\star},{\bf W}_{{q}}^{(1)}, \varnothing^l,\tilde{X}^\star_1, {\bf b}_1^\star).$
To find such a substructure $(\mathfrak{A}^{\star},R_2^{'\star},{\bf W}_{{q}}^{(2)},\varnothing^l, \tilde{X}^{\star}_2, {\bf b}_2^\star)$ of  $(\mathfrak{A},R,{\bf W}_{{q}},\varnothing^l,X),$
we have to ``shift'' from
$(\mathfrak{A}_2,R_2 ',{\bf W}_{{q}}^{(2)},\varnothing^l, \tilde{X}_2, {\bf b}_2)$
to $(\mathfrak{A}_2^{\star},R_2^{'\star},{\bf W}_{{q}}^{(2)},\varnothing^l, \tilde{X}^{\star}_2, {\bf b}_2^\star),$
by replacing $V(F_2)$ with $V(F^\star),$ and ``extending'' $R_2 '$ to $R_2^{'\star}$
so to contain all vertices in $R_1^{' \star} \setminus Y.$
This substructure $(\mathfrak{A}^{\star},R_2^{'\star},{\bf W}_{{q}}^{(2)},\varnothing^l, \tilde{X}^{\star}_2, {\bf b}_2^\star)$ will replace $(\mathfrak{A}^\star, R_1^{' \star},{\bf W}_{{q}}^{(1)} , \varnothing^l,\tilde{X}^\star_1, {\bf b}_1^\star)$ in~\eqref{@electronically},
thus providing a set $X'\subseteq Z'$ such that
$\partial_{\mathfrak{K}_2} (Z')\subseteq X'$ and
$(\mathfrak{A},R,{\bf W}_{{q}}^{(1)},\varnothing^l,X)\models θ^{\sf out}_q \iff (\mathfrak{A},R\setminus Y,{\bf W}_q^{(1)},\varnothing^l, X_{\rm out}\cup X')\models θ^{\sf out}_q.$

\myskip\paragraph{Defining a substructure of the initial structure with a different boundary.}
Let us now define this substructure $(\mathfrak{A}^{\star},R_2^{'\star},{\bf W}_{{q}}^{(2)}, \varnothing^l,\tilde{X}^{\star}_2, {\bf b}_2^\star)$ from $\big(\mathfrak{A}_2,R_2 ',{\bf W}_{{q}}^{(2)}, \varnothing^l,\tilde{X}_2, {\bf b}_2\big).$
We set ${\bf b}_2^\star$ to be
the tuple obtained from ${\bf b}_2$ after replacing each $v∈ V(F_2)$ with the corresponding $u∈ V(F^\star).$ See~\autoref{figure_boundariedgraph4} for an example.
We stress that the $h$-boundaried $τ$-structure $(\mathfrak{A}^{\star}, {\bf b}_2^\star)$ of~\autoref{figure_boundariedgraph4}
can also be defined as the one obtained from $(\mathfrak{A}^\star, {\bf b}_1^\star)$ of~\autoref{figure_boundariedgraph1} after replacing $\partial_{\mathfrak{K}_1}(Z)$ with $\partial_{\mathfrak{K}_2}(Z')$ in the boundary.
Also, let $R_2^{'\star} = R_1^{' \star}\setminus Y,$ and $\tilde{X}^{\star}_2 = (\tilde{X}_2\setminus V(F_2))\cup V(F^\star).$

\begin{figure}[ht]
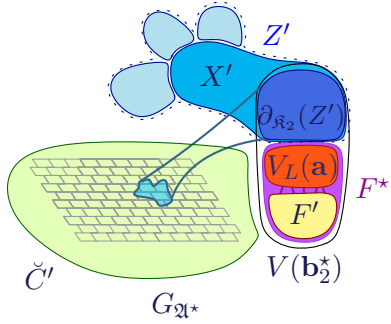

\centering
\tikzstyle{ipe stylesheet} = [
  ipe import,
  even odd rule,
  line join=round,
  line cap=butt,
  ipe pen normal/.style={line width=0.4},
  ipe pen heavier/.style={line width=0.8},
  ipe pen fat/.style={line width=1.2},
  ipe pen ultrafat/.style={line width=2},
  ipe pen normal,
  ipe mark normal/.style={ipe mark scale=3},
  ipe mark large/.style={ipe mark scale=5},
  ipe mark small/.style={ipe mark scale=2},
  ipe mark tiny/.style={ipe mark scale=1.1},
  ipe mark normal,
  /pgf/arrow keys/.cd,
  ipe arrow normal/.style={scale=7},
  ipe arrow large/.style={scale=10},
  ipe arrow small/.style={scale=5},
  ipe arrow tiny/.style={scale=3},
  ipe arrow normal,
  /tikz/.cd,
  ipe arrows, 
  <->/.tip = ipe normal,
  ipe dash normal/.style={dash pattern=},
  ipe dash dotted/.style={dash pattern=on 1bp off 3bp},
  ipe dash dashed/.style={dash pattern=on 4bp off 4bp},
  ipe dash dash dotted/.style={dash pattern=on 4bp off 2bp on 1bp off 2bp},
  ipe dash dash dot dotted/.style={dash pattern=on 4bp off 2bp on 1bp off 2bp on 1bp off 2bp},
  ipe dash normal,
  ipe node/.append style={font=\normalsize},
  ipe stretch normal/.style={ipe node stretch=1},
  ipe stretch normal,
  ipe opacity 10/.style={opacity=0.1},
  ipe opacity 30/.style={opacity=0.3},
  ipe opacity 50/.style={opacity=0.5},
  ipe opacity 75/.style={opacity=0.75},
  ipe opacity opaque/.style={opacity=1},
  ipe opacity opaque,
]
\definecolor{black}{rgb}{0,0,0}
\definecolor{white}{rgb}{1,1,1}
\definecolor{red}{rgb}{1,0,0}
\definecolor{blue}{rgb}{0,0,1}
\definecolor{green}{rgb}{0,1,0}
\definecolor{yellow}{rgb}{1,1,0}
\definecolor{orange}{rgb}{1,0.647,0}
\definecolor{gold}{rgb}{1,0.843,0}
\definecolor{purple}{rgb}{0.627,0.125,0.941}
\definecolor{gray}{rgb}{0.745,0.745,0.745}
\definecolor{brown}{rgb}{0.647,0.165,0.165}
\definecolor{navy}{rgb}{0,0,0.502}
\definecolor{pink}{rgb}{1,0.753,0.796}
\definecolor{seagreen}{rgb}{0.18,0.545,0.341}
\definecolor{turquoise}{rgb}{0.251,0.878,0.816}
\definecolor{violet}{rgb}{0.933,0.51,0.933}
\definecolor{darkblue}{rgb}{0,0,0.545}
\definecolor{darkcyan}{rgb}{0,0.545,0.545}
\definecolor{darkgray}{rgb}{0.663,0.663,0.663}
\definecolor{darkgreen}{rgb}{0,0.392,0}
\definecolor{darkmagenta}{rgb}{0.545,0,0.545}
\definecolor{darkorange}{rgb}{1,0.549,0}
\definecolor{darkred}{rgb}{0.545,0,0}
\definecolor{lightblue}{rgb}{0.678,0.847,0.902}
\definecolor{lightcyan}{rgb}{0.878,1,1}
\definecolor{lightgray}{rgb}{0.827,0.827,0.827}
\definecolor{lightgreen}{rgb}{0.565,0.933,0.565}
\definecolor{lightyellow}{rgb}{1,1,0.878}
\definecolor{aliceblue}{rgb}{0.941,0.973,1}
\definecolor{antiquewhite}{rgb}{0.98,0.922,0.843}
\definecolor{antiquewhite1}{rgb}{1,0.937,0.859}
\definecolor{antiquewhite2}{rgb}{0.933,0.875,0.8}
\definecolor{antiquewhite3}{rgb}{0.804,0.753,0.69}
\definecolor{antiquewhite4}{rgb}{0.545,0.514,0.471}
\definecolor{aquamarine}{rgb}{0.498,1,0.831}
\definecolor{aquamarine1}{rgb}{0.498,1,0.831}
\definecolor{aquamarine2}{rgb}{0.463,0.933,0.776}
\definecolor{aquamarine3}{rgb}{0.4,0.804,0.667}
\definecolor{aquamarine4}{rgb}{0.271,0.545,0.455}
\definecolor{azure}{rgb}{0.941,1,1}
\definecolor{azure1}{rgb}{0.941,1,1}
\definecolor{azure2}{rgb}{0.878,0.933,0.933}
\definecolor{azure3}{rgb}{0.757,0.804,0.804}
\definecolor{azure4}{rgb}{0.514,0.545,0.545}
\definecolor{beige}{rgb}{0.961,0.961,0.863}
\definecolor{bisque}{rgb}{1,0.894,0.769}
\definecolor{bisque1}{rgb}{1,0.894,0.769}
\definecolor{bisque2}{rgb}{0.933,0.835,0.718}
\definecolor{bisque3}{rgb}{0.804,0.718,0.62}
\definecolor{bisque4}{rgb}{0.545,0.49,0.42}
\definecolor{blanchedalmond}{rgb}{1,0.922,0.804}
\definecolor{blue1}{rgb}{0,0,1}
\definecolor{blue2}{rgb}{0,0,0.933}
\definecolor{blue3}{rgb}{0,0,0.804}
\definecolor{blue4}{rgb}{0,0,0.545}
\definecolor{blueviolet}{rgb}{0.541,0.169,0.886}
\definecolor{brown1}{rgb}{1,0.251,0.251}
\definecolor{brown2}{rgb}{0.933,0.231,0.231}
\definecolor{brown3}{rgb}{0.804,0.2,0.2}
\definecolor{brown4}{rgb}{0.545,0.137,0.137}
\definecolor{burlywood}{rgb}{0.871,0.722,0.529}
\definecolor{burlywood1}{rgb}{1,0.827,0.608}
\definecolor{burlywood2}{rgb}{0.933,0.773,0.569}
\definecolor{burlywood3}{rgb}{0.804,0.667,0.49}
\definecolor{burlywood4}{rgb}{0.545,0.451,0.333}
\definecolor{cadetblue}{rgb}{0.373,0.62,0.627}
\definecolor{cadetblue1}{rgb}{0.596,0.961,1}
\definecolor{cadetblue2}{rgb}{0.557,0.898,0.933}
\definecolor{cadetblue3}{rgb}{0.478,0.773,0.804}
\definecolor{cadetblue4}{rgb}{0.325,0.525,0.545}
\definecolor{chartreuse}{rgb}{0.498,1,0}
\definecolor{chartreuse1}{rgb}{0.498,1,0}
\definecolor{chartreuse2}{rgb}{0.463,0.933,0}
\definecolor{chartreuse3}{rgb}{0.4,0.804,0}
\definecolor{chartreuse4}{rgb}{0.271,0.545,0}
\definecolor{chocolate}{rgb}{0.824,0.412,0.118}
\definecolor{chocolate1}{rgb}{1,0.498,0.141}
\definecolor{chocolate2}{rgb}{0.933,0.463,0.129}
\definecolor{chocolate3}{rgb}{0.804,0.4,0.114}
\definecolor{chocolate4}{rgb}{0.545,0.271,0.075}
\definecolor{coral}{rgb}{1,0.498,0.314}
\definecolor{coral1}{rgb}{1,0.447,0.337}
\definecolor{coral2}{rgb}{0.933,0.416,0.314}
\definecolor{coral3}{rgb}{0.804,0.357,0.271}
\definecolor{coral4}{rgb}{0.545,0.243,0.184}
\definecolor{cornflowerblue}{rgb}{0.392,0.584,0.929}
\definecolor{cornsilk}{rgb}{1,0.973,0.863}
\definecolor{cornsilk1}{rgb}{1,0.973,0.863}
\definecolor{cornsilk2}{rgb}{0.933,0.91,0.804}
\definecolor{cornsilk3}{rgb}{0.804,0.784,0.694}
\definecolor{cornsilk4}{rgb}{0.545,0.533,0.471}
\definecolor{cyan}{rgb}{0,1,1}
\definecolor{cyan1}{rgb}{0,1,1}
\definecolor{cyan2}{rgb}{0,0.933,0.933}
\definecolor{cyan3}{rgb}{0,0.804,0.804}
\definecolor{cyan4}{rgb}{0,0.545,0.545}
\definecolor{darkgoldenrod}{rgb}{0.722,0.525,0.043}
\definecolor{darkgoldenrod1}{rgb}{1,0.725,0.059}
\definecolor{darkgoldenrod2}{rgb}{0.933,0.678,0.055}
\definecolor{darkgoldenrod3}{rgb}{0.804,0.584,0.047}
\definecolor{darkgoldenrod4}{rgb}{0.545,0.396,0.031}
\definecolor{darkgrey}{rgb}{0.663,0.663,0.663}
\definecolor{darkkhaki}{rgb}{0.741,0.718,0.42}
\definecolor{darkolivegreen}{rgb}{0.333,0.42,0.184}
\definecolor{darkolivegreen1}{rgb}{0.792,1,0.439}
\definecolor{darkolivegreen2}{rgb}{0.737,0.933,0.408}
\definecolor{darkolivegreen3}{rgb}{0.635,0.804,0.353}
\definecolor{darkolivegreen4}{rgb}{0.431,0.545,0.239}
\definecolor{darkorange1}{rgb}{1,0.498,0}
\definecolor{darkorange2}{rgb}{0.933,0.463,0}
\definecolor{darkorange3}{rgb}{0.804,0.4,0}
\definecolor{darkorange4}{rgb}{0.545,0.271,0}
\definecolor{darkorchid}{rgb}{0.6,0.196,0.8}
\definecolor{darkorchid1}{rgb}{0.749,0.243,1}
\definecolor{darkorchid2}{rgb}{0.698,0.227,0.933}
\definecolor{darkorchid3}{rgb}{0.604,0.196,0.804}
\definecolor{darkorchid4}{rgb}{0.408,0.133,0.545}
\definecolor{darksalmon}{rgb}{0.914,0.588,0.478}
\definecolor{darkseagreen}{rgb}{0.561,0.737,0.561}
\definecolor{darkseagreen1}{rgb}{0.757,1,0.757}
\definecolor{darkseagreen2}{rgb}{0.706,0.933,0.706}
\definecolor{darkseagreen3}{rgb}{0.608,0.804,0.608}
\definecolor{darkseagreen4}{rgb}{0.412,0.545,0.412}
\definecolor{darkslateblue}{rgb}{0.282,0.239,0.545}
\definecolor{darkslategray}{rgb}{0.184,0.31,0.31}
\definecolor{darkslategray1}{rgb}{0.592,1,1}
\definecolor{darkslategray2}{rgb}{0.553,0.933,0.933}
\definecolor{darkslategray3}{rgb}{0.475,0.804,0.804}
\definecolor{darkslategray4}{rgb}{0.322,0.545,0.545}
\definecolor{darkslategrey}{rgb}{0.184,0.31,0.31}
\definecolor{darkturquoise}{rgb}{0,0.808,0.82}
\definecolor{darkviolet}{rgb}{0.58,0,0.827}
\definecolor{deeppink}{rgb}{1,0.078,0.576}
\definecolor{deeppink1}{rgb}{1,0.078,0.576}
\definecolor{deeppink2}{rgb}{0.933,0.071,0.537}
\definecolor{deeppink3}{rgb}{0.804,0.063,0.463}
\definecolor{deeppink4}{rgb}{0.545,0.039,0.314}
\definecolor{deepskyblue}{rgb}{0,0.749,1}
\definecolor{deepskyblue1}{rgb}{0,0.749,1}
\definecolor{deepskyblue2}{rgb}{0,0.698,0.933}
\definecolor{deepskyblue3}{rgb}{0,0.604,0.804}
\definecolor{deepskyblue4}{rgb}{0,0.408,0.545}
\definecolor{dimgray}{rgb}{0.412,0.412,0.412}
\definecolor{dimgrey}{rgb}{0.412,0.412,0.412}
\definecolor{dodgerblue}{rgb}{0.118,0.565,1}
\definecolor{dodgerblue1}{rgb}{0.118,0.565,1}
\definecolor{dodgerblue2}{rgb}{0.11,0.525,0.933}
\definecolor{dodgerblue3}{rgb}{0.094,0.455,0.804}
\definecolor{dodgerblue4}{rgb}{0.063,0.306,0.545}
\definecolor{firebrick}{rgb}{0.698,0.133,0.133}
\definecolor{firebrick1}{rgb}{1,0.188,0.188}
\definecolor{firebrick2}{rgb}{0.933,0.173,0.173}
\definecolor{firebrick3}{rgb}{0.804,0.149,0.149}
\definecolor{firebrick4}{rgb}{0.545,0.102,0.102}
\definecolor{floralwhite}{rgb}{1,0.98,0.941}
\definecolor{forestgreen}{rgb}{0.133,0.545,0.133}
\definecolor{gainsboro}{rgb}{0.863,0.863,0.863}
\definecolor{ghostwhite}{rgb}{0.973,0.973,1}
\definecolor{gold1}{rgb}{1,0.843,0}
\definecolor{gold2}{rgb}{0.933,0.788,0}
\definecolor{gold3}{rgb}{0.804,0.678,0}
\definecolor{gold4}{rgb}{0.545,0.459,0}
\definecolor{goldenrod}{rgb}{0.855,0.647,0.125}
\definecolor{goldenrod1}{rgb}{1,0.757,0.145}
\definecolor{goldenrod2}{rgb}{0.933,0.706,0.133}
\definecolor{goldenrod3}{rgb}{0.804,0.608,0.114}
\definecolor{goldenrod4}{rgb}{0.545,0.412,0.078}
\definecolor{gray0}{rgb}{0,0,0}
\definecolor{gray1}{rgb}{0.012,0.012,0.012}
\definecolor{gray10}{rgb}{0.102,0.102,0.102}
\definecolor{gray100}{rgb}{1,1,1}
\definecolor{gray11}{rgb}{0.11,0.11,0.11}
\definecolor{gray12}{rgb}{0.122,0.122,0.122}
\definecolor{gray13}{rgb}{0.129,0.129,0.129}
\definecolor{gray14}{rgb}{0.141,0.141,0.141}
\definecolor{gray15}{rgb}{0.149,0.149,0.149}
\definecolor{gray16}{rgb}{0.161,0.161,0.161}
\definecolor{gray17}{rgb}{0.169,0.169,0.169}
\definecolor{gray18}{rgb}{0.18,0.18,0.18}
\definecolor{gray19}{rgb}{0.188,0.188,0.188}
\definecolor{gray2}{rgb}{0.02,0.02,0.02}
\definecolor{gray20}{rgb}{0.2,0.2,0.2}
\definecolor{gray21}{rgb}{0.212,0.212,0.212}
\definecolor{gray22}{rgb}{0.22,0.22,0.22}
\definecolor{gray23}{rgb}{0.231,0.231,0.231}
\definecolor{gray24}{rgb}{0.239,0.239,0.239}
\definecolor{gray25}{rgb}{0.251,0.251,0.251}
\definecolor{gray26}{rgb}{0.259,0.259,0.259}
\definecolor{gray27}{rgb}{0.271,0.271,0.271}
\definecolor{gray28}{rgb}{0.278,0.278,0.278}
\definecolor{gray29}{rgb}{0.29,0.29,0.29}
\definecolor{gray3}{rgb}{0.031,0.031,0.031}
\definecolor{gray30}{rgb}{0.302,0.302,0.302}
\definecolor{gray31}{rgb}{0.31,0.31,0.31}
\definecolor{gray32}{rgb}{0.322,0.322,0.322}
\definecolor{gray33}{rgb}{0.329,0.329,0.329}
\definecolor{gray34}{rgb}{0.341,0.341,0.341}
\definecolor{gray35}{rgb}{0.349,0.349,0.349}
\definecolor{gray36}{rgb}{0.361,0.361,0.361}
\definecolor{gray37}{rgb}{0.369,0.369,0.369}
\definecolor{gray38}{rgb}{0.38,0.38,0.38}
\definecolor{gray39}{rgb}{0.388,0.388,0.388}
\definecolor{gray4}{rgb}{0.039,0.039,0.039}
\definecolor{gray40}{rgb}{0.4,0.4,0.4}
\definecolor{gray41}{rgb}{0.412,0.412,0.412}
\definecolor{gray42}{rgb}{0.42,0.42,0.42}
\definecolor{gray43}{rgb}{0.431,0.431,0.431}
\definecolor{gray44}{rgb}{0.439,0.439,0.439}
\definecolor{gray45}{rgb}{0.451,0.451,0.451}
\definecolor{gray46}{rgb}{0.459,0.459,0.459}
\definecolor{gray47}{rgb}{0.471,0.471,0.471}
\definecolor{gray48}{rgb}{0.478,0.478,0.478}
\definecolor{gray49}{rgb}{0.49,0.49,0.49}
\definecolor{gray5}{rgb}{0.051,0.051,0.051}
\definecolor{gray50}{rgb}{0.498,0.498,0.498}
\definecolor{gray51}{rgb}{0.51,0.51,0.51}
\definecolor{gray52}{rgb}{0.522,0.522,0.522}
\definecolor{gray53}{rgb}{0.529,0.529,0.529}
\definecolor{gray54}{rgb}{0.541,0.541,0.541}
\definecolor{gray55}{rgb}{0.549,0.549,0.549}
\definecolor{gray56}{rgb}{0.561,0.561,0.561}
\definecolor{gray57}{rgb}{0.569,0.569,0.569}
\definecolor{gray58}{rgb}{0.58,0.58,0.58}
\definecolor{gray59}{rgb}{0.588,0.588,0.588}
\definecolor{gray6}{rgb}{0.059,0.059,0.059}
\definecolor{gray60}{rgb}{0.6,0.6,0.6}
\definecolor{gray61}{rgb}{0.612,0.612,0.612}
\definecolor{gray62}{rgb}{0.62,0.62,0.62}
\definecolor{gray63}{rgb}{0.631,0.631,0.631}
\definecolor{gray64}{rgb}{0.639,0.639,0.639}
\definecolor{gray65}{rgb}{0.651,0.651,0.651}
\definecolor{gray66}{rgb}{0.659,0.659,0.659}
\definecolor{gray67}{rgb}{0.671,0.671,0.671}
\definecolor{gray68}{rgb}{0.678,0.678,0.678}
\definecolor{gray69}{rgb}{0.69,0.69,0.69}
\definecolor{gray7}{rgb}{0.071,0.071,0.071}
\definecolor{gray70}{rgb}{0.702,0.702,0.702}
\definecolor{gray71}{rgb}{0.71,0.71,0.71}
\definecolor{gray72}{rgb}{0.722,0.722,0.722}
\definecolor{gray73}{rgb}{0.729,0.729,0.729}
\definecolor{gray74}{rgb}{0.741,0.741,0.741}
\definecolor{gray75}{rgb}{0.749,0.749,0.749}
\definecolor{gray76}{rgb}{0.761,0.761,0.761}
\definecolor{gray77}{rgb}{0.769,0.769,0.769}
\definecolor{gray78}{rgb}{0.78,0.78,0.78}
\definecolor{gray79}{rgb}{0.788,0.788,0.788}
\definecolor{gray8}{rgb}{0.078,0.078,0.078}
\definecolor{gray80}{rgb}{0.8,0.8,0.8}
\definecolor{gray81}{rgb}{0.812,0.812,0.812}
\definecolor{gray82}{rgb}{0.82,0.82,0.82}
\definecolor{gray83}{rgb}{0.831,0.831,0.831}
\definecolor{gray84}{rgb}{0.839,0.839,0.839}
\definecolor{gray85}{rgb}{0.851,0.851,0.851}
\definecolor{gray86}{rgb}{0.859,0.859,0.859}
\definecolor{gray87}{rgb}{0.871,0.871,0.871}
\definecolor{gray88}{rgb}{0.878,0.878,0.878}
\definecolor{gray89}{rgb}{0.89,0.89,0.89}
\definecolor{gray9}{rgb}{0.09,0.09,0.09}
\definecolor{gray90}{rgb}{0.898,0.898,0.898}
\definecolor{gray91}{rgb}{0.91,0.91,0.91}
\definecolor{gray92}{rgb}{0.922,0.922,0.922}
\definecolor{gray93}{rgb}{0.929,0.929,0.929}
\definecolor{gray94}{rgb}{0.941,0.941,0.941}
\definecolor{gray95}{rgb}{0.949,0.949,0.949}
\definecolor{gray96}{rgb}{0.961,0.961,0.961}
\definecolor{gray97}{rgb}{0.969,0.969,0.969}
\definecolor{gray98}{rgb}{0.98,0.98,0.98}
\definecolor{gray99}{rgb}{0.988,0.988,0.988}
\definecolor{green1}{rgb}{0,1,0}
\definecolor{green2}{rgb}{0,0.933,0}
\definecolor{green3}{rgb}{0,0.804,0}
\definecolor{green4}{rgb}{0,0.545,0}
\definecolor{greenyellow}{rgb}{0.678,1,0.184}
\definecolor{grey}{rgb}{0.745,0.745,0.745}
\definecolor{grey0}{rgb}{0,0,0}
\definecolor{grey1}{rgb}{0.012,0.012,0.012}
\definecolor{grey10}{rgb}{0.102,0.102,0.102}
\definecolor{grey100}{rgb}{1,1,1}
\definecolor{grey11}{rgb}{0.11,0.11,0.11}
\definecolor{grey12}{rgb}{0.122,0.122,0.122}
\definecolor{grey13}{rgb}{0.129,0.129,0.129}
\definecolor{grey14}{rgb}{0.141,0.141,0.141}
\definecolor{grey15}{rgb}{0.149,0.149,0.149}
\definecolor{grey16}{rgb}{0.161,0.161,0.161}
\definecolor{grey17}{rgb}{0.169,0.169,0.169}
\definecolor{grey18}{rgb}{0.18,0.18,0.18}
\definecolor{grey19}{rgb}{0.188,0.188,0.188}
\definecolor{grey2}{rgb}{0.02,0.02,0.02}
\definecolor{grey20}{rgb}{0.2,0.2,0.2}
\definecolor{grey21}{rgb}{0.212,0.212,0.212}
\definecolor{grey22}{rgb}{0.22,0.22,0.22}
\definecolor{grey23}{rgb}{0.231,0.231,0.231}
\definecolor{grey24}{rgb}{0.239,0.239,0.239}
\definecolor{grey25}{rgb}{0.251,0.251,0.251}
\definecolor{grey26}{rgb}{0.259,0.259,0.259}
\definecolor{grey27}{rgb}{0.271,0.271,0.271}
\definecolor{grey28}{rgb}{0.278,0.278,0.278}
\definecolor{grey29}{rgb}{0.29,0.29,0.29}
\definecolor{grey3}{rgb}{0.031,0.031,0.031}
\definecolor{grey30}{rgb}{0.302,0.302,0.302}
\definecolor{grey31}{rgb}{0.31,0.31,0.31}
\definecolor{grey32}{rgb}{0.322,0.322,0.322}
\definecolor{grey33}{rgb}{0.329,0.329,0.329}
\definecolor{grey34}{rgb}{0.341,0.341,0.341}
\definecolor{grey35}{rgb}{0.349,0.349,0.349}
\definecolor{grey36}{rgb}{0.361,0.361,0.361}
\definecolor{grey37}{rgb}{0.369,0.369,0.369}
\definecolor{grey38}{rgb}{0.38,0.38,0.38}
\definecolor{grey39}{rgb}{0.388,0.388,0.388}
\definecolor{grey4}{rgb}{0.039,0.039,0.039}
\definecolor{grey40}{rgb}{0.4,0.4,0.4}
\definecolor{grey41}{rgb}{0.412,0.412,0.412}
\definecolor{grey42}{rgb}{0.42,0.42,0.42}
\definecolor{grey43}{rgb}{0.431,0.431,0.431}
\definecolor{grey44}{rgb}{0.439,0.439,0.439}
\definecolor{grey45}{rgb}{0.451,0.451,0.451}
\definecolor{grey46}{rgb}{0.459,0.459,0.459}
\definecolor{grey47}{rgb}{0.471,0.471,0.471}
\definecolor{grey48}{rgb}{0.478,0.478,0.478}
\definecolor{grey49}{rgb}{0.49,0.49,0.49}
\definecolor{grey5}{rgb}{0.051,0.051,0.051}
\definecolor{grey50}{rgb}{0.498,0.498,0.498}
\definecolor{grey51}{rgb}{0.51,0.51,0.51}
\definecolor{grey52}{rgb}{0.522,0.522,0.522}
\definecolor{grey53}{rgb}{0.529,0.529,0.529}
\definecolor{grey54}{rgb}{0.541,0.541,0.541}
\definecolor{grey55}{rgb}{0.549,0.549,0.549}
\definecolor{grey56}{rgb}{0.561,0.561,0.561}
\definecolor{grey57}{rgb}{0.569,0.569,0.569}
\definecolor{grey58}{rgb}{0.58,0.58,0.58}
\definecolor{grey59}{rgb}{0.588,0.588,0.588}
\definecolor{grey6}{rgb}{0.059,0.059,0.059}
\definecolor{grey60}{rgb}{0.6,0.6,0.6}
\definecolor{grey61}{rgb}{0.612,0.612,0.612}
\definecolor{grey62}{rgb}{0.62,0.62,0.62}
\definecolor{grey63}{rgb}{0.631,0.631,0.631}
\definecolor{grey64}{rgb}{0.639,0.639,0.639}
\definecolor{grey65}{rgb}{0.651,0.651,0.651}
\definecolor{grey66}{rgb}{0.659,0.659,0.659}
\definecolor{grey67}{rgb}{0.671,0.671,0.671}
\definecolor{grey68}{rgb}{0.678,0.678,0.678}
\definecolor{grey69}{rgb}{0.69,0.69,0.69}
\definecolor{grey7}{rgb}{0.071,0.071,0.071}
\definecolor{grey70}{rgb}{0.702,0.702,0.702}
\definecolor{grey71}{rgb}{0.71,0.71,0.71}
\definecolor{grey72}{rgb}{0.722,0.722,0.722}
\definecolor{grey73}{rgb}{0.729,0.729,0.729}
\definecolor{grey74}{rgb}{0.741,0.741,0.741}
\definecolor{grey75}{rgb}{0.749,0.749,0.749}
\definecolor{grey76}{rgb}{0.761,0.761,0.761}
\definecolor{grey77}{rgb}{0.769,0.769,0.769}
\definecolor{grey78}{rgb}{0.78,0.78,0.78}
\definecolor{grey79}{rgb}{0.788,0.788,0.788}
\definecolor{grey8}{rgb}{0.078,0.078,0.078}
\definecolor{grey80}{rgb}{0.8,0.8,0.8}
\definecolor{grey81}{rgb}{0.812,0.812,0.812}
\definecolor{grey82}{rgb}{0.82,0.82,0.82}
\definecolor{grey83}{rgb}{0.831,0.831,0.831}
\definecolor{grey84}{rgb}{0.839,0.839,0.839}
\definecolor{grey85}{rgb}{0.851,0.851,0.851}
\definecolor{grey86}{rgb}{0.859,0.859,0.859}
\definecolor{grey87}{rgb}{0.871,0.871,0.871}
\definecolor{grey88}{rgb}{0.878,0.878,0.878}
\definecolor{grey89}{rgb}{0.89,0.89,0.89}
\definecolor{grey9}{rgb}{0.09,0.09,0.09}
\definecolor{grey90}{rgb}{0.898,0.898,0.898}
\definecolor{grey91}{rgb}{0.91,0.91,0.91}
\definecolor{grey92}{rgb}{0.922,0.922,0.922}
\definecolor{grey93}{rgb}{0.929,0.929,0.929}
\definecolor{grey94}{rgb}{0.941,0.941,0.941}
\definecolor{grey95}{rgb}{0.949,0.949,0.949}
\definecolor{grey96}{rgb}{0.961,0.961,0.961}
\definecolor{grey97}{rgb}{0.969,0.969,0.969}
\definecolor{grey98}{rgb}{0.98,0.98,0.98}
\definecolor{grey99}{rgb}{0.988,0.988,0.988}
\definecolor{honeydew}{rgb}{0.941,1,0.941}
\definecolor{honeydew1}{rgb}{0.941,1,0.941}
\definecolor{honeydew2}{rgb}{0.878,0.933,0.878}
\definecolor{honeydew3}{rgb}{0.757,0.804,0.757}
\definecolor{honeydew4}{rgb}{0.514,0.545,0.514}
\definecolor{hotpink}{rgb}{1,0.412,0.706}
\definecolor{hotpink1}{rgb}{1,0.431,0.706}
\definecolor{hotpink2}{rgb}{0.933,0.416,0.655}
\definecolor{hotpink3}{rgb}{0.804,0.376,0.565}
\definecolor{hotpink4}{rgb}{0.545,0.227,0.384}
\definecolor{indianred}{rgb}{0.804,0.361,0.361}
\definecolor{indianred1}{rgb}{1,0.416,0.416}
\definecolor{indianred2}{rgb}{0.933,0.388,0.388}
\definecolor{indianred3}{rgb}{0.804,0.333,0.333}
\definecolor{indianred4}{rgb}{0.545,0.227,0.227}
\definecolor{ivory}{rgb}{1,1,0.941}
\definecolor{ivory1}{rgb}{1,1,0.941}
\definecolor{ivory2}{rgb}{0.933,0.933,0.878}
\definecolor{ivory3}{rgb}{0.804,0.804,0.757}
\definecolor{ivory4}{rgb}{0.545,0.545,0.514}
\definecolor{khaki}{rgb}{0.941,0.902,0.549}
\definecolor{khaki1}{rgb}{1,0.965,0.561}
\definecolor{khaki2}{rgb}{0.933,0.902,0.522}
\definecolor{khaki3}{rgb}{0.804,0.776,0.451}
\definecolor{khaki4}{rgb}{0.545,0.525,0.306}
\definecolor{lavender}{rgb}{0.902,0.902,0.98}
\definecolor{lavenderblush}{rgb}{1,0.941,0.961}
\definecolor{lavenderblush1}{rgb}{1,0.941,0.961}
\definecolor{lavenderblush2}{rgb}{0.933,0.878,0.898}
\definecolor{lavenderblush3}{rgb}{0.804,0.757,0.773}
\definecolor{lavenderblush4}{rgb}{0.545,0.514,0.525}
\definecolor{lawngreen}{rgb}{0.486,0.988,0}
\definecolor{lemonchiffon}{rgb}{1,0.98,0.804}
\definecolor{lemonchiffon1}{rgb}{1,0.98,0.804}
\definecolor{lemonchiffon2}{rgb}{0.933,0.914,0.749}
\definecolor{lemonchiffon3}{rgb}{0.804,0.788,0.647}
\definecolor{lemonchiffon4}{rgb}{0.545,0.537,0.439}
\definecolor{lightblue1}{rgb}{0.749,0.937,1}
\definecolor{lightblue2}{rgb}{0.698,0.875,0.933}
\definecolor{lightblue3}{rgb}{0.604,0.753,0.804}
\definecolor{lightblue4}{rgb}{0.408,0.514,0.545}
\definecolor{lightcoral}{rgb}{0.941,0.502,0.502}
\definecolor{lightcyan1}{rgb}{0.878,1,1}
\definecolor{lightcyan2}{rgb}{0.82,0.933,0.933}
\definecolor{lightcyan3}{rgb}{0.706,0.804,0.804}
\definecolor{lightcyan4}{rgb}{0.478,0.545,0.545}
\definecolor{lightgoldenrod}{rgb}{0.933,0.867,0.51}
\definecolor{lightgoldenrod1}{rgb}{1,0.925,0.545}
\definecolor{lightgoldenrod2}{rgb}{0.933,0.863,0.51}
\definecolor{lightgoldenrod3}{rgb}{0.804,0.745,0.439}
\definecolor{lightgoldenrod4}{rgb}{0.545,0.506,0.298}
\definecolor{lightgoldenrodyellow}{rgb}{0.98,0.98,0.824}
\definecolor{lightgrey}{rgb}{0.827,0.827,0.827}
\definecolor{lightpink}{rgb}{1,0.714,0.757}
\definecolor{lightpink1}{rgb}{1,0.682,0.725}
\definecolor{lightpink2}{rgb}{0.933,0.635,0.678}
\definecolor{lightpink3}{rgb}{0.804,0.549,0.584}
\definecolor{lightpink4}{rgb}{0.545,0.373,0.396}
\definecolor{lightsalmon}{rgb}{1,0.627,0.478}
\definecolor{lightsalmon1}{rgb}{1,0.627,0.478}
\definecolor{lightsalmon2}{rgb}{0.933,0.584,0.447}
\definecolor{lightsalmon3}{rgb}{0.804,0.506,0.384}
\definecolor{lightsalmon4}{rgb}{0.545,0.341,0.259}
\definecolor{lightseagreen}{rgb}{0.125,0.698,0.667}
\definecolor{lightskyblue}{rgb}{0.529,0.808,0.98}
\definecolor{lightskyblue1}{rgb}{0.69,0.886,1}
\definecolor{lightskyblue2}{rgb}{0.643,0.827,0.933}
\definecolor{lightskyblue3}{rgb}{0.553,0.714,0.804}
\definecolor{lightskyblue4}{rgb}{0.376,0.482,0.545}
\definecolor{lightslateblue}{rgb}{0.518,0.439,1}
\definecolor{lightslategray}{rgb}{0.467,0.533,0.6}
\definecolor{lightslategrey}{rgb}{0.467,0.533,0.6}
\definecolor{lightsteelblue}{rgb}{0.69,0.769,0.871}
\definecolor{lightsteelblue1}{rgb}{0.792,0.882,1}
\definecolor{lightsteelblue2}{rgb}{0.737,0.824,0.933}
\definecolor{lightsteelblue3}{rgb}{0.635,0.71,0.804}
\definecolor{lightsteelblue4}{rgb}{0.431,0.482,0.545}
\definecolor{lightyellow1}{rgb}{1,1,0.878}
\definecolor{lightyellow2}{rgb}{0.933,0.933,0.82}
\definecolor{lightyellow3}{rgb}{0.804,0.804,0.706}
\definecolor{lightyellow4}{rgb}{0.545,0.545,0.478}
\definecolor{limegreen}{rgb}{0.196,0.804,0.196}
\definecolor{linen}{rgb}{0.98,0.941,0.902}
\definecolor{magenta}{rgb}{1,0,1}
\definecolor{magenta1}{rgb}{1,0,1}
\definecolor{magenta2}{rgb}{0.933,0,0.933}
\definecolor{magenta3}{rgb}{0.804,0,0.804}
\definecolor{magenta4}{rgb}{0.545,0,0.545}
\definecolor{maroon}{rgb}{0.69,0.188,0.376}
\definecolor{maroon1}{rgb}{1,0.204,0.702}
\definecolor{maroon2}{rgb}{0.933,0.188,0.655}
\definecolor{maroon3}{rgb}{0.804,0.161,0.565}
\definecolor{maroon4}{rgb}{0.545,0.11,0.384}
\definecolor{mediumaquamarine}{rgb}{0.4,0.804,0.667}
\definecolor{mediumblue}{rgb}{0,0,0.804}
\definecolor{mediumorchid}{rgb}{0.729,0.333,0.827}
\definecolor{mediumorchid1}{rgb}{0.878,0.4,1}
\definecolor{mediumorchid2}{rgb}{0.82,0.373,0.933}
\definecolor{mediumorchid3}{rgb}{0.706,0.322,0.804}
\definecolor{mediumorchid4}{rgb}{0.478,0.216,0.545}
\definecolor{mediumpurple}{rgb}{0.576,0.439,0.859}
\definecolor{mediumpurple1}{rgb}{0.671,0.51,1}
\definecolor{mediumpurple2}{rgb}{0.624,0.475,0.933}
\definecolor{mediumpurple3}{rgb}{0.537,0.408,0.804}
\definecolor{mediumpurple4}{rgb}{0.365,0.278,0.545}
\definecolor{mediumseagreen}{rgb}{0.235,0.702,0.443}
\definecolor{mediumslateblue}{rgb}{0.482,0.408,0.933}
\definecolor{mediumspringgreen}{rgb}{0,0.98,0.604}
\definecolor{mediumturquoise}{rgb}{0.282,0.82,0.8}
\definecolor{mediumvioletred}{rgb}{0.78,0.082,0.522}
\definecolor{midnightblue}{rgb}{0.098,0.098,0.439}
\definecolor{mintcream}{rgb}{0.961,1,0.98}
\definecolor{mistyrose}{rgb}{1,0.894,0.882}
\definecolor{mistyrose1}{rgb}{1,0.894,0.882}
\definecolor{mistyrose2}{rgb}{0.933,0.835,0.824}
\definecolor{mistyrose3}{rgb}{0.804,0.718,0.71}
\definecolor{mistyrose4}{rgb}{0.545,0.49,0.482}
\definecolor{moccasin}{rgb}{1,0.894,0.71}
\definecolor{navajowhite}{rgb}{1,0.871,0.678}
\definecolor{navajowhite1}{rgb}{1,0.871,0.678}
\definecolor{navajowhite2}{rgb}{0.933,0.812,0.631}
\definecolor{navajowhite3}{rgb}{0.804,0.702,0.545}
\definecolor{navajowhite4}{rgb}{0.545,0.475,0.369}
\definecolor{navyblue}{rgb}{0,0,0.502}
\definecolor{oldlace}{rgb}{0.992,0.961,0.902}
\definecolor{olivedrab}{rgb}{0.42,0.557,0.137}
\definecolor{olivedrab1}{rgb}{0.753,1,0.243}
\definecolor{olivedrab2}{rgb}{0.702,0.933,0.227}
\definecolor{olivedrab3}{rgb}{0.604,0.804,0.196}
\definecolor{olivedrab4}{rgb}{0.412,0.545,0.133}
\definecolor{orange1}{rgb}{1,0.647,0}
\definecolor{orange2}{rgb}{0.933,0.604,0}
\definecolor{orange3}{rgb}{0.804,0.522,0}
\definecolor{orange4}{rgb}{0.545,0.353,0}
\definecolor{orangered}{rgb}{1,0.271,0}
\definecolor{orangered1}{rgb}{1,0.271,0}
\definecolor{orangered2}{rgb}{0.933,0.251,0}
\definecolor{orangered3}{rgb}{0.804,0.216,0}
\definecolor{orangered4}{rgb}{0.545,0.145,0}
\definecolor{orchid}{rgb}{0.855,0.439,0.839}
\definecolor{orchid1}{rgb}{1,0.514,0.98}
\definecolor{orchid2}{rgb}{0.933,0.478,0.914}
\definecolor{orchid3}{rgb}{0.804,0.412,0.788}
\definecolor{orchid4}{rgb}{0.545,0.278,0.537}
\definecolor{palegoldenrod}{rgb}{0.933,0.91,0.667}
\definecolor{palegreen}{rgb}{0.596,0.984,0.596}
\definecolor{palegreen1}{rgb}{0.604,1,0.604}
\definecolor{palegreen2}{rgb}{0.565,0.933,0.565}
\definecolor{palegreen3}{rgb}{0.486,0.804,0.486}
\definecolor{palegreen4}{rgb}{0.329,0.545,0.329}
\definecolor{paleturquoise}{rgb}{0.686,0.933,0.933}
\definecolor{paleturquoise1}{rgb}{0.733,1,1}
\definecolor{paleturquoise2}{rgb}{0.682,0.933,0.933}
\definecolor{paleturquoise3}{rgb}{0.588,0.804,0.804}
\definecolor{paleturquoise4}{rgb}{0.4,0.545,0.545}
\definecolor{palevioletred}{rgb}{0.859,0.439,0.576}
\definecolor{palevioletred1}{rgb}{1,0.51,0.671}
\definecolor{palevioletred2}{rgb}{0.933,0.475,0.624}
\definecolor{palevioletred3}{rgb}{0.804,0.408,0.537}
\definecolor{palevioletred4}{rgb}{0.545,0.278,0.365}
\definecolor{papayawhip}{rgb}{1,0.937,0.835}
\definecolor{peachpuff}{rgb}{1,0.855,0.725}
\definecolor{peachpuff1}{rgb}{1,0.855,0.725}
\definecolor{peachpuff2}{rgb}{0.933,0.796,0.678}
\definecolor{peachpuff3}{rgb}{0.804,0.686,0.584}
\definecolor{peachpuff4}{rgb}{0.545,0.467,0.396}
\definecolor{peru}{rgb}{0.804,0.522,0.247}
\definecolor{pink1}{rgb}{1,0.71,0.773}
\definecolor{pink2}{rgb}{0.933,0.663,0.722}
\definecolor{pink3}{rgb}{0.804,0.569,0.62}
\definecolor{pink4}{rgb}{0.545,0.388,0.424}
\definecolor{plum}{rgb}{0.867,0.627,0.867}
\definecolor{plum1}{rgb}{1,0.733,1}
\definecolor{plum2}{rgb}{0.933,0.682,0.933}
\definecolor{plum3}{rgb}{0.804,0.588,0.804}
\definecolor{plum4}{rgb}{0.545,0.4,0.545}
\definecolor{powderblue}{rgb}{0.69,0.878,0.902}
\definecolor{purple1}{rgb}{0.608,0.188,1}
\definecolor{purple2}{rgb}{0.569,0.173,0.933}
\definecolor{purple3}{rgb}{0.49,0.149,0.804}
\definecolor{purple4}{rgb}{0.333,0.102,0.545}
\definecolor{red1}{rgb}{1,0,0}
\definecolor{red2}{rgb}{0.933,0,0}
\definecolor{red3}{rgb}{0.804,0,0}
\definecolor{red4}{rgb}{0.545,0,0}
\definecolor{rosybrown}{rgb}{0.737,0.561,0.561}
\definecolor{rosybrown1}{rgb}{1,0.757,0.757}
\definecolor{rosybrown2}{rgb}{0.933,0.706,0.706}
\definecolor{rosybrown3}{rgb}{0.804,0.608,0.608}
\definecolor{rosybrown4}{rgb}{0.545,0.412,0.412}
\definecolor{royalblue}{rgb}{0.255,0.412,0.882}
\definecolor{royalblue1}{rgb}{0.282,0.463,1}
\definecolor{royalblue2}{rgb}{0.263,0.431,0.933}
\definecolor{royalblue3}{rgb}{0.227,0.373,0.804}
\definecolor{royalblue4}{rgb}{0.153,0.251,0.545}
\definecolor{saddlebrown}{rgb}{0.545,0.271,0.075}
\definecolor{salmon}{rgb}{0.98,0.502,0.447}
\definecolor{salmon1}{rgb}{1,0.549,0.412}
\definecolor{salmon2}{rgb}{0.933,0.51,0.384}
\definecolor{salmon3}{rgb}{0.804,0.439,0.329}
\definecolor{salmon4}{rgb}{0.545,0.298,0.224}
\definecolor{sandybrown}{rgb}{0.957,0.643,0.376}
\definecolor{seagreen1}{rgb}{0.329,1,0.624}
\definecolor{seagreen2}{rgb}{0.306,0.933,0.58}
\definecolor{seagreen3}{rgb}{0.263,0.804,0.502}
\definecolor{seagreen4}{rgb}{0.18,0.545,0.341}
\definecolor{seashell}{rgb}{1,0.961,0.933}
\definecolor{seashell1}{rgb}{1,0.961,0.933}
\definecolor{seashell2}{rgb}{0.933,0.898,0.871}
\definecolor{seashell3}{rgb}{0.804,0.773,0.749}
\definecolor{seashell4}{rgb}{0.545,0.525,0.51}
\definecolor{sienna}{rgb}{0.627,0.322,0.176}
\definecolor{sienna1}{rgb}{1,0.51,0.278}
\definecolor{sienna2}{rgb}{0.933,0.475,0.259}
\definecolor{sienna3}{rgb}{0.804,0.408,0.224}
\definecolor{sienna4}{rgb}{0.545,0.278,0.149}
\definecolor{skyblue}{rgb}{0.529,0.808,0.922}
\definecolor{skyblue1}{rgb}{0.529,0.808,1}
\definecolor{skyblue2}{rgb}{0.494,0.753,0.933}
\definecolor{skyblue3}{rgb}{0.424,0.651,0.804}
\definecolor{skyblue4}{rgb}{0.29,0.439,0.545}
\definecolor{slateblue}{rgb}{0.416,0.353,0.804}
\definecolor{slateblue1}{rgb}{0.514,0.435,1}
\definecolor{slateblue2}{rgb}{0.478,0.404,0.933}
\definecolor{slateblue3}{rgb}{0.412,0.349,0.804}
\definecolor{slateblue4}{rgb}{0.278,0.235,0.545}
\definecolor{slategray}{rgb}{0.439,0.502,0.565}
\definecolor{slategray1}{rgb}{0.776,0.886,1}
\definecolor{slategray2}{rgb}{0.725,0.827,0.933}
\definecolor{slategray3}{rgb}{0.624,0.714,0.804}
\definecolor{slategray4}{rgb}{0.424,0.482,0.545}
\definecolor{slategrey}{rgb}{0.439,0.502,0.565}
\definecolor{snow}{rgb}{1,0.98,0.98}
\definecolor{snow1}{rgb}{1,0.98,0.98}
\definecolor{snow2}{rgb}{0.933,0.914,0.914}
\definecolor{snow3}{rgb}{0.804,0.788,0.788}
\definecolor{snow4}{rgb}{0.545,0.537,0.537}
\definecolor{springgreen}{rgb}{0,1,0.498}
\definecolor{springgreen1}{rgb}{0,1,0.498}
\definecolor{springgreen2}{rgb}{0,0.933,0.463}
\definecolor{springgreen3}{rgb}{0,0.804,0.4}
\definecolor{springgreen4}{rgb}{0,0.545,0.271}
\definecolor{steelblue}{rgb}{0.275,0.51,0.706}
\definecolor{steelblue1}{rgb}{0.388,0.722,1}
\definecolor{steelblue2}{rgb}{0.361,0.675,0.933}
\definecolor{steelblue3}{rgb}{0.31,0.58,0.804}
\definecolor{steelblue4}{rgb}{0.212,0.392,0.545}
\definecolor{tan}{rgb}{0.824,0.706,0.549}
\definecolor{tan1}{rgb}{1,0.647,0.31}
\definecolor{tan2}{rgb}{0.933,0.604,0.286}
\definecolor{tan3}{rgb}{0.804,0.522,0.247}
\definecolor{tan4}{rgb}{0.545,0.353,0.169}
\definecolor{thistle}{rgb}{0.847,0.749,0.847}
\definecolor{thistle1}{rgb}{1,0.882,1}
\definecolor{thistle2}{rgb}{0.933,0.824,0.933}
\definecolor{thistle3}{rgb}{0.804,0.71,0.804}
\definecolor{thistle4}{rgb}{0.545,0.482,0.545}
\definecolor{tomato}{rgb}{1,0.388,0.278}
\definecolor{tomato1}{rgb}{1,0.388,0.278}
\definecolor{tomato2}{rgb}{0.933,0.361,0.259}
\definecolor{tomato3}{rgb}{0.804,0.31,0.224}
\definecolor{tomato4}{rgb}{0.545,0.212,0.149}
\definecolor{turquoise1}{rgb}{0,0.961,1}
\definecolor{turquoise2}{rgb}{0,0.898,0.933}
\definecolor{turquoise3}{rgb}{0,0.773,0.804}
\definecolor{turquoise4}{rgb}{0,0.525,0.545}
\definecolor{violetred}{rgb}{0.816,0.125,0.565}
\definecolor{violetred1}{rgb}{1,0.243,0.588}
\definecolor{violetred2}{rgb}{0.933,0.227,0.549}
\definecolor{violetred3}{rgb}{0.804,0.196,0.471}
\definecolor{violetred4}{rgb}{0.545,0.133,0.322}
\definecolor{wheat}{rgb}{0.961,0.871,0.702}
\definecolor{wheat1}{rgb}{1,0.906,0.729}
\definecolor{wheat2}{rgb}{0.933,0.847,0.682}
\definecolor{wheat3}{rgb}{0.804,0.729,0.588}
\definecolor{wheat4}{rgb}{0.545,0.494,0.4}
\definecolor{whitesmoke}{rgb}{0.961,0.961,0.961}
\definecolor{yellow1}{rgb}{1,1,0}
\definecolor{yellow2}{rgb}{0.933,0.933,0}
\definecolor{yellow3}{rgb}{0.804,0.804,0}
\definecolor{yellow4}{rgb}{0.545,0.545,0}
\definecolor{yellowgreen}{rgb}{0.604,0.804,0.196}
\scalebox{1}{

}
\caption{The Gaifman graph of ${\mathfrak{A}^\star},$ when adding $\partial_{\mathfrak{K}_2}(Z')$ to the boundary. The set $X'$ is the set $\tilde{X}_2\setminus V(F^\star)$ and $\breve{C}'$ is the privileged connected component of $G_{\mathfrak{A}^\star}$ with respect to ${\bf W}_q^{(2)}$ and $X'.$}
\labels{figure_boundariedgraph4}
\end{figure}

For the tuple $(\mathfrak{A}^{\star},R_2^{'\star},{\bf W}_{{q}}^{(2)},\varnothing^l,\tilde{X}^{\star}_2, {\bf b}_2^\star)$ to be an $h$-boundaried $τ'$-structure, we need to show that
$R_2^{'\star}\subseteq V(\mathfrak{A}^\star),$
$\cupall {\bf W}_{{q}}^{(2)}\subseteq V(\mathfrak{A}^\star),$
$\tilde{X}^{\star}_2\subseteq V(\mathfrak{A}^\star),$ and
$V({\bf b}_2^\star)\subseteq V(\mathfrak{A}^\star).$
Notice that $R_2^{'\star}\subseteq V(\mathfrak{A}^\star),$ since  $R_2^{'\star} =  R_1^{' \star}\setminus Y$ and $R_1^{' \star}\subseteq V(\mathfrak{A}^\star).$
To show that $V({\bf b}_2^\star)\subseteq V(\mathfrak{A}^\star),$ we first notice that, since $\breve{C}$ respects ${\bf W}_q^{(1)}$ and $I_2^{(d)}\cap X = \emptyset,$ it holds that $I_2^{(d)}\subseteq \breve{C}.$
Therefore, since $Z'\subseteq I_2^{(d-r+1)},$ we have that $Z'\subseteq \breve{C}.$
The latter implies that $\partial_{\mathfrak{K}_2}(Z')$ is a subset of $V(\mathfrak{A}^\star).$
By the definition of ${\bf b}_2^\star$ and since $V({\bf b}_2) =  \partial_{\mathfrak{K}_2}(Z')\cup V(F_2),$ we have that $V({\bf b}_2^\star) = \partial_{\mathfrak{K}_2}(Z')\cup V(F^\star).$
Hence,  given that $\partial_{\mathfrak{K}_2}(Z')\subseteq V(\mathfrak{A}^{\star})$ and $V(F^\star)\subseteq  V(\mathfrak{A}^{\star}),$ $V({\bf b}_2^\star)\subseteq V(\mathfrak{A}^\star).$
Also, observe that the fact that $I_2^{(d)}\subseteq \breve{C}$ implies that $\cupall {\bf W}_q^{(2)} \subseteq V(\mathfrak{A}^\star),$ while $\tilde{X}^{\star}_2\subseteq V(\mathfrak{A}^\star),$ since $\tilde{X}^{\star}_2 = (\tilde{X}_2\setminus V(F_2))\cup V(F^\star),$ $\tilde{X}_2\setminus V(F_2)\subseteq Z',$ and $Z'\subseteq V(\mathfrak{A}^\star).$

\myskip\paragraph{All considered boundaried structures are $(θ^{\sf out}_q,h)$-equivalent.}
As a next step, we argue that the $h$-boundaried $τ'$-structures
$(\mathfrak{A}^{\star},R_2^{'\star},{\bf W}_{{q}}^{(2)},\varnothing^l,\tilde{X}^{\star}_2,{\bf b}_2^\star),$
$(\mathfrak{A}_2,R_2 ',{\bf W}_{{q}}^{(2)}, \varnothing^l,\tilde{X}_2, {\bf b}_2),$
and $(\mathfrak{A}^\star,R_1^{'\star},{\bf W}_{{q}}^{(1)},\varnothing^l,\tilde{X}^\star_1,{\bf b}_1^\star)$ are (pairwise) compatible.
To see why this holds, notice that, since $K_1^{\bf a}[\partial_{\mathfrak{K}_1} (Z) \cup V_L ({\bf a})]\cup F_1$
is strongly isomorphic to
$K_2^{\bf a}[\partial_{\mathfrak{K}_2} (Z') \cup V_L ({\bf a})]\cup F_2$
with respect to $({\cal V}_1, {\cal V}_2),$ it holds that $F_1$ and $F_2$ are isomorphic. This, together with the fact that $F_1$ is isomorphic to $F^\star,$ implies that $F_2,$ $F_1,$ and $F^\star$ are pairwise isomorphic graphs.
Therefore,
the structures $\mathfrak{A}^\star[V({\bf b}_2^\star)],$ $\mathfrak{A}_2 [V({\bf b}_2)],$ and $\mathfrak{A}^\star [V({\bf b}_1^\star)]$ are (pairwise) isomorphic.

By following the exactly symmetric arguments as in the proof of the subclaim above,
it is easy to show that
$(\mathfrak{A}^{\star},R_2^{'\star},{\bf W}_{{q}}^{(2)},\varnothing^l, \tilde{X}^{\star}_2, {\bf b}_2^\star)$ and $(\mathfrak{A}_2,R_2 ',{\bf W}_{{q}}^{(2)},\varnothing^l, \tilde{X}_2, {\bf b}_2)$ are $(θ^{\sf out}_q,h)$-equivalent. This implies that
\begin{eqnarray}
\big(\mathfrak{A}^\star, R_2^{' \star},{\bf W}_{{q}}^{(2)} , \varnothing^l,\tilde{X}^\star_2, {\bf b}_2^\star\big)\models \bar{φ}\iff\big(\mathfrak{A}_2,R_2 ',{\bf W}_{{q}}^{(2)},\varnothing^l, \tilde{X}_2, {\bf b}_2\big)\models \bar{φ}.\labels{@concordancia}
\end{eqnarray}

Therefore, combining~\eqref{@identificable},~\eqref{@photographic}, and~\eqref{@concordancia}, we conclude that
the $h$-boundaried $τ'$-structures $\big(\mathfrak{A}^{\star},R_2^{'\star},{\bf W}_{{q}}^{(2)},\varnothing^l, \tilde{X}^{\star}_2, {\bf b}_2^\star\big)$
and
$\big(\mathfrak{A}^\star,R_1^{'\star},{\bf W}_{{q}}^{(1)},\varnothing^l, \tilde{X}^\star_1, {\bf b}_1^\star\big)$
are $(θ^{\sf out}_q,h)$-equivalent.
Recall that, by~\eqref{@electronically},
$$(\mathfrak{A}_{\rm out}^\star, R\setminus (\breve{C}\cup Z),\emptyset^{2q} ,\varnothing^l,X_{\rm out}, {\bf b}_1^\star ) \oplus (\mathfrak{A}^\star, R_1^{' \star},{\bf W}_{{q}}^{(1)} ,\varnothing^l, \tilde{X}^\star_1, {\bf b}_1^\star)= (\mathfrak{A},R,{\bf W}_q^{(1)}, \varnothing^l,X)$$
and $(\mathfrak{A},R,{\bf W}_q^{(1)}, \varnothing^l,X)\models θ^{\sf out}_q.$
Since the $h$-boundaried $τ'$-structures $\big(\mathfrak{A}^\star,R_1^{'\star},{\bf W}_{{q}}^{(1)}, \varnothing^l,\tilde{X}^\star_1, {\bf b}_1^\star\big)$
and
$\big(\mathfrak{A}^{\star},R_2^{'\star},{\bf W}_{{q}}^{(2)},\varnothing^l, \tilde{X}^{\star}_2, {\bf b}_2^\star\big)$
are $(θ^{\sf out}_q,h)$-equivalent,
\begin{eqnarray}
(\mathfrak{A}_{\rm out}^\star, R\setminus (\breve{C}\cup Z),\emptyset^{2q} ,\varnothing^l,X_{\rm out}, {\bf b}_1^\star ) \oplus \big(\mathfrak{A}^{\star},R_2^{'\star},{\bf W}_{{q}}^{(2)},\varnothing^l, \tilde{X}^{\star}_2, {\bf b}_2^\star\big)\models θ^{\sf out}_q.\labels{@pseudorealism}
\end{eqnarray}

\myskip\paragraph{Another way to put a boundary in the initial structure.}
We set $X' := \tilde{X}_2^{\star}.$
To conclude the proof of~\autoref{claim_1}, it remains to prove that $$(\mathfrak{A}_{\rm out}^\star, R\setminus (\breve{C}\cup Z),\emptyset^{2q} ,\varnothing^l,X_{\rm out}, {\bf b}_1^\star ) \oplus \big(\mathfrak{A}^{\star},R_2^{'\star},{\bf W}_{{q}}^{(2)},\varnothing^l, \tilde{X}^{\star}_2, {\bf b}_2^\star\big) = (\mathfrak{A}, R\setminus Y, {\bf W}_q^{(2)},\varnothing^l, X_{\rm out} \cup X').$$
\begin{figure}[ht]
\vspace{-20mm}
\centering
\tikzstyle{ipe stylesheet} = [
  ipe import,
  even odd rule,
  line join=round,
  line cap=butt,
  ipe pen normal/.style={line width=0.4},
  ipe pen heavier/.style={line width=0.8},
  ipe pen fat/.style={line width=1.2},
  ipe pen ultrafat/.style={line width=2},
  ipe pen normal,
  ipe mark normal/.style={ipe mark scale=3},
  ipe mark large/.style={ipe mark scale=5},
  ipe mark small/.style={ipe mark scale=2},
  ipe mark tiny/.style={ipe mark scale=1.1},
  ipe mark normal,
  /pgf/arrow keys/.cd,
  ipe arrow normal/.style={scale=7},
  ipe arrow large/.style={scale=10},
  ipe arrow small/.style={scale=5},
  ipe arrow tiny/.style={scale=3},
  ipe arrow normal,
  /tikz/.cd,
  ipe arrows, 
  <->/.tip = ipe normal,
  ipe dash normal/.style={dash pattern=},
  ipe dash dotted/.style={dash pattern=on 1bp off 3bp},
  ipe dash dashed/.style={dash pattern=on 4bp off 4bp},
  ipe dash dash dotted/.style={dash pattern=on 4bp off 2bp on 1bp off 2bp},
  ipe dash dash dot dotted/.style={dash pattern=on 4bp off 2bp on 1bp off 2bp on 1bp off 2bp},
  ipe dash normal,
  ipe node/.append style={font=\normalsize},
  ipe stretch normal/.style={ipe node stretch=1},
  ipe stretch normal,
  ipe opacity 10/.style={opacity=0.1},
  ipe opacity 30/.style={opacity=0.3},
  ipe opacity 50/.style={opacity=0.5},
  ipe opacity 75/.style={opacity=0.75},
  ipe opacity opaque/.style={opacity=1},
  ipe opacity opaque,
]
\definecolor{black}{rgb}{0,0,0}
\definecolor{white}{rgb}{1,1,1}
\definecolor{red}{rgb}{1,0,0}
\definecolor{blue}{rgb}{0,0,1}
\definecolor{green}{rgb}{0,1,0}
\definecolor{yellow}{rgb}{1,1,0}
\definecolor{orange}{rgb}{1,0.647,0}
\definecolor{gold}{rgb}{1,0.843,0}
\definecolor{purple}{rgb}{0.627,0.125,0.941}
\definecolor{gray}{rgb}{0.745,0.745,0.745}
\definecolor{brown}{rgb}{0.647,0.165,0.165}
\definecolor{navy}{rgb}{0,0,0.502}
\definecolor{pink}{rgb}{1,0.753,0.796}
\definecolor{seagreen}{rgb}{0.18,0.545,0.341}
\definecolor{turquoise}{rgb}{0.251,0.878,0.816}
\definecolor{violet}{rgb}{0.933,0.51,0.933}
\definecolor{darkblue}{rgb}{0,0,0.545}
\definecolor{darkcyan}{rgb}{0,0.545,0.545}
\definecolor{darkgray}{rgb}{0.663,0.663,0.663}
\definecolor{darkgreen}{rgb}{0,0.392,0}
\definecolor{darkmagenta}{rgb}{0.545,0,0.545}
\definecolor{darkorange}{rgb}{1,0.549,0}
\definecolor{darkred}{rgb}{0.545,0,0}
\definecolor{lightblue}{rgb}{0.678,0.847,0.902}
\definecolor{lightcyan}{rgb}{0.878,1,1}
\definecolor{lightgray}{rgb}{0.827,0.827,0.827}
\definecolor{lightgreen}{rgb}{0.565,0.933,0.565}
\definecolor{lightyellow}{rgb}{1,1,0.878}
\definecolor{aliceblue}{rgb}{0.941,0.973,1}
\definecolor{antiquewhite}{rgb}{0.98,0.922,0.843}
\definecolor{antiquewhite1}{rgb}{1,0.937,0.859}
\definecolor{antiquewhite2}{rgb}{0.933,0.875,0.8}
\definecolor{antiquewhite3}{rgb}{0.804,0.753,0.69}
\definecolor{antiquewhite4}{rgb}{0.545,0.514,0.471}
\definecolor{aquamarine}{rgb}{0.498,1,0.831}
\definecolor{aquamarine1}{rgb}{0.498,1,0.831}
\definecolor{aquamarine2}{rgb}{0.463,0.933,0.776}
\definecolor{aquamarine3}{rgb}{0.4,0.804,0.667}
\definecolor{aquamarine4}{rgb}{0.271,0.545,0.455}
\definecolor{azure}{rgb}{0.941,1,1}
\definecolor{azure1}{rgb}{0.941,1,1}
\definecolor{azure2}{rgb}{0.878,0.933,0.933}
\definecolor{azure3}{rgb}{0.757,0.804,0.804}
\definecolor{azure4}{rgb}{0.514,0.545,0.545}
\definecolor{beige}{rgb}{0.961,0.961,0.863}
\definecolor{bisque}{rgb}{1,0.894,0.769}
\definecolor{bisque1}{rgb}{1,0.894,0.769}
\definecolor{bisque2}{rgb}{0.933,0.835,0.718}
\definecolor{bisque3}{rgb}{0.804,0.718,0.62}
\definecolor{bisque4}{rgb}{0.545,0.49,0.42}
\definecolor{blanchedalmond}{rgb}{1,0.922,0.804}
\definecolor{blue1}{rgb}{0,0,1}
\definecolor{blue2}{rgb}{0,0,0.933}
\definecolor{blue3}{rgb}{0,0,0.804}
\definecolor{blue4}{rgb}{0,0,0.545}
\definecolor{blueviolet}{rgb}{0.541,0.169,0.886}
\definecolor{brown1}{rgb}{1,0.251,0.251}
\definecolor{brown2}{rgb}{0.933,0.231,0.231}
\definecolor{brown3}{rgb}{0.804,0.2,0.2}
\definecolor{brown4}{rgb}{0.545,0.137,0.137}
\definecolor{burlywood}{rgb}{0.871,0.722,0.529}
\definecolor{burlywood1}{rgb}{1,0.827,0.608}
\definecolor{burlywood2}{rgb}{0.933,0.773,0.569}
\definecolor{burlywood3}{rgb}{0.804,0.667,0.49}
\definecolor{burlywood4}{rgb}{0.545,0.451,0.333}
\definecolor{cadetblue}{rgb}{0.373,0.62,0.627}
\definecolor{cadetblue1}{rgb}{0.596,0.961,1}
\definecolor{cadetblue2}{rgb}{0.557,0.898,0.933}
\definecolor{cadetblue3}{rgb}{0.478,0.773,0.804}
\definecolor{cadetblue4}{rgb}{0.325,0.525,0.545}
\definecolor{chartreuse}{rgb}{0.498,1,0}
\definecolor{chartreuse1}{rgb}{0.498,1,0}
\definecolor{chartreuse2}{rgb}{0.463,0.933,0}
\definecolor{chartreuse3}{rgb}{0.4,0.804,0}
\definecolor{chartreuse4}{rgb}{0.271,0.545,0}
\definecolor{chocolate}{rgb}{0.824,0.412,0.118}
\definecolor{chocolate1}{rgb}{1,0.498,0.141}
\definecolor{chocolate2}{rgb}{0.933,0.463,0.129}
\definecolor{chocolate3}{rgb}{0.804,0.4,0.114}
\definecolor{chocolate4}{rgb}{0.545,0.271,0.075}
\definecolor{coral}{rgb}{1,0.498,0.314}
\definecolor{coral1}{rgb}{1,0.447,0.337}
\definecolor{coral2}{rgb}{0.933,0.416,0.314}
\definecolor{coral3}{rgb}{0.804,0.357,0.271}
\definecolor{coral4}{rgb}{0.545,0.243,0.184}
\definecolor{cornflowerblue}{rgb}{0.392,0.584,0.929}
\definecolor{cornsilk}{rgb}{1,0.973,0.863}
\definecolor{cornsilk1}{rgb}{1,0.973,0.863}
\definecolor{cornsilk2}{rgb}{0.933,0.91,0.804}
\definecolor{cornsilk3}{rgb}{0.804,0.784,0.694}
\definecolor{cornsilk4}{rgb}{0.545,0.533,0.471}
\definecolor{cyan}{rgb}{0,1,1}
\definecolor{cyan1}{rgb}{0,1,1}
\definecolor{cyan2}{rgb}{0,0.933,0.933}
\definecolor{cyan3}{rgb}{0,0.804,0.804}
\definecolor{cyan4}{rgb}{0,0.545,0.545}
\definecolor{darkgoldenrod}{rgb}{0.722,0.525,0.043}
\definecolor{darkgoldenrod1}{rgb}{1,0.725,0.059}
\definecolor{darkgoldenrod2}{rgb}{0.933,0.678,0.055}
\definecolor{darkgoldenrod3}{rgb}{0.804,0.584,0.047}
\definecolor{darkgoldenrod4}{rgb}{0.545,0.396,0.031}
\definecolor{darkgrey}{rgb}{0.663,0.663,0.663}
\definecolor{darkkhaki}{rgb}{0.741,0.718,0.42}
\definecolor{darkolivegreen}{rgb}{0.333,0.42,0.184}
\definecolor{darkolivegreen1}{rgb}{0.792,1,0.439}
\definecolor{darkolivegreen2}{rgb}{0.737,0.933,0.408}
\definecolor{darkolivegreen3}{rgb}{0.635,0.804,0.353}
\definecolor{darkolivegreen4}{rgb}{0.431,0.545,0.239}
\definecolor{darkorange1}{rgb}{1,0.498,0}
\definecolor{darkorange2}{rgb}{0.933,0.463,0}
\definecolor{darkorange3}{rgb}{0.804,0.4,0}
\definecolor{darkorange4}{rgb}{0.545,0.271,0}
\definecolor{darkorchid}{rgb}{0.6,0.196,0.8}
\definecolor{darkorchid1}{rgb}{0.749,0.243,1}
\definecolor{darkorchid2}{rgb}{0.698,0.227,0.933}
\definecolor{darkorchid3}{rgb}{0.604,0.196,0.804}
\definecolor{darkorchid4}{rgb}{0.408,0.133,0.545}
\definecolor{darksalmon}{rgb}{0.914,0.588,0.478}
\definecolor{darkseagreen}{rgb}{0.561,0.737,0.561}
\definecolor{darkseagreen1}{rgb}{0.757,1,0.757}
\definecolor{darkseagreen2}{rgb}{0.706,0.933,0.706}
\definecolor{darkseagreen3}{rgb}{0.608,0.804,0.608}
\definecolor{darkseagreen4}{rgb}{0.412,0.545,0.412}
\definecolor{darkslateblue}{rgb}{0.282,0.239,0.545}
\definecolor{darkslategray}{rgb}{0.184,0.31,0.31}
\definecolor{darkslategray1}{rgb}{0.592,1,1}
\definecolor{darkslategray2}{rgb}{0.553,0.933,0.933}
\definecolor{darkslategray3}{rgb}{0.475,0.804,0.804}
\definecolor{darkslategray4}{rgb}{0.322,0.545,0.545}
\definecolor{darkslategrey}{rgb}{0.184,0.31,0.31}
\definecolor{darkturquoise}{rgb}{0,0.808,0.82}
\definecolor{darkviolet}{rgb}{0.58,0,0.827}
\definecolor{deeppink}{rgb}{1,0.078,0.576}
\definecolor{deeppink1}{rgb}{1,0.078,0.576}
\definecolor{deeppink2}{rgb}{0.933,0.071,0.537}
\definecolor{deeppink3}{rgb}{0.804,0.063,0.463}
\definecolor{deeppink4}{rgb}{0.545,0.039,0.314}
\definecolor{deepskyblue}{rgb}{0,0.749,1}
\definecolor{deepskyblue1}{rgb}{0,0.749,1}
\definecolor{deepskyblue2}{rgb}{0,0.698,0.933}
\definecolor{deepskyblue3}{rgb}{0,0.604,0.804}
\definecolor{deepskyblue4}{rgb}{0,0.408,0.545}
\definecolor{dimgray}{rgb}{0.412,0.412,0.412}
\definecolor{dimgrey}{rgb}{0.412,0.412,0.412}
\definecolor{dodgerblue}{rgb}{0.118,0.565,1}
\definecolor{dodgerblue1}{rgb}{0.118,0.565,1}
\definecolor{dodgerblue2}{rgb}{0.11,0.525,0.933}
\definecolor{dodgerblue3}{rgb}{0.094,0.455,0.804}
\definecolor{dodgerblue4}{rgb}{0.063,0.306,0.545}
\definecolor{firebrick}{rgb}{0.698,0.133,0.133}
\definecolor{firebrick1}{rgb}{1,0.188,0.188}
\definecolor{firebrick2}{rgb}{0.933,0.173,0.173}
\definecolor{firebrick3}{rgb}{0.804,0.149,0.149}
\definecolor{firebrick4}{rgb}{0.545,0.102,0.102}
\definecolor{floralwhite}{rgb}{1,0.98,0.941}
\definecolor{forestgreen}{rgb}{0.133,0.545,0.133}
\definecolor{gainsboro}{rgb}{0.863,0.863,0.863}
\definecolor{ghostwhite}{rgb}{0.973,0.973,1}
\definecolor{gold1}{rgb}{1,0.843,0}
\definecolor{gold2}{rgb}{0.933,0.788,0}
\definecolor{gold3}{rgb}{0.804,0.678,0}
\definecolor{gold4}{rgb}{0.545,0.459,0}
\definecolor{goldenrod}{rgb}{0.855,0.647,0.125}
\definecolor{goldenrod1}{rgb}{1,0.757,0.145}
\definecolor{goldenrod2}{rgb}{0.933,0.706,0.133}
\definecolor{goldenrod3}{rgb}{0.804,0.608,0.114}
\definecolor{goldenrod4}{rgb}{0.545,0.412,0.078}
\definecolor{gray0}{rgb}{0,0,0}
\definecolor{gray1}{rgb}{0.012,0.012,0.012}
\definecolor{gray10}{rgb}{0.102,0.102,0.102}
\definecolor{gray100}{rgb}{1,1,1}
\definecolor{gray11}{rgb}{0.11,0.11,0.11}
\definecolor{gray12}{rgb}{0.122,0.122,0.122}
\definecolor{gray13}{rgb}{0.129,0.129,0.129}
\definecolor{gray14}{rgb}{0.141,0.141,0.141}
\definecolor{gray15}{rgb}{0.149,0.149,0.149}
\definecolor{gray16}{rgb}{0.161,0.161,0.161}
\definecolor{gray17}{rgb}{0.169,0.169,0.169}
\definecolor{gray18}{rgb}{0.18,0.18,0.18}
\definecolor{gray19}{rgb}{0.188,0.188,0.188}
\definecolor{gray2}{rgb}{0.02,0.02,0.02}
\definecolor{gray20}{rgb}{0.2,0.2,0.2}
\definecolor{gray21}{rgb}{0.212,0.212,0.212}
\definecolor{gray22}{rgb}{0.22,0.22,0.22}
\definecolor{gray23}{rgb}{0.231,0.231,0.231}
\definecolor{gray24}{rgb}{0.239,0.239,0.239}
\definecolor{gray25}{rgb}{0.251,0.251,0.251}
\definecolor{gray26}{rgb}{0.259,0.259,0.259}
\definecolor{gray27}{rgb}{0.271,0.271,0.271}
\definecolor{gray28}{rgb}{0.278,0.278,0.278}
\definecolor{gray29}{rgb}{0.29,0.29,0.29}
\definecolor{gray3}{rgb}{0.031,0.031,0.031}
\definecolor{gray30}{rgb}{0.302,0.302,0.302}
\definecolor{gray31}{rgb}{0.31,0.31,0.31}
\definecolor{gray32}{rgb}{0.322,0.322,0.322}
\definecolor{gray33}{rgb}{0.329,0.329,0.329}
\definecolor{gray34}{rgb}{0.341,0.341,0.341}
\definecolor{gray35}{rgb}{0.349,0.349,0.349}
\definecolor{gray36}{rgb}{0.361,0.361,0.361}
\definecolor{gray37}{rgb}{0.369,0.369,0.369}
\definecolor{gray38}{rgb}{0.38,0.38,0.38}
\definecolor{gray39}{rgb}{0.388,0.388,0.388}
\definecolor{gray4}{rgb}{0.039,0.039,0.039}
\definecolor{gray40}{rgb}{0.4,0.4,0.4}
\definecolor{gray41}{rgb}{0.412,0.412,0.412}
\definecolor{gray42}{rgb}{0.42,0.42,0.42}
\definecolor{gray43}{rgb}{0.431,0.431,0.431}
\definecolor{gray44}{rgb}{0.439,0.439,0.439}
\definecolor{gray45}{rgb}{0.451,0.451,0.451}
\definecolor{gray46}{rgb}{0.459,0.459,0.459}
\definecolor{gray47}{rgb}{0.471,0.471,0.471}
\definecolor{gray48}{rgb}{0.478,0.478,0.478}
\definecolor{gray49}{rgb}{0.49,0.49,0.49}
\definecolor{gray5}{rgb}{0.051,0.051,0.051}
\definecolor{gray50}{rgb}{0.498,0.498,0.498}
\definecolor{gray51}{rgb}{0.51,0.51,0.51}
\definecolor{gray52}{rgb}{0.522,0.522,0.522}
\definecolor{gray53}{rgb}{0.529,0.529,0.529}
\definecolor{gray54}{rgb}{0.541,0.541,0.541}
\definecolor{gray55}{rgb}{0.549,0.549,0.549}
\definecolor{gray56}{rgb}{0.561,0.561,0.561}
\definecolor{gray57}{rgb}{0.569,0.569,0.569}
\definecolor{gray58}{rgb}{0.58,0.58,0.58}
\definecolor{gray59}{rgb}{0.588,0.588,0.588}
\definecolor{gray6}{rgb}{0.059,0.059,0.059}
\definecolor{gray60}{rgb}{0.6,0.6,0.6}
\definecolor{gray61}{rgb}{0.612,0.612,0.612}
\definecolor{gray62}{rgb}{0.62,0.62,0.62}
\definecolor{gray63}{rgb}{0.631,0.631,0.631}
\definecolor{gray64}{rgb}{0.639,0.639,0.639}
\definecolor{gray65}{rgb}{0.651,0.651,0.651}
\definecolor{gray66}{rgb}{0.659,0.659,0.659}
\definecolor{gray67}{rgb}{0.671,0.671,0.671}
\definecolor{gray68}{rgb}{0.678,0.678,0.678}
\definecolor{gray69}{rgb}{0.69,0.69,0.69}
\definecolor{gray7}{rgb}{0.071,0.071,0.071}
\definecolor{gray70}{rgb}{0.702,0.702,0.702}
\definecolor{gray71}{rgb}{0.71,0.71,0.71}
\definecolor{gray72}{rgb}{0.722,0.722,0.722}
\definecolor{gray73}{rgb}{0.729,0.729,0.729}
\definecolor{gray74}{rgb}{0.741,0.741,0.741}
\definecolor{gray75}{rgb}{0.749,0.749,0.749}
\definecolor{gray76}{rgb}{0.761,0.761,0.761}
\definecolor{gray77}{rgb}{0.769,0.769,0.769}
\definecolor{gray78}{rgb}{0.78,0.78,0.78}
\definecolor{gray79}{rgb}{0.788,0.788,0.788}
\definecolor{gray8}{rgb}{0.078,0.078,0.078}
\definecolor{gray80}{rgb}{0.8,0.8,0.8}
\definecolor{gray81}{rgb}{0.812,0.812,0.812}
\definecolor{gray82}{rgb}{0.82,0.82,0.82}
\definecolor{gray83}{rgb}{0.831,0.831,0.831}
\definecolor{gray84}{rgb}{0.839,0.839,0.839}
\definecolor{gray85}{rgb}{0.851,0.851,0.851}
\definecolor{gray86}{rgb}{0.859,0.859,0.859}
\definecolor{gray87}{rgb}{0.871,0.871,0.871}
\definecolor{gray88}{rgb}{0.878,0.878,0.878}
\definecolor{gray89}{rgb}{0.89,0.89,0.89}
\definecolor{gray9}{rgb}{0.09,0.09,0.09}
\definecolor{gray90}{rgb}{0.898,0.898,0.898}
\definecolor{gray91}{rgb}{0.91,0.91,0.91}
\definecolor{gray92}{rgb}{0.922,0.922,0.922}
\definecolor{gray93}{rgb}{0.929,0.929,0.929}
\definecolor{gray94}{rgb}{0.941,0.941,0.941}
\definecolor{gray95}{rgb}{0.949,0.949,0.949}
\definecolor{gray96}{rgb}{0.961,0.961,0.961}
\definecolor{gray97}{rgb}{0.969,0.969,0.969}
\definecolor{gray98}{rgb}{0.98,0.98,0.98}
\definecolor{gray99}{rgb}{0.988,0.988,0.988}
\definecolor{green1}{rgb}{0,1,0}
\definecolor{green2}{rgb}{0,0.933,0}
\definecolor{green3}{rgb}{0,0.804,0}
\definecolor{green4}{rgb}{0,0.545,0}
\definecolor{greenyellow}{rgb}{0.678,1,0.184}
\definecolor{grey}{rgb}{0.745,0.745,0.745}
\definecolor{grey0}{rgb}{0,0,0}
\definecolor{grey1}{rgb}{0.012,0.012,0.012}
\definecolor{grey10}{rgb}{0.102,0.102,0.102}
\definecolor{grey100}{rgb}{1,1,1}
\definecolor{grey11}{rgb}{0.11,0.11,0.11}
\definecolor{grey12}{rgb}{0.122,0.122,0.122}
\definecolor{grey13}{rgb}{0.129,0.129,0.129}
\definecolor{grey14}{rgb}{0.141,0.141,0.141}
\definecolor{grey15}{rgb}{0.149,0.149,0.149}
\definecolor{grey16}{rgb}{0.161,0.161,0.161}
\definecolor{grey17}{rgb}{0.169,0.169,0.169}
\definecolor{grey18}{rgb}{0.18,0.18,0.18}
\definecolor{grey19}{rgb}{0.188,0.188,0.188}
\definecolor{grey2}{rgb}{0.02,0.02,0.02}
\definecolor{grey20}{rgb}{0.2,0.2,0.2}
\definecolor{grey21}{rgb}{0.212,0.212,0.212}
\definecolor{grey22}{rgb}{0.22,0.22,0.22}
\definecolor{grey23}{rgb}{0.231,0.231,0.231}
\definecolor{grey24}{rgb}{0.239,0.239,0.239}
\definecolor{grey25}{rgb}{0.251,0.251,0.251}
\definecolor{grey26}{rgb}{0.259,0.259,0.259}
\definecolor{grey27}{rgb}{0.271,0.271,0.271}
\definecolor{grey28}{rgb}{0.278,0.278,0.278}
\definecolor{grey29}{rgb}{0.29,0.29,0.29}
\definecolor{grey3}{rgb}{0.031,0.031,0.031}
\definecolor{grey30}{rgb}{0.302,0.302,0.302}
\definecolor{grey31}{rgb}{0.31,0.31,0.31}
\definecolor{grey32}{rgb}{0.322,0.322,0.322}
\definecolor{grey33}{rgb}{0.329,0.329,0.329}
\definecolor{grey34}{rgb}{0.341,0.341,0.341}
\definecolor{grey35}{rgb}{0.349,0.349,0.349}
\definecolor{grey36}{rgb}{0.361,0.361,0.361}
\definecolor{grey37}{rgb}{0.369,0.369,0.369}
\definecolor{grey38}{rgb}{0.38,0.38,0.38}
\definecolor{grey39}{rgb}{0.388,0.388,0.388}
\definecolor{grey4}{rgb}{0.039,0.039,0.039}
\definecolor{grey40}{rgb}{0.4,0.4,0.4}
\definecolor{grey41}{rgb}{0.412,0.412,0.412}
\definecolor{grey42}{rgb}{0.42,0.42,0.42}
\definecolor{grey43}{rgb}{0.431,0.431,0.431}
\definecolor{grey44}{rgb}{0.439,0.439,0.439}
\definecolor{grey45}{rgb}{0.451,0.451,0.451}
\definecolor{grey46}{rgb}{0.459,0.459,0.459}
\definecolor{grey47}{rgb}{0.471,0.471,0.471}
\definecolor{grey48}{rgb}{0.478,0.478,0.478}
\definecolor{grey49}{rgb}{0.49,0.49,0.49}
\definecolor{grey5}{rgb}{0.051,0.051,0.051}
\definecolor{grey50}{rgb}{0.498,0.498,0.498}
\definecolor{grey51}{rgb}{0.51,0.51,0.51}
\definecolor{grey52}{rgb}{0.522,0.522,0.522}
\definecolor{grey53}{rgb}{0.529,0.529,0.529}
\definecolor{grey54}{rgb}{0.541,0.541,0.541}
\definecolor{grey55}{rgb}{0.549,0.549,0.549}
\definecolor{grey56}{rgb}{0.561,0.561,0.561}
\definecolor{grey57}{rgb}{0.569,0.569,0.569}
\definecolor{grey58}{rgb}{0.58,0.58,0.58}
\definecolor{grey59}{rgb}{0.588,0.588,0.588}
\definecolor{grey6}{rgb}{0.059,0.059,0.059}
\definecolor{grey60}{rgb}{0.6,0.6,0.6}
\definecolor{grey61}{rgb}{0.612,0.612,0.612}
\definecolor{grey62}{rgb}{0.62,0.62,0.62}
\definecolor{grey63}{rgb}{0.631,0.631,0.631}
\definecolor{grey64}{rgb}{0.639,0.639,0.639}
\definecolor{grey65}{rgb}{0.651,0.651,0.651}
\definecolor{grey66}{rgb}{0.659,0.659,0.659}
\definecolor{grey67}{rgb}{0.671,0.671,0.671}
\definecolor{grey68}{rgb}{0.678,0.678,0.678}
\definecolor{grey69}{rgb}{0.69,0.69,0.69}
\definecolor{grey7}{rgb}{0.071,0.071,0.071}
\definecolor{grey70}{rgb}{0.702,0.702,0.702}
\definecolor{grey71}{rgb}{0.71,0.71,0.71}
\definecolor{grey72}{rgb}{0.722,0.722,0.722}
\definecolor{grey73}{rgb}{0.729,0.729,0.729}
\definecolor{grey74}{rgb}{0.741,0.741,0.741}
\definecolor{grey75}{rgb}{0.749,0.749,0.749}
\definecolor{grey76}{rgb}{0.761,0.761,0.761}
\definecolor{grey77}{rgb}{0.769,0.769,0.769}
\definecolor{grey78}{rgb}{0.78,0.78,0.78}
\definecolor{grey79}{rgb}{0.788,0.788,0.788}
\definecolor{grey8}{rgb}{0.078,0.078,0.078}
\definecolor{grey80}{rgb}{0.8,0.8,0.8}
\definecolor{grey81}{rgb}{0.812,0.812,0.812}
\definecolor{grey82}{rgb}{0.82,0.82,0.82}
\definecolor{grey83}{rgb}{0.831,0.831,0.831}
\definecolor{grey84}{rgb}{0.839,0.839,0.839}
\definecolor{grey85}{rgb}{0.851,0.851,0.851}
\definecolor{grey86}{rgb}{0.859,0.859,0.859}
\definecolor{grey87}{rgb}{0.871,0.871,0.871}
\definecolor{grey88}{rgb}{0.878,0.878,0.878}
\definecolor{grey89}{rgb}{0.89,0.89,0.89}
\definecolor{grey9}{rgb}{0.09,0.09,0.09}
\definecolor{grey90}{rgb}{0.898,0.898,0.898}
\definecolor{grey91}{rgb}{0.91,0.91,0.91}
\definecolor{grey92}{rgb}{0.922,0.922,0.922}
\definecolor{grey93}{rgb}{0.929,0.929,0.929}
\definecolor{grey94}{rgb}{0.941,0.941,0.941}
\definecolor{grey95}{rgb}{0.949,0.949,0.949}
\definecolor{grey96}{rgb}{0.961,0.961,0.961}
\definecolor{grey97}{rgb}{0.969,0.969,0.969}
\definecolor{grey98}{rgb}{0.98,0.98,0.98}
\definecolor{grey99}{rgb}{0.988,0.988,0.988}
\definecolor{honeydew}{rgb}{0.941,1,0.941}
\definecolor{honeydew1}{rgb}{0.941,1,0.941}
\definecolor{honeydew2}{rgb}{0.878,0.933,0.878}
\definecolor{honeydew3}{rgb}{0.757,0.804,0.757}
\definecolor{honeydew4}{rgb}{0.514,0.545,0.514}
\definecolor{hotpink}{rgb}{1,0.412,0.706}
\definecolor{hotpink1}{rgb}{1,0.431,0.706}
\definecolor{hotpink2}{rgb}{0.933,0.416,0.655}
\definecolor{hotpink3}{rgb}{0.804,0.376,0.565}
\definecolor{hotpink4}{rgb}{0.545,0.227,0.384}
\definecolor{indianred}{rgb}{0.804,0.361,0.361}
\definecolor{indianred1}{rgb}{1,0.416,0.416}
\definecolor{indianred2}{rgb}{0.933,0.388,0.388}
\definecolor{indianred3}{rgb}{0.804,0.333,0.333}
\definecolor{indianred4}{rgb}{0.545,0.227,0.227}
\definecolor{ivory}{rgb}{1,1,0.941}
\definecolor{ivory1}{rgb}{1,1,0.941}
\definecolor{ivory2}{rgb}{0.933,0.933,0.878}
\definecolor{ivory3}{rgb}{0.804,0.804,0.757}
\definecolor{ivory4}{rgb}{0.545,0.545,0.514}
\definecolor{khaki}{rgb}{0.941,0.902,0.549}
\definecolor{khaki1}{rgb}{1,0.965,0.561}
\definecolor{khaki2}{rgb}{0.933,0.902,0.522}
\definecolor{khaki3}{rgb}{0.804,0.776,0.451}
\definecolor{khaki4}{rgb}{0.545,0.525,0.306}
\definecolor{lavender}{rgb}{0.902,0.902,0.98}
\definecolor{lavenderblush}{rgb}{1,0.941,0.961}
\definecolor{lavenderblush1}{rgb}{1,0.941,0.961}
\definecolor{lavenderblush2}{rgb}{0.933,0.878,0.898}
\definecolor{lavenderblush3}{rgb}{0.804,0.757,0.773}
\definecolor{lavenderblush4}{rgb}{0.545,0.514,0.525}
\definecolor{lawngreen}{rgb}{0.486,0.988,0}
\definecolor{lemonchiffon}{rgb}{1,0.98,0.804}
\definecolor{lemonchiffon1}{rgb}{1,0.98,0.804}
\definecolor{lemonchiffon2}{rgb}{0.933,0.914,0.749}
\definecolor{lemonchiffon3}{rgb}{0.804,0.788,0.647}
\definecolor{lemonchiffon4}{rgb}{0.545,0.537,0.439}
\definecolor{lightblue1}{rgb}{0.749,0.937,1}
\definecolor{lightblue2}{rgb}{0.698,0.875,0.933}
\definecolor{lightblue3}{rgb}{0.604,0.753,0.804}
\definecolor{lightblue4}{rgb}{0.408,0.514,0.545}
\definecolor{lightcoral}{rgb}{0.941,0.502,0.502}
\definecolor{lightcyan1}{rgb}{0.878,1,1}
\definecolor{lightcyan2}{rgb}{0.82,0.933,0.933}
\definecolor{lightcyan3}{rgb}{0.706,0.804,0.804}
\definecolor{lightcyan4}{rgb}{0.478,0.545,0.545}
\definecolor{lightgoldenrod}{rgb}{0.933,0.867,0.51}
\definecolor{lightgoldenrod1}{rgb}{1,0.925,0.545}
\definecolor{lightgoldenrod2}{rgb}{0.933,0.863,0.51}
\definecolor{lightgoldenrod3}{rgb}{0.804,0.745,0.439}
\definecolor{lightgoldenrod4}{rgb}{0.545,0.506,0.298}
\definecolor{lightgoldenrodyellow}{rgb}{0.98,0.98,0.824}
\definecolor{lightgrey}{rgb}{0.827,0.827,0.827}
\definecolor{lightpink}{rgb}{1,0.714,0.757}
\definecolor{lightpink1}{rgb}{1,0.682,0.725}
\definecolor{lightpink2}{rgb}{0.933,0.635,0.678}
\definecolor{lightpink3}{rgb}{0.804,0.549,0.584}
\definecolor{lightpink4}{rgb}{0.545,0.373,0.396}
\definecolor{lightsalmon}{rgb}{1,0.627,0.478}
\definecolor{lightsalmon1}{rgb}{1,0.627,0.478}
\definecolor{lightsalmon2}{rgb}{0.933,0.584,0.447}
\definecolor{lightsalmon3}{rgb}{0.804,0.506,0.384}
\definecolor{lightsalmon4}{rgb}{0.545,0.341,0.259}
\definecolor{lightseagreen}{rgb}{0.125,0.698,0.667}
\definecolor{lightskyblue}{rgb}{0.529,0.808,0.98}
\definecolor{lightskyblue1}{rgb}{0.69,0.886,1}
\definecolor{lightskyblue2}{rgb}{0.643,0.827,0.933}
\definecolor{lightskyblue3}{rgb}{0.553,0.714,0.804}
\definecolor{lightskyblue4}{rgb}{0.376,0.482,0.545}
\definecolor{lightslateblue}{rgb}{0.518,0.439,1}
\definecolor{lightslategray}{rgb}{0.467,0.533,0.6}
\definecolor{lightslategrey}{rgb}{0.467,0.533,0.6}
\definecolor{lightsteelblue}{rgb}{0.69,0.769,0.871}
\definecolor{lightsteelblue1}{rgb}{0.792,0.882,1}
\definecolor{lightsteelblue2}{rgb}{0.737,0.824,0.933}
\definecolor{lightsteelblue3}{rgb}{0.635,0.71,0.804}
\definecolor{lightsteelblue4}{rgb}{0.431,0.482,0.545}
\definecolor{lightyellow1}{rgb}{1,1,0.878}
\definecolor{lightyellow2}{rgb}{0.933,0.933,0.82}
\definecolor{lightyellow3}{rgb}{0.804,0.804,0.706}
\definecolor{lightyellow4}{rgb}{0.545,0.545,0.478}
\definecolor{limegreen}{rgb}{0.196,0.804,0.196}
\definecolor{linen}{rgb}{0.98,0.941,0.902}
\definecolor{magenta}{rgb}{1,0,1}
\definecolor{magenta1}{rgb}{1,0,1}
\definecolor{magenta2}{rgb}{0.933,0,0.933}
\definecolor{magenta3}{rgb}{0.804,0,0.804}
\definecolor{magenta4}{rgb}{0.545,0,0.545}
\definecolor{maroon}{rgb}{0.69,0.188,0.376}
\definecolor{maroon1}{rgb}{1,0.204,0.702}
\definecolor{maroon2}{rgb}{0.933,0.188,0.655}
\definecolor{maroon3}{rgb}{0.804,0.161,0.565}
\definecolor{maroon4}{rgb}{0.545,0.11,0.384}
\definecolor{mediumaquamarine}{rgb}{0.4,0.804,0.667}
\definecolor{mediumblue}{rgb}{0,0,0.804}
\definecolor{mediumorchid}{rgb}{0.729,0.333,0.827}
\definecolor{mediumorchid1}{rgb}{0.878,0.4,1}
\definecolor{mediumorchid2}{rgb}{0.82,0.373,0.933}
\definecolor{mediumorchid3}{rgb}{0.706,0.322,0.804}
\definecolor{mediumorchid4}{rgb}{0.478,0.216,0.545}
\definecolor{mediumpurple}{rgb}{0.576,0.439,0.859}
\definecolor{mediumpurple1}{rgb}{0.671,0.51,1}
\definecolor{mediumpurple2}{rgb}{0.624,0.475,0.933}
\definecolor{mediumpurple3}{rgb}{0.537,0.408,0.804}
\definecolor{mediumpurple4}{rgb}{0.365,0.278,0.545}
\definecolor{mediumseagreen}{rgb}{0.235,0.702,0.443}
\definecolor{mediumslateblue}{rgb}{0.482,0.408,0.933}
\definecolor{mediumspringgreen}{rgb}{0,0.98,0.604}
\definecolor{mediumturquoise}{rgb}{0.282,0.82,0.8}
\definecolor{mediumvioletred}{rgb}{0.78,0.082,0.522}
\definecolor{midnightblue}{rgb}{0.098,0.098,0.439}
\definecolor{mintcream}{rgb}{0.961,1,0.98}
\definecolor{mistyrose}{rgb}{1,0.894,0.882}
\definecolor{mistyrose1}{rgb}{1,0.894,0.882}
\definecolor{mistyrose2}{rgb}{0.933,0.835,0.824}
\definecolor{mistyrose3}{rgb}{0.804,0.718,0.71}
\definecolor{mistyrose4}{rgb}{0.545,0.49,0.482}
\definecolor{moccasin}{rgb}{1,0.894,0.71}
\definecolor{navajowhite}{rgb}{1,0.871,0.678}
\definecolor{navajowhite1}{rgb}{1,0.871,0.678}
\definecolor{navajowhite2}{rgb}{0.933,0.812,0.631}
\definecolor{navajowhite3}{rgb}{0.804,0.702,0.545}
\definecolor{navajowhite4}{rgb}{0.545,0.475,0.369}
\definecolor{navyblue}{rgb}{0,0,0.502}
\definecolor{oldlace}{rgb}{0.992,0.961,0.902}
\definecolor{olivedrab}{rgb}{0.42,0.557,0.137}
\definecolor{olivedrab1}{rgb}{0.753,1,0.243}
\definecolor{olivedrab2}{rgb}{0.702,0.933,0.227}
\definecolor{olivedrab3}{rgb}{0.604,0.804,0.196}
\definecolor{olivedrab4}{rgb}{0.412,0.545,0.133}
\definecolor{orange1}{rgb}{1,0.647,0}
\definecolor{orange2}{rgb}{0.933,0.604,0}
\definecolor{orange3}{rgb}{0.804,0.522,0}
\definecolor{orange4}{rgb}{0.545,0.353,0}
\definecolor{orangered}{rgb}{1,0.271,0}
\definecolor{orangered1}{rgb}{1,0.271,0}
\definecolor{orangered2}{rgb}{0.933,0.251,0}
\definecolor{orangered3}{rgb}{0.804,0.216,0}
\definecolor{orangered4}{rgb}{0.545,0.145,0}
\definecolor{orchid}{rgb}{0.855,0.439,0.839}
\definecolor{orchid1}{rgb}{1,0.514,0.98}
\definecolor{orchid2}{rgb}{0.933,0.478,0.914}
\definecolor{orchid3}{rgb}{0.804,0.412,0.788}
\definecolor{orchid4}{rgb}{0.545,0.278,0.537}
\definecolor{palegoldenrod}{rgb}{0.933,0.91,0.667}
\definecolor{palegreen}{rgb}{0.596,0.984,0.596}
\definecolor{palegreen1}{rgb}{0.604,1,0.604}
\definecolor{palegreen2}{rgb}{0.565,0.933,0.565}
\definecolor{palegreen3}{rgb}{0.486,0.804,0.486}
\definecolor{palegreen4}{rgb}{0.329,0.545,0.329}
\definecolor{paleturquoise}{rgb}{0.686,0.933,0.933}
\definecolor{paleturquoise1}{rgb}{0.733,1,1}
\definecolor{paleturquoise2}{rgb}{0.682,0.933,0.933}
\definecolor{paleturquoise3}{rgb}{0.588,0.804,0.804}
\definecolor{paleturquoise4}{rgb}{0.4,0.545,0.545}
\definecolor{palevioletred}{rgb}{0.859,0.439,0.576}
\definecolor{palevioletred1}{rgb}{1,0.51,0.671}
\definecolor{palevioletred2}{rgb}{0.933,0.475,0.624}
\definecolor{palevioletred3}{rgb}{0.804,0.408,0.537}
\definecolor{palevioletred4}{rgb}{0.545,0.278,0.365}
\definecolor{papayawhip}{rgb}{1,0.937,0.835}
\definecolor{peachpuff}{rgb}{1,0.855,0.725}
\definecolor{peachpuff1}{rgb}{1,0.855,0.725}
\definecolor{peachpuff2}{rgb}{0.933,0.796,0.678}
\definecolor{peachpuff3}{rgb}{0.804,0.686,0.584}
\definecolor{peachpuff4}{rgb}{0.545,0.467,0.396}
\definecolor{peru}{rgb}{0.804,0.522,0.247}
\definecolor{pink1}{rgb}{1,0.71,0.773}
\definecolor{pink2}{rgb}{0.933,0.663,0.722}
\definecolor{pink3}{rgb}{0.804,0.569,0.62}
\definecolor{pink4}{rgb}{0.545,0.388,0.424}
\definecolor{plum}{rgb}{0.867,0.627,0.867}
\definecolor{plum1}{rgb}{1,0.733,1}
\definecolor{plum2}{rgb}{0.933,0.682,0.933}
\definecolor{plum3}{rgb}{0.804,0.588,0.804}
\definecolor{plum4}{rgb}{0.545,0.4,0.545}
\definecolor{powderblue}{rgb}{0.69,0.878,0.902}
\definecolor{purple1}{rgb}{0.608,0.188,1}
\definecolor{purple2}{rgb}{0.569,0.173,0.933}
\definecolor{purple3}{rgb}{0.49,0.149,0.804}
\definecolor{purple4}{rgb}{0.333,0.102,0.545}
\definecolor{red1}{rgb}{1,0,0}
\definecolor{red2}{rgb}{0.933,0,0}
\definecolor{red3}{rgb}{0.804,0,0}
\definecolor{red4}{rgb}{0.545,0,0}
\definecolor{rosybrown}{rgb}{0.737,0.561,0.561}
\definecolor{rosybrown1}{rgb}{1,0.757,0.757}
\definecolor{rosybrown2}{rgb}{0.933,0.706,0.706}
\definecolor{rosybrown3}{rgb}{0.804,0.608,0.608}
\definecolor{rosybrown4}{rgb}{0.545,0.412,0.412}
\definecolor{royalblue}{rgb}{0.255,0.412,0.882}
\definecolor{royalblue1}{rgb}{0.282,0.463,1}
\definecolor{royalblue2}{rgb}{0.263,0.431,0.933}
\definecolor{royalblue3}{rgb}{0.227,0.373,0.804}
\definecolor{royalblue4}{rgb}{0.153,0.251,0.545}
\definecolor{saddlebrown}{rgb}{0.545,0.271,0.075}
\definecolor{salmon}{rgb}{0.98,0.502,0.447}
\definecolor{salmon1}{rgb}{1,0.549,0.412}
\definecolor{salmon2}{rgb}{0.933,0.51,0.384}
\definecolor{salmon3}{rgb}{0.804,0.439,0.329}
\definecolor{salmon4}{rgb}{0.545,0.298,0.224}
\definecolor{sandybrown}{rgb}{0.957,0.643,0.376}
\definecolor{seagreen1}{rgb}{0.329,1,0.624}
\definecolor{seagreen2}{rgb}{0.306,0.933,0.58}
\definecolor{seagreen3}{rgb}{0.263,0.804,0.502}
\definecolor{seagreen4}{rgb}{0.18,0.545,0.341}
\definecolor{seashell}{rgb}{1,0.961,0.933}
\definecolor{seashell1}{rgb}{1,0.961,0.933}
\definecolor{seashell2}{rgb}{0.933,0.898,0.871}
\definecolor{seashell3}{rgb}{0.804,0.773,0.749}
\definecolor{seashell4}{rgb}{0.545,0.525,0.51}
\definecolor{sienna}{rgb}{0.627,0.322,0.176}
\definecolor{sienna1}{rgb}{1,0.51,0.278}
\definecolor{sienna2}{rgb}{0.933,0.475,0.259}
\definecolor{sienna3}{rgb}{0.804,0.408,0.224}
\definecolor{sienna4}{rgb}{0.545,0.278,0.149}
\definecolor{skyblue}{rgb}{0.529,0.808,0.922}
\definecolor{skyblue1}{rgb}{0.529,0.808,1}
\definecolor{skyblue2}{rgb}{0.494,0.753,0.933}
\definecolor{skyblue3}{rgb}{0.424,0.651,0.804}
\definecolor{skyblue4}{rgb}{0.29,0.439,0.545}
\definecolor{slateblue}{rgb}{0.416,0.353,0.804}
\definecolor{slateblue1}{rgb}{0.514,0.435,1}
\definecolor{slateblue2}{rgb}{0.478,0.404,0.933}
\definecolor{slateblue3}{rgb}{0.412,0.349,0.804}
\definecolor{slateblue4}{rgb}{0.278,0.235,0.545}
\definecolor{slategray}{rgb}{0.439,0.502,0.565}
\definecolor{slategray1}{rgb}{0.776,0.886,1}
\definecolor{slategray2}{rgb}{0.725,0.827,0.933}
\definecolor{slategray3}{rgb}{0.624,0.714,0.804}
\definecolor{slategray4}{rgb}{0.424,0.482,0.545}
\definecolor{slategrey}{rgb}{0.439,0.502,0.565}
\definecolor{snow}{rgb}{1,0.98,0.98}
\definecolor{snow1}{rgb}{1,0.98,0.98}
\definecolor{snow2}{rgb}{0.933,0.914,0.914}
\definecolor{snow3}{rgb}{0.804,0.788,0.788}
\definecolor{snow4}{rgb}{0.545,0.537,0.537}
\definecolor{springgreen}{rgb}{0,1,0.498}
\definecolor{springgreen1}{rgb}{0,1,0.498}
\definecolor{springgreen2}{rgb}{0,0.933,0.463}
\definecolor{springgreen3}{rgb}{0,0.804,0.4}
\definecolor{springgreen4}{rgb}{0,0.545,0.271}
\definecolor{steelblue}{rgb}{0.275,0.51,0.706}
\definecolor{steelblue1}{rgb}{0.388,0.722,1}
\definecolor{steelblue2}{rgb}{0.361,0.675,0.933}
\definecolor{steelblue3}{rgb}{0.31,0.58,0.804}
\definecolor{steelblue4}{rgb}{0.212,0.392,0.545}
\definecolor{tan}{rgb}{0.824,0.706,0.549}
\definecolor{tan1}{rgb}{1,0.647,0.31}
\definecolor{tan2}{rgb}{0.933,0.604,0.286}
\definecolor{tan3}{rgb}{0.804,0.522,0.247}
\definecolor{tan4}{rgb}{0.545,0.353,0.169}
\definecolor{thistle}{rgb}{0.847,0.749,0.847}
\definecolor{thistle1}{rgb}{1,0.882,1}
\definecolor{thistle2}{rgb}{0.933,0.824,0.933}
\definecolor{thistle3}{rgb}{0.804,0.71,0.804}
\definecolor{thistle4}{rgb}{0.545,0.482,0.545}
\definecolor{tomato}{rgb}{1,0.388,0.278}
\definecolor{tomato1}{rgb}{1,0.388,0.278}
\definecolor{tomato2}{rgb}{0.933,0.361,0.259}
\definecolor{tomato3}{rgb}{0.804,0.31,0.224}
\definecolor{tomato4}{rgb}{0.545,0.212,0.149}
\definecolor{turquoise1}{rgb}{0,0.961,1}
\definecolor{turquoise2}{rgb}{0,0.898,0.933}
\definecolor{turquoise3}{rgb}{0,0.773,0.804}
\definecolor{turquoise4}{rgb}{0,0.525,0.545}
\definecolor{violetred}{rgb}{0.816,0.125,0.565}
\definecolor{violetred1}{rgb}{1,0.243,0.588}
\definecolor{violetred2}{rgb}{0.933,0.227,0.549}
\definecolor{violetred3}{rgb}{0.804,0.196,0.471}
\definecolor{violetred4}{rgb}{0.545,0.133,0.322}
\definecolor{wheat}{rgb}{0.961,0.871,0.702}
\definecolor{wheat1}{rgb}{1,0.906,0.729}
\definecolor{wheat2}{rgb}{0.933,0.847,0.682}
\definecolor{wheat3}{rgb}{0.804,0.729,0.588}
\definecolor{wheat4}{rgb}{0.545,0.494,0.4}
\definecolor{whitesmoke}{rgb}{0.961,0.961,0.961}
\definecolor{yellow1}{rgb}{1,1,0}
\definecolor{yellow2}{rgb}{0.933,0.933,0}
\definecolor{yellow3}{rgb}{0.804,0.804,0}
\definecolor{yellow4}{rgb}{0.545,0.545,0}
\definecolor{yellowgreen}{rgb}{0.604,0.804,0.196}

\caption{Left: The Gaifman graph of $(\mathfrak{A}_{\rm out}^\star, R\setminus (\breve{C}\cup Z),\emptyset^{2q} ,\varnothing^l,X_{\rm out}, {\bf b}_1^\star ) \oplus (\mathfrak{A}^\star, R_1^{' \star},{\bf W}_{{q}}^{(1)} ,\varnothing^l, \tilde{X}^\star_1, {\bf b}_1^\star).$
Right: The Gaifman graph of $\mathfrak{A}_{\rm out}^\star, R\setminus (\breve{C}\cup Z),\emptyset^{2q} ,\varnothing^l,X_{\rm out}, {\bf b}_1^\star ) \oplus \big(\mathfrak{A}^{\star},R_2^{'\star},{\bf W}_{{q}}^{(2)},\varnothing^l, \tilde{X}^{\star}_2, {\bf b}_2^\star\big).$}
\labels{figure_boundariedgraph5}
\end{figure}

To see why this holds, note that  the $h$-boundaried $τ'$-structures $(\mathfrak{A}_{\rm out}^\star, R\setminus (\breve{C}\cup Z),\emptyset^{2q} ,\varnothing^l,X_{\rm out}, {\bf b}_1^\star ) $ and $\big(\mathfrak{A}^{\star},R_2^{'\star},{\bf W}_{{q}}^{(2)}, \varnothing^l,\tilde{X}^{\star}_2, {\bf b}_2^\star\big)$ are compatible and that $R\setminus (\breve{C}\cup Z) \cup R_2^{' \star} = R\setminus (\breve{C}\cup Z) \cup (R_1^{' \star}\setminus Y) = R\setminus Y$ (the latter equality holds since $R\setminus (\breve{C}\cup Z) \cup R_1^{' \star} = R,$ $Y = V(\cupall{\sf influence}_{\tilde{\mathfrak{R}}_1} (W^\bullet))\subseteq I_1^{(d-r+1)},$ and $I_1^{(d-r+1)} \subseteq \breve{C} \cup Z$).
See~\autoref{figure_boundariedgraph5} for an example of how $\big(\mathfrak{A}^\star,R_1^{'\star},{\bf W}_{{q}}^{(1)},\varnothing^l, \tilde{X}^\star_1, {\bf b}_1^\star\big)$ is ``transformed'' to $\big(\mathfrak{A}^{\star},R_2^{'\star},{\bf W}_{{q}}^{(2)},\varnothing^l, \tilde{X}^{\star}_2, {\bf b}_2^\star\big).$

Finally, we have that
$(\mathfrak{A},R,{\bf W}_{{q}}^{(1)},\varnothing^l, X)\models θ^{\sf out}_q \iff (\mathfrak{A},R\setminus Y,{\bf W}_q^{(2)},\varnothing^l, X_{\rm out}\cup X')\models θ^{\sf out}_q.$
To conclude the proof of \autoref{claim_1}, it remains to prove that for every $V\subseteq Y$
that is also a subset of $V(\cupall{\sf influence}_{\breve{\mathfrak{R}}'}(\overline{W}),$ where $\overline{W}$ is the central $(j'-2)$-subwall of $W_1$
, $$(\mathfrak{A},R\setminus Y,{\bf W}_q^{(2)},\varnothing^l, X_{\rm out}\cup X')\models θ^{\sf out}_q\iff (\mathfrak{A}\setminus V,R\setminus Y,{\bf W}_q^{(2)},\varnothing^l, X_{\rm out}\cup X')\models θ^{\sf out}_q.$$

Let $V\subseteq Y$ that is also a subset of $V(\cupall{\sf influence}_{\breve{\mathfrak{R}}'}(\overline{W})),$ where $\overline{W}$ is the central $(j'-2)$-subwall of $W_1.$
Since $Y\subseteq I_1^{(w)},$ $(X_{\rm out}\cup X')\cap Y = \emptyset,$ and $\cupall {\bf W}_q^{(2)}\subseteq I_2^{(2)},$
$C''∈ {\sf pr}(G_{\mathfrak{A}},{\bf W}_q^{(2)}, X_{\rm out}\cup X') \iff C''\setminus V ∈ {\sf pr}(G_{\mathfrak{A}}\setminus V,{\bf W}_q^{(2)}, X_{\rm out}\cup X')$ and, if
${\cal C}$ (resp. ${\cal C}'$) is the set of all $C''∈{\sf cc}(G_{\mathfrak{A}}, X_{\rm out}\cup X')$ (resp. all $C''∈{\sf cc}(G_{\mathfrak{A}}\setminus V, X_{\rm out}\cup X')$)
that are not in ${\sf pr}(G_{\mathfrak{A}},{\bf W}_q^{(2)}, X_{\rm out}\cup X')$ (resp. $ {\sf pr}(G_{\mathfrak{A}}\setminus V,{\bf W}_q^{(2)}, X_{\rm out}\cup X')$), then ${\cal C}={\cal C}'.$
Therefore, to show that $(\mathfrak{A},R\setminus Y,{\bf W}_q^{(2)},\varnothing^l, X_{\rm out}\cup X')\models θ^{\sf out}_q\iff (\mathfrak{A}\setminus V,R\setminus Y,{\bf W}_q^{(2)},\varnothing^l, X_{\rm out}\cup X')\models θ^{\sf out}_q,$ it now suffices to prove
that $$(\mathfrak{A},X_{\rm out}\cup X')\models β|_{{\sf star}_{\sf X}} \iff
(\mathfrak{A}\setminus V,X_{\rm out}\cup X')\models β|_{{\sf star}_{\sf X}}.$$
Equivalently,
we want to prove that
${\sf star}_{\sf X} (\mathfrak{A},X_{\rm out}\cup X')\models β \iff {\sf star}_{\sf X} (\mathfrak{A}\setminus V,X_{\rm out}\cup X')\models β.$
Recall that $Y=V({\sf compass}_{\breve{\mathfrak{R}}'}(\breve{W}')),$ where $(\breve{W}', \breve{\mathfrak{R}}')$ is a $\breve{W}$-tilt of $(W,\mathfrak{R})$ and $\breve{W}$ is the central $j'$-subwall of $W_1.$
Note that, by the definition of a flatness pair and since $V\subseteq V(\cupall{\sf influence}_{\breve{\mathfrak{R}}'}(\overline{W})$ and $\overline{W}$ is the central $(j'-2)$-subwall of $W_1,$
no vertex in $V$ is adjacent to a vertex in $G_{\mathfrak{A}}\setminus (Y \cup V({\bf a})).$
Also, the fact that $(X_{\rm out}\cup X')\cap Y = \emptyset$
implies that every vertex in $X_{\rm out}\cup X'$ is either in $G_{\mathfrak{A}}\setminus Y$ or in $V({\bf a}).$
Therefore, if a vertex of $V$ is adjacent to a vertex $u$ in $X_{\rm out}\cup X',$ then $u∈ V({\bf a}).$
We will prove that every vertex $u$ in $X_{\rm out}\cup X'$ that is adjacent, in $G_{\mathfrak{A}},$ to a vertex in $V$ is also adjacent, in $G_{\mathfrak{A}},$ to a vertex in
$V({\sf compass}_{\mathfrak{R}}(W))\setminus V.$
Let $u$ be a vertex in $X_{\rm out}\cup X'$ that is adjacent, in $G_{\mathfrak{A}},$ to a vertex in $V.$
As observed above, $u∈ V({\bf a}).$
Let $\tilde{\cal Q}$ be a $(W,\mathfrak{R})$-canonical partition of $G\setminus V({\bf a}).$
Since by the hypothesis of the lemma ${\sf bid}_{(W,\mathfrak{R})}(N_{G_{\mathfrak{A}}}(V({\bf a})))≥(g+2)^2+1,$
we have that $u$ is adjacent to at least $(g+2)^2+1$ internal bags of $\tilde{\cal Q}.$
Notice that $V$ is a subset of the union of the vertex sets of all internal bags of $\tilde{\cal Q}$ that intersect the central $(g+2)$-subwall of $W_1.$
These bags are $(g+2)^2$ many.
This, in turn, implies that, since the vertex $u$ is adjacent to at least $(g+2)^2+1$ internal bags of $\tilde{\cal Q},$ $a$ is adjacent to a vertex $v$
in the vertex set of an internal bag of $\tilde{\cal Q}$ that
is disjoint from $V.$
By observing that $v∈ V({\sf compass}_{\mathfrak{R}}(W))\setminus V,$
we conclude that every vertex $u∈ (X_{\rm out}\cup X')\cap V({\bf a})$ is adjacent to a vertex in $V({\sf compass}_{\mathfrak{R}}(W))\setminus V.$
This implies that ${\sf star}_{\sf X} (\mathfrak{A}, X_{\rm out}\cup X') = {\sf star}_{\sf X} (\mathfrak{A}\setminus V, X_{\rm out}\cup X').$
\autoref{claim_1} follows.\hfill$\diamond$
\bigskip

Following~\autoref{claim_1}, let $X'\subseteq Z'$ such that $\partial_{\mathfrak{K}_2} (Z')\subseteq X'$ and $(\mathfrak{A},R,{\bf W}_q^{(1)}, \varnothing^l,X)\models θ^{\sf out}_q \iff (\mathfrak{A},R\setminus Y,{\bf W}_q^{(1)}, \varnothing^l,X_{\rm out}\cup X')\models θ^{\sf out}_q.$
Observe that since ${\bf W}_q^{(1)}\subseteq I_1^{(1)}$ and $I_1^{(d\cdot r)}\cap (X_{\rm out} \cup X')=\emptyset,$ we have that
${\bf W}_q^{(1)}\subseteq G\setminus (X_{\rm out} \cup X').$
Thus, there is a $\breve{C}'∈ {\sf cc}(G,X_{\rm out} \cup X')$ that respects ${\bf W}_q^{(1)}.$
Therefore, by~\autoref{@vuitcentistes}, $\{\breve{C}'\} = {\sf pr}(G,{\bf W}_{{q}}^{(1)},X_{\rm out}\cup X').$
Let $C'$ be the $w$-privileged set of $G$ with respect to ${\bf W}_q^{(1)}$ and $X_{\rm out}\cup X'$
and keep in mind that if $w=\circ,$ then $C'=\breve{C}',$ while, if $w=\bullet,$ then $C' = V(G)\setminus (X_{\rm out}\cup X')$ and $\breve{C}'\subseteq C'.$
Also, since every vertex in $V({\bf a})$ is adjacent, in $G,$ to at least $q$ internal bags of $\tilde{\cal Q},$
we have that
every $a∈ V({\bf a})$ is either in $V_L ({\bf a})$
 or belongs to both $\breve{C}$ and $\breve{C}'.$
Therefore, ${\bf a}\cap \breve{C} = {\bf a}\cap \breve{C}' = {\bf a}\cap C = {\bf a}\cap C'.$
We set ${\bf a}': = {\bf a}\cap \breve{C}.$
We aim to prove the following:\medskip

\setcounter{theotwo}{1}
\begin{theotwo}\label{claim_2}
It holds that
${\sf ap}_{{\bf c}}((\mathfrak{A}, R, {\bf a}')[C])\models \breve{ζ}_{\sf R}
\iff
{\sf ap}_{{\bf c}}((\mathfrak{A}, R\setminus Y, {\bf a}')[C'])\models \breve{ζ}_{\sf R}.$
\end{theotwo}

\noindent{\em \green{Proof of \autoref{claim_2}}:}
We will only prove that ${\sf ap}_{{\bf c}}((\mathfrak{A}, R, {\bf a}')[C])\models \breve{ζ}_{\sf R}
\implies
{\sf ap}_{{\bf c}}((\mathfrak{A}, R\setminus Y, {\bf a}')[C'])\models \breve{ζ}_{\sf R},$
since the other implication is trivial.
Suppose that
${\sf ap}_{{\bf c}}((\mathfrak{A}, R, {\bf a}')[C])\models \breve{ζ}_{\sf R}.$
Since $\breve{ζ}_{\sf R}$ is a Boolean combination of the basic local sentences $\breve{ζ}_1, \ldots, \breve{ζ}_p,$ there is a set $J\subseteq [p]$ such that for every $j∈ J$ it holds that ${\sf ap}_{{\bf c}}((\mathfrak{A}, R, {\bf a}')[C])\models \breve{ζ}_j$ and for every $j\notin J$ it holds that ${\sf ap}_{{\bf c}}((\mathfrak{A}, R, {\bf a}')[C])\models \neg \breve{ζ}_j.$
We will show that for every $j∈ J$ it holds that ${\sf ap}_{{\bf c}}((\mathfrak{A}, R\setminus Y, {\bf a}')[C'])\models\breve{ζ}_j$ and that for every $j\notin J$ it holds that ${\sf ap}_{{\bf c}}((\mathfrak{A}, R\setminus Y, {\bf a}')[C'])\models \neg \breve{ζ}_j.$
Therefore, we distinguish two cases.\bigskip

\noindent{\bf Case 1:} $j∈ J.$\medskip

We aim to prove that ${\sf ap}_{{\bf c}}((\mathfrak{A}, R,{\bf a}')[C])\models \breve{ζ}_j \iff {\sf ap}_{{\bf c}}((\mathfrak{A}, R\setminus Y, {\bf a}')[C'])\models \breve{ζ}_j.$
Suppose that ${\sf ap}_{{\bf c}}((\mathfrak{A}, R, {\bf a}')[C])\models \breve{ζ}_j.$
Recall that the constant-projection $(τ\cup\{R\})^{\bf c}$ of $(τ\cup\{R\}\cup{\bf c}),$ i.e.,
the vocabulary of the structure ${\sf ap}_{{\bf c}}((\mathfrak{A}, R ,{\bf a}')[C]),$ contains every unary relation symbol in $(τ\cup\{R\}\cup{\bf c})$
and note that in the structure ${\sf ap}_{{\bf c}}((\mathfrak{A}, R, {\bf a}')[C]),$ $R$ is interpreted as $R\cap C.$
We set $R_C:=R\cap C,$ $(\mathfrak{B}, R_C) := {\sf ap}_{{\bf c}}((\mathfrak{A}, R ,{\bf a}')[C]),$
and keep in mind that $\mathfrak{B}$ is a $τ^{\langle \bf c\rangle}$-structure.
Since the Gaifman graphs of $\mathfrak{B}$ and of $(\mathfrak{B},R_C)$ are the same, in the rest of the proof we will use $G_{\mathfrak{B}}$ to denote both of them.
Also, to get some intuition, notice that $G_{\mathfrak{B}}$ is obtained from $G[C]$ after removing some edges (namely, the edges of $G[C]$ that connect the vertices in $V({\bf a}')$ with $C\setminus V({\bf a}')$).

Since $\breve{ζ}_j$ is a basic local sentence with parameters $r_j$ and $\ell_j,$
we have that
$$(\mathfrak{B}, R_C)\models \breve{ζ}_j \iff \exists X_j\subseteq R_C \mbox{~that is $(\ell_j, r_j)$-scattered in $\mathfrak{B}$ and $\mathfrak{B}\models \bigwedge_{x∈ X_{j}} ψ_j (x)$}.$$

We prove the following, which intuitively states that, given the set $X_{j},$ we can find another set $X_j '$ that ``behaves'' in the same way as $X_j$ but also ``avoids'' some inner part of $K_2^{\bf a}.$
\medskip

\noindent{{\em Subclaim:}} There exists a $t ∈ [d-\frac{r}{2}+2\hat{r}+1, d- \hat{r}]$ and a set
$X_j^{\prime}$ that is $(\ell_j, r_j)$-scattered in $\mathfrak{B}$ such that $X_j \subseteq R_C,$ $\mathfrak{B}\models \bigwedge_{x∈ X_{j}}ψ_j (x)\iff \mathfrak{B}\models \bigwedge_{x∈ X_{j}^{\prime}}ψ_j (x),$ and $X_j^{\prime} \cap I_2^{(t)} = \emptyset.$\medskip

\noindent{\em Proof of Subclaim:}
Our goal is to find a flatness pair, say $(\tilde{W}_3, \tilde{\mathfrak{R}}_3),$ that is $θ$-equivalent to
$(\tilde{W}_2, \tilde{\mathfrak{R}}_2),$ and a proper ``buffer'' $t$ so as to replace the part of $X_j$ that is in $I_2^{(t)}$ to an ``equivalent'' one that is inside $I_3^{(t)}.$
For this replacement to be ``safe'', we first have to demand that the the influence of  $(\tilde{W}_3, \tilde{\mathfrak{R}}_3),$ i.e., the set $I_3^{(w)},$ is disjoint from both the modulator $X_{\rm out} \cup X'$ and the set $X_{j}.$
Recall that $\tilde{\cal W}''$ is a collection of $\hat{\ell} +2$ flatness pairs of $G\setminus V({\bf a})$ that are $θ$-equivalent to $(\tilde{W}_{1}, \tilde{\mathfrak{R}}_1)$ and the vertex sets of their influences are disjoint from $X_{\rm out} \cup X'.$
Therefore, since $X_j$ has size at most $\hat{\ell},$ there exists a flatness pair in $\tilde{\cal W}''\setminus \{(\tilde{W}_2, \tilde{\mathfrak{R}}_2)\},$ say $(\tilde{W}_3, \tilde{\mathfrak{R}}_3),$ such that $I_3^{(w)}\cap (X_{\rm out} \cup X'\cup  X_{j})=\emptyset.$

We now focus on the set $I_2^{(d)}\setminus I_2^{(d-r+1)}.$
Recall that for the set $X_{\rm out} \cup X'$ it holds that $X' \subseteq Z'\subseteq I_2^{(d-r)}$ and $X_{\rm out}\cap I_{2}^{(w)}=\emptyset.$
Therefore, $I_2^{(d)}\setminus I_2^{(d-r+1)}$ does not intersect the set $X_{\rm out}\cup X'.$
Since
$r=2\cdot(\hat{\ell}+3)\cdot\hat{r}$ and $|X_{j}|≤  \hat{\ell},$
there exists a $t ∈ [d-\frac{r}{2}+2\hat{r}+1, d- \hat{r}]$
such that $X_{j}$ does not intersect $I_2^{(t)}\setminus I_2^{(t-\hat{r}+1)}.$
Intuitively, we partition the $r$ layers of $\tilde{W}_2$ that are in
$I_2^{(d)}\setminus I_2^{(d-r+1)}$ into two parts,
the first $r/2$ layers and the second $r/2$ layers,
and then we find some layer among
the ``$\hat{r}$-central'' $(\hat{\ell} +1)\hat{r}$ layers of the second part.
This layer together with its preceding $\hat{r}-1$ layers define a ``buffer'' of size $\hat{r}$ that $X_j$ ``avoids'' - that is $I_2^{(t)}\setminus I_2^{(t-\hat{r}+1)}.$
Notice that $I_2^{(t)}\setminus I_2^{(t-\hat{r}+1)}$ is a subset of $I_2^{(d)}\setminus I_2^{(d-r+1)}$
and therefore $I_2^{(t)}\setminus I_2^{(t-\hat{r}+1)}$ intersects neither $X_j$ nor $X_{\rm out} \cup X'.$

We set $X_{j}^\star:=X_{j}\cap I_2^{(t -\hat{r}+1)}$
and $Y_{j}\subseteq [\ell_j]$  to be the set of indices of the vertices in $X_{j}^{\star}.$
Notice that $X_{j}^\star\subseteq R_2,$ given that $X_{j}^\star =  X_{j}\cap I_2^{(t -\hat{r}+1)}\subseteq R_C\cap I_2^{(t -\hat{r}+1)}$ and $R_C\cap I_2^{(t -\hat{r}+1)}\subseteq R_2.$
Therefore, since
$X_{j}^{\star}=X_{j}\cap I_{2}^{(t -\hat{r}+1)},$
$ψ_j (x)$ is an $r_j$-local formula (where ``$r_j$-local'' refers to distances in $G_{\mathfrak{B}}$),
and $\hat{r}≥ r_j,$ we have that  $\mathfrak{B}\models \bigwedge_{x∈ X_{j}^{\star}}ψ_j (x)\iff \mathfrak{B}[I_2^{(t)}]\models \bigwedge_{x∈ X_{j}^{\star}}ψ_{j}(x).$
To sum up, we observe that, since $\cupall  {\bf W}_q^{(2)} \subseteq I_2^{(t)},$ we have that ${\sf pr}(K_2^{\bf a}[I_2^{(t)}], {\bf W}_q^{(2)}, \emptyset) = \{I_2^{(t)}\}$ and also the set $X_{j}^{\star}$ is a subset of $I_{2}^{(t -\hat{r}+1)}\cap R_2$ that is $(|Y_j|, r_j)$-scattered in $\mathfrak{B}[I_2^{(t)}]$ (since $X_j$ is $(|Y_j|, r_j)$-scattered in  $\mathfrak{B}$) and
\begin{eqnarray}
\labels{eqeqeqeqeq3}
\mathfrak{B}\models \bigwedge_{x∈ X_{j}^{\star}}ψ_j (x)\iff \mathfrak{B}[I_2^{(t)}]\models \bigwedge_{x∈ X_{j}^{\star}}ψ_{j}(x).
\end{eqnarray}
Also, notice that ${\sf ap}_{{\bf c}}(\mathfrak{A}, {\bf a}')[I_2^{(t)}] = \mathfrak{B}[I_2^{(t)}].$

Using the fact that  $(\tilde{W}_2, \tilde{\mathfrak{R}}_2)$ is $θ$-equivalent to $(\tilde{W}_3, \tilde{\mathfrak{R}}_3),$
we now aim to find a set $\tilde{X}_j$ that is an ``equivalent'' (in $I_3^{(t)}$) set of $X_j^\star.$
Since 
 $(\tilde{W}_2, \tilde{\mathfrak{R}}_2)$ is $θ$-equivalent to $(\tilde{W}_3, \tilde{\mathfrak{R}}_3),$ we have that
${\sf in\mbox{-}sig}(\mathfrak{K}_2,R_2,t',L,\emptyset)=  {\sf in\mbox{-}sig}(\mathfrak{K}_3,R_3,t',L,\emptyset),$
for every $t'∈ [w].$
Therefore, we have that ${\sf in\mbox{-}sig}(\mathfrak{K}_2,R_2,t,L,\emptyset)=  {\sf in\mbox{-}sig}(\mathfrak{K}_3,R_3,t,L,\emptyset)$ for the particular value $t$ given above.
This implies that there exists a $\hat{C}∈ {\sf pr}(K_3^{\bf a}[I_3^{(t)}], {\bf W}_q^{(3)}, \emptyset)$ and a set $\tilde{X_{j}}\subseteq I_{3}^{(t -\hat{r}+1)}\cap R_3$ such that
$\tilde{X_{j}}$ is $(|Y_{j}|, r_{j})$-scattered in $\mathfrak{B}[I_3^{(t)}]$ and
${\sf ap}_{{\bf c}}(\mathfrak{A}, {\bf a}')[I_2^{(t)}]\models \bigwedge_{x∈ X_{j}^{\star}}ψ_{j}(x)\iff {\sf ap}_{{\bf c}}(\mathfrak{A}, {\bf a}')[\hat{C}]\models \bigwedge_{x∈ \tilde{X}_{j}}ψ_{j}(x).$
Observe that ${\sf ap}_{{\bf c}}(\mathfrak{A}, {\bf a}')[\hat{C}] = \mathfrak{B}[\hat{C}]$ and that ${\sf pr}(K_3^{\bf a}[I_3^{(t)}], {\bf W}_q^{(3)}, \emptyset) = \{I_3^{(t)}\}.$
Thus,
\begin{eqnarray}\labels{eqeeqeq2}
\mathfrak{B}[I_2^{(t)}]\models \bigwedge_{x∈ X_{j}^{\star}}ψ_{j}(x)\iff \mathfrak{B}[I_3^{(t)}]\models \bigwedge_{x∈ \tilde{X}_{j}}ψ_{j}(x).
\end{eqnarray}

Given that $X\cap I_3^{(w)}=\emptyset,$
$\breve{C}∈ {\sf pr}(G, {\bf W}_q^{(3)}, X),$ and $\cupall{\bf W}_q^{(3)}\subseteq I_3^{(w)},$
we have that $I_3^{(w)}\subseteq \breve{C}\subseteq C.$
We stress that the above holds, no matter which $q$-pseudogrid of $G$ we consider.
Also, since $\tilde{X_{j}}\subseteq I_{3}^{(t -\hat{r}+1)}\subseteq I_3^{(w-\hat{r})},$ for every $x∈\tilde{X_{j}}$ it holds that $N_{G_{\mathfrak{B}}}^{(≤ \hat{r})}(x)\subseteq C.$
Thus, since $ψ_{j}({\sf x})$ is $r_{j}$-local, it follows that
\begin{eqnarray}\labels{eqeqeq1}
\mathfrak{B}[I_3^{(t)}] \models \bigwedge_{x∈ \tilde{X}_{j}}ψ_{j}(x)\iff \mathfrak{B}\models \bigwedge_{x∈ \tilde{X}_{j}}ψ_{j}(x).
\end{eqnarray}

We now consider the set
\[{X}^{\prime}_{j}:=\left( X_{j}\setminus X_{j}^{\star}\right)\cup \tilde{X_{j}}.\]
Since $I_{3}^{(w)}\cap X_{j}=\emptyset$ and $\hat{r}≥ r_j,$ for every $x∈ X_{j},$ and thus, for every $x∈ X_{j}\setminus X_{j}^{\star},$ it holds that $N_{G_{\mathfrak{B}}}^{(≤ r_{j})}(x)\cap I_{3}^{(w-\hat{r}+1)}=\emptyset.$
Also, since $t ≤ w -\hat{r}$ and $\tilde{X_{j}}\subseteq I_{3}^{(t -\hat{r}+1)},$ for every $x∈\tilde{X_{j}}$ it holds that $N_{G_{\mathfrak{B}}}^{(≤ r_{j})}(x) \subseteq I_{3}^{(w-\hat{r}+1)}.$
Thus, for every $x∈ X_{j}\setminus X_{j}^{\star}$ and $x'∈ \tilde{X_{j}}$ we have that
$N_{G_{\mathfrak{B}}}^{(≤ r_{j})}(x)\cap N_{G_{\mathfrak{B}}}^{(≤ r_{j})}(x')=\emptyset.$
The latter, together with the fact that the set $X_{j}\setminus X_{j}^{\star}$ is $(\ell_{j}-|Y_{j}|, r_{j})$-scattered in $\mathfrak{B}$ and $\tilde{X_{j}}$ is $(|Y_{j}|, r_{j})$-scattered in $\mathfrak{B}[I_3^{(t)}],$
implies that ${X}^{\prime}_{j}$ is an $(\ell_{j}, r_{j})$-scattered set in $\mathfrak{B}.$
Moreover, by definition, we have that ${X}^{\prime}_{j}\subseteq R_C\cup R_3 = R_C$ (the latter equality holds since $I_3^{(w)} \subseteq C$) and ${X}^{\prime}_{j}$ does not intersect $I_2^{(t)},$ while, by~\eqref{eqeqeqeqeq3},~\eqref{eqeeqeq2}, and~\eqref{eqeqeq1}, we have that $\mathfrak{B} \models \bigwedge_{x∈ X_{j}}ψ_j (x)\iff \mathfrak{B} \models \bigwedge_{x∈ X_{j}^{\prime}}ψ_j (x).$ The subclaim follows.\hfill\blue{$\diamond$}
\bigskip

Following the above subclaim, let $t ∈ [d-\frac{r}{2}+2\hat{r}+1, d- \hat{r}]$ and let
$X_j^{\prime}$ be a set that is $(\ell_j, r_j)$-scattered in $\mathfrak{B}$ such that $X_j '\subseteq R_C,$ $\mathfrak{B} \models \bigwedge_{x∈ X_{j}}ψ_j (x)\iff \mathfrak{B} \models \bigwedge_{x∈ X_{j}^{\prime}}ψ_j (x),$ and $X_j^{\prime} \cap I_2^{(t)} = \emptyset.$

Since $r=2\cdot(\hat{\ell} +3)\cdot \hat{r}$ and $|{X}^{\prime}_{j}|≤ \hat{\ell},$ there exists a $t'∈ [d-r+2\hat{r}+1, d - \frac{r}{2}-\hat{r}]$ such that  ${X}^{\prime}_{j}$  does not intersect $I_1^{(t')}\setminus I_1^{(t'-\hat{r}+1)}.$
Intuitively, here, we partition the $r$ layers of $\tilde{W}_1$ that are in
$I_1^{(d)}\setminus I_1^{(d-r+1)}$ into two parts,
the first $r/2$ layers and the second $r/2$ layers,
and then we find some layer among
the ``$\hat{r}$-central'' $(\hat{\ell} +1)\hat{r}$ layers of the {\sl first} part.
This layer together with its preceding $\hat{r}-1$ layers define a ``buffer'' of size $\hat{r}$ that $X_j '$ ``avoids'' - that is $I_1^{(t')}\setminus I_1^{(t'-\hat{r}+1)}.$

Now, consider the set $U_1:={X}^{\prime}_{j}\cap (I_{1}^{(t'-\hat{r}+1)}\setminus  Z).$
Observe that $U_1\subseteq R_1$ and therefore $U_1\subseteq (I_{1}^{(t'-\hat{r}+1)}\setminus Z)\cap R_1.$
Recall that $Y =I_1^{(\hat{r})}$
and notice that, since $({X}^{\prime}_{j}\setminus U_1) \cap I_1^{(t')}=\emptyset$ and $t'>\hat{r},$
it holds that ${X}^{\prime}_{j}\setminus U_1\subseteq R\setminus Y.$

We now observe that
${\sf pr}(G,{\bf W}_q^{(1)}, X_{\rm out} \cup Z)  ={\sf pr}(G,{\bf W}_{{q}}^{{(1)}},X),$
 due to the fact that $\partial_{\mathfrak{K}_1} (Z)\subseteq X.$
Recall that ${\sf pr}(G,{\bf W}_{{q}}^{{(1)}},X) = \{\breve{C}\}.$
Τhus, ${\sf pr}(G,{\bf W}_q^{(1)}, X_{\rm out} \cup Z) = \{\breve{C}\}.$
Also, recall that
$X_{\rm out}\cap I_1^{(d)} = \emptyset.$
Therefore,
the fact that $\cupall{\bf W}_q^{(1)}\subseteq I_1^{(d)}$ implies that
 ${\sf pr}(K_1^{\bf a}[I_1^{(d)}], {\bf W}_q^{(1)}, Z)$ contains a unique element,
 say $C_1.$
Observe that $I_1^{(t')}\setminus Z\subseteq C_1.$
Also, note that $C_1\subseteq \breve{C}.$

Let $Y_{j}^{\prime}\subseteq [\ell_j]$ be the set of the indices of the vertices of
${X}^{\prime}_{j}$ in $U_1.$
Given that $U_1={X}^{\prime}_{j}\cap (I_{1}^{(t'-\hat{r}+1)}\setminus  Z)$ and
${X}^{\prime}_{j}$ is  $(\ell_j, r_j)$-scattered in $\mathfrak{B},$
 and
$\mathfrak{B} \models \bigwedge_{x∈ X_{j}^{\prime}}ψ_j (x),$
we get that
$U_1$ is $(|Y_{j}^{\prime}|,r_j)$-scattered in $\mathfrak{B}[I_1^{(t')}\setminus Z]$
and $\mathfrak{B}\models\bigwedge_{x∈ U_1}ψ_{j}(x).$
At this point, observe that,
since the formula $ψ_{j}({\sf x})$ is $r_j$-local, $U_1={X}^{\prime}_{j}\cap (I_{1}^{(t'-\hat{r}+1)}\setminus  Z),$
where $\hat{r}≥ r_j$ and $t'≤ d-\frac{r}{2}-\hat{r},$ for every $x∈ U_1$ we have that
$N_{G_{\mathfrak{B}}}^{(≤ r_j)} (x)\subseteq I_1^{(d)}\setminus Z.$
Thus, the fact that $I_1^{(t')}\setminus Z\subseteq C_1$ implies that
\begin{eqnarray}\labels{notasi1}
\mathfrak{B}\models\bigwedge_{x∈ U_1}ψ_{j}(x)\iff\mathfrak{B}[C_1]\models\bigwedge_{x∈ U_1}ψ_{j}(x).
\end{eqnarray}
Also, note that ${\sf ap}_{{\bf c}}(\mathfrak{A}, {\bf a}')[C_1] = \mathfrak{B}[C_1].$

As we mentioned before, ${\sf in\mbox{-}sig}(\mathfrak{K}_1,R_1,d,L,Z)=  {\sf in\mbox{-}sig}(\mathfrak{K}_2,R_2,d,L,Z').$
This implies the existence of a set $C_2∈ {\sf pr}(K_2^{\bf a}[I_2^{(d)}], {\bf W}_q^{(2)}, Z')$ and a set $U_2\subseteq (I_{2}^{(t'-\hat{r})}\setminus Z')\cap R_2\subseteq R\setminus Y$
such that $U_2$ is $(|Y_{j}^{\prime}|,r_{j})$-scattered in $\mathfrak{B}[I_2^{(t')}\setminus Z']$ and $\mathfrak{B}[C_1] \models\bigwedge_{x∈ U_1}ψ_{j}(x)\iff {\sf ap}_{{\bf c}}(\mathfrak{A}, {\bf a}')[C_2]\models\bigwedge_{x∈ U_2}ψ_{j}(x).$
We set $\tilde{\mathfrak{B}}: = {\sf ap}_{{\bf c}}(\mathfrak{A}, {\bf a}')[C_2].$
Therefore, we have that
\begin{eqnarray}\labels{romawi2}
\mathfrak{B}[C_1] \models\bigwedge_{x∈ U_1}ψ_{j}(x)\iff \tilde{\mathfrak{B}}\models\bigwedge_{x∈ U_2}ψ_{j}(x).
\end{eqnarray}
By~\eqref{notasi1} and~\eqref{romawi2}, we derive that
\begin{eqnarray}\labels{@corporations}
 \mathfrak{B}\models\bigwedge_{x∈ U_1}ψ_{j}(x)
& \iff &
\tilde{\mathfrak{B}}\models\bigwedge_{x∈ U_2}ψ_{j}(x).
\end{eqnarray}

We now observe that
${\sf pr}(G,{\bf W}_q^{(2)}, X_{\rm out} \cup Z')  ={\sf pr}(G,{\bf W}_{{q}}^{{(1)}},X_{\rm out}\cup X') .$
To see this, notice that ${\sf pr}(G,{\bf W}_{{q}}^{{(1)}},X_{\rm out}\cup X') = {\sf pr}(G,{\bf W}_{{q}}^{(2)},X_{\rm out}\cup X')$
and ${\sf pr}(G,{\bf W}_q^{(2)}, X_{\rm out} \cup Z') = {\sf pr}(G,{\bf W}_q^{(2)}, X_{\rm out} \cup X'),$ due to the fact that $\partial_{\mathfrak{K}_2} (Z')\subseteq X'.$
Τhus, ${\sf pr}(G,{\bf W}_q^{(2)}, X_{\rm out} \cup Z') = \{\breve{C}'\}.$
Recall that for the set $X_{\rm out} \cup X'$ it holds that $X' \subseteq Z'\subseteq I_2^{(d-r)}$ and $X_{\rm out}\cap I_{2}^{(w)}=\emptyset.$
Since
$C_2∈ {\sf pr}(K_2^{\bf a}[I_2^{(d)}], {\bf W}_q^{(2)}, Z'),$
$X_{\rm out}\cap I_{2}^{(w)}=\emptyset,$
and $\cupall {\bf W}_q^{(2)} \subseteq I_2^{(d)},$
it holds that
$I_2^{(t')}\setminus Z' \subseteq C_2$ and
$C_2\subseteq \breve{C}'\subseteq C'.$

We set $\mathfrak{B}':=\mathfrak{A}[C']$
and
$R_{C'}:= R\cap {C'}$
and we observe that, by construction, $V({\bf a}\cap C' )= V({\bf a}\cap C).$
Since $U_2$ is $(|Y_{j}^{\prime}|,r_{j})$-scattered in
$\mathfrak{B}[I_2^{(t')}\setminus Z'],$ where $U_2\subseteq I_{2}^{(t'-\hat{r}+1)}\setminus Z'$ and $t'< w - \hat{r},$ $U_2$ is also $(|Y_{j}^{\prime}|,r_{j})$-scattered in $\mathfrak{B}'.$
Moreover, the formula $ψ_{j}({\sf x})$ is $r_{j}$-local, so
\begin{eqnarray}\labels{@illusionistische}
\tilde{\mathfrak{B}}\models\bigwedge_{x∈ U_2}ψ_{j}(x)\iff
\mathfrak{B}'\models\bigwedge_{x∈ U_2}ψ_{j}(x).
\end{eqnarray}
Therefore, by~\eqref{@corporations} and~\eqref{@illusionistische}, it follows that $\mathfrak{B}\models\bigwedge_{x∈ U_1}ψ_{j}(x)\iff \mathfrak{B}'\models\bigwedge_{x∈ U_2}ψ_{j}(x).$

Consider the set $$X_j^\bullet:=({X}^{\prime}_{j}\setminus U_1)\cup U_2.$$

Notice that since ${X}^{\prime}_{j}\setminus U_1 \subseteq C$
and ${X}^{\prime}_{j}\setminus U_1$ does not intersect neither
$I_{2}^{(d-r+1)}$ (where $X'$ lies),
nor $I_1^{(d-r+1)}\subseteq I_{1}^{(t')}$ (where $Z$ lies),
it follows that ${X}^{\prime}_{j}\setminus U_1 \subseteq C\cap C'.$
This implies that ${X}^{\prime}_{j}\setminus U_1$ is an $(\ell_{j}-|Y_{j}^{\prime}|, r_{j})$-scattered set in $\mathfrak{B}$
and an $(\ell_{j}-|Y_{j}^{\prime}|, r_{j})$-scattered set in $\mathfrak{B}'.$
Since $U_2\subseteq I_{2}^{(t'-\hat{r}+1)}\setminus Z',$
${X}^{\prime}_{j}\cap  I_{2}^{(t)}=\emptyset,$
and $t'≤ t-2\hat{r},$
we have that for every $x∈ {X}^{\prime}_{j}\setminus U_1$ and $x'∈ U_2$ it holds that
$N_{G_{\mathfrak{B}'}}^{(≤ r_{j})}(x)\cap N_{G_{\mathfrak{B}'}}^{(≤ r_{j})}(x')=\emptyset.$
The latter, together with the fact that ${X}^{\prime}_{j}\setminus U_1$ is an $(\ell_{j}-|Y_{j}^{\prime}|,r_{j})$-scattered set in $\mathfrak{B}'$ and $U_2$ is a $(|Y_{j}^{\prime}|,r_{j})$-scattered set in $\mathfrak{B}',$
implies that  $X_j^\bullet$ is an $(\ell_{j}, r_{j})$-scattered set in $\mathfrak{B}'.$
Also, notice that $X_j^\bullet\subseteq R_{C'}\setminus Y.$
Furthermore, since the formula $ψ_{j}({\sf x})$ is $r_{j}$-local, it follows that
$\mathfrak{B}'\models\bigwedge_{x∈ X_{j}}ψ_{j}(x)\iff \mathfrak{B}'\models\bigwedge_{x∈ X_j^\bullet}ψ_{j}(x).$

Thus, assuming that there is a set  $X_j\subseteq R_C$  that is $(\ell_j, r_j)$-scattered in $\mathfrak{B}$ and $\mathfrak{B}\models \bigwedge_{x∈ X_j} ψ_j (x),$ we proved that there is a set $X_j^\bullet\subseteq R_{C'}\setminus Y\subseteq R\setminus Y$ that is $(\ell_j, r_j)$-scattered in $\mathfrak{B}'$ and $\mathfrak{B}'\models \bigwedge_{x∈ X_j^\bullet} ψ_j (x).$

To conclude Case 1, notice that we can prove the inverse implication,
i.e., by assuming the existence of a set $X_j^\bullet\subseteq R_{C'}\setminus Y\subseteq R\setminus Y$  that is $(\ell_j, r_j)$-scattered
in $\mathfrak{B}'$ and $\mathfrak{B}'\models \bigwedge_{x∈ X_j^\bullet} ψ_j (x)$
and, by using
the same arguments as above
(replacing $(\tilde{W}_{1}, \tilde{\mathfrak{R}}_1)$ with $(\tilde{W}_{2}, \tilde{\mathfrak{R}}_2),$ $Z$ with $Z'$ and $R$ with $R\setminus Y$),
we can prove the existence of a set $X_j\subseteq R$ that is  $(\ell_j, r_j)$-scattered  in $\mathfrak{B}$ such that $\mathfrak{B}\models \bigwedge_{x∈ X_j} ψ_j (x).$
\bigskip

\noindent{\bf Case 2:} $j\notin J.$\smallskip

We aim to prove that ${\sf ap}_{{\bf c}}((\mathfrak{A}, R, {\bf a}')[C])\models\neg
\breve{ζ}_j \iff {\sf ap}_{{\bf c}}((\mathfrak{A}, R\setminus Y, {\bf a}')[C'])\models\neg \breve{ζ}_j.$
In other words, we show that  for every set $X_{j}\subseteq R\cap C$ that is $(\ell_{j}, r_{j})$-scattered in $\mathfrak{B}$, $\mathfrak{B} \models \neg ψ_j (x),$ for some $x∈ X_{j}$ if and only if  for every set $X_{j}'\subseteq (R\setminus Y)\cap C'$ that is $(\ell_{j}, r_{j})$-scattered in $\mathfrak{B}'$, $\mathfrak{B}'\models \neg ψ_j (x),$ for some $x∈ X_{j}'.$
In Case 1, we showed that there is a set $X_j\subseteq R\cap C$ that is  $(\ell_j, r_j)$-scattered  in $\mathfrak{B}$ and $\mathfrak{B}\models \bigwedge_{x∈ X_j} ψ_j (x)$ if and only if there is a set $X_j^\bullet\subseteq (R\cap C')\setminus Y\subseteq R\setminus Y$ that is  $(\ell_j, r_j)$-scattered in $\mathfrak{B}'$ and $\mathfrak{B}'\models \bigwedge_{x∈ X_j^\bullet} ψ_j (x).$
This directly implies that  ${\sf ap}_{{\bf c}}((\mathfrak{A}, R, {\bf a}')[C])\models\neg\breve{ζ}_j \iff {\sf ap}_{{\bf c}}((\mathfrak{A}, R\setminus Y, {\bf a}')[C'])\models\neg \breve{ζ}_j.$
This concludes Case 2 and completes the proof of \autoref{claim_2}. \hfill$\diamond$
\bigskip

\setcounter{theothree}{2}
\begin{theothree}\label{claim_3}
It holds that $\mathfrak{A}[C]\models μ \iff \mathfrak{A}[C']\models μ.$
\end{theothree}

\noindent{\em \red{Proof of \autoref{claim_3}}}:
Observe that $\mathfrak{A}\models μ \iff G_{\mathfrak{A}}∈ \excl(\{K_{\hw(θ)}\}).$
Also, observe that $C\setminus D = C'\setminus D.$
If $Y = V(\mathfrak{A})\setminus C$ and $Y'= V(\mathfrak{A})\setminus  C',$ then note that $Y\setminus D = Y'\setminus D$ and $Y$ intersects at most $q$ bags of $\tilde{\cal Q}.$
Thus, by assumption, $G\setminus Y∈ \excl (\{K_{\hw(θ)}\}) \iff G\setminus (Y\setminus D)∈ \excl(\{K_{\hw(θ)}\}),$ which implies that
$G[C]∈ \excl (\{K_{\hw(θ)}\}) \iff G[C']∈ \excl (\{K_{\hw(θ)}\}).$\
Therefore,  $\mathfrak{A}[C]\models μ \iff \mathfrak{A}[C']\models μ.$
This concludes the proof of \autoref{claim_3}.\hfill$\diamond$
\bigskip

Recall that $W^\bullet$ be the central $g$-subwall of $W_1$
and let $(\tilde{W}', \tilde{\mathfrak{R}}')$ be the $W^\bullet$-tilt of $(W,\mathfrak{R})$ given by the algorithm ${\tt Find\_Equiv\_FlatPairs}$ in~\autoref{@inhumainement}.
We set $V:=V({\sf compass}_{\tilde{\mathfrak{R}}'}(\tilde{W}')).$
Also, recall that the algorithm ${\tt Find\_Equiv\_FlatPairs}$
outputs the set $Y= V({\sf compass}_{\breve{\mathfrak{R}}'}(\breve{W}')),$
where
$(\breve{W}',\breve{\mathfrak{R}}')$ is a $\breve{W}'$-tilt of $(W,\mathfrak{R})$ and $\breve{W}$ is the central $j'$-subwall of $W_1.$
Finally, recall that $j'=g+2\hat{r}+2.$
The definition of a tilt of a flatness pair implies that $V$ is a subset of $Y.$
By \autoref{claim_1}, we have that $(\mathfrak{A},R,{\bf W}_{{q}}^{(1)},\varnothing^l, X)\models θ^{\sf out}_q \iff (\mathfrak{A}\setminus V,R\setminus Y,{\bf W}_q^{(1)},\varnothing^l, X_{\rm out}\cup X')\models θ^{\sf out}_q.$

Recall that since $Y\subseteq I_1^{(w)},$ $(X_{\rm out}\cup X')\cap Y = \emptyset,$ and $\cupall {\bf W}_q^{(2)}\subseteq I_2^{(2)},$
$C''∈ {\sf pr}(G_{\mathfrak{A}},{\bf W}_q^{(2)}, X_{\rm out}\cup X') \iff C''\setminus V ∈ {\sf pr}(G_{\mathfrak{A}}\setminus V,{\bf W}_q^{(2)}, X_{\rm out}\cup X')$ and, if
${\cal C}$ (resp. ${\cal C}'$) is the set of all $C''∈{\sf cc}(G_{\mathfrak{A}}, X_{\rm out}\cup X')$ (resp. all $C''∈{\sf cc}(G_{\mathfrak{A}}\setminus V, X_{\rm out}\cup X')$)
that are not in ${\sf pr}(G_{\mathfrak{A}},{\bf W}_q^{(2)}, X_{\rm out}\cup X')$ (resp. $ {\sf pr}(G_{\mathfrak{A}}\setminus V,{\bf W}_q^{(2)}, X_{\rm out}\cup X')$), then ${\cal C}={\cal C}'.$
Also, recall
that all the basic Gaifman variables in $\breve{ζ}_{\sf R}$ are contained in $R$ and every $ψ_{i}({\sf x})$ is $r_{i}$-local.
The fact that $W^\bullet$ is the central $g$-subwall of $W_1,$ $\breve{W}$ has height $j'$ and $g = j'-2\hat{r} -2,$ and $R\cap Y = \emptyset$ implies that
 these no local formulas $ψ_{i}({\sf x})$ is evaluated using vertices in $V.$
  Therefore, ${\sf ap}_{{\bf c}}((\mathfrak{A}, R\setminus Y, {\bf a}')[C'])\models \breve{ζ}_{\sf R} \iff {\sf ap}_{{\bf c}}((\mathfrak{A}\setminus V, R\setminus Y, {\bf a}')[C'\setminus V])\models \breve{ζ}_{\sf R},$ and, by \autoref{claim_2},
${\sf ap}_{{\bf c}}((\mathfrak{A}, R, {\bf a}')[C])\models \breve{ζ}_{\sf R}
\iff
 {\sf ap}_{{\bf c}}((\mathfrak{A}\setminus V, R\setminus Y, {\bf a}')[C'\setminus V])\models \breve{ζ}_{\sf R}.$
Finally, we observe that
$\mathfrak{A}[C]\models μ \iff \mathfrak{A}[C'\setminus V]\models μ.$
Thus,
we get that $(\mathfrak{A},R, {\bf a})\models θ_{{\sf R},{\bf c}}\iff(\mathfrak{A}\setminus V,R\setminus Y, {\bf a})\models θ_{{\sf R},{\bf c}}.$

\myskip\section{Dealing with \texorpdfstring{$\bar{Θ}$}{bar-Theta}}
\label{the_second_level}

Let $τ$ be a vocabulary.
In this section we aim to present
how to generalize
the definition of $\tilde{θ}_q,$
given in~\autoref{@reconquistasen},
and the
definitions of
out-signature and in-signature in~\autoref{sec_out-sig_first-floor} and~\autoref{sec_in-sig_first-floor}
so as the algorithm ${\tt Fin\_Equiv\_FlatPairs}$ (given in~\autoref{@inhumainement})
outputs the correct answer
for sentences in $\bar{Θ}[τ].$
Given a sentence $θ∈\bar{Θ}[τ]$ and an enhanced version $θ_{{\sf R},{\bf c}}$
of $θ,$
in~\autoref{@arrangements},
we define the split version $\tilde{θ}_q$ of $θ_{{\sf R},{\bf c}}.$
Then, in~\autoref{@enthaltenden}, we generalize
the signatures given in~\autoref{sec_out-sig_first-floor} and~\autoref{sec_in-sig_first-floor},
for a sentence in $\bar{Θ}[τ]$
and we describe how to prove~\autoref{@desmembramientos}
for a sentence $θ∈\bar{Θ}[τ].$
The full proof of~\autoref{@desmembramientos}
for a sentence $θ∈\bar{Θ}[τ]$ can be found in~\autoref{second_level_more}.
\medskip

Let $h={\sf height}(θ).$
We fix ${\sf X}_1,\ldots,  {\sf X}_h$ to be a set of
second-order variables.
Let $σ∈ \FOL[τ]$ and $μ∈ \NTMC[τ]$
be the target sentences of $θ$ and,
for every $i∈ [r],$ let $β_i∈\MSOL^\tw[τ\cup \{{\sf X}_i\}]$
be a modulator sentence of $θ$ and a string $w = w_0\ldots w_{h-1}$ of size $h$ whose alphabet is $\{\circ, \bullet\},$
such that
if $θ_0=σ\wedge μ,$ and for every $i∈[h],$
%
\begin{eqnarray}
θ_i  & = & β_i \triangleright γ_{i-1},\labels{eq_bis}
\end{eqnarray}
where $γ_{i-1} = θ_{i-1}^{({\sf c})},$ if $w_i=\circ$ and $γ_{i-1} = θ_{i-1},$ if $w_i = \bullet,$
then $θ = θ_h.$
In other words, for every $i∈[h],$ $θ_i$ checks whether for a given $τ$-structure $\mathfrak{A},$
there exists a set $S_i\subseteq V(\mathfrak{A})$ such that
 \begin{itemize}
 \item ${\sf star}_{{\sf X}_i}(\mathfrak{A},S_i)\models β_i$  \mbox{~and~}
 \item depending whether $γ_{i-1} = θ_{i-1}^{({\sf c})}$ or $θ_{i-1}$:
 \begin{itemize}
 \item either for every connected component $C$ of $G_\mathfrak{A}\setminus S_i,$ it holds that
 $\mathfrak{A}[C]\models θ_{i-1},$ or
 \item ${\sf rm}_{{\sf X}_i}(\mathfrak{A}, S_i)\models θ_{i-1}.$
 \end{itemize}
\end{itemize}

We now generalize the notion of a $\circ/\bullet$-flag given in the beginning of~\autoref{sec_first_floor} to the notion of a $\circ/\bullet$-scenario.

\myskip\subsection{Scenarios}
\label{sec_scenarios}

\myskip\paragraph{The $\circ/\bullet$-scenarios.}
Let $θ∈ \bar{Θ}[τ]$ and let $h = {\sf height}(θ).$
We define the {\em $\circ/\bullet$-scenario} of $θ$ to be the string $w$ of size $h$ whose alphabet is $\{\circ, \bullet\}$ such that $w_i = \circ$ if $γ_i ∈ \bar{Θ}_i^{({\sf c})}$ and $w_i = \bullet$ if $γ_i ∈ \bar{Θ}_i.$
Every sentence $θ∈\bar{Θ}[τ]$
is associated with its $\circ/\bullet$-flag.
Intuitively, the $\circ/\bullet$-scenario of a sentence in $\bar{Θ}$ encodes whether each question, that is asked recursively in the structure obtained after removing a set $X_i,$ concerns every substructure that corresponds to a connected component of the Gaifman graph of the remaining substructure or the union of all these substructures, in each ``level'' of the recursion.

\myskip\paragraph{Choosing the privileged set given by a $\circ/\bullet$-scenario.}
Let $G$ be a graph, let $q,h∈ \mathbb{N},$ let ${\bf W}_q ∈ {(2^{V(G)})}^{2q},$ and let ${\cal X} = \{X_1,\ldots, X_h\}$ be a collection of pairwise disjoint subsets of $V(G).$
Also, let $w$ be a $\circ/\bullet$-scenario of size $h.$

A sequence of subsets $C_1, \ldots, C_{h+1}$ of $V(G)$ is called {\em $w$-privileged sequence of $G$ with respect to ${\bf W}_q$ and ${\cal X}$}, if
 $C_{h+1} = V(G)$ and
for every $i∈ [h],$ $X_i\subseteq C_{i+1}$ and

$$C_i=
\begin{cases}
\text{an element in }{\sf pr}(G, {\bf W}_q, X_{i}\cup \ldots \cup X_h), & \text{if } w_i = \circ \text{ and }{\sf pr}(G, {\bf W}_q, X_{i}\cup \ldots \cup X_h)\neq \emptyset,\\
\emptyset, & \text{if } w_i = \circ \text{ and }{\sf pr}(G, {\bf W}_q, X_{i}\cup \ldots \cup X_h)= \emptyset, \text{ and}\\
C_{i+1}\setminus X_i, & \text{if } w_i = \bullet.
\end{cases}
$$

By~\autoref{@vuitcentistes},
if ${\bf W}_q$ is a  ${q}$-pseudogrid of a graph $G$ and $X$ is a subset of $V(G),$ then ${\sf pr}(G,{\bf W}_{{q}},X)$ is either a singleton or the empty set.
Therefore, we get the following:
\begin{observation}
\labels{@imagelessness}
If ${\bf W}_q$ is a  ${q}$-pseudogrid of a graph $G,$ $X_1, \ldots, X_h$ are subsets of $V(G),$ and $w$ is a $\circ/\bullet$-scenario of size $h,$ then there is exactly one $w$-privileged sequence of $G$ with respect to ${\bf W}_q$ and $\{X_1, \ldots, X_h\}.$
\end{observation}

It is also easy to observe the following.
\begin{observation}\labels{obws_privi}
Let $G$ be a graph, let $q,h∈ \mathbb{N},$ let ${\bf W}_q ∈ {(2^{V(G)})}^{2q},$ and let ${\cal X} = \{X_1,\ldots, X_h\}$ be a collection of pairwise disjoint subsets of $V(G).$
Also, let $w$ be a $\circ/\bullet$-scenario of size $h.$
If $C_1, \ldots, C_{h+1}$ is a
$w$-privileged sequence of $G$ with respect to ${\bf W}_q$ and $\{X_1, \ldots, X_h\},$ then $C_1\subseteq \ldots \subseteq C_{h+1}$ and if for some $i∈[h]$ $C_i = \emptyset,$ then for every $j∈[i]$ $C_j = \emptyset.$
\end{observation}

A vertex set $C\subseteq V(G)$ is a {\em $w$-privileged set of $G$ with respect to ${\bf W}_q$ and ${\cal X}$}, if there is a $w$-privileged sequence $C_1,\ldots, C_{h+1}$ of $G$ with respect to ${\bf W}_q$ and ${\cal X}$ such that
$C_1 = C.$
Due to~\autoref{@imagelessness}, there is a unique $w$-privileged set of $G$ with respect to ${\bf W}_q$ and ${\cal X}.$
See~\autoref{@corazoncillo} for an example.

\begin{figure}[ht]
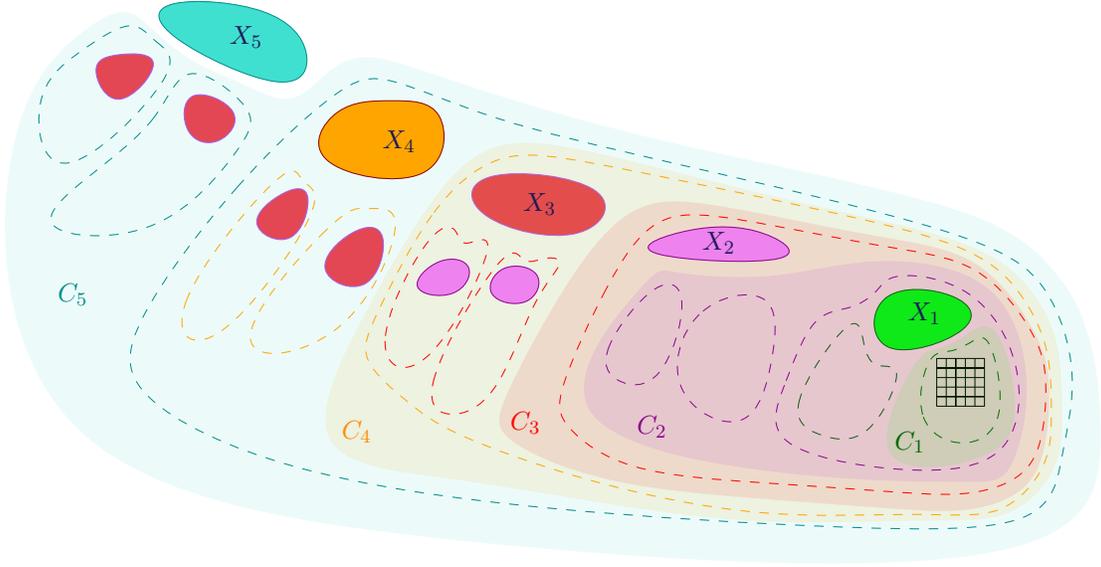

	\begin{center}
\tikzstyle{ipe stylesheet} = [
  ipe import,
  even odd rule,
  line join=round,
  line cap=butt,
  ipe pen normal/.style={line width=0.4},
  ipe pen heavier/.style={line width=0.8},
  ipe pen fat/.style={line width=1.2},
  ipe pen ultrafat/.style={line width=2},
  ipe pen normal,
  ipe mark normal/.style={ipe mark scale=3},
  ipe mark large/.style={ipe mark scale=5},
  ipe mark small/.style={ipe mark scale=2},
  ipe mark tiny/.style={ipe mark scale=1.1},
  ipe mark normal,
  /pgf/arrow keys/.cd,
  ipe arrow normal/.style={scale=7},
  ipe arrow large/.style={scale=10},
  ipe arrow small/.style={scale=5},
  ipe arrow tiny/.style={scale=3},
  ipe arrow normal,
  /tikz/.cd,
  ipe arrows, 
  <->/.tip = ipe normal,
  ipe dash normal/.style={dash pattern=},
  ipe dash dotted/.style={dash pattern=on 1bp off 3bp},
  ipe dash dashed/.style={dash pattern=on 4bp off 4bp},
  ipe dash dash dotted/.style={dash pattern=on 4bp off 2bp on 1bp off 2bp},
  ipe dash dash dot dotted/.style={dash pattern=on 4bp off 2bp on 1bp off 2bp on 1bp off 2bp},
  ipe dash normal,
  ipe node/.append style={font=\normalsize},
  ipe stretch normal/.style={ipe node stretch=1},
  ipe stretch normal,
  ipe opacity 10/.style={opacity=0.1},
  ipe opacity 30/.style={opacity=0.3},
  ipe opacity 50/.style={opacity=0.5},
  ipe opacity 75/.style={opacity=0.75},
  ipe opacity opaque/.style={opacity=1},
  ipe opacity opaque,
]
\definecolor{red}{rgb}{1,0,0}
\definecolor{blue}{rgb}{0,0,1}
\definecolor{green}{rgb}{0,1,0}
\definecolor{yellow}{rgb}{1,1,0}
\definecolor{orange}{rgb}{1,0.647,0}
\definecolor{gold}{rgb}{1,0.843,0}
\definecolor{purple}{rgb}{0.627,0.125,0.941}
\definecolor{gray}{rgb}{0.745,0.745,0.745}
\definecolor{brown}{rgb}{0.647,0.165,0.165}
\definecolor{navy}{rgb}{0,0,0.502}
\definecolor{pink}{rgb}{1,0.753,0.796}
\definecolor{seagreen}{rgb}{0.18,0.545,0.341}
\definecolor{turquoise}{rgb}{0.251,0.878,0.816}
\definecolor{violet}{rgb}{0.933,0.51,0.933}
\definecolor{darkblue}{rgb}{0,0,0.545}
\definecolor{darkcyan}{rgb}{0,0.545,0.545}
\definecolor{darkgray}{rgb}{0.663,0.663,0.663}
\definecolor{darkgreen}{rgb}{0,0.392,0}
\definecolor{darkmagenta}{rgb}{0.545,0,0.545}
\definecolor{darkorange}{rgb}{1,0.549,0}
\definecolor{darkred}{rgb}{0.545,0,0}
\definecolor{lightblue}{rgb}{0.678,0.847,0.902}
\definecolor{lightcyan}{rgb}{0.878,1,1}
\definecolor{lightgray}{rgb}{0.827,0.827,0.827}
\definecolor{lightgreen}{rgb}{0.565,0.933,0.565}
\definecolor{lightyellow}{rgb}{1,1,0.878}
\definecolor{black}{rgb}{0,0,0}
\definecolor{white}{rgb}{1,1,1}
\scalebox{0.9}{

}
	\end{center}
	\caption{An example where $γ_5∈ Θ_4,$ $γ_4∈ Θ_3^{({\sf c})},$ $γ_3∈ Θ_2^{({\sf c})},$ $γ_2∈ Θ_1,$ and $γ_1∈ Θ_0^{({\sf c})}.$ The sets $X_i,$ $i∈ [5]$ are the vertex sets that are recursively removed from the graph, while the colored areas correspond to the vertex sets in $G\setminus X_i$ that correspond to the ``privileged part'' of the graph.
	Connected components of each level of the construction are depicted by dashed boundaries.
	For example, since $γ_5∈ Θ_4,$ then $X_4$ is a vertex set in the graph $G\setminus X_5$ (which has three connected components - depicted by areas of dashed blue boundary), while, since $γ_4∈ Θ_3^{({\sf c})},$ we search for a set $X_3$ in every component of $G\setminus (X_4\cup X_5)$ (the privileged connected component of $G\setminus (X_4\cup X_5)$ is the orange area that has dashed orange boundary).}
	\labels{@corazoncillo}
\end{figure}

It is easy to see that the notion of $w$-privileged set can be expressed in $\MSOL.$

\begin{lemma}\labels{@assimilation}
Let $q,h∈ \mathbb{N},$ let $τ$ be a vocabulary, let ${\bf Q}\cup{\cal X}\cup\{{\sf C}\},$ where ${\cal X} = \{{\sf X}_1, \ldots, {\sf X}_h\},$ be a set of $2q+h+1$ new unary relation symbols,
and let $w$ be a $\circ/\bullet$-scenario of size $h.$
There is a sentence $η_{w\text{-}{\sf pr}_{{\cal X},{\sf C}}}∈ \MSOL[τ\cup{\bf Q}\cup{\cal X}\cup\{{\sf C}\}]$ such that for every $τ$-structure $\mathfrak{A},$ every ${\bf W}_q∈ {(2^{V(\mathfrak{A})})}^{2q},$ every ${\cal S} = \{S_1, \ldots, S_h\}$ of $h$ subsets of $V(\mathfrak{A}),$ and every $C\subseteq V(\mathfrak{A}),$
$(\mathfrak{A},{\bf W}_{q},{\cal S},C) \models η_{w\text{-}{\sf pr}_{{\cal X},{\sf C}}}$ (where every ${\sf X}_i$ is interpreted as $S_i$ and ${\sf C}$ is interpreted as $C$) if and only if ${\bf W}_q$ is a $q$-pseudogrid of $G_{\mathfrak{A}},$ the sets in ${\cal S}$ are pairwise disjoint, and $C$ is a $w$-privileged set of $G_{\mathfrak{A}}$ with respect to ${\bf W}_q$ and ${\cal S}.$
\end{lemma}


\myskip\subsection{The sentences \texorpdfstring{$θ^{\sf out}_{{q}}$}{theta-out-q} and  \texorpdfstring{$\tilde{θ}_{{q}}$}{tilde-theta-q}}\labels{@arrangements}
Let $τ$ be a vocabulary.
Let $θ∈\bar{Θ}[τ].$
We set $h: = {\sf height}(θ).$
Let $q,l∈\mathbb{N},$ let ${\bf c}$ be a collection of $l$ constant symbols not contained in $τ,$ and let $\{{\sf R}\}\cup{\bf Q}\cup{\cal X},$ where ${\cal X} = \{{\sf X}_1, \ldots, {\sf X}_h\},$ be a set of $2q+h+1$ unary symbols not contained in $τ.$
Let $θ_{{\sf R},{\bf c}}$ be an enhanced version of $θ$ and let $w$ be the $\circ/\bullet$-scenario of $θ.$

\myskip\paragraph{Τhe sentence $θ^{\sf out}_{{q}}.$}
We define the sentence $θ^{\sf out}_{{q}}∈ \MSOL[τ\cup\{{\sf R}\}\cup{\bf Q}\cup{\cal X}\cup{\bf c}]$ such that for every $τ$-structure $\mathfrak{A},$ every ${\bf W}_q∈ {(2^{V(\mathfrak{A})})}^{2q},$ every $R\subseteq V(\mathfrak{A}),$
every apex-tuple ${\bf a}$ of $\mathfrak{A}$ of size $l,$ and every collection ${\cal S} = \{S_1, \ldots, S_h\}$ of $h$ subsets of $V(\mathfrak{A}),$ $(\mathfrak{A},R,{\bf W}_{{q}},{\bf a},{\cal S})\models θ^{\sf out}_{{q}}$ if and only if  the following conditions are satisfied:
\begin{itemize}
\item ${\bf W}_q$ is a $q$-pseudogrid of $G_\mathfrak{A},$
\item the sets in ${\cal S}$ are pairwise disjoint, and \item there exists a sequence
$C_1, \ldots, C_h$ of subsets of $V(\mathfrak{A})$ that is a $w$-privileged sequence of $G_{\mathfrak{A}}$ with respect to ${\bf W}_q$ and ${\cal S}$ and
\begin{itemize}
\item for every $i∈[h],$ $S_{i}\subseteq C_{i+1},$
\item for every $i∈[h],$ $(\mathfrak{A}[C_{i+1}],S_{i})\models β_i|_{{\sf star}_{{\sf X}_i}},$ and
\item  for every $i∈[h]$ where $w_i = \circ,$ we have that for every $C∈ {\sf cc}(G_{\mathfrak{A}},S_i\cup \ldots \cup S_h)$ such that $C\subseteq C_{i+1}\setminus C_i,$ it holds that $(\mathfrak{A},R,{\bf a})[C]\models {θ_{i-1}}_{{\sf R}, {\bf c}}.$
\end{itemize}
\end{itemize}
Note that $θ^{\sf out}_{{q}}∈ \MSOL[τ\cup{\bf Q}\cup\{{\sf R}\}\cup{\cal X}\cup{\bf c}],$
since for every $i∈[h],$ $β_i |_{{\sf star}_{{\sf X}_i}}∈ \MSOL[τ\cup\{{\sf X}_i\}]$ and ${θ_{i-1}}_{{\sf R},{\bf c}}∈ \MSOL[τ\cup\{{\sf R}\}\cup{\bf c}].$

Intuitively, the sentence $θ^{\sf out}_q$ does everything that $θ_{{\sf R}, {\bf c}}$ does, except from the privileged component.

\myskip\paragraph{The sentence $\tilde{θ}_{{q}}.$}
We define the sentence $\tilde{θ}_{{q}}∈\MSOL[τ\cup{\bf Q}\cup\{{\sf R}\}\cup{\bf c}]$ such that
\begin{equation}
\tilde{θ}_{{q}}  = \exists {\sf X}_1,\ldots, {\sf X}_h \  \blue{\fbox{$θ^{\sf out}_{{q}}$}}\ \wedge
 \red{\fbox{$\exists {\sf C} \ η_{w\text{-}{\sf pr}_{{\cal X},{\sf C}}}\ \wedge  μ|_{{\sf ind}_{\sf C}}$}}\ \wedge   \green{\fbox{$\exists {\sf C} \ η_{w\text{-}{\sf pr}_{{\cal X},{\sf C}}}\ \wedge \breve{ζ}_{\sf R} |_{{\sf ap}_{\bf c}} |_{{\sf ind}_{\sf C}}$}}
\labels{@gaspilleront}
\end{equation}
(recall that $θ^{\sf out}_{q}∈ \MSOL[τ\cup\{{\sf R}\}\cup{\bf Q}\cup{\cal X}\cup{\bf c}],$ $η_{w\text{-}{\sf pr}_{{\cal X},{\sf C}}}∈ \MSOL[τ\cup {\bf Q}\cup{\cal X}\cup\{{\sf C}\}],$ $μ∈ \MSOL[τ],$ and $\breve{ζ}_{\sf R} |_{{\sf ap}_{\bf c}}∈ \FOL[τ\cup \{{\sf R}\}\cup{\bf c}]$).

Alternatively,  if $\mathfrak{A}$ is a $τ$-structure, $R\subseteq V(\mathfrak{A}),$ ${\bf W}_{{q}}∈ {(2^{V(\mathfrak{A})})}^{2q},$ and ${\bf a}$ is an apex-tuple of $\mathfrak{A}$ of size $l,$ then $(\mathfrak{A},R,{\bf W}_{{q}}, {\bf a})\models \tilde{θ}_{{q}} \iff \exists S_1, \ldots, S_h \subseteq V(\mathfrak{A})$ such that, if  ${\cal S} = \{S_1, \ldots, S_h\},$ then
\begin{itemize}
\item  $\blue{\fbox{$(\mathfrak{A},R,{\bf W}_{{q}},{\cal S})\models θ^{\sf out}_{{q}}$}},$
\item $\red{\fbox{\black{$\exists C\subseteq V(\mathfrak{A})$ that is $w$-privileged with respect to ${\bf W}_q$ and ${\cal S}$ such that $\mathfrak{A}[C] \models μ$}}},$ and
\item $\green{\fbox{\black{$\exists C\subseteq V(\mathfrak{A})$ that is $w$-privileged with respect to ${\bf W}_q$ and ${\cal S}$ such that $(\mathfrak{A}, R, {\bf a})[C]\models \breve{ζ}_{\sf R} |_{{\sf ap}_{\bf c}} .$}}}$
\end{itemize}

We stress that in the above sentence, the part ``$\exists {\sf C} \ η_{w\text{-}{\sf pr}_{{\cal X},{\sf C}}}$''
appears twice, but essentially refers to the same set since, by~\autoref{@imagelessness}, there is a unique
set in $V(\mathfrak{A})$ that is $w$-privileged
with respect to ${\bf W}_q$ and ${\cal S}.$

Intuitively, if the Gaifman graph contains a ``big enough'' wall, we can ``separate'' the questions that concern all $X_i$ and every non-privileged connected component of $G\setminus (X_i \cup \ldots \cup X_h)$ (expressed by $θ^{\sf out}_{{q}}$) and the question $σ\wedge μ$ that concerns the privileged part of $G\setminus (X_1 \cup \ldots \cup X_h).$

We call $\tilde{θ}_q$ the {\em split version} of $θ_{{\sf R},{\bf c}}.$
\bigskip

By using the same arguments as in the proof of~\autoref{lemma_equiva}, it is easy to prove the following:
\begin{lemma}\labels{@interruptions}
Let $τ$ be a vocabulary, ${\sf R}\notin τ$ be a unary relation symbol, and ${\bf c}$ be a collection of $l$ constant symbols, where $l∈\mathbb{N}_{≥ 1}.$
Let $θ∈ \bar{Θ}[τ],$ let ${q}={\sf height}(θ)\cdot(\tw(θ)+1)^2+1,$ let $θ_{{\sf R},{\bf c}}$ be an enhanced version of $θ,$ and let $\tilde{θ}_q$ be the split version of $θ_{{\sf R},{\bf c}}.$
If $\mathfrak{A}$ is a $τ$-structure, $R\subseteq V(\mathfrak{A}),$
${\bf W}_{{q}}$ is a $q$-pseudogrid of $G_{\mathfrak{A}},$
and ${\bf a}$ is an apex-tuple of $\mathfrak{A}$ of size $l,$
then $(\mathfrak{A},R, {\bf a})\models θ_{{\sf R},{\bf c}}\iff (\mathfrak{A},R,{\bf W}_{{q}}, {\bf a})\models \tilde{θ}_{{q}}.$
\end{lemma}

By~\autoref{@interruptions}, we also get the following:
\begin{corollary}\labels{corro_gen}
Let $τ$ be a vocabulary, ${\sf R}\notin τ$ be a unary relation symbol, and ${\bf c}$ be a collection of $l$ constant symbols, where $l∈\mathbb{N}_{≥ 1}.$
Let $θ∈ \bar{Θ}[τ],$ let ${q}={\sf height}(θ)\cdot(\tw(θ)+1)^2+1,$ let $θ_{{\sf R},{\bf c}}$ be an enhanced version of $θ,$ and let $\tilde{θ}_q$ be the split version of $θ_{{\sf R},{\bf c}}.$
If $\mathfrak{A}$ is a $τ$-structure, $R\subseteq V(\mathfrak{A}),$
${\bf W}_{{q}}^{(1)}, {\bf W}_q^{(2)}$ are two $q$-pseudogrids of $G_{\mathfrak{A}},$
and ${\bf a}$ is an apex-tuple of $\mathfrak{A}$ of size $l,$
then $(\mathfrak{A},R,{\bf W}_{{q}}^{(1)}, {\bf a})\models \tilde{θ}_{{q}}\iff (\mathfrak{A},R,{\bf W}_{{q}}^{(2)}, {\bf a})\models \tilde{θ}_{{q}}.$
\end{corollary}

\myskip\subsection{Modifying the signatures}\labels{@enthaltenden}
In this subsection
we generalize the definitions given in~\autoref{sec_out-sig_first-floor} and~\autoref{sec_in-sig_first-floor},
for a sentence in $\bar{Θ}[τ].$

Let $τ$ be a vocabulary, ${\sf R}\notin τ$ be a unary relation symbol, and ${\bf c}$ be a collection of $l$ constant symbols, where $l∈\mathbb{N}_{≥ 1}.$
Let $θ∈ \bar{Θ}_1[τ],$ let ${q}={\sf height}(θ)\cdot(\tw(θ)+1)^2+1,$ let $θ_{{\sf R},{\bf c}}$ be an enhanced version of $θ,$ and let $\tilde{θ}_q$ be the split version of $θ_{{\sf R},{\bf c}}.$
Recall that
$$\tilde{θ}_{{q}} =\exists {\sf X}_1,\ldots, {\sf X}_h\  \bigg(θ^{\sf out}_{{q}}\ \wedge \exists {\sf C} \Big( η_{w\text{-}{\sf pr}_{{\cal X},{\sf C}}}\ \wedge μ|_{{\sf ind}_{\sf C}} \Big) \wedge
\exists {\sf C} \Big( η_{w\text{-}{\sf pr}_{{\cal X},{\sf C}}}\ \wedge \breve{ζ}_{\sf R} |_{\sf ap_{\bf a}}|_{{\sf ind}_{\sf C}}\Big)\bigg)$$
where ${\cal X}=\{{\sf X}_1,\ldots, {\sf X}_h\},$ $θ^{\sf out}_{{q}}∈  \MSOL[τ\cup\{{\sf R}\}\cup{\bf Q}\cup{\cal X}\cup{\bf c}],$ $\breve{ζ}_{\sf R} |_{\sf ap_{\bf a}}∈ \FOL[τ\cup\{{\sf R}\}\cup{\bf c}],$ and $μ∈ \NTMC[τ].$

Recall that
there exist $p∈ \mathbb{N}_{≥ 1},$ $\ell_1, \ldots, \ell_p, r_1, \ldots, r_p∈ \mathbb{N}_{≥ 1},$ and sentences $\tilde{ζ}_1, \ldots, \tilde{ζ}_p∈ \FOL[τ^{\langle{\bf c}\rangle}\cup\{{\sf R}\}]$ such that $\breve{ζ}_{\sf R}$ is a Boolean combination of $\tilde{ζ}_1, \ldots, \tilde{ζ}_p$ and for every $h∈ [p],$ $\tilde{ζ}_h$ is a basic local sentence with parameters $\ell_h$ and $r_h,$ i.e.,
\begin{eqnarray*}
\tilde{ζ}_h = \exists {\sf x}_{1}\ldots\exists {\sf x}_{\ell_{h}}\big(\bigwedge_{i∈ [\ell_{h}]}{\sf x}_{i}∈ {\sf R}\wedge \bigwedge_{1≤ i<j≤ \ell_{h}} d({\sf x}_{i}, {\sf x}_{j})> 2r_{h}\wedge \bigwedge_{i∈ [\ell_{h}]}ψ_{h}({\sf x}_{i})\big),
\end{eqnarray*}
where $ψ_h$ is an $r_h$-local formula  in $\FOL[τ^{\langle{\bf c}\rangle}]$ with one free variable.
Let $\hat{r}:= \max_{h∈[p]}\{r_h\}$ and $\hat{\ell}:=\max_{h∈ [p]}\{\ell_h\}.$
\medskip

For the rest of this subsection, keep in mind that $q=(\tw(θ)+1)^2+1.$
Let $j'∈ \mathbb{N}.$
We set  \begin{eqnarray*}
j & = &  {\sf odd}(\max\{q/2,j'\}),\\
r & = &2\cdot (\hat{\ell}+ 3)\cdot \hat{r}, \mbox{~and~}\\
w & =& (r+2)\cdot q.\\
\end{eqnarray*}

\myskip\paragraph{Towards constructing a boundaried structure.}
Let $q,j',j,r,w,$ and $l$ as above.
Let $\mathfrak{A}$ be a $τ$-structure, let $G_{\mathfrak{A}}$ be its Gaifman graph, let ${\bf a}=(a_{1},\ldots,a_l)$ be an apex-tuple  of $G_{\mathfrak{A}},$ let $(W,\mathfrak{R})$ be a flatness pair of $G_{\mathfrak{A}}\setminus V({\bf a})$ of height $2w+j,$ and let ${\bf W}_{{q}}$ be the $q$-pseudogrid defined by the horizontal and vertical paths of the central ${q}$-subwall of $W.$
Also, let
$\mathfrak{K}=(\mathfrak{A}[V(K^{\bf a})],{\bf a}, {\bf I}, {\bf W}_{{q}})$  be the extended compass
of the flatness pair $(W,\mathfrak{R})$ of $G_{\mathfrak{A}}\setminus V({\bf a})$ and let
$\ell∈[0,h\cdot(\tw(θ)-1)].$
Given a $d∈[r,w],$ a collection ${\cal Z}$ of $h$ subsets of $I^{(d-r+1)},$ a collection of pairwise disjoint sets ${\cal L} = \{L_1, \ldots, L_h\}$ such that, for every $i∈[h],$ $L_i∈ 2^{[l]},$ and a collection of graphs ${\cal F} = \{F_1, \ldots, F_h\}$ such that for every $i∈[h],$ $F_i ∈ {\cal F}^{V_{L_i} ({\bf a})}_{\ell},$
we define the graph $K^{(d,{\cal Z},{\cal L},{\cal F})}$ as the one obtained from ${K^{\bf a}[I^{(d)}\cup V_{\cupall {\cal L}} ({\bf a})]\cup \cupall {\cal F}}$ by making every vertex in $V(\cupall {\cal F}\setminus V_{\cupall {\cal L}} ({\bf a}))$ adjacent to an arbitrarily chosen vertex in  $I^{(d)}\setminus \cupall {\cal Z}$ (see~\autoref{fig_contr_mult} for an example).
\begin{figure}[ht]
	\begin{center}
\tikzstyle{ipe stylesheet} = [
  ipe import,
  even odd rule,
  line join=round,
  line cap=butt,
  ipe pen normal/.style={line width=0.4},
  ipe pen heavier/.style={line width=0.8},
  ipe pen fat/.style={line width=1.2},
  ipe pen ultrafat/.style={line width=2},
  ipe pen normal,
  ipe mark normal/.style={ipe mark scale=3},
  ipe mark large/.style={ipe mark scale=5},
  ipe mark small/.style={ipe mark scale=2},
  ipe mark tiny/.style={ipe mark scale=1.1},
  ipe mark normal,
  /pgf/arrow keys/.cd,
  ipe arrow normal/.style={scale=7},
  ipe arrow large/.style={scale=10},
  ipe arrow small/.style={scale=5},
  ipe arrow tiny/.style={scale=3},
  ipe arrow normal,
  /tikz/.cd,
  ipe arrows, 
  <->/.tip = ipe normal,
  ipe dash normal/.style={dash pattern=},
  ipe dash dotted/.style={dash pattern=on 1bp off 3bp},
  ipe dash dashed/.style={dash pattern=on 4bp off 4bp},
  ipe dash dash dotted/.style={dash pattern=on 4bp off 2bp on 1bp off 2bp},
  ipe dash dash dot dotted/.style={dash pattern=on 4bp off 2bp on 1bp off 2bp on 1bp off 2bp},
  ipe dash normal,
  ipe node/.append style={font=\normalsize},
  ipe stretch normal/.style={ipe node stretch=1},
  ipe stretch normal,
  ipe opacity 10/.style={opacity=0.1},
  ipe opacity 30/.style={opacity=0.3},
  ipe opacity 50/.style={opacity=0.5},
  ipe opacity 75/.style={opacity=0.75},
  ipe opacity opaque/.style={opacity=1},
  ipe opacity opaque,
]
\definecolor{black}{rgb}{0,0,0}
\definecolor{white}{rgb}{1,1,1}
\definecolor{red}{rgb}{1,0,0}
\definecolor{blue}{rgb}{0,0,1}
\definecolor{green}{rgb}{0,1,0}
\definecolor{yellow}{rgb}{1,1,0}
\definecolor{orange}{rgb}{1,0.647,0}
\definecolor{gold}{rgb}{1,0.843,0}
\definecolor{purple}{rgb}{0.627,0.125,0.941}
\definecolor{gray}{rgb}{0.745,0.745,0.745}
\definecolor{brown}{rgb}{0.647,0.165,0.165}
\definecolor{navy}{rgb}{0,0,0.502}
\definecolor{pink}{rgb}{1,0.753,0.796}
\definecolor{seagreen}{rgb}{0.18,0.545,0.341}
\definecolor{turquoise}{rgb}{0.251,0.878,0.816}
\definecolor{violet}{rgb}{0.933,0.51,0.933}
\definecolor{darkblue}{rgb}{0,0,0.545}
\definecolor{darkcyan}{rgb}{0,0.545,0.545}
\definecolor{darkgray}{rgb}{0.663,0.663,0.663}
\definecolor{darkgreen}{rgb}{0,0.392,0}
\definecolor{darkmagenta}{rgb}{0.545,0,0.545}
\definecolor{darkorange}{rgb}{1,0.549,0}
\definecolor{darkred}{rgb}{0.545,0,0}
\definecolor{lightblue}{rgb}{0.678,0.847,0.902}
\definecolor{lightcyan}{rgb}{0.878,1,1}
\definecolor{lightgray}{rgb}{0.827,0.827,0.827}
\definecolor{lightgreen}{rgb}{0.565,0.933,0.565}
\definecolor{lightyellow}{rgb}{1,1,0.878}
\definecolor{aliceblue}{rgb}{0.941,0.973,1}
\definecolor{antiquewhite}{rgb}{0.98,0.922,0.843}
\definecolor{antiquewhite1}{rgb}{1,0.937,0.859}
\definecolor{antiquewhite2}{rgb}{0.933,0.875,0.8}
\definecolor{antiquewhite3}{rgb}{0.804,0.753,0.69}
\definecolor{antiquewhite4}{rgb}{0.545,0.514,0.471}
\definecolor{aquamarine}{rgb}{0.498,1,0.831}
\definecolor{aquamarine1}{rgb}{0.498,1,0.831}
\definecolor{aquamarine2}{rgb}{0.463,0.933,0.776}
\definecolor{aquamarine3}{rgb}{0.4,0.804,0.667}
\definecolor{aquamarine4}{rgb}{0.271,0.545,0.455}
\definecolor{azure}{rgb}{0.941,1,1}
\definecolor{azure1}{rgb}{0.941,1,1}
\definecolor{azure2}{rgb}{0.878,0.933,0.933}
\definecolor{azure3}{rgb}{0.757,0.804,0.804}
\definecolor{azure4}{rgb}{0.514,0.545,0.545}
\definecolor{beige}{rgb}{0.961,0.961,0.863}
\definecolor{bisque}{rgb}{1,0.894,0.769}
\definecolor{bisque1}{rgb}{1,0.894,0.769}
\definecolor{bisque2}{rgb}{0.933,0.835,0.718}
\definecolor{bisque3}{rgb}{0.804,0.718,0.62}
\definecolor{bisque4}{rgb}{0.545,0.49,0.42}
\definecolor{blanchedalmond}{rgb}{1,0.922,0.804}
\definecolor{blue1}{rgb}{0,0,1}
\definecolor{blue2}{rgb}{0,0,0.933}
\definecolor{blue3}{rgb}{0,0,0.804}
\definecolor{blue4}{rgb}{0,0,0.545}
\definecolor{blueviolet}{rgb}{0.541,0.169,0.886}
\definecolor{brown1}{rgb}{1,0.251,0.251}
\definecolor{brown2}{rgb}{0.933,0.231,0.231}
\definecolor{brown3}{rgb}{0.804,0.2,0.2}
\definecolor{brown4}{rgb}{0.545,0.137,0.137}
\definecolor{burlywood}{rgb}{0.871,0.722,0.529}
\definecolor{burlywood1}{rgb}{1,0.827,0.608}
\definecolor{burlywood2}{rgb}{0.933,0.773,0.569}
\definecolor{burlywood3}{rgb}{0.804,0.667,0.49}
\definecolor{burlywood4}{rgb}{0.545,0.451,0.333}
\definecolor{cadetblue}{rgb}{0.373,0.62,0.627}
\definecolor{cadetblue1}{rgb}{0.596,0.961,1}
\definecolor{cadetblue2}{rgb}{0.557,0.898,0.933}
\definecolor{cadetblue3}{rgb}{0.478,0.773,0.804}
\definecolor{cadetblue4}{rgb}{0.325,0.525,0.545}
\definecolor{chartreuse}{rgb}{0.498,1,0}
\definecolor{chartreuse1}{rgb}{0.498,1,0}
\definecolor{chartreuse2}{rgb}{0.463,0.933,0}
\definecolor{chartreuse3}{rgb}{0.4,0.804,0}
\definecolor{chartreuse4}{rgb}{0.271,0.545,0}
\definecolor{chocolate}{rgb}{0.824,0.412,0.118}
\definecolor{chocolate1}{rgb}{1,0.498,0.141}
\definecolor{chocolate2}{rgb}{0.933,0.463,0.129}
\definecolor{chocolate3}{rgb}{0.804,0.4,0.114}
\definecolor{chocolate4}{rgb}{0.545,0.271,0.075}
\definecolor{coral}{rgb}{1,0.498,0.314}
\definecolor{coral1}{rgb}{1,0.447,0.337}
\definecolor{coral2}{rgb}{0.933,0.416,0.314}
\definecolor{coral3}{rgb}{0.804,0.357,0.271}
\definecolor{coral4}{rgb}{0.545,0.243,0.184}
\definecolor{cornflowerblue}{rgb}{0.392,0.584,0.929}
\definecolor{cornsilk}{rgb}{1,0.973,0.863}
\definecolor{cornsilk1}{rgb}{1,0.973,0.863}
\definecolor{cornsilk2}{rgb}{0.933,0.91,0.804}
\definecolor{cornsilk3}{rgb}{0.804,0.784,0.694}
\definecolor{cornsilk4}{rgb}{0.545,0.533,0.471}
\definecolor{cyan}{rgb}{0,1,1}
\definecolor{cyan1}{rgb}{0,1,1}
\definecolor{cyan2}{rgb}{0,0.933,0.933}
\definecolor{cyan3}{rgb}{0,0.804,0.804}
\definecolor{cyan4}{rgb}{0,0.545,0.545}
\definecolor{darkgoldenrod}{rgb}{0.722,0.525,0.043}
\definecolor{darkgoldenrod1}{rgb}{1,0.725,0.059}
\definecolor{darkgoldenrod2}{rgb}{0.933,0.678,0.055}
\definecolor{darkgoldenrod3}{rgb}{0.804,0.584,0.047}
\definecolor{darkgoldenrod4}{rgb}{0.545,0.396,0.031}
\definecolor{darkgrey}{rgb}{0.663,0.663,0.663}
\definecolor{darkkhaki}{rgb}{0.741,0.718,0.42}
\definecolor{darkolivegreen}{rgb}{0.333,0.42,0.184}
\definecolor{darkolivegreen1}{rgb}{0.792,1,0.439}
\definecolor{darkolivegreen2}{rgb}{0.737,0.933,0.408}
\definecolor{darkolivegreen3}{rgb}{0.635,0.804,0.353}
\definecolor{darkolivegreen4}{rgb}{0.431,0.545,0.239}
\definecolor{darkorange1}{rgb}{1,0.498,0}
\definecolor{darkorange2}{rgb}{0.933,0.463,0}
\definecolor{darkorange3}{rgb}{0.804,0.4,0}
\definecolor{darkorange4}{rgb}{0.545,0.271,0}
\definecolor{darkorchid}{rgb}{0.6,0.196,0.8}
\definecolor{darkorchid1}{rgb}{0.749,0.243,1}
\definecolor{darkorchid2}{rgb}{0.698,0.227,0.933}
\definecolor{darkorchid3}{rgb}{0.604,0.196,0.804}
\definecolor{darkorchid4}{rgb}{0.408,0.133,0.545}
\definecolor{darksalmon}{rgb}{0.914,0.588,0.478}
\definecolor{darkseagreen}{rgb}{0.561,0.737,0.561}
\definecolor{darkseagreen1}{rgb}{0.757,1,0.757}
\definecolor{darkseagreen2}{rgb}{0.706,0.933,0.706}
\definecolor{darkseagreen3}{rgb}{0.608,0.804,0.608}
\definecolor{darkseagreen4}{rgb}{0.412,0.545,0.412}
\definecolor{darkslateblue}{rgb}{0.282,0.239,0.545}
\definecolor{darkslategray}{rgb}{0.184,0.31,0.31}
\definecolor{darkslategray1}{rgb}{0.592,1,1}
\definecolor{darkslategray2}{rgb}{0.553,0.933,0.933}
\definecolor{darkslategray3}{rgb}{0.475,0.804,0.804}
\definecolor{darkslategray4}{rgb}{0.322,0.545,0.545}
\definecolor{darkslategrey}{rgb}{0.184,0.31,0.31}
\definecolor{darkturquoise}{rgb}{0,0.808,0.82}
\definecolor{darkviolet}{rgb}{0.58,0,0.827}
\definecolor{deeppink}{rgb}{1,0.078,0.576}
\definecolor{deeppink1}{rgb}{1,0.078,0.576}
\definecolor{deeppink2}{rgb}{0.933,0.071,0.537}
\definecolor{deeppink3}{rgb}{0.804,0.063,0.463}
\definecolor{deeppink4}{rgb}{0.545,0.039,0.314}
\definecolor{deepskyblue}{rgb}{0,0.749,1}
\definecolor{deepskyblue1}{rgb}{0,0.749,1}
\definecolor{deepskyblue2}{rgb}{0,0.698,0.933}
\definecolor{deepskyblue3}{rgb}{0,0.604,0.804}
\definecolor{deepskyblue4}{rgb}{0,0.408,0.545}
\definecolor{dimgray}{rgb}{0.412,0.412,0.412}
\definecolor{dimgrey}{rgb}{0.412,0.412,0.412}
\definecolor{dodgerblue}{rgb}{0.118,0.565,1}
\definecolor{dodgerblue1}{rgb}{0.118,0.565,1}
\definecolor{dodgerblue2}{rgb}{0.11,0.525,0.933}
\definecolor{dodgerblue3}{rgb}{0.094,0.455,0.804}
\definecolor{dodgerblue4}{rgb}{0.063,0.306,0.545}
\definecolor{firebrick}{rgb}{0.698,0.133,0.133}
\definecolor{firebrick1}{rgb}{1,0.188,0.188}
\definecolor{firebrick2}{rgb}{0.933,0.173,0.173}
\definecolor{firebrick3}{rgb}{0.804,0.149,0.149}
\definecolor{firebrick4}{rgb}{0.545,0.102,0.102}
\definecolor{floralwhite}{rgb}{1,0.98,0.941}
\definecolor{forestgreen}{rgb}{0.133,0.545,0.133}
\definecolor{gainsboro}{rgb}{0.863,0.863,0.863}
\definecolor{ghostwhite}{rgb}{0.973,0.973,1}
\definecolor{gold1}{rgb}{1,0.843,0}
\definecolor{gold2}{rgb}{0.933,0.788,0}
\definecolor{gold3}{rgb}{0.804,0.678,0}
\definecolor{gold4}{rgb}{0.545,0.459,0}
\definecolor{goldenrod}{rgb}{0.855,0.647,0.125}
\definecolor{goldenrod1}{rgb}{1,0.757,0.145}
\definecolor{goldenrod2}{rgb}{0.933,0.706,0.133}
\definecolor{goldenrod3}{rgb}{0.804,0.608,0.114}
\definecolor{goldenrod4}{rgb}{0.545,0.412,0.078}
\definecolor{gray0}{rgb}{0,0,0}
\definecolor{gray1}{rgb}{0.012,0.012,0.012}
\definecolor{gray10}{rgb}{0.102,0.102,0.102}
\definecolor{gray100}{rgb}{1,1,1}
\definecolor{gray11}{rgb}{0.11,0.11,0.11}
\definecolor{gray12}{rgb}{0.122,0.122,0.122}
\definecolor{gray13}{rgb}{0.129,0.129,0.129}
\definecolor{gray14}{rgb}{0.141,0.141,0.141}
\definecolor{gray15}{rgb}{0.149,0.149,0.149}
\definecolor{gray16}{rgb}{0.161,0.161,0.161}
\definecolor{gray17}{rgb}{0.169,0.169,0.169}
\definecolor{gray18}{rgb}{0.18,0.18,0.18}
\definecolor{gray19}{rgb}{0.188,0.188,0.188}
\definecolor{gray2}{rgb}{0.02,0.02,0.02}
\definecolor{gray20}{rgb}{0.2,0.2,0.2}
\definecolor{gray21}{rgb}{0.212,0.212,0.212}
\definecolor{gray22}{rgb}{0.22,0.22,0.22}
\definecolor{gray23}{rgb}{0.231,0.231,0.231}
\definecolor{gray24}{rgb}{0.239,0.239,0.239}
\definecolor{gray25}{rgb}{0.251,0.251,0.251}
\definecolor{gray26}{rgb}{0.259,0.259,0.259}
\definecolor{gray27}{rgb}{0.271,0.271,0.271}
\definecolor{gray28}{rgb}{0.278,0.278,0.278}
\definecolor{gray29}{rgb}{0.29,0.29,0.29}
\definecolor{gray3}{rgb}{0.031,0.031,0.031}
\definecolor{gray30}{rgb}{0.302,0.302,0.302}
\definecolor{gray31}{rgb}{0.31,0.31,0.31}
\definecolor{gray32}{rgb}{0.322,0.322,0.322}
\definecolor{gray33}{rgb}{0.329,0.329,0.329}
\definecolor{gray34}{rgb}{0.341,0.341,0.341}
\definecolor{gray35}{rgb}{0.349,0.349,0.349}
\definecolor{gray36}{rgb}{0.361,0.361,0.361}
\definecolor{gray37}{rgb}{0.369,0.369,0.369}
\definecolor{gray38}{rgb}{0.38,0.38,0.38}
\definecolor{gray39}{rgb}{0.388,0.388,0.388}
\definecolor{gray4}{rgb}{0.039,0.039,0.039}
\definecolor{gray40}{rgb}{0.4,0.4,0.4}
\definecolor{gray41}{rgb}{0.412,0.412,0.412}
\definecolor{gray42}{rgb}{0.42,0.42,0.42}
\definecolor{gray43}{rgb}{0.431,0.431,0.431}
\definecolor{gray44}{rgb}{0.439,0.439,0.439}
\definecolor{gray45}{rgb}{0.451,0.451,0.451}
\definecolor{gray46}{rgb}{0.459,0.459,0.459}
\definecolor{gray47}{rgb}{0.471,0.471,0.471}
\definecolor{gray48}{rgb}{0.478,0.478,0.478}
\definecolor{gray49}{rgb}{0.49,0.49,0.49}
\definecolor{gray5}{rgb}{0.051,0.051,0.051}
\definecolor{gray50}{rgb}{0.498,0.498,0.498}
\definecolor{gray51}{rgb}{0.51,0.51,0.51}
\definecolor{gray52}{rgb}{0.522,0.522,0.522}
\definecolor{gray53}{rgb}{0.529,0.529,0.529}
\definecolor{gray54}{rgb}{0.541,0.541,0.541}
\definecolor{gray55}{rgb}{0.549,0.549,0.549}
\definecolor{gray56}{rgb}{0.561,0.561,0.561}
\definecolor{gray57}{rgb}{0.569,0.569,0.569}
\definecolor{gray58}{rgb}{0.58,0.58,0.58}
\definecolor{gray59}{rgb}{0.588,0.588,0.588}
\definecolor{gray6}{rgb}{0.059,0.059,0.059}
\definecolor{gray60}{rgb}{0.6,0.6,0.6}
\definecolor{gray61}{rgb}{0.612,0.612,0.612}
\definecolor{gray62}{rgb}{0.62,0.62,0.62}
\definecolor{gray63}{rgb}{0.631,0.631,0.631}
\definecolor{gray64}{rgb}{0.639,0.639,0.639}
\definecolor{gray65}{rgb}{0.651,0.651,0.651}
\definecolor{gray66}{rgb}{0.659,0.659,0.659}
\definecolor{gray67}{rgb}{0.671,0.671,0.671}
\definecolor{gray68}{rgb}{0.678,0.678,0.678}
\definecolor{gray69}{rgb}{0.69,0.69,0.69}
\definecolor{gray7}{rgb}{0.071,0.071,0.071}
\definecolor{gray70}{rgb}{0.702,0.702,0.702}
\definecolor{gray71}{rgb}{0.71,0.71,0.71}
\definecolor{gray72}{rgb}{0.722,0.722,0.722}
\definecolor{gray73}{rgb}{0.729,0.729,0.729}
\definecolor{gray74}{rgb}{0.741,0.741,0.741}
\definecolor{gray75}{rgb}{0.749,0.749,0.749}
\definecolor{gray76}{rgb}{0.761,0.761,0.761}
\definecolor{gray77}{rgb}{0.769,0.769,0.769}
\definecolor{gray78}{rgb}{0.78,0.78,0.78}
\definecolor{gray79}{rgb}{0.788,0.788,0.788}
\definecolor{gray8}{rgb}{0.078,0.078,0.078}
\definecolor{gray80}{rgb}{0.8,0.8,0.8}
\definecolor{gray81}{rgb}{0.812,0.812,0.812}
\definecolor{gray82}{rgb}{0.82,0.82,0.82}
\definecolor{gray83}{rgb}{0.831,0.831,0.831}
\definecolor{gray84}{rgb}{0.839,0.839,0.839}
\definecolor{gray85}{rgb}{0.851,0.851,0.851}
\definecolor{gray86}{rgb}{0.859,0.859,0.859}
\definecolor{gray87}{rgb}{0.871,0.871,0.871}
\definecolor{gray88}{rgb}{0.878,0.878,0.878}
\definecolor{gray89}{rgb}{0.89,0.89,0.89}
\definecolor{gray9}{rgb}{0.09,0.09,0.09}
\definecolor{gray90}{rgb}{0.898,0.898,0.898}
\definecolor{gray91}{rgb}{0.91,0.91,0.91}
\definecolor{gray92}{rgb}{0.922,0.922,0.922}
\definecolor{gray93}{rgb}{0.929,0.929,0.929}
\definecolor{gray94}{rgb}{0.941,0.941,0.941}
\definecolor{gray95}{rgb}{0.949,0.949,0.949}
\definecolor{gray96}{rgb}{0.961,0.961,0.961}
\definecolor{gray97}{rgb}{0.969,0.969,0.969}
\definecolor{gray98}{rgb}{0.98,0.98,0.98}
\definecolor{gray99}{rgb}{0.988,0.988,0.988}
\definecolor{green1}{rgb}{0,1,0}
\definecolor{green2}{rgb}{0,0.933,0}
\definecolor{green3}{rgb}{0,0.804,0}
\definecolor{green4}{rgb}{0,0.545,0}
\definecolor{greenyellow}{rgb}{0.678,1,0.184}
\definecolor{grey}{rgb}{0.745,0.745,0.745}
\definecolor{grey0}{rgb}{0,0,0}
\definecolor{grey1}{rgb}{0.012,0.012,0.012}
\definecolor{grey10}{rgb}{0.102,0.102,0.102}
\definecolor{grey100}{rgb}{1,1,1}
\definecolor{grey11}{rgb}{0.11,0.11,0.11}
\definecolor{grey12}{rgb}{0.122,0.122,0.122}
\definecolor{grey13}{rgb}{0.129,0.129,0.129}
\definecolor{grey14}{rgb}{0.141,0.141,0.141}
\definecolor{grey15}{rgb}{0.149,0.149,0.149}
\definecolor{grey16}{rgb}{0.161,0.161,0.161}
\definecolor{grey17}{rgb}{0.169,0.169,0.169}
\definecolor{grey18}{rgb}{0.18,0.18,0.18}
\definecolor{grey19}{rgb}{0.188,0.188,0.188}
\definecolor{grey2}{rgb}{0.02,0.02,0.02}
\definecolor{grey20}{rgb}{0.2,0.2,0.2}
\definecolor{grey21}{rgb}{0.212,0.212,0.212}
\definecolor{grey22}{rgb}{0.22,0.22,0.22}
\definecolor{grey23}{rgb}{0.231,0.231,0.231}
\definecolor{grey24}{rgb}{0.239,0.239,0.239}
\definecolor{grey25}{rgb}{0.251,0.251,0.251}
\definecolor{grey26}{rgb}{0.259,0.259,0.259}
\definecolor{grey27}{rgb}{0.271,0.271,0.271}
\definecolor{grey28}{rgb}{0.278,0.278,0.278}
\definecolor{grey29}{rgb}{0.29,0.29,0.29}
\definecolor{grey3}{rgb}{0.031,0.031,0.031}
\definecolor{grey30}{rgb}{0.302,0.302,0.302}
\definecolor{grey31}{rgb}{0.31,0.31,0.31}
\definecolor{grey32}{rgb}{0.322,0.322,0.322}
\definecolor{grey33}{rgb}{0.329,0.329,0.329}
\definecolor{grey34}{rgb}{0.341,0.341,0.341}
\definecolor{grey35}{rgb}{0.349,0.349,0.349}
\definecolor{grey36}{rgb}{0.361,0.361,0.361}
\definecolor{grey37}{rgb}{0.369,0.369,0.369}
\definecolor{grey38}{rgb}{0.38,0.38,0.38}
\definecolor{grey39}{rgb}{0.388,0.388,0.388}
\definecolor{grey4}{rgb}{0.039,0.039,0.039}
\definecolor{grey40}{rgb}{0.4,0.4,0.4}
\definecolor{grey41}{rgb}{0.412,0.412,0.412}
\definecolor{grey42}{rgb}{0.42,0.42,0.42}
\definecolor{grey43}{rgb}{0.431,0.431,0.431}
\definecolor{grey44}{rgb}{0.439,0.439,0.439}
\definecolor{grey45}{rgb}{0.451,0.451,0.451}
\definecolor{grey46}{rgb}{0.459,0.459,0.459}
\definecolor{grey47}{rgb}{0.471,0.471,0.471}
\definecolor{grey48}{rgb}{0.478,0.478,0.478}
\definecolor{grey49}{rgb}{0.49,0.49,0.49}
\definecolor{grey5}{rgb}{0.051,0.051,0.051}
\definecolor{grey50}{rgb}{0.498,0.498,0.498}
\definecolor{grey51}{rgb}{0.51,0.51,0.51}
\definecolor{grey52}{rgb}{0.522,0.522,0.522}
\definecolor{grey53}{rgb}{0.529,0.529,0.529}
\definecolor{grey54}{rgb}{0.541,0.541,0.541}
\definecolor{grey55}{rgb}{0.549,0.549,0.549}
\definecolor{grey56}{rgb}{0.561,0.561,0.561}
\definecolor{grey57}{rgb}{0.569,0.569,0.569}
\definecolor{grey58}{rgb}{0.58,0.58,0.58}
\definecolor{grey59}{rgb}{0.588,0.588,0.588}
\definecolor{grey6}{rgb}{0.059,0.059,0.059}
\definecolor{grey60}{rgb}{0.6,0.6,0.6}
\definecolor{grey61}{rgb}{0.612,0.612,0.612}
\definecolor{grey62}{rgb}{0.62,0.62,0.62}
\definecolor{grey63}{rgb}{0.631,0.631,0.631}
\definecolor{grey64}{rgb}{0.639,0.639,0.639}
\definecolor{grey65}{rgb}{0.651,0.651,0.651}
\definecolor{grey66}{rgb}{0.659,0.659,0.659}
\definecolor{grey67}{rgb}{0.671,0.671,0.671}
\definecolor{grey68}{rgb}{0.678,0.678,0.678}
\definecolor{grey69}{rgb}{0.69,0.69,0.69}
\definecolor{grey7}{rgb}{0.071,0.071,0.071}
\definecolor{grey70}{rgb}{0.702,0.702,0.702}
\definecolor{grey71}{rgb}{0.71,0.71,0.71}
\definecolor{grey72}{rgb}{0.722,0.722,0.722}
\definecolor{grey73}{rgb}{0.729,0.729,0.729}
\definecolor{grey74}{rgb}{0.741,0.741,0.741}
\definecolor{grey75}{rgb}{0.749,0.749,0.749}
\definecolor{grey76}{rgb}{0.761,0.761,0.761}
\definecolor{grey77}{rgb}{0.769,0.769,0.769}
\definecolor{grey78}{rgb}{0.78,0.78,0.78}
\definecolor{grey79}{rgb}{0.788,0.788,0.788}
\definecolor{grey8}{rgb}{0.078,0.078,0.078}
\definecolor{grey80}{rgb}{0.8,0.8,0.8}
\definecolor{grey81}{rgb}{0.812,0.812,0.812}
\definecolor{grey82}{rgb}{0.82,0.82,0.82}
\definecolor{grey83}{rgb}{0.831,0.831,0.831}
\definecolor{grey84}{rgb}{0.839,0.839,0.839}
\definecolor{grey85}{rgb}{0.851,0.851,0.851}
\definecolor{grey86}{rgb}{0.859,0.859,0.859}
\definecolor{grey87}{rgb}{0.871,0.871,0.871}
\definecolor{grey88}{rgb}{0.878,0.878,0.878}
\definecolor{grey89}{rgb}{0.89,0.89,0.89}
\definecolor{grey9}{rgb}{0.09,0.09,0.09}
\definecolor{grey90}{rgb}{0.898,0.898,0.898}
\definecolor{grey91}{rgb}{0.91,0.91,0.91}
\definecolor{grey92}{rgb}{0.922,0.922,0.922}
\definecolor{grey93}{rgb}{0.929,0.929,0.929}
\definecolor{grey94}{rgb}{0.941,0.941,0.941}
\definecolor{grey95}{rgb}{0.949,0.949,0.949}
\definecolor{grey96}{rgb}{0.961,0.961,0.961}
\definecolor{grey97}{rgb}{0.969,0.969,0.969}
\definecolor{grey98}{rgb}{0.98,0.98,0.98}
\definecolor{grey99}{rgb}{0.988,0.988,0.988}
\definecolor{honeydew}{rgb}{0.941,1,0.941}
\definecolor{honeydew1}{rgb}{0.941,1,0.941}
\definecolor{honeydew2}{rgb}{0.878,0.933,0.878}
\definecolor{honeydew3}{rgb}{0.757,0.804,0.757}
\definecolor{honeydew4}{rgb}{0.514,0.545,0.514}
\definecolor{hotpink}{rgb}{1,0.412,0.706}
\definecolor{hotpink1}{rgb}{1,0.431,0.706}
\definecolor{hotpink2}{rgb}{0.933,0.416,0.655}
\definecolor{hotpink3}{rgb}{0.804,0.376,0.565}
\definecolor{hotpink4}{rgb}{0.545,0.227,0.384}
\definecolor{indianred}{rgb}{0.804,0.361,0.361}
\definecolor{indianred1}{rgb}{1,0.416,0.416}
\definecolor{indianred2}{rgb}{0.933,0.388,0.388}
\definecolor{indianred3}{rgb}{0.804,0.333,0.333}
\definecolor{indianred4}{rgb}{0.545,0.227,0.227}
\definecolor{ivory}{rgb}{1,1,0.941}
\definecolor{ivory1}{rgb}{1,1,0.941}
\definecolor{ivory2}{rgb}{0.933,0.933,0.878}
\definecolor{ivory3}{rgb}{0.804,0.804,0.757}
\definecolor{ivory4}{rgb}{0.545,0.545,0.514}
\definecolor{khaki}{rgb}{0.941,0.902,0.549}
\definecolor{khaki1}{rgb}{1,0.965,0.561}
\definecolor{khaki2}{rgb}{0.933,0.902,0.522}
\definecolor{khaki3}{rgb}{0.804,0.776,0.451}
\definecolor{khaki4}{rgb}{0.545,0.525,0.306}
\definecolor{lavender}{rgb}{0.902,0.902,0.98}
\definecolor{lavenderblush}{rgb}{1,0.941,0.961}
\definecolor{lavenderblush1}{rgb}{1,0.941,0.961}
\definecolor{lavenderblush2}{rgb}{0.933,0.878,0.898}
\definecolor{lavenderblush3}{rgb}{0.804,0.757,0.773}
\definecolor{lavenderblush4}{rgb}{0.545,0.514,0.525}
\definecolor{lawngreen}{rgb}{0.486,0.988,0}
\definecolor{lemonchiffon}{rgb}{1,0.98,0.804}
\definecolor{lemonchiffon1}{rgb}{1,0.98,0.804}
\definecolor{lemonchiffon2}{rgb}{0.933,0.914,0.749}
\definecolor{lemonchiffon3}{rgb}{0.804,0.788,0.647}
\definecolor{lemonchiffon4}{rgb}{0.545,0.537,0.439}
\definecolor{lightblue1}{rgb}{0.749,0.937,1}
\definecolor{lightblue2}{rgb}{0.698,0.875,0.933}
\definecolor{lightblue3}{rgb}{0.604,0.753,0.804}
\definecolor{lightblue4}{rgb}{0.408,0.514,0.545}
\definecolor{lightcoral}{rgb}{0.941,0.502,0.502}
\definecolor{lightcyan1}{rgb}{0.878,1,1}
\definecolor{lightcyan2}{rgb}{0.82,0.933,0.933}
\definecolor{lightcyan3}{rgb}{0.706,0.804,0.804}
\definecolor{lightcyan4}{rgb}{0.478,0.545,0.545}
\definecolor{lightgoldenrod}{rgb}{0.933,0.867,0.51}
\definecolor{lightgoldenrod1}{rgb}{1,0.925,0.545}
\definecolor{lightgoldenrod2}{rgb}{0.933,0.863,0.51}
\definecolor{lightgoldenrod3}{rgb}{0.804,0.745,0.439}
\definecolor{lightgoldenrod4}{rgb}{0.545,0.506,0.298}
\definecolor{lightgoldenrodyellow}{rgb}{0.98,0.98,0.824}
\definecolor{lightgrey}{rgb}{0.827,0.827,0.827}
\definecolor{lightpink}{rgb}{1,0.714,0.757}
\definecolor{lightpink1}{rgb}{1,0.682,0.725}
\definecolor{lightpink2}{rgb}{0.933,0.635,0.678}
\definecolor{lightpink3}{rgb}{0.804,0.549,0.584}
\definecolor{lightpink4}{rgb}{0.545,0.373,0.396}
\definecolor{lightsalmon}{rgb}{1,0.627,0.478}
\definecolor{lightsalmon1}{rgb}{1,0.627,0.478}
\definecolor{lightsalmon2}{rgb}{0.933,0.584,0.447}
\definecolor{lightsalmon3}{rgb}{0.804,0.506,0.384}
\definecolor{lightsalmon4}{rgb}{0.545,0.341,0.259}
\definecolor{lightseagreen}{rgb}{0.125,0.698,0.667}
\definecolor{lightskyblue}{rgb}{0.529,0.808,0.98}
\definecolor{lightskyblue1}{rgb}{0.69,0.886,1}
\definecolor{lightskyblue2}{rgb}{0.643,0.827,0.933}
\definecolor{lightskyblue3}{rgb}{0.553,0.714,0.804}
\definecolor{lightskyblue4}{rgb}{0.376,0.482,0.545}
\definecolor{lightslateblue}{rgb}{0.518,0.439,1}
\definecolor{lightslategray}{rgb}{0.467,0.533,0.6}
\definecolor{lightslategrey}{rgb}{0.467,0.533,0.6}
\definecolor{lightsteelblue}{rgb}{0.69,0.769,0.871}
\definecolor{lightsteelblue1}{rgb}{0.792,0.882,1}
\definecolor{lightsteelblue2}{rgb}{0.737,0.824,0.933}
\definecolor{lightsteelblue3}{rgb}{0.635,0.71,0.804}
\definecolor{lightsteelblue4}{rgb}{0.431,0.482,0.545}
\definecolor{lightyellow1}{rgb}{1,1,0.878}
\definecolor{lightyellow2}{rgb}{0.933,0.933,0.82}
\definecolor{lightyellow3}{rgb}{0.804,0.804,0.706}
\definecolor{lightyellow4}{rgb}{0.545,0.545,0.478}
\definecolor{limegreen}{rgb}{0.196,0.804,0.196}
\definecolor{linen}{rgb}{0.98,0.941,0.902}
\definecolor{magenta}{rgb}{1,0,1}
\definecolor{magenta1}{rgb}{1,0,1}
\definecolor{magenta2}{rgb}{0.933,0,0.933}
\definecolor{magenta3}{rgb}{0.804,0,0.804}
\definecolor{magenta4}{rgb}{0.545,0,0.545}
\definecolor{maroon}{rgb}{0.69,0.188,0.376}
\definecolor{maroon1}{rgb}{1,0.204,0.702}
\definecolor{maroon2}{rgb}{0.933,0.188,0.655}
\definecolor{maroon3}{rgb}{0.804,0.161,0.565}
\definecolor{maroon4}{rgb}{0.545,0.11,0.384}
\definecolor{mediumaquamarine}{rgb}{0.4,0.804,0.667}
\definecolor{mediumblue}{rgb}{0,0,0.804}
\definecolor{mediumorchid}{rgb}{0.729,0.333,0.827}
\definecolor{mediumorchid1}{rgb}{0.878,0.4,1}
\definecolor{mediumorchid2}{rgb}{0.82,0.373,0.933}
\definecolor{mediumorchid3}{rgb}{0.706,0.322,0.804}
\definecolor{mediumorchid4}{rgb}{0.478,0.216,0.545}
\definecolor{mediumpurple}{rgb}{0.576,0.439,0.859}
\definecolor{mediumpurple1}{rgb}{0.671,0.51,1}
\definecolor{mediumpurple2}{rgb}{0.624,0.475,0.933}
\definecolor{mediumpurple3}{rgb}{0.537,0.408,0.804}
\definecolor{mediumpurple4}{rgb}{0.365,0.278,0.545}
\definecolor{mediumseagreen}{rgb}{0.235,0.702,0.443}
\definecolor{mediumslateblue}{rgb}{0.482,0.408,0.933}
\definecolor{mediumspringgreen}{rgb}{0,0.98,0.604}
\definecolor{mediumturquoise}{rgb}{0.282,0.82,0.8}
\definecolor{mediumvioletred}{rgb}{0.78,0.082,0.522}
\definecolor{midnightblue}{rgb}{0.098,0.098,0.439}
\definecolor{mintcream}{rgb}{0.961,1,0.98}
\definecolor{mistyrose}{rgb}{1,0.894,0.882}
\definecolor{mistyrose1}{rgb}{1,0.894,0.882}
\definecolor{mistyrose2}{rgb}{0.933,0.835,0.824}
\definecolor{mistyrose3}{rgb}{0.804,0.718,0.71}
\definecolor{mistyrose4}{rgb}{0.545,0.49,0.482}
\definecolor{moccasin}{rgb}{1,0.894,0.71}
\definecolor{navajowhite}{rgb}{1,0.871,0.678}
\definecolor{navajowhite1}{rgb}{1,0.871,0.678}
\definecolor{navajowhite2}{rgb}{0.933,0.812,0.631}
\definecolor{navajowhite3}{rgb}{0.804,0.702,0.545}
\definecolor{navajowhite4}{rgb}{0.545,0.475,0.369}
\definecolor{navyblue}{rgb}{0,0,0.502}
\definecolor{oldlace}{rgb}{0.992,0.961,0.902}
\definecolor{olivedrab}{rgb}{0.42,0.557,0.137}
\definecolor{olivedrab1}{rgb}{0.753,1,0.243}
\definecolor{olivedrab2}{rgb}{0.702,0.933,0.227}
\definecolor{olivedrab3}{rgb}{0.604,0.804,0.196}
\definecolor{olivedrab4}{rgb}{0.412,0.545,0.133}
\definecolor{orange1}{rgb}{1,0.647,0}
\definecolor{orange2}{rgb}{0.933,0.604,0}
\definecolor{orange3}{rgb}{0.804,0.522,0}
\definecolor{orange4}{rgb}{0.545,0.353,0}
\definecolor{orangered}{rgb}{1,0.271,0}
\definecolor{orangered1}{rgb}{1,0.271,0}
\definecolor{orangered2}{rgb}{0.933,0.251,0}
\definecolor{orangered3}{rgb}{0.804,0.216,0}
\definecolor{orangered4}{rgb}{0.545,0.145,0}
\definecolor{orchid}{rgb}{0.855,0.439,0.839}
\definecolor{orchid1}{rgb}{1,0.514,0.98}
\definecolor{orchid2}{rgb}{0.933,0.478,0.914}
\definecolor{orchid3}{rgb}{0.804,0.412,0.788}
\definecolor{orchid4}{rgb}{0.545,0.278,0.537}
\definecolor{palegoldenrod}{rgb}{0.933,0.91,0.667}
\definecolor{palegreen}{rgb}{0.596,0.984,0.596}
\definecolor{palegreen1}{rgb}{0.604,1,0.604}
\definecolor{palegreen2}{rgb}{0.565,0.933,0.565}
\definecolor{palegreen3}{rgb}{0.486,0.804,0.486}
\definecolor{palegreen4}{rgb}{0.329,0.545,0.329}
\definecolor{paleturquoise}{rgb}{0.686,0.933,0.933}
\definecolor{paleturquoise1}{rgb}{0.733,1,1}
\definecolor{paleturquoise2}{rgb}{0.682,0.933,0.933}
\definecolor{paleturquoise3}{rgb}{0.588,0.804,0.804}
\definecolor{paleturquoise4}{rgb}{0.4,0.545,0.545}
\definecolor{palevioletred}{rgb}{0.859,0.439,0.576}
\definecolor{palevioletred1}{rgb}{1,0.51,0.671}
\definecolor{palevioletred2}{rgb}{0.933,0.475,0.624}
\definecolor{palevioletred3}{rgb}{0.804,0.408,0.537}
\definecolor{palevioletred4}{rgb}{0.545,0.278,0.365}
\definecolor{papayawhip}{rgb}{1,0.937,0.835}
\definecolor{peachpuff}{rgb}{1,0.855,0.725}
\definecolor{peachpuff1}{rgb}{1,0.855,0.725}
\definecolor{peachpuff2}{rgb}{0.933,0.796,0.678}
\definecolor{peachpuff3}{rgb}{0.804,0.686,0.584}
\definecolor{peachpuff4}{rgb}{0.545,0.467,0.396}
\definecolor{peru}{rgb}{0.804,0.522,0.247}
\definecolor{pink1}{rgb}{1,0.71,0.773}
\definecolor{pink2}{rgb}{0.933,0.663,0.722}
\definecolor{pink3}{rgb}{0.804,0.569,0.62}
\definecolor{pink4}{rgb}{0.545,0.388,0.424}
\definecolor{plum}{rgb}{0.867,0.627,0.867}
\definecolor{plum1}{rgb}{1,0.733,1}
\definecolor{plum2}{rgb}{0.933,0.682,0.933}
\definecolor{plum3}{rgb}{0.804,0.588,0.804}
\definecolor{plum4}{rgb}{0.545,0.4,0.545}
\definecolor{powderblue}{rgb}{0.69,0.878,0.902}
\definecolor{purple1}{rgb}{0.608,0.188,1}
\definecolor{purple2}{rgb}{0.569,0.173,0.933}
\definecolor{purple3}{rgb}{0.49,0.149,0.804}
\definecolor{purple4}{rgb}{0.333,0.102,0.545}
\definecolor{red1}{rgb}{1,0,0}
\definecolor{red2}{rgb}{0.933,0,0}
\definecolor{red3}{rgb}{0.804,0,0}
\definecolor{red4}{rgb}{0.545,0,0}
\definecolor{rosybrown}{rgb}{0.737,0.561,0.561}
\definecolor{rosybrown1}{rgb}{1,0.757,0.757}
\definecolor{rosybrown2}{rgb}{0.933,0.706,0.706}
\definecolor{rosybrown3}{rgb}{0.804,0.608,0.608}
\definecolor{rosybrown4}{rgb}{0.545,0.412,0.412}
\definecolor{royalblue}{rgb}{0.255,0.412,0.882}
\definecolor{royalblue1}{rgb}{0.282,0.463,1}
\definecolor{royalblue2}{rgb}{0.263,0.431,0.933}
\definecolor{royalblue3}{rgb}{0.227,0.373,0.804}
\definecolor{royalblue4}{rgb}{0.153,0.251,0.545}
\definecolor{saddlebrown}{rgb}{0.545,0.271,0.075}
\definecolor{salmon}{rgb}{0.98,0.502,0.447}
\definecolor{salmon1}{rgb}{1,0.549,0.412}
\definecolor{salmon2}{rgb}{0.933,0.51,0.384}
\definecolor{salmon3}{rgb}{0.804,0.439,0.329}
\definecolor{salmon4}{rgb}{0.545,0.298,0.224}
\definecolor{sandybrown}{rgb}{0.957,0.643,0.376}
\definecolor{seagreen1}{rgb}{0.329,1,0.624}
\definecolor{seagreen2}{rgb}{0.306,0.933,0.58}
\definecolor{seagreen3}{rgb}{0.263,0.804,0.502}
\definecolor{seagreen4}{rgb}{0.18,0.545,0.341}
\definecolor{seashell}{rgb}{1,0.961,0.933}
\definecolor{seashell1}{rgb}{1,0.961,0.933}
\definecolor{seashell2}{rgb}{0.933,0.898,0.871}
\definecolor{seashell3}{rgb}{0.804,0.773,0.749}
\definecolor{seashell4}{rgb}{0.545,0.525,0.51}
\definecolor{sienna}{rgb}{0.627,0.322,0.176}
\definecolor{sienna1}{rgb}{1,0.51,0.278}
\definecolor{sienna2}{rgb}{0.933,0.475,0.259}
\definecolor{sienna3}{rgb}{0.804,0.408,0.224}
\definecolor{sienna4}{rgb}{0.545,0.278,0.149}
\definecolor{skyblue}{rgb}{0.529,0.808,0.922}
\definecolor{skyblue1}{rgb}{0.529,0.808,1}
\definecolor{skyblue2}{rgb}{0.494,0.753,0.933}
\definecolor{skyblue3}{rgb}{0.424,0.651,0.804}
\definecolor{skyblue4}{rgb}{0.29,0.439,0.545}
\definecolor{slateblue}{rgb}{0.416,0.353,0.804}
\definecolor{slateblue1}{rgb}{0.514,0.435,1}
\definecolor{slateblue2}{rgb}{0.478,0.404,0.933}
\definecolor{slateblue3}{rgb}{0.412,0.349,0.804}
\definecolor{slateblue4}{rgb}{0.278,0.235,0.545}
\definecolor{slategray}{rgb}{0.439,0.502,0.565}
\definecolor{slategray1}{rgb}{0.776,0.886,1}
\definecolor{slategray2}{rgb}{0.725,0.827,0.933}
\definecolor{slategray3}{rgb}{0.624,0.714,0.804}
\definecolor{slategray4}{rgb}{0.424,0.482,0.545}
\definecolor{slategrey}{rgb}{0.439,0.502,0.565}
\definecolor{snow}{rgb}{1,0.98,0.98}
\definecolor{snow1}{rgb}{1,0.98,0.98}
\definecolor{snow2}{rgb}{0.933,0.914,0.914}
\definecolor{snow3}{rgb}{0.804,0.788,0.788}
\definecolor{snow4}{rgb}{0.545,0.537,0.537}
\definecolor{springgreen}{rgb}{0,1,0.498}
\definecolor{springgreen1}{rgb}{0,1,0.498}
\definecolor{springgreen2}{rgb}{0,0.933,0.463}
\definecolor{springgreen3}{rgb}{0,0.804,0.4}
\definecolor{springgreen4}{rgb}{0,0.545,0.271}
\definecolor{steelblue}{rgb}{0.275,0.51,0.706}
\definecolor{steelblue1}{rgb}{0.388,0.722,1}
\definecolor{steelblue2}{rgb}{0.361,0.675,0.933}
\definecolor{steelblue3}{rgb}{0.31,0.58,0.804}
\definecolor{steelblue4}{rgb}{0.212,0.392,0.545}
\definecolor{tan}{rgb}{0.824,0.706,0.549}
\definecolor{tan1}{rgb}{1,0.647,0.31}
\definecolor{tan2}{rgb}{0.933,0.604,0.286}
\definecolor{tan3}{rgb}{0.804,0.522,0.247}
\definecolor{tan4}{rgb}{0.545,0.353,0.169}
\definecolor{thistle}{rgb}{0.847,0.749,0.847}
\definecolor{thistle1}{rgb}{1,0.882,1}
\definecolor{thistle2}{rgb}{0.933,0.824,0.933}
\definecolor{thistle3}{rgb}{0.804,0.71,0.804}
\definecolor{thistle4}{rgb}{0.545,0.482,0.545}
\definecolor{tomato}{rgb}{1,0.388,0.278}
\definecolor{tomato1}{rgb}{1,0.388,0.278}
\definecolor{tomato2}{rgb}{0.933,0.361,0.259}
\definecolor{tomato3}{rgb}{0.804,0.31,0.224}
\definecolor{tomato4}{rgb}{0.545,0.212,0.149}
\definecolor{turquoise1}{rgb}{0,0.961,1}
\definecolor{turquoise2}{rgb}{0,0.898,0.933}
\definecolor{turquoise3}{rgb}{0,0.773,0.804}
\definecolor{turquoise4}{rgb}{0,0.525,0.545}
\definecolor{violetred}{rgb}{0.816,0.125,0.565}
\definecolor{violetred1}{rgb}{1,0.243,0.588}
\definecolor{violetred2}{rgb}{0.933,0.227,0.549}
\definecolor{violetred3}{rgb}{0.804,0.196,0.471}
\definecolor{violetred4}{rgb}{0.545,0.133,0.322}
\definecolor{wheat}{rgb}{0.961,0.871,0.702}
\definecolor{wheat1}{rgb}{1,0.906,0.729}
\definecolor{wheat2}{rgb}{0.933,0.847,0.682}
\definecolor{wheat3}{rgb}{0.804,0.729,0.588}
\definecolor{wheat4}{rgb}{0.545,0.494,0.4}
\definecolor{whitesmoke}{rgb}{0.961,0.961,0.961}
\definecolor{yellow1}{rgb}{1,1,0}
\definecolor{yellow2}{rgb}{0.933,0.933,0}
\definecolor{yellow3}{rgb}{0.804,0.804,0}
\definecolor{yellow4}{rgb}{0.545,0.545,0}
\definecolor{yellowgreen}{rgb}{0.604,0.804,0.196}

	\end{center}
	\caption{The graph $K^{(d,{\cal Z},{\cal L},{\cal F})}.$}
	\labels{fig_contr_mult}
\end{figure}

We also define the structure $\mathfrak{A}^{(d,{\cal Z},{\cal L},{\cal F})}$ to be the one obtained from $\mathfrak{A}[I^{(d)}\cup V_{\cupall {\cal L}} ({\bf a})]$ after adding $\sum_{F∈ {\cal F}}|V(F)|$ new elements to its universe and a binary relation symbol that is interpreted as pairs of elements (corresponding to the additional edges of $\cupall {\cal F}$ and every edge between a vertex in  $V(\cupall {\cal F}\setminus V_{\cupall {\cal L}} ({\bf a}))$ and a vertex in  $I^{(d)}\setminus \cupall {\cal Z}$).
Notice that if $G$ is the Gaifman graph of $\mathfrak{A},$ then $K^{(d,{\cal Z},{\cal L},{\cal F})}$ is the Gaifman graph of  $\mathfrak{A}^{(d,{\cal Z},{\cal L},{\cal F})}.$

\myskip\paragraph{The out-signature of an extended compass.}
As in~\autoref{sec_out-sig_first-floor},
for every $\ell∈\mathbb{N},$ let
\begin{eqnarray*}
{\cal H}^{(\ell)} = \{ (H,{\cal V}) \mid H\mbox{~is a graph on $\ell$ vertices and ${\cal V}$ is a nice 3-partition of $H$}\}.
\end{eqnarray*}
We set $τ':=τ\cup{\bf Q}\cup\{{\sf R}\}\cup{\cal X}\cup{\bf c}$ and
\begin{eqnarray*}
\blue{{\sf SIG}_{\sf out}}=\{({\bf H}_1, \ldots, {\bf H}_h,\bar{φ})& \mid &\exists \ell_1, \ldots, \ell_h∈[0,h\cdot(\tw(θ)-1)]\mbox{\rm ~such that for every $i∈[h],$}\\
& &\hspace{4cm}{\bf H}_i∈ {\cal H}^{(\ell_i)}\mbox{~and~}\bar{φ}∈ {{\sf rep}^{(\sum_{i=1}^{r}\ell_i)}_{τ'}(θ^{\sf out}_{{q}})}\}.
\end{eqnarray*}

Let $\mathfrak{K}=(\mathfrak{A}[V(K^{\bf a})],{\bf a}, {\bf I}, {\bf W}_{{q}})$ be  the extended compass of a flatness pair $(W,\mathfrak{R})$ of $G\setminus V({\bf a})$ of height $2w+j,$  $R\subseteq V(K^{\bf a}),$ $d∈[r,w],$ ${\cal L} = \{L_1,\ldots, L_h\}∈ (2^{[l]})^h,$ and a collection ${\cal Z} = \{Z_1, \ldots, Z_h\}$ of subsets of $I^{(d-r+1)}.$
We define

\begin{eqnarray*}
\blue{{\sf out}\text{-}{\sf sig}}(\mathfrak{K},R,d,{\cal L},{\cal Z})=\{({\bf H}_1, \ldots, {\bf H}_h,\bar{φ})∈ \blue{{\sf SIG}_{\sf out}}& \mid
 & \mbox{for every $i∈[h],$~}\exists\  F_i∈ {\cal F}^{V_{L_i} ({\bf a})}_{|V(H_i)| - |\partial_{\mathfrak{K}}(Z_i)|},\\
 & & ~~~~\mbox{such that if ${\bf H}_i = (H_i,{\cal U}_i)$ and}\\
 & & ~~~~\mbox{${\cal V}_i = (\partial_{\mathfrak{K}}(Z_i),V_{L_i} ({\bf a}), V(F_i)\setminus V_{L_i} ({\bf a})),$}\\
& & ~~~~\mbox{then ${\cal V}_i$ is a nice 3-partition}\\
& & ~~~~\mbox{of $K^{\bf a}[\partial_{\mathfrak{K}}(Z_i)\cup V_{L_i} ({\bf a})]\cup F_i$ and}\\
& & ~~~~\mbox{$K^{\bf a}[\partial_{\mathfrak{K}}(Z_i)\cup V_{L_i} ({\bf a})]\cup F_i$ is strongly}\\
& & ~~~~\mbox{isomorphic to $H_i$ with respect to $({\cal V}_i, {\cal U}_i),$}\\
& & \exists\ \mbox{an ordering ${\bf b}$ of $ \bigcup_{i∈[h]} \big(\partial_{\mathfrak{K}} (Z_i)\cup V(F_i)\big),$ and }\\
& & \mbox{for every $i∈[h],$~}\exists\ \tilde{X}_i \subseteq Z_i\cup V(F_i),\\
& &  ~~~~~ \mbox{~such that~}  \partial_{\mathfrak{K}} (Z_i)\cup V(F_i)\subseteq \tilde{X}_{i} \mbox{~and}\\
& & ~~~~~ \mbox{if ${\cal X} = \{\tilde{X}_1,\ldots, \tilde{X}_h\}$ and}\\
& & ~~~~~\mbox{$R' = \bigcup_{i∈[h]} (Z_i\setminus \partial_{\mathfrak{K}}(Z_i))\cap R),$ then~}\\
& & ~~~~~ \big(\mathfrak{A}^{(d,{\cal Z},{\cal L},{\cal F})},R',{\bf W}_{q},\varnothing^l, {\cal X}, {\bf b}\big)\models \bar{φ}\}.
\end{eqnarray*}
Notice that if $h=1,$ then the out-signature given here is exactly the out-signature defined in~\autoref{sec_out-sig_first-floor}.

\myskip\paragraph{In-signature.}
We define:
$$\green{{\sf SIG}_{\sf in}} = 2^{[\ell_1]}\times\cdots\times 2^{[\ell_{q}]}\times[w].$$

Let $\mathfrak{K},R,d,{\cal L},{\cal Z}$ as in the previous paragraph.
We set
\begin{eqnarray*}
\green{{\sf in\mbox{-}sig}}(\mathfrak{K},R,d,{\cal L},{\cal Z}) &: = & \{(Y_{1},\ldots,Y_{p},t)∈ \green{{\sf SIG}_{\sf in}}\mid t≤ d\mbox{~and~  $\exists\  C∈ {\sf pr}(K^{\bf a}[I^{(d)}],{\bf W}_q, \cupall {\cal Z})$ ~and}\\
&  &~~~~~~~~~~~~~~~~~~~~~~~~~~~~~~~ \exists \ (\tilde{X}_{1},\ldots,\tilde{X}_p)\mbox{~such that~} \forall h∈[p]\  \\
& & ~~~~~~~~~~~~~~~~~~~~~~~~~~~~~~~~\tilde{X}_{h}=\{x_{i}^{h}\mid i∈ Y_{h}\},\\
& & ~~~~~~~~~~~~~~~~~~~~~~~~~~~~~~~~\mbox{$\tilde{X}_{h}\subseteq (I^{(t-\hat{r}+1)}\setminus \cupall {\cal Z})\cap R,$ and}\\
& &~~~~~~~~~~~~~~~~~~~~~~~~~~~~~~~~\text{if~}{\bf a}'={\bf a}\setminus V_{\cupall {\cal L}} ({\bf a}), \text{ then}\\
&  &~~~~~~~~~~~~~~~~~~~~~~~~~~~~~~~~\mbox{$\tilde{X}_h \text{~is~} (|Y_{h}|, r_{h})$-scattered in ${\sf ap}_{{\bf c}}(\mathfrak{A}, {\bf a}')[I^{(t)}\setminus \cupall {\cal Z}]$}\\
& &~~~~~~~~~~~~~~~~~~~~~~~~~~~~~~~\text{ and }{\sf ap}_{{\bf c}}(\mathfrak{A}, {\bf a}')[C]\models \bigwedge_{x∈ \tilde{X}_h} ψ_{h}(x)\}.\\
\end{eqnarray*}
%
%
We finally define
$${\sf CHAR}= [r,w] \times (2^{[l]})^h\times 2^{\blue{{\sf SIG}_{\sf out}}} \times 2^{\green{{\sf SIG}_{\sf in}}}$$
and
\begin{eqnarray*}
θ\text{-}{\sf char}(\mathfrak{K},R) & = &\{(d,{\cal L},\blue{{\sf sig}_{\sf out}},\green{{\sf sig}_{\sf in}})∈  {\sf CHAR}\mid \exists\ Z_1, \ldots, Z_h \subseteq I^{(d - r+1)}\mbox{~such that},\\
 & &\hspace{6cm} \text{if ${\cal Z}=\{Z_1, \ldots, Z_h \},$ then}\\
 & &\hspace{6cm}  \blue{{\sf out}\text{-}{\sf sig}}(\mathfrak{K},R,d,{\cal L},{\cal Z})= \blue{{\sf sig}_{\sf out}},  \mbox{~and~}\\
 & &\hspace{6cm} \green{{\sf in\mbox{-}sig}}(\mathfrak{K},R,d,{\cal L},{\cal Z})= \green{{\sf sig}_{\sf in}}\}.
\end{eqnarray*}

\myskip\paragraph{Summary of the algorithm of~\autoref{@desmembramientos} for sentences in $\bar{Θ}.$}

Let $\hat{r}:= \max_{h∈[p]}\{r_h\}$ and $\hat{\ell}:=\max_{h∈ [p]}\{\ell_h\}.$
We set $c$ to be the size of the \FOL-target sentence $σ$ of $θ,$
\begin{align*}
	q &: = {\sf height}(θ)\cdot(\tw(θ)+1)^2+1,\\
	\funref{@riguardavano}(|θ|,\tw(θ),g) & :=  \max\{q,(g+1)^2+1\},\\
	\funref{@carthaginoise}(|θ|,\tw(θ)) & := q-1,\\
	j' & := g+2\hat{r} +2,\\
	j & :=   {\sf odd}(\max\{q/2,j'\}),\\
	r &: = 2\cdot (\hat{\ell}+ 3)\cdot \hat{r}, \\
	w & := r\cdot q,\\
	m & := 2^{|{\sf CHAR}|} \cdot q\cdot (\hat{\ell} +3),\text{ and}\\
	\funref{@occidentales}(\hw(θ),\tw(θ),c,l, g)& := \lceil (2w+j)\cdot \sqrt{m}\rceil.
\end{align*}

We run the algorithm ${\tt Find\_Equiv\_FlatPairs}$ presented in~\autoref{@inhumainement},
which finds a collection $\tilde{\cal W}' =\{(\tilde{W}_1,\tilde{\mathfrak{R}}_1),\ldots, (\tilde{W}_{q\cdot (\hat{\ell} +3)}, \tilde{\mathfrak{R}}_{q\cdot (\hat{\ell} +3)})\}$
of $q\cdot (\hat{\ell} +3)$ flatness pairs that are $θ$-equivalent. i.e., their
extended compasses have the same characteristic.
Note that
the characteristic of each flatness pair, when dealing with a sentence $θ∈\bar{Θ}[τ]$ is defined using
the split version $\tilde{θ}_q$ of $θ_{{\sf R},{\bf c}},$ defined in~\autoref{@arrangements},
and the signatures defined in~\autoref{@enthaltenden}.
The algorithm
 ${\tt Find\_Equiv\_FlatPairs}$
also outputs
the set $Y:=V({\sf compass}_{\breve{\mathfrak{R}}'}(\breve{W}')),$ where
 $(\breve{W}',\breve{\mathfrak{R}}')$ is a $\breve{W}$-tilt
of $(W,\mathfrak{R})$ and $\breve{W}$ is the central $j'$-subwall of $\tilde{W}_1,$
and
a $W^\bullet$-tilt
$(\tilde{W}',\tilde{\mathfrak{R}}')$
of $(W,\mathfrak{R}),$
where $W^\bullet$ is be the central $g$-subwall of $\tilde{W}_1.$
The proof of why $(\mathfrak{A},R, {\bf a})\models θ_{{\sf R},{\bf c}}\iff (\mathfrak{A}\setminus V({\sf compass}_{\tilde{\mathfrak{R}}'}(\tilde{W}')),R\setminus Y, {\bf a})\models θ_{{\sf R},{\bf c}}$ can be found in~\autoref{second_level_more}.

Let us give a brief overview of the proof in~\autoref{second_level_more}.
The existence of a {\sl sequence} {\cal X} of modulators (that, however, still have ``small'' bidimensionality to the given flatness pair)
give rise to a non-empty $w$-privileged set with respect to ${\bf W}_q$ and ${\cal X}.$
Also, given that the bidimensionality of $\cupall{\cal X}$ with respect to the input flatness pair $(W,\mathfrak{R})$ is at most $q-1,$ there is
a ``buffer'' of $(\tilde{W}_1,\tilde{\mathfrak{R}}_1)$ that $\cupall {\cal X}$ does not intersect.
Again, we work with a flatness pair $(\tilde{W}_2,\tilde{\mathfrak{R}}_2)∈ \tilde{\cal W}'$ that is $θ$-equivalent to $(\tilde{W}_1,\tilde{\mathfrak{R}}_1)$ and
as in the case of $\bar{Θ}_1[τ]$ (\autoref{sec_proof_correctness}),
the proof is split into three claims (\autoref{claim_4},~\autoref{claim_5}, and~\autoref{claim_6}).

The main difference that appears is that, when dealing with a sentence $θ∈\bar{Θ},$ every $w$-privileged sequence gives rise to a {\sl sequence} of $Z_i$'s of parts of $I^{(d-r+1)}$ that
are being ``chopped off'' by the removal of the modulator sets in ${\cal X}$
and therefore all arguments in~\autoref{claim_1}, that deal with the satisfaction of $θ^{\sf out}_q,$ as encoded by the out-signature, have to be proved again, in \autoref{claim_4}, for their ``recursive'' analogue in $\bar{Θ}.$
For the in-signature, we still deal with the satisfaction of a Gaifman \FOL-sentence in the privileged component of $G_{\mathfrak{A}}$ with respect to ${\bf W}_q$ and ${\cal X}$ and therefore the proof of~\autoref{claim_5} is almost identical to the one of~\autoref{claim_2}.
For \autoref{claim_6},
the proof is again, as for~\autoref{claim_3}, a direct implication of the Assumption 5 of~\autoref{@desmembramientos}.

\myskip\section{Dealing with $Θ$ (the full story)}
\label{sec_final_combo}

In this section we describe the proof of~\autoref{@desmembramientos} for all sentences in $Θ.$
Our approach first considers a restricted version of $Θ,$ the logic $\hat{Θ}$ defined in~\autoref{@determinadas} where the only positive Boolean combination of sentences that we allow in each recursive level of a question expressed in $Θ$ is the {\sl conjunction} of a finite number of sentences.
Then, in~\autoref{@insinuations} we show how to insert also disjunctions to our arsenal of
positive Boolean combinations, in order to achieve the generality of~$Θ.$
This approach is based on the fact that every formula that is a positive Boolean combination of some set of formulas $Φ$ has an equivalent formula $φ'$ that is a disjunction of conjunctions of formulas in $Φ.$
More formally,
given a vocabulary $τ$ and a set ${\cal L}\subseteq \MSOL[τ],$
we define ${\bf CONJ}({\cal L})$ (resp. ${\bf DISJ}({\cal L})$) as the set of all sentences, in $\MSOL[τ],$ that are conjunctions (resp. disjunctions) of sentences of ${\cal L},$ and
we observe the following:

\begin{observation}\label{@inexpressibly}
Let ${\cal L}\subseteq \MSOL[τ],$ for some vocabulary $τ.$
For every formula $φ∈\bool({\cal L})$ there is a formula $φ'∈{\bf DISJ}({\bf CONJ}({\cal L}))$ such that
${\rm Mod}(φ) = {\rm Mod}(φ').$
Moreover, $|φ'|={\cal O}_{|φ|}(1).$
\end{observation}

For every formula $φ∈ {\bf CONJ}({\cal L})$ (resp. $φ∈{\bf DISJ}({\cal L})$), we define the {\em conjunction-width} (resp. {\em disjunction-width}) of $φ$ to be the minimum integer $\ell∈\mathbb{N}$ such that there exist formulas $φ_1,\ldots, φ_\ell ∈ {\cal L}$ such that $φ = φ_1\wedge \ldots \wedge φ_\ell$ (resp. $φ=φ_1\vee\ldots\veeφ_\ell$).

\myskip\paragraph{Rooted trees.}
Before we continue, let us give some definitions concerning rooted trees.
Given a tree $T,$ we denote by $L(T)$ the set of its leaves, i.e., the set of vertices of degree one in $T.$
Given a rooted tree $(T,r)$ and a vertex $v∈ V(T),$ we define the {\em height} of $v$ to be the number of edges in a minimum path in $T$ starting from a leaf of $T$ to $v.$
The \emph{height} of a rooted tree $(T,r)$ is the maximum height among all its vertices, and note that this maximum is always achieved by the root.
Given a rooted tree $(T,r)$ of height $h,$ we denote by $L_i,$ for $i∈ [h],$ the set of nodes of $T$ of height $i.$
We denote by ${\cal T}_{h,m}$ the set of all rooted trees of height $h$ and $|L(T)| = m.$

\myskip\subsection{Conjunctions}\label{@determinadas}
In this subsection we aim to sketch the proof of~\autoref{@desmembramientos} for all sentences in~$\hat{Θ},$ a restriction of~$Θ$ that deals only with conjunctions.
Our strategy is to ``reduce'' each $θ∈\hat{Θ}[τ]$ to a combination
of sentences in $\bar{Θ}[τ]$ and extend the definitions and the ideas presented in~\autoref{the_second_level} so to capture also $\hat{Θ}[τ].$
Every sentence $θ∈\hat{Θ}$ will be associated with a rooted tree expressing its recursive definition, where the root will correspond to $θ$ and every conjunction to a bifurcation of the tree.
Under the presence of a ``big enough'' pseudogrid in our given structure, we will use the aforementioned tree to define an equivalent version of the problem, where $θ$ is ``focused'' towards the privileged connected component occurring each time in the leaves of this tree.
Under this scope, every root-to-leaf path of the tree will correspond to a sentence in $\bar{Θ}.$
Performing this modification, we have to keep track of the bifurcations of the tree and ask the modulators that correspond to each such a bifurcation to be the same sets in all paths that contain this ``bifurcated'' node.
These equalities will have to be respected when searching for an equivalent ``solution-certificate'' that comes with the application of the irrelevant vertex technique.
For this reason, we have to (further) modify the definition of signatures and characteristics given in~\autoref{@enthaltenden} so as to ``add one dimension'' to them (corresponding to the shift from the ``path-like'' structure of sentences in $\bar{Θ}$ to the ``tree-like'' structure of sentences in $\hat{Θ}$), while respecting the equalities obtained from above.\medskip

Let us start by defining $\hat{Θ}.$
Let $τ$ be a vocabulary.
We define $\hat{Θ}[τ]:=\bigcup_{i∈\mathbb{N}}\hat{Θ}_i[τ],$ where
\begin{eqnarray*}
\hat{Θ}_{0}[τ] & = & \{σ\wedge μ\mid  σ∈ \FOL[τ]\mbox{~and~} μ∈ \NTMC[τ]\} \mbox{~and} \\
\mbox{for $i≥ 1,$ } \hat{Θ}_{i}[τ] & = & \{ β\triangleright γ\mid  β∈ \MSOL^\tw[τ\cup\{{\sf X}\}]  \mbox{~and~} γ∈ {\bf CONJ}({\hat{Θ}_{i-1}[τ]}^{({\sf c})})\}.
\end{eqnarray*}
Observe that $\hat{Θ}[τ]\subseteq Θ[τ].$

\myskip\paragraph{Conjunctive terms of sentences.}
For every $i∈\mathbb{N}_{≥ 1}$ and every $θ∈ \hat{Θ}_i[τ],$
there exist a sentence $β∈ \MSOL^{\tw}[τ]$ and a sentence $γ∈ {\bf CONJ}({\hat{Θ}_{i-1}[τ]}^{({\sf c})})$ such that
$$θ = β\triangleright γ.$$
Equivalently, there exist
$\ell$ sentences $θ_1, \ldots, θ_\ell∈ {\hat{Θ}_{i-1}[τ]}^{({\sf c})},$ where $\ell$ is the conjunction-width of $γ,$ such that
$$θ = β\triangleright
	 (θ_1\wedge \ldots \wedge θ_\ell ).$$
We call the sentences $θ_1, \ldots, θ_\ell$ the {\em conjunctive terms of $θ$}.
Observe that ${\bf CONJ}(\hat{Θ}_0) = \hat{Θ}_0.$
To see this, notice that the conjunction of a finite number of $σ_i\wedge μ_i$ can be expressed in a single sentence $σ^\star\wedge μ^\star,$ where $σ^\star$ is the conjunction of all $σ_i$'s and
$μ^\star$ is the sentence $μ_{\cal F}|_{\sf gf},$ where ${\cal F} = \{K_{{\bf hn}(θ)}\}.$
Therefore, every sentence $θ∈\hat{Θ}_1$ has exactly one conjunctive term.

We now recursively define a function ${\sf conj}:
\hat{Θ}[τ] \to \mathbb{N}$ as follows:
\begin{itemize}
\item For every $θ∈ \hat{Θ}_0[τ] \cup \hat{Θ}_1 [τ],$ we set ${\sf conj}(θ):=1.$
\item For every $θ∈ \hat{Θ}_i[τ], i≥ 2,$ we set
${\sf conj}(θ):=\sum_{i∈[\ell]}{\sf conj}(θ_i),$ where $θ_1, \ldots, θ_\ell$ are the conjunctive terms of $θ.$
\end{itemize}
Following the above definition, we stress that each sentence $θ∈\hat{Θ}$ has ${\sf conj}(θ)$ target sentences.

\myskip\paragraph{Tree-representations of sentences.}
We now describe an alternative way to view sentences in $\hat{Θ}$ that demonstrates the recursive structure of such a sentence.
Let $θ∈\hat{Θ},$ let $h={\sf height}(θ),$
and let $m={\sf conj}(θ).$
This means that $θ∈\hat{Θ}_h.$
Observe that each conjunctive term of $θ$ is a sentence in $\hat{Θ}_{h-1}.$
Then, each such a conjunctive term has its own conjunctive terms.
Following this recursive argument, we define a rooted tree $(T,r)∈ {\cal T}_{h,m}$ and a function $ρ:V(T)\to \hat{Θ}$
with the following properties:
\begin{itemize}
\item $ρ(r) = θ$ and
\item for every parent node $v∈ V(T),$ if $ρ(v) = θ,$ then the children nodes of $v$ are mapped, via $ρ,$ to the conjunctive terms of $θ.$
\end{itemize}
We stress that, if for a parent node $v∈ V(T),$  the sentence $ρ(v)$ has only one conjunctive term, then its only child node is mapped, via $ρ,$ to this single conjunctive term.
Therefore, every root-to-leaf path in $T$ has length $h$
and every leaf of $T$ is mapped, via $ρ,$ to a target sentence of $θ,$ which implies that $|L(T)| =m.$
We call $(T,r,ρ)$ the {\em tree-representation of $θ$}.

\myskip\paragraph{Linear projections of sentences.}
Let $θ∈\hat{Θ},$ let $h={\sf height}(θ),$ and let $m={\sf conj}(θ).$
Let also $(T,r,ρ)$ be the tree-representation of $θ.$

\begin{figure}[ht]
~~~\scalebox{1.04}{%
\tikzstyle{ipe stylesheet} = [
  ipe import,
  even odd rule,
  line join=round,
  line cap=butt,
  ipe pen normal/.style={line width=0.4},
  ipe pen heavier/.style={line width=0.8},
  ipe pen fat/.style={line width=1.2},
  ipe pen ultrafat/.style={line width=2},
  ipe pen normal,
  ipe mark normal/.style={ipe mark scale=3},
  ipe mark large/.style={ipe mark scale=5},
  ipe mark small/.style={ipe mark scale=2},
  ipe mark tiny/.style={ipe mark scale=1.1},
  ipe mark normal,
  /pgf/arrow keys/.cd,
  ipe arrow normal/.style={scale=7},
  ipe arrow large/.style={scale=10},
  ipe arrow small/.style={scale=5},
  ipe arrow tiny/.style={scale=3},
  ipe arrow normal,
  /tikz/.cd,
  ipe arrows, 
  <->/.tip = ipe normal,
  ipe dash normal/.style={dash pattern=},
  ipe dash dotted/.style={dash pattern=on 1bp off 3bp},
  ipe dash dashed/.style={dash pattern=on 4bp off 4bp},
  ipe dash dash dotted/.style={dash pattern=on 4bp off 2bp on 1bp off 2bp},
  ipe dash dash dot dotted/.style={dash pattern=on 4bp off 2bp on 1bp off 2bp on 1bp off 2bp},
  ipe dash normal,
  ipe node/.append style={font=\normalsize},
  ipe stretch normal/.style={ipe node stretch=1},
  ipe stretch normal,
  ipe opacity 10/.style={opacity=0.1},
  ipe opacity 30/.style={opacity=0.3},
  ipe opacity 50/.style={opacity=0.5},
  ipe opacity 75/.style={opacity=0.75},
  ipe opacity opaque/.style={opacity=1},
  ipe opacity opaque,
]
\definecolor{red}{rgb}{1,0,0}
\definecolor{blue}{rgb}{0,0,1}
\definecolor{green}{rgb}{0,1,0}
\definecolor{yellow}{rgb}{1,1,0}
\definecolor{orange}{rgb}{1,0.647,0}
\definecolor{gold}{rgb}{1,0.843,0}
\definecolor{purple}{rgb}{0.627,0.125,0.941}
\definecolor{gray}{rgb}{0.745,0.745,0.745}
\definecolor{brown}{rgb}{0.647,0.165,0.165}
\definecolor{navy}{rgb}{0,0,0.502}
\definecolor{pink}{rgb}{1,0.753,0.796}
\definecolor{seagreen}{rgb}{0.18,0.545,0.341}
\definecolor{turquoise}{rgb}{0.251,0.878,0.816}
\definecolor{violet}{rgb}{0.933,0.51,0.933}
\definecolor{darkblue}{rgb}{0,0,0.545}
\definecolor{darkcyan}{rgb}{0,0.545,0.545}
\definecolor{darkgray}{rgb}{0.663,0.663,0.663}
\definecolor{darkgreen}{rgb}{0,0.392,0}
\definecolor{darkmagenta}{rgb}{0.545,0,0.545}
\definecolor{darkorange}{rgb}{1,0.549,0}
\definecolor{darkred}{rgb}{0.545,0,0}
\definecolor{lightblue}{rgb}{0.678,0.847,0.902}
\definecolor{lightcyan}{rgb}{0.878,1,1}
\definecolor{lightgray}{rgb}{0.827,0.827,0.827}
\definecolor{lightgreen}{rgb}{0.565,0.933,0.565}
\definecolor{lightyellow}{rgb}{1,1,0.878}
\definecolor{black}{rgb}{0,0,0}
\definecolor{white}{rgb}{1,1,1}
\scalebox{0.9}{
\begin{tikzpicture}[ipe stylesheet]
  \draw[darkcyan]
    (112, 560)
     -- (128, 528);
  \draw[brown]
    (112, 560)
     -- (96, 528);
  \draw[purple]
    (144, 592)
     -- (112, 560);
  \pic[red]
     at (128, 528) {ipe disk};
  \draw[seagreen]
    (144, 592)
     -- (176, 560);
  \draw[purple]
    (176, 560)
     -- (160, 528);
  \pic[draw=darkorange, fill=gold]
     at (144, 592) {ipe fdisk};
  \pic[darkgreen]
     at (112, 560) {ipe disk};
  \pic[turquoise]
     at (96, 528) {ipe disk};
  \pic[green]
     at (160, 528) {ipe disk};
  \draw[red]
    (176, 560)
     -- (192, 528);
  \pic[orange]
     at (192, 528) {ipe disk};
  \pic[blue]
     at (176, 560) {ipe disk};
  \filldraw[draw=darkblue, fill=lightyellow]
    (200, 564)
     -- (212, 564)
     -- (212, 568)
     -- (220, 560)
     -- (212, 552)
     -- (212, 556)
     -- (200, 556)
     -- cycle;
  \draw[brown]
    (236, 560)
     -- (236, 528);
  \draw[purple]
    (236, 592)
     -- (236, 560);
  \pic[draw=darkorange, fill=gold]
     at (236, 592) {ipe fdisk};
  \pic[darkgreen]
     at (236, 560) {ipe disk};
  \pic[turquoise]
     at (236, 528) {ipe disk};
  \draw[darkcyan]
    (260, 560)
     -- (260, 528);
  \draw[purple]
    (260, 592)
     -- (260, 560);
  \pic[red]
     at (260, 528) {ipe disk};
  \pic[draw=darkorange, fill=gold]
     at (260, 592) {ipe fdisk};
  \pic[darkgreen]
     at (260, 560) {ipe disk};
  \node[ipe node]
     at (148, 592) {$a$};
  \node[ipe node]
     at (104, 560) {$b$};
  \node[ipe node]
     at (88, 528) {$d$};
  \node[ipe node]
     at (180, 560) {$c$};
  \node[ipe node]
     at (132, 528) {$e$};
  \node[ipe node]
     at (152, 528) {$f$};
  \node[ipe node]
     at (196, 528) {$g$};
  \draw[seagreen]
    (284, 592)
     -- (284, 560);
  \draw[purple]
    (284, 560)
     -- (284, 528);
  \pic[draw=darkorange, fill=gold]
     at (284, 592) {ipe fdisk};
  \pic[green]
     at (284, 528) {ipe disk};
  \pic[blue]
     at (284, 560) {ipe disk};
  \node[ipe node]
     at (288, 592) {$a$};
  \node[ipe node]
     at (288, 560) {$c$};
  \node[ipe node]
     at (288, 528) {$f$};
  \draw[seagreen]
    (308, 592)
     -- (308, 560);
  \pic[draw=darkorange, fill=gold]
     at (308, 592) {ipe fdisk};
  \draw[red]
    (308, 560)
     -- (308, 528);
  \pic[orange]
     at (308, 528) {ipe disk};
  \pic[blue]
     at (308, 560) {ipe disk};
  \node[ipe node]
     at (312, 592) {$a$};
  \node[ipe node]
     at (312, 560) {$c$};
  \node[ipe node]
     at (312, 528) {$g$};
  \node[ipe node]
     at (240, 592) {$a$};
  \node[ipe node]
     at (264, 592) {$a$};
  \node[ipe node]
     at (240, 560) {$b$};
  \node[ipe node]
     at (264, 560) {$b$};
  \node[ipe node]
     at (240, 528) {$d$};
  \node[ipe node]
     at (264, 528) {$e$};
\end{tikzpicture}
}
}
\caption{A rooted tree and its root-to-leaf paths.}
\labels{fig_conj_focus}
\end{figure}
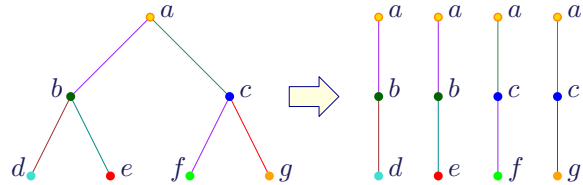

Let $P$ be a root-to-leaf path in $T.$
By definition, the vertices of $P$ are mapped, via $ρ,$ to a sequence $S_P:=(θ_h,\ldots, θ_0)$
of sentences, where, for each $i∈[0,h-1],$  $θ_{i}$ is a conjunctive term of $θ_{i+1}.$
Also by definition, $θ_h = θ$ and $θ_0$ is a sentence in $\bar{Θ}_0^{({\sf c})}.$
Notice that, as every sentence $θ_i ∈ \hat{Θ}_i,$
for every $i∈[h],$ there exists a sentence $β_i∈ \MSOL^\tw[[τ\cup\{{\sf X}_i\}]$
and an $\ell_i∈\mathbb{N}$ such that
$$θ_i = β_i\triangleright
	 (θ_{i,1}\wedge \ldots \wedge θ_{i,\ell_i} ),$$
where $θ_{i,1},\ldots, θ_{i,\ell}$ are the conjunctive terms of $θ_i,$ one of them being equal to $θ_{i-1}.$
Also, since every $θ_i, i∈[0,h-1],$ is a conjunctive term of $θ_{i+1},$ $θ_i$ is also a
sentence that belongs in
${\hat{Θ}_{i}[τ]}^{({\sf c})}.$
This allows us, for every $i∈[0,h-1],$ to consider a $\circ/\bullet$-flag corresponding to whether $θ_i∈ \hat{Θ}_{i}[τ]$ or $θ_i∈ {\hat{Θ}_{i}[τ]}^{({\sf c})}\setminus \hat{Θ}_{i}[τ].$
Therefore, we associate the sequence $S_P = (θ_h,\ldots, θ_0)$ with a $\circ/\bullet$-scenario $w$ of size $h,$
where for every $i∈[0,h-1],$ $w_i = \circ$ if $θ_i∈ {\hat{Θ}_{i}[τ]}^{({\sf c})}\setminus \hat{Θ}_{i}[τ]$ and $w_i = \bullet$ if $θ_i∈ \hat{Θ}_{i}[τ].$

We set $θ^P$ to be the sentence defined as follows:
\begin{itemize}
\item $θ^P_0: = θ_0,$
\item for every $i∈[h],$
 $θ^P_i:= β_i\triangleright θ^P_{i-1},$ and

\item $θ^P := θ^P_h.$
\end{itemize}
We call $θ^P$ the {\em linear projection of $θ$ with respect to $P$}.
It is now easy to see the following:
\begin{observation}\label{@insignifiante}
Let $θ∈\hat{Θ}$ and let $(T,r,ρ)$ be its tree-representation.
For every root-to-leaf path~$P$ of $T,$ the linear projection $θ^P$ οf $θ$ with respect to $P$ is a sentence in $\bar{Θ}_h,$ where $h={\sf height}(θ).$
\end{observation}

\myskip\paragraph{Focused linear projections.}
Let $θ∈\hat{Θ},$ let $h={\sf height}(θ),$ let $m={\sf conj}(θ),$ and let $(T,r,ρ)$ be the tree-representation of $θ.$
For each root-to-leaf path $P$ of $T,$ we consider  the linear projection $θ^P$ οf $θ$ with respect to $P.$
Due to~\autoref{@insignifiante}, every sentence $θ^P$ is
a sentence in $\bar{Θ}_h.$

To ease the readability, we consider an arbitrary ordering of the root-to-leaf paths of $T,$ say $P_1,\ldots, P_{m},$ and for every $i∈[m],$ we set $θ^{(i)}:=θ^{P_i}.$
The fact that, for every $i∈[m],$ $θ^{(i)}$ is a sentence in $\bar{Θ}_h,$ allows us to mimic the set of definitions in~\autoref{@arrangements}.
To do so,
first let $q,l∈\mathbb{N},$ let ${\bf c}$ be a collection of $l$ constant symbols not contained in $τ,$ and let $\{{\sf R}\}\cup{\bf Q}$ be a set of $2q+1$ unary relation symbols not contained in $τ.$
Also, for every $i∈[m],$ let
${\cal X}_i = \{{\sf X}_{i,1}, \ldots, {\sf X}_{i,h}\}$ be a set of $h$ unary relation symbols not contained in $τ,$
and let $w_i$ be the $\circ/\bullet$-scenario of $θ^{(i)}.$

For every $i∈[m],$
we let $σ_i$ and $μ_i$ be the target sentences of $θ^{(i)}$
and $ζ_i$  be the $l$-apex-projected sentence of $σ_i.$
Also, for every $i∈[m],$ we
define $θ_{q}^{(i)\sf out}$ to be the sentence as in~\autoref{@arrangements}.
Finally, for every $i∈[m],$ we define
the formula $\tilde{θ}_{q}^{(i)}({\sf X}_{i,1}, \ldots, {\sf X}_{i,h}),$ which is a formula with ${\sf X}_{i,1}, \ldots, {\sf X}_{i,h}$ as free variables and that is obtained from $θ^{(i)}$ as follows:
\begin{eqnarray*}
\tilde{θ}_{q}^{(i)}({\sf X}_{i,1}, \ldots, {\sf X}_{i,h})  = &\blue{\fbox{${θ^{(i)\sf out}_{{q}}}({\sf X}_{i,1}, \ldots, {\sf X}_{i,h})$}}\\
& \wedge &  \green{\fbox{$\exists {\sf C} \ η_{w\text{-}{\sf pr}_{\{{\sf X}_{i,1}, \ldots, {\sf X}_{i,h}\},{\sf C}}}\ \wedge {\breve{ζ}_{i{\sf R}} |_{{\sf ap}_{\bf c}}} |_{{\sf ind}_{\sf C}}$}}\\
& \wedge &
 \red{\fbox{$\exists {\sf C} \ η_{w\text{-}{\sf pr}_{\{{\sf X}_{i,1}, \ldots, {\sf X}_{i,h}\},{\sf C}}}\ \wedge  μ_i|_{{\sf ind}_{\sf C}}$}}.
\label{@desembanastasen}
\end{eqnarray*}
We stress that the only difference between the definition of a split sentence given in~\autoref{@gaspilleront} in~\autoref{@arrangements}  and the definition of $\tilde{θ}_{q}^{(i)}$ given above is the following:
in the definition of a split sentence in~\autoref{@gaspilleront},
the  second-order variables ${\sf X}_{i,1}, \ldots, {\sf X}_{i,h}$ are quantified by an existential quantifier inside the sentence,
while, in the the definition of $\tilde{θ}_{q}^{(i)}$ given here, ${\sf X}_{i,1}, \ldots, {\sf X}_{i,h}$ are considered as free variables of the formula.
We call $\tilde{θ}_{q}^{(i)}({\sf X}_{i,1}, \ldots, {\sf X}_{i,h}),$ $i∈ [m],$ a {\em focused linear projection of $θ$}.

\myskip\paragraph{Equality-checking formulas.}
Let $θ∈\hat{Θ},$ let $h={\sf height}(θ),$ let $m={\sf conj}(θ),$ and let $(T,r,ρ)$ be the tree-representation of $θ.$
We now define a binary relation ${\sf Eq}$ over  $[m]\times [h].$
Let $i,j∈[m]$ with $i\neq j.$
We consider the root-to-leaf paths $P_i$ and $P_j$ in $T.$
Let $z$ be the minimum height of a node in the intersection of $P_i$ and $P_j.$
It is easy to see that $z∈[h].$ We include in ${\sf Eq}$ the two pairs $(i,z)$ and $(j,z).$
In this way, we define the binary relation ${\sf Eq}\subseteq ([m]\times [h])^2.$
We call ${\sf Eq}$ the {\em equality relation corresponding to $(T,r,ρ)$}.

Let $a,b∈\mathbb{N}_{≥ 1}.$
Given a binary relation $Q\subseteq ([a]\times [b])^2,$ we define the formula $ξ_{Q}$ with $a\cdot b$ free second-order variables ${\sf X}_{1,1},\ldots, {\sf X}_{a,b}$ to be the following:
$$ξ_Q = \bigwedge_{\{(i,j),(i',j')\}∈ Q} {\sf X}_{i,j} = {\sf X}_{i',j'}.$$
Observe that  $ξ_{Q}∈ \MSOL[\{{\sf X}_{1,1},\ldots,{\sf X}_{a,b}\}].$
In the case that $Q = {\sf Eq},$ we call the formula $ξ_{\sf Eq}$ the {\em equality-checking formula of $(T,r,ρ)$}.

\myskip\paragraph{Tree-projected versions of sentences.}
Let $θ∈\hat{Θ},$ let $h={\sf height}(θ),$ let $m={\sf conj}(θ),$ let $(T,r,ρ)$ be the tree-representation of $θ,$ and let $θ^{(i)}, i∈[m],$ be the linear projections of $θ.$
Also, let $q,l∈\mathbb{N},$ let ${\bf c}$ be a collection of $l$ constant symbols not contained in $τ,$ and let $\{{\sf R}\}\cup{\bf Q}$ be a set of $2q+1$ unary relation symbols not contained in $τ.$
Also, for every $i∈[m],$ let
${\cal X}_i = \{{\sf X}_{i,1}, \ldots, {\sf X}_{i,h}\}$ be a set of $h$ unary symbols not contained in $τ,$
and let $w_i$ be the $\circ/\bullet$-scenario of $θ^{(i)}.$
Let $θ_{{\sf R}, {\bf c}}$ be an enhanced version of $θ.$
We consider the focused linear projections $\tilde{θ}_{q}^{(i)},$ $i∈ [m],$ of $θ$ and
let $ξ_{\sf Eq}$ be the equality-checking formula of $(T,r,ρ).$
Then, we define the sentence
$$\hat{θ}:=\exists {\bf X}\ ξ_{\sf Eq}({\bf X}) \wedge \bigwedge_{i∈[m]} \tilde{θ}_{q}^{(i)}({\sf X}_{i,1}, \ldots, {\sf X}_{i,h}),$$
where ${\bf X} = ({\sf X}_{1,1},\ldots, {\sf X}_{m,h}).$
Observe that $\hat{θ}∈ \MSOL[τ\cup\{{\sf R}\}\cup{\bf Q}\cup{\bf c}].$
We call $\hat{θ}$ the {\em tree-projected version of $θ_{{\sf R}, {\bf c}}$}.
By the definition of  $\hat{θ}$ and using~\autoref{@interruptions}, it is easy to prove the following:
\begin{lemma}\label{@individuation}
Let $θ_{{\sf R},{\bf c}}$ be an enhanced version of a sentence $θ∈ \hat{Θ}$ and let $l∈\mathbb{N}.$
If $\mathfrak{A}$ is a $τ$-structure, $R\subseteq V(\mathfrak{A}),$
${\bf W}_{{q}}$ is a $q$-pseudogrid in $G_{\mathfrak{A}},$ where ${q}={\sf height}(θ)\cdot (\tw(θ)+1)^2 +1,$ and
${\bf a}$ is an apex-tuple of $\mathfrak{A}$ of size $l,$
then  $(\mathfrak{A},R,{\bf a})\modelsθ_{{\sf R},{\bf c}}\iff (\mathfrak{A},R,{\bf W}_{{q}}, {\bf a})\models \hat{θ}.$
\end{lemma}

In what follows, we describe how to extend the signatures and characteristics defined in~\autoref{@enthaltenden} to capture sentences in $\hat{Θ}.$

\myskip\paragraph{Οut-signature.}
For every $i∈[m],$
we set $τ_i' :=τ\cup{\bf Q}\cup\{{\sf R}\}\cup\{{\sf X}_{i,1},\ldots, {\sf X}_{i,h}\}\cup{\bf c}$ and let
\begin{eqnarray*}
\blue{{\sf SIG}_{\sf out}}=\{({\bf H}_{1,1}, \ldots, {\bf H}_{m,h},\bar{φ}_1, \ldots, \bar{φ}_{m}) & \mid&\exists \ell_{1,1}, \ldots, \ell_{m,h}∈[0,\tw(θ)]\mbox{~such that}\\
& &\ \ \ \mbox{for every $i,j∈ [m]\times [h],$ ${\bf H}_{i,j}∈ {\cal H}^{(\ell_{i,j})}$~and~}\\
& &\ \ \ \mbox{for every $i∈[m],$~}\bar{φ}_i∈ {{\sf rep}^{(\sum_{j=1}^{h}\ell_{i,j})}_{τ_i'}(θ^{\sf out}_{i,q})}\}.
\end{eqnarray*}

Let $\mathfrak{K}=(\mathfrak{A}[V(K^{\bf a})],{\bf a}, {\bf I}, {\bf W}_{{q}})$ be  the extended compass of a flatness pair $(W,\mathfrak{R})$ of $G\setminus V({\bf a})$ of height $2w+j,$ let  $R\subseteq V(K^{\bf a}),$ let ${\bf d} = (d_1, \ldots, d_{p}),$ where for each $i∈[m],$ $d_i∈[r,w],$ and let, for $i∈[m],$ ${\cal L}_i = \{L_{i,1},\ldots, L_{i,h}\}∈ (2^{[l]})^h$ and  ${\cal Z}_i= \{Z_{i,1}, \ldots, Z_{i,h}\}$ be a collection of subsets of $I^{(d_i-r+1)}.$
We define
$\blue{{\sf out}\text{-}{\sf sig}}(\mathfrak{K},R,{\bf d},{\cal L}_1, \ldots, {\cal L}_{m},{\cal Z}_1, \ldots, {\cal Z}_{m})$ to be the following set:

\begin{eqnarray*}
\{({\bf H}_{1,1}, \ldots, {\bf H}_{m,h},\bar{φ}_1, \ldots, \bar{φ}_{m})∈ \blue{{\sf SIG}_{\sf out}}& \mid
 & \mbox{for every $i,j∈ [m]\times [h],$}\\
 & & \exists\  F_{i,j}∈ {\cal F}^{V_{L_{i,j}} ({\bf a})}_{|V(H_{i,j})| - |\partial_{\mathfrak{K}}(Z_{i,j})|}\mbox{~such that}\\
 & & ~~~~ξ_{\sf Eq}(F_{1,1},\ldots, F_{m, h})\mbox{ is true and}\\
 & & ~~~~\mbox{if ${\cal V}_{i,j} = (\partial_{\mathfrak{K}}(Z_{i,j}),V_{L_{i,j}} ({\bf a}), V(F_{i,j})\setminus V_{L_{i,j}} ({\bf a}))$}\\
 & & ~~~~\mbox{and ${\bf H}_{i,j} = (H_{i,j},{\cal U}_{i,j}),$}\\
 & & ~~~~\mbox{then ${\cal V}_{i,j}$ is a nice 3-partition}\\
& & ~~~~\mbox{of $K^{\bf a}[\partial_{\mathfrak{K}}(Z_{i,j})\cup V_{L_{i,j}} ({\bf a})]\cup F_{i,j}$ and}\\
& & ~~~~\mbox{$K^{\bf a}[\partial_{\mathfrak{K}}(Z_{i,j})\cup V_{L_{i,j}} ({\bf a})]\cup F_{i,j}$ is strongly}\\
& & ~~~~\mbox{isomorphic to $H_{i,j}$ with respect to $({\cal V}_{i,j}, {\cal U}_{i,j}),$}\\
& & \mbox{for every $i∈[m],$}\\
& & \exists\ \mbox{an ordering ${\bf b}_i$ of $ \bigcup_{j∈[h]} \big(\partial_{\mathfrak{K}} (Z_{i,j})\cup V(F_{i,j})\big)$ and }\\
& & \mbox{for every $i,j∈ [m]\times [h],$}\\
& & \exists\ \tilde{X}_{i,j} \subseteq Z_{i,j}\cup V(F_{i,j}) \mbox{~such that}\\
& & ~~~~ξ_{\sf Eq}(\tilde{X}_{1,1},\ldots, \tilde{X}_{m, h})\mbox{ is true,}\\
& &  ~~~~   \partial_{\mathfrak{K}} (Z_{i,j})\cup V(F_{i,j})\subseteq \tilde{X}_{i,j} \mbox{~and}\\
& & ~~~~ \mbox{if ${\cal X}_i = \{\tilde{X}_{i,1},\ldots, \tilde{X}_{i,h}\}$ and}\\
& & ~~~~ \mbox{$R_i' = \bigcup_{j∈[h]} R\cap (Z_{i,j}\setminus \partial_{\mathfrak{K}}(Z_{i,j})),$ then}\\
& & ~~~~ \big(\mathfrak{A}^{(d_i,{\cal Z}_i,{\cal L}_i,{\cal F}_i)},R_i',{\bf W}_{q},\varnothing^l, {\cal X}_i, {\bf b}_i\big)\models \bar{φ}_i\}.
\end{eqnarray*}

\myskip\paragraph{In-signature.}
For every $i∈[m],$
we let $σ_i$ and $μ_i$ be the target sentences of $θ^{(i)}$
and $ζ_i$  be the $l$-apex-projected sentence of $σ_i.$
Also, let $\breve{ζ}_{i\sf R}∈ \FOL[τ^{\langle \bf c\rangle}\cup\{{\sf R}\}]$
be the sentence obtained for a Gaifman sentence $\breve{ζ}_{i}$ that is equivalent to $ζ_i,$ after restricting its ``scattered'' variables of its basic local sentences to be contained in ${\sf R},$ as done in~\autoref{sec_equivalent-versions_first-floor}.
For every $i∈[m],$
let  $\breve{ζ}_{i,1}, \ldots, \breve{ζ}_{i,p_i}∈\FOL[τ^{\langle \bf c\rangle}\cup\{{\sf R}\}],$
be the corresponding basic local sentences of $\breve{ζ}_{i\sf R}$
with the corresponding integers $\ell_{i,1},\ldots,\ell_{i,p_i}$
and $r_{i,1},\ldots,r_{i,p_i}$ (all depending on $\breve{ζ}$).
We define
$$\green{{\sf SIG}_{\sf in}} = 2^{[\ell_{1,1}]}\times\cdots\times 2^{[\ell_{m,p_{m}}]}\times[w].$$
We define $\green{{\sf in\mbox{-}sig}}(\mathfrak{K},R,{\bf d},{\cal L}_1, \ldots, {\cal L}_{m},{\cal Z}_1, \ldots, {\cal Z}_{m})$ to be the following set:
\begin{eqnarray*}
 \{(Y_{1,1},\ldots,Y_{m,p_{m}},t_1, \ldots, t_{m})∈ \green{{\sf SIG}_{\sf in}}& \mid &
 \mbox{for every $i∈[m],$ }\\
&  &~~~~ t_i≤ d_i\mbox{~and~} \exists\  C_i∈ {\sf pr}(K^{\bf a}[I^{(d_i)}],{\bf W}_q, \cupall {\cal Z}_i) \mbox{~and}\\
&  &~~~~\exists \ (\tilde{X}_{i,1},\ldots,\tilde{X}_{i,p_i})\mbox{~such that~} \forall j∈[p_i]\  \\
& & ~~~~~~~\tilde{X}_{i,j}=\{x_{i}^{j}\mid i∈ Y_{j}\},\\
& & ~~~~~~~\tilde{X}_{i,j}\subseteq (I^{(t_i-\hat{r}+1)}\setminus \cupall {\cal Z}_i)\cap R, \text{ and}\\
& & ~~~~~~~\text{if~}{\bf a}_i '={\bf a}\setminus V_{\cupall {\cal L}_i} ({\bf a}), \text{ then}\\
& & ~~~~~~~\tilde{X}_h \text{~is~} (|Y_{h}|, r_{h})\text{-scattered}\\
& & ~~~~~~~\mbox{in~} {\sf ap}_{{\bf c}}(\mathfrak{A}, {\bf a}_i ')[I^{(t_i)}\setminus \cupall {\cal Z}_i]\text{ and }\\
& & ~~~~~~~{\sf ap}_{{\bf c}}(\mathfrak{A}, {\bf a}_i')[C_i\cup V({\bf a}_i')]\models \bigwedge_{x∈ \tilde{X}_{i,j}} ψ_{i,j}(x)\}.\\
\end{eqnarray*}

We finally define
$${\sf CHAR} = ([r,w])^{m} \times ((2^{[l]})^h)^{m}\times 2^{\blue{{\sf SIG}_{\sf out}}} \times 2^{\green{{\sf SIG}_{\sf in}}}$$
and

\begin{eqnarray*}
θ\text{-}{\sf char}(\mathfrak{K},R)& = &\{({\bf d},{\cal L}_1,\ldots, {\cal L}_{m},\blue{{\sf sig}_{\sf out}},\green{{\sf sig}_{\sf in}})∈  {\sf CHAR}\mid \\
& &\hspace{4cm}\mbox{for every $i∈[m]$}\\
& &\hspace{4cm} \exists\ Z_{i,1}, \ldots, Z_{i,h} \subseteq I^{(d_i - r+1)}\mbox{~such that}\\
& &\hspace{4.5cm} ξ_{\sf Eq}(Z_{1,1},\ldots, Z_{m, h})\mbox{ is true and}\\
& &\hspace{4.5cm} \mbox{if ${\cal Z}_i = \{Z_{i,1},\ldots, Z_{i,h}\},$ then}\\
& &\hspace{4.5cm}  \blue{{\sf out}\text{-}{\sf sig}}(\mathfrak{K},R,{\bf d},{\cal L}_1, \ldots, {\cal L}_{m},{\cal Z}_1, \ldots, {\cal Z}_{m})= \blue{{\sf sig}_{\sf out}},  \mbox{~and~}\\
& &\hspace{4.5cm} \green{{\sf in\mbox{-}sig}}(\mathfrak{K},R,{\bf d},{\cal L}_1, \ldots, {\cal L}_{m},{\cal Z}_1, \ldots, {\cal Z}_{m})= \green{{\sf sig}_{\sf in}}\}.
\end{eqnarray*}

Having stated the above definitions, we now sketch how to prove~\autoref{@desmembramientos} for a sentence in $\hat{Θ}[τ].$
Our argumentation is very similar to the case of $\bar{Θ}[τ].$
We run the algorithm ${\tt Find\_Equiv\_FlatPairs}$ presented in~\autoref{@inhumainement},
which finds a collection $\tilde{\cal W}' =\{(\tilde{W}_1,\tilde{\mathfrak{R}}_1),\ldots, (\tilde{W}_{p}, \tilde{\mathfrak{R}}_{p})\}$
of $p=q\cdot (\hat{\ell} +3)$ flatness pairs that are $θ$-equivalent. i.e., their
extended compasses have the same characteristic, as defined in the previous paragraph.
The algorithm
 ${\tt Find\_Equiv\_FlatPairs}$
also outputs
the set $Y:=V({\sf compass}_{\breve{\mathfrak{R}}'}(\breve{W}')),$ where
 $(\breve{W}',\breve{\mathfrak{R}}')$ is a $\breve{W}$-tilt
of $(W,\mathfrak{R})$ and $\breve{W}$ is the central $j'$-subwall of $\tilde{W}_1,$
and
a $W^\bullet$-tilt
$(\tilde{W}',\tilde{\mathfrak{R}}')$
of $(W,\mathfrak{R}),$
where $W^\bullet$ is be the central $g$-subwall of $\tilde{W}_1.$
We now provide a sketch of proof for the correctness of the above algorithm, i.e., we prove that $$(\mathfrak{A},R, {\bf a})\models θ_{{\sf R},{\bf c}}\iff (\mathfrak{A}\setminus V({\sf compass}_{\tilde{\mathfrak{R}}'}(\tilde{W}')),R\setminus Y, {\bf a})\models θ_{{\sf R},{\bf c}}.$$

We consider the tree-projected version $\hat{θ}$ of $θ_{{\sf R},{\bf c}}.$
Due to~\autoref{@individuation}, if  ${\bf W}_q$ is a $q$-pseudogrid of $G_{\mathfrak{A}},$ where ${q}={\sf height}(θ)\cdot (\tw(θ)+1)^2 +1,$
$(\mathfrak{A},R, {\bf a})\models θ_{{\sf R},{\bf c}}\iff (\mathfrak{A},R, {\bf W}_q, {\bf a})\models \hat{θ}.$

For each linear projection $θ^{(i)},i∈[m]$ of $θ,$
we treat the focused linear projection $\tilde{θ}_{q}^{(i)}({\sf X}_{i,1},\ldots,{\sf X}_{i,h})$ separately.
Intuitively,
for each sequence of disjoint sets $X_{i,1},\ldots, X_{i,h}$ assigned to the
free variables ${\sf X}_{i,1},\ldots,{\sf X}_{i,h}$ of $\tilde{θ}_{q}^{(i)},$
we find a ``buffer'' $d_i∈[r,w]$ in $(\tilde{W}_1, \tilde{\mathfrak{R}}_1)$ such that $\bigcup_j∈[h] X_{i,j}$ does not intersect $I_1^{(d_i)}\setminus I_1^{(d_i-r+1)}.$
We set ${\cal X}_{\rm in}^{(i)}$ to be the collection of the sets $X_{i,j}\cap I_1^{(d_i-r+1)}, j∈[h]$ and ${\cal X}_{\rm out}^{(i)} = \{X_{i,1}^{\rm out},\ldots, X_{i,h}^{\rm out}\},$ where $X_{i,j}^{\rm out} = X_{i,j}\cap I_1^{(d)}=\emptyset, j∈[h].$

For each $i∈[m],$ there exists some $k_i∈[p]$ such that
 $(\tilde{W}_{k_i}, \tilde{\mathfrak{R}}_i)∈ \tilde{\cal W}'\setminus\{(\tilde{W}_1, \tilde{\mathfrak{R}}_1)\}$ and $V({\sf compass}_{\tilde{\mathfrak{R}}_{k_i}})$ is disjoint from $\bigcup_j∈[h] X_{i,j}.$
We stress that for each $i∈[m],$ there is a possibly different $k_i∈[p]$ such that the above holds.
Our purpose is to ``replace'' ${\cal X}_{\sf in}^{(i)}$ with an other collection
${\cal X}^{(i)'} = \{X_{i,1}',\ldots, X_{i,h}'\}$ of sets that are inside $I_{k_i}^{(d_i-r+1)},$ that is an ``inner'' part of $(\tilde{W}_{k_i}, \tilde{\mathfrak{R}}_i).$

Notice that we can treat each focused linear projection $\tilde{θ}_{q}^{(i)}({\sf X}_{i,1},\ldots,{\sf X}_{i,h})$
and thus every sequence of sets $X_{i,1},\ldots, X_{i,h}$ separately, as
the equality-checking formula in the definition of $\hat{θ}$ is also incorporated inside the definition of characteristic and therefore, when
replacing ${\cal X}_{\rm in}^{(i)}$ with an other collection
${\cal X}^{(i)'},$ the satisfaction of equality-checking formula from
all ${\cal X}^{(i)'},i∈[m]$
will prove that $\hat{θ}$ is satisfied by interpreting ${\bf X}$ as
$X_{1,1},\ldots, X_{m,h}$ if and only if $\hat{θ}$ is satisfied by interpreting ${\bf X}$ as $X_{1,1}^{\rm out}\cup X_{1,1}',\ldots, X_{m,h}^{\rm out}\cup X_{m,h}'.$
We avoid to present a detailed proof for all the above, as it can be reproduced
by local adjustments in the proof in~\autoref{second_level_more}.

\myskip\subsection{Disjunctions of conjunctions}\label{@insinuations}
Let $τ$ be a vocabulary.
In this subsection we aim to provide the additional ideas that we need in order to complete the proof of~\autoref{@decendientes}, that is, dealing with the logic $Θ[τ].$
Recall that we defined $Θ =\bigcup_{i∈ \mathbb{N}}Θ_{i},$ where
\begin{eqnarray*}
Θ_{0}[τ] & = & \{σ\wedge μ\mid  σ∈ \FOL[τ]\mbox{~and~} μ∈ \NTMC[τ]\} \mbox{~and} \\
\mbox{for $i≥ 1,$ } Θ_{i}[τ] & = & \{ β\triangleright
	 γ \mid  β∈ \MSOL^\tw[τ\cup\{{\sf X}\}]  \mbox{~and~} γ∈ \bool(Θ_{i-1}[τ]^{({\sf c})})\}.
\end{eqnarray*}
Following~\autoref{@inexpressibly},
for each $i≥ 1,$ in the definition of $Θ_{i}[τ]$ we can replace the term
$γ∈\bool(Θ_{i-1}[τ]^{({\sf c})})$ by
$γ∈{\bf DISJ}({\bf CONJ}(Θ_{i-1}[τ]^{({\sf c})})).$
Having dealt with conjunctions in the previous subsection, we now aim to
resolve the case of disjunctions.
We first start by giving some additional definitions.

\myskip\paragraph{Disjunctive terms of sentences.}
Let $i∈\mathbb{N}_{≥ 1}$ and $θ∈ Θ_i [τ].$
Notice that there exist a sentence $β∈ \MSOL^{\tw}[τ \cup\{{\sf X}_i\}]$ and a sentence $γ∈{\bf DISJ}({\bf CONJ}({\hat{Θ}_{i-1}[τ]}^{({\sf c})}))$ such that
$$θ = β\triangleright γ.$$
Equivalently, there exist
$\ell$ sentences $ψ_1, \ldots, ψ_\ell∈ {\bf CONJ}(Θ_{i-1}[τ]^{({\sf c})}),$ where $\ell$ is the disjunction-width of $γ,$ such that
$$θ = β\triangleright
	 (ψ_1\vee \ldots \vee ψ_\ell ).$$
We call the sentences $ψ_1, \ldots, ψ_\ell$ the {\em disjunctive terms of $θ$}.
We stress that, since ${\bf DISJ}({\bf CONJ}(Θ_0)) = {\bf DISJ}(Θ_0)$ (this follows by the fact that
all target sentences are conjunctions), every sentence $θ∈Θ_1$ can have several disjunctive terms.

\myskip\paragraph{Boolean terms of sentences.}
As each disjunctive term $ψ_i$ of $θ$ is a sentence in ${\bf CONJ}(Θ_{i-1}[τ]^{({\sf c})}),$
 for each $i∈[\ell]$ there exist sentences $θ_{i,1}, \ldots, θ_{i,k_i} ∈ Θ_{i-1}[τ]^{({\sf c})},$ where $k_i$ is the conjunction-width of~$ψ_i,$
such that
$$ψ_i = θ_{i,1}\wedge \ldots \wedge θ_{i,k_i}.$$
Therefore, we can write
$$θ = β\triangleright \Big((θ_{1,1} \wedge \ldots \wedge θ_{1,k_1})\vee \ldots \vee (θ_{\ell, 1}\wedge \ldots \wedge θ_{\ell,k_\ell})\Big).$$
We call the sentences $θ_{1,1}, \ldots, θ_{1,k_1}, \ldots ,θ_{\ell, 1}, \ldots, θ_{\ell,k_\ell}$ the {\em Boolean terms of $θ$}.

We now recursively define a function ${\sf width}:
Θ[τ] \to \mathbb{N}$ as follows:
\begin{itemize}
\item For every $θ∈ {Θ}_0[τ],$ we set ${\sf width}(θ):=1.$
\item For every $i≥ 1$ and for every $θ∈ {Θ}_i [τ],$
we set
${\sf width}(θ):=\sum_{i∈ [\ell]}\sum_{j∈[k_i]}{\sf width}(θ_{i,j}),$ where $θ_{1,1}, \ldots, θ_{1,k_1}, \ldots ,θ_{\ell, 1}, \ldots, θ_{\ell,k_\ell}$ are the Boolean terms of $θ.$
\end{itemize}
Following the above definition, we stress that each sentence $θ∈Θ$ has ${\sf width}(θ)$ target sentences.
\bigskip

The ultimate goal of the rest of this subsection is to define a notion of a {\sl conjunctive scenario} of a sentence $θ∈ Θ,$ expressed in terms of a tree-representation as in the previous subsection,
and to prove that $θ$ is satisfied by a structure if and only if this structure satisfies at least one of the conjunctive scenarios of $θ.$

\myskip\paragraph{Representations of sentences in trees.}
Let $τ$ be a vocabulary and let $θ∈ Θ.$
Let $h:={\sf height}(θ)$ and $m:={\sf width}(θ).$
We define the {\em representation of $θ$} as the triple $(T,r,ρ),$ where $(T,r)∈ {\cal T}_{2h,m}$ and $ρ: V(T)\to Θ,$ such that
\begin{itemize}
\item $ρ(r) = θ,$
\item for every parent node $v∈ L_{2i}, i∈[h],$ if $ρ(v) = θ,$ then the children nodes of $v$ are mapped, via $ρ,$ to the disjunctive terms of $θ,$ and
\item for every parent node $v∈ L_{2i-1}, i∈[h],$ if $ρ(v) = ψ,$ then the children nodes of $v$ are mapped, via $ρ,$ to the conjunctive terms of $ψ.$
\end{itemize}
We stress that, if for a parent node $v$ in $L_{2i}, i∈[h]$  (resp.  in $L_{2i-1}, i∈[h]$), the sentence $ρ(v)$ has only one disjunctive (resp. conjunctive) term, then its only child node is mapped, via $ρ,$ to this single disjunctive (resp. conjunctive) term.
We can observe that, for every $i∈[h],$ every vertex $v∈ L_{2i}$ is mapped, via $ρ,$ to a sentence $θ∈ Θ_i$ and its grandchildren are nodes in $L_{2(i-1)}$ that are mapped, via $ρ,$ to the Boolean terms of $θ,$ which in turn are sentences in $Θ_{i-1}.$
The leaves of $(T,r)$ are mapped to the target sentences of $θ$ (thus, belonging to $Θ_0$).
Every root-to-leaf path of $T$ has length $2h.$

\myskip\paragraph{Conjunctive scenarios of sentences.}
Let $(T,r,ρ)$ be the representation of $θ.$
We now aim to define a collection ${\cal T}'$ of subtrees of $T.$
A subtree $T'$ of $T$ belongs in ${\cal T}'$
if and only if
$r∈ V(T'),$ all root-to-leaf paths of $T'$ have length $2h-1,$ and, for every $i∈[h],$ for every $v∈ L_{2i}(T'),$ there exists
a child $u_v$ of $v$ such that $u_v∈ V(T'),$ and for every $v∈ L_{2i-1}(T'),$ every child $u_v$ of $v$ belongs to $V(T').$

\begin{figure}[ht]
\!\!\!\!\!\!\!\!\!\!\!\!\!\!\!\!\!\!\!\!\!\!\scalebox{1.3}{%
\tikzstyle{ipe stylesheet} = [
  ipe import,
  even odd rule,
  line join=round,
  line cap=butt,
  ipe pen normal/.style={line width=0.4},
  ipe pen heavier/.style={line width=0.8},
  ipe pen fat/.style={line width=1.2},
  ipe pen ultrafat/.style={line width=2},
  ipe pen normal,
  ipe mark normal/.style={ipe mark scale=3},
  ipe mark large/.style={ipe mark scale=5},
  ipe mark small/.style={ipe mark scale=2},
  ipe mark tiny/.style={ipe mark scale=1.1},
  ipe mark normal,
  /pgf/arrow keys/.cd,
  ipe arrow normal/.style={scale=7},
  ipe arrow large/.style={scale=10},
  ipe arrow small/.style={scale=5},
  ipe arrow tiny/.style={scale=3},
  ipe arrow normal,
  /tikz/.cd,
  ipe arrows, 
  <->/.tip = ipe normal,
  ipe dash normal/.style={dash pattern=},
  ipe dash dotted/.style={dash pattern=on 1bp off 3bp},
  ipe dash dashed/.style={dash pattern=on 4bp off 4bp},
  ipe dash dash dotted/.style={dash pattern=on 4bp off 2bp on 1bp off 2bp},
  ipe dash dash dot dotted/.style={dash pattern=on 4bp off 2bp on 1bp off 2bp on 1bp off 2bp},
  ipe dash normal,
  ipe node/.append style={font=\normalsize},
  ipe stretch normal/.style={ipe node stretch=1},
  ipe stretch normal,
  ipe opacity 10/.style={opacity=0.1},
  ipe opacity 30/.style={opacity=0.3},
  ipe opacity 50/.style={opacity=0.5},
  ipe opacity 75/.style={opacity=0.75},
  ipe opacity opaque/.style={opacity=1},
  ipe opacity opaque,
]
\definecolor{red}{rgb}{1,0,0}
\definecolor{blue}{rgb}{0,0,1}
\definecolor{green}{rgb}{0,1,0}
\definecolor{yellow}{rgb}{1,1,0}
\definecolor{orange}{rgb}{1,0.647,0}
\definecolor{gold}{rgb}{1,0.843,0}
\definecolor{purple}{rgb}{0.627,0.125,0.941}
\definecolor{gray}{rgb}{0.745,0.745,0.745}
\definecolor{brown}{rgb}{0.647,0.165,0.165}
\definecolor{navy}{rgb}{0,0,0.502}
\definecolor{pink}{rgb}{1,0.753,0.796}
\definecolor{seagreen}{rgb}{0.18,0.545,0.341}
\definecolor{turquoise}{rgb}{0.251,0.878,0.816}
\definecolor{violet}{rgb}{0.933,0.51,0.933}
\definecolor{darkblue}{rgb}{0,0,0.545}
\definecolor{darkcyan}{rgb}{0,0.545,0.545}
\definecolor{darkgray}{rgb}{0.663,0.663,0.663}
\definecolor{darkgreen}{rgb}{0,0.392,0}
\definecolor{darkmagenta}{rgb}{0.545,0,0.545}
\definecolor{darkorange}{rgb}{1,0.549,0}
\definecolor{darkred}{rgb}{0.545,0,0}
\definecolor{lightblue}{rgb}{0.678,0.847,0.902}
\definecolor{lightcyan}{rgb}{0.878,1,1}
\definecolor{lightgray}{rgb}{0.827,0.827,0.827}
\definecolor{lightgreen}{rgb}{0.565,0.933,0.565}
\definecolor{lightyellow}{rgb}{1,1,0.878}
\definecolor{black}{rgb}{0,0,0}
\definecolor{white}{rgb}{1,1,1}
\begin{tikzpicture}[ipe stylesheet]
  \pic[ipe mark small]
     at (68, 400) {ipe disk};
  \pic[ipe mark small]
     at (68, 416) {ipe disk};
  \draw
    (68, 400)
     -- (68, 416);
  \draw
    (68, 416)
     -- (72, 432);
  \draw
    (72, 432)
     -- (80, 416);
  \draw
    (80, 416)
     -- (76, 400);
  \draw
    (80, 416)
     -- (84, 400);
  \draw
    (72, 432)
     -- (80, 448);
  \draw
    (80, 448)
     -- (88, 432);
  \draw
    (88, 432)
     -- (92, 416);
  \draw
    (92, 416)
     -- (88, 400);
  \draw
    (92, 416)
     -- (96, 400);
  \pic[ipe mark small]
     at (72, 432) {ipe disk};
  \pic[ipe mark small]
     at (80, 448) {ipe disk};
  \pic[ipe mark small]
     at (88, 432) {ipe disk};
  \pic[ipe mark small]
     at (92, 416) {ipe disk};
  \pic[ipe mark small]
     at (80, 416) {ipe disk};
  \pic[ipe mark small]
     at (76, 400) {ipe disk};
  \pic[ipe mark small]
     at (84, 400) {ipe disk};
  \pic[ipe mark small]
     at (88, 400) {ipe disk};
  \pic[ipe mark small]
     at (96, 400) {ipe disk};
  \draw
    (68, 416)
     -- (72, 400);
  \draw
    (64, 400)
     -- (68, 416);
  \pic[ipe mark small]
     at (64, 400) {ipe disk};
  \pic[ipe mark small]
     at (72, 400) {ipe disk};
  \pic[ipe mark small]
     at (148, 400) {ipe disk};
  \pic[ipe mark small]
     at (148, 416) {ipe disk};
  \draw
    (148, 400)
     -- (148, 416);
  \draw
    (148, 416)
     -- (152, 432);
  \draw
    (152, 432)
     -- (160, 448);
  \draw
    (160, 448)
     -- (168, 432);
  \draw
    (168, 432)
     -- (172, 416);
  \draw
    (172, 416)
     -- (168, 400);
  \draw
    (172, 416)
     -- (176, 400);
  \pic[ipe mark small]
     at (152, 432) {ipe disk};
  \pic[ipe mark small]
     at (160, 448) {ipe disk};
  \pic[ipe mark small]
     at (168, 432) {ipe disk};
  \pic[ipe mark small]
     at (172, 416) {ipe disk};
  \pic[ipe mark small]
     at (168, 400) {ipe disk};
  \pic[ipe mark small]
     at (176, 400) {ipe disk};
  \draw
    (148, 416)
     -- (152, 400);
  \draw
    (144, 400)
     -- (148, 416);
  \pic[ipe mark small]
     at (144, 400) {ipe disk};
  \pic[ipe mark small]
     at (152, 400) {ipe disk};
  \draw
    (184, 432)
     -- (192, 416);
  \draw
    (192, 416)
     -- (188, 400);
  \draw
    (192, 416)
     -- (196, 400);
  \draw
    (184, 432)
     -- (192, 448);
  \draw
    (192, 448)
     -- (200, 432);
  \draw
    (200, 432)
     -- (204, 416);
  \draw
    (204, 416)
     -- (200, 400);
  \draw
    (204, 416)
     -- (208, 400);
  \pic[ipe mark small]
     at (184, 432) {ipe disk};
  \pic[ipe mark small]
     at (192, 448) {ipe disk};
  \pic[ipe mark small]
     at (200, 432) {ipe disk};
  \pic[ipe mark small]
     at (204, 416) {ipe disk};
  \pic[ipe mark small]
     at (192, 416) {ipe disk};
  \pic[ipe mark small]
     at (188, 400) {ipe disk};
  \pic[ipe mark small]
     at (196, 400) {ipe disk};
  \pic[ipe mark small]
     at (200, 400) {ipe disk};
  \pic[ipe mark small]
     at (208, 400) {ipe disk};
  \draw
    (108, 432)
     -- (116, 416);
  \draw
    (116, 416)
     -- (116, 400);
  \draw
    (100, 432)
     -- (100, 416);
  \draw
    (100, 432)
     -- (108, 416);
  \draw
    (100, 416)
     -- (100, 400);
  \draw
    (108, 416)
     -- (104, 400);
  \draw
    (108, 416)
     -- (112, 400);
  \draw
    (100, 432)
     -- (100, 448)
     -- (100, 448);
  \draw
    (100, 448)
     -- (108, 432);
  \draw
    (80, 448)
     -- (92, 460);
  \draw
    (92, 460)
     -- (100, 448);
  \pic[ipe mark small]
     at (92, 460) {ipe disk};
  \pic[ipe mark small]
     at (100, 448) {ipe disk};
  \pic[ipe mark small]
     at (100, 432) {ipe disk};
  \pic[ipe mark small]
     at (108, 432) {ipe disk};
  \pic[ipe mark small]
     at (108, 416) {ipe disk};
  \pic[ipe mark small]
     at (100, 416) {ipe disk};
  \pic[ipe mark small]
     at (100, 400) {ipe disk};
  \pic[ipe mark small]
     at (104, 400) {ipe disk};
  \pic[ipe mark small]
     at (112, 400) {ipe disk};
  \pic[ipe mark small]
     at (116, 400) {ipe disk};
  \pic[ipe mark small]
     at (116, 416) {ipe disk};
  \draw
    (160, 448)
     -- (172, 460);
  \pic[ipe mark small]
     at (172, 460) {ipe disk};
  \draw
    (192, 448)
     -- (204, 460);
  \pic[ipe mark small]
     at (204, 460) {ipe disk};
  \draw
    (236, 432)
     -- (244, 416);
  \draw
    (244, 416)
     -- (244, 400);
  \draw
    (228, 432)
     -- (228, 416);
  \draw
    (228, 416)
     -- (228, 400);
  \draw
    (228, 432)
     -- (228, 448)
     -- (228, 448);
  \draw
    (228, 448)
     -- (236, 432);
  \draw
    (220, 460)
     -- (228, 448);
  \pic[ipe mark small]
     at (220, 460) {ipe disk};
  \pic[ipe mark small]
     at (228, 448) {ipe disk};
  \pic[ipe mark small]
     at (228, 432) {ipe disk};
  \pic[ipe mark small]
     at (236, 432) {ipe disk};
  \pic[ipe mark small]
     at (228, 416) {ipe disk};
  \pic[ipe mark small]
     at (228, 400) {ipe disk};
  \pic[ipe mark small]
     at (244, 400) {ipe disk};
  \pic[ipe mark small]
     at (244, 416) {ipe disk};
  \draw
    (264, 432)
     -- (272, 416);
  \draw
    (272, 416)
     -- (272, 400);
  \draw
    (256, 432)
     -- (264, 416);
  \draw
    (264, 416)
     -- (260, 400);
  \draw
    (264, 416)
     -- (268, 400);
  \draw
    (256, 432)
     -- (256, 448)
     -- (256, 448);
  \draw
    (256, 448)
     -- (264, 432);
  \draw
    (248, 460)
     -- (256, 448);
  \pic[ipe mark small]
     at (248, 460) {ipe disk};
  \pic[ipe mark small]
     at (256, 448) {ipe disk};
  \pic[ipe mark small]
     at (256, 432) {ipe disk};
  \pic[ipe mark small]
     at (264, 432) {ipe disk};
  \pic[ipe mark small]
     at (264, 416) {ipe disk};
  \pic[ipe mark small]
     at (260, 400) {ipe disk};
  \pic[ipe mark small]
     at (268, 400) {ipe disk};
  \pic[ipe mark small]
     at (272, 400) {ipe disk};
  \pic[ipe mark small]
     at (272, 416) {ipe disk};
  \filldraw[draw=blue, fill=lightyellow]
    (124, 432)
     -- (136, 432)
     -- (136, 436)
     -- (140, 428)
     -- (136, 420)
     -- (136, 424)
     -- (124, 424)
     -- cycle;
\end{tikzpicture}
}
\caption{A rooted tree and the collection ${\cal T}'.$}
\labels{fig_conj_scenarios}
\end{figure}
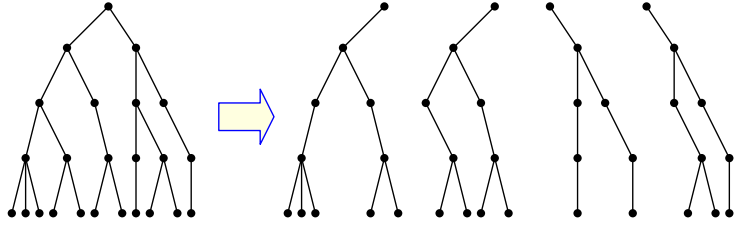

We denote by ${\cal S} = \{(L_{1}(T'),L_{3}(T'),\ldots, L_{2h-1}(T'))  \mid T'∈ {\cal T}'\}.$
Notice that for every $S∈ {\cal S},$ where $S=(V_1, \ldots, V_h),$ the
graph $T_S := T[\bigcup_{i∈ [h]} N_{(T,r)}[V_i]]$ is a tree of height $2h$ that is an induced subgraph of $T.$
By construction, for every $i∈[h]$ and for every $v∈ L_{2i}(T_S),$
there is exactly one $u_v∈ N_{(T,r)}(v)$ that is contained in $V(T_S)$
and
all the vertices in $N_{(T,r)}(u_v)$ are contained in $V(T_S).$
Following this, note that if $ρ(v)$ is a sentence $θ∈ Θ_i^{({\sf c})},$ then $ρ(u_v)$ is a disjunctive term of $θ$ (and the choice of a unique $u_v$ for every $v$ corresponds to the choice of a unique such a disjunctive term for every $ρ(v)$)
and $\{ρ(w)\mid w∈ N_{(T,r)}(u_v)\}$ is the set of all conjunctive terms of $ρ(u_v),$ that are sentences in $Θ_{i-1}^{({\sf c})}.$
For this reason, we call every $S∈ {\cal S}$ a \emph{conjunctive scenario} of $θ.$

\myskip\paragraph{Sentences corresponding to conjunctive scenarios.}
Let now $(T,r,ρ)$ be a representation of~$θ.$
Note that for each even $i∈[2h]$ and every $v∈ L_{i} (T),$
there exists a sentence $β∈ \MSOL^\tw[τ \cup \{{\sf X}\}]$
such that
$$ρ(v) = β\triangleright
	 (\bigvee_{u∈{N_{(T,r)}(v)}} ρ(u)),$$
while, for every odd $i∈[2h]$ and every $v∈ L_{i}(T),$
$$ρ(v) = \bigwedge_{u∈{N_{(T,r)}(v)}} ρ(u).$$
Let $S$ be a conjunctive scenario of $θ,$
and consider the subtree $T_S$ of $T.$
By the definition of $T_S,$ for every $i∈[h]$ and every vertex $v∈ L_{2i}(T_S),$ there is  a unique vertex $u_v∈ N_{(T,r)}(v) \cap V(T_S).$
We define $T^\star_S$ to be the tree obtained from $T_S$ after contracting, for every $i∈[h]$ and every $v∈ L_{2i}(T_S),$ the edge $\{v,u_v\}.$
Observe that $T^\star_S$ is a tree of height $h$ and, for every $i∈[0,h],$ the vertices of $L_i(T_S^\star)$ are the vertices of $L_{2i}(T_S).$
We define a function $ρ_S^\star:V(T^\star_S)\to Θ$
where
\begin{itemize}
\item for every $v∈ L_0 (T_S^\star),$ $ρ_S^\star(v) = ρ(v),$ and \item for every $i∈ [h]$ and for every $v∈ L_i(T_S^\star),$
if $ρ(v) =  β\triangleright
	 (\bigvee_{u∈{N_{(T_S,r)}(v)}} ρ(u)),$ for some $β∈ \MSOL^\tw[τ \cup \{{\sf X}\}],$ then $ρ_S^\star (v) = β\triangleright
	 (\bigwedge_{w∈{N_{(T_S^\star,r)}(v)}} ρ^\star_S(w)).$
\end{itemize}
Observe that for every $i∈ [0,h]$ and for every $v∈ L_i(T_S^\star),$
$ρ^\star_S(v)$ is a sentence in $\hat{Θ}_i.$
We call $ρ_S^\star(r)$ the {\em sentence corresponding to the scenario $S$} and we denote it by $θ_S.$
Since, for every $S∈ {\cal S},$ $θ_S$ is a sentence in $\hat{Θ}_h,$
we can consider its tree-representation and its tree-projected version $\hat{θ}_S,$ as defined in~\autoref{@determinadas}.

The next lemma says that, given a sentence $θ∈ Θ,$ its
enhanced version $θ_{{\sf R},{\bf c}}$ is equivalent to satisfying $\hat{θ}_S$ for at least one conjunctive scenario $S$ of $θ.$
Its proof can be easily derived using~\autoref{@interruptions}.

\begin{lemma}\label{@envelopperait}
Let $θ_{{\sf R},{\bf c}}$ be an enhanced version of a sentence $θ∈Θ$ and let $l∈\mathbb{N}.$
If $\mathfrak{A}$ is a $τ$-structure, $R\subseteq V(\mathfrak{A}),$
${\bf W}_{{q}}$ is  a $q$-pseudogrid in $G_{\mathfrak{A}},$ where ${q}={\sf height}(θ)\cdot (\tw(θ)+1),$ and
${\bf a}$ is an apex-tuple of $\mathfrak{A}$ of size $l,$
then  $(\mathfrak{A},R,{\bf a})\models θ_{{\sf R},{\bf c}}\iff (\mathfrak{A},R,{\bf W}_{{q}}, {\bf a})\models\bigvee_{S∈ {\cal S}} \hat{θ}_S.$
\end{lemma}

In what follows we sketch the proof of~\autoref{@desmembramientos}, for a general $θ∈ Θ.$
Let ${\cal S} = \{S_1,\ldots, S_{s}\}$ be the set of all conjunctive scenarios of $θ.$
First of all, we define
\begin{equation}\label{eq_product_char}
θ\text{-}{\sf char}(\mathfrak{K},R) = \Big(θ_{S_1}\text{-}{\sf char}(\mathfrak{K},R), \ldots, θ_{S_{s}}\text{-}{\sf char}(\mathfrak{K},R)\Big) .
\end{equation}
Then, we follow the same arguments as described at the end of~\autoref{@determinadas}.
In our current case, when considering a collection of $θ$-equivalent extended compasses, following the definition of the $θ$-characteristic given above, these extended compasses are
$θ_S$-equivalent for  every $S∈ {\cal S}.$
Therefore, when we find a part of a wall and declare it ``irrelevant'',
this part is ``irrelevant'' in any possible conjunctive scenario of $θ, $ and the proof is complete.

\myskip\section{Constructibility issues}\label{section_consrtr}

In this section we argue that our algorithm can be effectively constructed (see~\autoref{sec_constructing_meta_algo}) and present some consequences of our technique in the context of constructing the obstruction set of minor-closed graph classes (see~\autoref{@insuperables}).

\myskip\subsection{Constructing our (meta-)algorithm}
\label{sec_constructing_meta_algo}

All subroutines of our algorithm for deciding models of $Θ$-sentences can be effectively constructed, given a bound on the Hadwiger number of the corresponding sentence.
In order to do so, basic ingredients are (i) the quadratic algorithm for minor containment of~\cite{KawarabayashiKR12thedis} (ii) the linear algorithms from~\cite{SauST21kapiI,SauST21amor} (that are, in turn, based on the algorithm of~\cite{KawarabayashiTW18anew}) for finding a flat wall whose compass has bounded treewidth, and (iii) the computation of all the out- and in-signatures that is done in linear time, using the  treewidth bound and Courcelle's theorem, applied on sentences whose size effectively bounded by the constants of Gaifman's theorem. As the  dependencies in Gaifman's theorem as well as in the algorithm of Courcelle's theorem can be effectively (however non-elementarily) bounded (see e.g., \cite{FrickG04theco,DawarGKS07model,Buchi60weaks,Grohe07logi,HeimbergKS13,Gaifman82onloc}), it follows that our algorithm for  deciding models of $Θ$-sentences is also effectively constructible. We summarize this discussion in the following theorem.

 \begin{theorem}
 There is a Turing machine that
 receives as input a sentence $θ∈ Θ$ and an upper bound on $\hw(\Mod(θ))$ and returns as output the quadratic algorithm of \autoref{@decendientes}.
\end{theorem}

The above theorem permits us to state  \autoref{@decendientes} (and \autoref{@pertrechando} as well) so that the running time is bounded by
$f(|θ|)\cdot n^2$ (resp. $f(|θ|,\hw)\cdot n^2$) for some {\sl constructive} function $f.$
In this paper we did not focus on optimizing this function  or even giving explicit  upper bounds for it. We believe that for certain instantiations (or parameterizations) of $θ,$ providing reasonable upper  or lower bounds for the function $f$ is an interesting research direction.

\myskip\subsection{Constructibility horizon of Robertson-Seymour's  theorem}
\label{@insuperables}

Recall that the {\em (minor) obstrucion set} of a graph class  ${\cal G}$ is  the set $\obs({\cal G})$ of all minor-mininal graphs that are not contained in ${\cal G}.$
When ${\cal G}$ is minor-closed, the set
$\obs({\cal G})$ {\sl completely} characterizes ${\cal G},$ as ${\cal G}=\excl(\obs({\cal G})).$
By definition, no two elements of $\obs({\cal G})$ are comparable with respect to the minor relation, therefore, by Robertson-Seymour's theorem~\cite{RobertsonS04GMXX}, $\obs({\cal G})$ is always a finite set. Unfortunately, while we ``know'' the finiteness of $\obs({\cal G}),$ there is no general way to construct this set, given some  (finite) description of ${\cal G}$ \cite{FellowsL88nonc} (see also  \cite{FriedmanRS87them,krombholz2019upper}).
 This means that we may resort to a case  study of proving bounds\footnote{Notice that if we have a bound on $\obs({\cal G})$ one may use the ``finite description'' of ${\cal G}$ in order to identify all obstructions of
$\obs({\cal G}),$ by exhaustive search.} on the size of $\obs({\cal G})$
for particular instantiations of ${\cal G}$ (see~\cite{AdlerGK08comp,Lagergren98uppe,SauST21kapiI,AbrahamsonF93filit,CattellDDFL00onco,CourcelleDF97anote}).
As an attempt to enlarge the constructibility horizon of Robertson-Seymour's theorem, researchers have
considered several mechanisms to build minor-closed graphs classes from simpler ones. An interesting problem is whether it is possible to
{\sl construct} the obstruction of the new class given the obstructions of the simpler ones.
To detect the widest possible set of operations between graphs classes that maintains this constructibility is an interesting challenge.
We proceed with some definitions.

\myskip\paragraph{Constructive operations.}
Given an integer $r≥ 1,$  a {\em graph class operation of arity $r$} is
any function $\mathfrak{f}:(2^{{\cal G}_{\sf  all}})^r\rightarrow 2^{{\cal G}_{\sf all}}.$
Such an operation $\mathfrak{f}$ is {\em minor-invariant} if whenever ${\cal G}_{1},\ldots,{\cal G}_{r}$ are minor-closed graph classes, then so is $\mathfrak{f}({\cal G}_{1},\ldots,{\cal G}_{r}).$
%
%
We say that a minor-invariant graph class operation~$\mathfrak{f}$ is {\em explicitly constructive} if there is a {computable}  function~$f: \mathbb{N}^{r}\to\mathbb{N}$ such that if  ${\cal G}_{1},\ldots,{\cal G}_{r}$ are minor-closed graph classes, then
$\hw({\cal G})≤   f(\max\{\hw({\cal G}_{i})\mid i∈ [r]\}).$

It is easy to verify that the intersection  operation $\cap$ is explicitly constructive.
The case of the union  operation $\cup$ is more difficult and has been studied by Adler, Grohe, and Kreutzer~\cite{AdlerGK08comp} -- see also \cite{Lagergren98uppe} where the notion of {\sl intertwines}  has been introduced.
It has also been proved by Bulian and Dawar~\cite{BulianD16graph} that  the operation\! $^{\sf c}$\!  defined as
${\cal G}^{\sf c}=\{G\mid \forall C∈ {\sf cc}(G), G∈ {\cal G}\}$ is also explicitly constructive.
 The same was proven recently in~\cite{DinerGT21block}  for the  operation $^{\sf b}$
defined as ${\cal G}^{\sf b}=\{G\mid \forall C∈ {\sf bc}(G), G∈ {\cal G}\},$ where ${\sf bc}(G)$ is the set of all blocks of $G.$ We enlarge this set of operations by defining the
graph class  operation $\tritri$ as follows
\[{\cal B}\tritri {\cal G}=   \{G\mid \exists X\subseteq V(G), {\sf torso}(G,X)∈ {\cal B}\ \wedge\ G\setminus X∈ {\cal G}\}.\]
 It is easy to prove that $\tritri$ is minor-invariant.\footnote{Here we should stress that, alternatively, one might define ${\cal B}\tritri {\cal G}$ by replacing $\torso$ by $\torso^+$ as defined in~\autoref{ftmlpo}. The main statement of this subsection copies for this  definition as well.}
Notice that $\tritri$ is strongly related with $\triangleright$ and the above definition  closely imitates the definition of $\bar{Θ}_{1}.$
 Our results imply the following.

\begin{theorem}
\label{@proisionalmente} The operation
 $\tritri$  is explicitly constructive when restricted to ${\cal B}$'s where  $\obs({\cal B})$ contains some planar graph.
\end{theorem}
%

\begin{proof}[Proof (sketch).]
Let $G∈ \obs({\cal B}\tritri {\cal G}),$ $v∈ V(G),$ and $G'=G\setminus v.$
Let also $β\triangleright γ∈\bar{Θ}_{1},$
where $β$ expresses the fact that ${\sf torso}(G,X)∈ {\cal B}$ and $γ$ expresses the fact that $G\setminus X∈ {\cal G}.$ Notice that the fact that $\obs({\cal B})$ contains some planar graph, implies that ${\sf torso}(G,X)$
has bounded treewidth, therefore there is indeed some $β∈\MSOL^{\sf tw}$ such that  $\Mod(β\triangleright γ)={\cal B}\tritri {\cal G}.$
Since $G∈ \obs({\cal B}\tritri {\cal G}),$ we get $G'∈ {\cal B}\tritri {\cal G},$ therefore $\hw(G)≤ \hw(G')+1≤ \hw({\cal B}\tritri {\cal G})+1≤  \hw({\cal B})+\hw({\cal G})+1.$
We set $c:=\hw({\cal B})+\hw({\cal G})+1.$

We apply  the algorithm of the proof of~\autoref{@decendientes} using \autoref{corr_withoutR_irrele_flat} instead of \autoref{lemma_irrele_flat}
for  $(G,V(G),{\bf a}),$ no matter the  choice of ${\bf a},$  and the output is
either a report that  $\tw(G)≤ \funref{@interference}(c)\cdot r$
or  a non-empty vertex set $Z∈ V(G)$ such that $G\models θ\iff G\setminus Z\models θ.$ Here it is  important to note that, in  the second case,  no annotation vertices appear because  $β\triangleright γ$ has no \FOL-target sentences.
 As $G∈\obs({\cal B}\tritri {\cal G}),$ this second outcome of the proof of~\autoref{@decendientes}
is excluded, and therefore we obtain that
$\tw(G)$ is bounded by some explicit  function of $c.$

After bounding the treewidth of $G∈ \obs({\cal B}\tritri {\cal G}),$ it is possible to  bound its size as well,  just by using the
fact that $β\triangleright γ$ is a \MSOL-sentence.
For this, one may use the classic technique of Lagergren~\cite{Lagergren98uppe} (see also~\cite{Lagergren91anup,LagergrenA91mini} as well as the more recent application in~\cite{SauST21kapiI}) that combines the fact that
\MSOL-sentences have finite index with
the use of lean-decompositions (see \cite{GiannopoulouPRT19cutw,KanteK14anup,KanteK18line,GiannopoulouPRT17line} for further developments of this technique.
\end{proof}

As discussed above, the operation
$\tritri$ can be seen as a way to create minor-closed classes by ``composing together'' simpler ones.
The only previously known result about the constructibility  of $\tritri$ follows from~\cite{AdlerGK08comp,SauST21kapiI} (see also~\cite{FellowsL94onse}) for the case where $|X|≤ 1,$ that is,  when $\obs({\cal B})=\{K_{2}\}.$  \autoref{@proisionalmente} extends this for every obstruction set containing some planar graph.
It is an interesting question whether $\tritri$ remains explicitly constructive if we drop the planarity condition on $\obs({\cal B}).$  It is certainly desirable to give a reasonable estimation of an upper bound on  $\hw({\cal B}\tritri {\cal G})$ as a function
of $\hw({\cal B})$ and $\hw({\cal G}).$
We believe that this is aim is not ``out of reach'' as, at least,  one of the ``sources of non-elementarity'' in the dependencies of our main result, namely Gaifman's theorem, is  missing in the proof of~\autoref{@proisionalmente}.
\myskip


\myskip\section{From \textsf{\sf FOL} to \textsf{\sf FOL\!+DP}: the compound logic $\Theta^{\DP}$}
\label{sec_thetadp}

In the definition of $\Theta_0$,
the base case of $\Theta$, we consider compound sentences $\sigma\wedge\mu$, where $\sigma\in\FOL$ and
$\mu$ expresses minor-exclusion.
However,
one can consider extensions of \FOL in the compound sentences.
A possible candidate is First-order logic with disjoint-paths predicates defined in~\cite{SchirrmacherSV22first} (see the paragraph below for a formal definition).
This way we can define a more general logic $\Theta^{\DP}$ and prove an algorithmic meta-theorem that encompasses also the results in~\cite{GolovachST22model,GolovachST22model_arXiv}.
To ease reading, in this subsection we deal only with graphs and not with general structures. However, our results can be {straightforwardly} be extended to  general structures.


\subsection{The disjoint-paths logic}
We define the $2k$-ary predicate ${\sf dp}_{k}({\sf x}_1, {\sf y}_1, \ldots,{\sf x}_k, {\sf y}_k)$, which evaluates true in a  graph $G$  if and only if there
are paths $P_1,\ldots, P_k$ of $G$ of length at least two between (the interpretations of) ${\sf x}_i$ and ${\sf y}_i$ for all $i\in[k]$ such that for every $i,j\in[k]$, $i\neq j$, $V(P_i)\cap V(P_j)=\emptyset.$
We let $\FOL\!\!+\!\DP$ be the logic obtained from $\FOL$ after allowing
${\sf dp}_{k}({\sf x}_1, {\sf y}_1, \ldots,{\sf x}_k, {\sf y}_k), k\geq 1$ as atomic predicates.

\paragraph{The compound logic $\Theta^{\DP}$.}
We define an extension $\Theta^{\DP}$ of $Θ$  by considering, as the base
case, instead of  $Θ_{0},$ the logic
$$Θ_{0}^{\DP} = \{σ\wedge μ\mid  σ∈ \FOL\!+\!\DP\mbox{~and~} μ∈ \NTMC[\{{\sf E}\}]\}.$$

We sketch how to prove the following extension of~\autoref{@decendientes1}.

\begin{theorem}
\label{thm_thetadp}
For every $θ∈ \Theta^{\DP},$
there exists an algorithm that, given a graph $G$, outputs whether $G\models θ$
in time $\O_{|θ|}(n^2)$.
\end{theorem}

As we define the alternative $\tilde{\Theta}$ of $\Theta$,
we can also define $\tilde{\Theta}^{\DP}$ by taking $\tilde{Θ}_{0}^{\DP}=\FOL\!\!+\!\!\DP$ as the base case, i.e.,
by discarding the minor-exclusion   from the definition of ${Θ}_{0}^{\DP}.$
Notice that $\tilde{\Theta}^{\DP}$ contains $\FOL\!\!+\!\DP$ and can be seen as  a natural extension of it.
As a corollary of~\autoref{thm_thetadp}, we get the following analogue of \autoref{@pertrechando1}.

\begin{theorem}
\label{thm_tildethetadp}
For every $\tilde{θ}∈\tilde{\Theta}^{\DP},$ there exists an algorithm that, given a graph $G$, outputs whether $G\models θ$
in time $\O_{|θ|,\hw(G)}(n^2)$.
\end{theorem}

\autoref{thm_tildethetadp}  contains all results and applications of \cite{GolovachST22model,GolovachST22model_arXiv} as a (very) special case.
For a visualization of  the current meta-algorithmic landscape, see \autoref{@resolvesssremos}.
\newpage

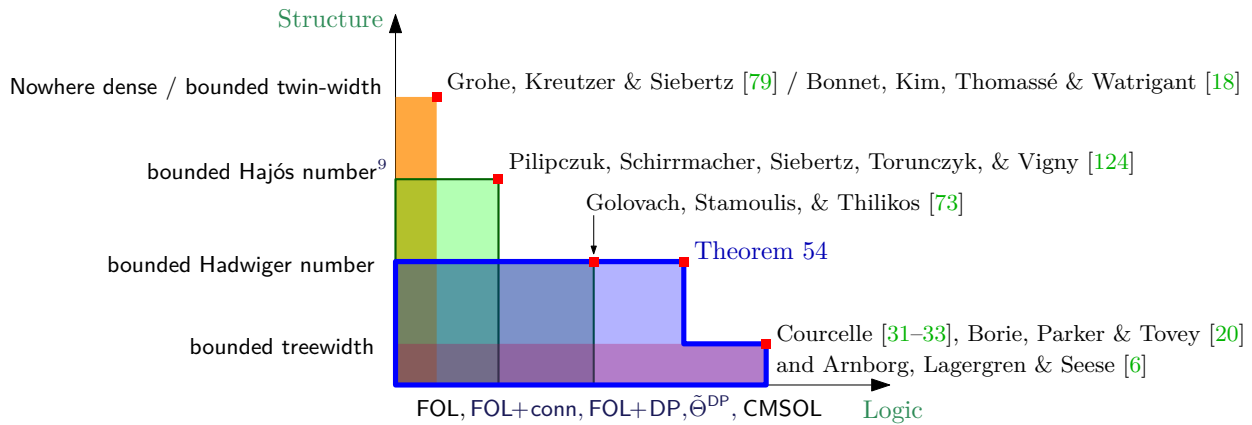
\begin{figure}[ht]
	\begin{center}
\tikzstyle{ipe stylesheet} = [
  ipe import,
  even odd rule,
  line join=round,
  line cap=butt,
  ipe pen normal/.style={line width=0.4},
  ipe pen heavier/.style={line width=0.8},
  ipe pen fat/.style={line width=1.2},
  ipe pen ultrafat/.style={line width=2},
  ipe pen normal,
  ipe mark normal/.style={ipe mark scale=3},
  ipe mark large/.style={ipe mark scale=5},
  ipe mark small/.style={ipe mark scale=2},
  ipe mark tiny/.style={ipe mark scale=1.1},
  ipe mark normal,
  /pgf/arrow keys/.cd,
  ipe arrow normal/.style={scale=7},
  ipe arrow large/.style={scale=10},
  ipe arrow small/.style={scale=5},
  ipe arrow tiny/.style={scale=3},
  ipe arrow normal,
  /tikz/.cd,
  ipe arrows, 
  <->/.tip = ipe normal,
  ipe dash normal/.style={dash pattern=},
  ipe dash dotted/.style={dash pattern=on 1bp off 3bp},
  ipe dash dashed/.style={dash pattern=on 4bp off 4bp},
  ipe dash dash dotted/.style={dash pattern=on 4bp off 2bp on 1bp off 2bp},
  ipe dash dash dot dotted/.style={dash pattern=on 4bp off 2bp on 1bp off 2bp on 1bp off 2bp},
  ipe dash normal,
  ipe node/.append style={font=\normalsize},
  ipe stretch normal/.style={ipe node stretch=1},
  ipe stretch normal,
  ipe opacity 10/.style={opacity=0.1},
  ipe opacity 30/.style={opacity=0.3},
  ipe opacity 50/.style={opacity=0.5},
  ipe opacity 75/.style={opacity=0.75},
  ipe opacity opaque/.style={opacity=1},
  ipe opacity opaque,
]
\definecolor{red}{rgb}{1,0,0}
\definecolor{blue}{rgb}{0,0,1}
\definecolor{green}{rgb}{0,1,0}
\definecolor{yellow}{rgb}{1,1,0}
\definecolor{orange}{rgb}{1,0.647,0}
\definecolor{gold}{rgb}{1,0.843,0}
\definecolor{purple}{rgb}{0.627,0.125,0.941}
\definecolor{gray}{rgb}{0.745,0.745,0.745}
\definecolor{brown}{rgb}{0.647,0.165,0.165}
\definecolor{navy}{rgb}{0,0,0.502}
\definecolor{pink}{rgb}{1,0.753,0.796}
\definecolor{seagreen}{rgb}{0.18,0.545,0.341}
\definecolor{turquoise}{rgb}{0.251,0.878,0.816}
\definecolor{violet}{rgb}{0.933,0.51,0.933}
\definecolor{darkblue}{rgb}{0,0,0.545}
\definecolor{darkcyan}{rgb}{0,0.545,0.545}
\definecolor{darkgray}{rgb}{0.663,0.663,0.663}
\definecolor{darkgreen}{rgb}{0,0.392,0}
\definecolor{darkmagenta}{rgb}{0.545,0,0.545}
\definecolor{darkorange}{rgb}{1,0.549,0}
\definecolor{darkred}{rgb}{0.545,0,0}
\definecolor{lightblue}{rgb}{0.678,0.847,0.902}
\definecolor{lightcyan}{rgb}{0.878,1,1}
\definecolor{lightgray}{rgb}{0.827,0.827,0.827}
\definecolor{lightgreen}{rgb}{0.565,0.933,0.565}
\definecolor{lightyellow}{rgb}{1,1,0.878}
\definecolor{black}{rgb}{0,0,0}
\definecolor{white}{rgb}{1,1,1}
\scalebox{0.97}{
\begin{tikzpicture}[ipe stylesheet]
 \draw[->]
    (304, 416)
     -- (368, 416);
       \draw[<-]
    (176, 560)
     -- (176, 416);
   \fill[darkgreen, ipe pen heavier, ipe opacity 30]
    (176, 464) rectangle (253, 416);
   \draw[darkgreen, ipe pen heavier]
    (176, 464) rectangle (253, 416);  
  \fill[darkorange, nonzero rule, ipe opacity 75]
    (176, 528)
     -- (192, 528)
     -- (192, 432)
     -- (192, 432)
     -- (192, 416)
     -- (176, 416)
     -- cycle;
  \fill[draw=darkgreen, ipe pen heavier, fill=green, ipe opacity 30]
    (176, 496) rectangle (216, 416);
   \draw[draw=darkgreen, ipe pen heavier]
    (176, 496) rectangle (216, 416);
  \fill[red, ipe opacity 30]
    (176, 432)
     -- (192, 432)
     -- (192, 432)
     -- (320, 432)
     -- (320, 416)
     -- (176, 416)
     -- cycle;
  \node[ipe node, font=\footnotesize]
     at (184, 404) {{\sf FOL},};
  \draw[orange, ipe pen fat]
    (188, 496) rectangle (188, 496);
  \draw
    (176, 416)
     -- (304, 416);
  \fill[blue, ipe pen ultrafat, ipe opacity 30]
    (176, 464)
     -- (288, 464)
     -- (288, 432)
     -- (320, 432)
     -- (320, 416)
     -- (176, 416)
     -- (176, 464)
     -- (288, 464);
     \draw[blue, ipe pen ultrafat]
    (176, 464)
     -- (288, 464)
     -- (288, 432)
     -- (320, 432)
     -- (320, 416)
     -- (176, 416)
     -- (176, 464)
     -- (288, 464);
  \node[ipe node, font=\footnotesize]
     at (324, 433.059) {Courcelle \cite{Courcelle90them,Courcelle97,Courcelle92}, Borie, Parker \& Tovey \cite{BoriePT92auto}};
  \node[ipe node, font=\footnotesize]
     at (324, 421.059) {and Arnborg, Lagergren \&  Seese \cite{ArnborgLS91easy}};
  \node[ipe node, font=\footnotesize]
     at (220, 500.059) {Pilipczuk, Schirrmacher, Siebertz, Torunczyk,
  \& Vigny \cite{PilipczukSSTV22algor}};
  \node[ipe node, font=\footnotesize]
     at (205, 404) {${\sf FOL\!\!+\!conn},$};
  \node[ipe node, font=\footnotesize]
     at (251, 404) {${\sf FOL\!\!+\!DP},$};
  \node[ipe node, font=\footnotesize]
     at (290, 404) {$\tilde{ \Theta}^{\sf DP},$};
  \node[ipe node, font=\footnotesize]
     at (311, 404) {{\sf CMSOL}};
  \pic[red]
     at (216, 496) {ipe square};
  \pic[red]
     at (253, 464) {ipe square};
  \pic[red]
     at (288, 464) {ipe square};
  \pic[red]
     at (320, 432) {ipe square};
  \node[ipe node, font=\small, text=blue]
     at (292, 465.059) {{\autoref{thm_tildethetadp}}};
  \node[ipe node, font=\footnotesize]
     at (250, 484.059) {Golovach, Stamoulis,
  \& Thilikos \cite{GolovachST22model}};
  
  \draw[-{>[ipe arrow tiny]}] (253,482) -- (253,467);
  
  \node[ipe node, font=\footnotesize]
     at (196, 531.059) {Grohe, Kreutzer \& Siebertz \cite{GroheKS17dec} /  Bonnet, Kim, Thomassé \& Watrigant \cite{Bonnet0TW20twinw}};
  \pic[red]
     at (192, 528) {ipe square};
  \node[ipe node, text=seagreen]
     at (130.311, 555.343) {{\small Structure}};
  \node[ipe node, text=seagreen]
     at (357.554, 403.343) {{\small Logic}};
  \node[ipe node, font=\footnotesize]
     at (64, 460) {{\sf bounded Hadwiger number}};
  \node[ipe node, font=\footnotesize]
     at (96, 428) {{\sf bounded treewidth}};
  \node[ipe node, font=\footnotesize]
     at (26, 529.059) {{\sf Nowhere dense /  bounded twin-width}};
  \node[ipe node, font=\footnotesize]
     at (79.399, 496.163) {{\sf bounded Hajós number}\footnotemark};
\end{tikzpicture}
}
	\end{center}
\myskip
	\caption{The current meta-algorithmic landscape and the position of \autoref{thm_tildethetadp} in it.}
	\label{@resolvesssremos}
	\myskip
\end{figure}
\footnotetext{The {\em Hajós number} of a graph $G$ is the maximum $k$ for which $G$ contains $K_{k}$ as a topological minor.}


\subsection{Bypassing Gaifman's Theorem}

We now explain how to use use the recent results of \cite{GolovachST22model,GolovachST22model_arXiv}
in order to modify the proofs of \autoref{@decendientes1} and \autoref{@pertrechando1}, towards proving the more general versions \autoref{thm_thetadp} and \autoref{thm_tildethetadp}. For this, we essentially show how
we may bypass the computation of the in-signature (based on the application of Gaifman's theorem)
by using instead the main result of \cite{GolovachST22model,GolovachST22model_arXiv}.

As a first step in the proof of~\autoref{thm_thetadp},
we modify the definition of the enhanced version of a sentence in $\Theta$ (see \autoref{sec_equivalent_version})
to the corresponding notion for sentences in $\Theta^{\DP}$ as follows:
we consider the enhanced version $σ_{{\sf R},{\bf c}}$ of every formula in $\FOLDP$ as defined in~\cite[Section 4]{GolovachST22model_arXiv}
and we define the {\em enhanced version} $θ_{{\sf R},{\bf c}}$ of $θ$
to be a sentence obtained
from $\theta$ after replacing each $\FOLDP$-target sentence $\sigma$ of $\theta$ with the enhanced version
$σ_{{\sf R},{\bf c}}$ of $σ$.
This way, it is easy to prove that both \autoref{obs_addingR} and \autoref{lem_no_matter_which_apex} hold in this setting.

The only missing ingredient for the proof of~\autoref{thm_thetadp} is an analogue of~\autoref{lemma_irrele_flat} for $\Theta^{\DP}$.
As described in \autoref{sec_reduce_instance},
to achieve this we inspect the three basic elements of our problem
(i.e., satisfaction of the modulator sentence in the modulator sets, satisfaction of the two target sentences in the ``remaining part'') and
we argue how to reduce the annotated set $R$ and characterize some non-annotated vertices as ``irrelevant''.
As explained in~\autoref{sec_reduce_instance}, for the ``irrelevancy'' for minor-exclusion we can use~\autoref{icalp_irrelevancy}, while the remaining two parts we design the algorithm ${\tt Find\_Equiv\_FlatPairs}$ (see~\autoref{@inhumainement}).
The proof of correctness of this algorithm is based on three Claims (see~\autoref{sec_proof_correctness} for $\bar{\Theta}_1$).
While \autoref{claim_1} and \autoref{claim_3} work for $\Theta^{\DP}$, this is not the case for \autoref{claim_2}.
The reason is that
dealing with  $\FOL\!\!+\!\DP$-target sentences, we can no longer use Gaifman's theorem and therefore we have to employ different techniques.
In fact, we use the following result from~\cite{GolovachST22model_arXiv} that intuitively says that,
given graph that contains a ``large enough'' flat wall,
we can finds an area inside the flat wall that is {\sl annotation-irrelevant} and
inside this area, the removal of any subset of vertices does not change the satisfiability of $σ_{{\sf R},{\bf c}}$.
To state it we also need to define the notion of {\em outer-compatibility}.

Let $G,G'$ be two graphs, let $R,R'$ be subsets of $V(G)$ and $V(G')$, respectively,
and let a partial function $ξ: V(G)\to V(G')$.
We say that the pairs $(G,R)$ and $(G',R')$ are {\em $ξ$-compatible} if for every $i\in[h]$ and every $v\in V(G)$,
$v\in X_i\iff ξ(v)\in X_i'$.
Let $(G,R)$ and $(G',R')$ be two annotated graphs and let a partial function $ξ: V(G)\to V(G')$ such that $(G,R)$ and $(G',R')$ are $ξ$-compatible.
We denote by $(G,R)\oplus_ξ (G',R')$ the colored graph obtained from the disjoint union of $(G,R)$ and $(G',R')$ after identifying vertices $v\in V(G)$ and $u\in V(G)$ if $ξ(v)=u$.
Let two annotated graphs $(G,R),(G',R')$, let ${\bf a}$ be an apex-tuple of $G$, and let $(W,\mathfrak{R})$ be a flatness pair of $G\setminus V({\bf a})$.
Given that $\mathfrak{R}=(X,Y,P,C,Γ,σ,π)$,
we call a partial function $ξ: V(X\cap Y)\cup V({\bf a})\to V(G')$ such that
$(G,R)$ and $(G',R')$ are $ξ$-compatible
an {\em outer-compatibility function of $(G,R)$ and $(G',R')$}.

Given an $r$-wall $W$ of some graph $G$ and an $r'\in\mathbb{N}$ such that $r'\leq r$,
we say that a subwall $W'$ of $W$ is {\em $r$-internal} if it is a subwall of $W^{(r-r')}$.

\begin{proposition}[\!\!\cite{GolovachST22model_arXiv}]
\label{prop_from_SODA}
There are two functions $\newfun{@individuated}:\mathbb{N}^3\to\mathbb{N}$ and $\newfun{fun_inte}:\mathbb{N}^2\to\mathbb{N}$
and an algorithm that, given
\begin{itemize}
\item $t,l,g\in\mathbb{N}$,
\item an $n$-vertex  graph ${G}$,
\item an apex-tuple ${\bf a}$ of ${G}$ of size $l$,
\item a regular flatness pair $(W,\mathfrak{R})$ of ${G}\setminus V({\bf a})$ of height at least $\funref{@individuated}(t,l,g)$ whose compass has treewidth at most ${\sf tw}$,
and
\item a set  $R\subseteq V(G)$,
\end{itemize}
outputs, in time $\O_{t,l,g,{\sf tw}}(n)$, a
flatness pair $(\tilde{W}',\tilde{\mathfrak{R}}')$ of $G\setminus V({\bf a})$
that is a $W'$-tilt of some subwall $W'$ of $W$ of height at least $\funref{fun_inte}(t,l)+g$
such that
\begin{itemize}
\item for every flatness pair $(\tilde{W}'',\tilde{\mathfrak{R}}'')$ of $G\setminus V({\bf a})$
that is a $W''$-tilt of some subwall $W''$ of $W'$ of height at least $\funref{fun_inte}(t,l)+g$
and for every flatness pair $(\tilde{W}''',\tilde{\mathfrak{R}}''')$ of $G\setminus V({\bf a})$
that is a $W'''$-tilt of some $\funref{fun_inte}(t,l)$-internal subwall $W'''$ of $W'$ of height $g$,
for every annotated graph $(F,R^\star)$,
every outer-compatibility function $ξ$ of $(G,{R})$ and $(F,R^\star)$,
every $σ\in\FOLDP$ of quantifier rank at most $t$,
and every $Y\subseteq V({\sf Compass}_{\tilde{\mathfrak{R}}'''}(\tilde{W}'''))$, it holds that
$$((F,R^\star)\oplus_{ξ} (G,R), {\bf a})\models σ_{{\sf R},{\bf c}} \iff ((F,R^\star)\oplus_{ξ} (G\setminus Y,R\setminus V({\sf Compass}_{\tilde{\mathfrak{R}}''}(\tilde{W}'')) ), {\bf a})\models σ_{{\sf R},{\bf c}}.$$
\end{itemize}
\end{proposition}

\paragraph{Modifying the definition of characteristics.}
To use~\autoref{prop_from_SODA} to prove \autoref{claim_2} for formulas in $\Theta^\DP$ (i.e., to show that $(G, R, {\bf a}')[C]\models  σ_{{\sf R},{\bf c}} \iff (G, R\setminus Y, {\bf a}')[C']\models  σ_{{\sf R},{\bf c}}$), we also have to slightly modify the definition of
the characteristic in the end of~\autoref{sec_in-sig_first-floor} and the algorithm ${\tt Find\_Equiv\_FlatPairs}$, as we proceed to describe.

Let $q =(\tw(θ)+1)^2+1$ and let $j'∈ \mathbb{N}$.
We set $j={\sf odd}(\max\{q/2,j'\})$.
We set $f(x,y,z,0) = z$ and for every $i\geq 1$ we set $f(x,y,z,q)= \funref{@individuated}(x,y,f(x,y,z,i-1))$.
Then, we set $w=f(t,l,g,q)$ and $r=\funref{@individuated}(t,l,g)$.
We set $${\sf CHAR} = [2,q]\times 2^{[l]}\times 2^{\blue{{\sf SIG}_{\sf out}}}.$$

Let $G$ be a graph, let ${\bf a}=(a_{1},\ldots,a_l)$ be an apex-tuple of $G,$ and let $(W,\mathfrak{R})$ be a flatness pair of $G\setminus V({\bf a})$ of height $2w+j$, and let $K:= {\sf compass}_{\mathfrak{R}}(W)$ and $K^{\bf a}:=G[V({\bf a})\cup V(K)].$
Also, let ${\bf W}_{{q}}$ be the $q$-pseudogrid defined by the horizontal and vertical paths of the central $q$-subwall of $W.$
We call {\em enhanced extended compass} of $(W,\mathfrak{R})$ a tuple $(\mathfrak{K},U_1,\ldots,U_q,R)$, where $\mathfrak{K} = (\mathfrak{A}[V(K^{\bf a})],{\bf a},{\bf W}_{{q}})$
and $\{U_1,\ldots,U_q,R\}\subseteq 2^{V(\mathfrak{K})}$.
We stress that in~\autoref{@verwechslungen}, in the definition of the extended compass of a flatness pair, we include to $\mathfrak{K}$ the tuple ${\bf I}$, while this is not anymore necessary in the definitions needed here.

Given a tuple $(i,L)\in  [2,q]\times 2^{[l]}$ and a set $Z\subseteq U_i$,
we define the {\em out-signature} of $(\mathfrak{K},R,U_i,L,Z)$,
denoted by $\blue{{\sf out}\text{-}{\sf sig}}(\mathfrak{K},R,U_i,L,Z)$, as in~\autoref{sec_out-sig_first-floor}, by replacing $\mathfrak{A}^{(d,Z,L,F)}$ by $G^{(U_i,Z,L,F)}$, where $G^{(U_i,Z,L,F)}$ is the graph obtained from $G[U_i\cup V_L ({\bf a})]$ after adding a set of $|V(F)|$ vertices and all edges corresponding to the additional edges of $F$ and every edge between a vertex in  $V(F\setminus V_L ({\bf a}))$ and a vertex in  $U_i\setminus Z$.

Given an enhanced extended compass $(\mathfrak{K},U_1,\ldots,U_q,R)$  of $(W,\mathfrak{R})$,
we define its {\em characteristic} to be
\begin{eqnarray}
\labels{eq_char_first-floor}θ\text{-}{\sf char}(\mathfrak{K},U_1,\ldots,U_q,R) & = &\{(i,L,\blue{{\sf sig}_{\sf out}})∈ {\sf CHAR} \mid
\exists\ Z\subseteq U_i\mbox{~such that}\\
\notag & &\hspace{5cm}  \blue{{\sf out}\text{-}{\sf sig}}(\mathfrak{K},R,U_i,L,Z)= \blue{{\sf sig}_{\sf out}}\}.
\end{eqnarray}
After defining the characteristic, we proceed to describe the modifications in the proof of \autoref{@desmembramientos}.

\paragraph{Modifying the algorithm ${\tt Find\_Equiv\_FlatPairs}$.}
The algorithm ${\tt Find\_Equiv\_FlatPairs}$ presented in~\autoref{@inhumainement},
in Step 2 computes a collection $\tilde{{\cal W}}$ of $2^{|{\sf CHAR}|}\cdot q$ flatness pairs of $\mathfrak{G}\setminus V({\bf a})$ of height $2w+j.$
We then add an extra step:
For each $(\tilde{W}_i, \tilde{\mathfrak{R}}_i)\in \tilde{{\cal W}}$,
we recursively apply the algorithm of \autoref{prop_from_SODA}
$q$-many times,
where, for every $i\in[2,q]$,
if $(W_{i-1}',\tilde{\mathfrak{R}}_i')$ is the output of the $(i-1)$-th recursive application,
the input of the algorithm in the $i$-th application is a tilt of an $f(k)$-internal subwall $W_{i-1}'''$ of $W_{i-1}'$.
This way,
for each $i\in[| \tilde{{\cal W}}|]$,
we compute flatness pairs
$(W_{i,1}',\mathfrak{R}_{i,1}'),\ldots, (W_{i,q}',\mathfrak{R}_{i,q}')$ and
$(W_{i,1}''',\mathfrak{R}_{i,1}'''),\ldots, (W_{i,q}''',\mathfrak{R}_{i,q}''')$ of $G\setminus V({\bf a})$
such that if we set
$(W_{i,0},\mathfrak{R}_{i,0}):=(\tilde{W}_i, \tilde{\mathfrak{R}}_i)$
and for every  $j\in[0,q]$, we set $U_{i,j}' := V({\sf Compass}_{\mathfrak{R}_{i,j}'}(W_{i,j}'))$ and $U_{i,j}''':= V({\sf Compass}_{\mathfrak{R}_{i,j}'''}(W_{i,j}'''))$,
then
\begin{itemize}
\item for every $j\in[q]$, $U_{i,j}'''\subseteq U_{i,j}'\subseteq U_{i,j-1}'''$, and
\item for every $j\in[q]$, every annotated graph $(F,R^\star)$,
every outer-compatibility function $ξ_{i,j}$ of $(G,R)[U_{i,j-1}]$ and $(F,R^\star)$,
every $σ\in\FOLDP$ of quantifier rank at most $t$,
and for every $S\subseteq U_{i,q}'''$, if $(G_{i,j-1},R_{i,j-1}):=(G,R)[U_{i,j-1}']$, then
\begin{eqnarray*}
& ((F,R^\star)\oplus_{ξ_{i,j}} (G_{i,j-1},R_{i,j-1}), {\bf a})\models σ_{{\sf R},{\bf c}},\\
& \iff\\
& ((F,R^\star)\oplus_{ξ_{i,j}} (G_{i,j-1}\setminus S,R_{i,j-1}\setminus U_{i,q}'), {\bf a})\models σ_{{\sf R},{\bf c}}.
\end{eqnarray*}
\end{itemize}
For every $(i,j)\in[|\tilde{\cal W}|]\times[q]$, we set $V_{i,j} = U_{i,j}'\setminus U_{i,j}'''$.
Observe that $V_{i,1},\ldots,V_{i,q}$ are pairwise disjoint subsets of $V({\sf Compass}_{\tilde{\mathfrak{R}}_i}(\tilde{W}_i))$.
Then, in Step 3, after
defining the enhanced extended compass $\mathfrak{K}_i,U_{i,1}''',\ldots,U_{i,q}''',R_i$ for every $i\in[|\tilde{{\cal W}}_i|]$,
we compute $θ\text{-}{\sf char}(\mathfrak{K}_i,U_{i,1}''',\ldots,U_{i,q}''',R_i)$ using Courcelle's theorem.
We output the set $Y:=U_{1,q}'$ and the flatness pair
$(W_{1,q}''',\mathfrak{R}_{1,q}''').$
\medskip

\paragraph{Proof of correctness of the modified ${\tt Find\_Equiv\_FlatPairs}$.}
What remains is to show that
$(G,R, {\bf a})\models θ_{{\sf R},{\bf c}}\iff (G\setminus U_{1,q}''',R\setminus U_{1,q}', {\bf a})\models θ_{{\sf R},{\bf c}}.$
We describe how to modify the proof of~\autoref{sec_proof_correctness} to work in our case.

We consider the simple case where $\theta = β\triangleright (σ\wedge μ)^{({\sf c})}$,
for some $β\in \MSOL^\tw[\{{\sf E},{\sf X}\}]$,
$σ\in \FOL\!+\!\DP$ and $\mu\in\NTMC$.
Suppose that
there is a set $X\subseteq V(G)$ such that
$(G,X)\models \beta$ and for every connected component $C$ of $G\setminus X$,
$G[C]\models μ$ and $(G,R, {\bf a})[C]\models σ_{{\sf R},{\bf c}}.$

Let $\breve{C}$ be the privileged connected component of $G$ with respect to a $q$-pseudowall ${\bf W}_q^{(1)}$ of $G$ and $X$.
Since $X$ intersects at most $q-1$ internal bags of any $(\tilde{W}_1,\tilde{\mathfrak{R}}_1)$-canonical partition of $G\setminus V({\bf a}),$
we know that there is a $j_0\in[q]$ such that $X\cap V_{1,j_0} = \emptyset$.
We set $X_{\sf in} := X\cap U_{1,j_0}'''$ and $X_{\sf out}:= X\setminus X_{\sf in}$.
Let $Z = U_{1,j_0}'''\setminus \breve{C}$.

Also, recall that
for every annotated graph $(F,R^\star)$ and
every outer-compatibility function $ξ_{1,j_0}$ of $(G_{1,j_0-1},R_{1,j_0-1})$ and $(F,R^\star)$,
it holds that
\begin{eqnarray}
& \notag((F,R^\star)\oplus_{ξ} (G_{1,j_0-1},R_{1,j_0-1}), {\bf a})\models σ_{{\sf R},{\bf c}},\\
& \iff \label{thetadpeq1}\\
& \notag((F,R^\star)\oplus_{ξ_{1,j_0}} (G_{1,j_0-1}\setminus Z,R_{1,j_0-1}\setminus U_{1,q}'), {\bf a})\models σ_{{\sf R},{\bf c}}.
\end{eqnarray}

The fact that $θ\text{-}{\sf char}(\mathfrak{K}_1,U_{1,1}''',\ldots,U_{1,q}''',R_1)= θ\text{-}{\sf char}(\mathfrak{K}_2,U_{2,1}''',\ldots,U_{2,q}''',R_2)$ implies that there is a $Z'\subseteq U_{2,j_0}'''$ such that
$\blue{{\sf out}\text{-}{\sf sig}}(\mathfrak{K}_1,R_1,U_{1,j_0}''',L,Z)= \blue{{\sf out}\text{-}{\sf sig}}(\mathfrak{K}_2,R_2,U_{2,j_0}''',L,Z')$
and by \autoref{claim_1} there is a set $X'$ satisfying $θ^{\sf out}_q$.
Also,
for every annotated graph $(F,R^\star)$ and
every outer-compatibility function $ξ_{2,j_0}$ of $(G_{2,j_0-1},R_{2,j_0-1})$ and $(F,R^\star)$,
it holds that
\begin{eqnarray}
& \notag((F,R^\star)\oplus_{ξ_{2,j_0}} (G_{2,j_0-1},R_{2,j_0-1}), {\bf a})\models σ_{{\sf R},{\bf c}},\\
& \iff \label{thetadpeq2}\\
& \notag((F,R^\star)\oplus_{ξ_{2,j_0}} (G_{2,j_0-1}\setminus Z',R_{2,j_0-1}\setminus U_{2,q}'), {\bf a})\models σ_{{\sf R},{\bf c}}.
\end{eqnarray}

We set $C'$ be the privileged connected component of $G$ with respect to ${\bf W}_q^{(1)}$ and $X_{\rm out}\cup X'$.
Note that $C\cup Z = C'\cup Z'$ (as both these sets are equal to the privileged
connected component of $G$ with respect to ${\bf W}_q^{(1)}$ and $X_{\rm out}$).

We also set $(F_1,R_{F_1}):=(G,R)[C\setminus U_{1,j_0 -1}']$ and $(F_2,R_{F_2}):= (G,R)[C'\setminus  U_{2,j_0 -1}']$.
Notice there is an outer-compatibility function $ξ_{1,j_0}$ of $(F_1,R_{F_1})$ and $(G_{1,j_0-1},R_{1,j_0-1})$
such that
$$(F_1,R_{F_1})\oplus_{ξ_{1,j_0}}(G_{1,j_0-1},R_{1,j_0-1}) = (G,R)[C\cup Z]$$
and there is an outer-compatibility function $ξ_{2,j_0}$ of $(F_2,R_{F_2})$ and $ (G_{2,j_0-1},R_{2,j_0-1})$
such that
$$(F_2,R_{F_2})\oplus_{ξ_{2,j_0}}  (G_{2,j_0-1},R_{2,j_0-1}) = (G,R)[C'\cup Z'].$$
Therefore, using the fact that $C\cup Z = C'\cup Z'$ and \eqref{thetadpeq1} and \eqref{thetadpeq2} we derive that
$(G,R, {\bf a})[C]\models σ_{{\sf R},{\bf c}} \iff (G,R\setminus U_{1,q}', {\bf a})[C']\models σ_{{\sf R},{\bf c}}.$

\myskip\section{Limitations, extensions, and further directions}
\label{sec_conclusions}
\myskip

To conclude the article, in  \autoref{nat_lim} we justify the necessity of the ingredients of our logic~$Θ$. Next, in \autoref{sec_extensions}, we discuss two additional extensions of our results. The one  is based on the notion of {\sl irrelevant-friendliness} and the other suggests an alternative meta-algorithmic trade off based on the {\sl scattered disjoint-paths predicate}. Finally, in \autoref{sec_further_research}
we   present several directions and open problems for further research.

\myskip
\myskip\subsection{Natural limitations}
\label{nat_lim}
\label{@achievements}

We now wish to comment on why the three basic
ingredients of the definition of our logic~$Θ$  are necessary for the statement and the
proof of a meta-algorithmic result such as \autoref{@decendientes}.

The first  ingredient of $Θ$ is that the modulator sentences belong in
$\MSOL^{\sf tw}[\{{\sf E},{\sf X}\}]$ which is  defined so
that the treewidth of $\torso(G,X)$ is bounded.~%
%
%
While it is known that bounding the treewidth is
necessary for \MSOL-model-checking~\cite{Kreutzer11algo,CourcelleMR00linea}, one may ask why it is not enough to just bound the treewidth of $G[X].$
To see why this unavoidable, consider a graph $G$ and let $G'$ be the graph obtained from $G$ by subdividing
 each edge once. Then, asking whether $G$ is Hamiltonian, which is a well-known \NP-complete problem~\cite{GJ79},
is equivalent to asking whether $G'$ has a vertex set $S'$ such that $G'[S']$
is a cycle and such that $G'\setminus S'$ is an edgeless graph, that is, a $K_{2}$-minor-free graph. Notice that, while $\tw(G'[S'])=2,$ $\torso(G',S')=G$
has unbounded treewidth.

The second ingredient of $Θ$ is minor-exclusion, that is materialized by the conjunction with $μ$ in the definition of $Θ_0.$ Notice first that expressing whether a graph $G$ contains a clique on $k$ vertices can be done by a \FOL-sentence, while the $k$-\textsc{Clique} problem is ${\sf W}[1]$-hard~\cite{cygan2015parameterized}. Therefore, the minor-exclusion condition cannot be dropped. Moreover, even if we consider a {\sl fixed} target \FOL-sentence, it was proved in~\cite{FominGT20onthe} that there exists a \FOL-sentence $σ$ such that
checking whether a graph $G$ has a set $S\subseteq V(G)$ with $|S|= k$
such that $G\setminus S\models σ$ is a ${\sf W}[1]$-hard problem, when parameterized by $k.$ This implies that, even for this restricted problem where the \FOL-sentence $σ$ is fixed, an algorithm in time $f(k)\cdot n^{\O(1)}$  cannot be expected.

 The third ingredient of $Θ$ is  the $\FOL$ demand, that is materialized by the conjunction with $σ$ in the definition of~$Θ_0.$ This is also necessary,
 as otherwise we may choose some property $σ$ not definable in \FOL, such as
 Hamiltonicity, which is  \MSOL-definable and {\sf NP}-complete
on
planar graphs~\cite{GJ79}. Without the restriction that $σ$ needs to be \FOL-definable, a void modulator and a sentence $μ$ expressing planarity would be able to model this  {\sf NP}-complete  problem. {Nevertheless, we may consider extensions of \FOL in the target sentence, as it is done in \autoref{sec_thetadp} and as  suggested in the next subsection.}

\myskip

\myskip\subsection{Extensions}
\label{sec_extensions}

\myskip\paragraph{Irrelevant-friendliness.}
In the definition of $Θ_{0}$ we include compound sentences $σ\wedge μ$
where $σ∈ \FOL$ and $μ$ expresses minor-exclusion.
Our proof is modulated so to allow  that  $σ$
may express a wider set of sentences, not definable in \FOL, which we proceed to discuss.
We call these sentences {\em irrelevant-friendly}, a technical concept whose formal definition is the following:

\begin{definition}[Irrelevant-friendliness]
\label{@disinfectants}
For a real number $\alpha ≥ 1,$ we say that a sentence $φ∈ {\rm \MSOL}[\{E\}]$ (evaluated on graphs) is  {\em $α$-irrelevant-friendly}
if $\Mod(φ)$ is hereditary\footnote{A graph class ${\cal G}$ is {\em hereditary} if every induced subgraph of a graph in ${\cal G}$ belongs in ${\cal G}.$} and there
 exists a function $\newfun{@reintroduced}:\mathbb{N}^3\to\mathbb{N}$ and an algorithm in time $\O(n^α)$ with the following specifications:
	
\medskip	
	\noindent{\sl Input:} two  integers $k,a∈ \mathbb{N},$ an odd integer $\ell∈ \mathbb{N}_{≥ 3},$  a graph $G,$ an $A\subseteq V(G)$ with $|A|≤ a,$ and
	a flatness pair $({W},{\mathfrak{R}})$
	of $G\setminus A$ of height $q≥ \funref{@reintroduced}(k,a,\ell).$
	
	\noindent{\sl Output:} A subwall $W'$ of $W$ of height $\ell$ such that if $K$ is the compass of some $W'$-tilt of $W$ and $X\subseteq V(G),$ where  ${\sf bid}(X\setminus {A},W,\mathfrak{R})≤ k,$ then $G\setminus X\models φ\Rightarrow  G\setminus (X\setminus  V(K))\models φ.$\end{definition}
	
(The definition of ${\sf bid}(X\setminus {A},W,\mathfrak{R})$ can be found in \autoref{sec_bidimensionality}.)  The above property is the abstraction of
what is required for a sentence in order to make our proof applicable. In fact, minor-exclusion is already an irrelevant-friendly property and
an important part of the proof is based on this fact.

For each  irrelevant-friendly sentence
$φ∈ \MSOL[\{{\sf E}\}],$ we define an extension $Θ^{φ}$ of $Θ$  by considering, as the base
case, instead of  $Θ_{0},$ the logic
$$Θ_{0}^{φ} = \{σ\wedge μ\wedge φ|_{\sf gf}\mid  σ∈ \FOL[τ]\mbox{~and~} μ∈ \NTMC[τ]\},$$
that is, we add $φ$ in the  conjunction  defining  $Θ_0.$

\begin{theorem}\label{@indudablemente}
If $\alpha≥ 1$ is a real number, $τ$ is some vocabulary, $φ∈ \MSOL[\{{\sf E}\}]$ is some   $α$-irrelevant-friendly sentence on graphs, and $θ∈ {Θ}^{φ}[τ],$ then $\Mod(θ)$ can be decided in time $\O(n^{α+1}).$
\end{theorem}

The  proof of~\autoref{@indudablemente}
is a direct consequence of the algorithm of \autoref{@decendientes}
and \autoref{@disinfectants}.
Indeed, the algorithm first finds an apex set $A$ and a flatness pair $(W,\mathfrak{R})$ of $G\setminus A$ as in the second step of proof of the algorithm of~\autoref{@decendientes} as presented in~\autoref{sec_gen_algo}.
Before proceeding
to the third step,
the algorithm uses the algorithm in \autoref{@disinfectants} in order to detect $W',$ and the compass $K'$ of some $W'$-tilt $W''$ of $W.$ According to the specifications of the algorithm in \autoref{@disinfectants}, the whole vertex set of $K'$  is already irrelevant with respect to the sentence $φ$ and, because
of the hypothesis that $\Mod(φ)$ is hereditary, we can now
proceed to the third step,
apply the algorithm of \autoref{lemma_irrele_flat}
inside the compass $K',$ and detect there the irrelevant vertices of $G$ as well as the vertices to be removed from the annotated set $R.$

The consequences of~\autoref{@indudablemente}
are open to investigate,  as it may permit to handle target sentences that are not necessarily \FOL-definable.
As an example, we mention the result
of Fiorini,  Hardy,  Reed, and Vetta in~\cite{FioriniHRV05plana} who designed an algorithm of time $\O_{k}(n)$ that  checks whether
a graph can be made planar and bipartite by removing $k$ vertices. Let  $φ∈\MSOL[{\sf E}]$ be the property of being bipartite. For planar graphs, using a simple parity argument, one can show that $φ$ is $1$-irrelevant-friendly.
Based on this and \autoref{@indudablemente},
one may  demand {\sl any} $\FOL$-property, apart from planarity and bipartiteness (which is not \FOL-definable)
and solve the corresponding problem in  time $\O_{k}(n^2)$ by \autoref{@indudablemente}.  This easy corollary
of \autoref{@indudablemente} can be further extended, as bipartiteness is $1$-irrelevant-friendly for other minor-closed graph classes as well. We prefer not to enter into details here, as we  mention  this problem  only as an example of  the potential of~\autoref{@indudablemente}. For other examples of irrelevant-friendly properties, one may consider, for instance, the exclusion of odd-minors~\cite{Huynh09theli} or of topological minors~\cite{FominLP0Z20hitti}.
%
\myskip

\myskip\myskip
\myskip\paragraph{From $\tilde{Θ}^{\sf DP}$ to its ``scattered'' extension $\tilde{Θ}^{\sf SDP}$.}

In \autoref{sec_thetadp} we defined the logic $Θ^{\sf DP}$ (and its counterpart $\tilde{Θ}^{\sf DP}$)
by using as target logic {\sf FOL\!+DP}, that is {\sf FOL} enhanced by {disjoint-path predicates}
${\sf dp}_{k}(\cdot)$ where
${\sf dp}_{k}({\sf x}_1, {\sf y}_1, \ldots,{\sf x}_k, {\sf y}_k)$  evaluates true in a  graph $G$  if and only if there    pairwise disjoint  paths $P_1,\ldots, P_k$ of $G$ of {length at least two}  between (the interpretations of) ${\sf x}_i$ and ${\sf y}_i$ for all $i\in[k].$ In the same section we proved extensions of \autoref{@decendientes1} and \autoref{@pertrechando1} for these more general logics by using the  recent results of \cite{GolovachST22model}. In  \cite{GolovachST22model}, a generalization  of  {\sf FOL\!+DP} was defined, namely  {\sf FOL\!+SDP},  by considering (for $s\in\mathbb{N}$) the   predicate
{\sf $s$-dp}$_{k}({\sf x}_1,{\sf y}_1,\ldots,{\sf x}_k,{\sf y}_k)$, where we now demand that the disjoint paths in question are pairwise $s$-scattered, i.e., there are no two vertices of two distinct
paths that are within distance at most  $s$.
 It is proved in \cite{GolovachST22model,GolovachST22model_arXiv} that model-checking for sentences in  {\sf FOL\!+SDP}  can be done in quadratic time for graphs of bounded Euler genus.
Similarly to the definition of   $\tilde{Θ}^{\sf DP},$
we may define the logic  $\tilde{Θ}^{\sf SDP}$  by using  {\sf FOL\!+SDP} as the target logic. By using the same arguments as in \autoref{sec_thetadp} for $\tilde{Θ}^{\sf SDP},$
it is possible to prove an analogue of \autoref{thm_tildethetadp} 
where   $\tilde{Θ}^{\sf DP}$  is replaced by the more expressive $\tilde{Θ}^{\sf SDP}$, while $\hw(G)$ is replaced by the more restrictive ${\bf eg}(G)$, that is the Euler genus of the graph $G$. The only essential change in the proof is that now, instead of~\autoref{prop_from_SODA}, we apply its ``{\sf FOL\!+SDP}-counterpart''  given by \cite[Lemma 12]{GolovachST22model_arXiv}.

\myskip\subsection{Further research}
\label{sec_further_research}

\myskip\paragraph{The minor-exclusion framework.} The graph-structural horizon in both \autoref{@decendientes}  and \autoref{@pertrechando} is delimited  by minor-exclusion.
In the case of \autoref{@decendientes}, this restriction is applied to the target
property defined by $μ$ in the logic $Θ,$ while in  \autoref{@pertrechando} this is the promise combinatorial restriction that yields efficient model-checking for  $\tilde{Θ}.$ This restriction is
hard-wired in our proof in the way it combines the Flat Wall theorem
with Gaifman's theorem. Recently, several efficient algorithms appeared
for modification problems targeting or assuming topological minor-freeness (see~\cite{PilipczukSSTV22algor,FominLP0Z20hitti,JansenK021verte,AgrawalKLPRSZ22delet} and the meta-algorithmic result in~{\cite{SchirrmacherSV22first}}). For such classes, to achieve efficient model-checking
for $Θ,$ or  some  fragment of it, is
an interesting open challenge.

\myskip\myskip

\myskip\paragraph{Quadratic time.}
The proof of \autoref{@decendientes} can be seen as a possible ``meta-algorithmization'' of the
{\sl irrelevant vertex technique} introduced by Robertson and Seymour~\cite{RobertsonS95XIII}, {going further than the two known recent attempts in this direction~\cite{GolovachST22model,FominGST20analgo}}.
The main routine of the algorithm
transforms the input of the problem
to a simpler graph by detecting territories in it that can be safely discarded, therefore producing a simpler instance. This routine is applied repetitively
until the graph  has ``small''  treewidth, so that the problem can
be solved in linear time by using Courcelle's theorem. This approach
gives an algorithm running in quadratic time. Any improvement of this running time
should rely on techniques escaping the above scheme of gradual simplification. The only results in this direction are the cases of making a graph planar by deleting at most $k$ vertices (resp. edges) in~\cite{JansenLS14anea} (resp. \cite{KawarabayashiR2007compu}) that run in time $\O_{k}(n).$

\myskip\myskip
\myskip\paragraph{Other modification operations.}
In \autoref{sec_applications} we gave a wide variety of problems that can be modeled by the logic $Θ,$ either directly or indirectly, via reductions. All these modifications concern problems involving edge or vertex removals. Is it possible to extend  $Θ$ so that it can also deal with other (local) operations such as edge contractions, edge additions, or others? This was done in~\cite{FominGST20analgo}
for the case of removing vertices to achieve planarity and a \FOL-definable property. Moreover, especially for
the part of~\autoref{thm_grammar} concerning the grammar ${\cal M},$
it is possible to make the following enhancement in order to include the contraction operation:
include the
production rule ${\sf M}\rightarrow ({\sf s}{{\sf M})}$ and add in the definition of $\tilde{M}$ that if $\tilde{\sf w}=({\sf s} {\sf w}'),$  then ${\cal G}_{\tilde{\sf w}}=\{G\mid \exists e∈ V(G),  G\,\slash\, e∈ {\cal G}_{\tilde{\sf w}'}\}.$\footnote{We use $G\,\slash\, e$ for the result of contracting the edge $e$ in $G.$} By using the more general result  in \autoref{thm_thetadp},  the proof of the second part of~\autoref{thm_grammar} can be easily adapted, using the expressive potential of the ${\sf dp}_{k}$ predicates,  so to work with this enhanced version of ${\cal M}$ by using one extra annotation set in order to mark the edges under contraction as ``specially colored'' vertices of $G'.$ We believe that using analogous enhancements it is possible to deal with other type of (not necessarily local) modification operations (see~\cite{FominGT19modi,GolovachST22model} for some previous steps in this direction).


\myskip\myskip
\myskip\paragraph{Further than connectivity closure.}

One of the key operations defining $Θ$
is the {connectivity extension} operation, that is, given a sentence $φ,$ to consider the (conjunctive) sentence $φ^{({\sf c})}.$ We incorporated this operation to our logic in order to express elimination distance modifications (such as those of tree-depth~\cite{BulianD17fixe} and bridge-depth~\cite{BougeretJS20bridg}) where, at each step, we remove some tree-like structure and then we apply the
current target sentence
to the connected components of the remaining graph.
In~\cite{DinerGT21block}, the notion of {\sl block elimination distance} has been
introduced, where the target property is applied to the biconnected components of the remaining graph (instead of the connected components). We are confident that our results can be adapted so to include the
biconnectivity extension -- or even the 3-conectivity extension, as defined by Tutte's decomposition~\cite{Tutte61athe}. However, we prefer to avoid this here as it would add undesirable burden
to the statement
of our results (and to the proofs as well).
Another direction is to consider different versions of $φ^{({\sf c})}.$ One of them might be a {\sl disjunctive} version, namely $φ^{\vee({\sf c})},$ where $G\models φ^{\vee({\sf c})}$ if
{\em at least one} of the connected  components of $G$
is a model of $φ.$ Another one is a {\sl selective} version, namely $φ^{\exists({\sf c})},$ where $G\models φ^{\wedge({\sf c})}$ if there is some subset of the connected  components of $G$ whose   union
is a model of $φ.$ Our proof fails if we wish to incorporate any of these two variants of $φ^{({\sf c})}$ in $Θ.$ However, it can be easily adapted so to  incorporate $φ^{\vee({\sf c})}$ in $\tilde{Θ}.$

\myskip\myskip
\myskip\paragraph{Descriptive complexity and the  $Θ$-hierarchy.}

Recall that $Θ=\bigcup_{i∈ \mathbb{N}}Θ_i,$
where each level of the sentence set $Θ_{i}$  is defined by adding an extra modulator sentence,  followed by some positive
Boolean combination of the connectivity closure
of the lower level. We extended our result from $Θ_1$ to every $Θ_{i}$
because $Θ$ is quite versatile and makes it easier to express more complex hierarchical modification problems, as we did in~\autoref{thm_grammar}.
However, it is an open problem whether this hierarchy is proper with respect to the descriptive complexity of the problems that it defines in each of its levels.
In simple cases where the modulator sentence asks for a set of bounded size,
and under the absence of positive Boolean combinations, it is
possible to express any $Θ$-definable problem using $Θ_{1}.$
For instance, elimination ordering to some
$Θ_{0}$-definable class can be straightforwardly expressed in $Θ,$ however with a  more technical proof one can also express it  in $Θ_1$ (see~\cite{FominGT22param}).
Is this collapse maintained
when we consider the full expressive power of $Θ$? We conjecture a negative answer to this question
for both $Θ$ and its extension $Θ^{\sf DP}$.

\myskip\myskip

\myskip\paragraph{Constructibility further than bounding treewidth.}

\autoref{@proisionalmente} extends the constructibility horizon of Robertson-Seymour's  theorem~\cite{RobertsonS04GMXX} to ${\cal B}\tritri{\cal G}$
when both ${\cal B}$ and ${\cal G}$ are  {minor-closed} and, moreover,  $\obs({\cal B})$ contains some planar graph (see \autoref{@insuperables}). Does this constructibility result still hold when we drop this latter planarity restriction? We are not in position to conjecture positively or negatively on this.

\bigskip

\noindent{\bf Acknowledgements.} We wish to thank \midnightblue{Stavros G. Kolliopoulos} and
\midnightblue{Christophe Paul} 
for their valuable remarks on earlier versions of this paper.

\newpage

{
\addcontentsline{toc}{section}{References}
}

\newpage

\appendix
\myskip\section{Flat walls framework}\label{@consideracions}

Here we present the framework on flat walls that was introduced in~\cite{SauST21amor}.
In~\autoref{label_administrando} we give some additional basic definitions and in~\autoref{label_domesticated} we define walls, subwalls, and other notions related to walls.
Next, in~\autoref{label_inappropriate}, we give the definitions of renditions and paintings, that are used in \autoref{label_exceptionalness} to define flatness pairs.
In~\autoref{label_exceptionalness}, apart from the definition of flatness pairs, we present notions like {\sl influence}, {\sl regularity}, and {\sl tilts}.
Then,
in~\autoref{sec_flt_b_comp}, we state~\autoref{label_proletarians} that is a critical
ingredient of our algorithm of~\autoref{@decendientes} in~\autoref{sec_gen_algo}.
In~\autoref{sec_homogeneouswalls} we present the definition of {\sl homogeneous flatness pairs} and state~\autoref{label_highlighting}, that is also important for the proof of~\autoref{@desmembramientos}.
Finally, in~\autoref{sec_canonical}, we give the definition of a canonical partition of a wall that allow us to define the bidimensionality of a set with respect to a flatness pair in~\autoref{sec_bidimensionality}.

\myskip\subsection{Basic definitions}\label{label_administrando}


Given a graph $G,$ we define the {\em detail} of $G,$ denoted by ${\sf detail}(G),$ to be the maximum among $|E(G)|$ and $|V(G)|.$
Given a finite collection ${\cal F}$ of graphs, we set $\ell_{\cal F}= \max\{{\sf detail}(H)\mid H∈ {\cal F}\}.$
\myskip\paragraph{Dissolutions and subdivisions.}
Given a vertex $v∈ V(G)$ of degree two with neighbors $u$ and $w,$ we define the {\em dissolution} of $v$
to be the operation of deleting $v$ and, if $u$ and $w$ are not adjacent, adding the edge $\{u,w\}.$
Given two graphs $H,G,$ we say that $H$ is a {\em dissolution} of $G$
if $H$ can be obtained from $G$ after dissolving vertices of $G.$
Given an edge $e=\{u,v\}∈ E(G),$ we define the {\em subdivision} of $e$
to be the operation of deleting $e,$ adding a new vertex $w$ and making it adjacent to $u$ and $v.$
Given two graphs $H,G,$ we say that $H$ is a {\em subdivision} of $G$
if $H$ can be obtained from $G$ after subdividing edges of $G.$

\myskip\paragraph{Contractions and minors.}
A graph $G'$ is a \emph{contraction} of a graph $G,$
if $G'$ can be obtained from $G$ by a sequence of edge contractions.
Given two graphs $H,G,$ if $H$ is a minor of $G$ then for every vertex $v∈ V(H)$ there is
a set of vertices in $G$ that are the endpoints of the edges of $G$ contracted towards creating $v.$
We call this set {\em model} of $v$ in $G.$

\myskip\subsection{Walls and subwalls}\label{label_domesticated}

\myskip\paragraph{Walls.}\label{label_mistreatment}
Let  $k,r∈\mathbb{N}.$ The
\emph{$(k\times r)$-grid} is the
graph whose vertex set is $[k]\times[r]$ and two vertices $(i,j)$ and $(i',j')$ are adjacent if and only if $|i-i'|+|j-j'|=1.$
An  \emph{elementary $r$-wall}, for some odd integer $r≥ 3,$ is the graph obtained from a
$(2 r\times r)$-grid
with vertices $(x,y)
	∈[2r]\times[r],$
after the removal of the
``vertical'' edges $\{(x,y),(x,y+1)\}$ for odd $x+y,$ and then the removal of
all vertices of degree one.
Notice that, as $r≥ 3,$  an elementary $r$-wall is a planar graph
that has a unique (up to topological isomorphism) embedding in the plane $\mathbb{R}^{2}$
such that all its finite faces are incident to exactly six
edges.
The {\em perimeter} of an elementary $r$-wall is the cycle bounding its infinite face,
while the cycles bounding its finite faces are called {\em bricks}.
Also, the vertices
in the perimeter of an elementary $r$-wall that have degree two are called {\em pegs},
while the vertices $(1,1), (2,r), (2r-1,1), (2r,r)$ are called {\em corners} (notice that the corners are also pegs).

An {\em $r$-wall} is any graph $W$ obtained from an elementary $r$-wall $\bar{W}$
after subdividing edges. A graph $W$ is a {\em wall} if it is an $r$-wall for some odd $r≥ 3$
and we refer to $r$ as the {\em height} of $W.$ Given a graph $G,$
a {\em wall of} $G$ is a subgraph of $G$ that is a wall.
We insist that, for every $r$-wall, the number $r$ is always odd.

We call the vertices of degree three of a wall $W$ {\em 3-branch vertices}.
A cycle of $W$ is a {\em brick} (resp. the {\em perimeter}) of $W$
if its 3-branch vertices are the vertices of a brick (resp. the perimeter) of $\bar{W}.$
We denote by ${\cal C}(W)$ the set of all cycles of $W.$
We  use $D(W)$ in order to denote the perimeter of the  wall $W.$
A brick of $W$ is {\em internal} if it is disjoint from $D(W).$

\myskip\paragraph{Subwalls.}
Given an elementary $r$-wall $\bar{W},$ some odd $i∈ \{1,3,\ldots,2r-1\},$ and $i'=(i+1)/2,$
the {\em $i'$-th  vertical path} of $\bar{W}$  is the one whose
vertices, in order of appearance, are $(i,1),(i,2),(i+1,2),(i+1,3),
	(i,3),(i,4),(i+1,4),(i+1,5),
	(i,5),\ldots,(i,r-2),(i,r-1),(i+1,r-1),(i+1,r).$
Also, given some $j∈[2,r-1]$ the {\em $j$-th horizontal path} of $\bar{W}$
is the one whose
vertices, in order of appearance, are $(1,j),(2,j),\ldots,(2r,j).$

A \emph{vertical} (resp. \emph{horizontal}) path of $W$ is one
that is a subdivision of a  vertical (resp. horizontal) path of $\bar{W}.$
Notice that the perimeter of an $r$-wall $W$
is uniquely defined regardless of the choice of the elementary $r$-wall $\bar{W}.$
A {\em subwall} of $W$ is any subgraph $W'$ of  $W$
that is an $r'$-wall, with $r' ≤ r,$ and such the vertical (resp. horizontal) paths of $W'$ are subpaths of the
	{vertical} (resp. {horizontal}) paths of $W.$

\myskip\paragraph{Layers.}
The {\em layers} of an $r$-wall $W$  are recursively defined as follows.
The first layer of $W$ is its perimeter.
For $i=2,\ldots,(r-1)/2,$ the $i$-th layer of $W$ is the $(i-1)$-th layer of the subwall $W'$
obtained from $W$ after removing from $W$ its perimeter and
removing recursively all occurring vertices of degree one.
We refer to the $(r-1)/2$-th layer as the {\em inner layer} of $W.$
The {\em central vertices} of an $r$-wall $W$ are its two branch vertices that do not belong to any of its layers
and are connected by a path of $W$  that does not intersect any layers of $W$.

\myskip\paragraph{Central walls.}
Given an $r$-wall $W$ and an odd $q∈\mathbb{N}_{≥ 3}$ where $q≤ r,$
we define the {\em central $q$-subwall} of $W,$ denoted by $W^{(q)},$
to be the $q$-wall obtained from $W$ after removing
its first $(r-q)/2$ layers and all occurring vertices of degree one.

\myskip\paragraph{Tilts.}
The {\em interior} of a wall $W$ is the graph obtained
from $W$ if we remove from it all edges of $D(W)$ and all vertices
of $D(W)$ that have degree two in $W.$
Given two walls $W$ and $\tilde{W}$ of a graph $G,$
we say that $\tilde{W}$ is a {\em tilt} of $W$ if $\tilde{W}$ and $W$ have identical interiors.

\myskip\subsection{Paintings and renditions}
\label{label_inappropriate}
In this subsection we present the notions of renditions and paintings, originating in the work of Robertson and Seymour \cite{RobertsonS95XIII}.
The definitions presented here were introduced by Kawarabayashi, Thomas, and Wollan \cite{KawarabayashiTW18anew} (see also~\cite{SauST21amor}).
\myskip\paragraph{Paintings.}
A {\em closed} (resp. {\em open}) {\em disk} is a set homeomorphic to the set
$\{(x,y)∈ \mathbb{R}^{2}\mid x^{2}+y^{2}≤ 1\}$ (resp. $\{(x,y)∈ \mathbb{R}^{2}\mid x^{2}+y^{2}< 1\}$).
Let $\Delta$ be a closed disk.
{Given a subset $X$ of $\Delta,$ we
denote its closure by $\bar{X}$ and its boundary by $\bd(X).$}
A {\em {$\Delta$}-painting} is a pair $Γ=(U,N)$
where
\begin{itemize}
	\item  $N$ is a finite set of points of $\Delta,$
	\item $N \subseteq U \subseteq \Delta,$ and
	\item $U \setminus  N$ has finitely many arcwise-connected  components, called {\em cells}, where, for every cell $c,$
	      \begin{itemize}
		      \item[$\circ$] the closure $\bar{c}$ of $c$
		            is a closed disk
		            and
		      \item[$\circ$]  $|\tilde{c}|≤ 3,$ where $\tilde{c}:=\bd(c)\cap N.$
	      \end{itemize}
\end{itemize}
We use the  notation $U(Γ) := U,$
$N(Γ) := N$  and denote the set of cells of $Γ$
by $C(Γ).$
For convenience, we may assume that each cell  of $Γ$ is an open disk of $\Delta.$
Notice that, given a $\Delta$-painting $Γ,$
the pair $(N(Γ),\{\tilde{c}\mid c∈ C(Γ)\})$  is a hypergraph whose hyperedges have cardinality at most three and  $Γ$ can be seen as a plane embedding of this hypergraph in $\Delta.$

\myskip\paragraph{Renditions.}
Let $G$ be a graph and let $\Omega$ be a cyclic permutation of a subset of $V(G)$ that we denote by $V(\Omega).$ By an {\em $\Omega$-rendition} of $G$ we mean a triple $(Γ, σ, π),$ where
\begin{itemize}
	\item[(a)] $Γ$ is a $\Delta$-painting for some closed disk $\Delta,$
	\item[(b)] $π: N(Γ)\to V(G)$ is an injection, and
	\item[(c)] $σ$ assigns to each cell $c ∈  C(Γ)$ a subgraph $σ(c)$ of $G,$ such that
	      \begin{enumerate}
		      \item[(1)] $G=\bigcup_{c∈ C(Γ)}σ(c),$
		      \item[(2)]  for distinct $c, c' ∈  C(Γ),$  $σ(c)$ and $σ(c')$  are edge-disjoint,
		      \item[(3)] for every cell $c ∈  C(Γ),$ $π(\tilde{c}) \subseteq V (σ(c)),$
		      \item[(4)]  for every cell $c ∈  C(Γ),$
		            $V(σ(c)) \cap \bigcup_{c' ∈  C(Γ) \setminus  \{c\}}V(σ(c')) \subseteq π(\tilde{c}),$ and
		      \item[(5)]  $π(N(Γ)\cap \bd(\Delta))=V(\Omega),$ such that the points
		            in $N(Γ)\cap \bd(\Delta)$ appear in $\bd(\Delta)$ in the same ordering
		            as their images, via $π,$ in $\Omega.$
	      \end{enumerate}
\end{itemize}

\myskip\subsection{Flatness pairs}
\label{label_exceptionalness}
In this subsection we define the notion of a flat wall, originating in the work of Robertson and Seymour \cite{RobertsonS95XIII} and later used in \cite{KawarabayashiTW18anew}.
Here, we define flat walls as in \cite{SauST21amor}.

\myskip\paragraph{Flat walls.}
Let $G$ be a graph and let $W$ be an $r$-wall  of $G,$ for some odd integer $r≥ 3.$
We say that a pair $(P,C)\subseteq D(W)\times D(W)$ is a {\em choice
		of pegs and corners for $W$} if $W$ is the subdivision of an  elementary $r$-wall $\bar{W}$
where $P$ and
$C$ are the pegs and the corners of $\bar{W},$ respectively (clearly, $C\subseteq P$).
To get more intuition, notice that a wall $W$ can occur in several ways from the elementary wall $\bar{W},$
depending on the way the vertices in the perimeter of $\bar{W}$ are subdivided.
Each of them gives a different selection $(P,C)$ of pegs and corners of $W.$

We say that $W$ is a {\em flat $r$-wall}
of $G$ if there is a separation $(X,Y)$ of $G$ and a choice  $(P,C)$
of pegs and corners for $W$ such that:
\begin{itemize}
	\item $V(W)\subseteq Y,$
	\item  $P\subseteq X\cap Y\subseteq V(D(W)),$ and
	\item  if $\Omega$ is the cyclic ordering of the vertices $X\cap Y$ as they appear in $D(W),$
	      then there exists an $\Omega$-rendition $(Γ,σ,π)$ of  $G[Y].$
\end{itemize}
We say that $W$ is a {\em flat wall}
of $G$ if it is a flat $r$-wall for some odd integer $r ≥ 3.$
%

\myskip\paragraph{Flatness pairs.}
Given the above, we  say that  the choice of the 7-tuple $\mathfrak{R}=(X,Y,P,C,Γ,σ,π)$
{\em certifies that $W$ is a flat wall of $G$}.
We call the pair $(W,\mathfrak{R})$ a {\em flatness pair} of $G$ and define
the {\em height} of the pair $(W,\mathfrak{R})$ to be the height of $W.$
We use the term {\em cell of} $\mathfrak{R}$ in order to refer to the cells of $Γ.$

We call the graph $G[Y]$ the {\em $\mathfrak{R}$-compass} of $W$ in $G,$
denoted by ${\sf compass}_{\mathfrak{R}}(W).$
It is easy to see that there is a connected component of ${\sf compass}_{\mathfrak{R}}(W)$ that contains the wall $W$ as a subgraph.
We can assume that ${\sf compass}_{\frR} (W)$ is connected, updating $\frR$ by removing from $Y$ the vertices of all the connected components of ${\sf compass}_\frR (W)$
except of the one that contains $W$ and including them in $X$ ($Γ, σ, π$ can also be easily modified according to the removal of the aforementioned vertices from $Y$).
We define the  {\em flaps} of the wall $W$ in $\mathfrak{R}$ as
${\sf flaps}_{\mathfrak{R}}(W):=\{σ(c)\mid c∈ C(Γ)\}.$
Given a flap $F∈ {\sf flaps}_{\mathfrak{R}}(W),$ we define its {\em base}
as $\partial F:=V(F)\cap π(N(Γ)).$
A  cell $c$ of ${\frR}$ is {\em untidy} if  $π(\tilde{c})$ contains a vertex
$x$ of ${W}$ such that two of the edges of ${W}$ that are incident to $x$ are edges of $σ(c).$ Notice that if $c$ is untidy then  $|\tilde{c}|=3.$
A cell $c$ of $\frR$ is {\em tidy} if it is not untidy.
The notion of tidy/untidy cell as well as the notions that we present in the rest of this subsection have been introduced in~\cite{SauST21amor}.

\myskip\paragraph{Cell classification.}
Given a cycle $C$ of ${\sf compass}_{\mathfrak{R}}(W),$ we say that
$C$ is {\em $\mathfrak{R}$-normal} if it is {\sl not} a subgraph of a flap $F∈ {\sf flaps}_{\mathfrak{R}}(W).$
Given an $\mathfrak{R}$-normal cycle $C$ of ${\sf compass}_{\mathfrak{R}}(W),$
we call a cell $c$ of $\mathfrak{R}$ {\em $C$-perimetric} if
$σ(c)$ contains some edge of $C.$
Since every $C$-perimetric cell $c$ contains some edge of $C$ and $|\partialσ(c)|≤ 3,$ we observe the following.
\begin{observation}\label{label_grundfalscher}
	For every pair $(C,C')$ of $\frR$-normal cycles of ${\sf compass}_{\frR} (W)$ such that $V(C)\cap V(C')=\emptyset,$ there is no cell of $\frR$ that is both $C$-perimetric and $C'$-perimetric.
\end{observation}
Notice that if $c$ is $C$-perimetric, then $π(\tilde{c})$ contains two points $p,q∈ N(Γ)$
such that  $π(p)$ and $π(q)$ are vertices of $C$ where one,
say $P_{c}^{\rm in},$ of the two $(π(p),π(q))$-subpaths of $C$ is a subgraph of $σ(c)$ and the other,
denoted by $P_{c}^{\rm out},$  $(π(p),π(q))$-subpath contains at most one internal vertex of $σ(c),$
which should be the (unique) vertex $z$ in $\partialσ(c)\setminus\{π(p),π(q)\}.$
We pick a $(p,q)$-arc $A_{c}$ in $\hat{c}:={c}\cup\tilde{c}$ such that  $π^{-1}(z)∈ A_{c}$ if and only if $P_{c}^{\rm in}$ contains
the vertex $z$ as an internal vertex.

We consider the circle  $K_{C}=\cupall\{A_{c}\mid \mbox{$c$ is a $C$-perimetric cell of $\mathfrak{R}$}\}$
and we denote by $\Delta_{C}$ the closed disk bounded by $K_{C}$  that is contained in  $\Delta.$
A cell $c$ of $\mathfrak{R}$ is called {\em $C$-internal} if $c\subseteq \Delta_{C}$
and is called {\em $C$-external} if $\Delta_{C}\cap c=\emptyset.$
Notice that  the cells of $\mathfrak{R}$ are partitioned into  $C$-internal,  $C$-perimetric, and  $C$-external cells.

Let $c$ be a tidy $C$-perimetric cell of $\mathfrak{R}$ where $|\tilde{c}|=3.$ Notice that $c\setminus A_{c}$ has two arcwise-connected components and one of them is an open disk $D_{c}$ that is a subset of $\Delta_{C}.$
If the closure $\overline{D}_{c}$  of $D_{c}$ contains only two points of $\tilde{c}$ then we call the cell $c$ {\em $C$-marginal}.
See~\autoref{label_exhalaciones} for a figure illustrating the above notions.
We refer the reader to \cite{SauST21amor} for more figures.

\begin{figure}[ht]
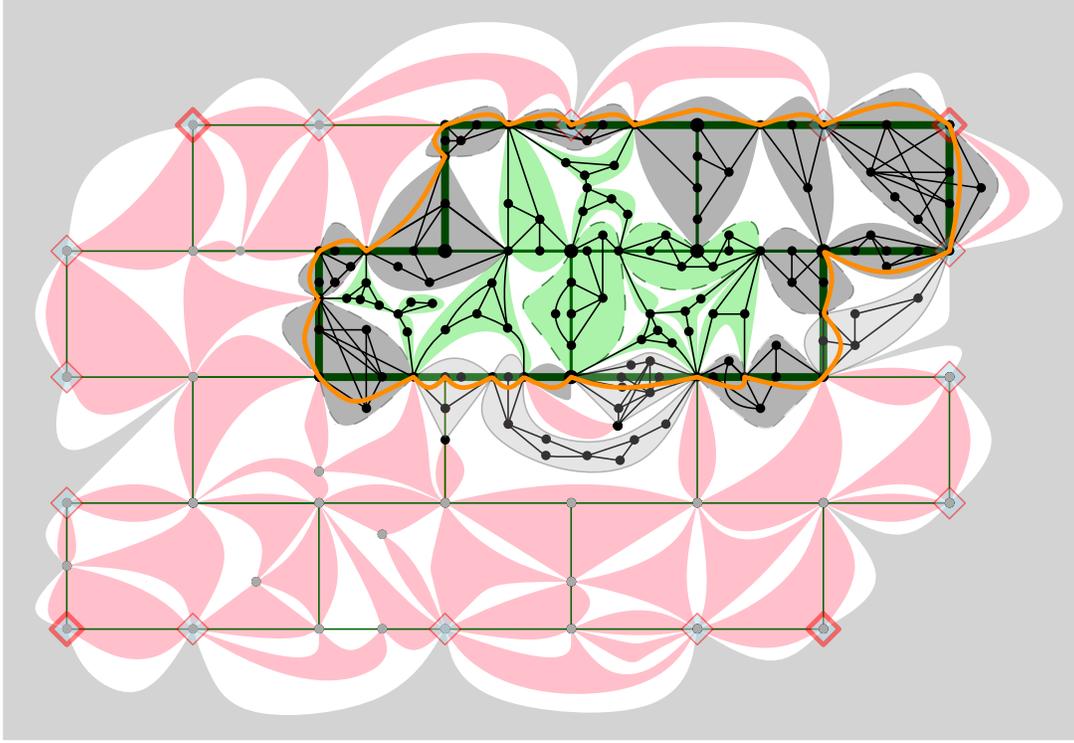

	\hspace{-4.6cm}\scalebox{1.485}{
\tikzstyle{ipe stylesheet} = [
 ipe import,
 even odd rule,
 line join=round,
 line cap=butt,
 ipe pen normal/.style={line width=0.4},
 ipe pen heavier/.style={line width=0.8},
 ipe pen fat/.style={line width=1.2},
 ipe pen ultrafat/.style={line width=2},
 ipe pen normal,
 ipe mark normal/.style={ipe mark scale=3},
 ipe mark large/.style={ipe mark scale=5},
 ipe mark small/.style={ipe mark scale=2},
 ipe mark tiny/.style={ipe mark scale=1.1},
 ipe mark normal,
 /pgf/arrow keys/.cd,
 ipe arrow normal/.style={scale=7},
 ipe arrow large/.style={scale=10},
 ipe arrow small/.style={scale=5},
 ipe arrow tiny/.style={scale=3},
 ipe arrow normal,
 /tikz/.cd,
 ipe arrows, 
 <->/.tip = ipe normal,
 ipe dash normal/.style={dash pattern=},
 ipe dash dotted/.style={dash pattern=on 1bp off 3bp},
 ipe dash dashed/.style={dash pattern=on 4bp off 4bp},
 ipe dash dash dotted/.style={dash pattern=on 4bp off 2bp on 1bp off 2bp},
 ipe dash dash dot dotted/.style={dash pattern=on 4bp off 2bp on 1bp off 2bp on 1bp off 2bp},
 ipe dash normal,
 ipe node/.append style={font=\normalsize},
 ipe stretch normal/.style={ipe node stretch=1},
 ipe stretch normal,
 ipe opacity 10/.style={opacity=0.1},
 ipe opacity 30/.style={opacity=0.3},
 ipe opacity 50/.style={opacity=0.5},
 ipe opacity 75/.style={opacity=0.75},
 ipe opacity opaque/.style={opacity=1},
 ipe opacity opaque,
]
\definecolor{red}{rgb}{1,0,0}
\definecolor{blue}{rgb}{0,0,1}
\definecolor{green}{rgb}{0,1,0}
\definecolor{yellow}{rgb}{1,1,0}
\definecolor{orange}{rgb}{1,0.647,0}
\definecolor{gold}{rgb}{1,0.843,0}
\definecolor{purple}{rgb}{0.627,0.125,0.941}
\definecolor{gray}{rgb}{0.745,0.745,0.745}
\definecolor{brown}{rgb}{0.647,0.165,0.165}
\definecolor{navy}{rgb}{0,0,0.502}
\definecolor{pink}{rgb}{1,0.753,0.796}
\definecolor{seagreen}{rgb}{0.18,0.545,0.341}
\definecolor{turquoise}{rgb}{0.251,0.878,0.816}
\definecolor{violet}{rgb}{0.933,0.51,0.933}
\definecolor{darkblue}{rgb}{0,0,0.545}
\definecolor{darkcyan}{rgb}{0,0.545,0.545}
\definecolor{darkgray}{rgb}{0.663,0.663,0.663}
\definecolor{darkgreen}{rgb}{0,0.392,0}
\definecolor{darkmagenta}{rgb}{0.545,0,0.545}
\definecolor{darkorange}{rgb}{1,0.549,0}
\definecolor{darkred}{rgb}{0.545,0,0}
\definecolor{lightblue}{rgb}{0.678,0.847,0.902}
\definecolor{lightcyan}{rgb}{0.878,1,1}
\definecolor{lightgray}{rgb}{0.827,0.827,0.827}
\definecolor{lightgreen}{rgb}{0.565,0.933,0.565}
\definecolor{lightyellow}{rgb}{1,1,0.878}
\definecolor{black}{rgb}{0,0,0}
\definecolor{white}{rgb}{1,1,1}

}
	\caption{This picture is taken from~\cite{SauST21amor}.
	It depicts a flat wall $W$ in a graph $G,$ the painting of a rendition $\mathfrak{R}$ certifying its flatness, a subwall $W'$ of $W,$ of height three, which is
		$\mathfrak{R}$-normal, and the $\mathfrak{R}$-flaps of $W,$ that correspond to  either $W'$-perimetric (depicted in grey) or $W'$-internal cells (depicted in green).
		The circle $K_{W'}$ is the fat orange cycle. The $W'$-marginal cells are depicted in light grey and the untidy cells are those with dashed boundary.
	}
	\label{label_exhalaciones}
\end{figure}

\myskip\paragraph{Influence.}
For every $\mathfrak{R}$-normal cycle $C$ of ${\sf compass}_{\mathfrak{R}}(W)$ we define the set
$${\sf influence}_{\mathfrak{R}}(C)=\{σ(c)\mid \mbox{$c$ is a cell of $\mathfrak{R}$ that is not $C$-external}\}.$$

A wall $W'$  of ${\sf compass}_{\mathfrak{R}}(W)$  is \emph{$\mathfrak{R}$-normal} if $D(W')$ is $\mathfrak{R}$-normal.
Notice that every wall of $W$ (and hence every subwall of $W$) is an $\mathfrak{R}$-normal wall of ${\sf compass}_{\mathfrak{R}}(W).$ We denote by ${\cal S}_{\mathfrak{R}}(W)$ the set of all $\mathfrak{R}$-normal walls of ${\sf compass}_{\mathfrak{R}}(W).$ Given a wall $W'∈ {\cal S}_{\mathfrak{R}}(W)$ and a cell $c$ of $\mathfrak{R},$
we say that $c$ is {\em $W'$-perimetric/internal/external/marginal} if $c$ is  $D(W')$-perimetric/internal/external/marginal, respectively.
We also use $K_{W'},$ $\Delta_{W'},$ ${\sf influence}_{\mathfrak{R}}(W')$ as shortcuts
for $K_{D(W')},$ $\Delta_{D(W')},$ ${\sf influence}_{\mathfrak{R}}(D(W')),$ respectively.

\myskip\paragraph{Regular flatness pairs.}
We call a  flatness pair $(W,\mathfrak{R})$ of a graph $G$ {\em regular}
if none of its cells is $W$-external, $W$-marginal, or untidy.

\myskip\paragraph{Tilts of flatness pairs.}
Let $(W,\mathfrak{R})$ and $(\tilde{W}',\tilde{\mathfrak{R}}')$  be two flatness pairs of a graph $G$
and let $W'∈ {\cal S}_{\mathfrak{R}}(W).$
We assume that ${\mathfrak{R}}=(X,Y,P,C,Γ,σ,π)$
and $\tilde{\mathfrak{R}}'=(X',Y',P',C',Γ',σ',π').$
We say that   $(\tilde{W}',\tilde{\mathfrak{R}}')$   is a {\em $W'$-tilt}
of $(W,\mathfrak{R})$ if
\begin{itemize}
	\item $\tilde{\mathfrak{R}}'$ does not have $\tilde{W}'$-external cells,
	\item  $\tilde{W}'$ is a tilt of $W',$
	\item  the set of $\tilde{W}'$-internal  cells of  $\tilde{\mathfrak{R}}'$ is the same as the set of $W'$-internal
	      cells of ${\mathfrak{R}}$ and their images via $σ'$ and ${σ}$ are also the same,
	\item ${\sf compass}_{\tilde{\mathfrak{R}}'}(\tilde{W}')$ is a subgraph of $\cupall{\sf influence}_{{\mathfrak{R}}}(W'),$ and
	\item if $c$ is a cell in $C(Γ') \setminus C(Γ),$ then $|\tilde{c}| ≤ 2.$
\end{itemize}

The next observation follows from the third item above and the fact that the cells corresponding to flaps
containing a central vertex of $W'$ are all internal (recall that the height of a wall is always at least three).

\begin{observation}\label{label_surreptitiously}
	Let $(W,\frR)$ be a flatness pair of a graph $G$ and $W'∈{\cal S}_{\frR}(W).$
	For every $W'$-tilt $(\tilde{W}',\tilde{\frR}')$ of $(W,\frR),$ the central vertices of $W'$ belong to the vertex set of ${\sf compass}_{\tilde{\frR}'}(\tilde{W}').$
\end{observation}

Also, given a regular flatness pair $(W,\frR)$ of a graph $G$ and a $W'∈ {\cal S}_{\mathfrak{R}}(W),$
for every $W'$-tilt $(\tilde{W}', \tilde{\frR}')$ of $(W,\mathfrak{R}),$ by definition, none of its cells is $\tilde{W}'$-external, $\tilde{W}'$-marginal, or untidy -- thus, $(\tilde{W}', \tilde{\frR}')$ is regular.
Therefore, regularity of a flatness pair is a property that its tilts ``inherit''.

\begin{observation}\label{label_expressionism}
	If $(W,\mathfrak{R})$ is a regular flatness pair of a graph $G,$ then for every $W'∈ {\cal S}_{\mathfrak{R}}(W),$ every $W'$-tilt of $(W,\mathfrak{R})$ is also regular.
\end{observation}

We next present one of the two main results of~\cite{SauST21amor} (see~\cite[Theorem 5]{SauST21amor}).

\begin{proposition}
\label{label_proporcionada}
There exists an algorithm that given a graph $G,$ a flatness pair $({W},{\mathfrak{R}})$ of $G,$ and a wall $W'∈ {\cal S}_{\mathfrak{R}}(W),$ outputs  a  $W'$-tilt of $({W},{\mathfrak{R}})$ in  time ${\cal O}(n+m).$
\end{proposition}

We conclude this subsection with the Flat Wall theorem and, in particular, the version proved by Chuzhoy \cite{Chuzhoy15impr}, restated in our framework (see \cite[Proposition 7]{SauST21amor}).

\begin{proposition}\label{label_aldobrandesco}
	There exist two functions  $\newfun{@donnescamente}:\mathbb{N}\to \mathbb{N}$  and
	$\newfun{@carlovingios}:\mathbb{N}\to \mathbb{N},$ where the images of $\funref{@donnescamente}$ are odd numbers, such that if $r ∈ \mathbb{N}_{≥ 3}$ is an odd integer, $t∈\mathbb{N}_{≥ 1},$
	$G$ is a graph that does not contain $K_t$ as a minor,  and  $W$ is an $\funref{@donnescamente}(t)\cdot r$-wall of $G,$
	then there is a set $A\subseteq V(G)$ with $|A|≤ \funref{@carlovingios}(t)$
	and a flatness pair $(\tilde{W}',\tilde{\mathfrak{R}}')$ of $G\setminus A$ of height $r.$
	Moreover, $\funref{@donnescamente}(t)={\cal O}(t^{2})$ and $\funref{@carlovingios}(t)=t-5.$
\end{proposition}

\myskip\subsection{Flat walls with compasses of bounded treewidth}\label{sec_flt_b_comp}

The following result was proved in~\cite[Theorem 8]{SauST21amor}.
It is a version of the Flat Wall theorem, originally proved in~\cite{RobertsonS95XIII}.
The proof in~\cite[Theorem 8]{SauST21amor} is strongly based on
the proof of an improved version of the Flat Wall theorem given by of Kawarabayashi, Thomas, and Wollan~\cite{KawarabayashiTW18anew} (see also~\cite{Chuzhoy15impr,GiannopoulouT13opti}).
%
\begin{proposition}\label{label_proletarians}
	There is a function $\newfun{@interference}:\mathbb{N}\to \mathbb{N}$    and
	an algorithm that receives as  input  a graph $G,$ an odd integer $r≥ 3,$ and a  $t∈\mathbb{N}_{≥ 1},$ and  outputs, in time $2^{{\cal O}_{t}(r^2)}\cdot n,$ one of the following:
	\begin{itemize}
		\item a report  that $K_{t}$ is a minor of $G,$
		\item a tree decomposition of $G$ of width at most $\funref{@interference}(t)\cdot r,$ or
		\item a set $A\subseteq V(G),$  where $|A|≤ \funref{@carlovingios}(t),$ a regular flatness pair $(W,\mathfrak{R})$ of $G\setminus A$ of height $r,$
		      and a tree decomposition of the $\mathfrak{R}$-compass of $W$ of width at most $\funref{@interference}(t)\cdot r.$  (Here $\funref{@carlovingios}(t)$ is the function of \autoref{label_aldobrandesco} and $\funref{@interference}(t)=2^{{\cal O}(t^2 \log t)}.$)
	\end{itemize}
\end{proposition}

\myskip\subsection{Homogeneous walls}
\label{sec_homogeneouswalls}

We first present some definitions on boundaried graphs and folios that will be used to define the notion of homogeneous walls.
Following this, we present some results concerning homogeneous walls that are key ingredients in our proofs.

\myskip\paragraph{Boundaried graphs.}
Let $t∈\mathbb{N}.$
A \emph{$t$-boundaried graph} is a triple $\bound{G} = (G,B,ρ)$ where $G$ is a graph, $B \subseteq V(G),$ $|B| = t,$ and
$ρ : B \to [t]$ is a bijection.
We call $B$ the {\em boundary} of ${\bf G}$ and the vertices of $B$ {\em the boundary vertices} of ${\bf G}.$
For $B'\subseteq B,$ we define the bijection $ρ[B']:B'\to [|B'|]$ such that for every $v∈ B',$ $ρ[B'](v) = |\{u∈ B' \mid ρ(u)≤ ρ(v)\}|.$
Also, for $S\subseteq V(G),$ we denote by ${\bf G}\setminus S$ the $t$-boundaried graph $(G\setminus S, B\setminus S, ρ[B\setminus S]).$
We  say that  $\textbf{G}_1=(G_1,B_1,ρ_1)$ and $\textbf{G}_{2}=(G_2,B_2,ρ_2)$
are {\em isomorphic} if there is an isomorphism from $G_{1}$ to $G_{2}$
that extends the bijection $ρ_{2}^{-1}\circ ρ_{1}.$
The triple $(G,B,ρ)$ is a {\em boundaried graph} if it is a $t$-boundaried graph for some $t∈\mathbb{N}.$
As in~\cite{RobertsonS95XIII} (see also \cite{BasteST20acom}), we define the {\em detail} of a boundaried graph
$\bound{G} = (G,B,ρ)$ as  ${\sf detail}(\bound{G}):=\max\{|E(G)|,|V(G)\setminus B|\}.$
We denote by ${\cal B}^{(t)}$ the set of all (pairwise non-isomorphic)  $t$-boundaried graphs and by ${\cal B}_{\ell}^{(t)}$ the set of all (pairwise non-isomorphic) $t$-boundaried graphs with detail at most $\ell.$
We also set ${\cal B}=\bigcup_{t∈\mathbb{N}}{\cal B}^{(t)}.$
%

\myskip\paragraph{Topological minors of boundaried graphs.}
We say that $(M,T)$   is a {\em {\sf tm}-pair} if $M$ is  a graph, $T\subseteq V(M),$ and  all vertices in
$V(M)\setminus T$ have degree two. We denote by ${\sf diss}(M,T)$ the graph obtained
from  $M$ by {dissolving} all vertices  in $V(M)\setminus T.$
A {\em {\sf tm}-pair} of a graph $G$  is a  {\em {\sf tm}-pair}  $(M,T)$ where $M$ is a subgraph of $G.$
We call the vertices in $T$ {\em branch} vertices of $(M,T).$
We need to deal with topological minors for the notion of homogeneity defined below, on which  the statement of~\cite[Theorem 5.2]{BasteST20acom} relies.

If $\textbf{M}=(M,B,ρ)∈{\cal B}$ and   $T\subseteq V(M)$ with $B\subseteq T,$ we  call  $(\textbf{M},T)$ a {\em {\sf btm}-pair}
and we  define  ${\sf diss}(\textbf{M},T)=({\sf diss}(M, T),B,ρ).$ Note that we do not permit dissolution of boundary vertices, as we consider all of them to be branch vertices. If $\textbf{G}=(G,B,ρ)$ is a boundaried graph and $(M,T)$ is a  {\sf tm}-pair of $G$
where $B\subseteq T,$  then we say that
$(\textbf{M},T),$ where $\textbf{M}=(M,B,ρ),$ is a   {\em {\sf btm}-pair} of $\textbf{G}=(G,B,ρ).$
Let $\textbf{G}_{1},{\bf G}_{2}$ be two boundaried graphs.
We say that $\textbf{G}_{1}$ is a {\em topological minor}
of $\textbf{G}_{2},$ denoted by $\textbf{G}_{1}\pretp\textbf{G}_{2},$ if
$\textbf{G}_{2}$ has a {\sf btm}-pair $(\textbf{M},T)$
such that  ${\sf diss}(\textbf{M},T)$ is isomorphic to $\textbf{G}_{1}.$

\myskip\paragraph{Folios.}
Given a $\textbf{G}∈ {\cal B}$ and a positive integer $\ell,$ we define the {\em $\ell$-folio} of ${\bf G}$
as
$${\ell}\mbox{\sf-folio}(\textbf{G})=\{\textbf{G}'∈ {\cal  B} \mid \textbf{G}'\pretp \textbf{G} \mbox{~and $\textbf{G}'$ has detail at most $\ell$}\}.$$

The number of distinct $\ell$-folios of $t$-boundaried graphs is indicated in the following result, proved first in~\cite{BasteST20acom} and used also in~\cite{BasteST20hitti}.
\begin{proposition}\label{label_objectivement}
	There exists a function $\newfun{label_dogmatizador}: \mathbb{N}^{2} \to \mathbb{N}$ such that for every $t,\ell∈ \mathbb{N},$ $|\{\ell\mbox{\sf-folio}({\bf G}) \mid {\bf G}∈ {\cal B}_{\ell}^{(t)}\}|≤ \funref{label_dogmatizador}(t,\ell).$ Moreover, $\funref{label_dogmatizador}(t,\ell)=2^{2^{{\cal O}((t+\ell)\cdot\log(t+\ell))}}.$\end{proposition}

\myskip\paragraph{Augmented flaps.}
Let $G$ be a graph, $A$ be a subset of $V(G)$ of size $a,$ and $(W,\frR)$ be a flatness pair of $G\setminus A.$
For each flap $F∈ {\sf flaps}_{\mathfrak{R}}(W)$ we consider a labeling  $\ell_{F}: \partial F\rightarrow\{1,2,3\}$ such that
the set of labels assigned by $\ell_{F}$ to $\partial F$ is  one of $\{1\},$ $\{1,2\},$ $\{1,2,3\}.$
Also, let $\tilde{a}∈[a].$
For every set ${\tilde{A}}∈ \binom{A}{\tilde{a}},$ we consider a bijection  $ρ_{\tilde{A}}: \tilde{A}\to [\tilde{a}].$
The labelings in ${\cal L}=\{\ell_{F} \mid F∈  {\sf flaps}_{\mathfrak{R}}(W)\}$ and the labelings in $\{ρ_{\tilde{A}} \mid \tilde{A} ∈ \binom{A}{\tilde{a}}\}$
will be useful for defining a set of boundaried graphs that we will call augmented flaps. We first need some more definitions.

Given a flap $F∈{\sf flaps}_{\mathfrak{R}}(W),$ we define an ordering
$\Omega(F)=(x_{1},\ldots,x_{q}),$ with $q≤ 3,$ of the vertices of $\partial{F}$
so that
\begin{itemize}
	\item $(x_{1},\ldots,x_{q})$ is a  counter-clockwise cyclic ordering of the vertices of $\partial F$ as they appear in the corresponding cell of $C(Γ).$ Notice that this cyclic ordering is significant  only when $|\partial F|=3,$
	      in the sense that $(x_{1},x_{2},x_{3})$ remains invariant under shifting, i.e., $(x_{1},x_{2},x_{3})$ is the same as $ (x_{2},x_{3},x_{1})$ but not  under inversion, i.e.,   $(x_{1},x_{2},x_{3})$ is not the same as $(x_{3},x_{2},x_{1}),$ and
	\item   for $i∈[q],$ $\ell_{F}(x_{i})=i.$
\end{itemize}
Notice that the second condition is necessary for completing the definition of the ordering $\Omega(F),$
and this is the reason why we set up the labelings in ${\cal L}.$\medskip\medskip

For each set $\tilde{A} ∈ \binom{A}{\tilde{a}}$ and each $F∈ {\sf flaps}_{\mathfrak{R}}(W)$ with $t_{F}:=|\partial F|,$
we fix $ρ_{F}: \partial F\to [\tilde{a}+1,\tilde{a}+t_F]$ such that
$(ρ^{-1}_{F}(\tilde{a}+1),\ldots,ρ^{-1}_{F}(\tilde{a}+t_F))= \Omega(F).$
Also, we define the boundaried graph $$\textbf{F}^{\tilde{A}}:=(G[\tilde{A}\cup F],\tilde{A}\cup \partial F,ρ_{\tilde{A}}\cup ρ_F)$$
and we denote by $F^{\tilde{A}}$ the underlying graph of $\textbf{F}^{\tilde{A}}.$ We call $\textbf{F}^{\tilde{A}}$ an {\em $\tilde{A}$-augmented flap} of the flatness pair $(W,\mathfrak{R})$ of $G\setminus A$
in $G.$

\myskip\paragraph{Palettes and homogeneity.}
For each $\frR$-normal cycle $C$ of ${\sf compass}_\frR (W)$ and each set $\tilde{A}∈ 2^A,$ we define $(\tilde{A},\ell)\mbox{\sf -palette}(C)=\{\ell\mbox{\sf -folio}({\bf F}^{\tilde{A}})\mid F∈ {\sf  influence}_{\mathfrak{R}}(C)\}.$
Given a set $\tilde{A}∈ 2^A,$ we say that the flatness pair $(W,\mathfrak{R})$  of $G\setminus A$ is {\em $λ$-homogeneous with respect to $\tilde{A}$} if every  {\sl internal} brick of ${W}$ has the {\sl same} $(\tilde{A},λ)$\mbox{\sf -palette} (seen as a cycle of ${\sf compass}_\frR (W)$).
Also, given a collection ${\cal S}\subseteq 2^A,$ we say that the flatness pair $(W,\mathfrak{R})$  of $G\setminus A$ is {\em $λ$-homogeneous
		with respect to ${\cal S}$}
if it is $λ$-homogeneous with respect to every $\tilde{A}∈ {\cal S}.$

The following observation is a consequence of the fact that, given a wall $W$ and a  subwall $W'$ of $W,$ every internal brick of a tilt $W''$ of $W'$ is also an internal brick of $W.$

\begin{observation}\label{label_convenciones}
	Let $λ∈\mathbb{N},$ $G$ be a graph, $A\subseteq V(G),$ ${\cal S}\subseteq 2^A,$ and $(W,\mathfrak{R})$  be a flatness pair of $G\setminus A.$ If $(W,\mathfrak{R})$ is  $λ$-homogeneous
	with respect to ${\cal S},$ then for every subwall $W'$ of $W,$ every $W'$-tilt of $(W,\mathfrak{R})$ is also $λ$-homogeneous
	with respect to ${\cal S}.$
\end{observation}

\medskip
%

Let $a,λ∈ \mathbb{N}.$
Also, let $G$ be a graph, $A$ be a subset of $V(G)$ of size at most $a,$ and $(W,\frR)$ be a flatness pair of $G\setminus A.$
For every flap  $F∈{\sf flaps}_\frR (W),$ we define the function
${\sf var}^{(A,λ)}_F:2^A\to \{λ\mbox{\sf-folio}({\bf G}) \mid {\bf G}∈ \bigcup_{i∈[|A|+3]}{\cal B}^{(i)}\}$
that maps each set $\tilde{A}∈2^A$ to the set  $λ\mbox{\sf -folio}({\bf F}^{\tilde{A}}).$

{The fact that there are ${\cal O}(2^{a})$ elements in $2^A$ together with
\autoref{label_objectivement} implies the existence of an upper bound to the number of different $λ$-folios of the augmented flaps of $(W,\frR),$ as indicated in the following result.  }

\begin{lemma}\label{label_anticipation}
	There exists a function $\newfun{label_exiieriences}:\mathbb{N}^2\to \mathbb{N}$ such that if $a,λ∈ \mathbb{N},$ $G$ is a graph, $A$ is a subset of $V(G)$ of size at most $a,$ and $(W,\frR)$ is a flatness pair of $G\setminus A,$ then $$|\{{\sf var}^{(A,λ)}_F\mid F∈ {\sf flaps}_\frR (W)\}|≤ \funref{label_exiieriences}(a,λ).$$
	Moreover, $ \funref{label_exiieriences}(a,λ)= 2^{2^{{\cal O}((a+λ )\cdot \log (a+λ))}}.$
\end{lemma}

\autoref{label_anticipation} allows us to define an injective function $σ: \{{\sf var}^{(A,λ)}_F\mid F∈ {\sf flaps}_\frR (W)\} \to [\funref{label_exiieriences}(a,λ)]$
that maps each function in $\{{\sf var}^{(A,λ)}_F\mid F∈ {\sf flaps}_\frR (W)\}$ to an integer in $[\funref{label_exiieriences}(a,λ)].$
Using $σ,$ we define a function $\zeta_{A,λ}:{\sf flaps}_\frR (W)\to [\funref{label_exiieriences}(a,λ)],$ that maps each flap $F∈{\sf flaps}_\frR (W)$ to the integer $σ({\sf var}^{(A,λ)}_F).$
In \cite{SauST21amor}, given a $w∈\mathbb{N},$ the notion of homogeneity is defined with respect to a {\sl flap-coloring $\zeta$ of $(W,\frR)$ with $w$ colors}, that is a function from ${\sf flaps}_\frR (W)$ to $[w].$
This function gives rise to the $\zeta\mbox{\sf -palette}$ of each $\frR$-normal cycle of ${\sf compass}_{\frR} (W)$ which, in turn, is used to define the notion of a {\em $\zeta$-homogeneous} flatness pair.
Hence, using the terminology of \cite{SauST21amor},  $\zeta_{A,λ}$ is a flap-coloring of $(W,\frR)$ with $\funref{label_exiieriences}(a,λ)$ colors, that ``colors''  each flap $F∈ {\sf flaps}_\frR (W)$ by mapping it to the integer $σ({\sf var}^{(A,λ)}_F),$ and the notion of $λ$-homogeneity
with respect to $2^A$ defined here can be alternatively interpreted as $\zeta_{A,λ}$-homogeneity.
The following result, which is the application of~\cite[Lemma 12]{SauST21amor} for the flap-coloring $\zeta_{A,λ},$ provides the conditions that guarantee the existence of a homogeneous flatness pair ``inside'' a given flatness pair of a graph.

\begin{proposition}\label{label_highlighting}
There is a function $\newfun{@poderosamente}:\mathbb{N}^3\to \mathbb{N},$ whose images are odd integers,
	and an algorithm that receives as  input  an odd integer $r≥ 3,$ $a,λ∈ \mathbb{N},$ a graph $G,$ a set $A\subseteq V(G)$ of size at most $a,$ and a flatness pair $(W,\frR)$ of $G\setminus A$ of height $\funref{@poderosamente}(r,a,λ),$
	and outputs
	a flatness pair $(\breve{W},\breve{\frR})$ of $G\setminus A$  of height $r$ that is $λ$-homogeneous with respect to $2^A$ and  is
	a $W'$-tilt of $(W,\frR)$ for some subwall $W'$ of $W.$
	Moreover, $\funref{@poderosamente}(r,a,λ) = {\cal O}(r^{\funref{label_exiieriences}(a,λ)})$ and the algorithm runs in time {$2^{{\cal O}(\funref{label_exiieriences}(a,λ)\cdot r \log r)}\cdot(n+m)$}.
\end{proposition}

\myskip\subsection{Canonical partitions}\label{sec_canonical}

%

\myskip\paragraph{Canonical partitions.}
Let $r≥ 3$ be an odd integer, let $W$ be an $r$-wall, and let $P_{1}, \ldots, P_{r}$ (resp. $L_{1},\ldots, L_{r}$) be its vertical (resp. horizontal) paths.
For every even  (resp. odd) $i∈[2,r-1]$ and every $j∈[2,r-1],$ we define ${A}^{(i,j)}$ to be the  subpath of $P_{i}$ that starts from a vertex of $P_{i}\cap L_{j}$ and finishes at a neighbor of a vertex in $L_{j+1}$ (resp. $L_{j-1}$), such that $P_{i}\cap L_{j}\subseteq A^{(i,j)}$ and $A^{(i,j)}$ does not intersect $L_{j+1}$ (resp. $L_{j-1}$).
Similarly, for every $i,j∈[2,r-1],$ we define $B^{(i,j)}$ to be the subpath of $L_{j}$ that starts from a vertex of $P_{i}\cap L_{j}$ and finishes at a neighbor of a vertex in $P_{i-1},$ such that $P_{i}\cap L_{j}\subseteq A^{(i,j)}$ and $A^{(i,j)}$ does not intersect $P_{i-1}.$

For every  $i,j∈[2,r-1],$ we denote by $Q^{(i,j)}$ the graph $A^{(i,j)}\cup B^{(i,j)}$ and by ${Q_{\rm ext}}$ the graph $W\setminus \bigcup_{i,j∈[2,r-1]} Q_{i,j}.$
Now consider the collection ${\cal Q}=\{Q_{\rm ext}\}\cup\{Q_{i,j}\mid i,j∈[2,r-1]\}$
and observe that the graphs in ${\cal Q}$ are connected subgraphs of $W$ and their vertex sets form a partition of $V(W).$
We call ${\cal Q}$ the {\em canonical partition} of $W.$ Also, we call every $Q_{i,j},$ for $i,j∈[2,r-1],$ an {\em internal bag} of ${\cal Q},$ while we refer to $Q_{\rm ext}$ as the {\em external bag} of ${\cal Q}.$ See \autoref{label_bitternesses} for an illustration of the notions defined above.

\begin{figure}[ht]
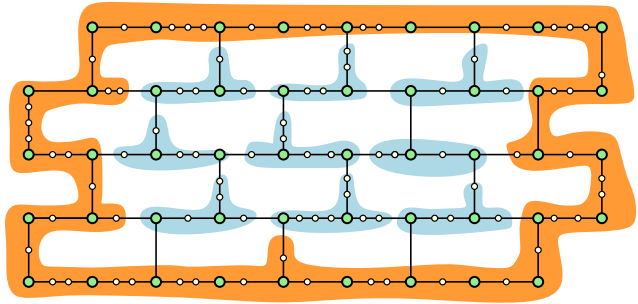

\hspace{1.35cm}\scalebox{0.75}{%
\tikzstyle{ipe stylesheet} = [
  ipe import,
  even odd rule,
  line join=round,
  line cap=butt,
  ipe pen normal/.style={line width=0.4},
  ipe pen heavier/.style={line width=0.8},
  ipe pen fat/.style={line width=1.2},
  ipe pen ultrafat/.style={line width=2},
  ipe pen normal,
  ipe mark normal/.style={ipe mark scale=3},
  ipe mark large/.style={ipe mark scale=5},
  ipe mark small/.style={ipe mark scale=2},
  ipe mark tiny/.style={ipe mark scale=1.1},
  ipe mark normal,
  /pgf/arrow keys/.cd,
  ipe arrow normal/.style={scale=7},
  ipe arrow large/.style={scale=10},
  ipe arrow small/.style={scale=5},
  ipe arrow tiny/.style={scale=3},
  ipe arrow normal,
  /tikz/.cd,
  ipe arrows, 
  <->/.tip = ipe normal,
  ipe dash normal/.style={dash pattern=},
  ipe dash dotted/.style={dash pattern=on 1bp off 3bp},
  ipe dash dashed/.style={dash pattern=on 4bp off 4bp},
  ipe dash dash dotted/.style={dash pattern=on 4bp off 2bp on 1bp off 2bp},
  ipe dash dash dot dotted/.style={dash pattern=on 4bp off 2bp on 1bp off 2bp on 1bp off 2bp},
  ipe dash normal,
  ipe node/.append style={font=\normalsize},
  ipe stretch normal/.style={ipe node stretch=1},
  ipe stretch normal,
  ipe opacity 10/.style={opacity=0.1},
  ipe opacity 30/.style={opacity=0.3},
  ipe opacity 50/.style={opacity=0.5},
  ipe opacity 75/.style={opacity=0.75},
  ipe opacity opaque/.style={opacity=1},
  ipe opacity opaque,
]
\definecolor{red}{rgb}{1,0,0}
\definecolor{blue}{rgb}{0,0,1}
\definecolor{green}{rgb}{0,1,0}
\definecolor{yellow}{rgb}{1,1,0}
\definecolor{orange}{rgb}{1,0.647,0}
\definecolor{gold}{rgb}{1,0.843,0}
\definecolor{purple}{rgb}{0.627,0.125,0.941}
\definecolor{gray}{rgb}{0.745,0.745,0.745}
\definecolor{brown}{rgb}{0.647,0.165,0.165}
\definecolor{navy}{rgb}{0,0,0.502}
\definecolor{pink}{rgb}{1,0.753,0.796}
\definecolor{seagreen}{rgb}{0.18,0.545,0.341}
\definecolor{turquoise}{rgb}{0.251,0.878,0.816}
\definecolor{violet}{rgb}{0.933,0.51,0.933}
\definecolor{darkblue}{rgb}{0,0,0.545}
\definecolor{darkcyan}{rgb}{0,0.545,0.545}
\definecolor{darkgray}{rgb}{0.663,0.663,0.663}
\definecolor{darkgreen}{rgb}{0,0.392,0}
\definecolor{darkmagenta}{rgb}{0.545,0,0.545}
\definecolor{darkorange}{rgb}{1,0.549,0}
\definecolor{darkred}{rgb}{0.545,0,0}
\definecolor{lightblue}{rgb}{0.678,0.847,0.902}
\definecolor{lightcyan}{rgb}{0.878,1,1}
\definecolor{lightgray}{rgb}{0.827,0.827,0.827}
\definecolor{lightgreen}{rgb}{0.565,0.933,0.565}
\definecolor{lightyellow}{rgb}{1,1,0.878}
\definecolor{black}{rgb}{0,0,0}
\definecolor{white}{rgb}{1,1,1}

}
	\caption{\small A $5$-wall and its canonical partition ${\cal Q}.$ The orange bag is the external bag $Q_{\rm ext}.$}
	\label{label_bitternesses}
\end{figure}

Let $(W,\mathfrak{R})$ be a flatness pair of a graph $G.$
Consider the canonical partition ${\cal Q}$ of $W.$ We enhance the graphs of ${\cal Q}$
so to include in them all the vertices of $G$ by applying the following procedure. We set $\tilde{\cal Q}:={\cal Q}$
and, as long as there is a vertex  $x∈ V({\sf compass}_{\mathfrak{R}}(W))\setminus V(\cupall \tilde{\cal Q})$ 
that is adjacent to a vertex of a graph $Q∈ \tilde{\cal Q},$  update $\tilde{\cal Q}:=\tilde{\cal Q}\setminus \{Q\}\cup \{\tilde{Q}\},$ where $\tilde{Q}={\sf compass}_{\mathfrak{R}}(W)[\{x\}\cup V(Q)].$ Since ${\sf compass}_{\frR}(W)$ is a connected graph, in this way we define a partition of the vertices of ${\sf compass}_{\mathfrak{R}}(W)$ into subsets inducing connected graphs.
We call the $\tilde{Q}∈\tilde{\cal Q}$ that contains $Q_{\rm ext}$ as a subgraph the {\em external bag} of $\tilde{\cal Q},$ and we denote it by $\tilde{Q}_{\rm ext},$ while we call {\em internal bags} of $\tilde{\cal Q}$ all graphs in $\tilde{\cal Q}\setminus \{\tilde{Q}_{\rm ext}\}.$
Moreover, we enhance $\tilde{\cal Q}$ by adding all vertices of $G\setminus V({\sf compass}_\frR (W)$ in its external bag, i.e., by updating $\tilde{Q}_{\rm ext}: = G[V(\tilde{Q}_{\rm ext})\cup V(G\setminus V({\sf compass}_\frR (W))].$
We call such a  partition $\tilde{\cal Q}$ a {\em $(W,\frR)$-canonical partition of $G.$}
Notice that a $(W,\frR)$-canonical partition of $G$ is  not unique (since the sets in ${\cal Q}$ can be ``expanded'' arbitrarily when introducing vertex $x$).
We stress that every internal bag of a $(W,\frR)$-canonical partition of $G$ contains vertices of at most three bricks of $W.$

\newpage
\myskip\section{{Details of dealing with \texorpdfstring{$\bar{Θ}$}{bar-Theta}}}
\labels{second_level_more}

Here we present the details of the proof of~\autoref{@desmembramientos} for sentences in $\bar{Θ}.$
In~\autoref{subsec_second_level_algo}
we present the algorithm ${\tt Find\_Equiv\_FlatPairs}$
for sentences in $\bar{Θ}$ that is a genaralization of the algorithm presented in~\autoref{@inhumainement} in the sense that
the characteristic of each flatness pair in the packing is defined as in~\autoref{@enthaltenden}.
Then, in~\autoref{sec_bar_theta_app}, we present the proof of correctness of the algorithm ${\tt Find\_Equiv\_FlatPairs}$
for sentences in $\bar{Θ}.$

\myskip\subsection{The algorithm for sentences in \texorpdfstring{$\bar{Θ}$}{bar-Theta}}
\label{subsec_second_level_algo}
\myskip\paragraph{The algorithm ${\tt Find\_Equiv\_FlatPairs}.$}

The algorithm has four steps.
First, recall that
there exist $p∈ \mathbb{N}_{≥ 1},$ $\ell_1, \ldots, \ell_p, r_1, \ldots, r_p∈ \mathbb{N}_{≥ 1},$ and sentences $\tilde{ζ}_1, \ldots, \tilde{ζ}_p∈ \FOL[τ^{\langle{\bf c}\rangle}\cup\{{\sf R}\}]$ such that $\breve{ζ}_{\sf R}$ is a Boolean combination of $\tilde{ζ}_1, \ldots, \tilde{ζ}_p$ and for every $h∈ [p],$ $\tilde{ζ}_h$ is a basic local sentence with parameters $\ell_h$ and $r_h,$ i.e.,
\begin{eqnarray*}
\tilde{ζ}_h = \exists {\sf x}_{1}\ldots\exists {\sf x}_{\ell_{h}}\big(\bigwedge_{i∈ [\ell_{h}]}{\sf x}_{i}∈ {\sf R}\wedge \bigwedge_{1≤ i<j≤ \ell_{h}} d({\sf x}_{i}, {\sf x}_{j})> 2r_{h}\wedge \bigwedge_{i∈ [\ell_{h}]}ψ_{h}({\sf x}_{i})\big),
\end{eqnarray*}
where $ψ_h$ is an $r_h$-local formula  in $\FOL[τ^{\langle {\bf c}\rangle}]$ with one free variable.

Let $\hat{r}:= \max_{h∈[p]}\{r_h\}$ and $\hat{\ell}:=\max_{h∈ [p]}\{\ell_h\}.$
We set $c$ to be the size of the \FOL-target sentence $σ$ of $θ,$
\begin{align*}
	q &: = {\sf height}(θ)\cdot(\tw(θ)+1)^2+1,\\
	\funref{@riguardavano}(|θ|,\tw(θ),g) & := \max\{q,(g+1)^2+1\},\\
	\funref{@carthaginoise}(|θ|,\tw(θ)) & := q-1,\\
	j' & := g+2\hat{r} +2,\\
	j & :=   {\sf odd}(\max\{q/2,j'\}),\\
	r &: = 2\cdot (\hat{\ell}+ 3)\cdot \hat{r}, \\
	w & := (r+2)\cdot q,\\
	m & := 2^{|{\sf CHAR}|} \cdot q\cdot (\hat{\ell} +3),\text{ and}\\
	\funref{@occidentales}(\hw(θ),\tw(θ),c,l, j')& := \lceil (2w+j)\cdot \sqrt{m}\rceil. \\
\end{align*}

\noindent{\bf Step 1}:
We first find a ``packing'' of subwalls of $W,$ i.e., a collection ${\cal W}$ of $m$ $(2w+j)$-subwalls of $W$ such that their influences are pairwise disjoint.
This collection exists because $W$ has height at least $\funref{@occidentales}(\hw(θ),\tw(θ),|σ|,l, j') = \lceil (2w+j)\cdot \sqrt{m}\rceil$
and because, due to~\autoref{label_grundfalscher}, for every distinct $W_i, W_j∈ {\cal W},$ there are no cells of $\mathfrak{R}$ that are both $W_i$-perimetric and $W_j$-perimetric.
This can be done in time ${\cal O}(n)$.\medskip

\noindent{\bf Step 2}:
Then, for every wall $W_i ∈ {\cal W},$ we compute a $W_i$-tilt of $(W,\mathfrak{R}),$ which we denote by $(\tilde{W}_i, \tilde{\mathfrak{R}}_i),$ and we consider the collection $\tilde{\cal W} := \{(\tilde{W}_i, \tilde{\mathfrak{R}}_i)\mid W_i ∈ {\cal W}\}$ of $m$ flatness pairs of $G_{\mathfrak{A}}\setminus V({\bf a})$ of height $2w+j.$
Note that $\tilde{\cal W}$ can be computed in time ${\cal O}(n),$ due to~\autoref{label_proporcionada}.
\medskip

\noindent{\bf Step 3}:
For every $i∈[m],$ let
$K_i := {\sf compass}_{\tilde{\mathfrak{R}}_i}(\tilde{W}_i),$
$K_i^{{\bf a}} := G_{\mathfrak{A}}[V({\bf a})\cup V(K_i)],$ and
${\bf W}_q^{(i)}$ be the $q$-pseudogrid defined by the horizontal and the vertical paths of the central $q$-subwall of $\tilde{W}_i.$
Also, for every $d∈ [w],$ let $I_i^{(d)} : = V(\cupall {\sf influence}_{\tilde{\mathfrak{R}}_i} (W_i^{(2d+j')}))$ and let ${\bf I}_i:=(I_i^{(1)},\ldots,I_i^{(w)}).$
Let $\mathfrak{K}_i := (\mathfrak{A}[V(K_i^{\bf a})],{\bf a}, {\bf I}_i, {\bf W}^{(i)}_{{q}})$ be the extended compass of $(\tilde{W}_i, \tilde{\mathfrak{R}}_i)$ in $G_{\mathfrak{A}}\setminus V({\bf a}),$ $R_i := R\cap V(K_i^{\bf a}),$ and observe that for every $i,j∈[m],$ $R_i \cap R_j = R\cap V({\bf a}).$
After defining the above collection $\{(\mathfrak{K}_1, R_1),\ldots, (\mathfrak{K}_m,R_m)\}$ of extended compasses of flatness pairs of $G_{\mathfrak{A}}\setminus V({\bf a}),$ we compute their characteristics:
Since, by the hypothesis of the lemma, $K_i, i∈[m]$ has treewidth at most $z,$ by Courcelle's theorem (\autoref{@originallypublishedin}), $θ\text{-}{\sf char}(\mathfrak{K}_i,R_i)$ can be computed in time ${\cal O}_{|θ|} (n).$
We say that two flatness pairs $(\tilde{W}_i, \tilde{\mathfrak{R}}_i), (\tilde{W}_j, \tilde{\mathfrak{R}}_j)∈ \tilde{\cal W}$ are {\em $θ$-equivalent} if $θ\text{-}{\sf char}(\mathfrak{K}_i,R_i)= θ\text{-}{\sf char}(\mathfrak{K}_j,R_j).$
\medskip

\noindent{\bf Step 4}:
Since $m = 2^{|{\sf CHAR}|} \cdot q\cdot (\hat{\ell} +3)$ and for every $i∈ [m],$ $θ\text{-}{\sf char}(\mathfrak{K}_i, R_i)\subseteq {\sf CHAR},$
we can find a collection $\tilde{\cal W}'\subseteq \tilde{\cal W}$ of pairwise
 $θ$-equivalent flatness pairs
such that
$|\tilde{\cal W}'| = q\cdot (\hat{\ell} +3).$
Without loss of generality, we assume that $(\tilde{W}_{1}, \tilde{\mathfrak{R}}_{1})∈ \tilde{\cal W}'.$
We set $\breve{W}$ to be the central $j'$-subwall of $\tilde{W}_1,$ $W^\bullet$ to be the central $g$-subwall of $\tilde{W}_1,$ and keep in mind that $j'=g+2\hat{r}+2.$
Note that $\breve{W}$ (resp. $W^\bullet$) is also the central $j'$-subwall (resp. $g$-subwall) of $W_1$ and, therefore, it is a subwall of $W$ of height $j'$ (resp. $g$).
Again, using~\autoref{label_proporcionada}, we compute, in time ${\cal O}(n),$
a $\breve{W}$-tilt
$(\breve{W}',\breve{\mathfrak{R}}')$
of $(W,\mathfrak{R})$
and
a $W^\bullet$-tilt
$(\tilde{W}',\tilde{\mathfrak{R}}')$
of $(W,\mathfrak{R}).$
We set $Y=V({\sf compass}(\breve{W}',\breve{\mathfrak{R}}')).$
We output the set $Y$ and the flatness pair
$(\tilde{W}',\tilde{\mathfrak{R}}').$

Οbserve that the overall algorithm runs in linear time.

\myskip\subsection{{Proof of correctness of the algorithm for sentences in \texorpdfstring{$\bar{Θ}$}{bar-Theta}}}
\label{sec_bar_theta_app}

In order to complete the proof of~\autoref{@desmembramientos} for a sentence $θ ∈ \bar{Θ}[τ],$
we have to prove that
$$(\mathfrak{A},R, {\bf a})\models θ_{{\sf R},{\bf c}}\iff (\mathfrak{A}\setminus V({\sf compass}_{\tilde{\mathfrak{R}}'}(\tilde{W}')),R\setminus Y, {\bf a})\models θ_{{\sf R},{\bf c}}.$$

For sake of  simplicity, we use $G$ to denote the Gaifman graph $G_{\mathfrak{A}}$ of $\mathfrak{A}.$

\myskip\paragraph{Observations on the collection $\tilde{\cal W}'.$}
Recall that, for every two $(\tilde{W}_i, \tilde{\mathfrak{R}}_i),(\tilde{W}_j, \tilde{\mathfrak{R}}_j)∈ \tilde{\cal W}',$
$(\tilde{W}_i, \tilde{\mathfrak{R}}_i)$ (resp. $(\tilde{W}_j, \tilde{\mathfrak{R}}_j)$) is a
$W_i$-tilt (resp. $W_j$-tilt) of $(W,\mathfrak{R}),$
where $W_i, W_j∈ {\cal W}$ and $V(\cupall{\sf influence}_{\mathfrak{R}} (W_i)) \cap V(\cupall {\sf influence}_{\mathfrak{R}} (W_j)) = \emptyset.$
This implies that
 $V({\sf compass}_{\tilde{\mathfrak{R}}_i}(\tilde{W}_i)) \cap V({\sf compass}_{\tilde{\mathfrak{R}}_j} (\tilde{W}_j)) = \emptyset.$
 Moreover, observe that
 if $\tilde{\cal Q}$ is a
 $(W, \mathfrak{R})$-canonical partition of $G\setminus V({\bf a}),$
 then no internal bag of $\tilde{\cal Q}$ intersects both $V(\cupall {\sf influence}_{\mathfrak{R}} (W_i))$ and $V(\cupall {\sf influence}_{\mathfrak{R}} (W_j)),$ for every $i,j∈[m], i\neq j.$

\myskip\paragraph{Shifting to the split version of $θ_{{\sf R},{\bf c}}.$}
Suppose that $(\mathfrak{A},R, {\bf a})\models θ_{{\sf R},{\bf c}}.$
Recall that
${\bf W}_q^{(1)}$ is the $q$-pseudogrid defined by the horizontal and the vertical paths of the central $q$-subwall of $\tilde{W}_1.$
Due to~\autoref{@interruptions}, we have that $(\mathfrak{A},R,{\bf W}_{{q}}^{(1)}, {\bf a})\models \tilde{θ}_{{q}},$
i.e.,
there are sets $X_1, \ldots, X_h \subseteq V(\mathfrak{A})$
such that
$(\mathfrak{A},R,{\bf W}_{{q}}^{(1)},{\bf a},X_1, \ldots, X_h)\models θ^{\sf out}_q$ and
there is a set $C\subseteq V(\mathfrak{A})$ that is a $w$-privileged set of
$\mathfrak{A}$ with respect to ${\bf W}_q^{(1)}$ and $\{X_1, \ldots, X_h\}$
and it holds that
$\mathfrak{A}[C]\models μ$
and $(\mathfrak{A},R, {\bf a})[C]\models\breve{ζ}_{\sf R} |_{\sf ap_{\bf c}}.$

We set ${\cal X} = \{X_1, \ldots, X_h\}.$
Let $C_1, \ldots, C_{h+1}$ be the collection of vertex sets certifying
that $C$ is a $w$-privileged set with respect to ${\bf W}_q^{(1)}$ and ${\cal X},$ i.e.,
\begin{itemize}
\item $C_1 = C,$ $C_{h+1} = V(G),$ and
\item for every $i∈ [h],$ if $w_i = \circ,$ then $C_{i}$ is the {\sl unique} element in ${\sf pr}(G, {\bf W}_q^{(1)}, X_{i}\cup \ldots \cup X_h),$ while if $w_i = \bullet,$ then $C_i = C_{i+1}\setminus X_i.$
\end{itemize}

By~\autoref{@imagelessness}, the sequence $C_1, \ldots, C_{h+1}$ (and therefore also $C$) is defined uniquely.

We fix $\tilde{\cal Q}$ to be a
$(\tilde{W}_1,\tilde{\mathfrak{R}}_1)$-canonical partition of $G\setminus V({\bf a}).$

\myskip\paragraph{The set $\cupall{\cal X}$ has bounded bidimensionality with respect to $(\tilde{W}_1,\tilde{\mathfrak{R}}_1).$}
Notice that, since $(\mathfrak{A},R,{\bf W}_q^{(1)},{\bf a},X_1, \ldots, X_h)\models θ^{\sf out}_q,$ for every $i∈ [h],$
it holds that $(\mathfrak{A}[C_{i+1}], X_i)\models β_i |_{{\sf star}_{{\sf X}_i}}.$
This, together with the fact that for every $i∈[h]$ $β_i∈ \MSOL^{\tw}[τ\cup\{{\sf X}_i\}]$
imply that
${\sf cl}_{{\sf X}_i}({\sf star}_{{\sf X}_i}(\mathfrak{A}[C_{i+1}], X_i))$ has treewidth at most $\tw(θ).$
By the definition of the $w$-privileged sequence $C_1,\ldots, C_{h+1}$ and due to~\autoref{@congregation}, for every $i∈[h],$
$X_i$ intersects at most $(\tw(θ)+1)^2$ internal bags  of every $(W, \mathfrak{R})$-canonical partition of $G\setminus V({\bf a}).$
Consequently, $\cupall {\cal X}$ intersects at most $h\cdot (\tw(θ)+1)^2 = q-1$ bags of $\tilde{\cal Q}.$

\myskip\paragraph{Finding a $θ$-equivalent extended compass that is disjoint from $X_1,\ldots, X_h.$}
Recall that $\tilde{\cal W}'$ is a collection of $q\cdot (\hat{\ell}+3)$ flatness pairs of $G\setminus V({\bf a})$
of height $2w+j$ that are $θ$-equivalent to $(\tilde{W}_1, \tilde{\mathfrak{R}}_1).$
The fact that $\cupall {\cal X}$ intersects at most $q-1$ bags of $\tilde{\cal Q},$
 $|\tilde{\cal W}'| = q\cdot (\hat{\ell} +3)$
and,
 if $\tilde{\cal Q}$ is a
 $(W, \mathfrak{R})$-canonical partition of $G\setminus V({\bf a}),$
 then no internal bag of $\tilde{\cal Q}$ intersects both $V(\cupall {\sf influence}_{\mathfrak{R}} (W_i))$ and $V(\cupall {\sf influence}_{\mathfrak{R}} (W_j)),$ for every $i,j∈[m], i\neq j,$
implies that there is a collection $\tilde{\cal W}''\subseteq \tilde{\cal W}'$ of size $\hat{\ell} +2$ such that $(\tilde{W}_{1}, \tilde{\mathfrak{R}}_{1})\notin \tilde{\cal W}'',$ every flatness pair in $\tilde{\cal W}''$ is $θ$-equivalent to  $(\tilde{W}_{1}, \tilde{\mathfrak{R}}_{1}),$ and the vertex set of its influence is disjoint from ${\cal X}.$
Assume, without loss of generality, that $(\tilde{W}_{2}, \tilde{\mathfrak{R}}_{2})∈ \tilde{\cal W}'',$ which implies that $θ\text{-}{\sf char}(\mathfrak{K}_1,R_1)= θ\text{-}{\sf char}(\mathfrak{K}_2,R_2)$ and $I_2^{(w)}\cap \cupall{\cal X} = \emptyset.$

\myskip\paragraph{Every modulator leaves an intact buffer.}
We fix $\tilde{\cal Q}$ to be a
$(\tilde{W}_1,\tilde{\mathfrak{R}}_1)$-canonical partition of $G\setminus V({\bf a}).$
Note that $\cupall {\cal X}$ intersects at most $q-1$ bags of $\tilde{\cal Q}.$
This implies that, given that $\tilde{W}_1$ has height $2w+j,$ and $w = (r+2)\cdot q,$
there is an $i∈ [q]$ such that no set in ${\cal X}$ intersects $I_1^{(i\cdot r-1)}\setminus I_1^{(i\cdot r-r)}.$ We set $d = i\cdot r-1$ and, for every $i∈[h],$ we set $X_i^{\rm in} = X_i \cap I_1^{(d-r+1)}$ and $X_i^{\rm out} = X_i\setminus I_1^{(d)}.$
Also, we set ${\cal X}_{\rm in} = \{X_1^{\rm in}, \ldots, X_h^{\rm in}\}$ and ${\cal X}_{\rm out} = \{X_1^{\rm out}, \ldots, X_h^{\rm out}\}.$

\myskip\paragraph{Picking the privileged component inside $G\setminus (X_1\cup\ldots\cup X_h).$}
Let $\breve{C}$ be the privileged connected component of $G$ with respect to ${\bf W}_q^{(1)}$ and $\cupall {\cal X}.$
We stress that, the target sentences are asked to be satisfied in $C$ but, depending on whether $w_1=\circ$ or $w_1=\bullet,$ $C$ is either equal to $\breve{C}$ or $C_2\setminus X.$
For every $i∈[2,h],$ let $Z_i = (I_1^{(d-r+1)}\setminus C_i)\cap C_{i+1}$ and let
$Z_1= (I_1^{(d-r+1)}\setminus \breve{C})\cap C_{2}.$
Observe that $\partial_{\mathfrak{K}_1}(Z_i)\subseteq X_i^{\rm in}$ and $X_i^{\rm in}\subseteq Z_i.$
We set ${\cal Z} = \{Z_1, \ldots, Z_h\}.$

\myskip\paragraph{All apices are adjacent to the privileged component.}
For every $i∈[h],$ we also set $V_{L_i}({\bf a}) = X_i\cap V({\bf a}),$ $L_i$ to be the set of indices of the vertices of ${\bf a}$ in $X_i,$ and ${\cal L} = \{L_1,\ldots, L_h\}.$

Observe that, for every $i∈[h],$ $V_{L_i}({\bf a}) \subseteq X_i^{\rm out}.$
We also claim that $\cup_{i∈[h]} V_{L_i}({\bf a}) = V({\bf a}) \cap N_{G} (\breve{C}).$
To see why this holds, recall that $\cupall {\cal X}$ intersects at most $q-1$ bags of $\tilde{\cal Q}$ and that, by assumption,
every vertex in $V({\bf a})$ is adjacent, in $G,$ to at least $q$ internal bags of $\tilde{\cal Q}.$
Therefore, for every $a∈ V({\bf a}),$ there is an internal
bag $Q$ of $\tilde{\cal Q}$ such that $V(Q)\subseteq V(G\setminus X)$ and $a$ is adjacent, in $G,$ to a vertex in $V(Q).$
For every such $Q,$ since $\breve{C}$ is the privileged component with respect to ${\bf W}_q^{(1)}$ and $\cupall{\cal X},$
it holds that $V(Q)\subseteq \breve{C}$ and therefore
every $a∈ V({\bf a})$ is adjacent, in $G,$ to a vertex in
$\breve{C}.$
This implies that every $a∈ V({\bf a})$ is either in $N_{G}(\breve{C})$ (that is a subset of $\cupall{\cal X}$) or belongs to $\breve{C}.$
Therefore $V_L ({\bf a}) = V({\bf a}) \cap N_{G} (\breve{C}).$
\bigskip

The fact that $θ\text{-}{\sf char}(\mathfrak{K}_1,R_1)= θ\text{-}{\sf char}(\mathfrak{K}_2,R_2)$ implies that there is a collection ${\cal Z}' = \{Z_1',\ldots, Z_h '\}$ of subsets of $I_2^{(d-r+1)},$ such that
\begin{itemize}
\item $\blue{{\sf out}\text{-}{\sf sig}}(\mathfrak{K}_1,R_1,d,{\cal L},{\cal Z})= \blue{{\sf out}\text{-}{\sf sig}}(\mathfrak{K}_2,R_2,d,{\cal L},{\cal Z}')$ and
\item $\green{{\sf in\mbox{-}sig}}(\mathfrak{K}_1,R_1,d,{\cal L},{\cal Z})=  \green{{\sf in\mbox{-}sig}}(\mathfrak{K}_2,R_2,d,{\cal L},{\cal Z}').$
\end{itemize}

We first prove the following:\medskip

\setcounter{theoone}{3}
\begin{theoone}\label{claim_4}
There is a collection ${\cal X}' =\{X_1',\ldots, X_h '\},$ such that for every $i∈[h],$ $X_i '\subseteq Z_i ',$ $\partial_{\mathfrak{K}_2} (Z_i')\subseteq X_i',$ and for every $V\subseteq Y$
that is also a subset of $V(\cupall{\sf influence}_{\breve{\mathfrak{R}}'}(\overline{W}),$ where $\overline{W}$ is the central $(j'-2)$-subwall of $W_1,$ it holds that $(\mathfrak{A},R,{\bf W}_q^{(1)},\varnothing^l,{\cal X})\models θ^{\sf out}_{q} \iff (\mathfrak{A}\setminus V,R\setminus Y,{\bf W}_q^{(1)},\varnothing^l, {\cal X}_{\rm out}\cup {\cal X}')\models θ^{\sf out}_{q}.$
\end{theoone}

\noindent{\em \blue{Proof of \autoref{claim_4}}:}
Let $t= |N_G (\breve{C})|.$
Since $\breve{C}∈{\sf cc}(G,\cupall{\cal X}),$ it holds that $N_G(\breve{C}) \subseteq \cupall {\cal X}.$
For every $i∈[h],$ let $t_i = |N_G (\breve{C}) \cap X_i|$ and observe that $\sum_{i∈[h]} t_i = t.$
Also,  for every $i∈[h],$ since ${\sf cl}_{{\sf X}_i}({\sf star}_{{\sf X}_i}(\mathfrak{A}[C_{i+1}], X_i))$ has treewidth at most $\tw(θ),$
and $N_G (\breve{C}) \cap X_i$ induces a complete graph on $t_i$ vertices in ${\sf cl}_{{\sf X}_i}({\sf star}_{{\sf X}_i}(\mathfrak{A}[C_{i+1}], X_i)),$
we have that $t_i∈[0,\tw(θ)-1].$
Therefore, since $\sum_{i∈[h]} t_i = t,$ $t∈[0,h\cdot(\tw(θ)-1)].$

\myskip\paragraph{Defining the boundary of our boundaried structure.}
For every $i∈[h],$ we set
\begin{itemize}
\item $F_i '$ to be the graph $G[(X_i^{\rm out}\setminus V_{L_i} ({\bf a}))\cap N_G (\breve{C})]$ and
\item $F_i^\star$ to be the graph obtained from $G[V_{L_i} ({\bf a})\cup V(F_i')]$ after removing every edge that has both endpoints in $V_{L_i} ({\bf a}).$
\end{itemize}
In other words, $F_i '$ is the subgraph of $G$ induced by the vertices of $X_i^{\rm out}$ that are not apices and are adjacent to vertices in $\breve{C}.$
Also, we extend $F_i '$ to  $F_i^\star$ by adding the vertices in $V_{L_i} ({\bf a})$ and the edges connecting vertices of  $V_{L_i} ({\bf a})$ and $V(F_i '),$ but not the edges that have both endpoints in $V_{L_i} ({\bf a}).$
This graph $F_i^\star$ will be later associated with a graph $F_i∈ {\cal F}_{t_i-|\partial_{\mathfrak{K}_1} (Z_i)|}^{V_{L_i}({\bf a})}.$

\myskip\paragraph{Separating $\mathfrak{A}$ into two boundaried structures.}
We use $Z$ and $Z'$ to denote $\cupall {\cal Z}$ and $\cupall {\cal Z}',$ respectively.
Also, we use $F^\star$ to denote the graph $\cupall F_i^\star.$
Let $$\mathfrak{A}_{\rm out}^\star = \mathfrak{A}\setminus (\breve{C}\cup (Z\setminus \partial_{\mathfrak{K}_1} (Z)))\mbox{~~and~~}\mathfrak{A}^\star = \mathfrak{A}[\breve{C}\cup Z \cup V(F^\star)].$$
Keep in mind that $V(\mathfrak{A}_{\rm out}^\star) \cap V(\mathfrak{A}^\star) = \partial_{\mathfrak{K}_1} (Z)\cup V(F^\star).$
Next, we will define two boundaried structures corresponding to $\mathfrak{A}_{\rm out}^\star$ and $\mathfrak{A}^\star,$ whose boundary will be the set $\partial_{\mathfrak{K}_1} (Z)\cup V(F^\star).$

\myskip\paragraph{An ordering on the (common) boundary of the two structures.}
We next claim that $\partial_{\mathfrak{K}_1} (Z)\cup V(F^\star) = N_G (\breve{C}),$ which directly implies that $|\partial_{\mathfrak{K}_1} (Z)\cup V(F^\star)| = t.$
To see why $\partial_{\mathfrak{K}_1} (Z)\cup V(F^\star) = N_G (\breve{C}),$ first recall that $N_G(\breve{C}) \subseteq \cupall {\cal X}$ and also notice that  $\cupall {\cal X}_{\rm in} \cap N_G (\breve{C}) = \partial_{\mathfrak{K}_1} (Z).$
Since $\cup_{i∈[h]} V_{L_i}({\bf a})=V({\bf a})\cap N_G (\breve{C})$ and
$\cup_{i∈[h]} V(F_i ') = (X^i_{\rm out}\setminus V_{L_i} ({\bf a}) )\cap N_G (\breve{C}),$ we have that $N_G (\breve{C}) = \partial_{\mathfrak{K}_1} (Z)\cup \bigcup_{i∈[h]} (V_{L_i} ({\bf a}) \cup V(F_i ')) = \partial_{\mathfrak{K}_1} (Z)\cup V(F^\star).$
Therefore, we can consider an ordering $v_1, \ldots, v_t$ of the vertices in $\partial_{\mathfrak{K}_1} (Z)\cup V(F^\star).$
Let ${\bf b}^\star_1 = (v_1, \ldots, v_t)$ and recall that $V(\mathfrak{A}_{\rm out}^\star) \cap V(\mathfrak{A}^\star) = \partial_{\mathfrak{K}_1} (Z)\cup V(F^\star).$
Now, consider the $t$-boundaried $τ$-structures $(\mathfrak{A}_{\rm out}^\star, {\bf b}_1^\star)$ and $(\mathfrak{A}^\star, {\bf b}_1^\star).$
Notice that $(\mathfrak{A}_{\rm out}^\star, {\bf b}_1^\star)$ and $(\mathfrak{A}^\star, {\bf b}_1^\star)$ are compatible
and that $(\mathfrak{A}_{\rm out}^\star, {\bf b}_1^\star)\oplus (\mathfrak{A}^\star, {\bf b}_1^\star)= \mathfrak{A}.$

\myskip\paragraph{Adding $V(F^\star)$ to each member of  ${\cal X}^{\rm in}.$}
Let ${\cal X}^\star_1 = \{\tilde{X}^\star_1, \ldots, \tilde{X}^\star_h\},$
where, for every $i∈[h],$ $\tilde{X}^\star_i = X_i^{\rm in} \cup V(F_i^\star).$
Since for every $i∈[h],$ $\partial_{\mathfrak{K}_1} (Z_i) \subseteq X_i^{\rm in},$ it holds that $V({\bf b}_1^\star)\subseteq \cupall {\cal X}^\star_1.$
Also, let ${\cal R}_1^{' \star} = \{R_1^{' \star}, \ldots, R_h^{' \star} \},$ where, for every $i∈[h],$ $R_i^{' \star} = R\cap (\breve{C}\cup Z_i).$
We have that $(R\setminus (\breve{C}\cup Z)) \cup  \cupall {\cal R}_1^{' \star} = R.$
Since $V(F^\star)\subseteq \cupall {\cal X}^{\rm out},$ it holds that ${\cal X}^{\rm out}\cup {\cal X}^\star_1 = {\cal X}.$

\myskip\paragraph{Separating $(\mathfrak{A},R,{\bf W}_q^{(1)}, \varnothing^l,{\cal X})$ into two boundaried structures.}
We consider the structure $(\mathfrak{A},R,{\bf W}_q^{(1)}, \varnothing^l, {\cal X}).$
We choose to include the $q$-pseudogrid ${\bf W}_q^{(1)}$ in the aforementioned structure (and not another $q$-pseudogrid,
even if, due to~\autoref{corro_gen}, they would yield equivalent instances)
in
order to be able to ``break'' $(\mathfrak{A},R,{\bf W}_q^{(1)}, \varnothing^l, {\cal X})$ into two $t$-boundaried structures corresponding to $\mathfrak{A}_{\rm out}$ and $\mathfrak{A}^\star,$ and $\mathfrak{A}^\star$ to contain $\cupall {\bf W}_q^{(1)},$ since $\cupall {\bf W}_q^{(1)}\subseteq \breve{C}\cup Z\subseteq V(\mathfrak{A}^\star).$

Thus, we have that
\begin{eqnarray}
(\mathfrak{A},R,{\bf W}_q^{(1)},\varnothing^l, X) = (\mathfrak{A}_{\rm out}^\star, R\setminus (\breve{C}\cup Z),\emptyset^{2q},\varnothing^l, X_{\rm out}, {\bf b}_1^\star ) \oplus (\mathfrak{A}^\star, R_1^{' \star},{\bf W}_{{q}}^{(1)},\varnothing^l,\tilde{X}^\star_1, {\bf b}_1^\star).\labels{neneneeletro}
\end{eqnarray}

Also, \autoref{cou_more} implies that there is a $\bar{φ}∈ {\sf rep}_{τ'}^{(t)} (θ^{\sf out}_{q}),$ where $τ' = τ \cup\{R\}\cup{\bf Q}\cup{\cal X}\cup{\bf c},$ such that
 $\big(\mathfrak{A}^\star, \cupall {\cal R}_1^{'\star},{\bf W}_{{q}}^{(1)}, \varnothing^l,{\cal X}^\star_1, {\bf b}_1^\star\big)\models \bar{φ}.$

\myskip\paragraph{Shifting from $\mathfrak{A}^\star$ to $\mathfrak{A}^{(d,Z,L,F_1)}.$}
Now, for every $i∈[h],$ consider a graph $F_i ∈ {\cal F}_{h-|\partial_{\mathfrak{K}_1} (Z_i)| - |L_i|}^{V_{L_i}({\bf a})}$ that is isomorphic\footnote{In the rest of the proof of the claim, we will usually consider a subgraph of $G,$ or a structure with universe $V(G)$ and isomorphic graphs/structures of them, and the latter will be ``abstract'' graphs/structures. For example, here we consider an ``abstract'' graph $F_1$ that is isomorphic to the graph $F^\star$ that is a subgraph of $G.$ We will always use superscript ``$^\star$'' in order to denote the subgraphs/structures that are being given by the graph, while the lack of superscript reflects to the corresponding isomorphic ``abstract'' graphs/structures.} to $F_i^\star,$ via a bijection $ξ_i: V(F_i) \leftrightarrow V(F_i^\star)$ that maps every $a∈ V_{L_i}({\bf a})$ to itself.

We set ${\cal V}_i = (\partial_{\mathfrak{K}_1} (Z_i), V_{L_i} ({\bf a}), V(F_i)\setminus V_{L_i} ({\bf a}))$ and observe that ${\cal V}_i$ is a nice 3-partition of $K_1^{\bf a}[\partial_{\mathfrak{K}_1} (Z_i) \cup V_{L_i} ({\bf a})]\cup F_i.$
Recall that, for every $i∈[h],$ let $t_i = |N_G (C) \cap X_i|$ and observe that $\sum_{i∈[h]} t_i = t.$
Also, observe that  the graph $V(K_1^{\bf a}[\partial_{\mathfrak{K}_1} (Z_i) \cup V_{L_i} ({\bf a})]\cup F_i)$ has $t_i$ vertices and therefore $(K_1^{\bf a}[\partial_{\mathfrak{K}_1} (Z_i) \cup V_{L_i} ({\bf a})]\cup F_i, {\cal V}_i)∈ {\cal H}^{(t_i)}.$
For every $i∈[h],$ let ${\bf H}_i = (K_1^{\bf a}[\partial_{\mathfrak{K}_1} (Z_i) \cup V_{L_i} ({\bf a})]\cup F_i,{\cal V}_i).$

\myskip\paragraph{A boundaried structure of bounded treewidth that satisfies $\bar{φ}.$}
Let ${\bf b}_1$ be the tuple obtained from ${\bf b}_1^\star$ after replacing, in ${\bf b}_1^\star,$ for every $i∈[h],$ each vertex $v∈ V(F_i^\star)$ with the vertex $ξ_i^{-1}(v)∈ V(F_i).$
Also, let $\mathfrak{A}_1 = \mathfrak{A}^{(d,{\cal Z},{\cal L},\{F_1,\ldots, F_h\})}.$
Observe that $\cupall{\bf W}_q^{(1)}\subseteq V(\mathfrak{A}_1)$ and $R_1\cap I_1^{(d)}\subseteq V(\mathfrak{A}_1).$
We set  ${\cal X}_1 = \{\tilde{X}_1, \ldots, \tilde{X}_h\},$ where, for every $i∈[h],$ $\tilde{X}_i = (\tilde{X}^\star_i\setminus V(F_i^\star))\cup V(F_i)$ and ${\cal R}_1^{'} =\{R_1 ', \ldots, R_h '\},$ where, for every $i∈[h],$ $R_i ' = R_1 \setminus  (Z_i\setminus \partial_{\mathfrak{K}_1}(Z_i))$ (recall that $R_1=R\cap V({\sf compass}_{\tilde{\mathfrak{R}}_1} (\tilde{W}_1)$).
Observe that, for every $i∈[h],$ $\tilde{X}_i\subseteq V(\mathfrak{A}_1),$
$R_i '\subseteq V(\mathfrak{A}_1),$
and
$R_i^{'}= R_i^{' \star} \setminus (\partial_{\mathfrak{K}_1}(Z)\cup \breve{C}).$
 At this point, we stress that while, for each $i∈[h],$
$\tilde{X}_i$ is obtained from $\tilde{X}^\star_i$
after replacing $V(F_i^\star)$ with $V(F_i),$ $R_i^{'}$ is obtained from $R_i^{' \star}$ after removing all elements in $V(\mathfrak{A}^\star)$ that are in $\partial_{\mathfrak{K}_1}(Z)$ and $\breve{C}.$

We aim to show that $({\bf H}_1,\ldots, {\bf H}_h, \bar{φ})∈ {\sf out}\text{-}{\sf sig}(\mathfrak{K}_1,R_1,d,{\cal L},{\cal Z}).$
To show this, by the definition of {\sf out}\text{-}{\sf sig} it remains to prove that $\big(\mathfrak{A}_1,\cupall {\cal R}_1 ',{\bf W}_{{q}}^{(1)}, \varnothing^l,{\cal X}_1, {\bf b}_1\big)\models \bar{φ}.$
To prove the latter, first notice that, since, for every $i∈[h],$ $F_i$ and $F_i^\star$ are isomorphic, we have that
$\mathfrak{A}_1[V({\bf b}_1)],$ $\mathfrak{A}^\star[V({\bf b}_1^\star)],$ and $\mathfrak{A}_{\rm out}^\star[V( {\bf b}_1^\star)]$ are (pairwise) isomorphic.
This implies that
 $(\mathfrak{A}_1, {\bf b}_1),$ $(\mathfrak{A}^\star, {\bf b}_1^\star),$ and $(\mathfrak{A}_{\rm out}^\star, {\bf b}_1^\star)$ are (pairwise) compatible.
We next consider the $t$-boundaried $τ'$-structures
 $\big(\mathfrak{A}^\star, \cupall {\cal R}_1^{' \star},{\bf W}_{{q}}^{(1)} ,\varnothing^l, {\cal X}^\star_1, {\bf b}_1^\star\big)$
and
$\big(\mathfrak{A}_1, \cupall {\cal R}_1 ',{\bf W}_{{q}}^{(1)}, \varnothing^l,{\cal X}_1, {\bf b}_1\big).$
These $t$-boundaried $τ'$-structures
are compatible.
We now prove that they are also $(θ^{\sf out}_{q},t)$-equivalent, which will imply that $\big(\mathfrak{A}^\star, \cupall {\cal R}_1^{' \star},{\bf W}_{{q}}^{(1)} , {\cal X}^\star_1, {\bf b}_1^\star\big)\models \bar{φ}\iff\big(\mathfrak{A}_1, \cupall {\cal R}_1 ',{\bf W}_{{q}}^{(1)}, \varnothing^l,{\cal X}_1, {\bf b}_1\big)\models \bar{φ}.$
\medskip

 \noindent{\em Subclaim:}
 $\big(\mathfrak{A}^\star, \cupall {\cal R}_1^{' \star},{\bf W}_{{q}}^{(1)},\varnothing^l, {\cal X}^\star_1, {\bf b}_1^\star\big)$
 and
 $\big(\mathfrak{A}_1, \cupall {\cal R}_1 ',{\bf W}_{{q}}^{(1)},\varnothing^l,  {\cal X}_1, {\bf b}_1\big)$
 are $(θ^{\sf out}_{q},t)$-equivalent.
 \medskip

\noindent{\em Proof of Subclaim:}
Let $\mathfrak{C}^\circ$ be a $τ$-structure, $R^\circ\subseteq V(\mathfrak{C}^\circ),$
${\bf W}_q^{\circ}∈ (2^{V(\mathfrak{C}^\circ)})^{2q},$
${\bf a}^\circ$ be an apex-tuple of  $\mathfrak{C}^\circ$ of size $l,$
${\cal X}^\circ$ be a collection of $h$ subsets of $V(\mathfrak{C}^\circ),$ and
${\bf b}^\circ$ be an apex-tuple of $\mathfrak{C}^\circ$ of size $t,$
such that
$(\mathfrak{C}^\circ,R^\circ,{\bf W}_q^{\circ},{\bf a}^\circ,{\cal X}^\circ, {\bf b}^\circ)$
is an  $t$-boundaried $τ'$-structure
that is compatible with $\big(\mathfrak{A}^\star, \cupall {\cal R}_1^{' \star},{\bf W}_{{q}}^{(1)} , {\cal X}^\star_1, {\bf b}_1^\star\big)$
and $\big(\mathfrak{A}_1, \cupall {\cal R}_1 ',{\bf W}_{{q}}^{(1)}, {\cal X}_1, {\bf b}_1\big).$
We aim to show that
$(\mathfrak{C}^\circ, R^\circ, {\bf W}_q^\circ,{\bf a}^\circ, {\cal X}^\circ, {\bf b}^\circ)\oplus \big(\mathfrak{A}^\star,R_1^{'\star},{\bf W}_{{q}}^{(1)},\varnothing^l, {\cal X}^\star_1, {\bf b}_1^\star\big)\models θ^{\sf out}_q
\iff
(\mathfrak{C}^\circ, R^\circ, {\bf W}_q^\circ, {\bf a}^\circ, {\cal X}^\circ, {\bf b}^\circ)\oplus \big( \mathfrak{A}_1,R_1 ',{\bf W}_{{q}}^{(1)}, \varnothing^l,{\cal X}_1, {\bf b}_1^\star\big)\models θ^{\sf out}_q.$
We set
$\mathfrak{B}^\star := (\mathfrak{C}^\circ, {\bf b}^\circ) \oplus (\mathfrak{A}^\star, {\bf b}_1^\star)$
and
$\mathfrak{B} := (\mathfrak{C}^\circ, {\bf b}^\circ)\oplus(\mathfrak{A}_1, {\bf b}_1).$
Equivalently, it suffices to prove that
$$(\mathfrak{B}^\star, R^\circ \cup R_1^{'\star}, {\bf W}_q^\circ \cup {\bf W}_{{q}}^{(1)},{\bf a}^\circ, {\cal X}^\circ \cup{\cal X}^\star_1)\models θ^{\sf out}_q\iff (\mathfrak{B}, R^\circ \cup R_1^{'}, {\bf W}_q^\circ \cup{\bf W}_{{q}}^{(1)},{\bf a}^\circ, {\cal X}^\circ \cup {\cal X}_1)\models θ^{\sf out}_q.$$

Let $\{\hat{X}_1^\star, \ldots, \hat{X}_h^\star\} = {\cal X}^\circ \cup {\cal X}^\star_1.$
Let $\{\hat{X}_1, \ldots, \hat{X}_h\} = {\cal X}^\circ \cup {\cal X}_1.$
By the definition of $ θ^{\sf out}_q$ (see~\autoref{@arrangements}), it will be enough to show that
\begin{enumerate}
\item
${\bf W}_q^\circ \cup {\bf W}_{{q}}^{(1)}$ is a $q$-pseudogrid of $G_{\mathfrak{B}^\star}\iff{\bf W}_q^\circ \cup {\bf W}_{{q}}^{(1)}$ is a $q$-pseudogrid of $G_{\mathfrak{B}},$

\item the sets $\hat{X}^\star_1, \ldots, \hat{X}^\star_h$ are pairwise disjoint if and only if  the sets $\hat{X}_1, \ldots, \hat{X}_h$ are pairwise disjoint, and

\item there exists a sequence $\hat{C}_1^\star, \ldots, \hat{C}^\star_{h+1}$ of subsets of $V(\mathfrak{B}^\star)$ that is a $w$-privileged sequence of $G_{\mathfrak{B}^\star}$ with respect to ${\bf W}_q^\circ \cup {\bf W}_{{q}}^{(1)}$ and $\{\hat{X}_1^\star, \ldots, \hat{X}_h^\star\}$ and
\begin{itemize}
\item[(i.)] for every $i∈ [h],$ $\hat{X}^\star_{i}\subseteq \hat{C}^\star_{i+1},$
\item[(ii.)] for every $i∈ [h],$ $(\mathfrak{B}^\star[\hat{C}^\star_{i+1}], \hat{X}^\star_{i})\models β_i|_{{\sf star}_{X_i}},$ and
\item[(iii.)]  for every $i∈ [h]$ where $w_i = \circ,$ we have that for every $C∈ {\sf cc}(G_{\mathfrak{B}^\star},\hat{X}^\star_i\cup \ldots \cup \hat{X}^\star_h)$ such that $C\subseteq \hat{C}^\star_{i+1}\setminus \hat{C}^\star_i,$ it holds that  $(\mathfrak{B}^\star,R^\circ \cup \cupall {\cal R}_1^{'\star})[C]\models  {θ_{i-1}}_{{\sf R},{\bf c}}.$
\end{itemize}
if and only if
there exists a sequence $\hat{C}_1, \ldots, \hat{C}_{h+1}$ of subsets of $V(\mathfrak{B})$ that is a $w$-privileged sequence of $G_{\mathfrak{B}}$ with respect to ${\bf W}_q^\circ \cup {\bf W}_{{q}}^{(1)}$ and $\{\hat{X}_1, \ldots, \hat{X}_h\}$ and
\begin{itemize}
\item[(i.)] for every $i∈ [h],$ $\hat{X}_{i}\subseteq \hat{C}_{i+1},$
\item[(ii.)] for every $i∈ [h],$ $(\mathfrak{B}[\hat{C}_{i+1}], \hat{X}_{i})\models β_i|_{{\sf star}_{{\sf X}_i}},$ and
\item[(iii.)]  for every $i∈ [h]$ where $w_i = \circ,$ we have that for every $C∈ {\sf cc}(G_{\mathfrak{B}},\hat{X}_i\cup \ldots \cup \hat{X}_h)$ such that $C\subseteq {\hat{C}_{i+1}\setminus \hat{C}_i},$ it holds that  $(\mathfrak{B},R^\circ \cup \cupall {\cal R}_1^{'})[C]\models  {θ_{i-1}}_{{\sf R},{\bf c}}.$
\end{itemize}

\item[(i).] $(\mathfrak{B}^\star,X^\circ \cup \tilde{X}^\star_1)\models β|_{{\sf star}_{\sf X}}\iff (\mathfrak{B}, X^\circ \cup\tilde{X}_1)\models β|_{{\sf star}_{\sf X}},$
\item[(iii).]  for every $C∈ {\sf cc}(\mathfrak{B}^\star,X^\circ \cup \tilde{X}^\star_1)\setminus {\sf pr}(G_{\mathfrak{B}^\star},{\bf W}_q^\circ \cup {\bf W}_{{q}}^{(1)},X^\circ \cup \tilde{X}^\star_1),$ $(\mathfrak{B}^\star,R^\circ \cup R_1^{'\star}, {\bf a}^0)[C]\models  \breve{ζ}_{\sf R} |_{\sf ap_{\bf c}}\wedge μ$
if and only if for every $C∈ {\sf cc}(\mathfrak{B},X^\circ \cup\tilde{X}_1)\setminus {\sf pr}(G_{\mathfrak{B}},{\bf W}_q^\circ \cup {\bf W}_{{q}}^{(1)},X^\circ \cup\tilde{X}_1),$
$(\mathfrak{B},R^\circ \cup R_1^{'}, {\bf a}^0)[C]\models  \breve{ζ}_{\sf R} |_{\sf ap_{\bf c}}\wedge μ.$
\end{enumerate}

Observe that, since $\cupall {\bf W}_q^{(1)} \subseteq V(\mathfrak{A}_1)\cap V(\mathfrak{A}_1^\star),$ it holds that $\cupall{\bf W}_q^\circ \cup {\bf W}_{{q}}^{(1)}\subseteq V(\mathfrak{B}^\star)\cap V(\mathfrak{B}).$
Thus, item~(1) above holds.

Recall that
${\cal X}_1 = \{\tilde{X}_1, \ldots, \tilde{X}_h\}$ and ${\cal X}_1^\star =\{\tilde{X}^\star_1, \ldots, \tilde{X}^\star_h\},$ where $\tilde{X}_i = (\tilde{X}^\star_i\setminus V(F_i^\star))\cup V(F_i)$ for every $i∈[h].$
This implies that if ${\cal X}^\circ \cup {\cal X}_1^\star= \{\hat{X}_1^\star, \ldots, \hat{X}_h^\star\}$ and
${\cal X}^\circ \cup {\cal X}_1= \{\hat{X}_1, \ldots, \hat{X}_h\},$
then the sets $\hat{X}^\star_1, \ldots, \hat{X}^\star_h$ are pairwise disjoint if and only if  the sets $\hat{X}_1, \ldots, \hat{X}_h$ are pairwise disjoint.
Thus, item~(2) above holds.

Βy~\autoref{@imagelessness}, there is a unique $w$-privileged sequence of $G_{\mathfrak{B}}$ with respect to ${\bf W}_q^\circ \cup {\bf W}_{{q}}^{(1)}$ and $\{\hat{X}_1, \ldots, \hat{X}_h\}$
and a unique
$w$-privileged sequence of $G_{\mathfrak{B}^\star}$ with respect to ${\bf W}_q^\circ \cup {\bf W}_{{q}}^{(1)}$ and $\{\hat{X}_1^\star, \ldots, \hat{X}_h^\star\}.$

Now notice that, for every $i∈[h]$ and for every $C∈ {\sf cc}(G_{\mathfrak{B}^\star},\hat{X}^\star_i\cup \ldots \cup \hat{X}^\star_h)$ such that $C\subseteq \hat{C}^\star_{i+1}\setminus \hat{C}^\star_i,$
it holds that $C\subseteq Z.$
Similarly,
for every $i∈[h]$ and for every $C∈ {\sf cc}(G_{\mathfrak{B}},\hat{X}_i\cup \ldots \cup \hat{X}_h)$ such that $C\subseteq {\hat{C}_{i+1}\setminus \hat{C}_i},$
it holds that $C\subseteq Z'.$
Since $G_{\mathfrak{A}^\star}[Z]$ is the same graph as $G_{\mathfrak{A}_1}[Z],$
we have that
 \begin{eqnarray*}
&\text{the set of all }C∈{\sf cc}(G_{\mathfrak{B}^\star},\hat{X}^\star_1\cup \ldots \cup \hat{X}^\star_h)\text{ that are not subsets of } \hat{C}^\star_1 =\\
&\text{the set of all }C∈{\sf cc}(G_{\mathfrak{B}},\hat{X}_1\cup \ldots \cup \hat{X}_h)\text{ that are not subsets of }\hat{C}_1.
\end{eqnarray*}
Also, notice that, by the definition of a $w$-privileged sequence, for every $i∈[h],$
$\hat{X}^\star_{i}\subseteq \hat{C}^\star_{i+1}$
and
$\hat{X}_{i}\subseteq \hat{C}_{i+1}.$
Therefore, it remains to prove that
for every $i∈[h],$
$(\mathfrak{B}[\hat{C}_{i+1}], \hat{X}_{i})\models β_i|_{{\sf star}_{{\sf X}_i}}\iff (\mathfrak{B}^\star[\hat{C}^\star_{i+1}], \hat{X}^\star_{i})\models β_i|_{{\sf star}_{{\sf X}_i}}.$
To prove this, we will argue that, for every $i∈[h],$ the structure ${\sf star}(\mathfrak{B}[\hat{C}_{i+1}], \hat{X}_{i})$ is isomorphic to the structure ${\sf star}(\mathfrak{B}^\star[\hat{C}^\star_{i+1}], \hat{X}^\star_{i}).$
To see why this holds, notice that
there is a bijection $ρ$ from $V({\sf star}(\mathfrak{B}[\hat{C}_{i+1}], \hat{X}_{i}))$ to $V({\sf star}(\mathfrak{B}^\star[\hat{C}^\star_{i+1}], \hat{X}^\star_{i})),$ such that for all ${\sf R}$ of arity $r≥ 1$ and all ${\bf x}∈ V({\sf star}(\mathfrak{B}[\hat{C}_{i+1}], \hat{X}_{i})),$ it holds that ${\bf x}∈ {\sf R}^{{\sf star}(\mathfrak{B}[\hat{C}_{i+1}], \hat{X}_{i})}$ if and only if ${\bf x}∈ {\sf R}^{{\sf star}(\mathfrak{B}^\star[\hat{C}^\star_{i+1}], \hat{X}^\star_{i})}.$
Therefore, $(\mathfrak{B}^\star, R^\circ \cup \cupall {\cal R}_1^{'\star},{\bf W}_q^\circ \cup {\bf W}_{{q}}^{(1)},{\bf a}^\circ,{\cal X}^\circ \cup {\cal X}^\star_1)\models θ^{\sf out}_{q} \iff (\mathfrak{B}, R^\circ \cup \cupall {\cal R}_1^{'}, {\bf W}_q^\circ \cup {\bf W}_{{q}}^{(1)},{\bf a}^\circ,{\cal X}^\circ \cup {\cal X}_1)\models θ^{\sf out}_{q}.$ The subclaim follows.\hfill$\diamond$
\bigskip

By the above subclaim, we have that  $\big(\mathfrak{A}^\star, \cupall {\cal R}_1^{' \star},{\bf W}_{{q}}^{(1)} ,\varnothing^l, {\cal X}^\star_1, {\bf b}_1^\star\big)$
 and
 $\big(\mathfrak{A}_1,\cupall {\cal R}_1 ',{\bf W}_{{q}}^{(1)}, \varnothing^l,{\cal X}_1, {\bf b}_1\big)$
 are $(θ^{\sf out}_{q},t)$-equivalent.
 Therefore,
 \begin{eqnarray}
 \big(\mathfrak{A}^\star, \cupall {\cal R}_1^{' \star},{\bf W}_{{q}}^{(1)} , \varnothing^l,{\cal X}^\star_1, {\bf b}_1^\star\big)\models \bar{φ}\iff\big(\mathfrak{A}_1, \cupall {\cal R}_1 ',{\bf W}_{{q}}^{(1)}, \varnothing^l,{\cal X}_1, {\bf b}_1\big)\models \bar{φ}.
 \labels{@menospreciado}
 \end{eqnarray}
Thus, we conclude that $F_1, \ldots, F_h,$ ${\bf b}_1,$ and ${\cal X}_1$
certify that $({\bf H}_1, \ldots, {\bf H}_h, \bar{φ})∈ {{\sf out}\text{-}{\sf sig}}(\mathfrak{K}_1,R_1,d,{\cal L}, {\cal Z}).$
Since ${{\sf out}\text{-}{\sf sig}}(\mathfrak{K}_1,R_1,d,{\cal L},{\cal Z})= {{\sf out}\text{-}{\sf sig}}(\mathfrak{K}_2,R_2,d,{\cal L},{\cal Z}'),$ we also have that $({\bf H}_1, \ldots, {\bf H}_h, \bar{φ})∈ {{\sf out}\text{-}{\sf sig}}(\mathfrak{K}_2,R_2,d,{\cal L},{\cal Z}').$
This implies that
\begin{itemize}
\item[(a).] for every $i∈[h],$ there is an $F_i '∈ {\cal F}_{t_i-|\partial_{\mathfrak{K}_2} (Z_i ')|}^{V_{L_i}({\bf a})}$ such that if ${\cal V}_i '= (\partial_{\mathfrak{K}_2} (Z_i '),V_{L_i} ({\bf a}), V(F_i ')\setminus V_{L_i} ({\bf a}))$ and ${\bf H}_i = (H_i, {\cal U}_i),$ then ${\cal V}_i '$ is a nice 3-partition of  $K_2^{\bf a}[\partial_{\mathfrak{K}_2} (Z_i ') \cup V_{L_i} ({\bf a})]\cup F_i '$  and $K_2^{\bf a}[\partial_{\mathfrak{K}_2} (Z_i ') \cup V_{L_i} ({\bf a})]\cup F_i '$ is strongly isomorphic to $H_i$ with respect to $({\cal V}_i ', {\cal U}')$ and
\item[(b).] there is an ordering ${\bf b}_2$ of the vertices in $ \bigcup_{i∈[h]} \big(\partial_{\mathfrak{K}} (Z_i ')\cup V(F_i ')\big)$
and for every $i∈[h],$ there is an $\tilde{X}_i '\subseteq Z_i '\cup V(F_i ')$
such that $\partial_{\mathfrak{K}_2}(Z_i ') \cup V(F_i ')\subseteq \tilde{X}_i '$ and,
if ${\cal X}_2 = \{\tilde{X}_1 ',\ldots, \tilde{X}_h '\}$ and $R'' = \bigcup_{i∈[h]} (Z_i'\setminus \partial_{\mathfrak{K}_2}(Z_i'))\cap R_2,$ then
$\big(\mathfrak{A}_2^{(d,{\cal Z}',{\cal L},\{F_1 ', \ldots, F_h '\})},R'',{\bf W}_{{q}}^{(2)}, \varnothing^l,{\cal X}_2, {\bf b}_2\big)\models \bar{φ}.$
\end{itemize}
Observe that, for every $i∈[h],$
since  $K_1^{\bf a}[\partial_{\mathfrak{K}_1} (Z_i) \cup V_{L_i} ({\bf a})]\cup F_i$
and $K_2^{\bf a}[\partial_{\mathfrak{K}_2} (Z_i ') \cup V_{L_i} ({\bf a})]\cup F_i '$
are strongly isomorphic to  $H_i$ with respect to
$({\cal V}_i, {\cal U}_i)$ and $({\cal V}_i ', {\cal U}_i),$ respectively,
we also have that  $K_1^{\bf a}[\partial_{\mathfrak{K}_1} (Z_i) \cup V_{L_i} ({\bf a})]\cup F_i$
is strongly isomorphic to
$K_2^{\bf a}[\partial_{\mathfrak{K}_2} (Z_i') \cup V_{L_i} ({\bf a})]\cup F_i '$
with respect to $({\cal V}_i, {\cal V}_i ').$
We now set $\mathfrak{A}_2 = \mathfrak{A}_2^{(d,{\cal Z}',{\cal L},\{F_1 ', \ldots, F_h '\})}.$
Notice that the fact that for every $i∈[h],$
$K_1^{\bf a}[\partial_{\mathfrak{K}_1} (Z_i) \cup V_{L_i} ({\bf a})]\cup F_i$
is strongly isomorphic to
$K_2^{\bf a}[\partial_{\mathfrak{K}_2} (Z_i') \cup V_{L_i} ({\bf a})]\cup F_i '$
with respect to $({\cal V}_i, {\cal V}_i ')$
implies that the $t$-boundaried $τ'$-structures
$\big(\mathfrak{A}_2,R'',{\bf W}_{{q}}^{(2)},\varnothing^l, {\cal X}_2, {\bf b}_2\big)$
and $\big(\mathfrak{A}_1,\cupall {\cal R}_1 ',{\bf W}_{{q}}^{(1)},\varnothing^l, {\cal X}_1, {\bf b}_1\big)$
are compatible.
Thus, given that
 $\big(\mathfrak{A}_2,R'',{\bf W}_{{q}}^{(2)},\varnothing^l, {\cal X}_2, {\bf b}_2\big)\models \bar{φ},$
 we have that  $\big(\mathfrak{A}_2,R'',{\bf W}_{{q}}^{(2)}, \varnothing^l,{\cal X}_2, {\bf b}_2\big)$
 and $\big(\mathfrak{A}_1,\cupall {\cal R}_1 ',{\bf W}_{{q}}^{(1)}, \varnothing^l,{\cal X}_1, {\bf b}_1\big)$ are
 $(θ^{\sf out}_{q},t)$-equivalent.
Therefore,
\begin{eqnarray}
\big(\mathfrak{A}_1,\cupall {\cal R}_1 ',{\bf W}_{{q}}^{(1)}, \varnothing^l,{\cal X}_1, {\bf b}_1\big)\models \bar{φ} \iff
\big(\mathfrak{A}_2,R'',{\bf W}_{{q}}^{(2)},\varnothing^l, {\cal X}_2, {\bf b}_2\big)\models \bar{φ}.
\labels{@descuartizada}
\end{eqnarray}

At this point, to give some intuition, we underline that even if
$\big(\mathfrak{A}_2,R'',{\bf W}_{{q}}^{(2)}, \varnothing^l,{\cal X}_2, {\bf b}_2\big)$
and $\big(\mathfrak{A}_1,\cupall {\cal R}_1 ',{\bf W}_{{q}}^{(1)}, \varnothing^l,{\cal X}_1, {\bf b}_1\big)$ are
$(θ^{\sf out}_{q},t)$-equivalent, we did not yet provide
a $t$-boundaried $τ'$-structure that is a \textsl{substructure} of $(\mathfrak{A},R,{\bf W}_q^{(1)},\varnothing^l,{\cal X})$ and that
is $(θ^{\sf out}_{q},t)$-equivalent to
$\big(\mathfrak{A}^\star, \cupall {\cal R}_1^{' \star},{\bf W}_{{q}}^{(1)} ,\varnothing^l, {\cal X}^\star_1, {\bf b}_1^\star\big).$
To find such a substructure $(\mathfrak{A}^{\star}, R^{'' \star},{\bf W}_{{q}}^{(2)}, \varnothing^l,{\cal X}^{\star}_2, {\bf b}_2^\star)$ of the structure $(\mathfrak{A},R,{\bf W}_{{q}},\varnothing^l,{\cal X}),$
we have to ``shift'' from
$\big(\mathfrak{A}_2,R'',{\bf W}_{{q}}^{(2)}, \varnothing^l,{\cal X}_2, {\bf b}_2\big)$
to $(\mathfrak{A}^{\star}, R^{'' \star},{\bf W}_{{q}}^{(2)}, \varnothing^l,{\cal X}^{\star}_2, {\bf b}_2^\star),$
by replacing, for every $i∈[h],$ $V(F_i ')$ with $V(F_i^\star),$ and ``extending'' $R''$ to $R_2^{'\star}$
so as to contain all vertices in $\cupall {\cal R}_1^{' \star} \setminus Y.$
This substructure $(\mathfrak{A}^{\star}, R^{'' \star},{\bf W}_{{q}}^{(2)}, \varnothing^l,{\cal X}^{\star}_2, {\bf b}_2^\star)$ will replace $\big(\mathfrak{A}^\star, \cupall {\cal R}_1^{' \star},{\bf W}_{{q}}^{(1)} ,\varnothing^l, {\cal X}^\star_1, {\bf b}_1^\star\big)$ in~\eqref{neneneeletro},
thus providing a collection ${\cal X}' =\{X_1',\ldots, X_h '\},$ such that for every $i∈[h],$ $X_i '\subseteq Z_i ',$ $\partial_{\mathfrak{K}_2} (Z_i')\subseteq X_i',$ and $(\mathfrak{A},R,{\bf W}_{{q}},\varnothing^l,{\cal X})\models {θ^{\sf out}_{q}} \iff (\mathfrak{A},R\setminus Y,{\bf W}_q,\varnothing^l, {\cal X}_{\rm out}\cup {\cal X}')\models {θ^{\sf out}_{q}}.$

\myskip\paragraph{Defining a substructure of the initial structure with a different boundary.}
Let us now define the substructure $(\mathfrak{A}^{\star}, R^{'' \star},{\bf W}_{{q}}^{(2)},\varnothing^l, {\cal X}^{\star}_2, {\bf b}_2^\star)$ from
$\big(\mathfrak{A}_2,R'',{\bf W}_{{q}}^{(2)}, \varnothing^l,{\cal X}_2, {\bf b}_2\big).$
We set ${\bf b}_2^\star$ to be
the tuple obtained from ${\bf b}_2$ after replacing, for every $i∈[h],$ each $v∈ V(F_i ')$ with the corresponding $u∈ V(F_i^\star).$
Also, let
$R^{'' \star}  = (\cupall {\cal R}_1^{' \star})\setminus Y,$ and ${\cal X}^{\star}_2 =\{\tilde{X}_1^{' \star}, \ldots, \tilde{X}_h^{' \star}\},$ where for every $i∈[h],$ $\tilde{X}_i^{' \star} = (\tilde{X}_i ' \setminus V(F_i '))\cup V(F_i^\star).$

For the tuple $(\mathfrak{A}^{\star}, R^{'' \star},{\bf W}_{{q}}^{(2)},\varnothing^l, {\cal X}^{\star}_2, {\bf b}_2^\star)$ to be a $t$-boundaried $τ'$-structure,
we need to show that $R^{'' \star}\subseteq V(\mathfrak{A}^{\star}),$ $\cupall{\bf W}_{{q}}^{(2)}\subseteq V(\mathfrak{A}^{\star}),$ $\cupall {\cal X}_2^\star \subseteq V(\mathfrak{A}^{\star}),$ and $V({\bf b}_2^\star)\subseteq V(\mathfrak{A}^{\star}).$
Notice that $R^{'' \star}\subseteq V(\mathfrak{A}^{\star}),$ since $R^{'' \star}  = (\cupall {\cal R}_1^{' \star})\setminus Y$ and $\cupall {\cal R}_1^{' \star}\subseteq V(\mathfrak{A}^{\star}).$
To show that $V({\bf b}_2^\star)\subseteq V(\mathfrak{A}^{\star}),$ we first notice that,
since $\breve{C}$ respects ${\bf W}_{{q}}^{(1)}$ and
$I_2^{(d)}\cap \cupall{\cal X} = \emptyset,$
it holds that
$I_2^{(d)}\subseteq \breve{C}.$
Therefore, since $Z'\subseteq I_2^{(d-r+1)},$ we also have that $Z'\subseteq C.$
The latter implies that $\partial_{\mathfrak{K}_2}(Z')$  is a subset of $V(\mathfrak{A}^{\star}).$
By the definition of ${\bf b}_2^\star$ and since $V({\bf b}_2) =  \bigcup_{i∈[h]} \big(\partial_{\mathfrak{K}} (Z_i ')\cup V(F_i ')\big) = \partial_{\mathfrak{K}_2} (Z') \cup  \bigcup_{i∈[h]} V(F_i '),$ we have that
$V({\bf b}_2^\star) = \partial_{\mathfrak{K}_2}(Z')\cup V(F^\star).$
Hence, given that $\partial_{\mathfrak{K}_2}(Z')\subseteq V(\mathfrak{A}^{\star})$ and $V(F^\star)\subseteq  V(\mathfrak{A}^{\star}),$ it holds that $V({\bf b}_2^\star)\subseteq V(\mathfrak{A}^\star).$
Also, observe that the fact that $I_2^{(d)}\subseteq \breve{C},$ implies that $\cupall {\bf W}_q^{(2)} \subseteq V(\mathfrak{A}^\star),$ while $\cupall {\cal X}^{\star}_2\subseteq V(\mathfrak{A}^\star),$ since ${\cal X}^{\star}_2 =\{\tilde{X}_1^{' \star}, \ldots, \tilde{X}_h^{' \star}\},$ where for every $i∈[h],$ $\tilde{X}_i^{' \star} = (\tilde{X}_i ' \setminus V(F_i '))\cup V(F_i^\star),$
$\tilde{X}_i '\setminus V(F_i ')\subseteq Z',$ and $Z'\subseteq V(\mathfrak{A}^\star).$

\myskip\paragraph{All considered boundaried structures are $(θ^{\sf out}_q,h)$-equivalent.}
As a next step, we argue that the $t$-boundaried $τ'$-structures
$(\mathfrak{A}^{\star}, R^{'' \star},{\bf W}_{{q}}^{(2)}, \varnothing^l,{\cal X}^{\star}_2, {\bf b}_2^\star),$
$(\mathfrak{A}_2,R'',{\bf W}_{{q}}^{(2)},\varnothing^l, {\cal X}_2, {\bf b}_2),$
and $\big(\mathfrak{A}^\star, \cupall {\cal R}_1^{' \star},{\bf W}_{{q}}^{(1)} , \varnothing^l,{\cal X}^\star_1, {\bf b}_1^\star\big)$ are (pairwise) compatible.
To see why this holds, notice that, for every $i∈[h],$ since
$K_1^{\bf a}[\partial_{\mathfrak{K}_1} (Z_i) \cup V_{L_i} ({\bf a})]\cup F_i$
is strongly isomorphic to
$K_2^{\bf a}[\partial_{\mathfrak{K}_2} (Z_i') \cup V_{L_i} ({\bf a})]\cup F_i '$
with respect to $({\cal V}_i, {\cal V}_i '),$ it holds that $F_i$ and $F_i '$ are isomorphic. This, together with the fact that, for every $i∈[h],$  $F_i$ is isomorphic to $F_i^\star$ implies that $\bigcup_{i∈[h]} F_i ',$ $\bigcup_{i∈[h]} F_i,$ and $F^\star$ are pairwise isomorphic graphs.
Therefore,
the structures $\mathfrak{A}^\star[V({\bf b}_2^\star)],$ $\mathfrak{A}_2 [V({\bf b}_2)],$ and $\mathfrak{A}^\star [V({\bf b}_1^\star)]$ are (pairwise) isomorphic.

By following the exactly symmetric arguments as in the proof of the subclaim above, it is easy to show that
$(\mathfrak{A}^{\star}, R^{'' \star},{\bf W}_{{q}}^{(2)},\varnothing^l, {\cal X}^{\star}_2, {\bf b}_2^\star)$ and $(\mathfrak{A}_2,R'',{\bf W}_{{q}}^{(2)}, \varnothing^l,{\cal X}_2, {\bf b}_2)$ are $(θ^{\sf out}_{q},t)$-equivalent. This implies that
\begin{eqnarray}
\big(\mathfrak{A}^\star, R^{'' \star},{\bf W}_{{q}}^{(2)} , \varnothing^l,\tilde{X}^\star_2, {\bf b}_2^\star\big)\models \bar{φ}\iff\big(\mathfrak{A}_2,R'',{\bf W}_{{q}}^{(2)}, \varnothing^l,{\cal X}_2, {\bf b}_2\big)\models \bar{φ}.\labels{@disviticchia}
\end{eqnarray}

Therefore, combining~\eqref{@menospreciado},~\eqref{@descuartizada}, and~\eqref{@disviticchia}, we conclude that
the $t$-boundaried $τ'$-structures $(\mathfrak{A}^{\star}, R^{'' \star},{\bf W}_{{q}}^{(2)}, \varnothing^l,{\cal X}^{\star}_2, {\bf b}_2^\star)$
and
$ \big(\mathfrak{A}^\star, \cupall {\cal R}_1^{' \star},{\bf W}_{{q}}^{(1)} ,\varnothing^l, {\cal X}^\star_1, {\bf b}_1^\star\big)$
are $(θ^{\sf out}_{q},t)$-equivalent.
Recall that, by~\eqref{neneneeletro},
$$(\mathfrak{A}_{\rm out}^\star, R\setminus (\breve{C}\cup Z),\emptyset^{2q} ,\varnothing^l,{\cal X}^{\rm out}, {\bf b}_1^\star ) \oplus (\mathfrak{A}^\star, \cupall {\cal R}_1^{' \star},{\bf W}_{{q}}^{(1)} ,\varnothing^l, {\cal X}^\star_1, {\bf b}_1^\star) = (\mathfrak{A},R,{\bf W}_q^{(1)}, \varnothing^l,{\cal X})$$
and $(\mathfrak{A},R,{\bf W}_q^{(1)}, {\cal X})\models θ^{\sf out}_{q}.$
Since the $t$-boundaried $τ'$-structures $(\mathfrak{A}^{\star}, R^{'' \star},{\bf W}_{{q}}^{(2)},\varnothing^l, {\cal X}^{\star}_2, {\bf b}_2^\star)$
and
$ \big(\mathfrak{A}^\star, \cupall {\cal R}_1^{' \star},{\bf W}_{{q}}^{(1)} , \varnothing^l,{\cal X}^\star_1, {\bf b}_1^\star\big)$
are $(θ^{\sf out}_{q},t)$-equivalent,
\begin{eqnarray}
(\mathfrak{A}_{\rm out}^\star, R\setminus (\breve{C}\cup Z),\emptyset^{2q} ,\varnothing^l,{\cal X}^{\rm out}, {\bf b}_1^\star ) \oplus \big(\mathfrak{A}^{\star},R^{'' \star},{\bf W}_{{q}}^{(2)},\varnothing^l, {\cal X}^{\star}_2, {\bf b}_2^\star\big)\models θ^{\sf out}_{q}.\labels{@easilyfinish}
\end{eqnarray}

\myskip\paragraph{Another way to put a boundary in the initial structure.}
We set ${\cal X}' = {\cal X}_2^{\star}.$
To conclude the proof of \autoref{claim_4}, it remains to prove that $$(\mathfrak{A}_{\rm out}^\star, R\setminus (\breve{C}\cup Z),\emptyset^{2q} ,\varnothing^l,{\cal X}^{\rm out}, {\bf b}_1^\star ) \oplus \big(\mathfrak{A}^{\star},R^{'' \star},{\bf W}_{{q}}^{(2)},\varnothing^l, {\cal X}', {\bf b}_2^\star\big) = (\mathfrak{A}, R\setminus Y, {\bf W}_q^{(2)}, \varnothing^l,{\cal X}^{\rm out} \cup {\cal X}').$$
To see why this holds, notice that  the $t$-boundaried $τ'$-structures $(\mathfrak{A}_{\rm out}^\star, R\setminus (\breve{C}\cup Z),\emptyset^{2q} ,\varnothing^l,{\cal X}^{\rm out}, {\bf b}_1^\star)$ and $\big(\mathfrak{A}^{\star},R^{'' \star},{\bf W}_{{q}}^{(2)}, \varnothing^l,{\cal X}', {\bf b}_2^\star\big)$
are compatible and that
$R\setminus (\breve{C}\cup Z) \cup R^{'' \star} = R\setminus (\breve{C}\cup Z) \cup ((\cupall {\cal R}_1^{' \star})\setminus Y) = R\setminus Y$ (the latter equality holds since $(R\setminus (\breve{C}\cup Z)) \cup  \cupall {\cal R}_1^{' \star} = R$ and $Y = V(\cupall{\sf influence}_{\tilde{\mathfrak{R}}_1} (W^\bullet))\subseteq I_1^{(d-r+1)} \subseteq \breve{C} \cup Z$).

Finally, we have that
$(\mathfrak{A},R,{\bf W}_{{q}}^{(1)},\varnothing^l, {\cal X})\models θ^{\sf out}_q \iff (\mathfrak{A},R\setminus Y,{\bf W}_q^{(2)},\varnothing^l, {\cal X}^{\rm out}\cup {\cal X}')\models θ^{\sf out}_q.$
To conclude the proof of \autoref{claim_4}, it remains to prove that for every $V\subseteq Y$
that is also a subset of $V(\cupall{\sf influence}_{\breve{\mathfrak{R}}'}(\overline{W})),$ where $\overline{W}$ is the central $(j'-2)$-subwall of $W_1$
, $(\mathfrak{A},R\setminus Y,{\bf W}_q^{(2)},\varnothing^l, {\cal X}_{\rm out}\cup {\cal X}')\models θ^{\sf out}_q\iff (\mathfrak{A}\setminus V,R\setminus Y,{\bf W}_q^{(2)},\varnothing^l, {\cal X}_{\rm out}\cup {\cal X}')\models θ^{\sf out}_q.$

Let $V\subseteq Y$ that is also a subset of $V(\cupall{\sf influence}_{\breve{\mathfrak{R}}'}(\overline{W}),$ where $\overline{W}$ is the central $(j'-2)$-subwall of $W_1.$
Since $Y\subseteq I_1^{(w)},$ $\cupall({\cal X}_{\rm out}\cup {\cal X}')\cap Y = \emptyset,$ and $\cupall {\bf W}_q^{(2)}\subseteq I_2^{(2)},$
$C''∈ {\sf pr}(G_{\mathfrak{A}},{\bf W}_q^{(2)}, \cupall({\cal X}_{\rm out}\cup {\cal X}')) \iff C''\setminus V ∈ {\sf pr}(G_{\mathfrak{A}}\setminus V,{\bf W}_q^{(2)}, \cupall({\cal X}_{\rm out}\cup {\cal X}'))$ and, if
${\cal C}$ (resp. ${\cal C}'$) is the set of all $C''∈{\sf cc}(G_{\mathfrak{A}}, \cupall({\cal X}_{\rm out}\cup {\cal X}'))$ (resp. all $C''∈{\sf cc}(G_{\mathfrak{A}}\setminus V, \cupall({\cal X}_{\rm out}\cup {\cal X}'))$)
that are not in ${\sf pr}(G_{\mathfrak{A}},{\bf W}_q^{(2)}, \cupall({\cal X}_{\rm out}\cup {\cal X}'))$ (resp. $ {\sf pr}(G_{\mathfrak{A}}\setminus V,{\bf W}_q^{(2)}, \cupall({\cal X}_{\rm out}\cup {\cal X}'))$), then ${\cal C}={\cal C}'.$
Therefore, in order to prove $(\mathfrak{A},R\setminus Y,{\bf W}_q^{(2)},\varnothing^l, \cupall({\cal X}_{\rm out}\cup {\cal X}'))\models θ^{\sf out}_q\iff (\mathfrak{A}\setminus V,R\setminus Y,{\bf W}_q^{(2)},\varnothing^l, \cupall({\cal X}_{\rm out}\cup {\cal X}'))\models θ^{\sf out}_q,$ it now suffices to prove
that for every $i∈ [h],$
it holds that $(\mathfrak{A}[C_{i+1}'], X_i^{\rm out}\cup X_i')\models β_i |_{{\sf star}_{{\sf X}_i}}\iff
(\mathfrak{A}[C_{i+1}'\setminus V], X_i^{\rm out}\cup X_i')\models β_i |_{{\sf star}_{{\sf X}_i}},$
where $C_1 ', \ldots, C_{h+1} '$ is the $w$-privileged sequence of $G$ with respect to ${\bf W}_q^{(1)}$ and ${\cal X}_{\rm out}\cup {\cal X}'.$
Equivalently,
for every $i∈ [h],$
we want to prove that
${{\sf star}_{{\sf X}_i}}(\mathfrak{A}[C_{i+1}'], X_i^{\rm out}\cup X_i')\models β_i\iff {{\sf star}_{{\sf X}_i}}(\mathfrak{A}[C_{i+1}'\setminus V], X_i^{\rm out}\cup X_i')\models β_i.$
Recall that $Y=V({\sf compass}_{\breve{\mathfrak{R}}'}(\breve{W}')),$ where $(\breve{W}', \breve{\mathfrak{R}}')$ is a $\breve{W}$-tilt of $(W,\mathfrak{R})$ and $\breve{W}$ is the central $j'$-subwall of $W_1.$
Note that, by the definition of a flatness pair and since $V\subseteq V(\cupall{\sf influence}_{\breve{\mathfrak{R}}'}(\overline{W})$ and $\overline{W}$ is the central $(j'-2)$-subwall of $W_1,$
no vertex in $V$ is adjacent to a vertex in $G_{\mathfrak{A}}\setminus (Y \cup V({\bf a})).$
Also, the fact that $\cupall({\cal X}_{\rm out}\cup {\cal X}')\cap Y = \emptyset$
implies that every vertex in $X_i^{\rm out}\cup X_i'$ is either in $G_{\mathfrak{A}}\setminus Y$ or in $V({\bf a}).$
Therefore, if a vertex of $V$ is adjacent to a vertex $u$ in $X_i^{\rm out}\cup X_i',$ then $u∈ V({\bf a}).$
We will prove that every vertex $u$ in $X_i^{\rm out}\cup X_i '$ that is adjacent, in $G_{\mathfrak{A}},$ to a vertex in $V$ is also adjacent, in $G_{\mathfrak{A}},$ to a vertex in
$V({\sf compass}_{\mathfrak{R}}(W))\setminus V.$
Let $u$ be a vertex in $X_{\rm out}\cup X'$ that is adjacent, in $G_{\mathfrak{A}},$ to a vertex in $V.$
As observed above, $u∈ V({\bf a}).$
Let $\tilde{\cal Q}$ be a $(W,\mathfrak{R})$-canonical partition of $G\setminus V({\bf a}).$
Since by the hypothesis of the lemma ${\sf bid}(N_{G_{\mathfrak{A}}}(V({\bf a})), W,\mathfrak{R})≥(g+2)^2+1,$
we have that $u$ is adjacent to at least $(g+2)^2+1$ internal bags of $\tilde{\cal Q}.$
Notice that $V$ is a subset of the union of the vertex sets of all internal bags of $\tilde{\cal Q}$ that intersect the central $(g+2)$-subwall of $W_1.$
These bags are $(g+2)^2$ many.
This, in turn, implies that, since the vertex $u$ is adjacent to at least $(g+2)^2+1$ internal bags of $\tilde{\cal Q},$ $a$ is adjacent to a vertex $v$
in the vertex set of an internal bag of $\tilde{\cal Q}$ that
is disjoint from $V.$
By observing that $v∈ V({\sf compass}_{\mathfrak{R}}(W))\setminus V,$
we conclude that every vertex $u∈ (X_i^{\rm out}\cup X_i')\cap V({\bf a})$ is adjacent to a vertex in $V({\sf compass}_{\mathfrak{R}}(W))\setminus V.$
This implies that ${{\sf star}_{{\sf X}_i}}(\mathfrak{A}[C_{i+1}'], X_i^{\rm out}\cup X_i')\models β_i\iff {{\sf star}_{{\sf X}_i}}(\mathfrak{A}[C_{i+1}'\setminus V], X_i^{\rm out}\cup X_i')\models β_i.$
\autoref{claim_4} follows.\hfill$\diamond$
\bigskip

Following \autoref{claim_4},
let ${\cal X}' =\{X_1',\ldots, X_h '\}$ be a collection of subsets of $V(\mathfrak{A}),$ such that for every $i∈[h],$ $X_i '\subseteq Z_i ',$ $\partial_{\mathfrak{K}_2} (Z_i')\subseteq X_i',$ and $(\mathfrak{A},R,{\bf W}_{{q}},\varnothing^l,{\cal X})\models {θ^{\sf out}_{q}} \iff (\mathfrak{A},R\setminus Y,{\bf W}_q, \varnothing^l,{\cal X}_{\rm out}\cup {\cal X}')\models {θ^{\sf out}_{q}}.$\bigskip

Let $C_1 ', \ldots, C_{h+1} '$ be the $w$-privileged sequence of $G$ with respect to ${\bf W}_q^{(1)}$ and ${\cal X}_{\rm out}\cup {\cal X}'.$
We set $C' = C_1 '.$
Observe that since ${\bf W}_q^{(1)}\subseteq I_1^{(1)}$ and $I_1^{(d\cdot r)}\cap \cupall({\cal X}^{\rm out} \cup {\cal X}')=\emptyset,$ we have that
${\bf W}_q^{(1)}\subseteq G\setminus \cupall({\cal X}_{\rm out} \cup {\cal X}').$
Thus, there is a $\breve{C}'∈ {\sf cc}(G,\cupall {\cal X}_{\rm out}\cup {\cal X}')$ that respects ${\bf W}_q^{(1)}.$
Keep in mind that if $w_1=\circ,$ then $C'=\breve{C}',$ while, if $w_1=\bullet,$ then $C' = C_2 '\setminus (X^{\rm out}_1\cup X_1')$ and $\breve{C}'\subseteq C'.$

Also, for every vertex $a∈ V({\bf a}),$ since it is adjacent, in $G,$ to at least $q$ internal bags of $\tilde{\cal Q},$ it holds that $a$ is either contained in some $X_i,$ $i∈[h],$ or it is adjacent to a vertex in $\breve{C}$ (and the same holds when replacing $X_i$ with $X_i '$ and $\breve{C}$ with $\breve{C}'$).
Therefore, every $a∈ V({\bf a})$ is either in $\bigcup_{i∈[h]} V_{L_i} ({\bf a})$ or belongs
 to both $\breve{C}$ and $\breve{C}'.$
Therefore, ${\bf a}\cap \breve{C} = {\bf a}\cap \breve{C}' = {\bf a}\cap C = {\bf a}\cap C'.$
We set ${\bf a}' = {\bf a}\cap \breve{C}.$
We aim to prove the following:\medskip

\setcounter{theotwo}{4}
\begin{theotwo}\label{claim_5}
It holds that
${\sf ap}_{{\bf c}}((\mathfrak{A}, R, {\bf a}')[C])\models \breve{ζ}_{\sf R}
\iff
{\sf ap}_{{\bf c}}((\mathfrak{A}, R\setminus Y, {\bf a}')[C'])\models \breve{ζ}_{\sf R}.$
\end{theotwo}

\noindent{\em \green{Proof of \autoref{claim_5}}:}
We will only prove that ${\sf ap}_{{\bf c}}((\mathfrak{A}, R, {\bf a}')[C])\models \breve{ζ}_{\sf R}
\implies
{\sf ap}_{{\bf c}}((\mathfrak{A}, R\setminus Y, {\bf a}')[C'])\models \breve{ζ}_{\sf R},$
since the other implication is trivial.
Suppose that
${\sf ap}_{{\bf c}}((\mathfrak{A}, R, {\bf a}')[C])\models \breve{ζ}_{\sf R}.$
Since $\breve{ζ}_{\sf R}$ is a Boolean combination of the sentences $\breve{ζ}_1, \ldots, \breve{ζ}_p,$ there is a set $J\subseteq [p]$ such that for every $j∈ J$ it holds that ${\sf ap}_{{\bf c}}((\mathfrak{A}, R, {\bf a}')[C])\models \breve{ζ}_j$ and for every $j\notin J$ it holds that ${\sf ap}_{{\bf c}}((\mathfrak{A}, R, {\bf a}')[C])\models \neg \breve{ζ}_j.$
We will show that for every $j∈ J$ it holds that ${\sf ap}_{{\bf c}}((\mathfrak{A}, R\setminus Y, {\bf a}')[C'])\models\breve{ζ}_j$ and that for every $j\notin J$ it holds that ${\sf ap}_{{\bf c}}((\mathfrak{A}, R\setminus Y, {\bf a}')[C'])\models \neg \breve{ζ}_j.$
Therefore, we distinguish two cases.\bigskip

\noindent{\bf Case 1:} $j∈ J.$\medskip

We aim to prove that ${\sf ap}_{{\bf c}}((\mathfrak{A}, R, {\bf a}')[C])\models \breve{ζ}_j \iff {\sf ap}_{{\bf c}}((\mathfrak{A}, R\setminus Y, {\bf a}')[C'])\models \breve{ζ}_j.$
Suppose that ${\sf ap}_{{\bf c}}((\mathfrak{A}, R, {\bf a}')[C])\models \breve{ζ}_j.$
Recall that the constant-projection $(τ\cup\{{\sf R}\})^{\bf c}$ of $(τ\cup\{{\sf R}\}\cup{\bf c}),$ i.e.,
the vocabulary of the structure ${\sf ap}_{{\bf c}}((\mathfrak{A}, R ,{\bf a}')[C]),$ contains every unary relation symbol in $(τ\cup\{{\sf R}\}\cup{\bf c})$
and note that in the structure ${\sf ap}_{{\bf c}}((\mathfrak{A}, R, {\bf a}')[C]),$ ${\sf R}$ is interpreted as $R\cap C.$
We set $R_C:=R\cap C,$ $(\mathfrak{B}, R_C) := {\sf ap}_{{\bf c}}((\mathfrak{A}, R ,{\bf a}')[C]),$
and keep in mind that $\mathfrak{B}$ is a $τ^{\bf c}$-structure.
Since the Gaifman graph of $\mathfrak{B}$ and of $(\mathfrak{B},R_C)$ are the same, in the rest of the proof we will use $G_{\mathfrak{B}}$ to denote both of them.
Also, notice that $G_{\mathfrak{B}}$ is obtained from $G[C]$ after removing some edges (namely, the edges of $G[C]$ that connect the vertices in $V({\bf a}')$ with $C\setminus V({\bf a}')$).
Since $\breve{ζ}_j$ is a basic local sentence with parameters $r_j$ and $\ell_j,$
we have that
$$(\mathfrak{B}, R_C)\models \breve{ζ}_j \iff \exists X_j\subseteq R_C \mbox{~that is $(\ell_j, r_j)$-scattered in $\mathfrak{B}$ and $\mathfrak{B}\models \bigwedge_{x∈ X_{j}} ψ_j (x)$}.$$

We prove the following, which intuitively states that, given the set $X_{j},$ we can find an other set $X_j '$ that ``behaves'' in the same way as $X_j$ but also ``avoids'' some inner part of $K_2^{\bf a}.$
\medskip

\noindent{{\em Subclaim:}} There exists a $t ∈ [d-\frac{r}{2}+2\hat{r}+1, d- \hat{r}]$ and a set
$X_j^{\prime}$ that is $(\ell_j, r_j)$-scattered in $\mathfrak{B}$ such that $X_j \subseteq R_C,$ $\mathfrak{B}\models \bigwedge_{x∈ X_{j}}ψ_j (x)\iff \mathfrak{B}\models \bigwedge_{x∈ X_{j}^{\prime}}ψ_j (x),$ and $X_j^{\prime} \cap I_2^{(t)} = \emptyset.$\medskip

\noindent{\em Proof of Subclaim:}	
Recall that $\tilde{\cal W}''$ is a collection of $\hat{\ell} +2$ flatness pairs of $G\setminus V({\bf a})$ that are $θ$-equivalent to $(\tilde{W}_{1}, \tilde{\mathfrak{R}}_1)$ and the vertex sets of their influences are disjoint from $\cupall {\cal X}_{\rm out} \cup {\cal X}'.$
Therefore, since $X_j$ has size at most $\hat{\ell},$ there exists a flatness pair in $\tilde{\cal W}''\setminus \{(\tilde{W}_2, \tilde{\mathfrak{R}}_2)\},$ say $(\tilde{W}_3, \tilde{\mathfrak{R}}_3),$ such that $I_3^{(w)}$ intersects neither $\cupall ({\cal X}_{\rm out} \cup {\cal X}')$ nor $X_{j}.$

We now focus on the set $I_2^{(d)}\setminus I_2^{(d-r+1)}.$
Recall that for the set $\cupall{\cal X}^{\rm out} \cup {\cal X}'$ it holds that $\cupall{\cal X}' \subseteq Z'\cap R_2\subseteq I_2^{(d-r)}\cap R_2$ and $\cupall{\cal X}^{\rm out}\cap I_{2}^{(w)}=\emptyset.$
Therefore, $I_2^{(d)}\setminus I_2^{(d-r+1)}$ does not intersect the set $\cupall {\cal X}_{\rm out}\cup {\cal X}'.$
Since
$r=2\cdot(\hat{\ell}+3)\cdot\hat{r}$ and $|X_{j}|≤  \hat{\ell},$
there exists a $t ∈ [d-\frac{r}{2}+2\hat{r}+1, d- \hat{r}]$
such that $X_{j}$ does not intersect $I_2^{(t)}\setminus I_2^{(t-\hat{r}+1)}.$
Intuitively, we partition the $r$ layers of $\tilde{W}_2$ that are in
$I_2^{(d)}\setminus I_2^{(d-r+1)}$ into two parts,
the first $r/2$ layers and the second $r/2$ layers,
and then we find some layer among
the ``$\hat{r}$-central'' $(\hat{\ell} +1)\hat{r}$ layers of the second part.
This layer together with its preceding $\hat{r}-1$ layers define a ``buffer'' of size $\hat{r}$ that $X_j$ ``avoids'' - that is $I_2^{(t)}\setminus I_2^{(t-\hat{r}+1)}.$
Notice that $I_2^{(t)}\setminus I_2^{(t-\hat{r}+1)}$ is a subset of $I_2^{(d)}\setminus I_2^{(d-r+1)}$
and therefore $I_2^{(t)}\setminus I_2^{(t-\hat{r}+1)}$ intersects neither $X_j$ nor $\cupall {\cal X}_{\rm out}\cup {\cal X}'.$

We set $X_{j}^\star:=X_{j}\cap I_2^{(t -\hat{r}+1)}$ and $Y_{j}\subseteq [\ell_j]$  to be the set of indices of the vertices in $X_{j}^{\star}.$
Notice that $X_{j}^\star\subseteq R_2,$ given that $X_{j}^\star =  X_{j}\cap I_2^{(t -\hat{r}+1)}\subseteq R_C\cap I_2^{(t -\hat{r}+1)}$
and $R_C\cap I_2^{(t -\hat{r}+1)}\subseteq R_2.$
Since $X_{j}^{\star}=X_{j}\cap I_{2}^{(t -\hat{r}+1)},$
$ψ_j ({\sf x})$ is an $r_j$-local formula  (where ``$r_j$-local'' refers to distances in $G_{\mathfrak{B}}$),
and $\hat{r}≥ r_j,$ we have that  $\mathfrak{B}\models \bigwedge_{x∈ X_{j}^{\star}}ψ_j (x)\iff \mathfrak{B}[I_2^{(t)}]\models \bigwedge_{x∈ X_{j}^{\star}}ψ_{j}(x).$
To sum up, we observe that, since $\cupall  {\bf W}_q^{(2)} \subseteq I_2^{(t)},$
we have that ${\sf pr}(K_2^{\bf a}[I_2^{(t)}], {\bf W}_q^{(2)}, \emptyset) = \{I_2^{(t)}\}$ and also the set $X_{j}^{\star}$ is a subset of $I_{2}^{(t -\hat{r}+1)}\cap R_2$ that is $(|Y_j|, r_j)$-scattered in $\mathfrak{B}[I_2^{(t)}]$ (since $X_j$ is $(|Y_j|, r_j)$-scattered in  $\mathfrak{B}$) and
\begin{eqnarray}
\labels{rigodle8}
\mathfrak{B}\models \bigwedge_{x∈ X_{j}^{\star}}ψ_j (x)\iff \mathfrak{B}[I_2^{(t)}]\models \bigwedge_{x∈ X_{j}^{\star}}ψ_{j}(x).
\end{eqnarray}
Also, notice that ${\sf ap}_{{\bf c}}(\mathfrak{A}, {\bf a}')[I_2^{(t)}] = \mathfrak{B}[I_2^{(t)}].$

Using the fact that  $(\tilde{W}_2, \tilde{\mathfrak{R}}_2)$ is $θ$-equivalent to $(\tilde{W}_3, \tilde{\mathfrak{R}}_3),$
we now aim to find a set $\tilde{X}_j$ that is an ``equivalent'' (in $I_3^{(t)}$) set of $X_j^\star.$
Since 
 $(\tilde{W}_2, \tilde{\mathfrak{R}}_2)$ is $θ$-equivalent to $(\tilde{W}_3, \tilde{\mathfrak{R}}_3),$ we have that
${{\sf in\mbox{-}sig}}(\mathfrak{K}_2,R_2,t',{\cal L},\emptyset^h)=  {{\sf in\mbox{-}sig}}(\mathfrak{K}_3,R_3,t',{\cal L},\emptyset^h),$
for every $t'∈ [w].$
Therefore, we have that ${{\sf in\mbox{-}sig}}(\mathfrak{K}_2,R_2,t,{\cal L},\emptyset^h)=  {{\sf in\mbox{-}sig}}(\mathfrak{K}_3,R_3,t,{\cal L},\emptyset^h)$ for the particular value $t$ given above.
The existence of the set $X_j^\star$ above and the fact that ${{\sf in\mbox{-}sig}}(\mathfrak{K}_2,R_2,t,{\cal L},\emptyset^h)=  {{\sf in\mbox{-}sig}}(\mathfrak{K}_3,R_3,t,{\cal L},\emptyset^h)$
imply that there exists a $\hat{C}∈ {\sf pr}(K_3^{\bf a}[I_3^{(t)}], {\bf W}_q^{(3)}, \emptyset)$ and a set $\tilde{X_{j}}\subseteq I_{3}^{(t -\hat{r}+1)}\cap R_3$ such that
$\tilde{X_{j}}$ is $(|Y_{j}|, r_{j})$-scattered in $\mathfrak{B}[I_3^{(t)}]$ and
${\sf ap}_{{\bf c}}(\mathfrak{A}, {\bf a}')[I_2^{(t)}]\models \bigwedge_{x∈ X_{j}^{\star}}ψ_{j}(x)\iff {\sf ap}_{{\bf c}}(\mathfrak{A}, {\bf a}')[\hat{C}]\models \bigwedge_{x∈ \tilde{X}_{j}}ψ_{j}(x).$
Observe that ${\sf ap}_{{\bf c}}(\mathfrak{A}, {\bf a}')[\hat{C}] = \mathfrak{B}[\hat{C}]$ and that ${\sf pr}(K_3^{\bf a}[I_3^{(t)}], {\bf W}_q^{(3)}, \emptyset) = \{I_3^{(t)}\}.$
Thus,
\begin{eqnarray}\labels{qopediru29}
\mathfrak{B}[I_2^{(t)}]\models \bigwedge_{x∈ X_{j}^{\star}}ψ_{j}(x)\iff \mathfrak{B}[I_3^{(t)}]\models \bigwedge_{x∈ \tilde{X}_{j}}ψ_{j}(x).
\end{eqnarray}

Since $I_3^{(w)}\cap \cupall {\cal X}=\emptyset,$ $I_3^{(w)}$ is a subset of the vertex set of a connected component of $G\setminus \cupall{\cal X}.$
Also, as $\breve{C}$ is the privileged component of $G$ with respect to ${\bf W}_q^{(3)}$ and $\cupall{\cal X}$ and
$\cupall{\bf W}_q^{(3)}\subseteq I_3^{(w)},$
 we have that
 $I_3^{(t)}\subseteq \breve{C}\subseteq C.$
We stress that the above holds no matter which $q$-pseudogrid of $G$ we consider.
Also, since $\tilde{X_{j}}\subseteq I_{3}^{(t -\hat{r}+1)}\subseteq I_3^{(w-\hat{r})},$
for every $x∈\tilde{X_{j}}$
it holds that $N_{G_{\mathfrak{B}}}^{(≤ \hat{r})}(x)\subseteq C.$
Thus, since $ψ_{j}({\sf x})$ is $r_{j}$-local, it follows that
\begin{eqnarray}\labels{joiospe3}
\mathfrak{B}[I_3^{(t)}] \models \bigwedge_{x∈ \tilde{X}_{j}}ψ_{j}(x)\iff \mathfrak{B}\models \bigwedge_{x∈ \tilde{X}_{j}}ψ_{j}(x).
\end{eqnarray}

We now consider the set
\[{X}^{\prime}_{j}:=\left( X_{j}\setminus X_{j}^{\star}\right)\cup \tilde{X_{j}}.\]
Since $I_{3}^{(w)}\cap X_{j}=\emptyset$ and $\hat{r}≥ r_j,$ for every $x∈ X_{j},$ and thus, for every $x∈ X_{j}\setminus X_{j}^{\star},$ it holds that $N_{G_{\mathfrak{B}}}^{(≤ r_{j})}(x)\cap I_{3}^{(w-\hat{r}+1)}=\emptyset.$
Also, since $t ≤ w -\hat{r}$ and $\tilde{X_{j}}\subseteq I_{3}^{(t -\hat{r}+1)},$ for every $x∈\tilde{X_{j}}$ it holds that $N_{G_{\mathfrak{B}}}^{(≤ r_{j})}(x) \subseteq I_{3}^{(w-\hat{r}+1)}.$
Thus, for every $x∈ X_{j}\setminus X_{j}^{\star}$ and $x'∈ \tilde{X_{j}}$ we have that
$N_{G_{\mathfrak{B}}}^{(≤ r_{j})}(x)\cap N_{G_{\mathfrak{B}}}^{(≤ r_{j})}(x')=\emptyset.$
The latter, together with the fact that the set $X_{j}\setminus X_{j}^{\star}$ is $(\ell_{j}-|Y_{j}|, r_{j})$-scattered in $\mathfrak{B}$ and $\tilde{X_{j}}$ is $(|Y_{j}|, r_{j})$-scattered in $\mathfrak{B}[I_3^{(t)}]$
implies that ${X}^{\prime}_{j}$ is an $(\ell_{j}, r_{j})$-scattered set in $\mathfrak{B}.$
Moreover, by definition, we have that ${X}^{\prime}_{j}\subseteq R_C\cup R_3 = R_C$ (the latter equality holds since $I_3^{(w)} \subseteq C$) and ${X}^{\prime}_{j}$ does not intersect $I_2^{(t)},$ while by~\eqref{rigodle8},~\eqref{qopediru29},~and~\eqref{joiospe3}, we have that $\mathfrak{B} \models \bigwedge_{x∈ X_{j}}ψ_j (x)\iff \mathfrak{B} \models \bigwedge_{x∈ X_{j}^{\prime}}ψ_j (x).$ The subclaim follows.\hfill$\diamond$
\bigskip

Following the above subclaim, let $t ∈ [d-\frac{r}{2}+2\hat{r}+1, d- \hat{r}]$ and let
$X_j^{\prime}$ be a set that is $(\ell_j, r_j)$-scattered in $\mathfrak{B}$ such that $X_j '\subseteq R_C,$ $\mathfrak{B} \models \bigwedge_{x∈ X_{j}}ψ_j (x)\iff \mathfrak{B} \models \bigwedge_{x∈ X_{j}^{\prime}}ψ_j (x),$ and $X_j^{\prime} \cap I_2^{(t)} = \emptyset.$

Since $r=2\cdot(\hat{\ell} +3)\cdot \hat{r}$ and $|{X}^{\prime}_{j}|≤ \hat{\ell},$ there exists a $t'∈ [d-r+2\hat{r}+1, d - \frac{r}{2}-\hat{r}]$ such that  ${X}^{\prime}_{j}$  does not intersect $I_1^{(t')}\setminus I_1^{(t'-\hat{r}+1)}.$
Intuitively, here, we partition the $r$ layers of $\tilde{W}_1$ that are in
$I_1^{(d)}\setminus I_1^{(d-r+1)}$ into two parts,
the first $r/2$ layers and the second $r/2$ layers,
and then we find some layer among
the ``$\hat{r}$-central'' $(\hat{\ell} +1)\hat{r}$ layers of the {\sl first} part.
This layer together with its preceding $\hat{r}-1$ layers define a ``buffer'' of size $\hat{r}$ that $X_j '$ ``avoids'' - that is $I_1^{(t')}\setminus I_1^{(t'-\hat{r}+1)}.$

Now, consider the set $U_1:={X}^{\prime}_{j}\cap (I_{1}^{(t'-\hat{r}+1)}\setminus  Z).$
Observe that $U_1\subseteq R_1$ and therefore $U_1\subseteq (I_{1}^{(t'-\hat{r}+1)}\setminus Z)\cap R_1.$
Recall that $Y =I_1^{(\hat{r})}$
and notice that, since $({X}^{\prime}_{j}\setminus U_1) \cap I_1^{(t')}=\emptyset$ and $t'>\hat{r},$
it holds that ${X}^{\prime}_{j}\setminus U_1\subseteq R\setminus Y.$

Let $Y_{j}^{\prime}\subseteq [\ell_j]$ be the set of the indices of the vertices of
${X}^{\prime}_{j}$ in $U_1.$
Given that $U_1={X}^{\prime}_{j}\cap (I_{1}^{(t'-\hat{r}+1)}\setminus  Z)$ and
${X}^{\prime}_{j}$ is  $(\ell_j, r_j)$-scattered in $\mathfrak{B},$
 and
$\mathfrak{B} \models \bigwedge_{x∈ X_{j}^{\prime}}ψ_j (x),$
we get that
$U_1$ is $(|Y_{j}^{\prime}|,r_j)$-scattered in $\mathfrak{B}[I_1^{(t')}\setminus Z]$
and $\mathfrak{B}\models\bigwedge_{x∈ U_1}ψ_{j}(x).$
At this point, observe that,
since the formula $ψ_{j}({\sf x})$ is $r_j$-local, $U_1={X}^{\prime}_{j}\cap (I_{1}^{(t'-\hat{r}+1)}\setminus  Z),$
where $\hat{r}≥ r_j$ and $t'≤ d-\frac{r}{2}-\hat{r},$ for every $x∈ U_1$ we have that
$N_{G_{\mathfrak{B}}}^{(≤ r_j)} (x)\subseteq I_1^{(d)}\setminus Z.$
Thus, the fact that $I_1^{(t')}\setminus Z\subseteq C_1$ implies that
\begin{eqnarray}\labels{jiro93}
\mathfrak{B}\models\bigwedge_{x∈ U_1}ψ_{j}(x)\iff\mathfrak{B}[C_1]\models\bigwedge_{x∈ U_1}ψ_{j}(x).
\end{eqnarray}
Also, note that ${\sf ap}_{{\bf c}}(\mathfrak{A}, {\bf a}')[C_1] = \mathfrak{B}[C_1].$

As we mentioned before, ${\sf in\mbox{-}sig}(\mathfrak{K}_1,R_1,d,L,Z)=  {\sf in\mbox{-}sig}(\mathfrak{K}_2,R_2,d,L,Z').$
This implies the existence of a set $C_2∈ {\sf pr}(K_2^{\bf a}[I_2^{(d)}], {\bf W}_q^{(2)}, Z')$ and a set $U_2\subseteq (I_{2}^{(t'-\hat{r})}\setminus Z')\cap R_2\subseteq R\setminus Y$
such that $U_2$ is $(|Y_{j}^{\prime}|,r_{j})$-scattered in $\mathfrak{B}[I_2^{(t')}\setminus Z']$ and $\mathfrak{B}[C_1] \models\bigwedge_{x∈ U_1}ψ_{j}(x)\iff {\sf ap}_{{\bf c}}(\mathfrak{A}, {\bf a}')[C_2]\models\bigwedge_{x∈ U_2}ψ_{j}(x).$
We set $\tilde{\mathfrak{B}} = {\sf ap}_{{\bf c}}(\mathfrak{A}, {\bf a}')[C_2].$
Therefore, we have that
\begin{eqnarray}\labels{opriwop4}
\mathfrak{B}[C_1] \models\bigwedge_{x∈ U_1}ψ_{j}(x)\iff \tilde{\mathfrak{B}}\models\bigwedge_{x∈ U_2}ψ_{j}(x).
\end{eqnarray}

By~\eqref{jiro93} and~\eqref{opriwop4}, we derive that
\begin{eqnarray}\labels{@desacralized}
 \mathfrak{B}\models\bigwedge_{x∈ U_1}ψ_{j}(x)
& \iff &
\tilde{\mathfrak{B}}\models\bigwedge_{x∈ U_2}ψ_{j}(x).
\end{eqnarray}

We now observe that
${\sf pr}(G,{\bf W}_q^{(2)}, \cupall({\cal X}^{\rm out} \cup {\cal Z}'))  ={\sf pr}(G,{\bf W}_{{q}}^{{(1)}},\cupall({\cal X}^{\rm out}\cup {\cal X}')) .$
To see this, notice that ${\sf pr}(G,{\bf W}_{{q}}^{{(1)}},\cupall({\cal X}^{\rm out}\cup {\cal X}')) = {\sf pr}(G,{\bf W}_{{q}}^{(2)},\cupall({\cal X}^{\rm out}\cup {\cal X}'))$
and ${\sf pr}(G,{\bf W}_q^{(2)},\cupall({\cal X}^{\rm out} \cup {\cal Z}')) = {\sf pr}(G,{\bf W}_q^{(2)}, \cupall({\cal X}^{\rm out}\cup {\cal X}')),$ due to the fact that $\partial_{\mathfrak{K}_2} (Z')\subseteq X'.$
Τhus, ${\sf pr}(G,{\bf W}_q^{(2)},\cupall({\cal X}^{\rm out} \cup {\cal Z}')) = \{\breve{C}'\}.$
Recall that for the set $\cupall{\cal X}^{\rm out} \cup {\cal X}'$ it holds that $\cupall{\cal X}' \subseteq Z'\cap R_2\subseteq I_2^{(d-r)}\cap R_2$ and $\cupall{\cal X}^{\rm out}\cap I_{2}^{(w)}=\emptyset.$
Since
$C_2∈ {\sf pr}(K_2^{\bf a}[I_2^{(d)}], {\bf W}_q^{(2)}, Z'),$
$\cupall {\cal X}^{\rm out}\cap I_{2}^{(w)}=\emptyset,$
and $\cupall {\bf W}_q^{(2)} \subseteq I_2^{(d)},$
it holds that
$I_2^{(t')}\setminus Z' \subseteq C_2$ and
$C_2\subseteq \breve{C}'\subseteq C'.$

We set $\mathfrak{B}':=\mathfrak{A}[C']$
and
$R_{C'}:= R\cap {C'}$
and we observe that, by construction, $V({\bf a}\cap C' )= V({\bf a}\cap C).$
Since $U_2$ is $(|Y_{j}^{\prime}|,r_{j})$-scattered in
$\mathfrak{B}[I_2^{(t')}\setminus Z'],$ where $U_2\subseteq I_{2}^{(t'-\hat{r}+1)}\setminus Z'$ and $t'< w - \hat{r},$ $U_2$ is also $(|Y_{j}^{\prime}|,r_{j})$-scattered in $\mathfrak{B}'.$
Moreover, the formula $ψ_{j}({\sf x})$ is $r_{j}$-local, so
\begin{eqnarray}\labels{@hullabalooing}
\tilde{\mathfrak{B}}\models\bigwedge_{x∈ U_2}ψ_{j}(x)\iff
\mathfrak{B}'\models\bigwedge_{x∈ U_2}ψ_{j}(x).
\end{eqnarray}
Therefore, by (\ref{@desacralized}) and (\ref{@hullabalooing}), it follows that $\mathfrak{B}\models\bigwedge_{x∈ U}ψ_{j}(x)\iff \mathfrak{B}'\models\bigwedge_{x∈ \tilde{U}}ψ_{j}(x).$

Consider the set $$X_j^\bullet:=({X}^{\prime}_{j}\setminus U_1)\cup U_2.$$

Notice that since ${X}^{\prime}_{j}\setminus U_1 \subseteq C$
and ${X}^{\prime}_{j}\setminus U_1$ does not intersect neither
$I_{2}^{(d-r+1)}$ (where $X'$ lies),
nor $I_1^{(d-r+1)}\subseteq I_{1}^{(t')}$ (where $Z$ lies),
it follows that ${X}^{\prime}_{j}\setminus U_1 \subseteq C\cap C'.$
This implies that ${X}^{\prime}_{j}\setminus U_1$ is an $(\ell_{j}-|Y_{j}^{\prime}|, r_{j})$-scattered set in $\mathfrak{B}$
and an $(\ell_{j}-|Y_{j}^{\prime}|, r_{j})$-scattered set in $\mathfrak{B}'.$
Since $U_2\subseteq I_{2}^{(t'-\hat{r}+1)}\setminus Z',$
${X}^{\prime}_{j}\cap  I_{2}^{(t)}=\emptyset,$
and $t'≤ t-2\hat{r},$
we have that for every $x∈ {X}^{\prime}_{j}\setminus U_1$ and $x'∈ U_2$ it holds that
$N_{G_{\mathfrak{B}'}}^{(≤ r_{j})}(x)\cap N_{G_{\mathfrak{B}'}}^{(≤ r_{j})}(x')=\emptyset.$
The latter, together with the fact that ${X}^{\prime}_{j}\setminus U_1$ is an $(\ell_{j}-|Y_{j}^{\prime}|,r_{j})$-scattered set in $\mathfrak{B}'$ and $U_2$ is a $(|Y_{j}^{\prime}|,r_{j})$-scattered set in $\mathfrak{B}',$
implies that  $X_j^\bullet$ is an $(\ell_{j}, r_{j})$-scattered set in $\mathfrak{B}'.$
Also, notice that $X_j^\bullet\subseteq R_{C'}\setminus Y.$
Furthermore, since the formula $ψ_{j}({\sf x})$ is $r_{j}$-local, it follows that
$\mathfrak{B}'\models\bigwedge_{x∈ X_{j}}ψ_{j}(x)\iff \mathfrak{B}'\models\bigwedge_{x∈ X_j^\bullet}ψ_{j}(x).$

Thus, assuming that there is a set  $X_j\subseteq R_C$  that is $(\ell_j, r_j)$-scattered in $\mathfrak{B}$ and $\mathfrak{B}\models \bigwedge_{x∈ X_j} ψ_j (x),$ we proved that there is a set $X_j^\bullet\subseteq R_{C'}\setminus Y\subseteq R\setminus Y$ that is $(\ell_j, r_j)$-scattered in $\mathfrak{B}'$ and $\mathfrak{B}'\models \bigwedge_{x∈ X_j^\bullet} ψ_j (x).$

To conclude Case 1, notice that we can prove the inverse implication,
i.e., by assuming the existence of a set $X_j^\bullet\subseteq R_{C'}\setminus Y\subseteq R\setminus Y$  that is $(\ell_j, r_j)$-scattered
in $\mathfrak{B}'$ and $\mathfrak{B}'\models \bigwedge_{x∈ X_j^\bullet} ψ_j (x)$
and, by using
the same arguments as above
(replacing $(\tilde{W}_{1}, \tilde{\mathfrak{R}}_1)$ with $(\tilde{W}_{2}, \tilde{\mathfrak{R}}_2),$ $Z$ with $Z'$ and $R$ with $R\setminus Y$),
we can prove the existence of a set $X_j\subseteq R$ that is  $(\ell_j, r_j)$-scattered  in $\mathfrak{B}$ such that $\mathfrak{B}\models \bigwedge_{x∈ X_j} ψ_j (x).$
\bigskip

\noindent{\bf Case 2:} $j\notin J.$\smallskip

We aim to prove that ${\sf ap}_{{\bf c}}((\mathfrak{A}, R, {\bf a}')[C])\models\neg
\breve{ζ}_j \iff {\sf ap}_{{\bf c}}((\mathfrak{A}, R\setminus Y, {\bf a}')[C'])\models\neg \breve{ζ}_j.$

In other words, we will prove that  for every set $X_{j}\subseteq R\cap C$ that is $(\ell_{j}, r_{j})$-scattered in $\mathfrak{B}$ and $\mathfrak{B} \models \neg ψ_j (x),$ for some $x∈ X_{j}$ if and only if  for every set $X_{j}'\subseteq (R\setminus Y)\cap C'$ that is $(\ell_{j}, r_{j})$-scattered in $\mathfrak{B}'$ and $\mathfrak{B}'\models \neg ψ_j (x),$ for some $x∈ X_{j}'.$
In Case 1, we argued that there is a set $X_j\subseteq R\cap C$ that is  $(\ell_j, r_j)$-scattered  in $\mathfrak{B}$ and $\mathfrak{B}\models \bigwedge_{x∈ X_j} ψ_j (x)$ if and only if there is a set $X_j^\bullet\subseteq (R\cap C')\setminus Y\subseteq R\setminus Y$ that is  $(\ell_j, r_j)$-scattered in $\mathfrak{B}'$ and $\mathfrak{B}'\models \bigwedge_{x∈ X_j^\bullet} ψ_j (x).$
This directly implies that  ${\sf ap}_{{\bf c}}((\mathfrak{A}, R, {\bf a}')[C])\models\neg\breve{ζ}_j \iff {\sf ap}_{{\bf c}}((\mathfrak{A}, R\setminus Y, {\bf a}')[C'])\models\neg \breve{ζ}_j.$
This concludes Case 2 and completes the proof of \autoref{claim_5}.\hfill$\diamond$
\bigskip

\setcounter{theothree}{5}
\begin{theothree}\label{claim_6}
It holds that $\mathfrak{A}[C]\models μ \iff \mathfrak{A}[C']\models μ.$\medskip
\end{theothree}

\noindent{\em \red{Proof of \autoref{claim_6}}}:
Observe that $\mathfrak{A}\models μ \iff G_{\mathfrak{A}}∈ \excl(\{K_{\hw(θ)}\}).$
Also, observe that $C\setminus D = C'\setminus D.$
If $Y = V(\mathfrak{A})\setminus C$ and $Y = V(\mathfrak{A})\setminus  C',$ then note that $Y\setminus D = Y'\setminus D$ and $Y$ intersects at most $q$ bags of $\tilde{\cal Q}.$
Thus, by assumption, $G\setminus Y∈ \excl (\{K_{\hw(θ)}\}) \iff G\setminus (Y\setminus D)∈ \excl(\{K_{\hw(θ)}\}),$ which implies that
$G[C]∈ \excl (\{K_{\hw(θ)}\}) \iff G[C']∈ \excl (\{K_{\hw(θ)}\}).$
Therefore, $\mathfrak{A}[C]\models μ \iff \mathfrak{A}[C']\models μ.$
This concludes the proof of \autoref{claim_6}.
\bigskip

Recall that $W^\bullet$ be the central $g$-subwall of $W_1$
and let $(\tilde{W}', \tilde{\mathfrak{R}}')$ be the $W^\bullet$-tilt of $(W,\mathfrak{R})$ given by the algorithm ${\tt Find\_Equiv\_FlatPairs}$ in~\autoref{subsec_second_level_algo}.
We set $V:=V({\sf compass}_{\tilde{\mathfrak{R}}'}(\tilde{W}')).$
Also, recall that the algorithm ${\tt Find\_Equiv\_FlatPairs}$
outputs the set $Y= V({\sf compass}_{\breve{\mathfrak{R}}'}(\breve{W}')),$
where
$(\breve{W}',\breve{\mathfrak{R}}')$ is a $\breve{W}'$-tilt of $(W,\mathfrak{R})$ and $\breve{W}$ is the central $j'$-subwall of $W_1.$
Finally, recall that $j'=g+2\hat{r}+2.$
The definition of a tilt of a flatness pair implies that $V$ is a subset of $Y.$
By \autoref{claim_4}, we have that $(\mathfrak{A},R,{\bf W}_{{q}}^{(1)},\emptyset^l,{\cal X})\models θ^{\sf out}_q \iff (\mathfrak{A}\setminus V,R\setminus Y,{\bf W}_q^{(1)},\varnothing^l, {\cal X}_{\rm out}\cup {\cal X}')\models θ^{\sf out}_q.$

Recall that since $Y\subseteq I_1^{(w)},$ $\cupall(X_{\rm out}\cup X')\cap Y = \emptyset,$ and $\cupall {\bf W}_q^{(2)}\subseteq I_2^{(2)},$
$C''∈ {\sf pr}(G_{\mathfrak{A}},{\bf W}_q^{(2)}, \cupall(X_{\rm out}\cup X')) \iff C''\setminus V ∈ {\sf pr}(G_{\mathfrak{A}}\setminus V,{\bf W}_q^{(2)}, \cupall(X_{\rm out}\cup X'))$ and, if
${\cal C}$ (resp. ${\cal C}'$) is the set of all $C''∈{\sf cc}(G_{\mathfrak{A}}, \cupall(X_{\rm out}\cup X'))$ (resp. all $C''∈{\sf cc}(G_{\mathfrak{A}}\setminus V, \cupall(X_{\rm out}\cup X'))$)
that are not in ${\sf pr}(G_{\mathfrak{A}},{\bf W}_q^{(2)}, \cupall(X_{\rm out}\cup X'))$ (resp. $ {\sf pr}(G_{\mathfrak{A}}\setminus V,{\bf W}_q^{(2)}, \cupall(X_{\rm out}\cup X'))$), then ${\cal C}={\cal C}'.$
Also, recall
that all the basic Gaifman variables in $\breve{ζ}_{\sf R}$ are contained in $R$ and every $ψ_{i}({\sf x})$ is $r_{i}$-local.
The fact that $W^\bullet$ is the central $g$-subwall of $W_1,$ $\breve{W}$ has height $j'$ and $g = j'-2\hat{r} -2,$ and $R\cap Y = \emptyset$ implies that
 these no local formulas $ψ_{i}({\sf x})$ is evaluated using vertices in $V.$
  Therefore, ${\sf ap}_{{\bf c}}((\mathfrak{A}, R\setminus Y, {\bf a}')[C'])\models \breve{ζ}_{\sf R} \iff {\sf ap}_{{\bf c}}((\mathfrak{A}\setminus V, R\setminus Y, {\bf a}')[C'\setminus V])\models \breve{ζ}_{\sf R},$ and, by \autoref{claim_5},
${\sf ap}_{{\bf c}}((\mathfrak{A}, R, {\bf a}')[C])\models \breve{ζ}_{\sf R}
\iff
 {\sf ap}_{{\bf c}}((\mathfrak{A}\setminus V, R\setminus Y, {\bf a}')[C'\setminus V])\models \breve{ζ}_{\sf R}.$
Finally, we observe that
$\mathfrak{A}[C]\models μ \iff \mathfrak{A}[C'\setminus V]\models μ.$
Thus,
we get that $(\mathfrak{A},R, {\bf a})\models θ_{{\sf R},{\bf c}}\iff(\mathfrak{A}\setminus V,R\setminus Y, {\bf a})\models θ_{{\sf R},{\bf c}}.$
\end{document}